\documentclass[a4paper,10pt]{book}
\usepackage{amsmath,amssymb,amsfonts,bbm, amscd}
\usepackage{diagrams}
\diagramstyle[labelstyle=\scriptstyle]

\usepackage[latin1]{inputenc}
\usepackage[T1]{fontenc}
\usepackage[francais]{babel}
\usepackage{fancyhdr}

\fancypagestyle{plain}{ 
\fancyhead{}
}

\usepackage[dvips]{graphicx}
\usepackage{hyperref}
\usepackage{lmodern}
\usepackage{epstopdf}
\usepackage{amsthm}
\usepackage{pstricks}

\DeclareMathOperator{\cyl}{Cyl}

\DeclareMathOperator{\Ad}{Ad}

\DeclareMathOperator{\Hom}{Hom}

\DeclareMathOperator{\id}{id}

\DeclareMathOperator{\tr}{tr}
\DeclareMathOperator{\Tr}{Tr}
\DeclareMathOperator{\Inv}{Inv}
\DeclareMathOperator{\diag}{diag}

\DeclareMathOperator{\Spin}{Spin}
\DeclareMathOperator{\SU}{SU}
\DeclareMathOperator{\SL}{SL}
\DeclareMathOperator{\SO}{SO}

\DeclareMathOperator{\ISU}{ISU}
\DeclareMathOperator{\U}{U}

\theoremstyle{definition}
\theoremstyle{definition}
\theoremstyle{definition}
\theoremstyle{definition}
\theoremstyle{definition}
\theoremstyle{definition}

\newcommand{\alg}{\mathfrak{g}}
\newcommand{\su}{\mathfrak{su}}

\newcommand{\so}{\mathfrak{so}}

\newcommand{\spin}{\mathfrak{spin}}

\def\unit{\mathbbm{1}}
\def\bra{\langle}
\def\ket{\rangle}
\def\f{\frac}

\def\mone{^{-1}}
\def\tl{\widetilde}
\def\pp{\partial}

\def\what{\widehat}

\def\eps{\epsilon}
\def\vareps{\varepsilon}
\def\thet{\vartheta}

\def\C{{\mathbbm C}}
\def\N{{\mathbbm N}}
\def\Z{{\mathbbm Z}}
\def\R{{\mathbbm R}}

\def\calA{{\mathcal A}}
\def\calC{{\mathcal C}}

\def\calG{{\mathcal G}}
\def\calH{{\mathcal H}}
\def\calI{{\mathcal I}}
\def\calL{{\mathcal L}}
\def\calM{{\mathcal M}}
\def\calN{{\mathcal N}}
\def\calO{{\mathcal O}}
\def\calT{{\mathcal T}}
\def\calV{{\mathcal V}}

\def\arr{\rightarrow}

\def\l({\left(}
\def\r){\right)}
\def\lv{\lvert}
\def\rv{\rvert}

\newcommand{\di}[1]{d_{j_{#1}}}

\def\beq{\begin{equation}}
\def\be{\begin{equation}}
\def\ee{\end{equation}}

\begin{document}



\thispagestyle{empty}
\sloppy\hbadness=5000\vbadness=100
{\center 

\fbox{%
\raisebox{.8cm}{\parbox{11cm}{%
\large \sc \center Universit\'e de la Méditerranée - Aix-Marseille II\\ Facult\'e des Sciences}

}}\\
\vspace{1cm}
{\huge \sc Th\`ese\\} 
\vspace{1.5cm}
Pr\'esent\'ee par\\
\vspace{1cm}
{\large \sc Valentin Bonzom} \\
\vspace{1cm}
Pour obtenir le grade de \\
\vspace{.5cm}
{\large \sc Docteur de l'universit\'e de la Méditerranée}\\
\'Ecole doctorale : Physique et Sciences de la mati\`ere\\
Sp\'ecialit\'e : Physique des particules, Physique mathématique\\
\vspace{1.2cm}
{\huge  \sc Géométrie quantique dans les mousses de spins }\\

\vspace{0.2cm}

{\Large De la théorie topologique BF vers la relativité générale}\\

\vspace{2.5cm}

Soutenue le 23 septembre 2010, devant la commission d'examen compos\'ee de\\
\begin{center}
\begin{tabular}[c]{ll}
{Mr Alejandro \sc Perez} & Pr\'esident du Jury\\
{Mr Carlo \sc Rovelli} & Directeur de th\`ese\\
{Mr Laurent \sc Freidel} & Rapporteur\\
{Mlle Bianca \sc Dittrich} & Rapporteur\\
{Mr Etera \sc Livine} & Examinateur\\
\end{tabular}
\end{center}
}
\newpage
~\newpage



\section*{Remerciements}

C'est un grand plaisir de remercier l'ensemble des personnes que j'ai pu cotoyer et qui m'ont entour\'e durant ces trois ann\'ees.

\noindent
Je dois remercier en premier lieu Etera Livine, qui fut mon directeur officieux durant une bonne partie de la th\`ese : toujours disponible, toujours pr\^et \`a discuter tout sujet scientifique et \`a proposer quatre projets de recherche simultan\'es ! Je le remercie pour son aide indispensable. Lui et Carlo Rovelli m'ont de plus laiss\'e une totale libert\'e aussi bien de penser que dans mon mode de travail. C'est quand m\^eme agr\'eable de faire ce que l'on a vraiment envie de faire !

\noindent
Je remercie l'ensemble du CPT et du groupe de gravit\'e quantique pour son accueil et son ambiance collaborative, p\^ele-m\^ele Carlo, Francesca, Eugenio, Antonino, Claudio, Elena et Roberto, toujours compr\'ehensifs \`a l'heure des pauses caf\'e, et plus particuli\`erement Simone Speziale avec qui j'ai le plaisir de travailler r\'eguli\`erement, et dont l'aide me fut pr\'ecieuse.

\noindent
L'accueil fut toujours chaleureux aupr\`es de Bianca Dittrich et de Daniele Oriti, qui furent les premiers \`a m'extirper de mon hermitage. J'esp\`ere pouvoir rendre un jour toutes les invitations que j'ai eu le plaisir de recevoir de leur part. Je remercie aussi Laurent Freidel et Lee Smolin pour leur enthousiasme lors de ma visite hivernale.

\noindent
Finalement, \c{c}a ne se voit pas sur les \'equations, mais trois ans de th\`ese, c'est trois ans dans la vie, avec une implication quotidienne. Content, pas content, trois fois par jour. Une pens\'ee profonde pour celle qui m'a support\'e tout en m'encourageant sans rel\^ache !

\newpage
\strut
\newpage

\section*{R\'esum\'e}

{\Large\sc Géométrie quantique dans les mousses de spins}

{\large De la théorie topologique BF vers la relativité générale}

\vspace{0.3cm}

La gravit\'e quantique \`a boucles a fourni un cadre d'\'etude particuli\`erement bien adapt\'e aux th\'eories de jauge d\'efinies sans m\'etrique fixe et invariantes sous diff\'eomorphismes. Les excitations fondamentales de cette quantification sont appel\'ees r\'eseaux de spins, et dans le contexte de la relativit\'e g\'en\'erale donnent un sens \`a la g\'eom\'etrie quantique au niveau canonique. Les mousses de spins constituent une sorte d'int\'egrale de chemins adapt\'ee aux r\'eseaux de spins, et donc destin\'ee \`a permettre le calcul des amplitudes de transition entre ces \'etats. Cette quantification est particuli\`erement efficace pour les th\'eories des champs topologiques, comme Yang-Mills 2d, la gravit\'e 3d ou les th\'eories BF, et des mod\`eles ont aussi \'et\'e propos\'es pour la gravit\'e quantique en dimension 4.

Nous discutons dans cette th\`ese diff\'erentes m\'ethodes pour l'\'etude des mod\`eles de mousses de spins. Nous pr\'esentons en particulier des relations de r\'ecurrence sur les amplitudes de mousses de spins. De mani\`ere g\'en\'erique, elles codent des sym\'etries classiques au niveau quantique, et sont susceptibles de permettre de faire le lien avec les contraintes hamiltoniennes. De telles relations s'interpr\`etent naturellement en termes de d\'eformations \'el\'ementaires sur des structures g\'eom\'etriques discr\`etes, telles que simplicielles. Une autre m\'ethode int\'eressante consiste \`a explorer la fa\c{c}on dont on peut r\'e\'ecrire les mod\`eles de mousses de spins comme des int\'egrales de chemins pour des syst\`emes de g\'eom\'etries sur r\'eseau, en s'inspirant \`a la fois des mod\`eles topologiques et du calcul de Regge. Cela aboutit \`a une vision tr\`es g\'eom\'etrique des mod\`eles, et fournit des actions classiques sur r\'eseau dont on \'etudie les points stationnaires.

Mots cl\'e : relativit\'e g\'en\'erale, th\'eorie des champs topologique, gravit\'e quantique, mousses de spins, r\'eseaux de spins.

\vspace{0.5cm}

{\Large\sc Quantum geometry in spin foams}

{\large From the topological BF theory towards general relativity}

\vspace{0.3cm}

Loop quantum gravity has provided us with a canonical framework especially devised for background independent and diffeomorphism invariant gauge field theories. In this quantization the fundamental excitations are called spin network states, and in the context of general relativity, they give a meaning to quantum geometry. Spin foams are a sort of path integral for spin network states, supposed to enable the computations of transition amplitudes between these states. The spin foam quantization has proved very efficient for topological field theories, like 2d Yang-Mills, 3d gravity or BF theories. Different models have also been proposed for 4-dimensional quantum gravity.

In this PhD manuscript, I discuss several methods to study spin foam models. In particular, I present some recurrence relations on spin foam amplitudes, which generically encode classical symmetries at the quantum level, and are likely to help fill the gap with the Hamiltonian constraints. These relations can be naturally interpreted in terms of elementary deformations of discrete geometric structures, like simplicial geometries. Another interesting method consists in exploring the way spin foam models can be written as path integrals for systems of geometries on a lattice, taking inspiration from topological models and Regge calculus. This leads to a very geometric view on spin foams, and gives classical action principles which are studied in details.

Keywords: general relativity, topological field theory, quantum gravity, spin networks, spin foams.\\

\vspace{0.4cm}

Centre de Physique Th\'eorique de Luminy - UMR 6207

\newpage
\strut
\newpage

\tableofcontents

{\renewcommand{\thechapter}{}\renewcommand{\chaptername}{}
\addtocounter{chapter}{-1}
\chapter{Introduction}\markboth{Introduction}{Introduction}}


For some hundred years, human beings enjoy two beautiful physical theories, that are very accurate descriptions of natural phenomena in their own domain of application, namely quantum mechanics and general relativity. Quantum mechanics turned out to be so powerful that people used it even before a correct interpretation arose. Its interpretation has been indeed challenging, and pushed us to reconsider interactions between systems, including observation and measuring processes. The greatest achievement is certainly quantum field theory, our ultimate description of microscopic physics, describing interactions of matter with fundamental forces of nature, except gravitation. On the other hand, general relativity is a theory of the dynamics of our universe, thus at very large scales, where it has been experimentally successful.

The attempt to include gravitation into the quantum field theory picture was unsuccessful, the theory being said to be perturbatively non-renormalizable, which makes it not predictible ! Still, a theory of \emph{quantum gravity} is expected to exist, and would then describe physics around the Planck scale. The latter is based on using both constants of quantum mechanics, i.e Planck constant $\hbar$, and Newton constant $G$ which is the coupling constant of gravitation, together with the maximal speed of propagation $c$. One can form the Planck length, $\ell_P = \sqrt{\hbar G/c^3}\approx 1.6\times 10^{-35}{\rm m}$, and we expect new physics to appear when probing objects at a few Planck lengths. Quantum gravity effects are typically expected to be relevant to describe the early stages of the universe and black hole physics.

Such scales look quite extreme to us, so that we do not have at our disposal quantum gravity experiments. So the failure of general relativity (GR) to be perturbatively renormalizable has led to various speculative roads aiming at unifying the quantum view of physics with gravitation, in which for instance GR is believed to be an effective theory, only valid at large scales. The most studied framework is that of string theory. It goes far beyond what is experimentally known in particle physics, adding new symmetries and extended objects instead of particles for instance. String theory is quite ambitious and would result in a great unification of all fundamental interactions.

However, it is worth looking closely at GR to realize it is not a standard gauge theory like electromagnetism for instance. Instead, it has brought new insights which make us drastically revise our picture of quantum field theory (QFT). In the latter, spacetime has a fixed, flat geometry, which hence enjoys a global symmetry coming from the principles of special relativity. This Lorentz symmetry is a foundation of the QFT framework, so that matter and interactions take place in this spacetime. This gives a picture of fundamental fields \emph{over} a spacetime background. But, the basic idea of GR is that spacetime geometry is the gravitational field, and is hence dynamical, and so typically does not enjoy the special symmetry used in QFT ! This means that spacetime geometry is \emph{not a background}, but evolves and interacts with other fields of particle physics. In agreement with this new picture, the fundamental symmetry of GR is the possibility of performing local changes of coordinates, called diffeomorphisms, which leave invariant the physical laws of nature. Consequently, there is physically speaking no intrinsic locations of phenomena with respect to a background, but instead phenomena are defined in relations to the dynamical geometry.

Thus, starting with the idea of merging GR with quantum physics, more modest than a great unification, one is led to reconsider the foundations of quantum field theory ! The aim is to build an intrinsically non-perturbative and background-independent framework which could apply to go beyond the perturbative non-renormalizability of GR. Many efforts, started in the eighties, have focused on this issue. The result is a mathematically rigourous and background-independent definition of quantum states on a canonical hypersurface (space). It is based on the same phase space as in Yang-Mills theory, and so potentially applies to any diffeomorphism invariant gauge theory with this phase space. The elementary excitations are called \emph{spin network states}, supported by embedded closed graphs, which can be made diffeomorphism invariant, and on which some operators depending on the connection and its momentum are well-defined. When applied to quantum gravity, it is called \emph{loop quantum gravity}, and gives a very nice picture of geometry at the quantum level ! A region of space containing a node of a spin network graph will get a non-trivial volume from this node. If a surface joins two such regions, with two nodes connected by a link of a graph, then this link carries an excitation of area for the surface. Moreover, these volumes and areas only take discrete values !

Next we expect these spin network states to evolve under the Hamiltonian. As it has been so far very hard to define a Hamiltonian operator for gravity on spin network states, understanding the corresponding dynamics is the purpose of a new formalism, the \emph{spin foam} formalism, which gives a general scheme to compute transition amplitudes between such states. In an analoguous way to spin networks, a spin foam is based on a discrete structure, a two-dimensional complex, which may be dual to a triangulation of spacetime. Summing over spin foams is supposed to give a clean implementation of the 'sum-over-histories', or path integral approach. Thus, spin foams applied to GR are expected to provide a picture of \emph{spacetime} geometry. Like for spin networks, we can distinguish the spin foam \emph{kinematics}, and their \emph{dynamics}. We will refer to kinematics to describe the structures of spin foams and their basic building blocks, while the dynamics is encoded in the sum over spin foams.

As there is no quantum gravity experiments and no well-established method to quantize a theory through spin foams, there exist different proposals referred to as \emph{spin foam models}. Also it seems bold to apply such new quantization processes to general relativity (and it is to some extend !). But we have two ideas to help us. The first one is a field theory used as a starting point to propose models, and the second consists in looking at the semi-classical limit. The starting point to propose spin foam models for quantum gravity is \emph{BF field theory}. It is a topological field theory of Schwarz's type, that is without local degrees of freedom, so that the latter come in finite numbers from the spacetime topology, and no metric is indeed to define it. It is closely related to other field theories of interest : in 2 and 4 dimensions, it is the zero coupling limit of Yang-Mills theory, and in 3 dimensions, it is gravity (but with degenerate metrics), describing flat spacetimes. It has been much studied in 2 and 3 dimensions, in connection with other topological field theories (like Chern-Simons theory) and with topological invariants of manifolds (like the Turaev-Viro invariant). Further, in 4 dimensions, it is possible to add constraints on the field variables to the topological action, so as to recover GR and its local degrees of freedom ! The nice point with this BF theory is that it is known: (i) that the kinematical framework can be described through spin network states, just like in LQG, (ii) how to quantize it with spin foams (which implies that its dynamics also is under control), (iii) how to relate this to a quantization of the Hamiltonian on spin network states, (iv) how to show the equivalence to other quantization schemes, at least in 3d ! Thus, this suggests to take advantage of the relation between 4d BF theory and GR to consequently modify the spin foam quantization of the former. The main idea is pictured on the diagram:
\begin{diagram}
 \text{spin network states} & \rTo   & \text{LQG Hamiltonian} & \lTo^? & \text{spin foams models}\\
                                          & \rdTo^{\text{flatness constraint}} &                                        &        & \uTo^{\text{constraints}} \\
                                          &           & \text{BF Hamiltonian}     & \rTo^{\text{topological invariance}}&\text{spin foams for BF theory}
\end{diagram}

Once a model for quantum gravity is proposed, the natural test is to investigate its semi-classical behaviour. A method has been proposed some years ago to compute correlation functions from spin foams so that they can be compared with standard QFT. This is a non-trivial task in a diffeomorphism invariant context, and relevant approximations have to be identified. So far, one considers a truncation of the sum over spin foams to a single 2-complex, and expand the summand around a given background with large, macroscopic area eigenvalues. This background is introduced to provide a mean geometry, necessary to make sense of correlation functions in real space. Since spin foams are usually considered on a dual complex to a spacetime triangulation, we expect their semi-classical limit to be related to an \emph{approximation} of GR on a triangulation. This is indeed what happens in the completely solved 3d gravity theory: there, the large area regime gives a semi-classical limit in which spin foam amplitudes are approximated by a form of quantum Regge calculus ! At this time, let us say that \emph{Regge calculus} is an approximation of GR on a triangulation, using piecewise flat metrics, where the basic variables are the lengths of edges, and curvature is encoded into the dihedral angles between simplices.

It is important to keep in mind that {\it a priori} there is more about spin foams than Regge calculus. Even in the asymptotics, corrections to the exponential of the Regge action are to be expected, and mainly spin foams are designed for the deep quantum regime. In addition, for the topological models, they provide a well-defined unambiguous measure for the path integral, and we expect this to hold for quantum gravity. Once this is stated, one may all the same ask for further connections with Regge calculus, beyond the semi-classical limit ! Indeed, there are strong pieces of evidence from the way spin networks and spin foams are built, as discrete structures of geometry, that their restrictions to a fixed graph and a fixed 2-complex can be naturally related and compared to Regge geometries. This means that the exploration of spin network and spin foam structures could be done by taking inspiration in Regge calculus.

The purpose is multifold. One is kinematical: to understand how much they differ in the canonical and spacetime formalisms, to get a deeper picture of the emergence of the so-called quantum 3-geometry and 4-geometry. At the dynamical level, it is somehow unknown what the Wheeler-de-Witt equation looks like on spin networks, while Regge equations of motion give an account of Einstein's dynamics on discrete structures. It is also commonly believed that the action of the quantized Hamiltonian of GR on spin network states could be geometrically pictured as elementary deformations of polyhedra. Such moves are known in the context of Regge calculus (and they enable to define a Hamiltonian formalism for these geometries), like tent moves for instance. Another interesting aspect is computional. To handle the kind of sums and integrals involved in sums over spin foams, it is really helpful to have at our disposal a nice and well-known geometric interpretation of the summand/integrand of the partition function. The scheme of the relations between LQG, spin foams and Regge calculus is depicted in the following diagram:
\begin{diagram}
 \text{spin networks on a single graph} & \rTo^{\text{deformation moves ?}} & \text{spin foams on a single graph} \\
 \dTo & & \dTo^{\text{semi-classical limit, only ?}} \\
 \text{canonical Regge/twisted geometries} & \rTo^{\text{tent moves}} & \text{Regge calculus}
\end{diagram}

\bigskip

The purpose of my doctoral work was to investigate in the most detailed way the content of the arrows of both the above diagrams. The directions of the arrows should not be taken too literally, as they may as well be reversed -- most of them contain speculative arguments. Another way to put it is to think of a 3-dimensional array, with two entries on each dimension. On one dimension, we have two field theories: BF and GR, which are physically totally different but surprisingly also technically related. The second dimension is about two quantization schemes: spin networks for the canonical quantization, and spin foams for the spacetime formalism. The third dimension describes two level of quantizations. First the kinematical level corresponds to the structure of states, or the type of geometries, encoded into spin networks and spin foams (they can be different for the two schemes), and the dynamical level which describes the evolution of spin network states (via a Hamiltonian operator or a sum over spin foams). Then, we are interested in filling the array, by precisely describing the geometries and their dynamics. This will be done partially taking inspiration from Regge calculus.

A typical result which will be presented here is the following. At the kinematical level, the variables which are quantized in spin foam models form genuine 4d Regge geometries, and it is possible to describe explicitly the quantum fluctuations around them within the partition function, while trying to construct similar geometries within the framework of LQG one rather arrives at some twisted, discontinuous 3d Regge geometries. Another example, coming from my work, is that spin foam evolution of spin network states and recurrence relations for these amplitudes can be derived from the action of the Hamiltonian operator in the topological BF case and should be interpreted as tent moves on simplices of a canonical hypersurface together with a nice geometric gauge fixing.

\medskip

I had the pleasure to begin my work exactly as the new spin foam models, from Engle-Pereira-Rovelli, Livine-Speziale and Freidel-Krasnov, were proposed. This has been an exciting period, with a lot to understand. Basically, how do they correct the previous attempts, what about their semi-classical limit, and how do they fit in the LQG picture ? Because both the formalism, devised for background-independent theories, and the quantum gravity physics are new and somehow speculative, there are two levels of reading of my work. On the one side, I tried to elaborate \emph{methods}, potentially useful for any spin foam models. On the other side, I obtained \emph{results} concerning specific models, namely the topological BF models in 3 and 4d, the old-fashioned Barrett-Crane model and the most recent ones.

I will focus along this manuscript on three methods. The first method is used to extract correlation functions from spin foams, proposed by Rovelli and collaborators some years ago. Actually, I will only illustrate this method in the context of a toy model in 3d, to show that it can be used to explicitly compute in the asymptotics at any desired order ! For this purpose, one has to find a way to include all corrections to the quantum Regge calculus (first order of the spin foam asymptotics) coming from the exact spin foam amplitude. It also answers interesting questions about the large spin limit of the 3d model beyond the first order, showing for instance that the Regge action is the only relevant frequency, and expliciting the full pattern of oscillations.

The second method is fully original, and aims at filling the gap between LQG and spin foams through recurrence relations on spin foam amplitudes. These relations are to be understood as Wheeler-de-Witt equations on boundary spin networks, or more generally as quantum implementations of symmetries. Interestingly, they admit geometric interpretations in terms of elementary deformations of simplices, such as tent moves, thus prefigurating possible Hamiltonian pictures of spin foams. I mainly focus on the topological models to develop the main ideas, but also found recurrence relations on non-topological models.

Finally, I looked at a third method to unfold the geometry of spin foams. The main idea is to rewrite a spin foam model on a triangulation as a path integral for a classical system of discrete geometries. For this, I developped a formalism to identify usual quantities of Regge calculus among the variables used to build spin foams, and applied it to the most well-known models. In particular for the new model, it shows how the BF discrete action reduces to a Regge-like action with the Immirzi parameter. The corresponding stationnary points are Regge geometries, and the formalism makes the path integral explicit and so the quantum fluctuations.

For a more complete r\'esum\'e of the methods and of the results, the reader is invited to have a look at the conclusion. The first part of the manuscript, \ref{part:classical}, presents the classical BF and general relativity field theories, and gives a brief introduction to the set-up of Regge calculus. Then, the second part, \ref{part:canonical geom}, is devoted to the Hamiltonian framework. There, I compare the Hamiltonian analyses of BF and GR, present the spin network quantization, relevant for both BF and GR, and I finally restrict attention to the geometric content of this quantization on a single graph, looking at the kinematical aspects, and at the dynamics in simplified situations. The third part, \ref{part:spinfoam1}, is a first contact with spin foams: the general heuristic ideas, and how they are explicitly carried out in BF theory. I also present in this part the way correlation functions can be recovered with spin foams, together with some explicit computations in the 3d toy model. The fourth part, \ref{sec:recurrence}, is dedicated to recurrence relations in the 3d and 4d topological spin foam models as quantization of the Hamiltonian constraint, and their geometric interpretations, with the help of spacetime Pachner moves and canonical tent moves. The last part of the manuscript, \ref{part:newmodel}, focuses on spin foam models for quantum gravity. It fully explores the classical setting, the different quantizations that have been proposed, from the Barrett-Crane model to the most recent ones. I also there develop the programme to build spin foam models out of path integrals for geometries on the lattice. It leads to geometrically clear derivations of all these models and more !



\part{Th\'eorie BF et relativit\'e g\'en\'erale classiques} \label{part:classical}

\chapter{Th\'eorie BF}

Notre pr\'esentation des th\'eories de type BF est fortement inspir\'ee de \cite{baez-bf-spinfoam} qui constitue une tr\`es jolie introduction aux mod\`eles de mousses de spins. Bien s\^ur, les articles originaux de Blau et Thompson \cite{blau-thompson-bf}, ainsi que d'Horowitz \cite{horowitz-bf} restent incontournables.

\section{\'Equations classiques}

Soit $P$ un fibr\'e principal de groupe de structure $G$, au-dessus d'une vari\'et\'e lisse $M$, notre espace-temps, de dimension $n$. Nous prendrons presque syst\'ematiquement dans cette th\`ese $G=\SU(2)$ ou $\Spin(4)$. Consid\'erons une connexion $A$ sur $P$, qui sera vue localement comme une 1-forme \`a valeurs dans l'alg\`ebre de Lie $\alg$ de $G$ ($A^{IJ}$ avec $I,J=0,1,2,3$ des indices euclidiens 4d, ou $A^i$, $i=1,2,3$ dans le cas de $\SU(2)$) et poss\'edant une loi de transformation particuli\`ere, que nous rappelons plus bas, sous l'effet des transformations de jauge. Nous avons \'egalement besoin d'une $(n-2)$-forme $B$ \`a valeurs dans le fibr\'e adjoint $\Ad(P)$, i.e. le fibr\'e vectoriel associ\'e \`a $P$ via l'action adjointe de $G$ sur son alg\`ebre $\alg$. Localement, nous verrons $B$ comme une $(n-2)$-forme sur $M$ \`a valeurs dans $\alg$ et se transformant sous la repr\'esentation adjointe de $G$. Finalement, l'action de la th\'eorie topologique BF se construit \`a l'aide d'une forme bilin\'eaire invariante non-d\'eg\'en\'er\'ee sur l'alg\`ebre $\alg$, que nous notons '$\tr$', et
\beq \label{BFaction}
S_{\mathrm{BF}}\bigl(B,A\bigr) = \int_M \tr\Bigl(B\wedge F(A)\Bigr),
\ee
o\`u $F(A)=dA+\f{1}{2}[A,A]$ est la 2-forme de courbure de $A$. Le groupe $\calG$ des transformations de jauge usuelles est constitu\'e des champs \`a valeurs dans $G$, et engendre les transformations suivantes
\beq \label{gauge transfo}
A \mapsto \Ad_g A + g\,dg\mone, \qquad B \mapsto \Ad_g B,
\ee
qui laisse l'action invariante. L'action adjointe s'\'ecrit matriciellement : $\Ad_g X = gX g\mone$. La lin\'earisation de ces transformations correspond aux variations : $\delta A=-d_A\phi$ and $\delta B = [\phi,B]$, telles que $\delta F = -[F,B]$. De m\^eme, \'etant formul\'ee comme l'int\'egrale d'une forme diff\'erentielle, $S_{\mathrm{BF}}$ est bien invariante sous les diff\'eomorphismes de $M$.

Les \'equations du mouvement sont particuli\`erement simples :
\begin{align}
d_A\, B &= 0, \label{daB}\\
F(A) &= 0, \label{flatness}
\end{align}
o\`u $d_A = d + [A,\cdot]$ est la diff\'erentielle covariante. Nous allons montrer qu'en fait toutes les solutions sont localement identiques \`a des transformations de jauge pr\`es ! La deuxi\`eme \'equation provient du fait que $B$ est un multiplicateur de Lagrange pour la courbure, et signifie simplement que $A$ est une connexion plate. Or toutes les connexions plates ne diff\`erent localement que par des transformations de jauge. Par ailleurs, puisque $A$ est plate, toute forme ferm\'ee $B$ satisfaisant $d_A B=0$ est localement exacte, et s'\'ecrit donc comme $d_A \psi$ pour une forme $\psi$. Mais, la sp\'ecificit\'e de cette th\'eorie est que l'on peut \'eliminer $\psi$ en un certain sens. En effet, l'action b\'en\'eficie d'une invariance locale particuli\`ere en plus de celle sous l'action de $\calG$ :
\beq \label{translation sym bf}
A \mapsto A, \qquad B \mapsto B + d_A\,\eta,
\ee
pour toute $(n-3)$-forme $\eta$ sur $M$ \`a valeurs dans $\alg$. Il s'agit en fait d'une sym\'etrie de jauge au sens usuel, parfois appel\'ee sym\'etrie de translation, en raison de son action sur $B$. En notant $\calT$ le groupe de ces sym\'etries de translation, le groupe complet des sym\'etries de la th\'eorie BF est le produit semi-direct $\calG \ltimes \calT$. Il est clair que de telles transformations permettent toujours de transformer une solution $B$ en la solution triviale localement. Ainsi, toutes les solutions des \'equations sur les champs sont localement \'equivalentes \`a la solution triviale. En termes physiques, cela signifie qu'il n'y a pas de degr\'es de libert\'e locaux.

Ces \'equations admettent une interpr\'etation cohomologique utile. En effet, l'action de la diff\'erentielle covariante au carr\'e sur une forme est proportionnelle \`a la courbure, $d^2_A\omega = [F(A),\omega]$. D\`es lors que $F=0$, nous pouvons utiliser $d_A$ pour d\'efinir des classes de cohomologie. L'ensemble des solutions non-\'equivalentes de jauge est donc form\'e par les connexions plates non-\'equivalentes de jauge avec les \'el\'ements du $(n-2)$-i\`eme groupe de cohomologie, ce dernier \'etant par dualit\'e de Poincar\'e isomorphe au deuxi\`eme groupe de cohomologie.

La th\'eorie BF et la relativit\'e g\'en\'erale telle que nous la pr\'esentons \`a la section suivante partagent des caract\'eristiques essentielles \`a l'id\'ee de formuler une g\'eom\'etrie de l'espace-temps qui soit dynamique. Toutes deux sont en effet formul\'ees en termes de formes diff\'erentielles, sans r\'ef\'erence \`a une m\'etrique, et sont ainsi invariantes sous les diff\'eomorphismes. Il est donc de premi\`ere importance de comprendre que c'est la sym\'etrie de jauge de translation \eqref{translation sym bf} qui distingue la th\'eorie topologique de la relativit\'e g\'en\'erale. La sym\'etrie de jauge standard avec la sym\'etrie de translation constituent l'ensemble des sym\'etries de la th\'eorie BF, et en particulier les diff\'eomorphismes. En effet, si $\xi$ est un champ de vecteurs sur $M$, en utilisant le produit int\'erieur $i_\xi$, nous pouvons former une fonction $\phi_\xi$ \`a valeurs dans le fibr\'e adjoint $\Ad(P)$, $\phi_\xi \equiv i_\xi A$, ainsi qu'une $(n-3)$-forme $\eta_\xi$ \`a valeurs dans $\Ad(P)$, $\eta_\xi \equiv i_\xi B$. Alors l'action de $\xi$ sur la connexion est
\begin{align}
\delta_\xi A &= \mathcal{L}_\xi A,\\
&= d_A\,\phi_\xi + i_\xi F,
\end{align}
o\`u $\mathcal{L}_\xi$ repr\'esente la d\'eriv\'ee de Lie selon $\xi$. Autrement dit, un diff\'eomorphisme infinit\'esimal s'\'ecrit comme une simple transformation de jauge, plus un terme proportionnel \`a la variation de l'action $\delta S/\delta B = F$. De m\^eme, l'action de $\xi$ sur $B$ est
\begin{align}
\delta_\xi B &= \mathcal{L}_\xi B,\\
&= d_A\, \eta_\xi + [\phi_\xi,B] + i_\xi\,d_A B.
\end{align}
Les deux premiers termes sont bien s\^ur des transformations de jauge, tandis que le troisi\`eme est proportionnel \`a la variation de l'action $\delta S/\delta A = d_AB$. Or comme le note Horowitz \cite{horowitz-bf}, on sait que deux sym\'etries qui diff\`erent par de telles variations de l'action, s'annulant sur l'espace des solutions des \'equations du mouvement, sont {\it de facto} \'equivalentes. La diff\'erence majeure entre la th\'eorie topologique et la relativit\'e g\'en\'erale est donc la possibilit\'e dans la th\'eorie topologique de r\'eexprimer les diff\'eomorphismes on-shell comme une sym\'etrie de jauge sur un espace \og interne\fg.

Signalons finalement le fait que les sym\'etries sp\'ecifiques de la th\'eorie BF ont \'et\'e interpr\'et\'ees dans le formalisme BRST (ou Batalin-Vilkovisky) comme une supersym\'etrie vectorielle \cite{maggiore-sorella-perturbative-4dbf, symmetries-bf-bv}, essentielle \`a la r\'esolution de la cohomology locale de l'op\'erateur BRST, et au final \`a la renormalisabilit\'e perturbative \cite{renormalisability-bf} ! Un aspect important de l'analyse de la th\'eorie BF par les m\'ethodes alg\'ebriques standards de la th\'eorie quantique des champs est le fait que l'alg\`ebre BRST n'est pas ferm\'ee \emph{off-shell}, l'op\'erateur BRST n'\'etant nilpotent que on-shell, et que la sym\'etrie de translation \eqref{translation sym bf} est r\'eductible en dimension $\geq4$. En effet, il y a des modes z\'ero pour la transformation de translation : si la $(n-3)$-forme $\eta$ est prise comme $\eta=d_A\,\psi$, alors $B$ est chang\'e en $B+[F_A, \psi]$, et cela n'a aucun effet on-shell ! Ce processus peut se poursuivre jusqu'\`a $d_A^{(n-2)}\kappa$, pour une fonction $\kappa$. Il faut donc \^etre prudent dans la formulation de l'int\'egrale de chemins, ce qui passe par l'introduction de fant\^omes de fant\^omes (\og ghosts of ghosts\fg) pour g\'erer les sym\'etries de jauge sur les fant\^omes.

\section{En dimensions 2 et 3} \label{sec:BFlowdim}

La th\'eorie topologique BF est intimement reli\'ee \`a d'autres th\'eories des champs de basses dimensions bien connues.

\subsection*{Yang-Mills en 2 dimensions}

A priori, la th\'eorie de Yang-Mills n\'ecessite d'\'equiper notre vari\'et\'e d'une m\'etrique. N\'eanmoins, pour une vari\'et\'e $M_2$ ferm\'ee et orient\'ee de dimension 2, une forme de volume $\vareps$ est suffisante. La courbure d'un connexion $A$ se d\'ecompose alors en $F(A) = f_A\vareps$, o\`u $f_A$ est une fonction sur $M_2$ \`a valeurs dans $\alg$ se transformant sous l'adjointe. L'action de Yang-Mills avec une constante de couplage not\'ee $e^2$ est alors
\beq
S_{\mathrm{YM2d}} = -\f 1{2e^2}\int_{M_2} \vareps\ \tr f_A^2.
\ee
Il est bien connu que la fonction de partition, d\'efinie formellement par
\beq
Z_{\mathrm{YM2d}}(e^2\rho) = \int DA\ e^{iS_{\mathrm{YM2d}}(A)},
\ee
ne d\'epend que la combinaison $e^2\rho$, o\`u $\rho = \int_{M_2}\vareps$ est le volume de $M_2$.

Pour voir le lien avec la th\'eorie BF, consid\'erons la fonctionnelle suivante
\beq
S_{\mathrm{BFYM2d}} = \f{e^2}{2} \int_{M_2} \vareps\ \tr B^2 + \int_{M_2} \tr\,BF(A).
\ee
$B$ est ici une fonction \`a valeurs dans $\alg$ se transformant sous l'adjointe. Le deuxi\`eme terme de l'action est bien s\^ur la fonctionnelle BF en dimension 2, auquel nous avons ajout\'e un terme quadratique en $B$ \`a l'aide de la forme de volume $\vareps$. La fonction de partition correspondante est
\beq
Z_{\mathrm{BFYM2d}}(e^2\rho) = \int DA\,DB\ e^{i\f{e^2}{2} \int_{M_2} \vareps\, \tr B^2}\ e^{i\int_{M_2}\vareps\, \tr\,B f_A}.
\ee
Il est possible d'effectuer formellement l'int\'egrale gaussienne sur $B$. Celle-ci se localise autour du point stationnaire 
\beq
B = -\f{2}{e^2}\,f_A,
\ee
et conduit \`a l'\'egalit\'e avec la fonction de partition de Yang-Mills,
\be
Z_{\mathrm{BFYM2d}}(e^2\rho) = Z_{\mathrm{YM2d}}(e^2\rho).
\ee
Cela nous permet de voir la limite de couplage faible (ou de petit volume de $M_2$) comme la th\'eorie topologique BF, ne d\'ependant plus de la forme de volume $\vareps$.

La m\'ethode la plus efficace pour calculer cette fonction de partition est certainement la r\'egularisation sur un r\'eseau, du fait qu'elle est invariante par subdivision du r\'eseau (\`a une renormalisation standard pr\`es) ! Ce calcul a d\'ej\`a \'et\'e utilis\'e par Witten \cite{witten-2d-YM} pour calculer le volume symplectique de l'espace de module des connexions plates sur le fibr\'e choisi au-dessus de $M_2$. Dans notre perspective, cela correspond pr\'ecis\'ement \`a un calcul de l'int\'egrale de chemins en forme de mousses de spins ! Dans ce cas, l'expression combinatoire fournie par la somme sur les mousses de spins est particuli\`erement simple, et c'est de plus la forme la plus simple que l'on peut donner \`a $Z_{\mathrm{YM2d}}(e^2\rho)$. Nous donnerons la formule explicite pour la limite topologique en \ref{sec:bf-spinfoam}.

\subsection*{La gravit\'e en dimension 3} 

Dans le formalisme dit du premier ordre, la gravit\'e riemannienne en trois dimensions fait intervenir une connexion de spin $A^i$ ($i=1,2,3$) \`a valeurs dans $\su(2)$, et une co-triade $e^i$, i.e. 1-forme \`a valeurs dans $\su(2)$, se transformant sous l'adjointe, et non-d\'eg\'en\'er\'ee. Les indices internes sont contract\'es gr\^ace \`a la forme de Killing sur $\su(2)$, i.e. le tenseur $\delta_{ij}$ (qui s'\'ecrit par ailleurs en fonction de la trace matricielle). L'action de la gravit\'e en 3d avec une constante cosmologique $\Lambda$ est
\begin{equation} \label{gravity action 3d}
S_{\mathrm{GR3d}}(e,A)=\,\f1{4\pi G}\int_{M_3} \left[e^i\wedge
F_i(A)\,+\,\frac{\Lambda}{6}\,\epsilon_{ijk}\,e^i\wedge e^j\wedge e^k\right].
\end{equation}
Nous omettrons syst\'ematiquement la constante de Newton $G$ en posant : $4\pi G\equiv 1$. L'action est bien invariante sous les diff\'eomorphismes de $M_3$, puisque formul\'ee comme int\'egrale d'une forme diff\'erentielle. Mise \`a part la condition de non-d\'eg\'en\'erescence sur $e$, nous reconnaissons dans le premier terme ci-dessus l'action BF pour $\SU(2)$, le deuxi\`eme terme correspondant au volume de $M_3$.

Les \'equations d'Einstein prennent une forme particuli\`erement simple en trois dimensions (ph\'enom\`ene que l'on ne retrouve \'evidemment pas en dimension 4),
\begin{align}
d_A e^i &= 0,\\
F^i+\f{\Lambda}{2}\,\epsilon^i_{\phantom{i}jk}\, e^j \wedge e^k &= 0.
\end{align}
La premi\`ere \'equation correspond \`a la condition usuelle de torsion nulle en relativit\'e g\'en\'erale. Elle permet d'exprimer la connexion $A$ comme une fonction de la co-triade d\`es lors que $e$ est bien inversible. La seconde \'equation quant \`a elle signifie que l'espace-temps a une courbure homog\`ene donn\'ee par $\Lambda$.

Il pourrait sembler {\it a priori} que la pr\'esence du terme de volume proportionnel \`a $\Lambda$ dans $S_{\mathrm{GR3d}}$ fasse dispara\^itre l'invariance de l'action sous la sym\'etrie de translation \eqref{translation sym bf}. N\'eanmoins, l'action poss\`ede toujours une sym\'etrie particuli\`ere qui fait de cette th\'eorie une th\'eorie topologique, ne poss\'edant pas de degr\'e de libert\'e locaux ! Cette sym\'etrie appara\^it comme une d\'eformation de la sym\'etrie de translation \eqref{translation sym bf},
\beq \label{translation symetry}
\delta A = \Lambda\, [e,\eta], \qquad \delta e=d_A \eta.
\ee
Notons que les diff\'eomorphismes peuvent \^etre g\'en\'er\'es \`a partir de ces transformations et des transformations de jauge usuelles. Pour une \'etude d\'etaill\'ee des sym\'etries dans le formalisme BRST, je renvoie par exemple \`a \cite{symmetries-bf-vs-gravity}.

La mani\`ere la plus efficace de traiter la gravit\'e 3d comme une th\'eorie des champs, et montrer en particulier sa renormalisabilit\'e, a \'et\'e donn\'ee par Witten dans \cite{witten-3d-gravity}. Il s'agit en fait de la formuler comme une th\'eorie de Chern-Simons pour un groupe de structure bien choisi. La th\'eorie de Chern-Simons est une th\'eorie des champs topologique qui a fait l'objet d'intenses recherches, en raison de ses liens avec la th\'eorie des invariants des 3-vari\'et\'es et des invariants de noeuds, comme l'a montr\'e Witten \cite{witten-jones-poly}. Cette th\'eorie n'est pas directement int\'eressante pour nous puisqu'on ne lui conna\^it pas de formulation de type somme sur les mousses de spins. Elle n'existe par ailleurs pas en dimension 4, mais seulement en dimensions impaires. N\'eanmoins, il s'agit d'une incontournable \`a laquelle il est souvent fait r\'ef\'erence. Un de ses aspects fascinants est qu'elle est tout \`a la fois reli\'ee au mod\`ele dit ``$G/G$-gauged Wess-Zumino-Witten`` en dimension 2, et \`a la th\'eorie BF en quatre dimensions, notamment via le fameux \'etat de Kodama, \cite{baez-knots-qg}. Par ailleurs, mon premier travail de (pr\'e-)th\`ese fut l'\'etude d'une ambiguit\'e dans l'action de la gravit\'e 3d, ambiguit\'e qui ressemble au param\`etre d'Immirzi de la gravit\'e 4d, et qui prend son sens plus ais\'ement dans le cadre de la th\'eorie de Chern-Simons. Nous en donnons donc une br\`eve exposition.

L'id\'ee est pour nous de combiner la sym\'etrie de jauge standard et la sym\'etrie de translation en une seule sym\'etrie de jauge pour un groupe de structure \'elargi, $\tl{G}$. Ce groupe \'elargi est $\Spin(4)$, $\ISU(2)$ ou $\SO(3,1)$ selon que la constante cosmologique est positive, nulle ou n\'egative. Les g\'en\'erateurs de l'alg\`ebre de Lie $\tl{\alg}$ satisfont aux relations de commutation :
\begin{equation}
[J_i,J_j]=\epsilon_{ij}^{\phantom{ij}k} J_k, \quad\quad [J_i,K_j]=\epsilon_{ij}^{\phantom{ij}k}
K_k, \quad\quad [K_i,K_j]=s\ \epsilon_{ij}^{\phantom{ij}k} J_k,
\end{equation}
ou $s=-1,0,1$ est le signe de la constante cosmologique. Nous pouvons maintenant consid\'erer une connexion pour le groupe \'elargi, $\omega = A^iJ_i + \sqrt{\lvert\Lambda\rvert}\ e^iK_i$. Dans le groupe des transformations de jauge correspondant, une rotation g\'en\'er\'ee par un \'el\'ement de la forme $u=u^iJ_i$ engendre une transformation de jauge $\SU(2)$, tandis qu'un boost $v=v^iK_i$ engendre une transformation \eqref{translation symetry}. Pour construire l'action de Chern-Simons, nous devons choisir une forme bilin\'eaire invariante non-d\'eg\'en\'er\'ee sur l'alg\`ebre,
\begin{equation} \label{forme killing}
\langle J_i,K_j\rangle=\delta_{ij} \qquad\qquad \langle J_i,J_j\rangle=\langle K_i,K_j\rangle=0.
\end{equation}
Alors, pour $\Lambda\neq 0$\footnote{Le cas $\Lambda=0$ est simplement d\'ecrit en prenant $s=0$ et $\sqrt{\lvert\Lambda\rvert}=1$ ou n'importe quelle constante.}, l'action de la gravit\'e en 3d s'\'ecrit aussi bien comme une action de Chern-Simons,
\begin{align} \label{reformulation witten}
S_{\mathrm{CS}}(\omega) &=\frac{1}{2\sqrt{\lvert\Lambda\rvert}}\int_{M_3} d^3x\ \epsilon^{\mu\nu\rho}\big(
\langle \omega_\mu,\partial_\nu \omega_\rho\rangle+\frac{1}{3}\langle \omega_\mu,[\omega_\nu,\omega_\rho]\rangle\big),\\
&= S_{\mathrm{GR3d}}(e,A).
\end{align}
L'\'equation du mouvement de Chern-Simons dit simplement que la courbure $R(\omega)$ de la connexion $\omega$ s'annule. Les composantes de la courbure $R(\omega)$ selon les g\'en\'erateurs $(J_i)_i$ et $(K_i)_i$ sont telles que nous retrouvons les \'equations d'Einstein,
\beq
R(\omega) = \bigl(F^i+\frac{\Lambda}{2}\,[e,e]^i\bigr)\, J_i + \bigl(d_\omega e\bigr)^i\,K_i = 0.
\ee

Introduisons d\'esormais une ambiguit\'e, caract\'eris\'ee par un param\`etre not\'e $\gamma$ dont l'origine est compl\`etement analogue \`a celle du param\`etre d'Immirzi en 4 dimensions. Cette ambiguit\'e fut remarqu\'ee par Witten, mais ne fut pas \`a ma connaissance \'etudi\'ee avant \cite{3d-imm}. Elle trouve en fait son origine dans l'existence, pour les cas $\Lambda\neq 0$, d'une deuxi\`eme forme bilin\'eaire invariante non-d\'eg\'en\'er\'ee sur l'alg\`ebre $\tl{\alg}$,
\begin{equation} \label{second bilinear form}
(J_i,J_j)=\delta_{ij},\qquad (K_i,K_j)=s\ \delta_{ij}\quad\mathrm{and}\quad(J_i,K_j)=0.
\end{equation}
L'existence d'une telle deuxi\`eme forme invariante est sp\'ecifique \`a ces groupes d'isom\'etries et peut se comprendre en observant que ceux-ci admettent une factorisation : $\Spin(4)=\SU(2)\times \SU(2)$, et de m\^eme pour $\SO(3,1)$ apr\`es complexification\footnote{Dans le cas lorentzien, c'est le groupe $\SO(2,2)$ qui intervient pour $\Lambda\leq0$ ; ce groupe s'\'ecrit de mani\`ere analogue, $\SO(2,2)=SL(2,\R)\times SL(2,\R)$.}. Ainsi, notre deuxi\`eme forme invariante n'est autre que la premi\`ere introduite, dans laquelle un des arguments se voit appliqu\'e l'op\'erateur de dualit\'e de Hodge $\star$ qui \'echangent les g\'en\'erateurs des rotations $(J_i)_i$ avec ceux des boosts $(K_i)_i$.

Cela siginifie que nous pouvons introduire une deuxi\`eme action de Chern-Simons,
\begin{equation} \label{2nd action 3d}
\tl{S}_{\mathrm{CS}}(\omega)
=\frac{1}{2\sqrt{\lvert\Lambda\rvert}}\int_{M_3} d^3x\ \epsilon^{\mu\nu\rho}\,\bigl( (\omega_\mu,\partial_\nu \omega_\rho)+\frac{1}{3}( \omega_\mu,[\omega_\nu,\omega_\rho])\bigr),
\end{equation}
qui conduit naturellement \`a la m\^eme \'equation du mouvement, $R(\omega) =0$. En termes de la co-triade et de la connexion de spin, cette action s'\'ecrit :
\beq
\tl{S}_{\mathrm{CS}}(\omega) = \frac{1}{2\sqrt{\lvert\Lambda\rvert}}S_{\mathrm{CS}}(A)+\f{s}{2}\sqrt{\lvert\Lambda\rvert}\ \int_{M_3} \,\tr \bigl( e\wedge d_A e\bigr).
\ee
$S_{\mathrm{CS}}(A)$ repr\'esente l'action de Chern-Simons $\SU(2)$ pour la connexion $A$, et le deuxi\`eme terme s'interpr\`ete comme un terme de torsion. De plus, il est possible de former une combinaison des deux actions \`a notre disposition \`a l'aide d'un param\`etre $\gamma$ r\'eel,
\beq
S_{\mathrm{GR3d}\gamma}(\omega) = S_{\mathrm{CS}}(\omega) + \gamma\mone\,\tl{S}_{\mathrm{CS}}(\omega).
\ee
Tant que la condition $\gamma^2\neq s$ est v\'erifi\'ee\footnote{Les cas o\`u $\gamma^2=s$ correspondent \`a des th\'eories de Chern-Simons pour la connexion self-duale ou anti-self-duale. Cela se v\'erifie gr\^ace \`a la d\'ecomposition
\beq
S_{\mathrm{GR3d}\gamma}(\omega) = \left(\gamma^{-1}+s\sigma\right)\, S_{CS}(\omega_+)\,+\,
\left(\gamma^{-1}-s\sigma\right)\, S_{CS}(\omega_-),
\ee
dans laquelle les connexions (anti-)self-duales sont $\omega_\pm^i = A^i\pm\sigma\sqrt{\lvert\Lambda\rvert}e^i$, avec : $\star\omega_\pm = \pm\sigma\omega_\pm$, le coefficient $\sigma$ \'etant 1 pour $\Lambda>0$ et $i$ pour $\Lambda<0$.}, les \'equations de stationarit\'e par rapport \`a $A$ et $e$ reproduisent bien les \'equations d'Einstein.

Ne disposant pas de mod\`ele type mousses de spins pour la th\'eorie de Chern-Simons, nous ne poursuivrons pas plus avant dans ce m\'emoire de th\`ese l'\'etude de cette ambiguit\'e. Le lecteur est renvoy\'e \`a \cite{3d-imm} pour plus de d\'etails. Dans cet article, j'ai \'etudi\'e avec E. R. Livine l'alg\`ebre des contraintes pour les crochets de Poisson modifi\'es par $\gamma$. Nous avons rapproch\'e cette structure symplectique de celle de la th\'eorie de Yang-Mills 3d avec masse topologique (induite par un terme de Chern-Simons), et \'etudi\'e la d\'eformation du spectre des longueurs par $\gamma$ dans la quantification par les r\'eseaux de spins. Une comparaison pr\'ecise avec la situation en 4d sugg\`ere que le param\`etre $\gamma$ introduit ici, bien qu'ayant la m\^eme origine que le param\`etre d'Immirzi en 4d, se rapproche plut\^ot des termes topologiques-$\theta$, du point de vue de la structure symplectique. L'\'etude de cette ambiguit\'e a \'et\'e reprise dans \cite{meusburger-immirzi-chern-simons}, en lien avec le r\^ole des sym\'etries dites kappa-Poincar\'e en gravit\'e 3d.

\chapter{Relativit\'e g\'en\'erale \`a la Palatini et Plebanski}

Nous nous int\'eressons d\'esormais \`a la relativit\'e g\'en\'erale \emph{riemannienne} (comme dans le reste de cette th\`ese), sous sa formulation utilis\'ee en gravit\'e dite \`a boucles qui a pour action l'action de Holst. Nous expliciterons le lien entre l'action de Holst et celle dite de Plebanski qui \'etablit un lien pr\'ecis avec la th\'eorie BF. C'est de plus laction de Plebanski, \`a travers son lien \`a la th\'eorie BF, qui sert de base pour la quantification en mousses de spins.

\section{L'action de Holst}

Nous choisissons un fibr\'e not\'e $\calT M$, sur notre vari\'et\'e d'espace-temps $M$ de dimension 4, qui soit isomorphe au fibr\'e tangent $TM$ mais pas de mani\`ere canonique. L'id\'ee est que chaque fibre de $\calT M$ peut \^etre vue comme une sorte d'imitation de l'espace tangent \`a $M$, pour reprendre le langage de Baez dans \cite{baez-book}, qu'on appelle en th\'eorie de jauge l'espace interne. Sur ces fibres, on introduit une m\'etrique euclidienne (ou minkowskienne) et on consid\'erera que la physique s'y comporte et s'y formule de mani\`ere euclidienne (ou minkowskienne). Ainsi, le champ gravitationnel n'est autre que l'isomorphisme entre $\calT M$ et $TM$ dont la t\^ache est de transposer la dynamique du fibr\'e d'imitation au fibr\'e tangent ! Sur une telle interpr\'etation du champ gravitationnel, j'invite \`a lire pour le plus grand profit les sections consacr\'ees du livre de Rovelli \cite{rovelli-book}. Cette isomorphisme prend donc la forme une 1-forme \`a valeurs dans $\calT M$, ou autrement dit d'une cot\'etrade not\'ee $e$.

Si $A$ est une connexion sur le fibr\'e $P$ des rep\`eres orthonormaux de $M$, sa courbure $F$ est une 2-forme \`a valeurs dans $\Ad(P)$ ou de mani\`ere \'equivalente dans $\Lambda^2\calT M$. Localement, nous verrons $A = A_\mu^{IJ} J_{IJ} dx^\mu$ comme une 1-forme sur $M$ \`a valeurs dans l'alg\`ebre de Lie $\so(4)$, dont les g\'en\'erateurs sont not\'es $J_{IJ}$. De m\^eme, la cot\'etrade a pour composantes $e^I = e_\mu^I dx^\mu$. Les indices internes, i.e. sur les fibres de $\calT M$, sont manipul\'es \`a l'aide de la m\'etrique euclidienne $\delta_{IJ}$. Avec ces pr\'eliminaires, nous sommes en mesure d'\'ecrire l'action de Holst pour un param\`etre d'Immirzi $\gamma$,
\beq \label{holst action}
S_{\mathrm{H}}(e,A) = \f 12 \int_M \eps_{IJKL}\ e^I\wedge e^J\wedge F(A)^{KL} + \f 1\gamma \int_M e^I\wedge e^J\wedge F(A)_{IJ}.
\ee
Cette action ne fait pas intervenir de m\'etrique sur $M$ mais se formule plut\^ot en termes de formes diff\'erentielles et est ainsi naturellement invariante sous les diff\'eomorphismes de $M$. Pour retrouver le formalisme d'origine d'Einstein-Hilbert, il faut former une m\'etrique sur $M$ via : $g = \delta_{IJ}\,(e^I\otimes e^J)$, ce qui fonctionne uniquement si la cot\'etrade est non-d\'eg\'en\'er\'ee. Notons \`a cet effet que $S_{\mathrm{H}}$ reste bien d\'efinie m\^eme lorsque $e$ est d\'eg\'en\'er\'ee, $\det e=0$. Par ailleurs, en plus d'appara\^itre comme pr\'ef\'erable ou attirant du point de vue physique \cite{rovelli-book}, l'utilisation de la cot\'etrade \`a la place de la m\'etrique est n\'ecessaire pour coupler la gravit\'e aux fermions.

La th\'eorie est ici dans un formalisme du premier ordre, ce qui signifie que $e$ (ou la m\'etrique) et $A$ sont des variables ind\'ependentes. Pour retrouver une connexion compatible avec la m\'etrique, il faut utiliser les \'equations du mouvement. Oublions pour cela dans un premier temps le second terme de l'action de Holst et formons simplement l'action de dite de Hilbert-Palatini,
\beq \label{palatini action}
S_{\mathrm{HP}}(e,A) = \f 12 \int_M \eps_{IJKL}\ e^I\wedge e^J\wedge F(A)^{KL} .
\ee
En extremisant cette action par rapport \`a $A$, on obtient les \'equations suivantes,
\beq \label{torsion nulle}
d_A (e\wedge e) = 0 \quad \overset{\det e\neq 0}{\Longrightarrow} \quad d_A\, e = 0.
\ee
Toujours pour $e$ non-d\'eg\'en\'er\'ee, cette derni\`ere relation permet d'exprimer la connexion comme une fonction de la cot\'etrade, $A(e)$. Celle-ci peut ensuite \^etre ramen\'ee sur l'espace tangent (en utilisant $e$) pour donner la connexion de Levi-Civita. C'est pour cette raison que l'\'equation : $d_A e =0$, est appel\'ee la condition de torsion nulle. Par ailleurs, la variation de l'action par rapport \`a la cot\'etrade fournit l'\'equation
\beq \label{einstein palatini eq}
\eps_{IJKL}\ e^J\wedge F^{KL} = 0.
\ee
En ramenant l'\'equation sur l'espace tangent et en utilisant $A(e)$, cela correspond pr\'ecis\'ement aux \'equations d'Einstein imposant \`a la courbure de Ricci de s'annuler.

Revenons maintenant \`a l'action de Holst \eqref{holst action} pour consid\'erer l'influence du deuxi\`eme terme. Ce terme, pris isol\'ement, n'a \'et\'e \'etudi\'e que tr\`es r\'ecemment \cite{perez-liu-topo}. Les auteurs ont montr\'e qu'il s'agit d'un terme topologique, i.e. ne poss\'edant pas de degr\'es de libert\'e localement. Ce r\'esultat \'etait attendu dans la mesure o\`u son effet au sein de l'action de Holst \'etait d\'ej\`a connu : les \'equations du mouvement pour l'action de Holst sont exactement les m\^emes que celles de l'action de Hilbert-Palatini \eqref{palatini action}. Cela signifie en particulier que le param\`etre d'Immirzi $\gamma$ est classiquement invisible en gravit\'e pure. En fait, le passage de l'action de Palatini-Cartan \`a l'action de Holst se traduit en termes hamiltoniens par une transformation canonique sur l'espace des phases. N\'eanmoins, les r\'esultats de \cite{perez-liu-topo} peuvent para\^itre surprenants d'un autre point de vue. En effet, le terme en $e^I\wedge e^J\wedge F_{IJ}$ dans \eqref{holst action} ne conduit qu'\`a l'\'equation du mouvement \eqref{torsion nulle}, du fait que l'\'equivalent de \eqref{einstein palatini eq}, 
\beq
e^I\wedge F_{IJ}=0,
\ee
est satisfaite d\`es lors que $A$ est la connexion de torsion nulle $A(e)$. On pourrait alors penser que l'ensemble des solutions classiques est beaucoup plus grand que celui associ\'e \`a $S_{\mathrm{H}}$. Cette assertion se r\'ev\`ele en fait fausse, car le terme suppl\'ementaire de l'action de Holst jouit d'un groupe de sym\'etrie de jauge beaucoup plus large que $S_{\mathrm{HP}}$, ce qui permet d'identifier des solutions a priori non-\'equivalentes. Les auteurs de \cite{perez-liu-topo} obtiennent ce r\'esultat par une analyse hamiltonienne compl\`ete. Mais la formulation lagrangienne de ces sym\'etries de jauge suppl\'ementaires semble compliqu\'ee et n'est toujours pas connue.

D'autres travaux ont \'et\'e men\'es pr\'ec\'edemment pour apporter des \'eclaircissements sur le param\`etre d'Immirzi qui reste malgr\'e tout aujourd'hui mal compris. Classiquement, il d\'efinit une famille \`a un param\`etre de th\'eories \'equivalentes pour la gravit\'e pure. N\'eanmoins, le couplage \`a la mati\`ere est susceptible d'apporter des effets physiques particuli\`erement int\'eressants. En effet, il est connu que le couplage minimal de fermions dans la th\'eorie d'Einstein-Cartan apporte de la torsion, sous la forme : $d_A e =$\og courant fermionique\fg. En prenant pour base l'action de Holst, les auteurs de \cite{perez-torsion} trouvent alors que l'action effective comporte un terme d'interactions \`a quatre fermions dont la constante de couplage d\'epend explicitement de $\gamma$,
\begin{equation} \label{int einstein cartan}
S_{\mathrm{int}}(e,\psi)=-\frac{3}{2}\pi G \frac{\gamma^2}{\gamma^2+1}\int_M d^4x \sqrt{g}\
\big(\overline{\psi}\gamma_5\gamma_I\psi\big)\ \big(\overline{\psi}\gamma_5\gamma^I\psi\big).
\end{equation}
Dans \cite{freidel-torsion} sont aussi \'etudi\'es le couplage non-minimal en pr\'esence de torsion et du param\`etre d'Immirzi, et l'influence sur la violation de la parit\'e.

Du point de vue math\'ematique le param\`etre d'Immirzi tire son origine de la pr\'esence de deux formes bilin\'eaires invariantes non-d\'eg\'en\'er\'ees sur l'alg\`ebre $\spin(4)$, exactement de la m\^eme mani\`ere que le param\`etre $\gamma$ introduit en section \ref{sec:BFlowdim} dans la formulation \`a la Chern-Simons de la gravit\'e 3d. Ces deux formes permettent deux contractions diff\'erentes entre $e^I\wedge e^J$ et la courbure $F$ et sont reli\'ees par dualit\'e de Hodge comme le montre l'\'ecriture de l'action,
\beq \label{holst2}
S_{\mathrm{H}}(e,A) = \f 12 \int_M \eps_{IJKL}\ e^I\wedge e^J \wedge \bigl( F + \gamma\mone \star F\bigr)^{KL}.
\ee
Il s'agit bien s\^ur des formes invariantes pr\'ec\'edemment d\'efinie, en \eqref{forme killing} et \eqref{second bilinear form}. \'Etant donn\'e la possibilit\'e de construire ces deux termes de l'action, le fait essentiel reste que la deuxi\`eme forme invariante ne modifie pas les \'equations du mouvement.

\section{L'action de Plebanski}

Nous pr\'esentons ici l'action de Plebanski $\SO(4)$, qui fut initialement introduite en relation avec les solutions self-duales de la relativit\'e g\'en\'erale complexifi\'ee. Pour de r\'ecentes pr\'esentations en liens avec les mod\`eles de mousses de spins, nous renvoyons \`a \cite{freidel-plebso4} et \cite{reisenberger-pleb}.

Comme le sugg\`ere l'\'ecriture \eqref{holst2} de l'action de Holst, celle-ci peut \^etre vue comme une action BF,
\beq \label{BFimmirzi}
\int_M \tr\ B\wedge \bigl(F + \gamma\mone \star F\bigr),
\ee
pour laquelle on ne regarde que les configurations de $B$ prenant la forme
\beq
B = \star (e\wedge e).
\ee
Ici, nous avons normalis\'e la trace pour que $\tr (AB) = A^{IJ} B_{IJ}$. Plus g\'en\'eralement, on a l'habitude de dire que $B$ est simple s'il s'\'ecrit $\pm(e\wedge e)$ ou $\pm\star(e\wedge e)$ pour une 1-forme $e$. Ainsi, si les configurations simples $\pm\star(e\wedge e)$ correspondent \`a l'action de Holst pour un param\`etre d'Immirzi $\gamma$, les configurations simples $\pm(e\wedge e)$ donnent l'action de Holst pour un param\`etre d'Immirzi $\gamma\mone$. Par ailleurs, soulignons que l'op\'erateur $(\id +\gamma\mone\star)$ \'etant inversible sur l'alg\`ebre $\so(4)$, une simple red\'efinition de $B$ dans l'action \eqref{BFimmirzi} montre qu'il s'agit bien d'une th\'eorie BF authentique.

Bien s\^ur, il ne s'agit pas pour nous de restreindre l'espace des solutions de la th\'eorie BF aux 2-formes simples, mais bien plut\^ot d'introduire cette condition comme une contrainte dans l'action, \`a respecter lorsque l'on regarde les variations de celle-ci. Cela nous conduit \`a l'action de Plebanski $\SO(4)$ dont les variations conduisent aux \'equations du mouvement voulues. Nous introduisons donc un multiplicateur de Lagrange $\phi$ qui force $B$ \`a \^etre simple :
\beq \label{pleb action1}
S_{\mathrm{Pl}}(B,A,\phi) = \f12 \int_M \tr\ B\wedge \bigl(F + \gamma\mone \star F\bigr) + \f12 \int_M \phi_{IJKL}\ B^{IJ}\wedge B^{KL}.
\ee
Le multiplicateur $\phi$ doit en outre \^etre antisym\'etrique dans l'\'echange des indices $I$ et $J$, et $K$ et $L$, $\phi_{IJKL} = \phi_{[IJ][KL]}$, sym\'etrique dans l'\'echange des paires $[IJ]$ et $[KL]$, et de trace nulle, $\eps^{IJKL}\phi_{IJKL} =0$. Il contient donc 20 composantes ind\'ependantes, qui conduisent \`a 20 \'equations alg\'ebriques sur les 36 composantes de $B$,
\beq \label{simplicity pleb}
\eps^{\mu\nu\rho\sigma}\ B_{\mu\nu}^{IJ}\,B_{\rho\sigma}^{KL} = \f\calV {4!}\ \eps^{IJKL},
\ee
pour
\beq \label{4-volume}
\calV = \eps^{\mu\nu\rho\sigma}\,\eps_{IJKL}\,B_{\mu\nu}^{IJ}\,B_{\rho\sigma}^{KL}.
\ee
Les contraintes \eqref{simplicity pleb} sont appel\'ees contraintes de simplicit\'e, et lorsqu'elles sont satisfaites $\calV$ repr\'esente le d\'eterminant de la cot\'etrade associ\'ee \`a $B$. Plus pr\'ecis\'ement, les solutions des contraintes de simplicit\'e ont \'et\'e \'etudi\'ees et classifi\'ees par diff\'erents auteurs. Je renvoie \`a Freidel-De Pietri \cite{freidel-plebso4} et \`a Reisenberger \cite{reisenberger-pleb} pour les d\'etails des preuves (et sp\'ecifiquement \`a \cite{reisenberger-pleb} pour l'\'etude des cas d\'eg\'en\'er\'es), et \`a \cite{pleb-high-dim} pour la g\'en\'eralisation des contraintes de simplicit\'e aux dimensions sup\'erieures.

Lorsque la condition de non-d\'eg\'en\'erescence $\calV\neq0$ est satisfaite, les contraintes \eqref{simplicity pleb} sont v\'erifi\'ees si et seulement $B$ est simple, c'est-\`a-dire s'il existe une cot\'etrade $e$ reli\'ee \`a $B$ par l'une des \'egalit\'es suivantes :
\begin{align}
&\textrm{secteur topologique}\ (I\pm)\qquad &B &= \pm (e\wedge e),\\
&\textrm{secteur gravitationnel}\ (II\pm)\qquad &B &= \pm \star(e\wedge e).
\end{align}
Nous avons rapport\'e les appellations de secteur topologique/gravitationnel qui restent couramment utilis\'ees, bien que toutes les solutions aboutissent \`a l'action de Holst en pr\'esence du param\`etre d'Immirzi dans \eqref{pleb action1} (\`a une red\'efinition pr\`es de ce param\`etre).

De plus, toujours pour $\calV\neq0$, il est possible de troquer l'ensemble des contraintes \eqref{simplicity pleb} pour un autre en \'echangeant le r\^ole des indices internes et tangents,
\beq \label{simplicity pleb2}
\eps_{IJKL}\ B_{\mu\nu}^{IJ}\,B_{\rho\sigma}^{KL} = \f\calV {4!}\ \eps_{\mu\nu\rho\sigma}.
\ee
Celles-ci s'obtiennent \`a partir de l'action suivante :
\beq \label{pleb action2}
\tl{S}_{\mathrm{Pl}}(B,A,\phi) = \f12 \int_M \tr\Bigl( B\wedge \bigl(F + \gamma\mone \star F\bigr)\Bigr) + \int_M d^4x\ \phi^{\mu\nu\rho\sigma}\ \tr\bigl(B_{\mu\nu}\, (\star B_{\rho\sigma})\bigr),
\ee
o\`u $\tr(A(\star B)) = \f12\eps_{IJKL}A^{IJ}B^{KL}$. Le nouveau multiplicateur de Lagrange doit satisfaire aux m\^emes sym\'etries que dans la formulation pr\'ec\'edente, $\phi^{\mu\nu\rho\sigma} = -\phi^{\nu\mu\rho\sigma} = -\phi^{\mu\nu\sigma\rho} = \phi^{\rho\sigma\mu\nu}$ et $\eps_{\mu\nu\rho\sigma}\phi^{\mu\nu\rho\sigma} =0$. C'est en fait cette formulation qui est implicitement utilis\'ee pour construire la plupart des mod\`eles de mousses de spins en gravit\'e quantique. En fait, seul le mod\`ele de Reisenberger \cite{reisenberger-spinfoam}, qui ne sera pas \'etudi\'e dans cette th\`ese, s'appuie sur la formulation \eqref{simplicity pleb} des contraintes de simplicit\'e (mais dans une version self-duale).

Une autre description des contraintes de simplicit\'e est int\'eressante en ce qu'elle permet d'appr\'ecier l'information g\'eom\'etrique contenue dans la th\'eorie BF. On utilise pour cela la d\'ecomposition de l'alg\`ebre $\so(4)$ en deux facteurs $\su(2)$, self-dual et anti-self-dual, pour \'ecrire $B = B_+ \oplus B_-$, avec $\star B_+ = B_+$ et $\star B_- = -B_-$. A partir de chacune de ces formes \`a valeurs dans $\su(2)$, il est possible de construire une m\'etrique sur $M$, donn\'ee par la formule d'Urbantke,
\beq
g_{\pm\mu\nu} = -\f2{3\calV}\,\eps_{ijk}\,\eps^{\alpha\beta\gamma\delta}\ B_{\pm\mu\alpha}^i\,B_{\pm\beta\gamma}^j\,B_{\pm\delta\nu}^k.
\ee
Cela nous donne donc deux m\'etriques pour un champ $B$ \`a valeurs dans $\so(4)$. Lorsque les contraintes de simplicit\'e sont satisfaites, ces expressions se simplifient et les secteurs topologique et gravitationnel se distinguent du fait de la relation induite par la cot\'etrade entre les m\'etriques $g_+$ et $g_-$,
\begin{align}
&\textrm{secteur topologique}\ (I\pm)\qquad &g_{+\mu\nu} &= -g_{-\mu\nu} = \pm \delta_{IJ}\ e^I_\mu e^J_\nu,\\
&\textrm{secteur gravitationnel}\ (II\pm)\qquad &g_{+\mu\nu} &= g_{-\mu\nu} = \pm \delta_{IJ}\ e^I_\mu e^J_\nu.
\end{align}

\chapter{Calcul de Regge} \label{sec:regge calculus}

Nous verrons en discutant le formalisme canonique de la LQG que les graphes qui constituent les excitations du champ gravitationnel portent une certaine forme de g\'eom\'etrie associ\'ee aux liens et noeuds de chaque graphe. Cela se transmet en mousses de spins o\`u la quantification se fait directement \`a partir d'une triangulation fix\'ee de l'espace-temps, dont les propri\'et\'es g\'eom\'etriques doivent \^etre quantifi\'ees. Il est donc essentiel de comparer, voire de s'inspirer d'une approche \`a la relativit\'e g\'en\'erale d\'ej\`a connue, fond\'ee sur la g\'eom\'etrie de structures discr\`etes : le calcul de Regge \cite{regge-calculus}. A la vue des r\'esultats existants, notamment en 3d, on peut s'attendre \`a ce que la limite semi-classique des mousses de spins soit en effet donn\'ee par une forme de calcul de Regge quantique ! Cela va nous motiver pour pr\'esenter une formulation r\'ecente \cite{dittrich-speziale-aarc} des g\'eom\'etries de Regge utilisant les aires et angles dih\'edraux 3d comme variables fondamentales.

Le calcul de Regge est une approximation de la relativit\'e g\'en\'erale dans laquelle l'espace-temps est une vari\'et\'e triangul\'ee dont les blocs sont des $n$-simplexes. Le champ dynamique, i.e. la m\'etrique, est d\'ecrit par la donn\'ee des longueurs $(\ell_e)$ de toutes les ar\^etes de la triangulation. On peut voir une g\'eom\'etrie de Regge comme un cas de vari\'et\'e riemannienne continue dite plate par morceaux : les $n$-simplexes sont plats, et la courbure est concentr\'ee autour des $(n-2)$-simplexes. Nous nous en tiendrons ici \`a une pr\'esentation des bases du calcul de Regge (classique), et d'une de ces variantes r\'ecemment introduite, le but \'etant d'y avoir recours lorsqu'il sera pertinent dans l'\'etude des mousses de spins, ou de toute g\'eom\'etrie discr\`ete. N\'eanmoins, il s'agit d'une approche offrant une multitude de facettes int\'eressantes, du point de vue math\'ematique, sur la g\'eom\'etrie des vari\'et\'es plates par morceaux, analytiquement mais aussi num\'eriquement. Je renvoie le lecteur au tr\`es bon expos\'e \cite{frohlich-regge}, aux revues \cite{williams-discrete-qg, williams-tuckey-biblio, loll-discrete-approaches} sur le calcul de Regge et d'autres approches discr\`etes, \`a \cite{regge-williams-discrete-structures, gentle-miller-numerical-regge} sur le formalisme classique, et \cite{rocek-williams-quantization-regge, rocek-willliams-quantum-regge, hamber-williams-3d, hamber-williams-lattice-graviton} pour des r\'esultats au niveau quantique.

Toutes les grandeurs g\'eom\'etriques sont d\'etermin\'ees par l'ensemble des longueurs. Celles-ci formant une m\'etrique plate par morceaux, il s'agit donc d'un formalisme du second ordre. Les angles dih\'edraux entre deux $(n-1)$-simplexes (t\'etra\`edres en 4d), $\thet_{tt'}(\ell_e)$, sont des fonctions explicites des longueurs, et permettent de d\'efinir une notion de courbure discr\`ete. Pour cela, par exemple en 4d, on colle un ensemble de 4-simplexes le long de t\'etra\`edres partag\'es deux \`a deux, et ayant tous un triangle $f$ en commun, et on mesure l'angle de d\'eficit :
\beq \label{deficit angle}
\vareps_f(\ell_e) \,=\, 2\pi - \sum_{(tt')} \thet_{tt'}(\ell_e),
\ee
par la somme des angles dih\'edraux autour du triangle. De m\^eme, les longueurs donnent acc\`es aux volumes $A_f(\ell_e)$ des $(n-2)$-simplexes (les aires des triangles en 4d, et en 3d les longueurs elles-m\^emes). Ces volumes discr\'etisent la $(n-2)$-forme dans le continu form\'ee par le produit ext\'erieur du vielbein (cot\'etrade en 4d) $(n-2)$ fois avec lui-m\^eme, $\bigwedge_{}^{n-2} e$. On peut alors former l'action de Regge se d\'ecomposant sur les $(n-2)$-simplexes de l'int\'erieur de la triangulation :
\beq \label{regge action}
S_{\rm R}(\ell_e) \,=\, \sum_f A_f(\ell_e)\,\vareps_f(\ell_e).
\ee
Les \'equations de Regge associ\'ees s'obtiennent par variations de l'action par rapport aux longueurs. Un ph\'enom\`ene important est que les termes de variation des angles dih\'edraux n'interviennent pas. La raison en est l'identit\'e de Schl\H{a}fli, \'etablissant que dans chaque $n$-simplexe $v$ :
\beq \label{schlafli}
\sum_{f\subset \pp v} A_f\, \delta\theta_{tt'} = 0,
\ee
pour toute variation de la g\'eom\'etrie du $n$-simplexe ($(t, t')$ repr\'esentent la paire de $(n-1)$-simplexes dans $v$ partageant $f$). Ainsi les \'equations de Regge sont :
\beq \label{regge equation}
\f{\pp S_{\rm R}}{\pp \ell_e} \,=\, \sum_{f\,/\,\pp f \supset e} \f{\pp A_f}{\pp \ell_e}\,\vareps_f \,=\, 0.
\ee
En 3d, cela dit simplement que les angles de d\'eficit s'annulent.

\medskip

Il nous sera en fait surtout utile de conna\^itre la structure d\'etaill\'ee des g\'eom\'etries de Regge, et les relations entre les diff\'erentes grandeurs. En 3 dimensions, o\`u la th\'eorie continue est topologique, la quantification est bien connue -- via Cherns-Simons \cite{witten-3d-gravity}, en boucles \cite{noui-perez-ps3d} ou en mousses de spins \cite{ooguri-3d} -- et on sait relier la quantification exacte donn\'ee par le mod\`ele de mousses de spins de Ponzano-Regge au calcul de Regge dans la limite semi-classique : il s'agit d'une forme de calcul de Regge quantique, dans lequel les longueurs prennent des valeurs discr\`etes. Cela fut pressenti dans  le travail fondateur de Ponzano-Regge \cite{PR}, et je renvoie \`a plus \cite{dowdall-asym-PR} pour les travaux les plus r\'ecents sur l'asymptotique du mod\`ele pour une certaine classe de vari\'et\'es. Cette quantification des longueurs, initialement un ansatz, fut comprise par Rovelli, \cite{rovelli-PRTVO}, \`a la lumi\`ere de quantification en boucles. Cela entre autres indications incita \`a suivre une voie similaire en 4d, \`a savoir de se fonder sur la quantification du mod\`ele topologique 4d \cite{ooguri4d}, offrant une mesure bien d\'efinie. Les nombres quantiques qui \'emergent sont alors plus naturellement interpr\'et\'es comme des aires de triangles, ce qui coincide avec la description de la g\'eom\'etrie quantique en gravit\'e \`a boucles, o\`u les aires sont plus naturelles que les longueurs. Cela sugg\`ere donc \cite{rovelli-PRTVO} que le r\'egime asymptotique des grandes distances pourrait mieux s'appr\'ehender en termes d'un calcul de Regge utilisant plut\^ot les aires que les longueurs. Malgr\'e des efforts certains dans cette direction, le probl\`eme reste ouvert. La principale difficult\'e r\'eside dans le fait qu'il y a plus de triangles que d'ar\^etes dans une triangulation r\'eguli\`ere \footnote{Chaque triangle a exactement trois ar\^etes, tandis que chaque ar\^ete est partag\'ee par au moins triangles.}, de sorte que toutes les aires ne sont pas ind\'ependantes. Il en r\'esulte des contraintes sur les aires, qui sont non-locales et difficiles \`a contr\^oler.

Pour surmonter ces difficult\'es, il a r\'ecemment \'et\'e propos\'e par Dittrich et Speziale, \cite{dittrich-speziale-aarc}, d'introduire des variables suppl\'ementaires, qui mettent en avant le r\^ole des angles dih\'edraux dans la construction des g\'eom\'etries de Regge. C'est un pas en avant certain dans la compr\'ehension des articulations entre calcul de Regge et mousses de spins, car les angles dih\'edraux, comme nous le verrons, peuvent \^etre rassembl\'es pour former des objets vivant dans des groupes de Lie (chapitre \ref{sec:bf-regge}). De telles structures alg\'ebriques se pr\`etent particuli\`erement bien \`a un d\'eveloppement en mousses de spins. Ainsi nous pourrons obtenir une vision compl\`etement g\'eom\'etrique des mod\`eles de mousses de spins (chapitre \ref{sec:pathint fk}) ! Mais ces structures alg\'ebriques forment aussi le cadre naturel du formalisme canonique de la LQG, de sorte que Dittrich et Ryan \cite{dittrich-ryan-simplicial-phase} ont pu explorer la g\'eom\'etrie de l'espace des phases de la LQG au regard du calcul de Regge \og aires - angles 3d\og (voir section \ref{sec:twisted-simplicial}) !

Nous oublions donc les longueurs, et nous concentrons sur les angles dih\'edraux 3d, $\phi_{ff'}^t$, entre les triangles $f, f'$ dans le t\'etra\`edre $t$. Ceux-ci permettent de d\'ecrire la forme de chaque t\'etra\`edre de mani\`ere unique (i.e. \`a l'\'echelle pr\`es). Mais dans $t$, les six angles ne sont pas ind\'ependants ; il n'y a que cinq degr\'es de libert\'e. Il nous faut donc une contrainte s'exer\c{c}ant sur chaque t\'etra\`edre en reliant ces six angles. Une telle contrainte est bien connue et consiste \`a former la matrice $G_{ff'}^t = \cos\phi_{ff'}^t$ (avec par convention $\phi_{ff} = \pi$, et les entr\'ees \'etant les triangles de $t$). Alors, le t\'etra\`edre $t$ est ferm\'e si et seulement si :
\beq \label{gram tet}
\det\, G_{ff'}^t \,=\,0.
\ee
Cela revient \`a interpr\'eter $G_{ff'}$ comme la matrice de Gram d'une famille de quatre vecteurs $\vec{n}_f\in\R^3$, repr\'esentants les directions orthogonales aux triangles dans l'espace 3d engendr\'e par $t$. Alors :
\beq
\cos\phi_{ff'}^t = -\f{\vec{n}_f\cdot\vec{n}_{f'}}{\lv \vec{n}_{f}\rv\,\lv\vec{n}_{f'}\rv}.
\ee
En formant les produits vectoriels de ces vecteurs, on a acc\`es \`a des vecteurs d\'ecrivant l'orientation des ar\^etes dans l'espace 3d, et par suite aux angles dih\'edraux 2d, entre les ar\^etes de chaque triangle $f$. Dans le t\'etra\`edre $t$, l'angle entre les ar\^etes $e, e'$ du triangle $f$, qui se rencontrent en un sommet partag\'e avec les triangles $f_1, f_2$, est donn\'e en fonction des trois angles 3d autour de ce sommet :
\beq \label{relation 2d-3d angles}
\cos\alpha_{ee'}^{ft} \,=\, \f{\cos\phi_{f_1f_2}^t \,+\, \cos\phi_{ff_1}^t \cos\phi_{ff_2}^t}{\sin\phi_{ff_1}^t\,\sin\phi_{ff_2}^t}.
\ee
N\'eanmoins, lorsque l'on colle deux t\'etra\`edres $t, t'$, il n'est pas du tout clair que leur fronti\`ere commune, le triangle disons $f$, ait la m\^eme forme selon qu'il soit vu de $t$ ou de $t'$ ! En fait, sa forme se caract\'erise par les angles dih\'edraux 2d entre ses trois ar\^etes, qui peuvent alors \^etre calcul\'es en fonction des angles 3d de $t$, soit : $\cos\alpha_{ee'}^{ft}(\phi_{ff'}^t)$, ou des angles 3d de $t'$ : $\cos\alpha_{ee'}^{ft'}(\phi_{ff'}^{t'})$. La consistence g\'eom\'etrique requiert alors :
\beq \label{2dconstraints-aarc}
C_{ee'}^f \equiv \cos\alpha_{ee'}^{ft} \,-\, \cos\alpha_{ee'}^{ft'} \,=\, 0.
\ee
Suivant \cite{dittrich-speziale-aarc}, les contraintes \eqref{gram tet} et \eqref{2dconstraints-aarc} permettent d'utiliser les angles 3d pour d\'ecrire les g\'eom\'etries de Regge. Les contraintes \eqref{2dconstraints-aarc} ont une jolie interpr\'etation g\'eom\'etrique et comme nous le verrons en \ref{sec:twisted-simplicial}, selon \cite{dittrich-ryan-simplicial-phase}, forment la cl\'e pour comprendre la diff\'erence entre le type de g\'eom\'etrie d\'ecrit par la LQG et le calcul de Regge.

N\'eanmoins, le formalisme des mousses de spins, triangulant l'espace-temps et utilisant les holonomies d'une connexion entre t\'etra\`edres, pousse plus naturellement \`a consid\'erer les angles dih\'edraux 4d, entre t\'etra\`edres adjacents, que les angles 2d. Aussi, nous ferons souvent r\'ef\'erence \`a la section \ref{sec:gluing-aarc} \`a un jeu de contraintes diff\'erent mais \'equivalent \`a \eqref{2dconstraints-aarc}, d\'ej\`a pr\'esent\'e dans \cite{dittrich-speziale-aarc}. L'id\'ee est que dans un 4-simplexe, il est possible de calculer de plusieurs mani\`eres l'angle dih\'edral $\thet_{t_1t_2}$ en fonctions des angles 3d. Pour cela, on consid\`ere que $t_1, t_2$ se rencontrent le long de $f$. Dans le 4-simplexe correspondant, nous choisissons un des trois t\'etra\`edres restants, disons $t$. Celui-ci partage avec $t_1$ le triangle $f_1$, et avec $t_2$ le triangle $f_2$. Alors, on calcule :
\beq
\cos\,\thet_{t_1 t_2}(t) \,=\, \f{\cos\phi_{f_1f_2}^t - \cos\phi_{f_1f}^{t_1}\cos\phi_{ff_2}^{t_2}}{\sin\phi_{f_1f}^{t_1}\sin\phi_{ff_2}^{t_2}}.
\ee
Cette formule fait donc intervenir les angles dih\'edraux du t\'etra\`edre $t$. Ainsi, en choisissant un autre t\'etra\`edre interm\'ediaire $t'$, rencontrant $t_1, t_2$ le long des triangles $f_1', f_2'$, nous formons une autre expression pour le m\^eme angle, $\thet_{t_1 t_1}(t')$, faisant intervenir les angles 3d $\phi_{f_1' f_2'}^{t'}$. Le jeu de contraintes que nous cherchons prend donc la forme :
\beq \label{constraints4dangles-aarc}
\cos\thet_{t_1 t_2}(t) - \cos\thet_{t_1 t_2}(t') = \f{\cos\phi_{f_1f_2}^t - \cos\phi_{f_1f}^{t_1}\cos\phi_{ff_2}^{t_2}}{\sin\phi_{f_1f}^{t_1}\sin\phi_{ff_2}^{t_2}} - \f{\cos\phi_{f'_1f'_2}^{t'} - \cos\phi_{f'_1f}^{t_1}\cos\phi_{ff'_2}^{t_2}}{\sin\phi_{f'_1f}^{t_1}\sin\phi_{ff'_2}^{t_2}} = 0.
\ee
Bien s\^ur un tel jeu de contraintes semble peu ais\'e \`a manipuler$\dotsc$ Mais nous verrons \`a la section \ref{sec:gluing-aarc} comment \'ecrire cela avec de simples produits d'\'el\'ements du groupe $\SU(2)$, et formuler un ensemble de contraintes covariantes \'equivalentes \`a \eqref{gram tet} et \eqref{constraints4dangles-aarc} !

Ces contraintes d\'ecrivent donc la forme de la triangulation, mais nous n'avons ici pas mis d'\'echelle, et plut\^ot mis en avant le r\^ole des angles. Il reste que les aires sont des variables naturelles en LQG et en mousses de spins, que l'on souhaite prendre en compte dans cette formulation du calcul de Regge. Nous consid\'erons donc des variables $A_f$, a priori ind\'ependantes. Pour qu'elles puissent s'identifier aux aires des triangles, il faut revenir aux contraintes \eqref{gram tet} d\'ecrivant la g\'eom\'etrie d'un t\'etra\`edre isol\'e. L'annulation du d\'eterminant de la matrice de Gram $G_{ff'}$ traduit simplement une relation de d\'ependance lin\'eaire entre les quatre vecteurs $\vec{n}_{f}$, dont on peut facilement se convaincre que les coefficients sont pr\'ecis\'ement (proportionnels) aux aires. On forme ainsi la relation de fermeture du t\'etra\`edre :
\beq \label{3dclosure}
\sum_{f\subset \pp t} A_f\,\vec{n}_f \,=\,0.
\ee
Cette relation s'exprime en fonction des angles 3d simplement en la projetant sur chacun des quatre $\vec{n}_f$ :
\beq \label{gauge invariant closure}
A_f - \sum_{f'\neq f} A_{f'}\,\cos\phi_{ff'}^t =0.
\ee

\part{G\'eom\'etrie dans l'approche canonique} \label{part:canonical geom}


\chapter{Description hamiltonienne}

\section{Th\'eorie BF}

L'action BF \'etant lin\'eaire en les d\'eriv\'ees des champs, elle se pr\^ete ais\'ement \`a une analyse hamiltonienne, qui permet un traitement complet des sym\'etries de jauge. Pour simplifier les notations nous choisissons une vari\'et\'e $M$ de dimension 4, et demandons \`a pouvoir la d\'ecomposer en $M = \Sigma\times\R$. Nous prenons des coordonn\'ees adapt\'ees, et notons $t$ celle sur $\R$. L'action se r\'e\'ecrit alors :
\begin{align}
S_{\mathrm{BF}} &= \int d^4x\ \eps^{\mu\nu\rho\sigma}\ \tr\,\bigl(B_{\mu\nu}\,F_{\rho\sigma}\bigr) \\
&= \int dt \int d^3x\ \tr \bigl(B^a\ \pp_t A_a\bigr) - \tr \bigl(A_t\ D_a B^a\bigr) + \tr\bigl(B_{at}\ \eps^{abc}\,F_{bc}\bigr), \label{hamiltonian bf}
\end{align}
o\`u les indices $a,b,c$ repr\'esentent les indices des coordonn\'ees sur $\Sigma$, et $\eps^{abc}$ est d\'efini \`a partir du tenseur compl\`etement antisym\'etrique $\eps^{\mu\nu\rho\sigma}$ par : $\eps^{abc} = \eps^{abct}$. L'op\'erateur $D$ repr\'esente la d\'eriv\'ee covariante pour la restriction de la connexion $A$ \`a $\Sigma$. On constate facilement que le moment conjugu\'e \`a la partie spatiale de la connexion est
\beq
B^a = \eps^{abc}\,B_{bc},
\ee
ce qui conduit aux crochets de Poisson fondamentaux,
\beq
\bigl\{ B^a_i(x), A_b^j(y)\bigr\} = \delta^a_b\,\delta^j_i\,\delta^{(3)}(x-y),
\ee
que nous avons exprim\'e \`a l'aide d'une base orthonorm\'ee de $\alg$, indic\'ee par des lettres du milieu de l'alphabet $i,j$.

De mani\`ere \'equivalente, si $P_{\vert\Sigma}$ est la restriction du fibr\'e $P$ \`a la surface $\Sigma$, l'espace des phases est constitu\'e par l'ensemble des connexions $\calA$ sur ce fibr\'e et des $(n-2)$-formes \`a valeurs dans $\Ad(P_{\vert\Sigma})$. Il s'identifie \`a l'espace cotangent $T^*\calA$ muni de la structure symplectique naturelle
\beq
\omega\bigl( (\delta A,\delta B),(\delta A',\delta B')\bigr) = \int \tr\bigl(\delta A\wedge\delta B' - \delta A'\wedge \delta B\bigr).
\ee
A ce stade, c'est donc le m\^eme espace des phases que dans la th\'eorie de Yang-Mills !

Par ailleurs, il appara\^it clairement dans \eqref{hamiltonian bf} que les variables $A_t$ et $B_{at}$ ne sont pas dynamiques. Ce sont des multiplicateurs de Lagrange imposant les parties spatiales des \'equations du mouvement,
\begin{align} \label{gauss law bf}
D_aB^a &= 0,\\
F_{ab} &= 0. \label{flat constraint bf}
\end{align}
Ces \'equations n'impliquant pas de d\'eriv\'ees temporelles, on dit que ce sont des contraintes (primaires en l'occurence) sur l'espace des phases $T^*\calA$. Il appara\^it de plus que le hamiltonien, issu de de la d\'ecomposition \eqref{hamiltonian bf}, n'est qu'une combinaison lin\'eaire de ces contraintes, et doit donc s'annuler ! Il s'agit d'un ph\'enom\`ene assez g\'en\'eral pour les th\'eories invariantes sous les reparam\'etrisations des cordonn\'ees \cite{henneaux-teitelboim} ; c'est en particulier aussi le cas en relativit\'e g\'en\'erale.
Lorsque la d\'ecomposition hamiltonienne g\'en\`ere des contraintes, il faut s'assurer que celles-ci sont pr\'eserv\'ees par l'\'evolution sous l'action du hamiltonien. Puisque le hamiltonien est une combinaison des contraintes, cela revient \`a calculer l'alg\`ebre des contraintes pour les crochets de Poisson. Celle-ci est particuli\`erement simple, car elle refl\`ete les sym\'etries de la th\'eorie. Int\'egrons tout d'abord les contraintes contre des fonctions quelconques pour former
\begin{align}
\calC_G(\Lambda) &= \int dx\ \tr\bigl(\Lambda(x)\,D_aB^a(x)\bigr),\\
\calC_F(N) &= \int dx\ \eps^{abc}\ \tr\bigl(N_c(x)\,F_{ab}(x)\bigr).
\end{align}
Un calcul simple montre que ces contraintes g\'en\`erent, via les crochets de Poisson, des transformations sur les variables canoniques qui correspondent aux parties spatiales des transformations covariantes standards \eqref{gauge transfo},
\beq
\bigl\{ \calC_G(\Lambda), A_a^i(x)\bigr\} = D_a\Lambda^i(x),\qquad \bigl\{ \calC_G(\Lambda), B^a_i(x)\bigr\} = -[\Lambda,B^a]_i(x),
\ee
et des transformations de translation \eqref{translation sym bf},
\beq
\bigl\{ \calC_F(N), A_a^i(x)\bigr\} = 0,\qquad \bigl\{\calC_F(N),B^a_i(x)\bigr\} = -2\,\eps^{abc}\,D_bN_{ci}(x).
\ee
L'ensemble des transformations de jauge covariantes est retrouv\'e en prenant en compte les lois de transformations des multiplicateurs de Lagrange. Le fait que ces contraintes g\'en\`erent les sym\'etries de jauge est confirm\'e par le calcul de l'alg\`ebre des contraintes qui prend la structure attendue,
\begin{align}
\bigl\{ \calC_G(\Lambda), \calC_G(\Lambda')\bigr\} &= \calC_G([\Lambda,\Lambda']),\\
\bigl\{ \calC_G(\Lambda), \calC_F(N)\bigr\} &= \calC_F([\Lambda,N]),\\
\bigl\{ \calC_F(N), \calC_F(N')\bigr\} &= 0,
\end{align}
o\`u la notation $[\cdot,\cdot]$ repr\'esente le commutateur dans $\alg$ (en particulier $[\Lambda,N]_c = [\Lambda,N_c]$). Il s'agit donc de contraintes de premi\`ere classe dans la terminologie de Dirac, i.e. que leurs crochets s'annulent sur la surface d\'efinie par \eqref{gauss law bf} et \eqref{flat constraint bf}. En particulier, les crochets du hamiltonien avec $\calC_G(\Lambda)$ et $\calC_F(N)$ ne font pas appara\^itre de contraintes secondaires.


La prise en compte de contraintes de premi\`ere classe en g\'eom\'etrie symplectique se fait en deux \'etapes. Tout d'abord, il faut restreindre l'espace des phases, ici $T^*\calA$, \`a la portion satisfaisant les contraintes, \eqref{gauss law bf} et \eqref{flat constraint bf}. Puisque les contraintes commutent au sens des crochets de Poisson, leurs flots laissent ce sous-espace invariant. Finalement, l'espace des phases r\'eduit est form\'e par les orbites correspondantes : autrement dit, une configuration physique de l'espace des phases r\'eduit est une solution des contraintes \`a des transformations de jauge pr\`es.

La r\'eduction par rapport \`a la contrainte dite de Gau\ss, $D_aB^a = 0$, conduit comme dans les th\'eories de jauge usuelles (Yang-Mills) \`a l'espace $T^*(\calA/\calG)$. Puis la contrainte $F_{ab}=0$ nous demande de ne regarder que l'ensemble des connexions plates $\calA_0$, tout en identifiant les moments conjugu\'es $B^a$ qui diff\`erent d'un terme du type $\eps^{abc}D_b\eta_c$. Puisque pour une connexion plate l'op\'erateur $D$ permet de d\'efinir des classes de cohomologie, ces derniers apparaissent comme des \'el\'ements de la seconde classe de cette cohomologie. On montre finalement \cite{horowitz-bf} que l'espace des phases r\'eduit, form\'e des connexions plates modulo l'action de $\calG$ et des \'el\'ements de la seconde classe de cohomologie, poss\`ede la structure d'un espace cotangent, $T^*(\calA_0/\calG)$.

Notons que la d\'efinition pr\'ecise de l'espace quotient $\calA/\calG$ fait intervenir quelques subtilit\'es qui naturellement se retrouvent dans les approches discr\`etes \`a la quantification. En particulier, dans les mod\`eles topologiques de mousses de spins de type BF, ces subtilit\'es doivent \^etre prises en compte dans l'\'etude des propri\'et\'es de finitude/divergence des amplitudes \cite{barrett-PR, cell-homology, twisted-homology} (visant \`a renormaliser les mousses de spins) ! L'id\'ee est que toutes les connexions n'ont pas le m\^eme type d'orbite sous l'action de $\calG$. En regardant les connexions ayant le plus petit type de groupe d'isotropie, on peut voir qu'il existe un ensemble ouvert et dense de $\calA/\calG$ qui est une vari\'et\'e lisse de dimension infinie. Les points manquants sont dus aux connexions ayant plus de sym\'etries et dites r\'eductibles. N\'eanmoins, on peut montrer que l'ensemble des connexions admet une stratification en sous-vari\'et\'es lisses ayant chacune un type d'orbite donn\'e.

Il est possible de donner, assez qualitativement au moins, une description de l'espace r\'eduit $\calA_0/\calG$. Une connexion plate se caract\'erise par le fait que ses holonomies le long de courbes ferm\'ees ne d\'ependent que des classes d'homotopie des courbes. Autrement dit, une connexion plate d\'efinit un morphisme du groupe fondamental $\pi_1(\Sigma)$ dans le groupe de structure $G$. Sous une transformation de jauge, ce morphisme se trouve conjugu\'e par la valeur de la transformation au point utilis\'e pour d\'efinir $\pi_1(\Sigma)$. Ainsi, $G$ agissant par conjugaison sur l'espace des homomorphismes de $\pi_1(\Sigma)$ dans $G$, nous avons,
\beq
\calA_0/\calG \subseteq \Hom\bigl(\pi_1(\Sigma),G\bigr)/G.
\ee
L'espace $\Hom(\pi_1(\Sigma),G)/G$ est connu comme l'espace de modules des $G$-fibr\'es plats sur $\Sigma$, ou l'espace des classes de conjugaison des repr\'esentations de $\pi_1(\Sigma)$ dans $G$. Pour $G$ un groupe matriciel et $\Sigma$ compacte, c'est une vari\'et\'e alg\'ebrique r\'eelle. Malgr\'e tout, comme pour $\calA/\calG$, il y a g\'en\'eriquement des singularit\'es, mais il admet une structure stratifi\'ee en sous-vari\'et\'es lisses (de m\^eme pour $\calA_0/\calG$). Cet espace a \'et\'e tr\`es largement \'etudi\'e lorsque $\Sigma$ est une surface ferm\'ee orientable, o\`u il admet une structure symplectique naturelle (voir le travail fondateur de Goldman \cite{goldman}). Cette approche est particuli\`erement efficace dans la r\'esolution de Yang-Mills en 2d (par int\'egrales de chemins \cite{witten-2d-YM}), et de la th\'eorie de Chern-Simons/gravit\'e 3d (quantification canonique, \cite{witten-3d-gravity}). Il semble aussi que ce soit une bonne approche pour \'evaluer les divergences dans les mod\`eles de mousses de spins topologiques, comme le sugg\`ere fortement le travail de Barrett et Naish-Guzman \cite{barrett-PR}. C'est le point de vue que j'ai r\'ecemment d\'evelopp\'e avec M. Smerlak dans \cite{cell-homology, twisted-homology}.

\subsection*{En dimension 3}

En trois dimensions, o\`u cette th\'eorie d\'ecrit la gravit\'e, il est possible d'extraire du multiplicateur imposant les contraintes $F_{ab} = 0$ les variables de lapse et de shift. Cela permet de d\'ecomposer ces contraintes en une contrainte qui g\'en\`erent les diff\'eomorphismes de $\Sigma$ et une contrainte dite scalaire \cite{thiemann-3d}. Cela rend les contraintes de la relativit\'e g\'en\'erale en dimension 3 et en dimension 4 formellement identiques. Naturellement il n'est pas possible de faire de m\^eme \`a partir de la th\'eorie BF en dimension 4 dans la mesure o\`u les contraintes de la th\'eorie topologique $F_{ab}=0$ sont alors plus fortes que celles associ\'ees aux diff\'eomorphismes en relativit\'e g\'en\'erale.

Lorsque le champ $B$ est non-d\'eg\'en\'er\'e, il s'interpr\^ete comme une co-triade que nous noterons $e$. Le multiplicateur imposant la courbure nulle est $e_t$, et le moment conjugu\'e \`a la connexion est
\beq \label{3d moment}
B^a_i = \eps^{ab}\,\delta_{ij}\, e_b^j.
\ee
Introduisons le vecteur normal dans l'espace interne :
\beq
n^i = \eps^{ijk}\,\eps_{ab}\,B^a_j\,B^b_k = \bigl(\vec{e}_1\times \vec{e}_2\bigr)^i,
\ee
en utilisant la notation vectorielle pour les indices internes. On d\'efinit \'egalement le vecteur de courbure, $F^i = \eps^{ab}F^i_{ab}$. On peut alors remarquer que la norme du vecteur $\vec{n}$ est pr\'ecis\'ement le d\'eterminant de la 2-m\'etrique induite par $e$ sur $\Sigma$, $\lvert \vec{n}\rvert^2 = \det({^2}g)$. Par ailleurs, l'orthogonalit\'e avec $B^a$, i.e. $n^iB^a_i = 0$, permet de d\'ecomposer la contrainte de courbure nulle en deux types de termes :
\beq
\vec{e}_t\cdot\vec{F} = N^a\,V_a + NH,
\ee
o\`u $N^a$ et $N$ sont respectivement le shift et le lapse, d\'efinis par
\beq
N^a = \f{1}{\det({^2}g)}\eps^i_{\phantom{i}jk}\,B^a_i\,e_t^j\,n^k, \qquad N = \f{1}{\sqrt{\det({^2}g)}} \vec{e}_t\cdot\vec{n}.
\ee
Cette r\'e\'ecriture fait appara\^itre la contrainte vectorielle $V_a$, qui impose l'invariance sous les diff\'eomorphismes de $\Sigma$, et la contrainte scalaire $H$ qui n'est autre que la projection de $F$ le long de la direction normale $\vec{n}$,
\begin{align}
V_a &= B^b_i\,F_{ab}^i = \vec{e}_a\cdot\vec{F},\\
H &= \f{1}{2\sqrt{\det({^2}g)}}\eps^{i}_{\phantom{i}jk}\,B^a_j\,B^b_k\,F^i_{ab} = \f{\vec{n}}{\lvert\vec{n}\rvert}\cdot\vec{F}.
\end{align}

\section{Relativit\'e g\'en\'erale}

L'analyse hamiltonienne des actions de Holst ou Plebanski est relativement lourde, bien que directe, du fait de l'algorithme des contraintes de Dirac. En particulier, cela n'a \'et\'e fait qu'assez r\'ecemment pour l'action de Plebanski $\tl{S}_{\mathrm{Pl}}$ dans \cite{henneaux-pleb} et pour $S_{\mathrm{Pl}}$ dans \cite{alexandrov-hamiltonian-pleb}. L'analyse hamiltonienne de l'action de Holst a \'et\'e effectu\'ee par Barros dans \cite{barros-hamiltonian} et approfondie dans un formalisme compl\`etement $\SO(4)$-covariant par Alexandrov \cite{alexandrov-covariant-holst}. Ces approches sont naturellement \'equivalentes entre elles, voir en particulier \cite{alexandrov-roche}. Par ailleurs l'approche initiale au param\`etre d'Immirzi s'est faite par une transformation canonique sur l'espace des phases de l'action de Hilbert-Palatini, par Barbero \cite{barbero-immirzi1,barbero-immirzi2}. Bien s\^ur, l'approche de Barbero fut d\'ecisive pour la gravit\'e \`a boucles \emph {lorentzienne} en ce sens qu'elle permet de rester dans un formalisme r\'eel, par opposition au formalisme self-dual d'Ashtekar qui introduisait une complexification de la th\'eorie, tout en gardant la plupart des avantages offerts par les variables d'Ashtekar (le formalisme self-dual dans la th\'eorie riemanienne que nous pr\'esentons ne n\'ecessite naturellement pas de complexification). Nous ne pr\'esentons pas ici les d\'etails de l'analyse hamiltonienne qui se trouvent dans les r\'ef\'erences ci-dessus, mais seulement les points essentiels. Notons enfin les travaux connexes de Perez et ses collaborateurs, qui ont \'etudi\'e l'ajout de termes topologiques \`a l'action de Holst dans \cite{perez-rezende} et l'action topologique obtenue dans la limite $\gamma\rightarrow0$ \cite{perez-liu-topo}.

Rappelons \`a toutes fins utiles les bases usuelles du formalisme hamiltonien pour ce type d'action, telles que l'on peut les trouver dans \cite{ashtekar-book-lectures}. Supposons que $M$ est de la forme $\Sigma\times\R$, et prenons une coordonn\'ee $t$ de \og temps \fg{} sur $\R$ qui indexe les surfaces $\Sigma_t\sim\Sigma\times\{t\}$. Notons $\pp_t$ le champ de vecteurs associ\'e sur $M$, $n$ le champ de vecteurs orthonormal aux surfaces $\Sigma_t$ (pour la m\'etrique riemanienne provenant de la cot\'etrade, $g(n,n)=1$). On peut alors d\'ecomposer de mani\`ere unique le champ de vecteurs $\pp_t$ et sa contraction avec la cot\'etrade,
\begin{equation}
\pp_t = N n + \vec{N}\qquad \Rightarrow\qquad e_t^I = N n^I + e_a^I\,N^a,
\end{equation}
en prenant le vecteur $\vec{N}$ orthogonal \`a $n$. Les quantit\'es $N$ et $\vec{N}$ sont respectivement appel\'es le \emph{lapse} et le \emph{shift}. Ils permettent de d\'ecrire l'\'evolution \og temporelle \fg{} selon $t$ de mani\`ere adapt\'ee \`a la m\'etrique, en une \'evolution normale et une \'evolution tangentielle. Le vecteur $n^I$ est le vecteur interne d'imitation de $n$, $n^I = e^I(n)$. Les indices latins $a,b,c$ du d\'ebut de l'alphabet font r\'ef\'erence \`a un jeu de coordonn\'ees quelconques sur $\Sigma$.

\subsection*{Contraintes de simplicit\'e hamiltonienne}

La premi\`ere \'etape importante est de voir qu'il existe des contraintes sur le moment conjugu\'e \`a la partie spatiale de la connexion $A_a^{IJ}$ pour l'action de Holst \eqref{holst action}. En effet, on trouve facilement que celui-ci est donn\'e par :
\beq \label{gamma-moment}
P^{(\gamma)a}_{IJ} = P^a_{IJ} + \gamma\mone(\star P^a)_{IJ},
\ee
o\`u $P^a_{IJ}$ est le moment conjugu\'e \`a la connexion lorsque $\gamma\rightarrow\infty$,
\beq \label{momentumP}
P^a_{IJ} = \f12\eps^{abc}\,\eps_{IJKL}\,e_b^K\,e_c^L = \sqrt{q} \bigl(n_I^{\phantom{a}} e^a_J - n_J^{\phantom{a}} e^a_{I}\bigr).
\ee
On utilise ici l'inverse de la cot\'etrade, et la racine carr\'ee du d\'eterminant de la 3-m\'etrique sur $\Sigma$, $\sqrt{q}$. Mais il est clair que l'on ne peut pas utiliser $P^a_{IJ}$ comme une variable libre et ind\'ependante, car tous les vecteurs \`a valeurs dans $\so(4)$, du type $X^a_{IJ}$, ne sont pas de la forme \eqref{momentumP}. A la vue de la section pr\'ec\'edente, on comprend que le moment doit venir de la projection d'une 2-forme \emph{simple} sur $\Sigma$, ce qui nous permet, par projection de \eqref{simplicity pleb2}, de deviner les contraintes \`a satisfaire :
\beq \label{hamiltonian simplicity}
C^{ab} = \eps^{IJKL}\ P^a_{IJ}\,P^b_{KL}.
\ee
Il s'agit de contraintes primaires \`a prendre en compte dans le hamiltonien. On trouve par ailleurs par un calcul standard \cite{ashtekar-book-lectures} que celui-ci est une combinaison lin\'eaire de contraintes,
\beq
H = \int d^3x\ \bigl( -A_t^{IJ}\,G_{IJ} + N^a\,V_a - N\,S + \psi_{ab}\,C^{ab}\bigr),
\ee
avec
\begin{align}
G_{IJ} &= D_a P^{(\gamma)a}_{IJ},\\
V_a &= P^{(\gamma)b}_{IJ}\,F_{ab}^{IJ},\\
S &= \f1{\sqrt{q}}\,P^{(\gamma)a}_{IJ}\,P^{bJ}_{\phantom{bJ}K}\,F_{ab}^{IK},
\end{align}
du fait que le lapse, le shift et les composantes $A_t^{IJ}$ ne sont pas des variables dynamiques. En calculant les crochets de Poisson, on montre que les contraintes $G_{IJ}$ constituent la loi de Gau\ss : elles engendrent les transformations de jauge $\so(4)$. On peut \'egalement montrer que sur la surface des contraintes, $V_a$ et $S$ g\'en\`erent respectivement les diff\'eomorphismes de $\Sigma$ et les diff\'eomorphismes dans la direction normale $n$ (modulo l'action de $G_{IJ}$), de la m\^eme mani\`ere que dans le formalisme ADM. Elles sont appel\'ees contraintes vectorielle et scalaire (ou hamiltonienne pour cette derni\`ere, par un l\'eger abus de langage qui nous montre bien o\`u se situent les difficult\'es). Ces trois ensembles de contraintes forment entre elles un syst\`eme de \emph{premi\`ere classe} (et il n'y en a pas d'autres). Autrement dit, le flot de ces contraintes sur la surface des contraintes est pure jauge, et identifie diff\'erentes solutions sous l'action de $\SO(4)$ et des diff\'eomorphismes.

L'alg\`ebre des contraintes (que nous donnons plus bas avec le fixage de jauge usuel de la gravit\'e \`a boucles) montre en outre qu'il existe des contraintes secondaires, dues \`a la stabilisation des contraintes de simplicit\'e \eqref{hamiltonian simplicity} sous l'effet de $C$, de la forme $PP(DP)$. Ces contraintes secondaires forment un syst\`eme de contraintes de \emph{seconde classe} avec les $C^{ab}$ \cite{barros-hamiltonian, alexandrov-covariant-holst} \footnote{Dans le formalisme complexe d'Ashtekar, il est int\'eressant que de noter ces contraintes secondaires se retrouvent sous la forme de contraintes de r\'ealit\'e.}. L'\'elimination des contraintes de seconde classe de l'alg\`ebre peut se faire en passant aux crochets de Dirac. Ceux-ci ont \'et\'e explicitement calcul\'es dans \cite{alexandrov-covariant-holst}. Dans cette situation, les variables canoniques initiales perdent leur statut privil\'egi\'e. En particulier, la principale difficult\'e de cette approche est que la connexion ne commute plus avec elle-m\^eme pour les crochets de Dirac. Cela est assez g\^enant, et m\^eme r\'edhibitoire jusqu'\`a pr\'esent, pour une quantification de type boucles, qui utilise les r\'eseaux de spins comme repr\'esentation de l'alg\`ebre des holonomies d'une connexion.

\subsection*{Dans la jauge \og temps\fg}

Afin de comprendre l'origine de la non-commutativit\'e de la connexion et la solution propos\'ee \`a l'origine de la gravit\'e \`a boucles, il nous faut \'etudier l'influence des contraintes de seconde classe sur les variables canoniques et la structure symplectique. Pour cela, on utilise le fait que l'on conna\^it les solutions de $C^{ab}$, qui ne sont autres que \eqref{momentumP} ! En introduisant une quantit\'e $E^a_i$ d\'efinie comme la partie \og boost\fg{} du moment $P$ :
\beq
E^a_i := P^a_{0i} = \f12\eps^{abc}\,\eps_{ijk}\, e_b^j\,e_c^k,
\ee
la partie \og rotation spatiale\fg{} de $P$ s'\'ecrit
\beq
P^a_{ij} = n_i^{\phantom{a}}E^a_j - n_j^{\phantom{a}} E^a_i.
\ee
Les indices $i,j=1,2,3$ sont ici les indices des rotations 3d de $\SO(4)$ selon la d\'ecomposition standard. S'il est possible de r\'eexprimer la structure symplectique et les contraintes en fonction de $E^a_i$ et du vecteur $n^i$ \cite{barros-hamiltonian}, il est tr\`es avantageux d'effectuer une fixation de jauge partielle, en consid\'erant $n^I\in S^3$ fix\'e, disons \`a $n^I =(1,0,0,0)$. Ce choix de jauge est d\'enomm\'e jauge \og temps\fg{} (en anglais, time gauge) et est \'equivalent \`a $e_a^0 = 0$. Avec une telle fixation de jauge, les seules composantes non-nulles du moment sont les $P^a_{0i} = E^a_i$ qui forment de la sorte une triade pour la 3-m\'etrique $q_{ab}$ sur $\Sigma$,
\beq \label{triade 3metrique}
q\,q^{ab} = \delta^{ij}\ E^a_i\,E^b_j.
\ee
Nous devons alors r\'esoudre la partie \og boost\fg{}, i.e. relative \`a notre choix de $n^I$, des contraintes $G_{IJ}$, ce qui nous permet dans le m\^eme temps de r\'esoudre les contraintes secondaires de seconde classe. Le r\'esultat est que la partie boost de la connexion, $A^{0i}_a$, reste libre, tandis que la partie rotation 3d, $A^{ij}_a$ est compl\`etement d\'etermin\'ee par la triade $E^a_i$. On d\'efinit 
\beq
K^i_a := A^{0i}_a,
\ee
qui s'interpr\^ete g\'eom\'etriquement comme la courbure extrins\`eque d\`es lors que la connexion $A$ sur $M$ satisfait la condition de torsion nulle, $d_A e=0$. Quant \`a la partie rotationnelle de la connexion, les contraintes secondaires et la contrainte de Gau\ss{} la contraignent \`a \^etre la connexion compatible avec la triade sur $\Sigma$, au sens suivant :
\beq
A_a^{ij} = \eps^{ij}_{\phantom{ij}k}\ \Gamma_a^k(E),
\ee
o\`u $\Gamma(E)$ est d\'efinie par l'\'equation
\beq
\eps^{abc}\ \bigl(\pp^{\phantom{i}}_{a} e_{b}^i + \eps^i_{\phantom{i}jk}\ \Gamma_{a}^j\,e_{b}^k\bigr) = 0.
\ee
Ainsi, les termes de l'action de Holst lin\'eaires en la d\'eriv\'ee temporelle se r\'e\'ecrivent :
\beq
P^{(\gamma)a}_{IJ}\,\pp_t A_a^{IJ} = \gamma\mone\ E^a_i\,\pp_tA_a^i,
\ee
$A_a^i$ \'etant une connexion $\SU(2)$ sur $\Sigma$, appel\'ee la \emph{connexion d'Ashtekar-Barbero} et d\'efinie comme
\beq
A_a^i := \Gamma_a^i(E) +\gamma K_a^i.
\ee
Comme on le voit ais\'ement, les variables canoniques de l'action de Hilbert-Palatini, i.e. lorsque $\gamma\rightarrow\infty$, sont les couples $(E^a_i,K_a^i)$. De ce point de vue, le passage de $K_a^i$ \`a la connexion d'Ashtekar-Barbero (\`a un scaling par $\gamma$ pr\`es) est une transformation canonique \cite{barbero-immirzi2}. Le fait remarquable est naturellement que les crochets $\{A_a^i,A_b^j\}$ s'annulent \'etant donn\'ee la d\'efinition de la connexion d'Ashtekar-Barbero. Cela nous donne comme seuls crochets non-triviaux :
\beq
\{ E^a_i(x), A_b^j(y)\} = \gamma\,\delta^a_b\,\delta^j_i\,\delta^{(3)}(x-y).
\ee
Dans ces nouvelles variables, les contraintes de Gau\ss{} $\SU(2)$, vectorielle et scalaire gardent une forme relativement similaire \`a la pr\'ec\'edente ($\SO(4)$) malgr\'e l'utilisation de la nouvelle connexion,
\begin{align}
&G_i = D_aE^a_i := \pp_aE^a_i + \eps_{ij}^{\phantom{IJ}k} A_a^j E^a_k,\\
&V_a = E^b_i F_{ab}^i + \textrm{termes proportionnels \`a la contrainte de Gau\ss},\\
&\begin{aligned} S =&\ \f{E^a_i E^b_j}{\sqrt{\det E}} \bigl(\eps^{ij}_{\phantom{ij}k} F_{ab}^k + 2(1-\gamma^2) K_{[a}^i K^j_{b]}\bigr)  \\
&\qquad+ \textrm{termes proportionnels \`a la contrainte de Gau\ss}. \end{aligned}
\end{align}
Nous avons not\'e : $F_{ab}^i = \pp_aA_b^i - \pp_bA_a^i + \eps^i_{\phantom{i}jk}A_a^j A_b^k$, la courbure de la connexion d'Ashtekar-Barbero, et utilis\'e des crochets droits pour l'antisym\'etrisation. Ces contraintes forment une alg\`ebre de premi\`ere classe. Pour le voir, on les int\`egre sur $\Sigma$ contre des fonctions test,
\begin{gather}
\calC_G(\Lambda) = \int_{\Sigma} d^3x\ \Lambda^i\,G_i,\qquad \calC_{\mathrm{Diff}} (\vec{N})= \int_{\Sigma} d^3x\ N^a\bigl(E^b_i F_{ab}^i - A_a^i G_i\bigr),\\
\textrm{et}\qquad \calC_S(N) = \int_{\Sigma} d^3x\ N\,\f{E^a_i E^b_j}{\sqrt{\det E}} \bigl(\eps^{ij}_{\phantom{ij}k} F_{ab}^k + 2(1-\gamma^2) K_{[a}^i K^j_{b]}\bigr).
\end{gather}
On v\'erifie que l'alg\`ebre est bien celles des transformations de jauge $\su(2)$ et des diff\'eomorphismes,
\begin{align}
&\{\calC_G(\Lambda), \calC_G(\Lambda')\} = \calC_G([\Lambda,\Lambda']),\\
&\{\calC_G(\Lambda), \calC_{\mathrm{Diff}}(\vec{N})\} = -\calC_G(\mathcal{L}_{\vec{N}}\Lambda),\\
&\{\calC_{\mathrm{Diff}}(\vec{N}), \calC_{\mathrm{Diff}}(\vec{N}')\} = \calC_{\mathrm{Diff}}(\mathcal{L}_{\vec{N}}\vec{N}'),\\
&\{\calC_G(\Lambda), \calC_S(N)\} = 0,\\
&\{\calC_{\mathrm{Diff}}(\vec{N}), \calC_S(M)\} = -\calC_S(\mathcal{L}_{\vec{N}}M).
\end{align}
Le dernier crochet est responsable du fait que l'alg\`ebre est ouverte au sens BRST, car il fait appara\^itre des \emph{fonctions} de structure,
\beq
\{\calC_S(N),\calC_S(M)\} = \calC_{\mathrm{Diff}}(\vec{S}) + \textrm{termes proportionnels \`a la contrainte de Gau\ss},
\ee
pour un vecteur $S^a = E^a_i E^{bi} (N\pp_bM-M\pp_bN)/\lvert\det(E)\rvert$. Pour finir cette pr\'esentation des contraintes, notons que la triade et la connexion d'Ashtekar-Barbero se transforment de mani\`ere naturelle sous la contrainte de Gau\ss,
\beq
\{A_a^i,\calC_G(\Lambda)\} = -D_a\Lambda^i,\qquad \{E^a_i,\calC_G(\Lambda)\} = \eps_{ij}^{\phantom{ij}k}\Lambda^j E^a_k,
\ee
et sous les 3-diff\'eomorphismes,
\beq
\{A_a^i,\calC_{\mathrm{Diff}}(\vec{N})\} = \calL_{\vec{N}}A_a^i,\qquad \{E^a_i,\calC_{\mathrm{Diff}}(\vec{N})\} = \calL_{\vec{N}}E^a_i.
\ee
Les \'equations de Hamilton d\'ecrivent l'\'evolution des variables canoniques en fonction du param\`etre arbitraire $t$, en posant que les d\'eriv\'ees $\pp_t A_a^i$ et $\pp_t E^a_i$ sont obtenues en prenant les crochets de $A^i_a$ et $E^a_i$ avec le hamiltonien, $\calC_G(\Lambda) + \calC_{\mathrm{Diff}}(\vec{N}) + \calC_S(N)$.

Pour plus de d\'etails sur le formalisme r\'eel $\SU(2)$, je renvoie aux excellentes introductions \`a la gravit\'e quantique en boucles \cite{perez-intro-lqg, ashtekar-status-report, thiemann-lectures-lqg, thiemann-book}.

R\'esumons les \'etapes cl\'es de cette analyse, pour comparaison avec la th\'eorie BF. Les contraintes de simplicit\'e de Plebanski se retrouvent en formalisme hamiltonien comme des contraintes de seconde classe ; elles contraignent la forme du moment conjugu\'e \`a la connexion et n'engendrent pas des transformations de jauge. Par une fixation de jauge ad\'equate, on voit que les contraintes de seconde classe obligent \`a exprimer la partie \og rotation 3d\fg{} interne de la connexion en termes de la triade sur $\Sigma$. Comme cette triade est conjugu\'ee est la partie \og boost\fg{} de la connexion, cela explique que la partie rotation et la partie boost de la connexion $\SO(4)$ ne commutent plus pour les crochets de Dirac\footnote{Les crochets de Dirac donnent les crochets dans l'espace des phases r\'eduit par les contraintes de seconde classe, sans avoir \`a r\'esoudre explicitement celles-ci, et donc sans avoir \`a chercher de param\'etrisation de cet espace r\'eduit.}. Notons \`a ce propos les travaux d'Alexandrov, souvent cit\'es comme {\it covariant loop quantum gravity}, qui replace les discussions ci-dessus dans un contexte compl\`etement $\SO(4)$-covariant. Cela permet d'adopter un point de vue qui va au-del\`a de la fixation de jauge usuelle, et ainsi d'\'etudier diff\'erents choix possibles de connexions \cite{alexandrov-choice-connection} aux niveaux classique et quantique. Pour une revue de cette approche et des r\'ef\'erences suppl\'ementaires, voir \cite{livine-clqg}. Dans le cadre de travail discret et $\SO(4)$-covariant des mousses de spins, nous verrons des choses similaires dues \`a des contraintes de simplicit\'e discr\`etes. En particulier nous trouverons un analogue au champ de normales internes $n^I$ qui jouera un r\^ole important.

Pour l'action de Hilbert-Palatini, nous trouvons apr\`es r\'esolution des contraintes de seconde classe que c'est la courbure extrins\`eque $K^i_a$ qui se retrouve conjugu\'e \`a la triade \cite{ashtekar-book-lectures}. Puisque la courbure extrins\`eque se transforme sous l'adjointe de $\SU(2)$, il semble que tout contact avec l'espace des phases de Yang-Mills, et {\it a fortiori} de la th\'eorie BF, soit rompu. C'est ici que l'introduction du param\`etre d'Immirzi est fondamentale, en ce qu'elle permet de se ramener \`a une variable canonique qui soit une connexion. Cela est tout \`a fait essentiel \`a la gravit\'e en boucles qui utilise le sens g\'eom\'etrique profond d'une connexion comme outil de transport parall\`ele. 

Malgr\'e tout, il serait dommage d'avoir obtenu des simplifications au niveau cin\'ematique, i.e. de l'espace des phases non-r\'eduit, mais de se retrouver avec des contraintes davantage compliqu\'ees. Notons donc que les contraintes sont compl\`etement polynomiales except\'e le terme $(\det(E))\mone$ dans la contrainte scalaire. Pour une discussion sur la pertinence de ce terme qui est tel que $S$ est de densit\'e 1, je renvoie au livre de Thiemann, \cite{thiemann-book}.

\chapter{Quantifier des th\'eories de jauge}

Face \`a des th\'eories avec contraintes, nous disposons de deux strat\'egies diff\'erentes pour la quantification. Il est possible de quantifier l'espace des phases r\'eduit. C'est potentiellement l'approche la plus simple, \`a condition que les contraintes et leur sens physique soit bien ma\^itris\'es. Pour la th\'eorie BF, cela revient \`a une quantification sur $T^*(\calA_0/\calG)$, $\calA_0$ \'etant le sous-espace des connexions plates, qui peut \^etre plus ou moins explicit\'ee selon $\Sigma$, \cite{horowitz-bf} (et particuli\`erement efficace en gravit\'e 3d, \cite{witten-3d-gravity}, o\`u l'on voit tout de suite que la th\'eorie a un nombre fini de degr\'es de libert\'e). Mais il y a peu de chances que cela nous aide et puisse servir de bases pour des th\'eories plus compliqu\'ees comme la relativit\'e g\'en\'erale en dimension 4, ayant un nombre infini de degr\'es de libert\'e. Nous allons donc suivre le programme de Dirac, consistant \`a quantifier l'espace des phases non-r\'eduit, puis contraindre la th\'eorie au niveau quantique (voir par exemple \cite{matschull-dirac-program} pour une introduction).

Nous nous contenterons de traiter le cas o\`u le groupe de structure est $G=\SU(2)$ (le cas $\Spin(4)$ s'obtient imm\'ediatement en prenant deux copies de $\SU(2)$), le lecteur pourra trouver g\'en\'eralisations et d\'etails dans la th\`ese de Livine \cite{livine-these}, en particulier pour le cas physiquement aussi int\'eressant de $\SL(2,\C)$. Nous avons obtenu dans l'analyse hamiltonienne de la th\'eorie BF $\SU(2)$ et de la relativit\'e g\'en\'erale le m\^eme espace des phases (au param\`etre d'Immirzi pr\`es), dont les variables canoniques consistent en une connexion $\SU(2)$ sur $\Sigma$, $A_a^i$, et son moment conjugu\'e, qu'on notera maintenant $E^a_i$, se transformant sous l'adjointe. C'est exactement le m\^eme espace des phases que dans la th\'eorie de Yang-Mills $\SU(2)$ ! Par analogie avec le cas ab\'elien, l'\'electrodynamique, on appelle parfois le champ $E^a_i$ champ \'electrique. De m\^eme, dans l'analyse de l'action de Holst avant la prise en compte des contraintes de seconde classe et la fixation de jauge partielle, l'espace des phases est similaire mais pour le groupe $\SO(4)$, avec les paires canoniques $(A_a^{IJ}, P^{a(\gamma)}_{IJ})$. Ainsi, nous souhaitons b\'en\'eficier d'une proc\'edure de quantification commune \`a ces situations (permettant notamment d'\'etudier leurs diff\'erences -- dues aux contraintes de simplicit\'e, \`a la fixation de jauge, et bien s\^ur la dynamique ! -- au niveau quantique).

Pour cela nous devrons former une alg\`ebre de fonctions sur l'espace des phases, \`a repr\'esenter sur un espace de Hilbert dit cin\'ematique\footnote{On utilisera aussi le terme cin\'ematique pour d\'esigner l'espace de Hilbert obtenu apr\`es imposition de l'invariance de jauge due \`a $G$.}. Nous choisirons une polarisation o\`u la variable de configuration est la connexion, \`a la Schr\H{o}dinger. L'espace de Hilbert cin\'ematique sera muni d'un produit scalaire, gr\^ace \`a la mesure d'Ashtekar-Lewandowski $\mu_{\rm AL}$,
\beq
\langle \phi \,|\, \psi\rangle = \int d\mu_{\rm AL}(A)\ \overline{\phi}(A)\,\psi(A).
\ee
Sur cet espace (ou son dual), on doit alors chercher les \'etats physiques de la th\'eorie consid\'er\'ee. Ils sont d\'efinis comme les \'etats annihil\'es par la promotion en op\'erateurs des contraintes de premi\`ere classe (nous discuterons les contraintes de seconde classe dans les parties mousses de spins). Pour la relativit\'e g\'en\'erale, cela donne formellement :
\begin{alignat}{3}
&\widehat{G_i}&\,&\psi(A) &\,=\,& 0,\\
&\widehat{V_a}&\,&\psi(A) &\,=\,& 0,\\
&\widehat{S}&\,&\psi(A) &\,=\,& 0.
\end{alignat}
Au niveau quantique, la prise en compte des sym\'etries de jauge se fait donc en une seule \'etape, en comparaison de la r\'eduction au niveau classique qui implique de se placer sur la surface des contraintes puis de quotienter par les orbites des transformations. La contrainte de Gau\ss{} et la contrainte vectorielle jouissent d'une interpr\'etation simple, comme transformations de jauge $\SU(2)$ et diff\'eomorphismes de $\Sigma$. Cela permet de les imposer en moyennant les \'etats cin\'ematiques le long de leurs orbites. En revanche, la contrainte scalaire r\'esiste toujours a une description satisfaisante.

N\'eanmoins, la prise en compte des contraintes de Gau\ss{} et vectorielle est une r\'eussite, rigoureuse, et de plus pertinente pour toute th\'eorie de jauge invariante sous les diff\'eomorphismes. Un des r\'esultats majeurs \cite{lost} est que le choix qui est fait de l'alg\`ebre \`a repr\'esenter conduit \`a une unique repr\'esentation invariante sous les diff\'eomorphismes ! L'approche d\'evelopp\'ee a en particulier l'avantage de ne pas faire intervenir de m\'etrique privil\'egi\'ee et fix\'ee, et respecte de mani\`ere naturelle l'invariance fondamentale de la th\'eorie sous les diff\'eomorphismes. Un des aspects essentiels de cette quantification est, par contraste avec la th\'eorie quantique des champs usuelle o\`u les \'etats (de Fock) sont particulaires donc ponctuels, que les excitations \'el\'ementaires sont port\'ees par des \emph{graphes}, donc de dimension 1, appel\'es \emph{r\'eseaux de spins}. Dans la suite de cette th\`ese, nous travaillerons la plupart du temps sur un graphe fix\'e. Mais il est fondamental de garder \`a l'esprit que la th\'eorie compl\`ete (exacte) contient l'ensemble des (classes d'\'equivalence sous diff\'eomorphismes) des graphes ! Ainsi, tout r\'esultat obtenu sur un graphe donn\'e doit : soit se comprendre en termes d'une troncature des degr\'es de libert\'e de la th\'eorie au graphe consid\'er\'e, soit \^etre prolong\'e \`a la th\'eorie compl\`ete en \og sommant\fg{} sur tous les graphes, ce qui peut se r\'ev\'eler difficile. Le cas de la th\'eorie BF est particulier : il s'agit d'une th\'eorie topologique, dont on peut donc capter l'ensemble des degr\'es de libert\'e (fini) \`a l'aide d'un seul graphe, comme une triangulation.

Je pr\'esente la quantification en r\'eseaux de spins \`a la section \ref{sec:spinnet}. Cela a permis l'av\`enement d'une g\'eom\'etrie quantique 3d ! C'est-\`a-dire que l'on peut d\'efinir sur cet espace des \'etats cin\'ematiques des op\'erateurs dont les valeurs propres sont reli\'ees directement aux nombres quantiques caract\'erisant les \'etats de r\'eseaux de spins. Ces op\'erateurs s'interpr\`etent en gravit\'e 4d comme des op\'erateurs d'aire de surfaces, et de 3-volumes, et prennent des valeurs propres discr\`etes ! Je pr\'esente les r\'esultats essentiels \`a la section \ref{sec:operatorsLQG}. Je m'attarde aussi sur le formalisme du t\'etra\`edre quantique, bien adapt\'e \`a l'\'etude des noeuds (4-valents) des r\'eseaux de spins, et qui permet de retrouver tr\`es facilement la 3-g\'eom\'etrie quantique. Cela donne une approche intuitive que j'utilise pour introduire des op\'erateurs d'angles, sans entrer dans les d\'etails de la construction de \cite{major-3dangles}, angles que j'ai \'etudi\'e comme observables d'angles dih\'edraux dans les mousses de spins, section \ref{sec:insertion obsBF}.

\section{Les r\'eseaux de spins} \label{sec:spinnet}

Pour construire l'alg\`ebre des observables cin\'ematiques, au lieu d'int\'egrer les champs sur tout $\Sigma$ contre des fonctions test comme c'est l'usage en th\'eorie des champs, nous utilisons le sens g\'eom\'etrique profond (au sens de la ge\'om\'etrie diff\'erentielle et des th\'eories de jauge) des objets impliqu\'es, la connexion et la 2-forme qui lui est conjugu\'ee. La connexion, en tant que 1-forme localement sur $\Sigma$ s'int\`egre naturellement sur des courbes. Etant donn\'ee une courbe $e$, cela permet de d\'efinir l'holonomie,
\beq
U_e(A) \,=\, P e^{-\int_e A},
\ee
o\`u $P$ d\'esigne le fait que l'exponentielle est ordonn\'ee. Pour une param\'etrisation $x(t)$ de la courbe, $t\in[0,1]$, on a : $\int_e A = \int_0^1 A_a(x(t))\,\f{dx^a}{dt}(x(t))\,dt$. L'holonomie, vue comme un \'el\'ement du groupe $G$, d\'efinit une notion de transport parall\`ele, en permettant de transporter un objet vivant dans une fibre au-dessus de $x(0)$ jusqu'\`a la fibre au-dessus de $x(1)$ par l'action de $U_e(A)$. Dans une transformation de jauge telle que $A^g = gAg\mone + g\,dg\mone$, l'holonomie se transforme \`a gauche au point d'arriv\'ee et \`a droite au point de d\'epart de $e$ :
\beq \label{gauge transfo hol}
U_e(A^g) \,=\, g(x(1))\,U_e(A)\,g(x(0))\mone.
\ee

Avec ces objets, nous allons construire non plus une quantification \`a la Fock, pour des \'etats particulaires (de dimension 0), mais une quantification dite en r\'eseaux de spins, dans laquelle les excitations \'el\'ementaires ont pour support des courbes ! Pour cela nous formons l'alg\`ebre $\cyl$ des fonctions dites \emph{cylindriques} de la connexion $A$. Soit un graphe orient\'e et ferm\'e $\Gamma$, avec $E$ liens ($e_1,\dotsc, e_E$) et $V$ noeuds. L'id\'ee est de sonder la g\'eom\'etrie de la connexion \`a travers un nombre fini de variables : les holonomies le long des liens du graphe. Une fonction cylindrique $\psi_{\Gamma,f}\in\cyl_\Gamma$ est donc form\'ee \`a l'aide d'une fonction $f$ continue \`a valeurs complexes sur $\SU(2)^E$ :
\beq
\psi_{\Gamma,f}(A) \,=\, f\bigl( U_{e_1}(A),\dotsc, U_{e_E}(A)\bigr).
\ee
L'alg\`ebre de ces fonctions poss\`ede un produit scalaire naturel, provenant de la mesure de Haar sur $\SU(2)^E$, $d\mu_\Gamma = \prod_e dg_e$, soit
\beq \label{ps graphe}
\langle \phi_{\Gamma,h} | \psi_{\Gamma,f}\rangle \,=\, \int \prod_{e=1}^E dg_e\ \bar{h}(g_1,\dotsc,g_E)\,f(g_1,\dotsc,g_E).
\ee
En prenant la compl\'etion de cet espace pour la norme associ\'ee \`a ce produit, nous arrivons naturellement \`a l'espace $L^2(\SU(2)^E, d\mu_{\Gamma})$.

Pour construire l'espace de Hilbert \og cin\'ematique\fg{} de la LQG, et en fait adapt\'e \`a toute th\'eorie de jauge invariante sous les diff\'eomorphismes et sans m\'etrique (\og background independent\fg), il faut consid\'erer les fonctions cylindriques sur tous les graphes possibles sur $\Sigma$, $\cyl=\cup_\Gamma\cyl_\Gamma$ ! Cela peut se faire rigoureusement en consid\'erant une limite projective des espaces sur chaque graphe. L'espace de Hilbert qui en r\'esulte peut \^etre vu comme l'espace $L^2(\bar{\calA},d\mu_{\rm AL})$ des fonctionnelles de carr\'e int\'egrable sur $\bar{\calA}$. Ce dernier espace est appel\'e l'espace des \og connexions g\'en\'eralis\'ees\fg, qui sont arbitrairement discontinues mais assignent \`a chaque courbe une holonomie ! La mesure $\mu_{\rm AL}$ est la mesure d'Ashtekar-Lewandowski, provenant naturellement de l'ensemble des mesures $d\mu_\Gamma$, et qui rend orthogonaux les fonctionnelles cylindriques associ\'ees \`a des graphes diff\'erents.

Nous n'aurons pas vraiment besoin dans cette th\`ese de consid\'erer l'int\'egralit\'e de cet espace de Hilbert, m\^eme si c'est bien s\^ur son existence qui justifie en partie l'int\'er\^et des travaux que j'ai pu effectuer ! En fait, dans la perspective de d\'evoiler la g\'eom\'etrie cod\'ee dans les mousses et r\'eseaux de spins, il sera suffisant de tronquer cet espace \`a un graphe fix\'e, soit un nombre fini de variables ! Ce n'est toutefois bien s\^ur qu'en consid\'erant $L^2(\bar{\calA},d\mu_{\rm AL})$ dans sa totalit\'e que l'on peut pleinement appr\'ecier le travail fourni depuis des ann\'ees pour traiter un nombre infini de degr\'es de libert\'e, dans un cadre \og background independent\fg. A cette fin, je renvoie aux excellentes revues \cite{ashtekar-status-report, perez-intro-lqg, thiemann-lectures-lqg, baez-bf-spinfoam}, et au livre \cite{thiemann-book} pour plus de d\'etails, avec force r\'ef\'erences. Aussi, je me concentre maintenant sur la description de l'espace $L^2(\SU(2)^E, d\mu_{\Gamma})$ et de son secteur invariant de jauge $\calH_\Gamma$.

\bigskip

Sur un graphe $\Gamma$ fix\'e, nous invoquons la d\'ecomposition de $L^2(\SU(2))$ sur chaque lien, par le th\'eor\`eme de Peter-Weyl. Cela permet de consid\'erer une base hilbertienne form\'ee de tous les \'el\'ements de matrices des repr\'esentations irr\'eductibles. Celles-ci sont indic\'ees par un spin $j\in\N/2$, et les \'el\'ements de matrices sont les fonctions : $g\mapsto D^{(j)}_{mn}(g) = \langle j, m\lv g\rv j, n\rangle$ en notation bra-ket. Les vecteurs de base sont ainsi des r\'eseaux de spins non-invariants de jauge, caract\'eris\'es par la donn\'ee d'un spin $j_e$ sur chaque lien, et de deux nombres magn\'etiques, $m_e, n_e$, un pour chaque extr\'emit\'es du lien :
\beq \label{non-inv spinnet}
s^{\{j_e\}}_{\Gamma\{m_e,n_e\}} \,:\, (g_1,\cdots,g_E)\,\mapsto\, \prod_{e=1}^E D^{(j_e)}_{m_e n_e}(g_e).
\ee
Ces fonctions sont orthogonales pour le produit \eqref{ps graphe}, et peuvent \^etre normalis\'ees en multipliant chaque \'el\'ement de matrice par $\sqrt{d_{j_e}}$, o\`u $d_j = (2j+1)$ est la dimension de la repr\'esentation de spin $j$.

La contrainte de Gau\ss{} impose classiquement l'invariance de jauge $\SU(2)$ pour l'espace interne. Lorsque que l'on sonde la connexion sur le graphe $\Gamma$, nous avons vu que les transformations de jauge agissent uniquement aux extr\'emit\'es des liens, \eqref{gauge transfo hol}. Autrement dit, sur le graphe $\Gamma$, il s'agit d'une action de $\SU(2)^V$. Ainsi, il est possible de se restreindre au secteur invariant de jauge sur le graphe $\Gamma$ en moyennant simplement sur $\SU(2)^V$ \`a l'aide de la mesure de Haar correspondante (qui est bien invariante par translation),
\beq
\int \prod_{v=1}^V dh_v\ f\bigl(h_{t(e)}\,g_e\,h_{s(e)}\mone\bigr),
\ee
$s(e)$ et $t(e)$ d\'esignant le noeud source et le noeud d'arriv\'ee de chaque lien. De telles fonctions forment l'espace invariant $\calH_\Gamma$,
\beq
\calH_\Gamma = L^2\bigl(\SU(2)^E/\SU(2)^V, d\mu_\Gamma\bigr).
\ee
On peut de plus en exhiber une base orthogonale, dite des r\'eseaux de spins ! Pour cela regardons un noeud, avec des liens entrants $(e_{\rm in})$, et des liens sortants $(e_{\rm out})$. En utilisant le processus de moyenne ci-dessus sur les vecteurs de base non-invariants \eqref{non-inv spinnet}, on peut voir que les int\'egrales entrainent l'insertion \`a chaque noeud d'un tenseur entrelaceur entre les repr\'esentations entrantes et sortantes, i.e. $\otimes_{e_{\rm in}}\calH_{j_e} \rightarrow \otimes_{e_{\rm out}}\calH_{j_e}$, si $\calH_j$ est l'espace de Hilbert de dimension $d_j$ portant la repr\'esentation de spin $j$. Par dualit\'e, cet entrelaceur peut \^etre vu comme un objet dans la repr\'esentation triviale : $\otimes_{e_{\rm in}}\calH_{j_e} \otimes_{e_{\rm out}}\overline{\calH}_{j_e}\rightarrow \C$.

Pour former une base des r\'eseaux de spins, nous avons donc besoin d'une base d'entrelaceurs \`a chaque noeud. Pour un noeud trivalent, il n'existe qu'un seul vecteur invariant $\lv \iota_{j_1 j_2 j_3}\rangle$ dans le produit tensoriel des trois repr\'esentations : c'est le symbole $3mj$ de Wigner (ou le coefficient de Clebsch-Gordan) \`a la normalisation pr\`es, soit $\langle j_1, m_1 ; j_2, m_2 ; j_3, m_3 | \iota_{j_1 j_2 j_3}\rangle = \bigl(\begin{smallmatrix} j_1 &j_2 &j_3 \\m_1 &m_2 &m_3 \end{smallmatrix}\bigr)$ (plus de d\'etails en appendice, \ref{sec:app}). Dans le cas d'un noeud de valence $n$, il faut d\'eplier celui-ci en un graphe, avec $n$ pattes externes, ne contenant que des noeuds trivalents. A chaque lien virtuel, i.e. interne, on assigne un spin (virtuel), et chaque noeud interne porte alors un symbole $3mj$ (\`a dualisation pr\`es). Un entrelaceur $\iota_v$ est donc d\'etermin\'e par le choix d'une d\'ecomposition en un arbre trivalent et la donn\'ee de spins virtuels sur les liens internes. On forme les r\'eseaux de spins par contraction des nombres magn\'etiques :
\beq
s_\Gamma^{\{j_e,\iota_v\}}(g_1,\dotsc,g_E) \,=\, \sum_{\{m_e, n_e\}} \prod_{e=1}^E \langle j_e, m_e\lv g_e\rv j_e, n_e\rangle\ \prod_{v=1}^V \langle \otimes_{e\,{\rm in}}\, j_e,m_e \lv \iota_v \rv \otimes_{e\,{\rm out}}\, j_e, n_e\rangle.
\ee
En sommant sur tous les graphes $\Gamma$, l'ensemble des r\'eseaux de spins, $(s_\Gamma^{\{j_e,\iota_v\}})_{\Gamma, (j_e), (\iota_v)}$, constitue une base de l'espace des \'etats cin\'ematiques invariants de jauge,
\beq
\calH_{\rm{kin}} = \bigoplus_\Gamma \calH_\Gamma = L^2(\bar{\calA}/\calG, d\mu_{\rm{AL}}).
\ee
Il faut prendre garde \`a exclure la repr\'esentation triviale dans les r\'eseaux de spins, ce qui est n\'ecessaire pour que des r\'eseaux de spins associ\'es \`a des graphes diff\'erents soient orthogonaux,
\beq
\langle s_\Gamma^{c}\,\vert s_{\Gamma'}^{c'}\rangle = \delta_{\Gamma,\Gamma'}\ \delta_{c,c'}\ N_c,
\ee
en notant $c=\{j_e, \iota_v\}$ le coloriage des graphes (et $N_c$ \'etant la normalisation des vecteurs de base).

\begin{figure}
\begin{center}
\includegraphics[width=7cm]{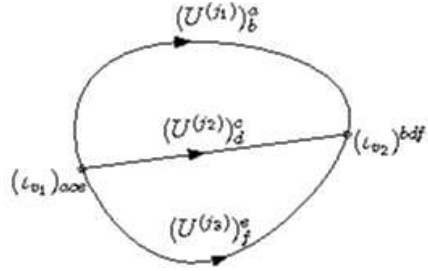}
\end{center}\caption{ \label{fig:theta-spinnet} Un graphe de r\'eseau de spins en \og theta\fg{}. Les liens sont colori\'es par des repr\'esentations et les noeuds par des entrelaceurs. A la normalisation pr\`es (et \`a dualisation pr\`es des repr\'esentations incidentes), il n'existe qu'un seul entrelaceur entre 3 repr\'esentations de $\SU(2)$, qui est donc sp\'ecifi\'e par les trois spins. La fonctionelle de r\'eseau de spins est obtenue en contractant les \'el\'ements matriciels des holonomies d'une connexion le long des 3 courbes dans les repr\'esentations consid\'er\'ees avec les entrelaceurs, ce qui donne bien un r\'esultat invariant sous les transformations de jauge agissant aux noeuds.
}
\end{figure}

Ce sont ces objets que nous consid\'ererons, \`a la partie \ref{sec:recurrence}, pour traduire l'action de la contrainte hamiltonienne de la th\'eorie topologique en relations de r\'ecurrence sur les amplitudes de mousses de spins. Avant cela il nous faut un peu plus de pratique. Le graphe ferm\'e le plus simple est une boucle. Alors nous retrouvons la boucle de Wilson bien connue et bien invariante de jauge, $\tr(D^{(j)}(g))$, la trace \'etant prise dans la repr\'esentation de spin $j$. Regardons ensuite le graphe \og theta\fg, figure \ref{fig:theta-spinnet}. Pour vraiment d\'efinir la fonction d'onde de r\'eseau de spins, nous avons besoin $(i)$ sur le graphe lui-m\^eme, d'une orientation des liens, et d'une orientation (positive-sens trigo, n\'egative sinon, nous utilisons les conventions graphiques de \cite{varshalovich-book}) autour des noeuds, $(ii)$ d'un coloriage des liens par des spins $j_1, j_2, j_3$ (les vertexes sont trivalents, donc pas de spins virtuels). Alors :
\begin{multline}
s^{\{j_e\}}_\Gamma(g_1, g_2, g_3) = \sum_{\substack{n_1, n_2, n_3 \\ m_1, m_2, m_3}} D^{(j_1)}_{m_1 n_1}(g_1)\,D^{(j_2)}_{m_2 n_2}(g_2)\,D^{(j_3)}_{m_3 n_3}(g_3)\ \begin{pmatrix} j_1 &j_2 &j_3 \\m_1 & m_2 &m_3 \end{pmatrix} \\
\times (-1)^{\sum_{e=1}^3 (j_e-n_e)}\begin{pmatrix} j_1 &j_2 &j_3 \\ -n_1 &-n_2 &-n_3 \end{pmatrix}
\end{multline}
Dans l'\'etude des relations de r\'ecurrence associ\'ees \`a la contrainte de courbure nulle de la th\'eorie topologique, nous serons amen\'e \`a \'evaluer les r\'eseaux de spins sur la connexion triviale. Cela donne ici :
\begin{align}
s^{\{j_e\}}_{\Gamma|g_e=\mathbbm{1}} &= \sum_{m_1, m_2, m_3} \begin{pmatrix} j_1 &j_2 &j_3 \\m_1 & m_2 &m_3 \end{pmatrix}
(-1)^{\sum_{e=1}^3 (j_e-m_e)}\begin{pmatrix} j_1 &j_2 &j_3 \\ -m_1 &-m_2 &-m_3 \end{pmatrix},\\
&= \sum_{m_1, m_2, m_3} \begin{pmatrix} j_1 &j_2 &j_3 \\m_1 & m_2 &m_3 \end{pmatrix}
\begin{pmatrix} j_1 &j_2 &j_3 \\ m_1 &m_2 &m_3 \end{pmatrix},\\
&= 1,
\end{align}
pourvu que les in\'egalit\'es triangulaires soient satisfaites entre les trois spins, et $0$ sinon.

Nous avons construit l'espace de Hilbert $\calH_\Gamma$ des fonctions $L^2$ de la connexion sur le graphe, invariantes de jauge. Cet espace porte une repr\'esentation de l'alg\`ebre des holonomies sur $\Gamma$. En effet, on peut v\'erifier que n'importe quel r\'eseau de spins sur $\Gamma$ devient un op\'erateur hermitien en agissant par multiplication sur les r\'eseaux de spins. Ce sont les observables de r\'eseaux de spins, qui g\'en\'eralisent la boucle de Wilson. Mais nous souhaitons aussi repr\'esenter la triade, i.e. le champ $E^a_i$ ! Des observables dites de \emph{flux} se construisent en gardant \`a l'esprit que $E^a_i$ est une densit\'e, i.e. provient d'une 2-forme sur $\Sigma$, et s'int\`egre donc naturellement sur des 2-surfaces. Soit $S\subset \Sigma$ une telle surface, et une fonction $f : S\rightarrow \SU(2)$. Le flux de $E^a_i$ est alors d\'efini comme :
\beq
E_{S,f} = \int_S E_i\,f^i = \int_S f^i\,E^a_i\ \eps_{abc}\,dx^b\wedge dx^c.
\ee
Il se trouve que la $*$-alg\`ebre fondamentale de cette quantification, voir par exemple \cite{lost}, est l'alg\`ebre des holonomies et des flux ! On peut notamment v\'erifier que l'action de ce flux non-ab\'elien sur une surface ferm\'ee au voisinage d'un noeud d'un r\'eseau de spins portant un entrelaceur est nulle. Je ne d\'evelopperai pas plus avant ce formalisme, mais je pr\'esenterai des observables proches des flux, dont le sens physique n'est autre que l'aire de la surface $S$.

Apr\`es avoir trait\'e la contrainte de Gau\ss, il est possible de construire un espace de Hilbert invariant sous les diff\'eomorphismes de $\Sigma$. L'id\'ee est que les diff\'eomorphismes agissent de fa\c{c}on simple sur les fonctions cylindriques, en d\'epla\c{c}ant leur graphe. En effet, si $\phi$ est un diff\'eomorphisme, alors : $\psi_{\Gamma,f}(\phi^* A) = \psi_{\phi\Gamma, f}(A)$. Ainsi, alors que l'espace $L^2(\bar{\calA}/\calG)$ voyait des fonctions cylindriques ayant des graphes diff\'erents comme orthogonales, une base de l'espace invariant est obtenue en consid\'erant les r\'eseaux de spins (et leur donn\'ees) pour les classes d'\'equivalence de graphes sous les diff\'eomorphismes -- la mesure d'Ashtekar-Lewandowski passant au quotient. Ces \'etats sont appel\'es \emph{s-knots}, soulignant bien le fait qu'ils d\'ependent de la fa\c{c}on dont ils sont nou\'es (et donc de la topologie de $\Sigma$). Je renvoie aux r\'ef\'erences donn\'ees pr\'ec\'edemment pour plus de d\'etails. Travaillant le plus souvent sur un graphe fix\'e, je n'utiliserai pas cette construction ici. En accord avec les r\'esultats de la LQG et des mousses de spins en 3d, on s'attend \`a ce qu'un \og bon\fg{} mod\`ele de mousses de spins prenne soin cette sym\'etrie de jauge, tout comme il est sens\'e impl\'ementer la projection sur le noyau de la contrainte scalaire !

Pour compl\'eter la th\'eorie, il faudrait ensuite impl\'ementer les sym\'etries de la contrainte scalaire $S$, et projeter sur les \'etats invariants ou les classes d'\'equivalence de cette action pour obtenir enfin lespace des \'etats physiques de gravit\'e quantique. Apr\`es d'importants efforts de la part de nombreux auteurs, Thiemann a pu proposer une d\'efinition d'un op\'erateur $\hat{S}$, \cite{thiemann-qsd}, sans anomalies, agissant sur les noeuds des r\'eseaux de spins, et en modifiant les spins mais aussi le graphe par la cr\'eation de nouveaux noeuds. Une id\'ee importante \`a signaler en LQG est que la limite dans laquelle on \'elimine la r\'egularisation de la contrainte hamiltonienne doit \^etre trivialis\'ee par l'invariance sous les diff\'eomorphismes, donnant une dynamique finie \`a la LQG. N\'eanmoins, un certain nombre d'aspects reste difficile \`a aborder avec cette quantification de $S$. C'est en partie pour cela que le formalisme des mousses de spins s'est d\'evelopp\'e, pour construire d'une mani\`ere \og propre\fg{} l'exponentielle de la contrainte scalaire (et donc la r\'ealisation de l'int\'egrale de chemins sur les r\'eseaux de spins), et avoir acc\`es \`a la bonne limite semi-classique. Des progr\`es ont permis dans les derni\`eres ann\'ees une convergence int\'eressante avec la LQG, notamment exploit\'ee dans \cite{alesci-noui-sardelli, alesci-rovelli-hamiltonian} pour construire des op\'erateurs susceptibles d'engendrer la m\^eme dynamique que les r\'ecents mod\`eles de mousses de spins.

\section{Op\'erateurs g\'eom\'etriques} \label{sec:operatorsLQG}

Les r\'eseaux de spins portent une repr\'esentation de l'alg\`ebre des holonomies et flux. Il est aussi possible de d\'efinir des op\'erateurs qui en LQG ont un sens g\'eom\'etrique fort : les aires de surfaces, et les 3-volumes. Nous allons voir que ce sont les excitations de r\'eseaux de spins qui fournissent des valeurs non-nulles de ces quantit\'es. Ainsi, cette quantification aboutit \`a une vision pr\'ecise et math\'ematiquement rigoureuse de la 3-g\'eom\'etrie quantique ! Notons n\'eanmoins que ces op\'erateurs ne commutent pas avec la contrainte hamiltonienne (ce ne sont pas des observables de Dirac, invariantes de jauge au sens de l'ensemble des sym\'etries de jauge de la th\'eorie). Leur quantification reposant sur la construction de l'espace cin\'ematique en repr\'esentation de connexions, il est aussi possible de d\'efinir ces op\'erateurs dans le contexte de la th\'eorie BF, ou toute th\'eorie ayant un espace des phases param\'etr\'e par une connexion et son moment conjugu\'e vivant dans l'adjointe. L'interpr\'etation g\'eom\'etrique est alors toutefois diff\'erente (et il faut enlever le param\`etre d'Immirzi des formules que nous allons donner).

Ces r\'esultats sont valides dans la th\'eorie compl\`ete, form\'ee par l'ensemble des graphes de r\'eseaux de spins. Il est toutefois utile d'en donner une vision simplifi\'ee, comme nous le faisons dans une sous-section sur le t\'etra\`edre quantique \ref{sec:quantum tet}. L'espace des phases d'un t\'etra\`edre correspond en fait \`a l'alg\`ebre des observables associ\'ee \`a un noeud 4-valent d'un r\'eseau de spins en LQG. Ainsi, une quantification g\'eom\'etrique de cet espace des phases conduit de mani\`ere simplifi\'ee aux r\'esultats de la th\'eorie compl\`ete. Cela offre une vision intuitive et g\'eom\'etrique utile, que nous utiliserons pour comprendre pr\'ecis\'ement les nombres quantiques caract\'erisant un noeud 4-valent, et pour discuter la quantification d'op\'erateurs d'angles dih\'edraux sans se lancer dans la quantification au sein de la th\'eorie compl\`ete. Ajoutons que cette approche du t\'etra\`edre quantique, mais avec le groupe $\Spin(4)$ \`a la place de $SU(2)$, a eu une influence majeure sur la construction des mod\`eles de mousses de spins et le lien avec la th\'eorie de Plebanski !

\subsection{G\'eom\'etrie quantique : Aires, volumes, longueurs}

L'observable invariante de jauge $\SU(2)$ la plus simple est l'aire d'une surface. Nous formons d'abord pour une surface $S$, param\'etr\'ee par des cooordonn\'ees $x(\sigma)$, $\sigma=(\sigma^1, \sigma^2)$, la quantit\'e :
\beq
E_i(S) = \int_S d\sigma^1 d\sigma^2\ \eps_{abc}\,\f{\pp x^a}{\pp\sigma_1}(\sigma)\, \f{\pp x^b}{\pp\sigma_2}(\sigma)\ E^c_i(x(\sigma)),
\ee
qui n'est pas invariante de jauge. Malgr\'e cela, l'action de l'op\'erateur correspondant sur les r\'eseaux de spins est particuli\`erement simple. Supposons que $S$ ne croise un lien $e$ d'un r\'eseau de spins, portant le spin $j$, qu'en un point $p$, alors $\widehat{E}_i(S)$ ins\`ere dans l'holonomie le long de $e$ un g\'en\'erateur $\tau_i\in\su(2)$ dans la repr\'esentation de spin $j$ au point $p$. Cela provient du crochet fondamental,
\beq
\bigl\{ E_i(S), U_e(A)\bigr\} = \gamma\ U_{e>p}(A)\,\tau_i\,U_{e<p}(A).
\ee
Cela signifie que l'action de $E_i(S)$ est locale : elle n'agit qu'\`a l'intersection de $S$ et $e$. De plus, on en d\'eduit que $E_i(S)E^i(S)$ produit le Casimir du spin $j$, i.e. $\tau_i\tau_j\delta^{ij}$, de sorte que les r\'eseaux de spins sont des vecteurs propres de cet op\'erateur !

Nous pouvons alors calculer l'aire d'une surface $S$ par sa somme de Riemann. D\'ecoupons $S$ en $N$ petites plaquettes $S_n$ de taille d'ordre $1/N$, et formons (classiquement) :
\beq \label{area riemann sum}
A(S) = \lim_{N\rightarrow \infty} \sum_{n=1}^N \sqrt{ E_i(S_n)\,E^i(S_n)}.
\ee
Cela correspond bien la quantit\'e recherch\'ee, comme on le voit par exemple en prenant des coordonn\'ees telles que $S$ correspond \`a $x^3=0$, et $\sigma^1=x^1, \sigma^2=x^2$. Alors le membre de droite de \eqref{area riemann sum} est la somme de Riemann de :
\beq
\int_S d\sigma^1 d\sigma^2\ \sqrt{E_i^3(x)\,E^{3i}(x)} = \int_S d\sigma^1 d\sigma^2\ \sqrt{q^{33}(x)\,\det({^3}q)} = \int_S d\sigma^1 d\sigma^2\ \sqrt{\det({^2}q)} = A(S),
\ee
par d\'efinition de la triade, \eqref{triade 3metrique}. Au niveau quantique, imaginons que $S$ croise le graphe $\Gamma$ d'un r\'eseau de spins en un nombre fini de point, et pas de mani\`ere tagentielle. Alors la r\'egularisation \eqref{area riemann sum} ne fait agir $\widehat{E}_i(S_n)$ qu'en ces intersections, et l'aire ne re\c{c}o\^it de contributions qu'en croisant le graphe d'un r\'eseau de spins ! De plus, elle est diagonale sur ceux-ci :
\beq \label{area spectrum lqg}
\widehat{A(S)}\ s^{\{j_e\}}_\Gamma = \gamma\ \biggl(\sum_{e, e\cap S\neq\emptyset} \sqrt{j_e(j_e+1)}\biggr)\ s^{\{j_e\}}_\Gamma.
\ee
Nous avons tenu compte ici de l'\'echelle de la structure symplectique de la gravit\'e, donn\'ee par le param\`etre d'Immirzi. N\'eanmoins, toutes ces constructions et les r\'esultats associ\'es tiennent toujours pour une th\'eorie d'une connexion et de son moment conjugu\'e (sans le param\`etre $\gamma$) ou la th\'eorie BF, \cite{baez-bf-spinfoam}. Seule l'interpr\'etation g\'eom\'etrique doit \^etre chang\'ee. A nouveau les r\'ef\'erences pr\'ecit\'ees contiennent tous les d\'etails sur l'action g\'en\'erique des aires. Mentionnons tout de m\^eme un aspect technique : une ambiguit\'e dans la quantification des op\'erateurs quadratiques en la triade \'evalu\'es au m\^eme point, qui peut conduire \`a des spectres diff\'erant du Casimir en $j(j+1)$ par une constante et/ou un terme lin\'eaire. Parmi les choix naturels, on trouve $j^2$ et $(j+\f12)^2$, qui donnent des spectres ayant toujours le m\^eme \'ecart, discut\'es \`a propos de l'entropie des trous noirs \cite{alekseev-area-entropy}, et que l'on retrouvera dans les mod\`eles des mousses de spins \cite{eprl}.

Des calculs similaires peuvent \^etre conduits \`a partir de la structure symplectique $\SO(4)$ covariante. Mais pour interpr\'eter les flux de la norme du moment $P^{a}_{IJ}$ comme des aires, il faut prendre en compte les contraintes de seconde classe, autrement dit utiliser les crochets de Dirac pour la quantification. Etant donn\'e qu'il n'y a plus de connexions privil\'egi\'ees pour les crochets de Dirac, plusieurs choix sont possibles. Un choix naturel (par rapport aux sym\'etries) conduit alors \`a un spectre des aires donn\'e par la partie boost du Casimir de $\SO(4)$ (par rapport au champ de normales $n^I$ d\'efinissant la direction temps de l'espace interne)\footnote{Cela donne un spectre continu dans la th\'eorie lorentzienne, contrastant avec le spectre discret de la LQG, qui reste associ\'e aux Casimir de $\SU(2)$ dans la th\'eorie lorentzienne.} \cite{livine-clqg}, et ind\'ependant du param\`etre d'Immirzi. De mani\`ere remarquable, la quantification en mousses de spins des contraintes de simplicit\'e, voir section \ref{sec:epr-quantisation}, permet de relier directement ce spectre \`a celui de la LQG, \eqref{area spectrum lqg}, voir \cite{eprl}.

Ainsi, les liens des r\'eseaux de spins portent des excitations qui s'interpr\`etent directement comme les aires de surfaces duales aux liens. On peut donc s'attendre \`a ce qu'un noeud porte une excitation de volume de la r\'egion 3d duale contenue dans les surfaces duales aux liens qui s'y rencontrent. C'est pr\'ecis\'ement ce qui se passe ! Classiquement, le volume d'une r\'egion $R$ s'exprime en fonction de la triade par :
\beq \label{3volume}
V(R) = \int_R d^3x\ \sqrt{\det({^3}q)} = \int_R d^3x\ \sqrt{\f{1}{3!}\,\lvert \eps_{abc}\,\eps^{ijk}\ E^a_i(x)\,E^b_j(x)\,E^c_k(x)\rvert}.
\ee
Cette observable peut \^etre quantifi\'ee, basiquement en rempla\c{c}ant les triades par des d\'erivations (insertions de g\'en\'erateurs de l'alg\`ebre dans les holonomies), en un op\'erateur hermitien $\widehat{V(R)}$ bien d\'efini \cite{thiemann-book}. Il n'agit qu'au niveau des noeuds des r\'eseaux de spins, sans modifier le graphe, ni les spins des liens. Il annihile les noeuds de valence inf\'erieurs \`a trois. Cela se comprend intuitivement par le fait que les surfaces duales aux liens accroch\'es \`a un tel noeud ne peuvent former que des poly\`edres d\'eg\'en\'er\'es. En revanche, $\widehat{V(R)}$ agit de mani\`ere non-triviale \`a partir des noeuds de valence 4 : il fait intervenir l'entrelaceur $\iota_v$ en plus des spins $j_e$ sur les pattes du noeud, et prend des valeurs propres discr\`etes. Mais il n'est pas diagonal sur la base des r\'eseaux de spins construites ci-dessus (qui demande de sp\'ecifier une d\'ecomposition en vertexes trivalents et des spins virtuels), et son spectre n'est pas compl\`etement connu.

\bigskip

Nous avons donc le tableau suivant d'une 3-g\'eom\'etrie quantique. Un r\'eseau de spins correspond \`a un graphe $\Gamma$ immerg\'e dans une 3-vari\'et\'e, dont les liens portent des repr\'esentations (spins) de $\SU(2)$ et les noeuds des entrelaceurs entre les repr\'esentations qui s'y rencontrent. Un noeud permet alors de d\'efinir une 3-cellule duale ayant un volume non-trivial, prenant des valeurs discr\`etes. Un lien est dual \`a une surface qui d\'elimite les deux r\'egions de l'espace duales aux extr\'emit\'es. L'aire de cette surface est bien d\'efinie et donn\'ee par le casimir de la repr\'esentation port\'ee par le lien.

Dans cette vision duale, un noeud connect\'e \`a deux autres noeuds permet de consid\'erer le lieu de l'intersection des deux surfaces duales qui est une courbe. On peut alors se demander s'il existe un op\'erateur de longueur associ\'e \`a cette courbe. Cela permettrait d'approfondir la g\'eom\'etrie quantique, mais aussi de r\'epondre \`a une question naturelle : est-il possible d'assigner \`a un r\'eseau de spins des quanta de longueurs duales, et de les interpr\'eter comme la donn\'ee d'une m\'etrique \`a la Regge sur le complexe dual ?
Une construction d'un op\'erateur de longueur a \'et\'e donn\'ee par Bianchi, \cite{bianchi-lengthLQG}, (voir aussi Thiemann \cite{thiemann-lengthLQG}) bien adapt\'ee au complexe dual au r\'eseau de spins.

Pour d\'efinir correctement ces courbes duales, nous supposons que $\Gamma$ est le 1-squelette dual \`a une d\'ecomposition cellulaire de la 3-vari\'et\'e $\Sigma$. Les courbes qui nous int\'eressent sont alors les 1-cellules de cette d\'ecomposition, duales aux faces du 2-squelette qui contient $\Gamma$. Une telle courbe $\gamma$ est alors caract\'eris\'ee par deux liens se rencontrant en un noeud, ce qui d\'efinit une portion, appel\'ee {\it wedge}, d'une 2-cellule bord\'ee par un cycle ferm\'e de $\Gamma$. Mais il faut identifier les courbes associ\'ees aux diff\'erents wedges d'une 2-cellule. Classiquement, la longueur d'une courbe $\gamma : s\in [0,1] \mapsto \gamma^a(s) \in\Sigma$ est donn\'ee par :
\beq
L(\gamma) = \int_0^1 ds\ \sqrt{\delta_{ij}\,G^i(s)\,G^j(s)},
\ee
o\`u $G^i(s)$ est une fonction de la triade :
\beq
G^i(s) = \f1{2\sqrt{\det({^3}q)}}\,\eps^{ijk}\,\eps_{abc}\,\dot{\gamma}^a(s)\,E^b_j(\gamma(s))\,E^c_k(\gamma(s)).
\ee
L'expression du d\'eterminant de la 3-m\'etrique en fonction de la triade est donn\'ee en \eqref{3volume}.

Je renvoie \`a \cite{bianchi-lengthLQG} pour les d\'etails de la r\'egularisation et l'\'etude des propri\'et\'es de l'op\'erateur construit. Il existe un op\'erateur de longueur $L_w(\gamma)$ par wedge $w$, mesurant la longueur de la courbe $\gamma$, qui agit alors sur le noeud du wedge, sans modifier le graphe, ni les spins coloriant les liens. Ainsi, ses \'el\'ements de matrices sont indic\'es par les entrelaceurs sur le noeud. Il n'est pas diagonal sur les r\'eseaux de spins, mais on sait que ces valeurs propres sont discr\`etes ! Il commute avec les op\'erateurs d'aires des surfaces duales aux liens, et avec les op\'erateurs de volume des r\'egions duales aux noeuds diff\'erents de celui du wedge consid\'er\'e. Il poss\`ede la limite semi-classique attendue : pour un \'etat piqu\'e sur la g\'eom\'etrie classique d'un t\'etra\`edre, il mesure les longueurs des ar\^etes de celui-ci.

Par ailleurs, il est possible de diagonaliser simultan\'ement les op\'erateurs d'aires et les op\'erateurs de longueurs associ\'es \`a diff\'erents wedges d'une m\^eme 2-cellule. Comme le montre Bianchi, les nombres quantiques de longueurs correspondants, caract\'erisant une seule courbe, sont g\'en\'eriquement diff\'erents ! Cela signifie que la longueur d'une courbe d\'epend du wedge, ou encore de la 3-cellule duale au noeud de $\Gamma$ dans laquelle elle est vue. Cela sugg\`ere fortement d'aller regarder la m\^eme situation mais au niveau classique, dans le but de voir si de mani\`ere g\'en\'erique, l'espace des phases de la LQG, restreint \`a un graphe dual \`a une d\'ecomposition cellulaire ou simplicielle, donne naissance \`a des \emph{m\'etriques discontinues} d'une 3-cellule \`a l'autre, i.e. assignant des longueurs qui d\'ependent de la cellule dans laquelle elles sont consid\'er\'ees. Nous verrons, comme le montrent Dittrich et Ryan, \cite{dittrich-ryan-simplicial-phase}, \`a la section \ref{sec:twisted-simplicial}, que c'est en effet le cas.

Mais auparavant, pour pr\'eciser le sens g\'eom\'etrique des entrelaceurs, regardons un noeud 4-valent. A celui-ci, on peut associer un t\'etra\`edre dual, tel que les pattes du noeud sont duaux aux triangles. Les spins port\'es par les liens sont alors les aires quantifi\'ees de ces quatre triangles. N\'eanmoins, la g\'eom\'etrie d'un t\'etra\`edre n'est pas compl\`etement sp\'ecifi\'ee par ces aires. Certes, l'action de l'op\'erateur de volume, tenant compte de l'entrelaceur, apporte une information suppl\'ementaire. Mais on sait que classiquement, il nous suffit de prendre deux angles dih\'edraux non-oppos\'es en plus des aires pour caract\'eriser le t\'etra\`edre. Cela sugg\`ere de s'int\'eresser \`a une quantification des observables d'angles dih\'edraux entre les surfaces duales aux liens des r\'eseaux de spins. Une proposition dans ce sens a \'et\'e faite par Major, \cite{major-3dangles}, qui s'ins\`ere tr\`es bien dans la vision de la 3-g\'eom\'etrie duale. Major donne des d\'efinitions pr\'ecises d'op\'erateurs d'angles et de leur cosinus, et discute les conditions qui permettent de reproduire la bonne limite semi-classique.

\subsection{Le t\'etra\`edre quantique et les op\'erateurs d'angles} \label{sec:quantum tet}

Il sera int\'eressant pour nous de comprendre les id\'ees principales de la construction des observables d'angles du point de vue de la quantification de l'espace des g\'eom\'etries d'un t\'etra\`edre. Cette approche a \'et\'e initi\'ee par Barbieri \cite{barbieri}, et proc\`ede, de fa\c{c}on math\'ematiquement rigoureuse, par quantification g\'eom\'etrique \cite{baez-barrett-quantum-tet}. Etant proche de la LQG, dont elle d\'ecrit les observables cin\'ematiques, fonctions de la triade, agissant au niveau d'un vertex 4-valent, elle peut servir de guide pour la d\'efinition rigoureuse d'op\'erateurs en LQG. Mais il faut prendre garde \`a ce qu'elle ne traite qu'un simplexe \emph{isol\'e} ! Cela signifie que les holonomies ne sont pas prises en compte, elles qui sont fondamentales \`a la LQG bien s\^ur, et comme nous le verrons tr\`es largement par la suite, tout aussi fondamentales pour d\'efinir les th\'eories de jauge sur r\'eseau et assurer la \emph{consistence} des g\'eom\'etries discr\`etes. (En particulier dans cette derni\`ere approche, nous devrons prendre soin du collage des simplexes, en utilisant le transport parall\`ele, pour \'etendre les r\'esultats de la quantification g\'eom\'etrique du t\'etra\`edre \`a une triangulation et in fine construire des int\'egrales de chemins sur des g\'eom\'etries discr\`etes consistentes.)

On d\'ecrit un t\'etra\`edre par les normales aux triangles, not\'ees\footnote{Nous utilisons ici les notations associ\'ees au graphe $\Gamma$ du r\'eseau de spins. Dans les sections suivantes, nous travaillerons avec un graphe \emph{fix\'e} dual \`a une triangulation. Nous aurons alors recours aux notations d\'esignant les simplexes de la triangulation. Le sens des variables sera tout \`a fait similaire.} $\vec{b}_e\in\R^3$. Ces vecteurs satisfont une relation de fermeture,
\beq \label{closure 3d tet}
\vec{b}_1+\vec{b}_2+\vec{b}_3+\vec{b}_4 = 0.
\ee
Ces sont en fait des bivecteurs, \'el\'ements de $\Lambda^2\R^3$, form\'es par les produits vectoriels des vecteurs d\'ecrivant les ar\^etes du triangle dans $\R^3$. Une structure de Poisson sur ces bivecteurs est obtenue en les regardant comme des \'el\'ements de l'alg\`ebre duale $\so(3)^*$,
\beq
\{ b_e^i, b_{e}^j\} \,=\, \eps^{ij}_{\phantom{ij}k}\, b_e^k,
\ee
les crochets s'annulant entre liens diff\'erents\footnote{Puisque les vecteurs $\vec{b}_e$ sont sens\'es correspondre \`a la triade, ou \`a ses flux \`a travers les triangles, il peut para\^itre surprenant qu'ils ne commutent pas comme c'est le cas dans la th\'eorie compl\`ete. En fait, en passant des variables de connexion et triade \`a l'alg\`ebre des holonomies et des flux, il est n\'ecessaire que les flux ne commutent plus entre eux pour que les crochets de Poisson satisfassent l'identit\'e de Jacobi (voir \cite{corichi-zapata-fluxes}) ! Cela est tout \`a fait standard lorsque l'on travaille avec le fibr\'e cotangent \`a un groupe de Lie, et sera repris dans les sections suivantes.}. La contrainte de fermeture \eqref{closure 3d tet} est le g\'en\'erateur pour ces crochets des rotations dans $\R^3$, i.e. la contrainte de Gau\ss. On peut former des observables invariantes,
\beq
A_e = \lv \vec{b}_e \rv, \quad \text{et} \qquad A_{ee'} = \lv \vec{b}_e + \vec{b}_{e'}\rv.
\ee
$A_e$ correspond \`a l'aire du triangle dual \`a $e$, tandis que $A_{ee'}$ repr\'esente l'aire du parall\'elogramme dont les sommets sont les milieux des ar\^etes des triangles duaux \`a $e$ et $e'$, sauf l'ar\^ete partag\'ee par ces deux triangles. Cela revient \`a couper en deux le t\'etra\`edre. Remarquons que seules trois de ces variables sont distinctes, correspondants aux trois fa\c{c}ons de couper le t\'etra\`edre en deux. On peut s'assurer qu'un ensemble complet d'observables qui commutent pour les crochets de Poisson est form\'ee par les quatre aires $A_e$ plus une variable $A_{ee'}$. En effet, les aires des parall\'elogrammes ne commutent pas entre elles, typiquement : $\{A_{12}^2, A_{13}^2\} = 4 V_{\rm tet}$, $V_{\rm tet}$ \'etant le volume du t\'etra\`edre. Cela nous indique que le t\'etra\`edre quantique est caract\'eris\'e par cinq nombres quantiques, la diff\'erence avec les six degr\'es de libert\'e classiques provenant des relations d'incertitude !

\begin{figure}
\begin{center}
\includegraphics[width=4cm]{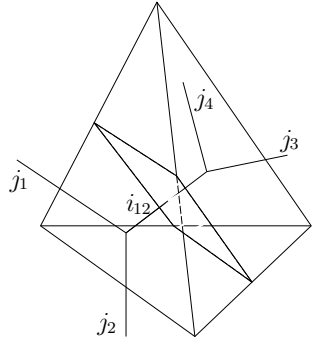}
\end{center}\caption{ \label{fig:quantum-tet} Un t\'etra\`edre et une repr\'esentation graphique des \'etats du t\'etra\`edre quantique. De l'ext\'erieur du t\'etra\`edre, on voit un noeud 4-valent portant un entrelaceur. Les pattes externes sont duales aux triangles et sont colori\'ees par des repr\'esentations $(j_e)_{e=1,\dotsc,4}$ qui sont les valeurs propres, discr\`etes, des aires des triangles. Une base d'entrelaceurs, et donc des \'etats quantiques, est d\'efinie en diagonalisant un des op\'erateurs $\hat{A}_{ee'}$, ici $\hat{A}_{12}$, dont les valeurs propres $i_{12}$ repr\'esentent l'aire du parall\'elogramme dessin\'e, joignant les milieux de 4 ar\^etes. Alg\'ebriquement, on peut d\'ecomposer cet entrelaceur en deux noeuds trivalents reli\'es par un spin $i_{12}$ virtuel, comme en \eqref{tree 4-node}, alors port\'e par un lien dual au parall\'elogramme !
}
\end{figure}

Mais on peut aussi quantifier avant d'imposer la contrainte de Gau\ss, comme en LQG. La quantification g\'eom\'etrique \cite{baez-barrett-quantum-tet}, associe \`a chaque triangle un espace de Hilbert form\'e par l'ensemble des repr\'esentations irr\'eductibles de $\SU(2)$, et donc au t\'etra\`edre l'espace de Hilbert
\beq
\calH_{\rm tet} = \bigotimes_{e=1,2,3,4}\ \bigoplus_{j_e\in\f{\N}{2}} \calH_{j_e}.
\ee
Les composantes $b_e^i$ des vecteurs normaux sont alors promues en g\'en\'erateurs $J_e^i$ de $\su(2)$ agissant sur $\calH_{\rm tet}$, et satisfaisant les relations de commutation : $[J^i_e, J^j_e] = i\eps^{ij}_{\phantom{ij}k} J^k_e$. La contrainte de Gau\ss{} restreint alors les \'etats physiques aux vecteurs invariants sous $\SU(2)$ :
\beq
\Inv\bigl(\calH_{\rm tet}\bigr) = \bigoplus_{j_1,j_2, j_3, j_4} \Inv\bigl(\calH_{j_1}\otimes \calH_{j_2}\otimes \calH_{j_3}\otimes \calH_{j_4}\bigr),
\ee
autrement dit, aux entrelaceurs. Une base se forme donc en choisissant quatre spins $(j_e)$ et un entrelaceur entre ces repr\'esentations. Cela se fait par le choix d'un appariement de ces spins deux \`a deux, disons $j_1$ avec $j_2$, et d'un spin virtuel $i_{12}$ dans la d\'ecomposition $\calH_{j_1}\otimes\calH_{j_2} = \oplus_{i_{12}}\calH_{i_{12}}$. Ainsi, nous avons nos cinq degr\'es de libert\'e quantiques, qui s'av\`erent \^etre les valeurs propres des quatre aires $\widehat{A}_e$ et de $\widehat{A}_{12}$ :
\begin{alignat}{5} \label{eigenstate area tet}
&\widehat{A}_e&\ &\lv (j_1 j_2), i_{12}, (j_3 j_4)\rangle &\,=\,& \sqrt{j_e(j_e+1)}&\ &\lv (j_1 j_2), i_{12}, (j_3 j_4)\rangle,&\qquad &e=1,2,3,4,\\ \label{eigenstate A12 tet}
&\widehat{A}_{12}&\ &\lv (j_1 j_2), i_{12}, (j_3 j_4)\rangle &\,=\,& \sqrt{i_{12}(i_{12}+1)}&\ &\lv (j_1 j_2), i_{12}, (j_3 j_4)\rangle.&&
\end{alignat}
Remarquons que les trois fa\c{c}ons de couper le t\'etra\`edre en deux pour mesurer l'aire d'un parall\'elogramme se traduit au niveau quantique par les trois choix d'appariements possibles entre les quatre spins $j_e$. Le d\'eveloppement de tels entrelaceurs 4-valents en arbre trivalent est donn\'e alg\'ebriquement en appendice, \eqref{tree 4-node}, et se traduit graphiquement en consid\'erant le lien virtuel portant $i_{12}$ comme dual au parall\'elogramme dont il donne l'aire, voir figure \ref{fig:quantum-tet}.

La donn\'ee des variables $A_{ee'}$ est naturellement \'equivalente classiquement \`a celle des angles dih\'edraux $\phi_{ee'}$ entre les triangles, car :
\beq
A_{ee'}^2 = A_e^2 + A_{e'}^2 - 2\,A_e A_{e'}\,\cos\phi_{ee'}.
\ee
On extrait alors le spectre :
\beq
\widehat{\cos\phi_{12}}\ \lv (j_1 j_2), i_{12}, (j_3 j_4)\rangle = \f{j_1(j_1+1)+j_2(j_2+1)-i_t(i_t+1)}{2\sqrt{j_1(j_1+1)\,j_2(j_2+1)}}\ \lv (j_1 j_2), i_{12}, (j_3 j_4)\rangle,
\ee
et l'on comprend bien qu'on ne peut donner en m\^eme temps les valeurs de deux angles non-oppos\'es du fait de la non-commutativit\'e des $A_{ee'}$.

Nous avons donc expos\'e l'id\'ee principale \`a l'origine de la quantification des op\'erateurs d'angles dans \cite{major-3dangles}. De mani\`ere assez remarquable, nous retrouverons ce r\'esultat, section \ref{sec:insertion obsBF}, par une route totalement diff\'erente et originale. Nous n'utiliserons que des grandeurs classiques, et consid\'ererons des insertions d'observables g\'eom\'etriques dans une int\'egrale sur les g\'eom\'etries d'une triangulation. En plus de la m\'ethode (quantification canonique versus int\'egrale de chemins), une diff\'erence essentielle est que pour passer d'un simplexe isol\'e \`a une triangulation, des relations de collage, telles que \eqref{constraints4dangles-aarc} en calcul de Regge aires - angles 3d doivent \^etre prises en compte, et ce dans l'int\'egrale sur les g\'eom\'etries, ce qui est non-trivial !

L'approche que nous avons pr\'esent\'ee, la quantification d'un simplexe isol\'e, est \`a la base du mod\`ele de mousses de spins de Barrett-Crane (BC), qui d\'ecrit un 4-simplexe quantique. Nous avons au cours de cette th\`ese formul\'e une d\'erivation de ce mod\`ele, \cite{BCpaper-val}, qui met en \'evidence des soucis au niveau du collage des simplexes. Et c'est en prenant soin de ce collage, i.e. en impl\'ementant les contraintes sur les angles 3d \eqref{constraints4dangles-aarc} au niveau des mousses de spins que j'ai pu d\'eriver une vision de type int\'egrale de chemins pour les nouveaux mod\`eles de mousses de spins EPR/FK \cite{bf-aarc-val, new-model-val}. L'id\'ee de la d\'emarche est assez similaire au travail initial de Freidel-Krasnov \cite{new-model-fk}, mais formul\'ee tr\`es diff\'eremment.

Pour terminer cette section, notons que les r\'esultats obtenus ici peuvent aussi \^etre vus du point de vue du calcul sur les invariants de $\SU(2)$ (comme les symboles 6j de Wigner) comme des r\`egles dites de grasping, permettant notamment d'\'evaluer des insertions d'observables en mousses de spins \cite{speziale-grasping-rules}.

\chapter{G\'eom\'etries sur un graphe fix\'e}

Puisque la quantification en r\'eseaux de spins met en \'evidence des graphes comme excitations fondamentales, nous nous int\'eressons plus en d\'etails aux aspects cin\'ematiques et dynamiques de la th\'eorie sur un graphe fix\'e. En fait, nous allons regarder et comparer de mani\`ere tr\`es pr\'ecise diff\'erents types de g\'eom\'etrie sur un graphe, au niveau canonique : celles correspondant aux variables de la LQG, et les g\'eom\'etries de Regge. Une importante partie de mon travail de th\`ese ayant consist\'e \`a extraire la g\'eom\'etrie des mousses de spins, qui donnent l'\'evolution des r\'eseaux de spins, il est important pour comprendre l'action des mousses de spins sur les \'etats cin\'ematiques de cerner la diff\'erence entre la th\'eorie compl\`ete (la LQG), et l'approximation qu'est le calcul de Regge. C'est ce que je d\'ecris dans la section \ref{sec:twisted-simplicial}, suivant \cite{dittrich-ryan-simplicial-phase}, avec la notion de g\'eom\'etries tordues pour caract\'eriser la LQG, r\'ecemment approfondie dans \cite{freidel-speziale-twisted-geom}.

Une des difficult\'es pour traiter la contrainte scalaire en relativit\'e g\'en\'erale est que l'on ne dispose pas d'une interpr\'etation claire de son effet g\'eom\'etrique. Ainsi conna\^itre le lien entre la g\'eom\'etrie de la LQG sur un graphe et le calcul de Regge sugg\`ere l'utilisation de ce dernier comme source d'inspiration. Nous discutons alors bri\`evement \`a la section \ref{sec:flatness} une contrainte de courbure nulle pr\'esent\'ee dans \cite{dittrich-ryan-simplicial-phase}. Pour prolonger la discution, nous utilisons le mod\`ele topologique de gravit\'e en 3 dimensions, o\`u la LQG sur un graphe et le calcul de Regge coincident. Nous d\'erivons dans ce cadre explicitement une forme g\'eom\'etrique, reliant courbure intrins\`eque et courbure extrins\`eque, de la contrainte de courbure nulle de la th\'eorie BF. Cela nous sera important pour relier les relations de r\'ecurrence sur les mousses de spins en 3d aux contraintes classiques de la th\'eorie canonique.

Puis dans la section \ref{sec:classical-tent}, je d\'ecris le formalisme des mouvements \og tente\fg{} qui permettent d'obtenir une description hamiltonienne de la dynamique de l'action de Regge. En plus de s'int\'eresser aux sym\'etries de cette derni\`ere, par comparaison avec les sym\'etries du continu, qui s'av\`erent \^etre g\'en\'eralement bris\'ees au niveau discret comme le montre \cite{bahr-dittrich-broken-symmetries}, cela d\'efinit un cadre d'\'etude adapt\'e aux actions discr\`etes, potentiellement utile pour la th\'eorie compl\`ete. Comme application de cette id\'ee, nous verrons aux chapitres suivants comment, dans le mod\`ele topologique en 3d, la contrainte hamiltonienne agit sur les r\'eseaux de spins en g\'en\'erant ces mouvements \og tente\fg{} et conduit dans la base des spins \`a des \'equations au diff\'erences finies sur les amplitudes de mousses de spins !

\section{G\'eom\'etries tordues et \`a la Regge} \label{sec:twisted-simplicial}

Nous avons vu que la quantification d'une th\'eorie de jauge par les r\'eseaux de spins fournit des excitations fondamentales sous forme de graphes, et des op\'erateurs g\'eom\'etriques associ\'es \`a des nombres quantiques sur les liens et les noeuds. Nous nous int\'eressons ici aux espaces des phases classiques de diff\'erents types de g\'eom\'etries discr\`etes, vivant sur le graphe dual \`a une triangulation d'une 3-vari\'et\'e, avec la logique suivante :
\begin{itemize}
 \item Tout d'abord, l'espace des phases issu de la th\'eorie BF $\Spin(4)$ tronqu\'e au graphe dual \`a la triangulation.
 \item Puis sa r\'eduction \`a l'espace des phases $\SU(2)$ de la LQG tronqu\'e au m\^eme graphe, conduisant \`a des g\'eom\'etries dites \emph{tordues}.
 \item Puis une nouvelle r\'eduction conduisant aux g\'eom\'etries de Regge sur la triangulation.
\end{itemize}
On peut \'egalement consid\'erer le premier point comme la discr\'etisation naturelle de la th\'eorie BF $\Spin(4)$ canonique sur une triangulation. Comme dans le continu, nous verrons qu'il existe des contraintes dites de \emph{simplicit\'e} qui permettent de se ramener \`a un espace des phases $\SU(2)$ comme en LQG. Une question naturelle est de savoir si cela aboutit \`a une notion de t\'etrade discr\`ete, par analogie avec le continu o\`u les contraintes de simplicit\'e font passer d'une 2-forme $B$ \`a une cot\'etrade $e$. La r\'eponse est qu'il existe alors une notion de m\'etrique sur la triangulation, mais \emph{discontinue} d'un t\'etra\`edre \`a un autre ! C'est-\`a-dire que de telles m\'etriques n'accordent pas les m\^emes longueurs aux ar\^etes de la triangulation selon le t\'etra\`edre dans lequel elles sont calcul\'ees : on parle, suivant Freidel et Speziale, \cite{freidel-speziale-twisted-geom}, de g\'eom\'etries \emph{tordues}. Ainsi, il ne s'agit pas de g\'eom\'etries de Regge. Puis nous reproduisons des contraintes propos\'ees dans \cite{dittrich-ryan-simplicial-phase} qui r\'eduisent l'espace des phases $\SU(2)$ aux g\'eom\'etries de Regge. Il s'agit en fait de contraintes propos\'ees dans la formulation aires - angles 3d du calcul de Regge dans \cite{dittrich-speziale-aarc}. Au final, cela donne une vision tr\`es pr\'ecise des liens existants entre ces types de g\'eom\'etries discr\`etes. De plus, cela permet de relier le formalisme utilis\'e dans les r\'eseaux de spins et celui des variables scalaires typiques du calcul de Regge. Nous verrons dans les chapitres suivants une analyse similaire dans l'approche covariante, et nous montrerons que les contraintes n\'ecessaires si l'on souhaite des g\'eom\'etries de Regge sont naturellement pr\'esentes dans le formalisme des mousses de spins !

Nous d\'ecrivons donc le cadre d'\'etude et quelques r\'esultats essentiels de Dittrich et Ryan, \cite{dittrich-ryan-simplicial-phase}. Il s'agit d'une discr\'etisation de la th\'eorie \emph{canonique}, reposant sur une triangulation 3d. Le d\'epart s'effectue avec l'espace des phases des holonomies et flux de $\Spin(4)$ sur la triangulation, en pr\'esence du param\`etre d'Immirzi. Autrement dit, l'espace des phases issu de l'action de Holst, avant r\'eduction par les contraintes de seconde classe, est tronqu\'e \`a un nombre fini de degr\'es de libert\'e, vivant sur la triangulation. Les triangles et t\'etra\`edres sont indiqu\'es par les lettres $f$ et $t$. Un triangle peut aussi \^etre d\'etermin\'e par la paire de t\'etra\`edres $(tt')$ qui se le partagent ; nous utiliserons plut\^ot cette notation pour d\'esigner un lien dual \`a un triangle, qui relie les t\'etra\`edres $t$ et $t'$. Ainsi, les holonomies sont associ\'ees au 1-squelette dual, not\'ees $G_{tt'}\in\Spin(4)$, et repr\'esentent le transport parall\`ele entre les deux t\'etra\`edres $t$ et $t'$. Le changement d'orientation d'un lien se traduit par une inversion, $G_{t't} = G_{tt'}\mone$.

Ce transport parall\`ele relie des rep\`eres locaux de $\R^4$ associ\'es \`a chaque t\'etra\`edre. Ceux-ci sont donc consid\'er\'es comme \emph{plats}, et comme en calcul de Regge (3d), la courbure se concentre autour des ar\^etes de la triangulation. Elle est d\'etermin\'ee par le produit des $G_{tt'}$ autour de ces ar\^etes. Du point de vue dual, il s'agit du produit le long du bord de la face duale \`a une ar\^ete $\ell$, $G_\ell = G_{t_0t_n}\dotsm G_{t_2 t_1}G_{t_1t_0}$. Les triangles permettent de construire les flux, et donc de \og discr\'etiser\fg{} la 2-forme $B$ en $B_f\in\spin(4)$. Nous avons bien s\^ur besoin du r\'ef\'erentiel local d'un t\'etra\`edre pour d\'efinir chacune de ces variables, appel\'ee \emph{bivecteur}, $B_f(t)$. Les bivecteurs d'un triangle donn\'e consid\'er\'es dans des t\'etra\`edres adjacents $t_1, t_2$ partageant $f$ ne sont pas ind\'ependants mais reli\'es par du transport parall\`ele, $B_f(t_1) = - G_{t_1 t_2}\,B_f(t_2)\,G_{t_1 t_2}\mone$, si $G_{t_1 t_2}$ va de $t_2$ \`a $t_1$.

Nous pouvons alors consid\'erer la structure symplectique h\'erit\'ee de l'action,
\beq
\int_M \Tr\,\bigl( B\wedge F + \gamma\mone\star B\wedge F\bigr) = (1+\gamma\mone) \int_M \tr\, \bigl(b_+\wedge f_+\bigr) + (1-\gamma\mone) \int_M \tr\, \bigl(b_-\wedge f_-\bigr),
\ee
que nous avons d\'ecompose\'ee en parties self-duale et anti-self-duale, selon $\spin(4)=\su(2)_+\oplus\su(2)_-$. Du fait du param\`etre d'Immirzi, $B_f$ n'est pas le moment canonique de l'holonomie $G_{tt'}$, comme nous l'avons vu en \eqref{gamma-moment}. Celui-ci est en effet,
\beq
J_f(t) = B_f(t) +\f1\gamma \star B_f(t) \quad \Leftrightarrow \quad B_f(t) = \f{\gamma^2}{\gamma^2-1}\Bigl( J_f(t) -\f1\gamma \star J_f(t)\Bigr),
\ee
et la structure symplectique associ\'ee se traduit dans les crochets de Poisson en fonction des $B_e$ par :
\begin{alignat}{3}
&\{ g_{\pm tt'}, g_{\pm t''t'''}\}& &=&\ &0,\\
&\{ b_{\pm f}^i(t), g_{\pm tt'}\}& &=&\ &\f{\gamma}{\gamma\pm 1}\ \tau_\pm^i\,g_{\pm tt'},\\
&\{ b_{\pm f}^i(t), b_{\pm f}^j(t)\}& &=&\ &\f{\gamma}{\gamma\pm 1}\ \eps^{ij}_{\phantom{ij}k}\,b_{\pm f}^k(t).
\end{alignat}
Ces crochets ne sont autres que les crochets issus des variables canoniques $(A_a^{IJ}, P^{(\gamma)a}_{IJ})$, \eqref{gamma-moment}, de la th\'eorie continue, qui donnent naissance \`a l'alg\`ebre des holonomies et des flux, celle-ci \'etant alors tronqu\'ee \`a un seul graphe. (Quant au fait que les moments commutent dans la th\'eorie continue, mais pas ici, on consultera \cite{corichi-zapata-fluxes}. Un tel ph\'enom\`ene est naturel sur le fibr\'e cotangent d'un groupe de Lie, et n\'ecessaire pour satisfaire l'identit\'e de Jacobi). Nous avons utilis\'e les g\'en\'erateurs anti-hermitiens $\tau_\pm^i = -i\sigma^i/2$ des secteurs $\su(2)_\pm$, satisfaisant $[\tau^i_\pm, \tau^j_\pm] = \eps^{ij}_{\phantom{ij}k}\tau^k_\pm$, tandis que $t(e)$ repr\'esente le vertex d'arriv\'ee du lien $e$. Les cas limites $\gamma\rightarrow0$ et $\gamma\rightarrow\infty$ sont int\'eressants car ils correspondent respectivement aux secteurs topologique et gravitationnel. Les moments conjugu\'es aux holonomies sont :
\beq
J_f(t) = \gamma\mone\star B_f(t), \quad \text{et}\qquad J_f(t) = B_f(t).
\ee
Les deux structures symplectiques associ\'ees, remarqu\'ees dans \cite{baez-barrett-quantum-tet}, sont en un sens les structures symplectiques naturelles de $T^\star \Spin(4)$. Elles sont reli\'ees par un \og flip\fg, provoqu\'e par l'op\'erateur de dualit\'e de Hodge : celui-ci change les crochets anti-self-duaux en leurs oppos\'es (\`a un rescaling pr\`es par $\gamma$ au voisinage de 0). Ainsi la non-unicit\'e de la structure symplectique a la m\^eme origine que celle de l'existence de deux formes bilin\'eaires invariantes sur l'alg\`ebre $\spin(4)$, qui permet d'introduire le param\`etre d'Immirzi dans l'action de Palatini-Hilbert, \`a savoir la d\'ecomposition $\spin(4) = \su(2)\oplus\su(2)$. Comme pour former l'action de Holst, la bonne structure symplectique consiste \`a prendre une combinaison des deux situations particuli\`eres ci-dessus avec comme param\`etre $\gamma$.

Les transformations de jauge changent les r\'ef\'erentiels locaux des t\'etra\`edres. Une transformation $K$ agit par une famille d'\'el\'ements $(K_t\in\Spin(4))$ selon :
\begin{align}
&K\,\cdot\, G_{tt'} = K_{t}\,G_{tt'}\,K_{t'}\mone, \\
&K\,\cdot\, B_f(t) = \Ad\bigl(K_t\bigr)\,B_f(t),
\end{align}
en notation matricielle, $\Ad$ \'etant l'adjointe (ici sur $\su(2)\oplus\su(2)$). L'invariance de jauge se traduit au niveau canonique par la contrainte de Gau\ss,
\beq
\sum_{f\subset \pp t} B_f(t) = 0, \label{gauss closure}
\ee
qui g\'en\`ere via les crochets de Poisson les transformations lin\'earis\'ees. Physiquement, cette contrainte n'est autre que la relation de fermeture du t\'etra\`edre $t$ !

Comme dans le continu, des choix de $B_f$ ou des moments $J_f$ quelconques ne peuvent d\'ecrire la g\'eom\'etrie de la triangulation, y compris celle d'un t\'etra\`edre plat ! Autrement dit, nous souhaitons contraindre ces variables, avec comme objectif de retrouver deux types de configurations, et de comprendre en quoi elles diff\`erent : l'espace des phases $\SU(2)$ de la LQG sur le graphe, et les g\'eom\'etries de Regge de la triangulation. Dans ce dernier cas, les variables $B_f$ sont contraintes de mani\`ere \`a provenir de 4-vecteurs $E^I_\ell(t)$ d\'ecrivant la direction (dans l'espace interne) et la longueur des ar\^etes $\ell$ de la triangulation. Cela revient donc \`a obtenir une projection consistente de la cot\'etrade sur la triangulation 3d, de sorte que : $(\star B_f(t))^{IJ} = E_{\ell_1}^{[I} E_{\ell_2}^{J]}$, si $\ell_1, \ell_2$ sont deux ar\^etes du triangle dual \`a $e$ (et les crochets indiquant l'antisym\'etrisation).

Les contraintes permettant cela peuvent se trouver directement au niveau discret \cite{baez-barrett-quantum-tet} ou \^etre tir\'ees d'une discr\'etisation directe des contraintes de simplicit\'e utilis\'ees dans le formalisme hamiltonien \eqref{hamiltonian simplicity}, $C^{ab} = \Tr(\star P^a\,P^b) = 0$. Lorsque ces contraintes sont discr\'etis\'ees sur un unique triangle, elles donnent les contraintes \emph{diagonales}, et sur deux triangles d'un m\^eme t\'etra\`edre les contraintes dites \emph{crois\'ees} :
\begin{alignat}{2}
&\text{Simplicit\'e diagonale}&&\qquad\qquad\qquad \Tr\,\bigl(\star B_f(t)\, B_f(t)\bigr) = 0, \label{diag3d}\\
&\text{Simplicit\'e crois\'ee}&&\qquad\qquad\qquad \Tr\,\bigl(\star B_f(t)\, B_{f'}(t)\bigr) = 0. \label{off-diag3d}
\end{alignat}
La r\'esolution de ces contraintes souffre de la m\^eme ambiguit\'e que dans le continu : des bivecteurs d'un t\'etra\`edre les satisfaisant, ainsi que la relation de fermeture, s'\'ecrivent \cite{baez-barrett-quantum-tet} : $B_f^{IJ}(t) = \pm E_{\ell_1}^{[I}(t) E_{\ell_2}^{J]}(t)$ ou bien $B_f^{IJ}(t) = \pm \f12 \eps^{IJ}_{\phantom{IJ}KL}E_{\ell_1}^{[I}(t) E_{\ell_2}^{J]}(t)$, ce qui correspond \`a nouveau aux secteurs topologique et gravitationnel. Des bivecteurs de cette forme sont appel\'es des bivecteurs \emph{simples}.

Mais ces contraintes qui permettent de reconstruire la g\'eom\'etrie de chaque t\'etra\`edre ne sont pas suffisantes pour recoller ces derniers de mani\`ere consistente ! Cela peut para\^itre surprenant car les contraintes du continu fonctionnent tr\`es bien. En revanche, on sait que la stabilisation des contraintes de simplicit\'e par le hamiltonien engendre des contraintes secondaires, de la forme $B\,B\,\star(D_AB)=0$. On peut imaginer que des contraintes similaires vont permettre dans le discret d'assurer la g\'eom\'etricit\'e (avant m\^eme de regarder la dynamique). C'est ce que montre Dittrich et Ryan \cite{dittrich-ryan-simplicial-phase}, en compl\'etant \eqref{diag3d} et \eqref{off-diag3d} par :
\beq \label{edge simplicity}
\Tr\,\bigl(\star B_{f}(t)\ \Ad\bigl(G_{tt'}\bigr)\,B_{f'}(t')\bigr) = 0.
\ee
Cette contrainte concerne des triangles $f$ et $f'$ ayant une ar\^ete commune et dans le bord de t\'etra\`edres voisins $t$ et $t'$. Notons que comme les contraintes secondaires de la th\'eorie continue, celle-ci n\'ecessite l'intervention de la connexion/transport parall\`ele, pour \og transporter \fg{} le triangle $f'$ dans le r\'ef\'erentiel de $t$. Cette condition assure que tous les bivecteurs partageant une ar\^ete de la triangulation engendrent un sous-espace (de l'espace interne $\R^4$) de dimension 3 (au plus), du fait que tous ces bivecteurs doivent \^etre orthogonaux \`a l'ar\^ete en commun. Ce type d'interpr\'etation sera tr\`es utile pour les mousses de spins, section \ref{sec:classical sf}.

Pour \'etudier l'alg\`ebre des contraintes et l'espace des phases r\'eduit, i.e. des g\'eom\'etries simplicielles, en d\'etails (ce que nous ne pr\'esenterons pas), les auteurs de \cite{dittrich-ryan-simplicial-phase} introduisent des variables scalaires, invariantes de jauge. Formons d'abord :
\beq
A_{\pm f} = \lv \vec{b}_{\pm f}\rv, \quad \text{et} \qquad \cos\phi_{\pm ff'} = \f{\vec{b}_{\pm f}(t)\cdot\vec{b}_{\pm f'}(t)}{A_{\pm f}\,A_{\pm f'}},
\ee
en utilisant la notation vectorielle sur $\su(2)\simeq\R^3$. Les contraintes \eqref{diag3d} et \eqref{off-diag3d} se r\'e\'ecrivent exactement
\beq
A_{+f} - A_{-f} = 0, \quad\text{et}\qquad \cos\phi_{+ff'} - \cos\phi_{-ff'} = 0.
\ee
$A_{\pm f}$ et $\phi_{\pm ff'}$ s'interpr\`etent alors comme les aires des triangles, et les angles dih\'edraux 3d entre les triangles $f$ et $f'$. Pour prendre en compte les holonomies, on forme le produit scalaire entre les \og normales\fg{} $\vec{N}_{\pm ff'}(t) = \vec{b}_{\pm f}(t)\wedge \vec{b}_{\pm f'}(t)$ (toujours avec la notation vectorielle, et le produit vectoriel) :
\beq \label{4d angle momentum}
\cos\theta_{\pm t't''} = \f{\vec{N}_{\pm ff'}(t')\cdot R(g_{\pm t't''}) \vec{N}_{\pm f''f}(t'')}{\sqrt{\vec{N}_{\pm ff'}^2\,\vec{N}_{\pm f''f}^2}},
\ee
o\`u $R(g)$ est simplement la repr\'esentation vectorielle de $\SU(2)$. Les triangles $f, f'$ (et $f, f''$) sont dans le t\'etra\`edre $t'$ ( $t''$). Ces deux t\'etra\`edres partagent le triangle $f$. A priori, cette quantit\'e d\'epend du choix des triangles utilis\'es pour la calculer, mais ce n'est en fait pas le cas dans le secteur gravitationnel lorsque les contraintes sont satisfaites. La contrainte additionnelle \eqref{edge simplicity} \emph{implique, mais n'est pas \'equivalente \`a} :
\beq
\cos\theta_{+ tt'} - \cos\theta_{-tt'} = 0,
\ee
La variable $\theta_{\pm tt'}$ s'interpr\`ete alors naturellement comme l'angle dih\'edral entre $t$ et $t'$ \cite{dittrich-ryan-simplicial-phase} (mais c'est \`a ce stade une variable ind\'ependante des aires et des angles $\phi_{ff'}$).

Ainsi, nous sommes partis de l'espace des phases des holonomies-flux de $\Spin(4)$ sur un graphe, r\'eexprim\'e avec les variables invariantes de jauge $A_{\pm f}, \phi_{\pm ff'}$ et $\theta_{\pm tt'}$. Alors les contraintes de simplicit\'e r\'eduisent cet espace \`a un seul secteur ind\'ependant, disons le secteur self-dual. C'est pr\'ecis\'ement l'espace des phases, en variables scalaires, provenant des holonomies et flux de $\SU(2)$ sur le graphe. Autrement dit, nous arrivons au m\^eme espace des phases en partant de l'alg\`ebre holonomies-flux de la th\'eorie BF $\SU(2)$, celle qu'utilise la LQG non-r\'eduite, tronqu\'ee par le graphe fix\'e. Avec ces variables, $b_f(t)\in\su(2), g_{tt'}\in\SU(2)$, nous pouvons en effet former directement les aires $A_f$, angles 3d $\phi_{ff'}$, et angles 4d $\theta_{tt'}$.

La question est alors de savoir si tout le contenu des contraintes \eqref{edge simplicity} a bien \'et\'e pris en compte. Ce n'est pas le cas, m\^eme pour une triangulation de cinq t\'etra\`edres formant le bord d'un 4-simplexe, comme on peut le voir par un simple argument de comptage ! On s'attend dans cette situation \`a un espace des phases correspondant aux g\'eom\'etries de Regge, form\'ee de 20 variables, typiquement les longueurs des 10 ar\^etes et leurs moments conjugu\'es. Avec les variables pr\'ec\'edentes, nous avons 10 aires (pour les 10 triangles), 10 angles dih\'edraux 4d, ainsi que 10 angles dih\'edraux 3d ind\'ependants. En effet, la g\'eom\'etrie d'un t\'etra\`edre est d\'etermin\'ee par six variables, typiquement les 4 aires, et 2 angles dih\'edraux non-oppos\'es (ce qui fait bien 10 angles 3d au bord du 4-simplexe). Cela fait donc 10 variables en trop ! L'id\'ee principale est la m\^eme que dans la proposition de calcul de Regge en variables aires-angles 3d \cite{dittrich-speziale-aarc}, expos\'ee \`a la section \ref{sec:regge calculus}. Consid\'erons deux t\'etra\`edres voisins. Les variables $A_f$ et $\phi_{ff'}$ permettent de calculer toutes les longueurs dans chaque t\'etra\`edre ind\'ependamment, mais rien ne garantit qu'elles attribuent les m\^emes longueurs aux trois ar\^etes du triangle en commun ! C'est en ce sens qu'on parlera de g\'eom\'etrie \emph{tordue}, comme \'etant la g\'eom\'etrie cod\'ee par les holonomies-flux de $\SU(2)$ sur un graphe fix\'e. Ce type de g\'eom\'etrie correspond \`a une m\'etrique plate par morceaux, comme en calcul de Regge, mais discontinue d'un t\'etra\`edre \`a l'autre !

Pour passer d'une g\'eom\'etrie tordue \`a une g\'eom\'etrie simplicielle authentique, avec une m\'etrique de Regge continue, des contraintes suppl\'ementaires sont n\'ecessaires, assurant que les triangles ont la m\^eme forme dans les deux t\'etra\`edres voisins. Si l'on peut pour cela essayer de se ramener aux variables de longueurs, d'autres choix sont possibles. En particulier, dans \cite{dittrich-speziale-aarc}, il est propos\'e de demander que les angles dih\'edraux 2d entre les ar\^etes des triangles, comme fonctions des angles 3d, prennent les m\^emes valeurs quelque soit le t\'etra\`edre utilis\'e pour le calcul. L'expression standard des angles 2d en fonction des angles 3d conduit \`a choisir 10 contraintes du type \eqref{2dconstraints-aarc}. Un autre choix, peut-\^etre plus naturel, consiste \`a se dire que les contraintes de recollement recherch\'ees doivent permettre de calculer toute la g\'eom\'etrie intrins\`eque de mani\`ere consistente en fonction des longueurs, ou bien des aires \footnote{Une triangulation a g\'en\'eriquement plus de triangles que d'ar\^etes, car si chaque triangle a exactement trois ar\^etes, chaque ar\^ete est partag\'ee par au moins triangles.}, de sorte que l'on peut choisir des contraintes
\beq \label{constraint 3dangle-area}
\cos\phi_{ff'} = F_{ff'}(A),
\ee
qui expriment les angles 3d en termes des aires. Malheureusement, il n'existe pas de d'expressions explicites pour ces fonctions $F_{ff'}$. Elles sont par ailleurs \emph{non-locales}, par opposition \`a \eqref{2dconstraints-aarc} qui n'implique que les variables de deux t\'etra\`edres voisins, et au sens o\`u elles n\'ecessitent a priori l'utilisation des 10 aires du 4-simplexe pour calculer chaque angle 3d.

Terminons cette section en mentionnant la structure des crochets de Dirac issus de la r\'eduction par les contraintes de simplicit\'e \cite{dittrich-ryan-simplicial-phase}. Comme en LQG dans la jauge \og temps\fg, les crochets entre la triade (ici les aires) et le moment conjugu\'e, ici les angles dih\'edraux 4d, sont proportionnels au param\`etre d'Immirzi,
\beq
\{ A_f, \theta_{tt'} \} = \sqrt{2}\,\gamma\ \delta_{f, (tt')},
\ee
La notation $\delta_{f, (tt')}$ signifie que les crochets sont nuls sauf si $f$ est le triangle commun \`a $t$ et $t'$. De plus, les aires commutent entre elles, de m\^eme que les angles 3d du fait des contraintes \eqref{constraint 3dangle-area} qui les expriment comme fonctions des aires. Quant aux crochets entre les angles 4d, ils nont pas \'et\'e calcul\'es explicitement, mais les auteurs de \cite{dittrich-ryan-simplicial-phase} conclut \`a leur annulation par un argument indirect. Un tel ph\'enom\`ene serait tr\`es int\'eressant \`a comparer avec l'approche $\Spin(4)$ covariante de l'analyse hamiltonienne, dans la th\'eorie continue, dans laquelle la connexion ne commute plus avec elle-m\^eme pour les crochets de Dirac.

\section{La contrainte de courbure nulle} \label{sec:flatness}

Ainsi, nous disposons d'une \'etude d\'etaill\'ee de l'espace des phases des g\'eom\'etries discr\`etes de la LQG, au niveau cin\'ematique. Et il nous faudrait maintenant des \'equivalents des contraintes vectorielles et scalaire, imposant l'invariance sous les diff\'eomorphismes, et qui soient de premi\`ere classe. Mais de telles contraintes ne sont pas connues et il n'est pas clair qu'elles existent ! A dire vrai, il manque une bonne compr\'ehension des diff\'eomorphismes sur les espace-temps discrets. C'est ce qui emp\^eche de raisonner g\'eom\'etriquement pour construire la contrainte scalaire sur les r\'eseaux de spins en LQG, et c'est \'egalement une difficult\'e pour comprendre quelle dynamique engendre les mod\`eles de mousses de spins. Dans ce dernier cadre, de plus, le contr\^ole des sym\'etries a un r\^ole crucial pour discuter la pr\'esence d'anomalies \cite{bojowald-perez-anomalies} et les divergences, associ\'ees aux bulles. Par exemple, dans le mod\`ele topologique de Ponzano-Regge pour la gravit\'e 3d, des divergences proviennent d'une sym\'etrie de jauge au niveau discret, qui d\'eplace les sommets de la triangulation, ce qui se comprend bien du fait que l'espace-temps est strictement plat. Cette sym\'etrie peut \^etre vue comme une discr\'etisation de la sym\'etrie de jauge de translation \eqref{translation sym bf} de la th\'eorie BF continue \cite{freidel-louapre-diffeo}, dont on sait qu'elle est \'equivalente aux diff\'eomorphismes !

On pourrait tenter d'utiliser le calcul de Regge comme guide, \`a nouveau, mais aucune alg\`ebre ferm\'ee n'est connue en calcul de Regge. Il semble m\^eme que les sym\'etries de jauge soient g\'en\'eriquement bris\'ees en calcul de Regge, comme le montre les travaux de Dittrich et ses collaborateurs \cite{bahr-dittrich-broken-symmetries, bahr-dittrich-perfect-regge, dittrich-hohn-linearized-rc}(nous y reviendrons \`a la section suivante) ! N\'eanmoins, pour des solutions particuli\`eres des \'equations de Regge, on peut s'attendre \`a trouver une sym\'etrie de jauge. Typiquement, une solution plate reste inchang\'ee si on modifie les longueurs des ar\^etes autour d'un sommet en d\'epla\c{c}ant celui-ci. La sym\'etrie \'etant intuitive, il reste \`a trouver des contraintes associ\'ees \`a cette g\'eom\'etrie plate.

A partir de travaux similaires en 3d de Dittrich et Freidel (non parus), Dittrich et Ryan \cite{dittrich-ryan-simplicial-phase} proposent un ensemble de contraintes pour la courbure nulle applicable aux triangulations de la 3-sph\`ere pouvant \^etre vues comme le bord d'une triangulation 4d ne poss\'edant pas de triangles internes. Pour voir l'id\'ee principale, regardons simplement la triangulation de bord d'un unique 4-simplexe. On souhaite \`a partir des variables d'aires $A_f$ et d'angles 4d $\theta_{tt'}$ construire un 4-simplexe \emph{plat}. Un tel simplexe est bien connu, et construit en demandant que les angles $\theta_{tt'}$ soient bien les angles dih\'edraux $\thet_{tt'}(A)$ d\'etermin\'es par les longueurs ou les aires,
\beq \label{flat dynamics 4-simplex}
\theta_{tt'} \,-\, \thet_{tt'}(A_f) \,=\, 0.
\ee
Pour comprendre les transformations engendr\'ees par ces relations, les auteurs de \cite{dittrich-ryan-simplicial-phase} utilisent d'abord la contrainte de courbure nulle du mod\`ele topologique, $G_\ell =\unit$, pour les holonomies $\Spin(4)$ autour des ar\^etes. Ils construisent des combinaisons de cette contrainte dont les flots laissent les contraintes de simplicit\'e invariantes sur l'espace des connexions plates. Ces combinaisons engendrent des d\'eplacements des sommets de la triangulation, et sont de premi\`ere classe pour les crochets de Dirac. Ensuite, ils montrent que \eqref{flat dynamics 4-simplex} est form\'e de combinaisons invariantes de jauge de ces g\'en\'erateurs de transation des sommets.

Cela laisse envisager la possibilit\'e d'un traitement complet de la dynamique des g\'eom\'etries plates, du classique au quantique, et du formalisme canonique pour les g\'eom\'etries de Regge au formalisme plus covariant des mousses de spins. Cela nous am\`enera \`a consid\'erer des relations de r\'ecurrence sur les amplitudes de mousses de spins comme traduisant les sym\'etries classiques au niveau quantique dans les mod\`eles topologiques, partie \ref{sec:recurrence}. Pour aller plus loin et voir vraiment, i.e. alg\'ebriquement, ces relations de r\'ecurrence comme des quantifications des contraintes g\'eom\'etriques \eqref{flat dynamics 4-simplex}, ce que nous allons faire pour la gravit\'e 3d, nous avons besoin d'\'etablir une correspondance directe entre les contraintes en variables scalaires, \eqref{flat dynamics 4-simplex} et la contrainte de courbure nulle en variables covariantes (holonomies).

Pour initier un tel programme, nous simplifions les choses en regardant la th\'eorie topologique en 3d, et nous pr\'esentons un r\'esultat simple mais efficace (pas encore publi\'e). Nous consid\'erons une triangulation 2d de la surface canonique $\Sigma$. Le graphe dual est form\'e de faces duales aux sommets, de liens duaux aux ar\^etes, et de vertexes duaux aux triangles. Les holonomies $g_e$ permettent le transport parall\`ele $SU(2)$ entre triangles adjacents, i.e. le long des liens duaux $e$. La courbure se concentre autour des sommets de la triangulation, et est cod\'ee par l'holonomie $g_f = \prod_{e\subset f} g_e$ le long du bord de la face duale $f$. La 1-forme $e_a^i$, \eqref{3d moment}, se discr\'etise en un vecteur $\vec{l}_e(v)$ sur les ar\^etes, dans le r\'ef\'erentiel d'un triangle dual au vertex $v$, qui d\'ecrit la direction de l'ar\^ete dans $\R^3$, et dont la norme est la longueur de l'ar\^ete $l_e$. La relation de fermeture des triangles s'\'ecrit :
\beq
\vec{l}_{e_1}(v) + \vec{l}_{e_2}(v) + \vec{l}_{e_3}(v) \,=\,0,
\ee
pour les trois liens attach\'es au vertex $v$ (et \`a des signes d'orientation pr\`es). La structure symplectique est celle du fibr\'e cotangent $T^*\SU(2)$ sur chaque lien, pour laquelle la relation de fermeture est la contrainte de Gau\ss. Il faut bien s\^ur relier les directions d'une m\^eme ar\^ete vue dans les r\'ef\'erentiels de triangles voisins. C'est le r\^ole du transport parall\`ele :
\beq \label{transport 3d vectors}
\vec{l}_e(v) \,=\, R(g_e)\,\vec{l}_e(v'),
\ee
si le lien $e$ va du vertex $v'$ \`a $v$. $R(g)$ d\'esigne la repr\'sentation vectorielle de $\SU(2)$ (qui se confond avec l'adjointe en regardant les vecteurs $\vec{l}$ dans l'alg\`ebre $\su(2)$). Il n'y a bien s\^ur pas de contraintes de seconde classe (type simplicit\'e) en 3d. La contrainte de courbure nulle (hamiltonienne) est :
\beq \label{3d canonical simplicial flatness}
g_f = \prod_{e\subset f} g_e = \mathbbm{1}.
\ee
Elle est de premi\`ere classe : elle engendre des translations des sommets de la triangulation dans $\R^3$.

Nous souhaitons r\'eexprimer cette contrainte en termes de variables g\'eom\'etriques \`a la Regge. Pour cela, nous allons introduire une param\'etrisation particuli\`ere, qui sera aussi utilis\'ee dans les mousses de spins (et qui d\'ecrit bien les g\'eom\'etries tordues en 4d, voir \cite{freidel-speziale-twisted-geom, rovelli-speziale-lqg-regge}). Techniquement, l'id\'ee est de trouver un cadre interm\'ediaire entre les variables $(g_e, \vec{l}_e)$ et les variables scalaires invariantes de jauge, en \og rentrant\fg{} ces derni\`eres dans des \'el\'ements de groupe. On distingue explicitement les normes des directions des vecteurs $\vec{l}_e(v)$. Soit $n_{ev}\in\SU(2)$ une rotation qui envoie l'axe de r\'ef\'erence $\hat{z}=(0,0,1)$ de $\R^3$ sur la direction de $\vec{l}_e$,
\beq
\vec{l}_e(v) \,=\, l_e\ R(n_{ev})\,\hat{z}.
\ee
On voit bien que seuls deux des trois param\`etres de $n_{ev}$ sont d\'efinis par $\vec{l}_e(v)$ : $n_{ev}$ n'est d\'efinie qu'\`a une rotation pr\`es autour de $\hat{z}$. Dans la param\'etrisation d'Euler, $n=e^{-\f i2\alpha\sigma_z}\,e^{-\f i2\beta\sigma_y}\,e^{-\f i2\gamma\sigma_z}$, l'angle $\gamma$ n'est pas physique. Il faudra donc prendre soin de cette ambiguit\'e. En particulier, la relation de transport parall\`ele \eqref{transport 3d vectors} peut se r\'e\'ecrire en utilisant les rotations $n_{ev}$ au prix de l'introduction d'angles suppl\'ementaires $\xi_e$,
\beq
n_{ev} \,=\, g_e\,n_{ev'}\,e^{\f i2\xi_e\sigma_z}.
\ee
Autrement dit, la r\'esolution des \'equations de transport parall\`ele pour les holonomies en fonction des vecteurs $\vec{l}_e(v), \vec{l}_e(v')$ laisse un degr\'e de libert\'e, $\xi_e$ :
\beq
g_e \,=\, n_{ev}\,e^{-\f i2\xi_e\sigma_z}\,n_{ev'}\mone.
\ee
Ce formalisme sera plus largement d\'evelopp\'e aux sections \ref{sec:gluing-aarc} et \ref{sec:pathint fk}, en 4d. Avant d'injecter ces expressions des holonomies dans la contrainte \eqref{3d canonical simplicial flatness}, regardons les variables scalaires desquelles nous aimerions nous rapprocher.

Les variables invariantes de jauge $\SU(2)$ sont les longueurs $l_e$, et leur moment conjugu\'e $\theta_e$, construit comme en \eqref{4d angle momentum} : on forme une normale au triangle (dual \`a $v$) par le produit vectoriel des deux ar\^etes, $\vec{N}_{ee'}(v) = \vec{l}_e(v)\wedge \vec{l}_{e'}(v)$. L'angle $\theta_e$ est donn\'e par le produit scalaire des normales \`a deux triangles voisins, mais pour comparer ces normales, il faut en transporter une dans le r\'ef\'erentiel de l'autre triangle,
\beq
\cos\theta_e = -\f{\vec{N}_{ee'}(v')\,\cdot\,R(g_e)\,\vec{N}_{ee''}(v'')}{\sqrt{\vec{N}^2_{ee'}\,\vec{N}^2_{ee''}}}.
\ee
Ici $e$ est le lien joignant les vertexes $v'$ et $v''$. Ces moments ne sont pas fix\'es par les longueurs : ils d\'ependent des holonomies. Les angles dih\'edraux 2d, entre ar\^etes d'un triangle, en revanche, sont compl\`etement d\'etermin\'es par les longueurs via la relation de fermeture,
\beq
\cos\phi_{ee'}^v \,=\, -\eps_{ee'}\, \f{\vec{l}_e(v)\cdot\vec{l}_{e'}(v)}{l_e\,l_{e'}}.
\ee
Le signe $\eps_{ee'}=\pm 1$ est positif si les liens duaux $e,e'$ sont tous deux entrants ou sortant du vertex $v$, et n\'egatif sinon. A ce stade, remarquons que les rotations $n_{ev}$ qui codent les directions des ar\^etes fournissent une expression tr\`es simple des angles 2d,
\beq
\cos\phi_{ee'}^v \,=\, -\eps_{ee'}\ \langle 1,0\lv n_{ev}\mone\,n_{e'v}^{\phantom{}}\rv 1, 0\rangle,
\ee
via la repr\'esentation de spin 1, dans l'\'etat $\lv 1, 0\rangle$ invariant sous le sous-groupe $\U(1)$ selon $\hat{z}$. Ainsi, si les rotations $n_{ev}$ ne sont bien s\^ur pas d\'etermin\'ees par les longueurs, l'angle $\beta$ d'Euler dans la d\'ecomposition des produits $n_{ev}\mone\,n_{e'v}$ l'est. Nous \'ecrivons :
\beq
n_{ev}\mone\,n_{e'v}^{\phantom{}} = e^{-\f i2 \alpha_{ee'}^v\sigma_z}\, e^{-\f i2 \phi_{ee'}^v\sigma_z}\, e^{-\f i2 \gamma_{ee'}^v\sigma_z},
\ee
ce qui d\'efinit les angles $\alpha_{ee'}^v$ et $\gamma_{ee'}^e$. Une question est de savoir comment ces angles, et les angles $\xi_e$, interviennent dans les quantit\'es physiques. Nous allons pr\'esenter une r\'eponse.

\begin{figure}
\begin{center}
\includegraphics[width=7cm]{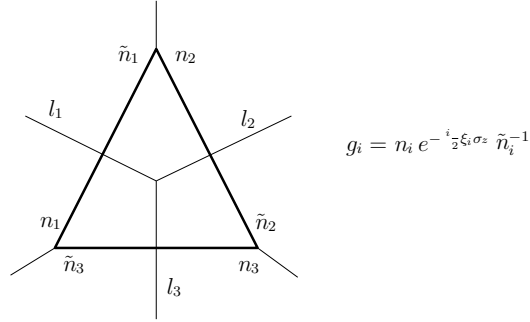}
\end{center}\caption{ \label{fig:3-valent-flatness} Un bout de triangulation 2d et de la d\'ecomposition duale. Les trois ar\^etes se rencontrant au centre font partie de la triangulation, et ont des longueurs $l_1, l_2, l_3$. En gras sont dessin\'es les liens duaux qui portent des holonomies faisant le lien entre les r\'ef\'erentiels de diff\'erents triangles. Comme on le voit, un sommet de la triangulation est dual \`a une 2-cellule, ici triangulaire car le sommet est trivalent. La contrainte de courbure nulle s'applique ainsi \`a cette plaquette. Les variables $n_e$ et $\tilde{n}_e$ sont attach\'ees aux extr\'emit\'es des liens duaux et codent la direction des ar\^etes dans les diff\'erents r\'ef\'erentiels. Elles peuvent aussi param\'etrer les holonomies selon la formule donn\'ee \`a droite.
}
\end{figure}

Nous regardons maintenant la contrainte de courbure nulle en fonction de ces variables pour en d\'egager une interpr\'etation (de la contrainte et des variables \`a la fois !). On ne s'attend bien s\^ur pas \`a ce que cette contrainte impose une courbure nulle sur la triangulation 2d, mais plut\^ot qu'elle nous donne des relations entre courbures intrins\`eque et extrins\`eque. Concentrons-nous sur une face duale triangulaire, i.e. que trois triangles se rencontrent deux \`a deux le long d'ar\^etes, et tous les trois autour d'un sommet. Notons les rotations $n_{ev}$ aux deux extr\'emit\'es d'un lien $n_e$ et $\tl{n}_e$. Nous avons alors selon la figure \ref{fig:3-valent-flatness} :
\beq
n_1\,e^{-\f i2 \xi_1\sigma_z}\,\tl{n}_1\mone\,n_2\,e^{-\f i2 \xi_2\sigma_z}\,\tl{n}_2\mone\,n_3\,e^{-\f i2 \xi_3\sigma_z}\,\tl{n}_3\mone\,=\,\mathbbm{1}.
\ee
Nous r\'e\'ecrivons cela comme :
\beq
e^{\f i2 \xi_1\sigma_z}\,\tl{n}_1\mone\,n_2\,e^{\f i2 \xi_2\sigma_z} \,=\, \bigl(n_1\mone \tl{n}_3\bigr)\, e^{\f i2 \xi_3\sigma_z}\,\bigl(n_3\mone \tl{n}_2\bigr).
\ee
Il suffit alors de projeter cette \'equation sur l\'el\'ement de matrice diagonal dans l'\'etat $\lv 1, 0\rangle$, et d'utiliser explicitement la repr\'esentation de spin $1$ (voire appendice, \ref{sec:app}) pour aboutir \`a :
\beq
\cos\phi_{12} = \cos\phi_{23} \cos\phi_{31} - \sin\phi_{23} \sin\phi_{31}\,\cos\bigl(\xi_3 +\alpha_{31}+\gamma_{23}\bigr).
\ee
Pour des triangles non-d\'eg\'en\'er\'es, on peut alors extraire l'angle $(\xi_3 +\alpha_{31}+\gamma_{23})$. La surprise est que cela produit la forme de la relation donnant les angles dih\'edraux 3d en fonction des angles dih\'edraux 2d ! De plus, on sait que les angles $\phi_{ee'}$  sont vraiment les angles 2d, c'est-\`a-dire fonctions des longueurs gr\^ace \`a la relation de fermeture. De la sorte, nous voyons que cette contrainte impose \`a la combinaison $(\xi_3 +\alpha_{31}+\gamma_{23})$ d'\^etre l'angle dih\'edral entre les triangles duaux aux vertexes $v_{23}$ et $v_{31}$ (cette paire de triangles est caract\'eris\'ee par l'ar\^ete qu'ils partagent, duale au lien $e_3$), d\'etermin\'e par les longueurs ! Explicitement :
\beq
\cos\bigl(\xi_3 +\alpha_{31}+\gamma_{23}\bigr) = - \f{\cos\phi_{12} - \cos\phi_{23}\,\cos\phi_{31}}{\sin\phi_{23}\ \sin\phi_{31}} = -\cos\thet_{3}(l).
\ee
Ainsi, la contrainte des g\'eom\'etries plates imposent une relation entre la courbure intrins\`eque, donn\'ee par les angles $\phi_{ee'}$, et la courbure extrins\`eque, cod\'ee par les angles dih\'edraux 3d : ces derniers sont d\'etermin\'es par les premiers, selon la formule permettant de former des 3-simplexes plats (des t\'etra\`edres, voir plus bas). Cela confirme donc l'id\'ee de la contrainte \eqref{flat dynamics 4-simplex} (en 4d) pour les g\'eom\'etries plates, en termes des variables scalaires, tout en sugg\'erant une forme l\'eg\`erement diff\'erente, avec les cosinus :
\beq \label{flat dynamics 3d scalar}
\cos\theta_e \,-\,\cos\thet_e(l) \,=\,0.
\ee
Il faudrait alors \'etudier ces contraintes plus pr\'ecis\'ement (comme cela a apparemment \'et\'e fait par Dittrich et Freidel dans des travaux encore non parus).

\medskip

Ce r\'esultat nous permet de discuter l'influence de la triangulation d'une fa\c{c}on plus g\'eom\'etrique que dans \cite{ooguri-3d}, dans un mod\`ele topologique certes. Cette question en contient en fait deux : le choix de la triangulation de la surface canonique, et si celle-ci est un bord, le choix de la triangulation \`a l'int\'erieur. Pour le premier point, nous pouvons voir dans le mod\`ele en gravit\'e 3d qu'en fait les contraintes \eqref{flat dynamics 3d scalar} permettent d'effectuer pour une face duale triangulaire un mouvement de Pachner 3-1 ! En effet, ce mouvement consiste \`a changer un morceau d'une triangulation 2d pour obtenir une triangulation de la m\^eme vari\'et\'e topologique, en transformant trois triangles autour d'un sommet en un seul triangle, dont les ar\^etes sont celles des triangles initiaux non-connect\'ees au sommet \'elimin\'e. A un tel sommet, nous avons vu que les angles dih\'edraux de la courbure extrins\`eque se calculent, via la contrainte de courbure nulle, comme les angles dih\'edraux d'un t\'etra\`edre plat en fonction des longueurs. Pour visualiser le t\'etra\`edre correspondant, il suffit de consid\'erer les trois triangles initiaux, formant le \og haut\fg{} du t\'etra\`edre et d'y ajouter une \og base\fg{} : un triangle form\'e des trois ar\^etes des triangles initiaux ne touchant pas le sommet. Alors, on ne change clairement pas la g\'eom\'etrie et la physique en \'eliminant les trois triangles et le sommet qui les attache, et en consid\'erant \`a la place le triangle formant la base du t\'etra\`edre plat. On s'attend de la m\^eme fa\c{c}on \`a pouvoir faire des mouvements de Pachner 2-2 en transformant des deux triangles voisins. De plus amples recherches sur le sujet sont en cours.

Revenons \`a une surface canonique telle que la 3-sph\`ere, comme dans \cite{dittrich-ryan-simplicial-phase}. Se pose alors la question du changement de la triangulation int\'erieure. Du point de vue des variables scalaires, combien de solutions aux \'equations de Regge y a-t-il pour des longueurs fix\'ees au bord, et comment se comporte les solutions dans les changements de triangulations ? On peut typiquement effectu\'e un mouvement de Pachner 1-5 (en 4d), divisant un 4-simplexe (plat) en cinq 4-simplexes. Une solution est bien s\^ur la solution plate, mais il est possible que des solutions avec courbure existent.

Les mod\`eles topologiques, comme celui pr\'esent\'e ci-dessus sont des terrains int\'eressants pour \'elaborer des m\'ethodes en vue d'illustrer ces questions et pour \'etudier des dynamiques simplifi\'ees. Il s'agit bien s\^ur de situations particuli\`eres : dire que le mod\`ele est topologique, c'est dire qu'il ne d\'epend que de la topologie de la vari\'et\'e, et est donc invariant sous les changements de la triangulation. Cette sym\'etrie est bien s\^ur le coeur de ces mod\`eles. Mais en faisant le lien avec les variables g\'eom\'etriques du calcul de Regge, nous pouvons esp\'erer aller plus loin. A l'heure actuelle, nous pouvons pr\'esenter diff\'erents r\'esultats, partie \ref{sec:recurrence} sur le mod\`ele de gravit\'e 3d. En ce qui concerne l'invariance de la triangulation int\'erieure, nous verrons que le mouvement de Pachner 2-3 (transformant deux t\'etra\`edres en trois) se traduit par une relation de r\'ecurrence qui s'interpr\`ete directement comme une quantification des contraintes \eqref{flat dynamics 3d scalar} ! De plus, le mouvement de Pachner 1-4 est g\'en\'er\'e par l'action des modes de Fourier $\SU(2)$ de la contrainte \eqref{3d canonical simplicial flatness} sur un r\'eseau de spins associ\'e \`a un t\'etra\`edre. En prenant le point de vue du bord, nous pourrons g\'en\'eraliser cela en \'ecrivant des relations de r\'ecurrence exactes sur les amplitudes correspondant \`a des \'evolutions sous des mouvements \og tente\fg. Avant cela, nous allons pr\'esenter ces mouvements \og tente\fg{} et montrer comment ils permettent de d\'efinir une formulation canonique du calcul de Regge !



\section{Contraintes hamiltoniennes en calcul de Regge} \label{sec:classical-tent}

Je pr\'esente ici, de mani\`ere succinte quelques \'el\'ements du travail de Bahr et Dittrich, \cite{bahr-dittrich-broken-symmetries}, dont le but est d'\'etudier les sym\'etries du calcul de Regge, avec en toile de fond la recherche de la sym\'etrie sous les diff\'eomorphismes au niveau discret. Un point que nous retrouverons dans la suite est l'\'evolution de la surface canonique triangul\'ee sous des transformations appel\'ees \og mouvement tente\fg. Ces mouvements d\'efinissent un formalisme hamiltonien qui reproduit exactement la dynamique de l'action. Les auteurs argumentent pour montrer que le probl\`eme bien connu de trouver une alg\`ebre des contraintes ferm\'ee et consistente dans les th\'eories de gravit\'e discr\`ete revient \`a trouver une action b\'en\'eficiant d'une notion d'invariance exacte sous les diff\'eomorphismes. Plus concr\`etement, il est montr\'e qu'une solution des \'equations de Regge impliquant de la courbure ne jouit g\'en\'eralement pas de sym\'etries exactes. Dans le formalisme hamiltonien, cela signifie qu'il n'existe pas de contraintes, mais plut\^ot des pseudo-contraintes !

Les mouvements de type tente, introduits par Sorkin, permettent de d\'efinir une \'evolution \emph{locale et en temps discret} d'une hypersurface triangul\'ee, dont l'un des aspects est que les triangulations g\'en\'er\'ees sont identiques \`a celles de d\'epart. Ainsi, le nombre de variables canoniques est inchang\'e, ce qui repr\'esente un avantage par rapport \`a d'autres outils autorisant ce type d'\'evolution en temps discret comme les mouvements de Pachner. Consid\'erons donc une triangulation $\Delta_{(d-1)}$, $(d-1)$-dimensionnelle. Nous allons d\'ecrire un mouvement \'el\'ementaire, i.e. n'agissant que sur un seul sommet de $\Delta_{(d-1)}$, disons $s$ de valence $n$. Les mouvements tente sont locaux au sens o\`u ils agissent de mani\`ere ind\'ependante sur chaque vertex de la triangulation, ou plut\^ot sur l'\'etoile autour d'un vertex, i.e. ici l'ensemble des ar\^etes se rencontrant en $s$. Ajoutons alors un vertex $s^*$ (dans le \og futur\fg{} de $s$), que l'on connecte \`a $s$. Cette ar\^ete est le \og pole\fg{} de la tente. Si les $(s_i), i=1,\dotsc,n$, sont les sommets de $\Delta_{(d-1)}$ connect\'es \`a $s$, on les relie maintenant \`a $s^*$. Nous avons alors une nouvelle triangulation $\Delta_{(d-1)}^*$, dont les simplexes sont obtenus \`a partir de ceux de $\Delta_{(d-1)}$ en rempla\c{c}ant $s$ par $s^*$.

Pour \'ecrire des \'equations d'\'evolution, d\'ecrivant la dynamique entre les deux hypersurfaces, il faut consid\'erer le morceau de triangulation $d$-dimensionnelle, coll\'e entre $\Delta_{(d-1)}$ et $\Delta_{(d-1)}^*$, dont les $d$-simplexes s'obtiennent \`a partir des $(d-1)$-simplexes de $\Delta_{(d-1)}$ contenant $s$ par l'ajout du sommet $s^*$ et des ar\^etes ad\'equates (si $ss_{i_1}\dotsm s_{i_{d-2}}$ est un $(d-1)$-simplexe de $\Delta_{(d-1)}$, alors $ss^*s_{i_1}\dotsm s_{i_{d-2}}$ est un $d$-simplexe). Notons \'egalement que cette transformation de $\Delta_{(d-1)}$ peut s'effectuer par des combinaisons de mouvements de Pachner.

On peut proc\'eder ainsi sur les sommets de notre choix pour cr\'eer des morceaux de plus en plus larges de triangulations $d$-dimensionnelle. Le choix des sommets utilis\'es peut s'interpr\'eter comme le choix d'un \og lapse\fg{} discr\'etis\'e. L'\'evolution de $\Delta_{(d-1)}$ se d\'ecompose en temps discret, par un certain nombre de ces \'etapes \'el\'ementaires. Cette d\'ecomposition permet d'\'etudier un nombre r\'eduit d'\'equations \`a chaque \'etape, par opposition \`a un sch\'ema d'\'evolution qui ferait \'evoluer l'hypersurface en une seule fois. Si $T$ est un morceau de triangulation $d$-dimensionnelle, la dynamique est g\'er\'ee par l'action de Regge correspondante, fonction des longueurs des ar\^etes, avec termes de bord,
\beq
S_{\rm R} = -\sum_{f\subset \mathrm{int}T} A_f(l_e)\ \vareps_f(l_e) - \sum_{f\subset \pp T} A_f(l_e)\ \psi_f^T(l_e) + \Lambda \sum_v V_v.
\ee
Les angles de d\'eficit autour des $(d-2)$-simplexes, not\'es $f$, de l'int\'erieur et du bord sont respectivement :
\beq
\vareps_f(l_e) = \bigl(2\pi-\sum_{v\supset f}\thet_{fv}(l_e)\bigr), \qquad\text{et}\qquad \psi_f^T(l_e) = \bigl(\pi - \sum_{v\supset f} \thet_{fv}(l_e)\bigr),
\ee
tandis que $A_f$ d\'esigne le volume de ces $(d-2)$-simplexes, $V_v$ le $d$-volume des $d$-simplexes, et $\Lambda$ la constante cosmologique.

Imaginons donc une s\'erie de mouvements tente \`a partir d'un sommet $s$ et appliqu\'es successivement au sommet $s^n$ g\'en\'er\'e \`a l'issue de la $n$-i\`eme \'etape. Notons $s_i$ les sommets auxquels $s$ est initialement reli\'e, $l_i^n$ les longueurs des ar\^etes $(s^n s_i)$, et $t^n$ la longueur du pole $(s^n s^{n+1})$. Soit $S_{n}$ l'action de Regge pour la $d$-triangulation cr\'e\'ee autour du pole $(s^n s^{n+1})$, d\'elimit\'ee par les ar\^etes $(s_i s_j)$ (non-dynamiques), $(s^n s_i)$ et $(s^{n+1} s_i)$. On introduit des \og moments\fg{} qui vont permettrent de donner une vision canonique des \'equations de Regge,
\begin{align}
p^n_i &= -\f{\pp S_n}{\pp l^n_i},\\
\tl{p}_i^n &=\ \f{\pp S_n}{\pp l_i^{n+1}},\\
p^n_t &= -\f{\pp S_n}{\pp t^n}.
\end{align}
L'\'equation du mouvement obtenue par variation de l'action par rapport \`a la longueur du pole donne bien s\^ur :
\beq
p^n_t = 0.
\ee
Par ailleurs, en effectuant deux mouvements cons\'ecutifs, nous obtenons deux moments pour les longueurs $l_i^{n+1}$. Mais les \'equations de Regge correspondantes donnent tout juste :
\beq
\tl{p}^n_i = p^{n+1}_i,
\ee
de sorte qu'il y a bien un moment par longueur. Cette r\'e\'ecriture permet de construire des solutions des \'equations avec courbure, dans le langage des mouvements tente, pour des triangulations ayant des sommets internes. L'id\'ee est de partir avec des donn\'ees initiales $l^1_i, l^{2}_i$ et d'utiliser l'\'equation $p_t^n=0$ pour obtenir une solution pour la longueur du pole $t^1$. Par suite, cela d\'etermine les moments $p^{2}_i$. On a ainsi un  jeu complet de donn\'ees canoniques initiales, qui permettent de faire un deuxi\`eme mouvement tente, et de trouver des solutions pour $l^3_i, t^2$.

Les sym\'etries d'une solution sont cod\'ees dans les valeurs propres nulles de la matrice hessienne de l'action \'evalu\'ee sur cette solution. Les auteurs de \cite{bahr-dittrich-broken-symmetries} construisent d'abord une solution plate, qui poss\`edent naturellement des valeurs propres nulles : celles-ci correspondent aux degr\'es de libert\'e de lapse et shift, i.e. la possibilit\'e de d\'eplacer le vertex interne $v^1$ sans changer les donn\'ees de bord ni le fait que la g\'eom\'etrie est plate. Puis, les auteurs s'int\'eressent \`a des solutions avec courbure \og proches\fg{} de la g\'eom\'etrie plate. La question est bien s\^ur de savoir si la sym\'etrie de cette derni\`ere persiste, ou bien si elle est bris\'ee (m\^eme faiblement) par la courbure. Il est alors montr\'e que celle-ci est g\'en\'eralement bris\'ee.

En suivant \cite{bahr-dittrich-broken-symmetries}, il est int\'eressant de regarder le cas 3d. Le mod\`ele est alors strictement plat lorsque la constante cosmologique $\Lambda$ est nulle. On peut donc utiliser celle-ci pour cr\'eer des solutions de courbure homog\`ene faible, de fac\c{c}on \`a observer voire contr\^oler la brisure de la sym\'etrie de la solution plate. Dans ce cas, les $(d-2)$-simplexes sont simplement les ar\^etes, et les \'equations pr\'ec\'edentes prennent une forme assez simples :
\begin{alignat}{2}
&p^n_i &\,=\,& \psi^{n+}_i - \Lambda \f{\pp V_n}{\pp l^n_i},\\
&p^{n+1}_i &\,=\,& -\psi^{(n+1)-}_i + \Lambda \f{\pp V_n}{\pp l^{n+1}_i},\\
&p^n_t &\,=\,& \vareps_t^n - \Lambda \f{\pp V_n}{\pp t^n} = 0.
\end{alignat}
Ici, $\psi^{n+}_i$ est l'angle dih\'edral autour de l'ar\^ete $(s^n s_i)$ dans le morceau de triangulation entre les \'etapes $n$ et $(n+1)$, et $\psi^{(n+1)-}_i$ l'angle dih\'edral autour de $(s^{n+1} s_i)$ dans cette m\^eme triangulation dont le 3-volume est $V_n$. 

Cosid\'erons pour simplifier que $s^n$ est trivalent, et regardons d'abord le cas $\Lambda=0$. Alors, on peut facilement construire une solution, plate, par des consid\'erations g\'eom\'etriques : $t^n$ est la longueur entre les sommets de deux toits de pyramides ayant la m\^eme base triangulaire -- celle qui serait form\'ee par les trois ar\^etes $(s_i s_j)$ de longueur $l_{ij}$. Cette solution permet de r\'eexprimer $p^n_i$ de mani\`ere simple :
\beq
p^n_i = -\pi + \thet_{s^n s_i}(l_i^n, l_{ij}),
\ee
o\`u $\thet_{s^n s_i}$ est l'angle dih\'edral en $(s^n s_i)$ pour un t\'etra\`edre (plat) constitu\'e par les quatre sommets $s^n$ et $(s_i)$, $i=1,2,3$. Autrement dit, il n'y a pas de d\'ependance en les longueurs $l^{n+1}_i$, et on aboutit \`a des relations n'impliquant que les longueurs et les moments de la triangulation 2d au temps $n$, i.e. une contrainte ! Le lecteur des sections pr\'ec\'edentes aura bien s\^ur remarquer qu'il s'agit de la contrainte que l'on avait d\'ej\`a propos\'ee et d\'eriv\'ee \`a la section pr\'ec\'edente !

Mais ce formalisme permet d'aller un peu plus loin que dans les sections pr\'ec\'edentes, en introduisant un peu de courbure via $\Lambda\neq 0$. On peut alors chercher une solution de $p^n_t=0$ pour $t^n$, perturbativement en $\Lambda$ autour de la solution plate. Les auteurs de \cite{bahr-dittrich-broken-symmetries} montrent que la contrainte d\'eriv\'ee dans le cas plat est affect\'ee de sorte que la d\'ependance en les longueurs $l^{n+1}_i$ ne dispara\^it plus. Ainsi, la nouvelle relation fait intervenir les donn\'ees canoniques des deux hypersurfaces cons\'ecutives, aux temps $n$ et $(n+1)$, traduisant une d\'ependance en le lapse et le shift ! On peut donc former une \og pseudo-contrainte\fg{}, i.e. en fait une \'equation du mouvement, du type
\beq
p^n_i - p^n_i(l^n_j,l^{n+1}_j) = 0.
\ee

En fait, on sait que la gravit\'e 3d avec constante cosmologique est topologique, car il existe une transformation de jauge exacte qui g\'en\'eralise \eqref{translation sym bf} au cas $\Lambda\neq 0$. L'intuition physique est naturellement que comme dans le cas plat, une courbure homog\`ene ne propage pas de degr\'e de libert\'e locaux. Ainsi, on peut comprendre que les contraintes de l'analyse en mouvement tente, qui ne font que traduire les sym\'etries de l'action autour de certaines solutions, sont bris\'ees par la constante cosmologique car l'action de Regge utilis\'ee n'est pas adapt\'ee \`a la physique de la courbure homog\`ene. Ceci vient du fait que la discr\'etisation de Regge utilise comme blocs \'el\'ementaires des t\'etra\`edres plats, plut\^ot adapt\'es aux solutions plates. Mais pour les solutions avec courbure, il est clair que l'on ne peut plus d\'eplacer les sommets de ces t\'etra\`edres plats sans changer la physique. Dans une telle situation, il est en fait beaucoup plus naturel de discr\'etiser la th\'eorie sur des simplexes ayant eux-m\^emes la courbure homog\`ene $\Lambda$ ! De cette fa\c{c}on la courbure n'est plus cod\'ee dans un collage sp\'ecial de simplexes plats, mais dans les blocs \'el\'ementaires eux-m\^emes. Et il est clair qu'une telle formulation autorise \`a d\'eplacer les sommets de la triangulation, comme dans le cas plat, traduisant une sym\'etrie de jauge retrouv\'ee.

On doit donc consid\'erer des t\'etra\`edres b\'en\'eficiant de la courbure homog\`ene de l'espace que l'on veut d\'ecrire. Les \'equations des mouvements tente (pour des sommets trivalents) permettent de former d'authentiques contraintes :
\beq
p^n_i = -\pi + \thet^{(\Lambda)}_{s^n s_i}(l_i^n, l_{ij}),
\ee
o\`u l'angle dih\'edral $\thet^{(\Lambda)}_{s^n s_i}$ est maintenant calcul\'e pour un t\'etra\`edre \`a la g\'eom\'etrie courbe homog\`ene. Il est possible de d\'evelopper cette expression en puissance de $\Lambda$, pour obtenir des contraintes tronqu\'ees \`a un ordre souhait\'e. Par exemple, au premier ordre en $\Lambda$,
\beq
p^n_i = -\pi +\thet_{s^n s_i}(l_i^n, l_{ij}) + \Lambda \f{\pp V}{\pp l^n_i} + \calO(\Lambda^2),
\ee
$V$ \'etant le volume du t\'etra\`edre.

Ainsi, nous voyons qu'il est possible en gravit\'e discr\`ete d'identifier des sym\'etries de jauge, et de les d\'ecrire par des contraintes sur des variables canoniques, m\^emes pour des g\'eom\'etries courbes. La pr\'esence de ces sym\'etries est en fait intimement li\'ee au choix de l'action discr\`ete. Ce point de vue a \'et\'e affin\'e dans \cite{bahr-dittrich-perfect-regge}, pour r\'epondre \`a la question : comment construire une action ayant les sym\'etries recherch\'ees \`a partir d'une discr\'etisation na\"ive ? L'id\'ee est de mettre en place un processus de \og coarse-graining\fg classique. On consid\`ere une triangulation initiale $\Delta_1$, dont les ar\^etes sont not\'ees $E$, et form\'ee de blocs de simplexes d'une subdivision $\Delta_2$ de celle-ci, dont les ar\^etes sont not\'ees $e$.  A partir de l'action de Regge standard, pour des simplexes plats, sur la triangulation plus fine, on peut former une action dite \emph{am\'elior\'ee} pour l'autre triangulation. L'id\'ee est de fixer les longueurs $L_E$ et de r\'esoudre les \'equations du mouvement pour une g\'eom\'etrie courbe homog\`ene sur $\Delta_2$, d\'etermin\'ee par $\Lambda$, pour les longueurs $l_e$, avec la contrainte $\sum_{e\subset E}l_e =L_E$. L'action am\'elior\'ee est alors 
\beq
S_{\Delta_2}(L_E) = S_{\Delta_1}(l_e^*)_{|\sum_{e\subset E}l^*_e =L_E},
\ee
pour des longueurs $l_e^*$ d\'ecrivant une courbure homog\`ene $\Lambda$ sur $\Delta_2$. Autrement dit, elle prend en compte la dynamique de la triangulation plus fine. En r\'ep\'etant le processus un grand nombre de fois, on comprend que les simplexes de $\Delta_1$ voit leur g\'eom\'etrie interne peu \`a peu courb\'ee par la r\'esolution des \'equations de Regge pour $\Lambda\neq 0$. Dans cette limite, on construit alors l'action de Regge pour des simplexes ayant la courbure homog\`ene voulue ! Je renvoie naturellement \`a \cite{bahr-dittrich-perfect-regge} pour une discussion plus compl\`ete du proc\'ed\'e.


\part{Les mousses de spins : introduction et premiers calculs} \label{part:spinfoam1}

\chapter{Les mousses de spins}

\section{Dynamique des r\'eseaux de spins}

Nous avons d\'ecrit, en termes de r\'eseaux de spins, un espace de Hilbert s'interpr\'etant comme $L^2(\calA/\calG)$. Celui-ci est form\'e des \'etats \og cin\'ematiques\fg, fonctionnelles de la connexion (g\'en\'eralis\'ee) et invariants de jauge, pour une th\'eorie de jauge background independent, bien adapt\'es \`a l'invariance sous les diff\'eomorphismes. La dynamique de ces \'etats est ensuite g\'er\'ee par le hamiltonien, sp\'ecifique \`a la th\'eorie consid\'er\'ee.En particulier, en relativit\'e g\'en\'erale, si l'invariance sous les 3-diff\'eomorphismes de la surface canonique est ma\^itris\'ee, la contrainte scalaire, formant la derni\`ere partie du hamiltonien, reste myst\'erieuse en gravit\'e quantique \`a boucles. Et c'est pr\'ecis\'ement pour contourner et surmonter les obstacles de la quantification canonique que se sont d\'evelopp\'es les mod\`eles de mousses de spins pour la gravit\'e quantique, depuis une dizaine d'ann\'ees (la premi\`ere proposition \'etant le mod\`ele BC, \cite{BCpaper}), l'id\'ee \'etant de passer \`a une repr\'esentation de type int\'egrales de chemins sur les g\'eom\'etries de l'espace-temps. Cette interpr\'etation g\'eom\'etrique (comme int\'egrales de chemins) est un r\'esultat essentiel et subtil, qui permet de distinguer les mod\`eles topologiques des autres, et qu'il convient d'analyser en d\'etails.

L'un des principaux objectifs de la th\'eorie quantique est de parvenir au calcul d'amplitudes (puis de probabilit\'e) de transition entre diff\'erents \'etats quantiques. En m\'ecanique quantique, cela revient \`a conna\^itre les \'el\'ements de matrices de l'op\'erateur d'\'evolution. En th\'eorie quantique des champs standard, il faut calculer les \'el\'ements de la matrice $S$ de diffusion entre \'etats \og in\fg{} et \og out\fg. Ces calculs se font typiquement par un d\'eveloppement en diagrammes de Feynman. Dans ce cadre, les \'etats de Fock sont particulaires, ce sont des excitations de dimension 0, et les diagrammes de Feynman qui repr\'esentent des transitions entre ces excitations sont des graphes. Ici, nos \'etats de base sont des excitations port\'ees par des graphes. On s'attend donc \`a ce que les transitions entre \'etats de r\'eseaux de spins prennent la forme de surfaces reliant les graphes consid\'er\'es !

On peut voir cela qualitativement de la mani\`ere suivante. Imaginons que l'on souhaite imposer une contrainte $H(x)=0$ et que l'op\'erateur correspondant soit bien d\'efini sur les r\'eseaux de spins. Dans une th\'eorie o\`u le hamiltonien est une contrainte de premi\`ere classe, comme dans BF ou la relativit\'e g\'en\'erale, l'op\'erateur d'\'evolution est remplac\'e par un op\'erateur de projection sur le sous-espace de Hilbert (ou un sous-espace du dual) satisfaisant la contrainte au niveau quantique :
\beq
P = \prod_x \delta(\hat{H}(x)) = \int DN\ e^{i\int_\Sigma N(x) \hat{H}(x) dx}.
\ee
Ici $N$ est un multiplicateur de Lagrange -- typiquement le lapse $N$ et le shift $\vec{N}$ en relativit\'e g\'en\'erale. L'amplitude de transition entre deux \'etats cin\'ematiques $s, s'$ est alors donn\'ee par l'\'el\'ement de matrice de $P$ entre $s$ et $s'$,
\beq
\langle s\lvert P\rvert s'\rangle = \langle s\,\vert s'\rangle_{\rm{phys}}.
\ee
Cette quantit\'e est appel\'ee le \emph{produit scalaire physique} entre les \'etats cin\'ematiques. D\'eveloppons (au moins formellement) le projecteur selon :
\beq
\langle s\,\vert s'\rangle_{\rm{phys}} = \sum_{n\geq0} \f1{n!} \int DN\ \langle s\lvert \biggl(\int_\Sigma N(x)\,\hat{H}(x)\,dx\biggr)^n\rvert s'\rangle.
\ee
Pour un terme de cette somme \`a $n$ fix\'e, nous pouvons introduire dans le produit scalaire cin\'ematique une r\'esolution de l'identit\'e, typiquement dans la base des r\'eseaux de spins, \'ecrite formellement comme :
\beq \label{id hkin}
\id_{\calH_{\rm kin}} \,=\, \sum_{\Gamma,c} \, \lvert s_{\Gamma}^{c}\rangle\ \langle s_{\Gamma}^{c}\rvert,
\ee
entre chacune des $n$ actions de l'op\'erateur hamiltonien. Pour simplifier la notation, \'ecrivons $\int_\Sigma N(x)\hat{H}(x)dx = NH$. Il vient
\beq
\langle s\,\vert s'\rangle_{\rm{phys}} = \sum_{n\geq0} \f1{n!} \sum_{\substack{\Gamma_1,\dotsc,\Gamma_{n-1} \\ c_1,\dotsc, c_{n-1}}} \int DN\ \langle s\lvert NH \rvert s_{\Gamma_{n-1}}^{c_{n-1}}\rangle\ \dotsm\ \langle s_{\Gamma_2}^{c_2}\lvert NH \rvert s_{\Gamma_1}^{c_1}\rangle\ \langle s_{\Gamma_1}^{c_1}\lvert NH \rvert s'\rangle.
\ee
Imaginons que $NH$ envoie un r\'eseau de spins sur une combinaison lin\'eaire de r\'eseaux de spins en modifiant le graphe initial par l'ajout de noeuds et de liens colori\'es en nombre fini. Par d\'efinition du produit scalaire cin\'ematique, sur $L^2(\calA/\calG)$, l'\'el\'ement $\langle s_j \lvert NH\rvert s_i\rangle$ est non nul uniquement si le graphe du r\'eseau $\lvert s_j\rangle$ est le m\^eme que celui de $NH\vert s_i\rangle$. Ainsi, en regardant une famille de r\'eseaux $(s_j)$, \`a $n$ fix\'e, telle que les amplitudes sont non-nulles, nous avons en fait d\'epli\'e la transition de $s$ \`a $s'$ en mettant en \'evidence une \og histoire\fg{}, ou une s\'erie d'\'evolutions \'el\'ementaires transformant $s$ en $s'$. Cela se visualise par un 2-complexe cellulaire reliant les graphes de $s$ et $s'$. Dans cette \'evolution, les ar\^etes de $s$ donnent naissance \`a des faces, i.e. des 2-cellules, et les noeuds du graphe \'evoluent en des liens, i.e. les 1-cellules du 2-complexe. Si l'on prend une tranche de ce 2-complexe, on trouve pr\'ecis\'ement un des r\'eseaux de spins interm\'ediaires. Ainsi, le complexe h\'erite sur ces faces des repr\'esentations de $G$ coloriant les liens des graphes, et sur ces liens des entrelaceurs des graphes. Quant aux vertexes du 2-complexe, ils sont le lieu des \'evolutions des graphes et coloriages des r\'eseaux de spins ! 

On appelle ces 2-complexes, dont les coloriages par des repr\'esentations et entrelaceurs sont somm\'es pour produire une amplitude, des \emph{mousses de spins}. Et le produit scalaire physique s'\'ecrit comme une somme sur les mousses de spins reliant les r\'eseaux de spins au bord, $F:s\rightarrow s'$,
\beq \label{spin foam transition}
\langle s\,\vert s'\rangle_{\rm{phys}} = \sum_{F:s\rightarrow s'} Z(F).
\ee
Bien s\^ur la quantit\'e qui d\'efinit un mod\`ele de mousses de spins est l'amplitude $Z(F)$ qui repr\'esente une histoire de r\'eseaux de spins avec un nombre fini de transitions (i.e. de vertexes). On peut r\'esumer cela par comparaison avec les diagrammes de Feynman. Un diagramme de Feynman correspond \`a un op\'erateur sur l'espace de Fock, qui est d\'etermin\'e par une s\'equence de cr\'eations et d'annihilations de particules virtuelles. De mani\`ere similaire, une mousse de spins d\'ecrit une s\'erie de cr\'eations et d'annihilations de r\'eseaux de spins virtuels, permettant de passer de $s$ \`a $s'$ !

En s'inspirant de la th\'eorie quantique des champs topologique, il est possible de formaliser ces id\'ees \cite{baez-spin-foam, baez-bf-spinfoam}. L'id\'ee est d'associer \`a chaque vari\'et\'e $\Sigma$ (compacte orient\'ee) de dimension $(n-1)$ un espace de Hilbert cin\'ematique $\calH_\Sigma$, engendr\'e par les r\'eseaux de spins invariants de jauge. Puis \`a chaque cobordisme $M: \Sigma\rightarrow\Sigma'$ repr\'esentant un espace-temps, on associe une mousse de spins $F : \calH_\Sigma\rightarrow \calH_{\Sigma'}$. N\'eanmoins, en tentant de faire cela, on voit poindre une difficult\'e qui distingue les mod\`eles de mousses de spins non-topologiques. En effet, pour former un 2-complexe sur lequel vit la mousse de spins $F$, on prend typiquement le 2-squelette dual \`a une triangulation de $M$. Dans un mod\`ele topologique, on s'attend bien s\^ur \`a ce que l'amplitude $Z(F)$ ne d\'epende pas de la triangulation, et donc du 2-complexe utilis\'e. Mais, dans une th\'eorie poss\'edant des degr\'es de libert\'e localement, la formule \eqref{spin foam transition} sugg\`ere que pour propager ceux-ci, il est n\'ecessaire de sommer sur toutes les s\'equences de cr\'eation et annihilation de r\'eseaux de spins virtuels, i.e. de sommer sur toutes les mousses de spins reliant les \'etats cin\'ematiques choisis sur $\Sigma$ et $\Sigma'$.

Par ailleurs, nous n'avons pas pr\'ecis\'ement mentionn\'e le choix de l'espace-temps associ\'e aux mousses de spins. On peut en effet imaginer que l'on se contente de sommer sur les mousses de spins vivant dans un espace-temps fix\'e, et en particulier avec une topologie fix\'ee, comme dans la d\'efinition de l'action de Hilbert-Einstein. Mais il peut \^etre plus satisfaisant d'imaginer la topologie comme une donn\'ee non-fix\'ee a priori et \'emergeant de la dynamique quantique, comme c'est le cas en 2d gr\^ace aux mod\`eles de matrices. Dans ce cas, la somme sur les mousses de spins peut \^etre comprise dans un sens g\'en\'eral, o\`u l'on somme sur tous les 2-complexes cellulaires reliant les r\'eseaux de spins consid\'er\'es. Cette approche permet de pousser l'analogie entre mousses de spins et diagrammes de Feynman : en effet, il est possible de voir les amplitudes de mousses de spins $Z(F)$ comme des diagrammes de Feynman pour des th\'eories d'un champ scalaire formul\'ee sur le groupe de structure $G$ et non-locale. Ainsi, la th\'eorie des champs sur le groupe, appel\'ee \og group field theory\fg (GFT), permet de construire la somme sur les amplitudes de mousses de spins, \cite{freidel-gft, oriti-gft-review-06}. 

\bigskip

La r\'ealisation non-triviale la plus compl\`ete du programme des mousses de spins comme dynamique de la LQG a \'et\'e obtenue en gravit\'e 3d, en grande partie par Noui et Perez, \cite{noui-perez-ps3d}. Cela permet de comprendre comment la repr\'esentation en mousses de spins peut se construire explicitement, et comment elle s'articule avec le formalisme canonique de la LQG. La gravit\'e 3d se formule au niveau hamiltonien par une paire de variables canoniques sur une surface $\Sigma$, $(A_a^i, E^a_i)$, form\'ee d'une connexion $\SU(2)$ et d'une 1-forme conjugu\'ee \`a valeurs dans l'adjointe, plus pr\'ecis\'ement $E^a_i = \eps^{ab}\delta_{ij} e_b^j$.

Le formalisme de la LQG s'applique donc, et un espace de Hilbert cin\'ematique est obtenu par les fonctionnelles cylindriques, ou les r\'eseaux de spins non-invariants de jauge, sur l'ensemble des graphes possibles. L'espace des phases est soumis \`a la contrainte de Gau\ss{}, $D_aE^a_i=0$, impos\'ee au niveau quantique en se restreignant aux r\'eseaux de spins portant des entrelaceurs aux noeuds. De plus, les spins $j_l$ coloriant les liens des \'etats de r\'eseaux de spins sont les nombres quantiques associ\'es aux valeurs propres de l'op\'erateur de longueur pour des liens duaux \`a ceux du graphe, \cite{livine-3dlength}. Si $l^*$ est une courbe qui intersecte en un point un lien $l$ portant le spin $j_l$, elle acquiert une longueur $L^2_{l^*} = j_l(j_l+1)$, ou \`a un ordering pr\`es (comme le sugg\`ere les \'etudes asymptotiques) $L^2_{l^*} = (j_l+\f12)^2$. C'est bien s\^ur l'\'equivalent de l'op\'erateur d'aire en LQG.

Il reste alors une contrainte, de premi\`ere classe (ab\'elienne), caract\'erisant l'absence de degr\'es de libert\'e locaux, $F(A)^i =0$, qui exprime le fait que la connexion doit \^etre plate sur $\Sigma$. Il nous faut donc trouver l'espace des solutions de cette contrainte quantifi\'ee et le produit scalaire physique. On forme donc le projecteur :
\beq
P = \int DN\ \exp \biggl(i\int_\Sigma \tr\, \bigl[N \widehat{F(A)}\bigr]\biggr),
\ee
qui permet d'identifier l'espace des \'etats physiques comme :
\beq
\calH_{\rm{phys}} = \overline{\cyl/\calI},\qquad \text{pour}\qquad \calI = \{ \vert s\rangle,\ \langle s\, \vert s\rangle_{\rm{phys}} =0\}.
\ee
$\calI$ est l'ensemble des \'etats de norme nulle, et la barre repr\'esente la compl\'etion de Cauchy\footnote{Pour simplifier les notations, nous entendons par $\cyl$ les fonctionnelles cylindriques de la connexion invariantes sous l'action de $G$.}.

Il nous faut bien s\^ur d\'efinir l'op\'erateur de courbure sur les fonctionnelles cylindriques, ou de mani\`ere \'equivalente sur les r\'eseaux de spins. Pour cela, nous utilisons une r\'egularisation sur un r\'eseau, celui-ci \'etant choisi bien adapt\'e aux graphes des r\'eseaux de spins dont on veut calculer le produit scalaire. Soit $\Gamma$ et $\Gamma'$ les graphes des r\'eseaux de spins $s, s'$. On prend donc un r\'eseau dont les liens passent le long de ceux de $\Gamma, \Gamma'$. Sur ce r\'eseau, la courbure s'interpr\`ete comme l'holonomie autour de chaque plaquette \'el\'ementaire, via l'approximation :
\beq
W_p(A) = \unit + \eps^2\, F_p(A) + o(\eps^2).
\ee
Ici, $W_p(A)$ est la boucle de Wilson autour de la plaquette $p$, et $\eps$ la taille typique des liens du r\'eseau. Ainsi, on r\'egularise le projecteur $P$ comme :
\beq
P = \prod_p \delta(W_p(A)).
\ee
Mais on peut d\'ej\`a intuiter que $P$ ne d\'epend pas du r\'egulateur ! Cela tient \`a deux points importants : le fait que $P$ projette sur les connexions plates, et le fait que les r\'eseaux de spins ne d\'ependent que d'un nombre fini d'holonomies. Ainsi, il est possible de fusionner les plaquettes adjacentes en \'eliminant les liens communs. Pour le dire simplement, comme $P$ contient les conditions de trivialit\'e autour de chaque plaquette, il impose la m\^eme condition sur la plaquette r\'esultant de leur fusion \footnote{Cela est naturellement reli\'e au fait que les bords des plaquettes sont homotopes \`a un point, en m\^eme temps que l'holonomie d'une connexion \emph{plate} le long d'une courbe ne d\'epend que du type d'homotopie de la courbe}. Cette transformation, effectu\'ee sur les liens qui ne sont pas contenus dans $\Gamma$ ni $\Gamma'$, n'affecte pas le produit scalaire.



Bien s\^ur, comme on sait que la boucle de Wilson est un op\'erateur bien d\'efini sur les fonctionnelles cylindriques, nous pouvons \'ecrire le produit scalaire physique :
\beq
\langle s\,\vert s'\rangle_{\rm{phys}} = \langle s\,\vert \prod_{p=1}^{N} \delta(W_p(A))\, \rvert s'\rangle.
\ee
Pour obtenir la repr\'esentation en mousses de spins, il faut d\'evelopper l'op\'erateur de plaquette sur les repr\'esentations de $\SU(2)$,
\beq
\delta(W_p(A)) = \sum_{j_p} (2j_p+1)\,\chi_{j_p}(W_p(A)),
\ee
et introduire la r\'esolution de l'identit\'e \eqref{id hkin} sur $\calH_{\rm kin}$ entre chacune des plaquettes (notons que l'ordre des plaquettes n'importe pas). On a donc :
\beq \label{phys ps noui-perez}
\langle s\,\vert s'\rangle_{\rm{phys}} = \sum_{\{j_p\}} \sum_{\substack{\Gamma_1,\dotsc, \Gamma_N \\ c_1,\dotsc, c_N}}\,\prod_{p=1}^N (2j_p+1)\ \langle s_{\Gamma_p}^{c_p} \lvert \chi_{j_p}(W_p(A)) \rvert s_{\Gamma_{p-1}}^{c_{p-1}}\rangle.
\ee
Par d\'efinition de la mesure d'Ashtekar-Lewandowski, seuls les r\'eseaux de spins $\rvert s_{\Gamma_p}^{c_p}\rangle$ dont le graphe $\Gamma_p$ diff\`erent du graphe $\Gamma_{p-1}$ par l'ajout de la boucle autour de la plaquette $p$ contribuent. Ainsi, la somme sur les graphes interm\'ediaires dispara\^it et leur choix est d\'etermin\'e par l'action successive des plaquettes. On peut bien s\^ur visualiser les transitions de fa\c{c}on continue, ce qui correspondrait \`a faire tendre la taille des plaquettes vers z\'ero et $N$ vers l'infini. On obtient alors une sorte d'\'evolution \og temporelle\fg{} de $s$ vers $s'$, i.e. un 2-complexe dont les faces sont colori\'ees par des spins et les liens par des entrelaceurs. Une tranche \og spatiale\fg{} de ce complexe correspond \`a un graphe $\Gamma_p$, colori\'e par $(c_p)$.

Un 2-complexe sans vertex correspond \`a une transition entre des \'etats ayant des graphes hom\'eomorphes. Cela se voit bien en consid\'erant deux graphes qui ne diff\`erent qu'entre deux vertexes par le choix de chemins $\gamma$ et $\gamma'$ diff\'erents mais dans la m\^eme classe d'homotopie. Un vertex dans un 2-complexe signifie la cr\'eation/annihilation de liens et noeuds du graphe du r\'eseau de spins. En fait, comme le montre pr\'ecis\'ement \cite{noui-perez-ps3d}, cette construction permet de montrer que des r\'eseaux de spins dont les graphes sont dans la m\^eme classe d'homotopie sont physiquement \'equivalents ! Je ne d\'etaille pas ici les amplitudes de transition explicitement, car elles sont donn\'ees, pour des r\'eseaux \`a noeuds trivalents, par le mod\`ele de Ponzano-Regge que je pr\'esente \`a la section suivante.

Notons que la forme du 2-complexe d\'epend de l'ordre des plaquettes. N\'eanmoins, le produit scalaire physique ne fait pas intervenir ici de sommes sur les 2-complexes reliant $s\rightarrow s'$ ! C'est bien s\^ur une manifestation attendue du caract\`ere topologique de la th\'eorie. Ainsi, \eqref{phys ps noui-perez} est une r\'ealisation de la forme g\'en\'erique \eqref{spin foam transition}, dans laquelle la somme sur les mousses de spins est juste une somme sur les coloriages d'un 2-complexe, et avec :
\beq
Z(F) = \prod_{p=1}^N (2j_p+1)\ \langle s_{\Gamma_p}^{c_p} \lvert \chi_{j_p}(W_p(A)) \rvert s_{\Gamma_{p-1}}^{c_{p-1}}\rangle.
\ee
La mousse de spins $F:s\rightarrow s'$ est ici d\'etermin\'ee par les spins des plaquettes $j_p$, et les coloriages \`a chaque transition $c_p$. Quant au 2-complexe, il se construit de proche en proche via le fait que $\Gamma_p$ est issu de l'ajout de la plaquette $p$ \`a $\Gamma_{p-1}$.

\medskip

Nous avons d\'ej\`a vu de quelle fa\c{c}on les r\'eseaux de spins d\'ecrivent la 3-g\'eom\'etrie quantique, en tant qu'\'etats propres des op\'erateurs d'aires, et via les op\'erateurs de volume associ\'es aux noeuds (en gravit\'e 3d, ce sont plut\^ot les op\'erateurs de longueurs pour des courbes duales aux liens des r\'eseaux et des op\'erateurs d'aires autour des noeuds). Puisque les mousses de spins fournissent l'\'evolution hamiltonienne des r\'eseaux de spins, on peut naturellement penser qu'elles offrent une vision 4-dimensionnelle de la g\'eom\'etrie quantique ! Cela est vrai, mais de mani\`ere assez subtile. Vrai, car on s'attend \`a pouvoir \'ecrire le produit scalaire physique comme une int\'egrale de chemins sur les g\'eom\'etries, pour l'exponentielle de l'action classique. Vrai \'egalement par dualit\'e cellulaire, car les repr\'esentations associ\'ees aux faces des 2-complexes s'interpr\`etent comme les aires de surfaces duales (en 4d), les entrelaceurs des liens sont associ\'es \`a des 3-volumes duaux, et il est donc naturel de voir des 4-volumes duaux aux vertexes des mousses de spins. Dans notre exemple en 3d, les spins $j_p$ ne sont autres que les nombres quantiques d\'eterminant les longueurs d'ar\^etes duales aux faces du 2-complexe, et les entrelaceurs sont associ\'es aux aires de surfaces duales aux liens du 2-complexe.

La subtilit\'e de cette interpr\'etation tient au fait que le hamiltonien est dans les th\'eories qui nous int\'eressent une contrainte. La dynamique engendr\'ee est donc une dynamique de transformations de jauge. Et, ainsi, les histoires des r\'eseaux de spins que d\'ecrivent les mousses de spins sont des histoires de jauge, \cite{perez-spin-foam-representation} ! Comme nous l'avons vu ci-dessus dans le mod\`ele en 3d, les mousses de spins \'emergent du d\'epliement de l'action de la contrainte de courbure nulle, $\delta(W_p(A))$, r\'ep\'et\'ee sur chaque plaquette, et permettent d'identifier comme physiquement \'equivalents les graphes homotopes.

Nous allons maintenant pr\'esenter les mod\`eles de mousses de spins pour la th\'eorie topologique BF, appel\'es en dimensions 3 et 4 mod\`eles de Ponzano-Regge et d'Ooguri, en les d\'erivant \`a partir de l'int\'egrale fonctionnelle discr\'etis\'ee, \`a la mani\`ere de ce qui a \'et\'e fait dans Yang-Mills 2d pour calculer le volume de l'espace des modules des connexions plates \cite{witten-2d-YM}. Plus tard, nous pr\'esenterons les mod\`eles de gravit\'e quantique, issus du m\'elange des mod\`eles topologiques et de l'impl\'ementation des contraintes de simplicit\'e de Plebanski au niveau quantique. Autant dire tout de suite que l'interpr\'etation g\'eom\'etrique des donn\'ees des mousses de spins, est tr\`es importante dans cette construction. A la suite de cela, je pr\'esenterai une partie importante de mes travaux de th\`ese qui a vis\'e \`a pousser le plus loin possible l'interpr\'etation des mousses de spins comme une int\'egrale sur les g\'eom\'etries d'un r\'eseau dual au 2-complexe, en \'ecrivant explicitement cette int\'egrale en termes de l'action classique discr\'etis\'ee.

\section{Mod\`ele pour la th\'eorie BF} \label{sec:bf-spinfoam}

Pour appliquer la d\'erivation heuristique pr\'ec\'edente des mousses de spins concr\`etement, il faut bien s\^ur un op\'erateur hamiltonien bien d\'efini sur les r\'eseaux de spins. Si un tel op\'erateur existe en LQG, d\'evelopp\'e par Thiemann, des doutes subsistent quant \`a sa pertinence, pour des raisons intuitives (comme le fait que cet op\'erateur agit de mani\`ere tr\`es locale). Aussi, pour contourner les difficult\'es habituelles de la quantification canonique, on peut esp\'erer obtenir des mod\`eles de mousses de spins \`a partir de l'approche covariante, par int\'egrales de chemins. C'est du moins ce que nous a appris la th\'eorie des champs, pour le calcul des amplitudes de transition. Nous nous concentrons ici sur les mod\`eles les plus simples, associ\'es \`a la th\'eorie topologique BF.

La quantit\'e la plus simple form\'ee \`a partir d'une int\'egrale fonctionnelle est la fonction de partition qui int\`egre sur toutes les configurations des champs sans donn\'ees externes. Pour obtenir des amplitudes de transition, il faut en revanche fixer des donn\'ees de bord dans l'int\'egrale fonctionnelle. N\'eanmoins, les ressorts du calcul sont les m\^emes que ceux du calcul de la fonction de partition, de sorte que nous nous concentrons sur celle-ci. De mani\`ere formelle, elle prend la forme suivante dans la th\'eorie BF, sur une vari\'et\'e $M$ de dimension $n$,
\begin{align} \label{formal bf path int}
Z_{\rm BF}(M) &= \int DA\,DB\ \exp\bigl(i\int_M \tr(B\wedge F_A)\bigr),\\
&= \int DA\quad \delta(F_A).
\end{align}
L'int\'egrale sur les $(n-2)$-formes $B$ est interpr\'et\'ee formellement comme une projection sur les connexions plates. Ainsi, on s'attend \`a ce que cette int\'egrale calcule le \og volume\fg{} de l'espace des connexions plates ! Ainsi, nous avons d'un c\^ot\'e un objet math\'ematique, l'espace des connexions plates, dont on souhaite calculer le volume, celui-ci s'exprimant a priori comme une int\'egrale fonctionnelle (donc un objet assez complexe). Un tel volume n'est par ailleurs pas d\'efini sans une mesure d\'etermin\'ee. D'un autre c\^ot\'e, nous allons montrer que la repr\'esentation en mousses de spins fournit une formulation combinatoire de ce volume (et donc une mesure appropri\'ee) qui respecte l'invariance topologique du mod\`ele.

Il est important de savoir que cette fonction de partition a \'et\'e largement \'etudi\'ee par les m\'ethodes standards de la th\'eorie des champs ! Les travaux les plus connus sont peut-\^etre ceux de Witten, en 2d \cite{witten-2d-YM} et en 3d \cite{witten-3d-gravity,witten-amplitude-3d}, et ceux de Blau et Thompson, \cite{blau-thompson-bf,blau-thompson-bf-torsion}. Signalons \'egalement l'approche de Gegenberg et Kunstatter par d\'ecomposition de Hodge des connexions \cite{gegenberg-partition-function-bf}, et la preuve de la renormalisabilit\'e, \`a la BRST et en exhibant une supersym\'etrie vectorielle, par Lucchesi, Piguet et Sorella, \cite{renormalisability-bf}, en toutes dimensions. Ces approches ont permis de montrer que la fonction de partition peut s'\'ecrire comme une int\'egrale sur l'espace des connexions plates modulo les transformations de jauge usuelles, la mesure \'etant donn\'ee par la torsion analytique de Ray-Singer.  Mais ce traitement n\'ecessite l'introduction d'une m\'etrique sur la vari\'et\'e (typiquement pour la fixation de jauge), ce qui n'est pas compatible avec notre ambition de d\'evelopper un formalisme background independent\footnote{M\^eme si la torsion au final ne d\'epend pas de la m\'etrique choisie.}.

L'avantage du formalisme des mousses de spins est donc d'\^etre totalement background independent, et donc bien adapt\'e pour traiter une g\'eom\'etrie pleinement dynamique. De plus, comme nous l'avons expliqu\'e, les mousses de spins \'emergent naturellement du formalisme canonique, en tant qu'\'evolution (de jauge) des excitations fondamentales de r\'eseaux de spins ! En particulier, les donn\'ees combinatoires des sommes sur les mousses de spins s'interpr\`etent naturellement comme les valeurs propres des op\'erateurs g\'eom\'etriques de la th\'eorie canonique.

\begin{figure}
\begin{center}
\includegraphics[width=3.5cm]{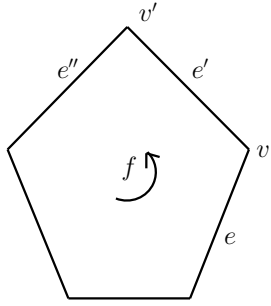}
\end{center}\caption{ \label{fig:dualface} Repr\'esentation typique d'une face duale du complexe cellulaire \`a la triangulation. C'est autour de ces faces qu'est concentr\'ee la courbure. La face $f$ est duale \`a un $(n-2)$-simplexe (ar\^ete en 3d, triangle en 4d), et son bord est form\'e de liens duaux \`a des $(n-1)$-simplexes (triangle en 3d, t\'etra\`edre en 4d), et de vertex duaux aux $n$-simplexes (t\'etra\`edre en 3d, et 4-simplexe en 4d).
}
\end{figure}

Etant donn\'e que la th\'eorie BF ne poss\`ede pas de degr\'es de libert\'e locaux, on peut s'attendre \`a calculer la fonction de partition de mani\`ere \emph{exacte} \`a partir d'une discr\'etisation de la th\'eorie. On consid\`ere pour cela une triangulation de la vari\'et\'e d'espace-temps $M$. Le r\^ole cl\'e est en fait jou\'e par le 2-squelette dual $\Gamma$ \`a cette triangulation. Celui-ci est form\'e de $V$ vertexes, duaux aux $n$-simplexes, de $E$ liens duaux aux $(n-1)$-simplexes et de $F$ faces duales aux $(n-2)$-simplexes. Notons $\Gamma_i$, $i=0,1,2$, l'ensemble des $i$-cellules duales de $\Gamma$. L'id\'ee est ensuite, comme en calcul de Regge de concentrer la courbure autour des $(n-2)$-simplexes. Pour cela, on discr\'etise la connexion et sa courbure de la mani\`ere suivante. Une connexion sur $\Gamma$ est une collection d'\'el\'ements de groupe sur les liens,
\beq
A \,=\, \bigl( g_e\bigr)_{e\in\Gamma_1}\, \in\, G^E.
\ee
Les cellules doivent \^etre orient\'ees, et le changement d'orientation d'un lien se traduit par le changement de $g_e$ en $g_e\mone$. Une notion de courbure est alors d\'efinie par le produit orient\'e des $g_e$ le long du bord de chaque face,
\beq
H_f(A) = \prod_{e\in \pp f} g_e^{[f:e]},
\ee
ce qui n\'ecessite un vertex de r\'ef\'erence et une orientation pour chaque face, $[f:e]$ \'etant le nombre d'incidence de la face $f$ sur le lien $e$. G\'eom\'etriquement, on peut comprendre les \'el\'ements $g_e$ comme des op\'erateurs de transport parall\`ele entre les $n$-simplexes duaux aux extr\'emit\'es des liens de $\Gamma_1$, ou comme les holonomies d'une connexion sur $M$ le long d'un nombre fini de liens. Ainsi, $H_f(A)$ peut \^etre vue comme l'holonomie autour d'une boucle ferm\'ee d\'elimitant la face $f$, de sorte que l'on d\'efinit une connexion plate telle que pour toute face,
\beq
H_f(A) = \unit.
\ee
Il est alors possible de donner un sens \`a la fonction de partition de la th\'eorie BF en consid\'erant
\beq \label{partition bf lattice}
Z_{\rm BF}(M) = \int_{G^E} dA\ \prod_{f\in\Gamma_2} \delta(H_f(A)),
\ee
o\`u la mesure $dA$ est la mesure de Haar sur $G^E$, $dA = \prod_{e\in\Gamma_1} dg_e$. Etant construit avec la mesure de Haar $dg_e$ sur chaque lien, on s'attend \`a ce que $dA$ s'accorde naturellement avec la mesure d'Ashtekar-Lewandowski dans le calcul des amplitudes de transition.

Les transformations de jauge changent le r\'ef\'erentiel local interne sur chaque $n$-simplexe (plat au sens de l'espace interne). Cela d\'efinit une action de $G^V$ sur les connexions, selon
\be
h \cdot A = \bigl(h_{t(e)}\,g_e\,h^{-1}_{s(e)}\bigr)_{e\in\Gamma_1}.
\ee
Ici $h=(h_v)_{v\in\Gamma_0}$ est un ensemble de $V$ \'el\'ements du groupe, et $t(e)$ (respectivement $s(e)$) est le vertex d'arriv\'ee (respectivement de d\'epart) du lien $e$. L'\'el\'ement de courbure $H_f(A)$ se transforme par conjugaison sous $h_{v_f}$, si $v_f$ est le vertex de r\'ef\'erence de $f$. On voit donc la fonction de partition est invariante sous ces transformations de jauge.

Bien s\^ur, il y a de forts risques pour que le produit de delta sur $G$ soit mal d\'efini, conduisant \`a des divergences. Mais ignorons ce souci et calculons cette quantit\'e formellement par transform\'ee de Fourier. Cela commence par le d\'eveloppement
\beq
\delta(g) = \sum_\rho \dim(\rho)\ \chi_\rho(g),
\ee
sur les repr\'esentations irr\'eductibles de $G$, $\chi_\rho$ \'etant le caract\`ere de la repr\'esentation $\rho$, pour chaque face. Notons $\calH_{\rho}$ l'espace de repr\'esentation de $\rho$. Nous sommes conduits \`a :
\beq
Z_{\rm BF}(M) = \sum_{(\rho_f)_{f\in\Gamma_2}}\,\int dA\ \prod_{f\in\Gamma_2} \dim(\rho_f)\ \chi_{\rho_f}\bigl(H_f(A)\bigr).
\ee
Dans l'expression formelle \eqref{formal bf path int} de la fonction de partition, le champ $B$ \'etait conjugu\'e \`a la courbure par transform\'ee de Fourier fonctionnelle en un sens. Ici, ce r\^ole est jou\'e par l'ensemble des repr\'esentations $(\rho_f)_{f\in\Gamma_2}$. Celles-ci colorent les faces de $\Gamma$, c'est-\`a-dire que ce sont les modes de Fourier de la courbure sur les $(n-2)$-simplexes de la triangulation. C'est bien s\^ur les m\^emes donn\'ees que celles port\'ees par les liens des r\'eseaux de spins dans le formalisme canonique.

L'avantage de travailler sur le complexe cellulaire dual \`a une triangulation est que le nombre de faces se rencontrant sur chaque lien est constant. Ce nombre n'est autre que le nombre de $(n-2)$-simplexes formant le bord d'un $(n-1)$-simplexe, \'egal \`a $n$. Cela signifie que chaque holonomie $g_e$ intervient exactement une fois dans $n$ holonomies de courbure $H_f$. Pour des repr\'esentations $(\rho_f)$ fix\'es, on doit donc int\'egrer sur $G$ le produit de $n$ \'el\'ements de matrices de $g_e$. Cela donne :
\beq
\int_G dg\ D^{(\rho_1)}(g)\otimes \dotsm\otimes D^{(\rho_n)}(g) = \id_{\Inv(\calH_{\rho_1}\otimes\dotsm\otimes\calH_{\rho_n})}.
\ee
Ici $\Inv(\calH_{\rho_1}\otimes\dotsm\otimes\calH_{\rho_n})$ est le sous-espace invariant sous $G$ du produit tensoriel des repr\'esentations se rencontrant en $e$. Il s'agit bien s\^ur d'un processus de moyenne sur le groupe qui produit l'identit\'e sur le sous-espace invariant. Sans surprise, cette identit\'e se d\'eveloppe sur une base orthogonale d'entrelaceurs $\iota : \otimes_{i=1}^n \calH_{\rho_i}\rightarrow \C$,
\beq
\id_{\Inv(\calH_{\rho_1}\otimes\dotsm\otimes\calH_{\rho_n})} = \sum_{\iota: \otimes_{i=1}^n \calH_{\rho_i}\rightarrow \C} \iota\,\iota^*,
\ee
chacun d'eux assurant l'invariance \`a gauche et \`a droite sous $G$. On peut donc remplacer les int\'egrales sur $G$ par des coloriages des $(n-1)$-simplexes de la triangulation par des entrelaceurs. On obtient ainsi :
\beq
Z_{\rm BF}(M) = \sum_{(\rho_f)_{f\in\Gamma_2}} \sum_{(\iota_e)_{e\in\Gamma_1}} \prod_{f\in\Gamma_2} \dim(\rho_f)  \prod_{v\in\Gamma_0} W_v(\rho_f, \iota_e).
\ee
La fonction de partition est donc constitu\'ee par l'assemblage d'une amplitude de r\'ef\'erence sur chaque $(n-2)$-simplexe, $W_f = \dim(\rho_f)$, appel\'ee \emph{amplitude de face}, et d'une amplitude de r\'ef\'erence sur chaque $n$-simplexe, \'egalement appel\'ee \emph{amplitude de vertex} par dualit\'e, $W_v$. Cette derni\`ere est construite de la mani\`ere suivante. Un $n$-simplexe poss\`ede $(n+1)$ $(n-1)$-simplexes, portant des entrelaceurs $\iota_e$. Il faut alors contracter les indices de ces entrelaceurs deux \`a deux, avec une contraction par $(n-2)$-simplexe dans la repr\'esentation $\rho_f$. En fait, un $(n-1)$-simplexe porte un entrelaceur qui est donc contract\'e avec $n$ entrelaceurs des $(n-1)$-simplexes voisins. Remarquons que l'amplitude des liens duaux, i.e. sur les $(n-1)$-simplexes, peut \^etre absorb\'ee de mani\`ere g\'en\'erique dans l'amplitude de vertex (car un $(n-1)$-simplexe est toujours partag\'e par exactement deux $n$-simplexes).

Ces amplitudes de vertex se repr\'esentent par des graphes de r\'eseaux de spins (et correspondent \`a l'\'evaluation de fonctionnelles de r\'eseaux de spins sur la connexion triviale). Les entrelaceurs $\iota_e$ sont des vertexes de valence $n$, et leurs pattes correspondent aux indices libres des repr\'esentations $(\rho_1,\dotsc, \rho_n)$. Ces vertexes peuvent \^etre \og d\'epli\'es\fg{} en arbres trivalents, avec des liens virtuels portant les repr\'esentations virtuelles qui caract\'erisent $\iota_e$. L'amplitude est obtenue par contraction deux \`a deux des indices libres des entrelaceurs, ce qui se traduit graphiquement par des liens entre vertexes portant les repr\'esentations $\rho_f$. Ainsi, les vertexes du graphe sont associ\'es aux $(n-1)$-simplexes et les liens aux $(n-2)$-simplexes. Cela forme un graphe qui combinatoirement parlant a pr\'ecis\'ement la structure d'un $n$-simplexe.

\subsection{En deux dimensions}

Nous d\'etaillons maintenant ces amplitudes de mani\`ere plus pr\'ecise en basses dimensions, $n=2, 3, 4$. En dimension 2, notre d\'emarche n'est autre que celle de Witten, \cite{witten-2d-YM}, pour la th\'eorie de Yang-Mills et sa limite de couplage faible qu'est la th\'eorie BF. L'int\'egrale sur $G$ du produit de deux \'el\'ements de matrices peut se lire comme la relation d'orthogonalit\'e des \'el\'ements matriciels de repr\'esentations,
\beq
\int_G dg\ D^{(\rho_1)}_{ab}(g)\ \overline{D^{(\rho_2)}_{cd}(g)} = \f{1}{\dim\rho_1}\,\delta_{\rho_1\rho_2}\ \delta_{ac}\,\delta_{bd}.
\ee
La barre correspond \`a la conjugaison complexe. Ainsi, en 2 dimensions, toutes les repr\'esentations sont \'egales ! Cela explique la \og trivialit\'e\fg{} des th\'eories de jauge en 2d de notre point de vue. Les seules amplitudes qui interviennent sont donc les dimensions des espaces de repr\'esentations. Le facteur $\dim\rho$ est pr\'esent $F$ fois dans le d\'eveloppement des delta de Dirac, puis $-E$ fois du fait de la relation d'orthogonalit\'e ci-dessus, et finalement la contraction des indices restants le donne encore $V$ fois, ce qui fait $F-E+V = \chi(M)$, c'est-\`a-dire la caract\'eristique d'Euler de la vari\'et\'e, qui est un invariant topologique.

Cela donne une expression particuli\`erement simple de la fonction de partition. Par ailleurs, on sait par Goldman \cite{goldman} que l'espace $\calM$ des modules des connexions plates sur un fibr\'e au-dessus d'une vari\'et\'e $M$ ferm\'ee et orient\'ee admet une structure symplectique naturelle $\Omega$. Witten a montr\'e que la fonction de partition calcule en fait ce volume symplectique de l'espace des connexions plates (modulo les transformations de jauge), et en \'egalisant avec l'expression en mousses de spins, a obtenu l'\'el\'egante correspondance
\beq \label{2d ym partition function}
Z_{\rm BF} = \int_{\calM} \Omega \,=\, \sum_\rho \bigl(\dim(\rho)\bigr)^{\chi(M)},
\ee
\`a des coefficients inessentiels (pour nous) pr\`es. Pour une surface ferm\'ee orient\'ee de genre $g$, $\chi(M) = 2-2g$. Comme $\chi(M)$ est un invariant topologique, il est clair que l'expression obtenue est \emph{ind\'ependante de la triangulation}.

\subsection{Mod\`ele de Ponzano-Regge, en 3d} \label{sec:pr model}

\begin{figure}
\begin{center}
\includegraphics[width=8cm]{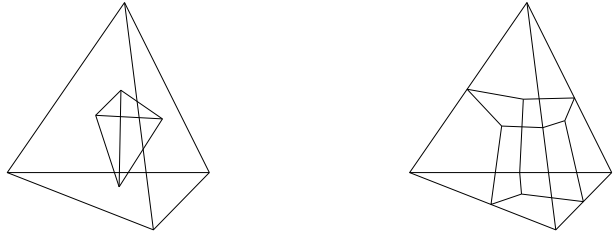}
\end{center}\caption{ \label{fig:6j} L'amplitude d'un t\'etra\`edre du mod\`ele PR, i.e. le symbole 6j, est obtenue en mettant des entrelaceurs trivalents sur les triangles et des spins sur les ar\^etes, et en contractant les entrelaceurs, selon la structure du t\'etra\`edre dual dessin\'e \`a gauche. On peut aussi le voir comme un r\'eseau de spins au bord du t\'etra\`edre, dessin\'e \`a droite.
}
\end{figure}

En dimension 3, un raisonnement similaire conduit au \emph{mod\`ele de Ponzano-Regge}, qui pour $G=\SU(2)$ est une quantification de la gravit\'e 3d riemannienne\footnote{Ce n'est en fait pas exactement la gravit\'e, du fait de la pr\'esence de m\'etriques d\'eg\'en\'er\'ees.}. Chaque \'el\'ement de groupe $g_e$ appara\^it maintenant dans trois faces duales, correspondant aux trois ar\^etes de chaque triangle. Les int\'egrales produisent donc des entrelaceurs trivalents. Concentrons-nous sur le cas $G=\SU(2)$, la g\'en\'eralisation \'etant \'evidente. Dans ce cas, les repr\'esentations irr\'eductibles sont donn\'ees par des spins demi-entiers $(j_f)_{f\in\Gamma_2}$ coloriant les faces. A spins fix\'es, il existe un unique entrelaceur \`a la normalisation pr\`es entre $\calH_{j_1}\otimes\calH_{j_2}\otimes\calH_{j_3}$ et $\C$, qui est le symbole 3mj de Wigner, voir appendice \ref{sec:app},
\beq
\int_{\SU(2)} dg\ \prod_{a=1}^3 D^{(j_a)}_{m_an_a}(g) = \begin{pmatrix} j_1 &j_2 &j_3 \\ m_1 & m_2 & m_3\end{pmatrix} \begin{pmatrix} j_1 &j_2 &j_3 \\ n_1 & n_2 & n_3\end{pmatrix}.
\ee
Dans une triangulation 3d, un triangle est partag\'e par exactement deux t\'etra\`edres. Les deux symboles 3mj ci-dessus d'un triangle sont alors contract\'es avec les symboles 3mj port\'es par les autres triangles de chacun de ces deux t\'etra\`edres. Cela donne comme amplitude fondamentale d'un t\'etra\`edre :
\beq
W_{v}^{\rm PR}(j_f) = (-1)^{\sum_{i=1}^6 j_i}\ \begin{Bmatrix} j_1 &j_2 &j_3\\ j_4 &j_5 &j_6 \end{Bmatrix},
\ee
o\`u la quantit\'e entre accolades est le symbole 6j de Wigner, d\'efini en appendice, qui est issu de la contraction des 3mj invariants de $\SU(2)$. Ce type de symboles se repr\'esente graphiquement comme le graphe colori\'e d'un r\'eseau de spins. Un symbole 3mj est un vertex trivalent, dont les pattes portent des indices libres vivant dans les repr\'esentations $(j_1, j_2, j_3)$. Les contractions de ces indices correspondent \`a l'\'etablissement de liens entre ces vertexes dans les repr\'esentations concern\'ees. Pour l'amplitude de vertex de Ponzano-Regge, il faut voir que chaque triangle porte un entrelaceur trivalent, de sorte que le graphe du symbole 6j est t\'etra\'edral, comme \`a la figure \ref{fig:6j}.

On peut aussi comprendre cette amplitude comme l'\'evaluation des fonctionnelles de r\'eseaux de spins vivant au bord de chaque t\'etra\`edre. En effet, le graphe du symbole 6j peut \^etre vu comme \emph{dual \`a la triangulation du bord} d'un t\'etra\`edre. Aussi on peut associer \`a ce graphe (fix\'e) un espace de Hilbert cin\'ematique, engendr\'e par les r\'eseaux de spins. Les spins induits par le mod\`ele de Ponzano-Regge forment les m\^emes donn\'ees que celles de la quantification en r\'eseaux de spins.

Pour terminer la pr\'esentation du mod\`ele, il faut donner l'amplitude de face. Pour les repr\'esentations irr\'eductibles de $\SU(2)$, on a : $\dim \calH_j = (2j+1)$, et :
\beq
Z_{\rm PR} = \sum_{(j_f)_{f\in\Gamma_2}} \prod_f (-1)^{2j_f} \bigl( 2j_f+1\bigr) \prod_v W_v^{\rm PR}(j_f).
\ee
L'interpr\'etation, provenant du formalisme canonique, des spins $j_f$ comme longueurs quantifi\'ees des ar\^etes de la triangulation, duales aux faces de $\Gamma$, est renforc\'ee par l'analyse asymptotique du symbole 6j ! On consid\`ere un t\'etra\`edre ayant pour longueurs $\ell_i = (j_i+\f12)$, et on note $V(\ell)$ son volume. Introduisons aussi l'action de Regge correspondante, $S_{\rm R}(\ell_i) = \sum_{i=1}^6 \ell_i\,\thet_i(\ell_j)$, comme fonction des longueurs des ar\^etes, et $\thet_i(\ell)$ d\'esignant les angles dih\'edraux. Alors, Ponzano et Regge ont conjectur\'e, \cite{PR}, \`a l'aide d'analyses num\'eriques, la formule asymptotique suivante, lorsque les spins sont envoy\'es vers l'infini de mani\`ere homog\`ene, $j_i\rightarrow \lambda j_i$ pour $\lambda$ tr\`es grand\footnote{En g\'en\'eral, la solution est une combinaison de fonctions d'Airy, qui se r\'eduit \`a \eqref{6jasymp} quand $V(\ell_e)^2>0$.} \cite{PR, schulten-gordon2} (voir plus r\'ecemment \cite{roberts-asym-6j,freidel-louapre-6j, gurau-asym-6j, dowdall-asym-PR}),
\be \label{6jasympPR}
\begin{Bmatrix} j_1 &j_2 &j_3 \\ j_4 &j_5 &j_6 \end{Bmatrix} \sim
\f1{\sqrt{12\pi V(\ell_i)}}\cos\left( S_{\rm R}(\ell_i) + \f\pi4\right),
\ee
Le r\'egime des grands spins fournit ainsi une notion de limite semi-classique des mousses de spins sous la forme d'un calcul de Regge quantique, i.e. avec des longueurs quantifi\'ees. Cette quantification des longueurs, qui \'etait initialement un ansatz de Ponzano et Regge, se trouve justifi\'ee par la quantification en boucles, comme l'a remarqu\'e Rovelli \cite{rovelli-PRTVO}.

Mais le mod\`ele de Ponzano-Regge offre plusieurs avantages sur le calcul de Regge. Il nous fournit en effet une mesure d'int\'egration, compatible avec les in\'egalit\'es triangulaires qui assurent que la m\'etrique de Regge, d\'efinie par les longueurs $(\ell_f)$, est bien d\'efinie positive, et ce gr\^ace au symbole 6j issu de la th\'eorie des repr\'esentations de $\SU(2)$. De plus, la mesure d\'etermin\'ee par l'amplitude de face, est telle que le mod\`ele est topologique, i.e ne d\'epend pas de la triangulation choisie !

Mais avant de voir cela, il faut s'attarder un peu sur les divergences. Les divergences les plus simples, \'etudi\'ees dans la litt\'erature, proviennent de distributions delta sur le groupe $G$ \emph{redondantes}. De telles redondances se produisent d\`es que la triangulation poss\`ede un sommet interne. Car alors, ce sommet est duale \`a une 3-cellule $b$ appel\'ee \emph{bulle}, et il est possible de trouver un ordre des faces formant le bord de la bulle telle que :
\be \label{bubble group variables}
\prod_{f\in \pp b} H_f^{\epsilon_f} = \unit,
\ee
o\`u $\epsilon_f=\pm1$ et \`a une conjugaison pr\`es de certaines $H_f$. Cela signifie que ces faces ne sont pas ind\'ependantes : il est possible d'exprimer une des holonomies autour d'une face en fonction des autres, de sorte qu'une des fonctions delta sur $G$ est redondante \`a chaque bulle. Cela donne une divergence formelle du type $\delta(\unit)$ pour chaque sommet de la triangulation, comme l'avait conjectur\'e Ponzano et Regge, proposant \'egalement une r\'egularisation adapt\'ee \`a cette situation. L'\'equation \eqref{bubble group variables} peut se comprendre g\'eom\'etriquement comme une version discr\`ete de l'identit\'e de Bianchi du continu, $d_A F_A =0$, sur chaque 3-cellule duale. Cette interpr\'etation a \'et\'e renforc\'ee par Freidel et Louapre, \cite{freidel-louapre-diffeo}, qui ont donn\'e une explication de ces divergences par l'existence d'une sym\'etrie de jauge, \'equivalent discret de la sym\'etrie de translation de la th\'eorie BF \eqref{translation sym bf}. En cons\'equence, les auteurs proposent une fixation de jauge qui \'eliminent ces divergences ! Cette fixation consiste \`a \'eliminer les fonctions delta associ\'ees aux ar\^etes le long d'un arbre maximal de la triangulation (c'est-\`a-dire touchant tous les sommets sans former de boucles), ou de mani\`ere \'equivalente \`a ne selectionner que le mode trivial $j=0$ dans la d\'ecomposition $\delta(g) = \sum_j (2j+1)\chi_j(g)$, le long de cet arbre.

Nous verrons aux chapitres consacr\'es aux relations de r\'ecurrence l'int\'er\^et de s\'electionner un mode $j\neq0$ de la contrainte de courbure nulle, et comment cela permet de relier l'action de la contrainte hamiltonienne sur les r\'eseaux de spins aux amplitudes de mousses de spins. Signalons aussi que les divergences pr\'esent\'ees ci-dessus sont les divergences les plus simples et na\"ives, la conjecture de Ponzano et Regge, \'etant fausse de mani\`ere g\'en\'erale. Pour une \'etude plus compl\`ete des divergences dans les mod\`eles de mousses de spins pour la th\'eorie BF, et plus g\'en\'eralement sur un 2-complexe arbitraire, je renvoie aux travaux que j'ai r\'ecemment effectu\'e en collaboration avec M. Smerlak, \cite{cell-homology, twisted-homology} (et \`a \cite{barrett-PR} plus sp\'ecifiquement pour le mod\`ele PR), et que je ne pr\'esenterai pas ici par manque de temps et de place.

\begin{figure}
\begin{center}
\includegraphics[width=6cm]{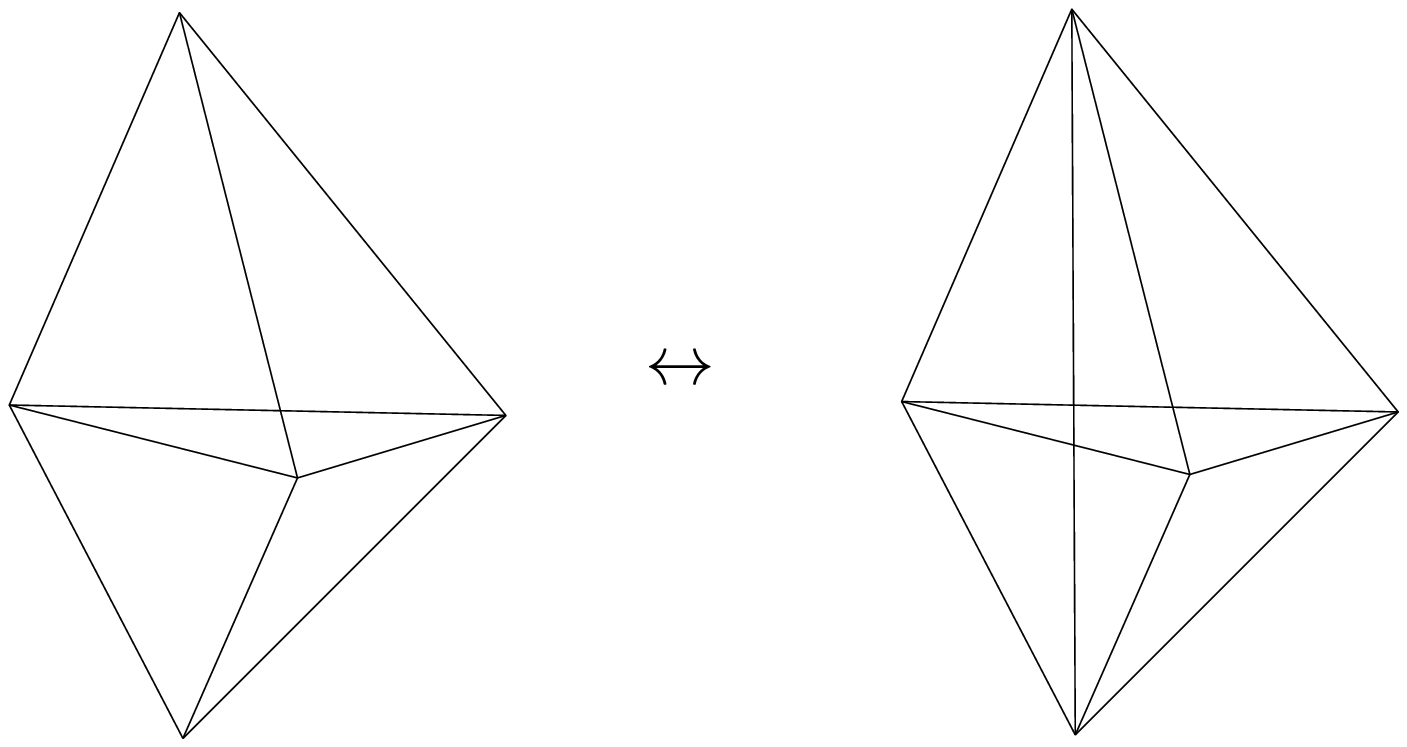}\qquad\qquad\qquad\qquad\includegraphics[width=6cm]{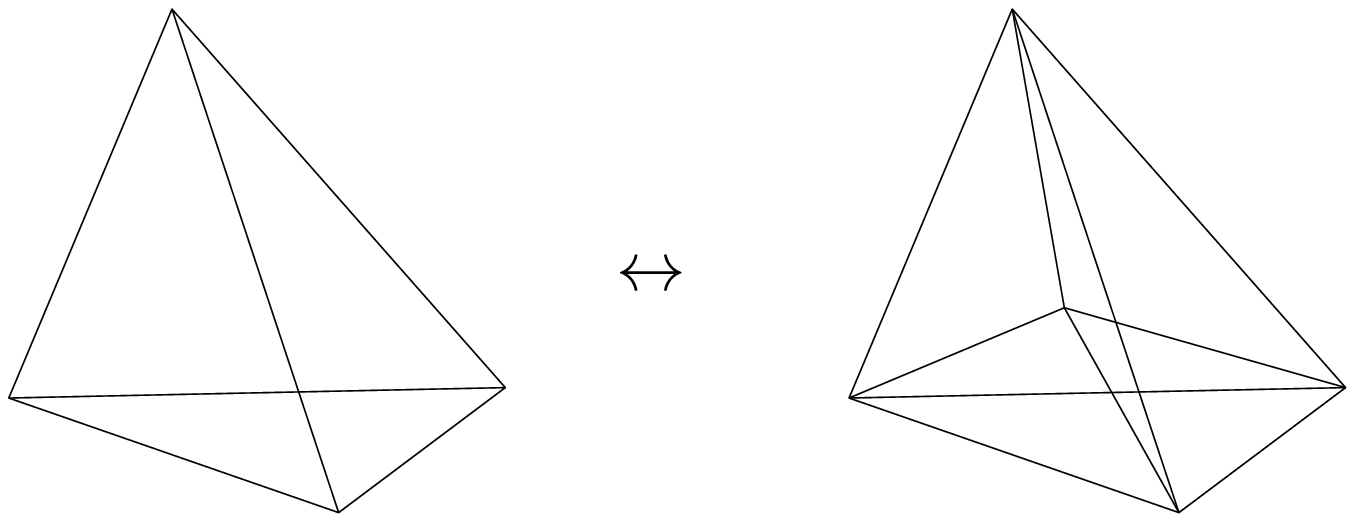}
\end{center}\caption{ \label{fig:3dpachner-moves} Les mouvements de Pachner relient les triangulations de vari\'et\'es hom\'eomorphes. En 3d, les mouvements \'el\'ementaires sont le mouvement 2-3, \`a gauche, et le mouvement 1-4, \`a droite. Sur la gauche, deux t\'etra\`edres sont coll\'es le long d'un triangle, et le mouvement consiste \`a enlever ce triangle, ajouter une ar\^ete reliant les deux sommets oppos\'es et trois triangles partageant cette ar\^ete. Le mouvement 1-4 consiste \`a ajouter un sommet au centre d'un t\'etra\`edre en le reliant aux 4 autres sommets pour former 4 t\'etra\`edres.
}
\end{figure}

L'invariance topologique du mod\`ele provient de l'invariance sous des transformations de la triangulations, appel\'es mouvements de Pachner. Ces mouvements relient les triangulations de vari\'et\'es hom\'eomorphes. En dimension 3, ces transformations s'\'ecrivent comme des s\'equences de deux mouvements \'el\'ementaires locaux et de leur inverse. Le mouvement 3-2 relie l'amplitude de Ponzano-Regge pour une configuration \`a trois t\'etra\`edres partageant une ar\^ete commune \`a l'amplitude d'une configuration \`a deux t\'etra\`edres voisins, comme illustr\'e en figure \ref{fig:3dpachner-moves}. L'\'egalit\'e entre les deux amplitudes est imm\'ediate lorsque l'on \'ecrit le mod\`ele comme une int\'egrale sur les connexions discr\`etes $A=(g_e)_{e\in\Gamma_1}$. Mais, cela appara\^it moins trivial dans la repr\'esentation en mousses de spins ! L'identit\'e qui correspond \`a ce mouvement est n\'eanmoins bien connue en th\'eorie des repr\'esentations : il s'agit de l'identit\'e de Biedenharn-Elliott (aussi appel\'ee dans un contexte plus g\'en\'eral identit\'e du pentagone),
\begin{equation} \label{biedenharn-elliott}
\sum_l (-1)^{S+l}(2l+1)\,\begin{Bmatrix} k &f &l\\ d &a &g\end{Bmatrix}\,\begin{Bmatrix} a &d &l\\ c &b &j\end{Bmatrix}\,\begin{Bmatrix} b &c &l\\ f &k &h\end{Bmatrix} = \begin{Bmatrix} j &h &g\\ k &a &b \end{Bmatrix}\,\begin{Bmatrix} j &h &g\\ f &d &c \end{Bmatrix},
\end{equation}
o\`u $S$ est la somme des neuf spins fix\'es. On a bien trois 6j pour trois t\'etra\`edres dans le membre de gauche, et deux \`a droite. La somme sur $l$ est pr\'ecis\'ement la somme sur le spin, et donc la longueur de l'ar\^ete commune aux trois t\'etra\`edres. Ainsi, cette identit\'e est intrins\`equement \emph{dynamique} : elle permet de faire explicitement la somme sur les spins pour des ar\^etes partag\'ees par trois t\'etra\`edres !

Le deuxi\`eme mouvement \'el\'ementaire est le mouvement 1-4, qui remplace un t\'etra\`edre par quatre, en ajoutant un sommet \`a l'int\'erieur du t\'etra\`edre initial, \ref{fig:3dpachner-moves}. Ce nouveau sommet est alors reli\'e aux quatre autres sommets. N\'eanmoins, une telle configuration contient un sommet interne, de sorte que le mouvement est en fait divergent. L'invariance est donc purement formelle : l'amplitude asosci\'ee aux quatre t\'etra\`edres est un simple 6j, fois un facteur $\delta(\unit)$. Notons que la proc\'edure de fixation de jauge de \cite{freidel-louapre-diffeo} qui impose $j_f=0$ le long d'un arbre maximal transforme directement l'amplitude divergente \`a quatre t\'etra\`edres en un 6j.

Nous verrons en d\'etails aux chapitres sur les relations de r\'ecurrence comment l'invariance sous ces mouvements de Pachner impl\'emente la contrainte hamiltonienne de la th\'eorie. La possibilit\'e de consid\'erer le mod\`ele de Ponzano-Regge comme un projecteur sur les \'etats physiques, satisfaisant la contrainte hamiltonienne au niveau quantique a \'et\'e montr\'ee par Ooguri, \cite{ooguri-3d}. Pour cela, il faut \'etendre le mod\`ele PR de sorte qu'il traite les vari\'et\'es \`a bord, et identifier des \'etats de la gravit\'e sur r\'eseau au bord. Dans ce cas, une triangulation de la 3-vari\'et\'e $M$ induit une triangulation $\Delta$ du bord. De m\^eme la d\'ecomposition cellulaire duale $\Gamma$ induit la d\'ecomposition cellulaire duale \`a la triangulation du bord. Nous pouvons donc consid\'erer au bord les donn\'ees des fonctionnelles de r\'eseaux de spins, permettant de d\'efinir un espace de Hilbert cin\'ematique. Ces donn\'ees sont des coloriages des liens duaux \`a $\Delta$, ou de mani\`ere \'equivalente des ar\^etes de $\Delta$, par des spins $c=(j_e)$, et l'espace naturel est :
\beq
\calH_{\rm kin}(\Delta) = \Bigl\{ \sum_c \phi_\Delta(c)\,\vert s^c_\Delta\rangle\Bigr\},
\ee
muni du produit scalaire $\langle s^{c'}_\Delta \vert s^c_\Delta\rangle = \delta_{c'c}$. Cela coincide bien s\^ur avec l'espace cin\'ematique de la LQG mais sur un graphe fix\'e ! Les fonctions des coloriages, $\phi_\Delta(c)$, sont les fonctions d'onde de la gravit\'e sur r\'eseau.

On se souvient aussi qu'une tranche du 2-squelette duale \`a la triangulation s'interpr\`ete naturellement comme un graphe de r\'eseaux de spins, ce qui est ici le cas pour le 1-squelette dual \`a la triangulation du bord. Pour finir, un ensemble de spins coloriant les faces duales de la 3-vari\'et\'e induit un coloriage par ces spins du 1-squelette dual \`a $\Delta$, ou par dualit\'e des ar\^etes de $\Delta$. On peut alors calculer la fonction de partition \`a la Ponzano-Regge en fixant un coloriage $c = (j_e)$ des ar\^etes au bord, ce qui fixe le coloriage des faces duales de $\Gamma$ touchant le bord,
\beq
Z_\Delta(c) = \prod_{e\in\Delta} i^{j_e}\sqrt{(2j_e+1)}\ \sum_{(j_f)} \prod_{\rm faces\ internes} (-1)^{2j_f}(2j_f+1) \prod_{\rm vertexes\ internes} W_v^{\rm PR}(j_f).
\ee

On cherche maintenant \`a voir le mod\`ele PR comme un projecteur. Consid\'erons donc une surface $\Sigma$ ferm\'ee orientable, qui sera notre surface canonique initiale et finale. Soit $N$ une 3-vari\'et\'e ayant la topologie de $\Sigma\times [0,1]$. On peut alors calculer le mod\`ele PR pour des coloriages $c_1, c_2$ fix\'es sur les deux composantes du bord de $N$, triangul\'ees par $\Delta_1, \Delta_2$. Cela fournit la quantit\'e $Z_{\Delta_1\Delta_2}(c_1, c_2)$. On peut alors composer de telles fonctions en ajoutant un autre cobordisme $\Sigma\times [0, 1]$. Gr\^ace \`a l'ind\'ependance en la triangulation, on met en \'evidence la propri\'et\'e remarquable suivante :
\beq
\Lambda(\Delta_2) \sum_{c_2\ {\rm de}\ \Delta_2} Z_{\Delta_1\Delta_2}(c_1, c_2)\ Z_{\Delta_2\Delta_3}(c_2, c_3) = Z_{\Delta_1\Delta_3}(c_1, c_3),
\ee
$\Lambda(\Delta_2)$ \'etant une constante qui ne d\'epend que du nombre de sommets de $\Delta_2$. Cela sugg\`ere de d\'efinir comme op\'erateur sur l'espace cin\'ematique $\calH_{\rm kin}(\Delta)$ d'une triangulation $\Delta$,
\beq
P_\Delta (\phi_\Delta)(c) = \Lambda(\Delta)\, \sum_{c'\ {\rm de}\ \Delta} Z_{\Delta \Delta}(c, c') \  \phi_\Delta(c'),
\ee
qui est bien un projecteur, puisque $P_\Delta\circ P_\Delta = P_\Delta$ ! On d\'efinit les \'etats physiques de la sorte :
\beq
\phi_\Delta\in\calH_{\rm phys}(\Delta) \quad \text{si}\qquad \phi_\Delta(c) = P_\Delta (\phi_\Delta)(c).
\ee
Il faut voir cette condition comme l'\'equation de Wheeler-De-Witt pour la th\'eorie discr\`ete ! Le produit scalaire physique est donn\'e par la contraction du noyau $Z_{\Delta \Delta}$ avec deux \'etats sur le graphe dual \`a $\Delta$,
\beq
\langle \phi_\Delta \vert \psi_\Delta \rangle_{\rm phys} = \Lambda(\Delta)^2 \sum_{c,c'} \phi_\Delta(c)\ Z_{\Delta \Delta}(c, c') \  \psi_\Delta(c').
\ee
Il s'agit pr\'ecis\'ement d'une r\'ealisation de l'\'equation g\'en\'erique \eqref{spin foam transition}, dans laquelle le 2-complexe est fix\'e, comme 2-squelette dual \`a la triangulation de $N$.

Ce formalisme a l'avantage de pouvoir traiter des situations inaccessibles au formalisme hamiltonien et \`a la quantification canonique. Une restriction \'evidente du formalisme canonique est la seule prise en compte de la topologie du type $\Sigma \times [0, 1]$. Par contraste, le formalisme du mod\`ele PR conduit naturellement \`a consid\'erer des amplitudes de transition entre des surfaces de topologie diff\'erentes, $\Sigma_1, \Sigma_2$, reli\'ees par des cobordismes, $(M,\Sigma_1, \Sigma_2)$, comme dans \cite{witten-amplitude-3d}. Il est aussi possible de construire des \'etats \`a la Hartle-Hawking. Supposons que $\Sigma$ soit hom\'eomorphe au bord d'une 3-vari\'et\'e $M$. Fixons une triangulation de $M$ et $\Delta$ la triangulation induite sur le bord. Alors on peut consid\'erer en repr\'esentations de spins l'\'etat
\beq
\phi_{M, \Delta} (c) = Z_{M, \Delta}(c),
\ee
ou en repr\'esentations de connexion,
\beq
\phi_{M, \Delta}(A) = \sum_c Z_{M, \Delta}(c)\ s_{\Delta}^c(A).
\ee
Ces aspects ont \'et\'e \'etudi\'es en d\'etails par Witten, \cite{witten-amplitude-3d}, pour $M$ un \og handlebody\fg, dans la th\'eorie continue, et l'\'equivalence avec la th\'eorie sur r\'eseau est donn\'ee dans \cite{ooguri-3d}.

Le lien avec la quantification de la contrainte hamiltonienne par Noui et Perez, \cite{noui-perez-ps3d}, tient d'abord \`a la repr\'esentation de cette quantification en mousses de spins, comme \'ebauch\'ee \`a la section pr\'ec\'edente. En restreignant l'action de la contrainte aux r\'eseaux de spins trivalents, les amplitudes correspondant au produit scalaire physique sont pr\'ecis\'ement les amplitudes PR. Il nous reste donc \`a expliquer pourquoi une telle restriction n'en est en fait pas une, ou de mani\`ere \'equivalente, pourquoi l'espace physique $\calH_{\rm phys}(\Delta)$ ne d\'epend pas de $\Delta$. Cela vient de la contrainte de courbure nulle, $F_{ab}^i =0$, qui pousse \`a chercher les fonctionnelles des connexions plates (invariantes de jauge). Or les holonomies d'une connexion plate ne d\'ependent que des classes d'homotopie des courbes. On peut donc se restreindre \`a un nombre fini de degr\'es de libert\'e, sur une triangulation, qui contient toutes les informations topologiques de la vari\'et\'e. Au niveau canonique, il suffit en fait de consid\'erer les r\'eseaux de spins construits le long d'un ensemble de g\'en\'erateurs de $\pi_1(\Sigma)$, la contrainte de courbure nulle se r\'eduisant aux relations qui existent entre ces g\'en\'erateurs.

\subsection{Mod\`ele d'Ooguri, en 4d} \label{sec:ooguri model}

\begin{figure}[ht]
\begin{center}
\includegraphics[width=10cm]{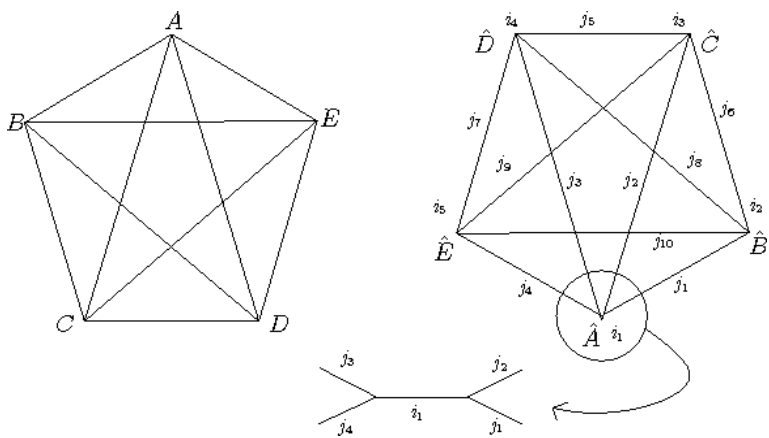}
\end{center}\caption{ \label{fig:15j} Sur la figure de gauche est repr\'esent\'e un 4-simplexe, de sommets $A, B, C, D, E$ (les t\'etra\`edres \'etant form\'e par des collections de 4 de ces points et ainsi de suite). Sur la droite, il s'agit du r\'eseau dual au bord du 4-simplexe. Ainsi, un t\'etra\`edre, disons $(ABCD)$ (donc obtenu en \'eliminant le sommet $E$), est repr\'esent\'e par un vertex, appel\'e $\hat{E}$. Les liens sont de m\^eme duaux aux triangles, et donc color\'es par des spins $j_f$. Pour \'ecrire le mod\`ele d'Ooguri en termes de mousses de spins, on d\'eveloppe chaque entrelaceur 4-valent en deux vertex trivalents reli\'es par un spin virtuel, dont les caract\'eristiques ont \'et\'e donn\'es \`a la section \ref{sec:quantum tet}.
}
\end{figure}

Le mod\`ele similaire en dimension 4 est appel\'e mod\`ele d'Ooguri, \cite{ooguri4d}. Il se pr\'esente sous la forme d'une amplitude associ\'ee \`a chaque type de cellules, via des coloriages des faces par des spins et des liens par des entrelaceurs. La fonction de partition est obtenue en faisant le produit de ces amplitudes sur toute la triangulation, puis en sommant sur tous les coloriages -- et au final en sommant sur tous les 2-complexes. En 4d, chaque lien est partag\'e par quatre faces duales, correspondant aux quatre triangles d'un t\'etra\`edre. Cela signifie que l'on a \`a int\'egrer un produit de quatre \'el\'ements matriciels de repr\'esentations, aboutissant \`a l'identit\'e sur l'espace invariant du produit tensoriel de quatre repr\'esentations, $\Inv(\calH_{j_1}\otimes\calH_{j_2}\otimes\calH_{j_3}\otimes\calH_{j_4})$. Il nous faut donc exhiber une base des entrelaceurs 4-valents, en d\'eveloppant ceux-ci en deux entrelaceurs trivalents. On choisit un des trois appariements possibles entre les quatre repr\'esentations $j_f$, ainsi qu'une repr\'esentation interm\'ediaire $i$. Un entrelaceur s'\'ecrit par exemple $\lvert (j_1j_2),(j_3j_4);i\rangle$, o\`u $i$ est choisi dans la d\'ecomposition en repr\'esentations irr\'eductibles de $\calH_{j_1}\otimes\calH_{j_2} = \oplus_{i_{12}} \calH_{i_{12}}$ et de $\calH_{j_3}\otimes\calH_{j_4} = \oplus_{i_{34}} \calH_{i_{34}}$. On a alors :
\begin{align}
\int_{\SU(2)} dg\ \prod_{a=1}^4 D^{(j_a)}_{m_an_a}(g) &= \langle (j_a, m_a)\, \vert\ \id_{\Inv(\calH_{j_1}\otimes\dotsm\otimes\calH_{j_4})}\ \vert (j_a, n_a)\rangle,\\
&= \sum_i (2i+1)\ \langle (j_a, m_a)\, \vert (j_1j_2),(j_3j_4);i\rangle\ \langle (j_1j_2),(j_3j_4);i\,\vert (j_a, n_a)\rangle.  \label{int g4}
\end{align}
Les \'etats $\vert (j_a, m_a)\rangle$ repr\'esentent les \'etats de moments magn\'etiques bien d\'efinis, $\otimes_{a=1}^4 \vert j_a, m_a\rangle$. Leur contraction avec les entrelaceurs $\lvert (j_1j_2),(j_3j_4);i\rangle$ sont donn\'ees explicitement en appendice \ref{sec:app}. L'amplitude de vertex s'obtient en contractant les indices libres des quantit\'es $\langle (j_a, m_a)\, \vert (j_1j_2),(j_3j_4);i\rangle$ au sein de chaque 4-simplexe. Il y a alors 10 spins sur les faces, et cinq entrelaceurs (caract\'eris\'e par un appariement et un spin virtuel) sur les t\'etra\`edres. La structure combinatoire peut \^etre comprise en regardant le graphe dual \`a la triangulation du bord. Les entrelaceurs sur les 5 t\'etra\`edres sont repr\'esent\'es par 5 noeuds 4-valents, et les 10 faces deviennent des liens reliant ces noeuds, comme sur la figure \ref{fig:15j}. Le graphe correspondant a bien s\^ur la m\^eme structure que le 4-simplexe, et l'amplitude r\'esultante est appel\'ee symbole 15j.

L'interpr\'etation canonique de ce 15j, en termes d'\'etats de bord du 4-simplexe, est imm\'ediate : les \'etats $\lvert (j_1j_2),(j_3j_4);i\rangle$ ne sont autres que les \'etats du t\'etra\`edre quantique, pr\'esent\'es en section \ref{sec:quantum tet}, et \'etats propres des op\'erateurs d'aires\footnote{Les \'etats cin\'ematiques de la LQG et de la th\'eorie BF 4d $\SU(2)$ sont les m\^emes, et sauf \`a consid\'erer la gravit\'e compl\`ete, il faudrait plut\^ot parler des flux du moment conjugu\'e \`a la connexion \`a travers les triangles \`a la place d'aires physiques.}, et de l'op\'erateur d'aire d'un parall\'elogramme coupant le t\'etra\`edre en deux, \eqref{eigenstate area tet}, \eqref{eigenstate A12 tet} (ou d'un angle dih\'edral) !

Le lecteur attentif aura not\'e que la structure de chaque 15j d\'ependra du choix des appariements pour chacun des cinq entrelaceurs d'un 4-simplexe. Il est n\'eanmoins possible d'\'ecrire le mod\`ele en n'utilisant qu'un seul type de 15j, mais cela requiert g\'en\'eralement des choix d'appariements diff\'erents pour les deux entrelaceurs impliqu\'es dans \eqref{int g4}. L'invariance de jauge est alors garantie par l'apparition de symboles 6j appropri\'es \`a chaque t\'etra\`edre pour recoupler correctement les deux appariements diff\'erents. Sans pr\'eciser le d\'etails des 15j, nous pouvons donc \'ecrire la fonction de partition comme :
\beq \label{oog model}
Z_{\rm BF4d} = \sum_{(j_f)_{f\in\Gamma_2}} \sum_{(i_e)_{e\in\Gamma_1}} \prod_f (2j_f+1)\ \prod_e (2i_e+1)\ \prod_v \{15j\}(j_f, i_e).
\ee
Comme en dimension 3, des mouvements de Pachner permettent de relier deux triangulations d'une m\^eme vari\'et\'e topologique. Il existe trois mouvements \'el\'ementaires (et leur inverse): le mouvement 1-5, qui se construit comme le mouvement 1-4 en 3d, en ajoutant un sommet au centre d'un 4-simplexe, le mouvement 2-4 que nous \'etudierons en d\'etails \`a la section \ref{sec:move4-2}, et le mouvement 3-3. Seul ce dernier est une invariance rigoureuse du mod\`ele tel que nous l'avons d\'efini. En effet, il ne poss\`ede qu'une invariance formelle sous les autres mouvements, du fait de divergences. Comme en 3d, ces divergences sont associ\'ees \`a la pr\'esence de bulles, i.e. de 3-cellules duales, dans les configurations mises en sc\`ene dans ces mouvements. A nouveau, ces bulles sont le lieu de fonctions delta sur le groupe redondantes, mais il n'y a pas eu jusqu'ici de proposition de fixation de jauge. Une des difficult\'es est la suivante : le mouvement 1-5, typiquement, produit cinq 3-cellules, de sorte que l'argument de Ponzano et Regge incite \`a attendre 5 delta redondants. Mais, seulement 4 le sont. Cela vient du fait que les cinq 3-cellules ne sont pas ind\'ependantes, mais forment le bord d'une 4-cellule duale au sommet ajout\'e. Bien s\^ur ce ph\'enom\`ene est reli\'e \`a la r\'eductibilit\'e de la sym\'etrie de jauge de translation \eqref{translation sym bf} de la th\'eorie continue, se traduisant dans le formalisme BRST par la n\'ecessit\'e d'introduire des fant\^omes de fant\^omes (\og ghosts of ghosts\fg), \cite{renormalisability-bf}. Il serait certainement int\'eressant d'\'etudier pr\'ecis\'ement les sym\'etries de jauge au niveau discret en dimension quelconque, pour g\'en\'eraliser \cite{freidel-louapre-diffeo}, et comprendre la structure des divergences associ\'ees dans la fonction de partition !

\medskip

Comme cela devient clair, l'invariance sous les mouvements de Pachner du mod\`ele topologique est la cl\'e de la dynamique du mod\`ele. Cette invariance (formelle) est facile \`a v\'erifier dans la formulation du mod\`ele en termes d'int\'egrales sur les connexions discr\`etes, \eqref{partition bf lattice}. En revanche, cela est nettement moins trivial dans la repr\'esentation en mousses de spins. Une autre fa\c{c}on de le dire est que l'action de la contrainte hamiltonienne $F_{ab}^i=0$ reste simple sur les \'etats en repr\'esentations de Schr\H{o}dinger, fonctionnelles de la connexion, mais prend une forme moins \'evidente en repr\'esentations de coloriages par des spins,
\beq
\widehat{F_A}\,\phi_\Delta(c) = 0.
\ee
Naturellement, cela doit donner des \'equations au diff\'erences finies sur les coloriages des \'etats physiques, contenues dans l'invariance du mod\`ele de mousses de spins sous les mouvements de Pachner ! Ce sont ces \'equations que nous \'etudions \`a la partie \ref{sec:recurrence}.



\chapter{Calculer avec les mousses de spins} \label{sec:semiclass}

\section{Les fonctions de corr\'elations}

Si nous avons pr\'ec\'edemment discut\'e les amplitudes de transition entre r\'eseaux de spins, i.e. leur produit scalaire physique, nous n'avons encore pas parl\'e des observables standards en th\'eorie quantique des champs que sont les fonctions de Green/de corr\'elations. Elles sont pourtant essentielles pour d\'eriver la limite semi-classique, de basse \'energie, et ainsi v\'erifier son bon comportement, mais aussi si l'on veut pouvoir comparer nos mod\`eles, pour l'op\'erateur hamiltonien ou de mousses de spins, \`a d'autres approches. Le r\'egime semi-classique peut aussi nous permettre de d\'evelopper des approximations et m\'ethodes perturbatives originales, diff\'erentes de celles de la gravit\'e lin\'earis\'ee, \`a partir de nos mod\`eles de mousses de spins intrins\`equement non-perturbatifs. Nous discutons donc dans cette section une approche g\'en\'erique pour d\'efinir proprement les fonctions de corr\'elations dans un contexte g\'en\'eralement covariant et sans m\'etrique d'espace-temps privil\'egi\'ee a priori. Cette approche a \'et\'e initi\'ee dans \cite{rovelli-modesto-scattering}, et est \`a la base de tous les calculs d'observables type graviton dans les mod\`eles de mousses de spins, en particulier celui que je d\'eveloppe en gravit\'e 3d \`a la section suivante. Pour une pr\'esentation plus compl\`ete de la m\'ethode, je renvoie le lecteur \`a \cite{gravitonpropinlqg}.

En th\'eorie quantique des champs conventionnelle, les fonctions de Green \`a $n$ points sont formellement les insertions de $n$ champs dans l'int\'egrale de chemins,
\beq \label{greenqft}
W(x_1,\dotsc,x_n) = Z\mone \int D\phi\ \phi(x_1)\dotsm\phi(x_n)\ e^{iS(\phi)},
\ee
o\`u $S(\phi)$ est l'action de la th\'eorie, et $Z$ l'int\'egrale de chemins sans insertion de champs $\phi(x)$. Les points $x_1,\dotsc,x_n$ sont des points de l'espace-temps. De tels quantit\'es se calculent explicitement, en th\'eorie des perturbations \`a la Feynman, ou avec une r\'egularisation sur r\'eseau, \`a l'aide d'une renormalisation appropri\'ee. Il faut cependant remarquer que l'espace-temps est alors muni d'une m\'etrique physiquement privil\'egi\'ee et fix\'ee, i.e. non-dynamique, comme la m\'etrique de Minkowski. Celle-ci nous permet de mesurer la distance entre les points $x_i$, et est essentielle \`a l'invariance de $W(x_1,\dotsc,x_n)$ sous le groupe de Poincar\'e typiquement. Par ailleurs, si la mesure et l'action, $D\phi\,e^{iS}$, sont invariantes sous les diff\'eomorphismes, alors il est clair que $W$ est localement ind\'ependants des  points $x_i$.

Il nous faut donc une strat\'egie pour traiter ces difficult\'es, et \^etre capable de traduire \eqref{greenqft} en termes de mousses de spins. Si $R$ est une r\'egion de l'espace-temps dont le bord, $\Sigma=\pp R$ abrite les points $x_1,\dotsc,x_n$, notons $\varphi$ la restriction du champ $\phi$ \`a $\Sigma$, et r\'e\'ecrivons \eqref{greenqft} comme :
\beq \label{backindptgreen}
W(x_1,\dotsc,x_n) = Z\mone \int D\varphi\ \varphi(x_1)\dotsm\varphi(x_n)\ W(\varphi,\Sigma)\ \Psi_0(\varphi),
\ee
o\`u $W(\varphi,\Sigma)$ n'est autre que l'int\'egrale de chemins sur $R$,
\beq
W(\varphi,\Sigma) = \int \left.D\phi_R\right|_{\phi_{R|\Sigma}=\varphi}\ e^{iS_R(\phi_R)},
\ee
avec la condition de champ fixe, $\phi_{R|\Sigma}=\varphi$, sur $\Sigma$, et $S_R(\phi_R)$ \'etant l'action restreinte \`a $R$. Du fait de la covariance g\'en\'erale, $W(\varphi,\Sigma)$ est invariant sous les petites d\'eformations de $\Sigma$. On peut donc simplement noter le noyau de propagation du champ \`a l'int\'erieur de $R$, $W(\varphi)$. En gravit\'e quantique, la g\'eom\'etrie sur $\Sigma$ utilis\'ee pour calculer ce noyau n'est donc naturellement pas li\'ee \`a une m\'etrique privil\'egi\'ee, mais d\'etermin\'ee par $\varphi$ qui n'est autre que le champ gravitationnel au bord. Disons tout de suite qu'un int\'er\^et pour nous de ce d\'ecoupage est la possibilit\'e de d\'ecrire les configurations de bord, $\varphi$, en termes des r\'eseaux de spins qui forment une base de l'espace des \'etats cin\'ematique, et d'observer alors que $W(\varphi)$ est pr\'ecis\'ement ce que les mod\`eles de mousses de spins cherchent \`a calculer !

Quant \`a $\Psi_0(\varphi)$, il s'agit de l'int\'egrale de chemins sur l'ext\'erieur de $R$. Si l'on peut supposer que la th\'eorie est (approximativement) libre en dehors de $R$, alors $\Psi_0(\varphi)$ r\'esulte d'une int\'egrale gaussienne (avec des conditions \`a l'infini bien s\^ur) et n'est autre que l'\'etat du vide de la th\'eorie (libre) en repr\'esentation de Schr\H{o}dinger. En gravit\'e lin\'earis\'ee autour d'une m\'etrique de fond, $g_{\mu\nu} = g_{\mu\nu}^0+h_{\mu\nu}$, cet \'etat d\'ecrit une g\'eom\'etrie quantique piqu\'ee autour de $h_{ab}=0$ et $\pi^{ab}=0$, ce dernier \'etant le moment conjugu\'e \`a $h$. Cela sugg\`ere une g\'en\'eralisation \`a des \'etats de type coh\'erent sur $\Sigma$, i.e. de $\Psi_0(\varphi)$ \`a un \'etat $\Psi_q(\varphi)$ qui serait piqu\'e autour d'une 3-g\'eom\'etrie $q$, d\'ecrivant la g\'eom\'etrie intrins\`eque \emph{et} la g\'eom\'etrie extrins\`eque de $\Sigma$. De telles donn\'ees permettent alors de situer les entr\'ees $x_1,\dotsc,x_n$ des fonctions de Green en lien avec la m\'etrique introduite sur $\Sigma$ issue de $q$, et donc en lien avec les valeurs que prend le champ gravitationnel sur $\Sigma$.

En r\'egime quelconque de la gravit\'e quantique, on souhaite bien s\^ur consid\'erer des surfaces $\Sigma$ arbitraires. Mais, pour \'etudier la limite semi-classique des mousses de spins, en comparaison avec la gravit\'e quantique lin\'earis\'ee, il nous est suffisant de regarder sur deux hyperplans de genre espace de petites perturbations $h'_{ab}$ et $h''_{ab}$ qui s'annulent asymptotiquement. Dans ce cadre, le noyau de propagation $W[h',h'',T]$ en relativit\'e g\'en\'erale lin\'earis\'ee a \'et\'e \'etudi\'e (voir \cite{mattei-transition-graviton} pour un calcul explicite en jauge temporelle et pour plus de d\'etails et r\'ef\'erences sur le sujet), $T$ \'etant le temps propre asymptotique entre les deux hypersurfaces. Alors, la fonction \`a deux points est donn\'ee par :
\beq
W_{abcd}(\vec{x},\vec{y};T) = \f1\calN \int Dh'\,Dh''\ W[h',h'',T]\ \Psi_0(h')\,h'_{ab}(\vec{x},0)\ \Psi_0(h'')\,h''_{cd}(\vec{y},T),
\ee
o\`u $\calN$ est une constante de normalisation. Notons que si le noyau et l'\'etat du vide sont bien invariants de jauge, ce n'est pas le cas des insertions, qui n\'ecessitent donc des choix de fixations de jauge spatiales. Nous pouvons alors donner une formulation en termes de mousses de spins,
\beq \label{loop prop}
W_{\mu\nu\rho\sigma}(x,y) = \f1N \sum_s W[s]\ \Psi_0(s)\ \langle s\lvert h_{\mu\nu}(x)\rvert s\rangle\ \langle s\lvert h_{\rho\sigma}(y)\rvert s\rangle.
\ee
Ici, la somme est prise sur l'ensemble des r\'eseaux de spins. Le noyau $W[s]$ est fourni explicitement par un mod\`ele de mousses de spins choisi, et prend en compte la somme sur tous les 2-complexes (ou sur toutes les mousses) compatibles avec le r\'eseau $s$. $\Psi_0(s)$ est une fonctionnelle sur les r\'eseaux de spins qui doit \^etre piqu\'e sur ceux d\'ecrivant l'espace plat dans la limite semi-classique.

Une telle formule (avec des approximations comme nous le verrons) a permis dans tous les mod\`eles \'etudi\'es pour linstant de reproduire correctement le comportement du propagateur avec la distance, en $1/\ell^2$. Mais ces calculs sont aussi \`a l'origine de la d\'ech\'eance du mod\`ele de Barrett-Crane, dont on a pu montrer \cite{alesci-BC1, alesci-BC2} qu'il ne produisait pas la structure tensorielle de $W_{\mu\nu\rho\sigma}$ attendue, dans le cas d'un 4-simplex isol\'e. Je renvoie notamment \`a \cite{gravitonEPR-bianchi} pour des r\'esultats encourageants sur le nouveau mod\`ele dit EPR, toujours pour un unique 4-simplex, mais avec le param\`etre d'Immirzi, et une comparaison \emph{explicite} \`a la m\^eme quantit\'e calcul\'ee par le calcul de Regge (dans \cite{modesto-perturbativeregge}) et par la relativit\'e g\'en\'erale lin\'earis\'ee. Signalons \'egalement une approche assez similaire concernant aussi la limite semi-classique du mod\`ele EPR, mais \`a travers la cosmologie, pour retrouver les \'equations de Friedmann \cite{vidotto-sfcosmo}.

Certains de ces calculs utilisent les formes asymptotiques de $W[s]$, tronqu\'e \`a un seul vertex, i.e. un seul 4-simplex, et \`a l'ordre dominant dans la limite des grands spins. Je renvoie \`a \cite{barrett-asym15j} pour le symbole 15j, \cite{barrett-asymEPR, pereira-asymEPR} pour le mod\`ele EPR riemannien et lorentzien, et \`a \cite{freidel-conrady-semiclass} pour le mod\`ele FK, qui montrent tous la proximit\'e du d\'eveloppement semi-classique (asymptotique \`a grands spins) des mousses de spins avec le calcul de Regge !

\section{Corr\'elations entre deux longueurs dans un mod\`ele en 3d} \label{sec:3dgraviton}

Nous regardons ici un mod\`ele simplificateur, en gravit\'e 3d, fond\'e sur le mod\`ele de Ponzano-Regge avec un unique t\'etra\`edre. Il fut introduit par S. Speziale \cite{graviton3d-simone}, puis d\'evelopp\'e aux premiers ordres dans \cite{graviton3d-corrections}, avant d'obtenir un d\'eveloppement complet dans \cite{graviton3d-val}. Bien que la th\'eorie ne poss\`ede aucun degr\'e de libert\'e localement, il est possible de calculer la fonction \`a deux points dans une jauge o\`u elle ne s'annule pas (m\^eme si les gravitons sont alors de purs effets de jauge). Cela va nous permettre de discuter plus ais\'ement le choix de l'\'etat de bord, et de parvenir \`a un d\'eveloppement asymptotique complet dans la limite des grandes distances. Ce d\'eveloppement reproduit \`a l'ordre le plus bas le comportement en $1/\ell$ du propagateur en 3d dans la jauge ad\'equate. De plus, la m\^eme m\'ethode pourra s'appliquer au symbole 6j, dans un cas particulier dit isoc\`ele, et nous donnera un d\'eveloppement asymptotique complet ! Celui-ci a par la suite \'et\'e confirm\'e et g\'en\'eralis\'e dans \cite{maite-etera-6jcorr, maite-etera-6jasym} par Dupuis et Livine, notamment en utilisant des relations de r\'ecurrence sur le symbole 6j, telles que celles que nous regardons \`a la partie \ref{sec:recurrence} dans diff\'erents mod\`eles de mousses de spins !

Ce mod\`ele en 3d permet de constater explicitement comment le calcul de Regge intervient. A l'ordre le plus bas, il en effet possible, comme mentionn\'e en fin de section pr\'ec\'edente, de se contenter de l'asymptotique du mod\`ele, ici du symbole 6j, qui est basiquement\footnote{En g\'en\'eral, la solution est une combinaison de fonctions d'Airy, qui se r\'eduit \`a \eqref{6jasymp} quand $V(\ell_e)^2>0$.} le cosinus de l'action de Regge 3d $S_{\rm R}$ \cite{PR, schulten-gordon2},
\be \label{6jasymp}
\begin{Bmatrix} j_1 &j_2 &j_3 \\ j_4 &j_5 &j_6 \end{Bmatrix} \sim
\f1{\sqrt{12\pi V(\ell_e)}}\cos\left( S_{\rm R}[\ell_e] + \f\pi4\right),
\ee
o\`u $V(\ell_e)$ est le volume du t\'etra\`edre dont les longueurs sont donn\'ees par $\ell_e = j_e+\f12$, et poss\'edant des angles dih\'edraux $\theta_e(\ell)$ fonctions des longueurs. L'action de Regge est $S_{\rm R} = \sum_e\ell_e\,\theta_e(\ell)$. Le r\'egime des grands spins fournit ainsi une notion de limite semi-classique des mousses de spins sous la forme d'un calcul de Regge quantique (avec des longueurs quantifi\'ees). Mais le mod\`ele de Ponzano-Regge offre plusieurs avantages sur le calcul de Regge. Il nous fournit en effet une mesure d'int\'egration, compatible avec les in\'egalit\'es triangulaires qui assurent que la m\'etrique de Regge est bien d\'efinie positive, et ce gr\^ace au symbole 6j issu de la th\'eorie des repr\'esentations de $\SU(2)$. Cette mesure est par ailleurs fix\'ee par la condition de faire un mod\`ele topologique, ne d\'ependant pas de la triangulation.

Ce mod\`ele topologique est bien s\^ur bas\'e sur l'utilisation du 6j complet, et pas seulement de son asymptotique \eqref{6jasymp}, comme en atteste les relations de r\'esurrence \emph{exactes} sur le 6j qui caract\'erisent les sym\'etries de la th\'eorie au niveau quantique. En ce qui concerne le calcul du propagateur, il a \'et\'e montr\'e dans \cite{graviton3d-corrections} que les d\'eviations du 6j \`a l'asymptotique de Ponzano et Regge engendrent des corrections \`a partir du deuxi\`eme ordre sous-dominant (le next-to-next-to-leading order, NNLO) aux grands spins. La m\'ethode que nous utilisons ici repose sur une forme int\'egrale du 6j et du propagateur, suivie d'une approximation du point col. Cette repr\'esentation int\'egrale, exacte, prend donc en compte les corrections \`a l'asymptotique \eqref{6jasymp} de mani\`ere naturelle !

\subsection{Pr\'esentation du mod\`ele}

\begin{figure}[ht]
\begin{center}
\includegraphics[width=4cm]{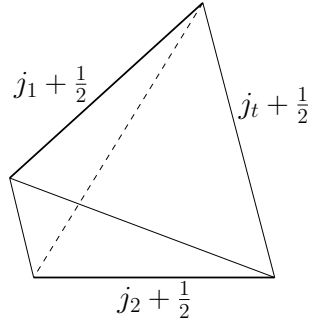}
\end{center}\caption{ \label{graviton setting} Le cadre du calcul de la fonction \`a 2 points. Les deux liens pour lesquels nous allons calculer les corr\'elations de fluctuations de longueur son en gras, et ont pour longueur moyenne $(j_1+\f{1}{2})$ et $(j_2+\f{1}{2})$. Ces donn\'ees sont prises en compte \`a travers l'\'etat de bord du t\'etra\`edre. Dans la jauge temps, les quatre autres liens ont des longueurs impos\'ees, $(j_t+\f12)$, \`a interpr\'eter comme le temps propre d'une particule se propageant le long de ces liens. De mani\`ere \'equivalente, le temps entre deux plans contenant les liens $e_1, e_2$ a \'et\'e mesur\'e : $T=(j_t+\f{1}{2})/\sqrt{2}$.}
\end{figure}

Le mod\`ele est pr\'esent\'e en d\'etails dans \cite{graviton3d-simone}, nous en donnons ici les grandes lignes. Nous consid\'erons un t\'etra\`edre dans l'espace(-temps) 3d euclidien, et souhaitons calculer les corr\'elations entre les fluctuations des longueurs $\ell_1',\ell_2'$ de deux ar\^etes oppos\'ees, autour de valeurs moyennes $\ell_1, \ell_2$, suivant la figure \ref{graviton setting}. Comme fixation de jauge, nous imaginons que le temps propre $T$ entre ces deux ar\^etes a \'et\'e mesur\'e, fixant les longueurs des quatre autres ar\^etes \`a une valeur commune, $\ell_t = (j_t+\f12)\ell_P = \sqrt{2}T$, o\`u $\ell_P$ est la longueur de Planck. En consid\'erant que les vecteurs $\ell_1^\mu$ et $\ell_2^\mu$ sont dans les directions $x^1$ et $x^2$, nous projetons le propagateur $W_{\mu\nu\rho\sigma}(x,y)$ selon
\beq
\ell_1^2\ell_2^2 W_{1122}(T) := \ell^{\mu}_1\ell^\nu_1\ell^\rho_2\ell^\sigma_2\ W_{\mu\nu\rho\sigma}(x,y).
\ee
Du fait de la mesure du \og temps\fg{} $T$, la somme sur les r\'eseaux de spins dans \eqref{loop prop} est ici r\'eduite \`a une somme sur les coloriages $j_1', j_2'$ des ar\^etes $e_1, e_2$, qui sont reli\'es aux longueurs par $\ell_e'=(j_e'+\f12)\ell_P$, ces longueurs \'etant quantifi\'ees en gravit\'e 3d en boucles, en accord avec le mod\`ele de Ponzano-Regge. On a donc :
\beq
\ell_1^2\ell_2^2 W_{1122}(T) = \f1\calN \sum_{j_1',j_2'} W[j_1',j_2',j_t]\ \Psi_0(j_1',j_2')\ \langle j_1'\lvert \ell_1^\mu\ell_1^\nu h_{\mu\nu}(x)\rvert j_1'\rangle\,\langle j_2'\lvert \ell_2^\rho\ell_2^\sigma h_{\rho\sigma}(y)\rvert j_2'\rangle.
\ee
Le noyau est donn\'e par l'amplitude de Ponzano-Regge pour le t\'etra\`edre avec quatre spins fix\'es,
\beq
W[j_1',j_2',j_t] = \begin{Bmatrix} j_1' &j_t &j_t \\ j_2' &j_t &j_t \end{Bmatrix}.
\ee

L'id\'ee principale pour calculer la fonction \`a deux points est d'effectuer un d\'eveloppement vis-\`a-vis de la g\'eom\'etrie du bord, qui se comporte donc comme une g\'eom\'etrie de fond admettant des perturbations. Pour cela nous devons sp\'ecifier les g\'eom\'etries intrins\`eque et extrins\`eque du bord, ici donn\'ees par $\ell_t$, les longueurs moyennes $\ell_1,\ell_2$, et leurs moments conjugu\'es. Ces derniers sont des variables angulaires $\eta_e$ sur chaque ar\^ete. Ils doivent satisfaire les \'equations du mouvement, qui comme nous l'avons vu \'etablissent que ces variables angulaires sont les angles dih\'edraux associ\'es aux longueurs. 
Notre \'etude montre par ailleurs, de fa\c{c}on claire, que les d\'eveloppements s'\'ecrivent beaucoup plus naturellement avec les variables de longueur qu'avec les spins eux-m\^emes. Nous utiliserons en fait des variables adimensionn\'ees \'equivalentes aux longueurs, $d_j = 2\ell_j/\ell_P = 2j+1$, soit la dimension de la repr\'esentation de spin $j$ de $\SU(2)$. Ce choix peut para\^itre physiquement plus naturel dans la mesure o\`u nous nous int\'eressons \`a un d\'eveloppement asymptotique aux grandes longueurs. Cet aspect avait auparavant \'et\'e peu consid\'er\'e car il s'agit de la premi\`ere \'etude qui allant au del\`a de l'ordre dominant aux grandes longueurs n\'ecessite d'avoir vraiment les bons param\`etres pour simplifier les formules au possible. Posons
\beq
k_e=\f{\ell_e}{2\ell_t},
\ee
pour $e=e_1,e_2$, qui seront gard\'es \emph{fix\'es}, i.e. que $\ell_t$, ou de mani\`ere \'equivalente $d_{j_t}$, fixe l'\'echelle du t\'etra\`edre. Les angles dih\'edraux $\thet_1, \thet_2$ et $\thet_t$ sont donn\'es en fonction des longueurs par :
\be
\thet_{1,2}=2 \arccos \Big(\f{k_{2,1}}{\sqrt{1-k_{1,2}^2}} \Big) \quad\mathrm{and}\quad \thet_t=\arccos\Big( \f{-k_1 k_2}{\sqrt{1-k_1^2}\sqrt{1-k_2^2}} \Big),
\ee
pourvu que $k_e<1$, condition assur\'ee par les in\'egalit\'es triangulaires. Notons la relation pariculi\`erement simple qui existe entre ces angles : $\cos\thet_t= -\cos(\f{\thet_1}{2})\cos(\f{\thet_2}{2})$.

\subsubsection*{L'\'etat de bord}

Nous avons maintenant besoin d'expliciter un \'etat de bord, $\Psi_0(j_1',j_2')$, que nous prenons factoris\'e, c'est-\`a-dire : $\Psi_0(j_1',j_2') = \Psi_{e_1}(j_1')\Psi_{e_2}(j_2')$. Le r\^ole de cet \'etat est de piquer l'amplitude sur la g\'eom\'etrie classique du t\'etra\`edre telle que d\'ecrite ci-dessus. On s'inspire pour cela des \'etats coh\'erents de la m\'ecanique quantique standard, en proposant une forme gaussienne sur les spins, et une phase sur les moments conjugu\'es,
\beq \label{gaussian state}
\Psi_e^{\rm Gauss}(j) \propto e^{-\f{(j-j_e)^2}{\gamma_e}}\ e^{\f{i}{2}d_j\thet_e}.
\ee
Il s'agit de l'ansatz utilis\'e initialement par Rovelli dans \cite{rovelli-gravitonBC} (premier calcul de ce type dans le cadre du mod\`ele de Barrett-Crane). Le coefficient $\gamma_e$ doit se comporter en fonction de $j_e$ de sorte que les incertitudes relatives tendent vers z\'ero aux grandes distances,
\beq
\f{\langle \Delta j_e'\rangle}{\langle j_e'\rangle}\sim \f{\sqrt{\gamma_e}}{j_e} \rightarrow 0,\qquad \f{\langle \Delta \phi_e\rangle}{\langle \phi_e\rangle}\sim \f{1}{\sqrt{\gamma_e}\thet_e} \rightarrow 0,
\ee
ce qui autorise une gamme de scaling pour $\gamma_e$. Mais, il a \'et\'e montr\'e \cite{graviton3d-simone} que le bon choix correspond \`a $\gamma_e\propto j_e$. Comme nous l'avons d\'ej\`a mentionn\'e, il sera plus ais\'e d'utiliser les longeurs (ou les dimensions $d_j$) que les spins. Une autre petite modification de $\gamma_e$ s'av\`ere commode, que l'on prend
\beq
\gamma_e=d_{j_t}(1-k_e^2).
\ee

On s'attend \`a ce que la phase ait un r\^ole essentiel, comme en m\'ecanique quantique, dans l'\'emergence des trajectoires classiques. Cela se r\'ealise par le fait que cette phase compense dans l'analyse au point col une phase oppos\'ee provenant du noyau, ph\'enom\`ene d\'ej\`a observ\'e dans \cite{rovelli-gravitonBC}. On souhaiterait aussi comprendre cette phase par dualit\'e de Fourier, i.e. en exprimant l'\'etat de bord en repr\'esentation \og moments\fg. On s'attend alors \`a voir une inversion des r\^oles entre longueurs et angles dans \eqref{gaussian state}. La structure naturelle de la th\'eorie provenant du groupe $\SU(2)$, on utilise la transform\'ee de Fourier sur ce groupe, ou plut\^ot sur $\SO(3)$,
\beq
\what{\Psi}_e(\phi) = \sum_{j\in\N} \Psi_e(j)\,\chi_j(\phi),\qquad \Psi_e(j) = \int dg\ \what{\Psi}_e(\phi)\,\chi_j(\phi).
\ee
La mesure de Haar normalis\'ee est donn\'e par $dg=\sin^2\phi\, d\phi\,d^2\vec{n}$ pour $\vec{n}\in S^2$. Avec cette param\'etrisation, les caract\`eres sont : $\chi_j(\phi) = \sin d_j\phi/\sin\phi$. Notre volont\'e d'utiliser pleinement la structure math\'ematique \`a disposition montre alors une limite de l'\'etat gaussien \eqref{gaussian state}. La transform\'ee de celui-ci se comportant en $e^{-\gamma_e(\phi-\thet_e)^2}$, elle n'est pas bien d\'efinie sur le groupe. Un choix plus naturel est de consid\'erer des fonctions du type : $\exp(-\gamma_e\sin^2(\phi-\thet_e))$. C'est ce qui a \'et\'e fait \`a partir de \cite{graviton3d-corrections}, et que nous faisons aussi en prenant \`a la place de l'ansatz gaussien :
\beq \label{bound state bessel}
\Psi_e(j) = \f{e^{-\gamma_e/2}}{N}\ \Big[ I_{\lvert j-j_e\rvert}(\f{\gamma_e}{2}) - I_{j+j_e+1}(\f{\gamma_e}{2})\Big]\ \cos(d_j\alpha_e), \quad\text{pour}\quad \alpha_e = \thet_e/2.
\ee
$N$ est une normalisation qui d\'epend de $\gamma_e$. Les fonctions $I_n(z)$ sont les fonctions de Bessel modifi\'ees du premier type, d\'efinies par : $I_n(z) = \f{1}{\pi}\int_0^\pi d\phi\ e^{z\cos\phi}\cos(n\phi)$. Quant \`a $\alpha_e = \thet_e/2$, ce sont les demi-angles dih\'edraux\footnote{On utilisera conjointement les quantit\'es $\alpha_e$ et $d_{j_e}$. On aurait tr\`es bien pu prendre \`a la place $\thet_e=2\alpha_e$ et $\ell_e/\ell_P = d_{j_e}/2$, d'interpr\'etation peut-\^etre encore plus imm\'ediate.}. Le comportement asymptotique reproduit une gaussienne, centr\'ee sur $j_e$, de largeur au carr\'e $\gamma_e$,
\be \label{asymptotics bound state}
\Psi_e(j) \sim \f{1}{N}\sqrt{\f{4}{\pi\gamma_e}}\ e^{-\f{(j-j_e)^2}{\gamma_e}}\ \cos(d_j\alpha_e).
\ee
Le r\^ole du cosinus est similaire \`a celui de la phase dans \eqref{gaussian state}, tout en fournissant un \'etat de bord r\'eel. Il permet de piquer la variable duale \`a $j$, i.e. l'angle $\phi$, sur l'angle dih\'edral classique. Cela se voit maintenant tr\`es bien par dualit\'e :
\begin{gather} \label{boundary state}
\widehat{\Psi}_e (\phi) = \f{1}{2}\ \sum_{\eta=\pm 1} \widehat{\Psi}_e^{(\eta)}(\phi) \\
\text{avec}\quad\ \widehat{\Psi}_e^{(\eta)} = \f{1}{N\sin(\phi)}\ \sin\big(d_{j_e}(\phi+\eta\ \alpha_e)\big)\ e^{-\gamma_e\sin^2(\phi+\eta\ \alpha_e)}
\end{gather}
Notons qu'il s'agit d'une fonction de classe sur $\SU(2)$, gr\^ace \`a l'invariance sous $\phi\rightarrow \pi-\phi$ (et \`a l'invariance sous $\phi\rightarrow -\phi$), ce qui n'est pas le cas de $\widehat{\Psi}^{(\eta)}$ seule. Cet \'etat est donc centr\'e autour de $\phi=\alpha_e$ et $\phi=\pi-\alpha_e$.

Insistons finalement sur le fait que nos \'etats de bord sont des ansatz, calqu\'es sur les \'etats coh\'erents standards. Une autre approche, physiquement plus satisfaisante, voire n\'ecessaire, est de demander \`a ces \'etats de bord d'\^etre des \'etats physiques de la th\'eorie, comme c'est le cas dans \eqref{backindptgreen}. Un crit\`ere d'\'evolution physique dans notre situation avec quatre spins fix\'es a \'et\'e propos\'e dans \cite{3dphys-state}, fond\'e sur le 6j isoc\`ele comme noyau de propagation, et les auteurs montrent que ce crit\`ere est bien satisfait dans l'asymptotique par les \'etats gaussiens.

\subsubsection{Analyse des points col du noyau}

Les insertions de champs s'obtiennent en \'ecrivant $\ell_1^\mu\ell_1^\nu h_{\mu\nu} = \ell_1^\mu\ell_1^\nu g_{\mu\nu} - \ell_1^\mu\ell_1^\nu\delta_{\mu\nu}$. Le deuxi\`eme terme correspond \`a la g\'eom\'etrie moyenne fix\'ee comme fond, $\ell_1^\mu\ell_1^\nu\delta_{\mu\nu} = \ell_P^2 d_{j_1}^2/4$, o\`u $\ell_P$ est la longueur de Planck. Le premier terme agit simplement comme l'op\'erateur de longueur sur le r\'eseau de spins dual au t\'etra\`edre. Celui-ci \'etant un \'etat propre, on aboutit \`a \cite{graviton3d-simone}
\beq
\langle j_1'\lvert \ell_1^\mu\ell_1^\nu\,h_{\mu\nu}(x) \rvert j_1'\rangle = \f{\ell_P^2}{4}\bigl(d_{j_1'}^2 - d_{j_1}^2\bigr),
\ee
et de m\^eme pour la deuxi\`eme insertion. Cela est g\'eom\'etriquement assez clair : la contraction diagonale de $h_{\mu\nu}$ mesure bien les fluctuations de longueur dans la direction de contraction. Ces longueurs apparaissent ici comme les valeurs propres des op\'erateurs correspondants en LQG.

Le propagateur $W_{1122}$ s'\'ecrit donc :
\begin{align} \label{exact propagator}
&W_{1122}=\f{1}{\calN}\sum_{j'_1,j'_2}\ \begin{Bmatrix} j'_1 & j_t & j_t \\ j'_2 & j_t & j_t \end{Bmatrix}\ \calO_{j_1}(j'_1) \Psi_{e_1}(j'_1)\ \calO_{j_2}(j'_2) \Psi_{e_2}(j'_2), \\
&\text{avec}\quad\ \calO_{j_e}(j')=\f{1}{\di{e}^2}\Big(d_{j'}^2-\di{e}^2\Big).
\end{align}
Le facteur de normalisation $\calN$ est donn\'e par la m\^eme somme, sans les insertions d'observables $\calO_{j_e}(j_e')$. La cl\'e de notre formulation r\'eside dans l'existence d'une formule int\'egrale simple pour le 6j isoc\`ele\footnote{Pour une configuration g\'en\'erique, seul le 6j au carr\'e admet une telle repr\'esentation sous forme int\'egrale, \`a moins d'utiliser plus de donn\'ees au bord, comme avec les \'etats coh\'erents $\SU(2)$. Pour les mod\`eles en 4d, l'amplitude du 4-simplex est bien donn\'ee par une int\'egrale, avec des donn\'ees de bord un peu plus \'evolu\'ees qu'ici, notamment ces m\^emes \'etats coh\'erents de $\SU(2)$.}, dont nous prenons les spins entiers,
\be \label{6jiso so3}
\begin{Bmatrix}
j_1 & j_t & j_t \\
j_2 & j_t & j_t
\end{Bmatrix} = \int_{\SU(2)^2} dg_1 dg_2\ \chi_{j_t}(g_1g_2)\ \chi_{j_t}(g_1g_2^{-1})\ \chi_{j_1}(g_1)\ \chi_{j_2}(g_2).
\ee
Gr\^ace \`a cette formule et la transform\'ee de Fourier de l'\'etat de bord, nous pourrons \'egalement \'ecrire le propagateur comme une int\'egrale sur le groupe. Cela va nous permettre d'effectuer un d\'eveloppement asymptotique, via la m\'ethode du  point col/phase stationnaire, pour les grands spins. Par rapport \`a \cite{graviton3d-corrections} sur le m\^eme sujet, notre m\'ethode offre l'avantage d'inclure naturellement les corrections \`a l'ordre dominant provenant de l'action de Regge, mais aussi les corrections du 6j \`a l'action de Regge.

Commen\c{c}ons par regarder les points col de la formule \eqref{6jiso so3} pour le 6j isoc\`ele. Le calcul pour $W_{1122}$ offrira en fait une structure tr\`es similaire. Nous avons d'abord besoin des angles de classe des \'el\'ements de groupe $g_1g_2$ et $g_1g_2\mone$,
\be
\phi_{12}^\pm=\arccos\big( \cos\phi_1 \cos\phi_2 \mp u\ \sin\phi_1 \sin\phi_2 \big)
\ee
avec la notation $u=\vec{n}_1\cdot\vec{n}_2$. Les phases oscillant rapidement se d\'eveloppent sous forme exponentielle pour amener
\be \label{6j exp form}
\begin{Bmatrix}
j_1 & j_t & j_t \\
j_2 & j_t & j_t
\end{Bmatrix}
=\f{1}{8\pi^2}\sum_{\eps_1, \eps_2, \eps_{12}^+, \eps_{12}^- =\pm 1} \eps_1\ \eps_2\ \eps_{12}^+\
\eps_{12}^-\ \int d\phi_1 d\phi_2 du\ f(\phi_1,\phi_2,u)\ e^{id_{j_t}\Phi_{\{\eps\}}},
\ee
o\`u la \og mesure\fg{} $f(\phi_1,\phi_2,u)$ et la phase $\Phi_{\{\eps\}}$, ind\'ependantes de l'\'echelle ($j_t, j_1$ et $j_2$ ayant la m\^eme \'echelle) sont donn\'ees par
\begin{align} \label{measure f}
f(\phi_1,\phi_2,u) &= \f{\sin(\phi_1)\sin(\phi_2)}{\sin(\phi_{12}^+)\sin(\phi_{12}^-)}, \\
\Phi_{\{\eps\}}(\phi_1,\phi_2,u) &= (\eps_{12}^+\phi_{12}^+ + \eps_{12}^-\phi_{12}^-)\ +\ 2k_1\eps_1\ \phi_1\ +\ 2k_2\eps_2\ \phi_2. \label{6j phases}
\end{align}

Recherchons maintenant les points stationnaires de $\Phi_{\{\eps\}}$. La variable $u$ n'intervient que dans $\phi_{12}^\pm$, et l'\'equation de stationnarit\'e associ\'ee, $\eps_{12}^+\pp_u \phi_{12}^+ + \eps_{12}^-\pp_u \phi_{12}^-=0$, se r\'esout par : $u=\vec{n}_1\cdot\vec{n}_2=0$ et $\eps_{12}^+=\eps_{12}^-=\eps_{12}$. Les variations de $\phi_1$ et $\phi_2$ conduisent aux \'equations :
\be
d_{j_t}\bigl(\eps_{12}^+\pp_{\phi_e}\phi_{12}^+ + \eps_{12}^-\pp_{\phi_e}\phi_{12}^-\bigr)+d_{j_e}\eps_e=0, \qquad e=1,2,
\ee
dont les solutions dans $[0,\pi[$ sont :
\be \label{6j stationary points}
\left\{ \begin{array}{l}
\bar{\phi}_1=-\eps_{12}\ \eps_2\arccos \Big(\f{k_2}{\sqrt{1-k_1^2}}\Big)+ (1+\eps_{12}\ \eps_2) \f{\pi}{2} \\
\bar{\phi}_2=-\eps_{12}\ \eps_1\arccos \Big(\f{k_1}{\sqrt{1-k_2^2}}\Big)+ (1+\eps_{12}\ \eps_1) \f{\pi}{2}
\end{array}
\right.
\ee
Cela nous permet d'acc\'eder \`a une interpr\'etation g\'eom\'etrique simple des angles de classe intervenant dans l'int\'egrale \eqref{6jiso so3}, semblable \`a celle, connue, pour les variables de l'int\'egrale d\'efinissant le 6j au carr\'e. Le symbole 6j isoc\`ele provient d'une amplitude piqu\'ee aux grands spins sur les (demi-)angles dih\'edraux, internes ou externes, de la g\'eom\'etrie classique. En effet, par exemple, lorsque $\eps_{12}\eps_2=-1$, l'angle stationnaire $\bar{\phi}_1$ est $\bar{\phi}_1=\alpha_1=\thet_1/2$, tandis que pour $\eps_{12}\eps_2=1$, c'est : $\bar{\phi}_1=\pi-\alpha_1$. Cette interpr\'etation n'est pas une surprise dans la mesure o\`u le symbole 6j, g\'en\'erique, est le seul \'etat physique pour une triangulation form\'ee par le bord d'un t\'etra\`edre, dans la base des r\'eseaux de spins. Il satisfait la contrainte de courbure nulle au niveau quantique, et c'est lui qui doit \^etre utilis\'e comme \'etat de bord dans le cadre d'un bord g\'en\'erique, sans la fixation de jauge sur $T$ \cite{3dphys-state}.

Nous utilisons en compl\'ement cette analyse pour d\'eriver le d\'eveloppement asymptotique complet du 6j isoc\`ele, plus bas en section \ref{sec:asymp6jiso}.

\subsubsection*{Une int\'egrale pour le propagateur}

Nous ins\'erons l'expression \eqref{6jiso so3} dans \eqref{exact propagator}, ce qui nous oblige \`a constater que les sommes sur $j_1', j_2'$ forment dor\'enavant et pr\'ecis\'ement les transform\'ees de Fourier des \'etats de bord ! Regardons d'abord la normalisation $\calN$,
\begin{align}
\calN &= \int dg_1 dg_2\ \chi_{j_t}(g_1 g_2^{-1})\ \chi_{j_t}(g_1 g_2)\ \Bigl[\sum_{j'_1}\Psi_{e_1}(j'_1)\chi_{j'_1}(g_1)\Bigr]\ \Bigl[\sum_{j'_2}\Psi_{e_2}(j'_2)\chi_{j'_2}(g_2)\Bigr], \\
&= \int dg_1 dg_2\ \chi_{j_t}(g_1 g_2^{-1})\ \chi_{j_t}(g_1 g_2)\ \widehat{\Psi}_{e_1} (g_1)\ \widehat{\Psi}_{e_2} (g_2). \label{norm group integral}
\end{align}
On d\'eveloppe les phases oscillant rapidement sous formes exponentielles,
\begin{gather} \label{normalisation exp form}
\calN = \f{1}{32\pi^2} \sum_{\substack{\eps_1, \eps_2, \eps_{12}^+, \eps_{12}^-,\\ \eta_1, \eta_2 =\pm 1}} \eps_1\ \eps_2\ \eps_{12}^+\ \eps_{12}^-\ \int d\phi_1 d\phi_2 du\ f(\phi_1,\phi_2,u)\ e^{d_{j_t} S_{\{\eps,\eta\}}(\phi_1,\phi_2,u)}, \quad \text{pour}\\
S_{\{\eps,\eta\}}(\phi_1,\phi_2,u) = i\bigl(\eps_{12}^+\phi_{12}^+ + \eps_{12}^-\phi_{12}^-\bigr) + \sum_{e=1,2} 2ik_e\eps_e(\phi_e+\eta_e\alpha_e) - \big(1-k_e^2\big)\sin^2(\phi_e+\eta_e\alpha_e). \label{action}
\end{gather}
Le label $\{\eps,\eta\}$ fait \'etat de la d\'ependance en les signes, et $f$ est donn\'e par \eqref{measure f}. Le point remarquable est que la partir imaginaire de $S_{\{\eps,\eta\}}$ n'est autre que $\Phi_{\{\eps\}}$, \eqref{6j phases}, \`a des termes constants en $\alpha_{1,2}$ pr\`es qui ne jouent aucun r\^ole dans la m\'ethode de la phase stationnaire. En particulier, les points stationnaires de $S_{\{\eps,\eta\}}$ sont \`a rechercher parmi ceux de $\Phi_{\{\eps\}}$.

Le num\'erateur de $W_{1122}$ se pr\^ete \`a la m\^eme analyse. Pour prendre en compte les observables $\calO_{j_e}$, remarquons que le caract\`ere de $\SU(2)$ est une fonction propre du laplacien sur la sph\`ere $S^3$, ayant le Casimir de la repr\'esentation pour valeur propre,
\be
\Delta_{S^3}\chi_j(\phi)=\f{1}{\sin^2\phi}\ \pp_\phi\bigl(\sin^2\phi\ \pp_\phi\ \chi_j\bigr)=-(d_j^2-1)\ \chi_j(\phi).
\ee
Cela permet \`a nouveau de faire les sommes sur $j_1', j_2'$ en introduisant les transform\'ees de Fourier $\widehat{\Psi}_e$. Le r\'esultat de ces op\'erations est \'egalement d\'evelopp\'e en exponentielles,
\begin{multline}  \label{prop exp form}
W_{1122} = \f{1}{32\pi^2\ \calN}\ \f{k_1 k_2}{4\cos^2\thet_t} \sum_{\substack{\eps_1, \eps_2, \eps_{12}^+, \eps_{12}^-,\\ \eta_1, \eta_2 =\pm 1}} \eps_1\ \eps_2\ \eps_{12}^+\ \eps_{12}^-\ \int d\phi_1 d\phi_2 du\ f(\phi_1,\phi_2,u)\ e^{d_{j_t} S_{\{\eps,\eta\}}(\phi_1,\phi_2,u)} \\
\times \Bigl( a_{\{\eps,\eta\}}(\phi_1,\phi_2)+\f{b_{\{\eps,\eta\}}(\phi_1,\phi_2)}{d_{j_t}}+\f{c_{\{\eps,\eta\}}(\phi_1,\phi_2)}{d_{j_t}^2} \Bigr).
\end{multline}
Les fonctions  $a_{\{\eps,\eta\}}$, $b_{\{\eps,\eta\}}$ et $c_{\{\eps,\eta\}}$ repr\'esentent les insertions des observables, et sont donn\'ees explicitement par :
\begin{equation}
a_{\{\eps,\eta\}}(\phi_1,\phi_2) = \prod_{e=1,2} \Bigl( \f{1-k_e^2}{2k_e}\sin^2 2(\phi_e+\eta_e\ \alpha_e) - 2i\eps_e \sin2(\phi_e+\eta_e\ \alpha_e)\Bigr),
\end{equation}
\begin{multline}
b_{\{\eps,\eta\}}(\phi_1,\phi_2) = -\f{1}{k_2}\cos 2(\phi_2+\eta_2\ \alpha_2)\ \Bigl( \f{1-k_1^2}{2k_1}\sin^2 2(\phi_1+\eta_1\ \alpha_1) - 2i\eps_1 \sin2(\phi_1+\eta_1\ \alpha_1)\Bigr) \\ +( e_1\leftrightarrow e_2),
\end{multline}
\beq
c_{\{\eps,\eta\}}(\phi_1,\phi_2) = \f{1}{k_1 k_2}\ \cos 2(\phi_1+\eta_1\ \alpha_1)\ \cos 2(\phi_2+\eta_2\ \alpha_2).
\ee
Tout comme $f(\phi_1,\phi_2,u)$, elles sont ind\'ependantes de l'\'echelle.

L'\'ecriture \eqref{prop exp form} envisage donc $W_{1122}$ comme la valeur moyenne de la fonction $a_{\{\eps,\eta\}}+\f{b_{\{\eps,\eta\}}}{d_{j_t}}+\f{c_{\{\eps,\eta\}}}{d_{j_t}^2}$ pour une th\'eorie non-lin\'eaire d\'efinie par l'action $S_{\{\eps,\eta\}}$ et la mesure d'int\'egration $f$. La strat\'egie est donc de calculer s\'epar\'ement $\calN$ et le num\'erateur, en perturbations, \`a l'aide d'un d\'eveloppement autour des points stationnaires de l'action $S_{\{\eps,\eta\}}$.

Comme remarqu\'e ci-dessus, les points stationnaires de la partie imaginaire de $S_{\{\eps,\eta\}}$ sont les m\^emes que ceux du 6j isoc\`ele, i.e $u=0$ et $\bar{\phi}_e$ donn\'e par \eqref{6j stationary points}, ind\'ependamment des signes $\eta_1$ et $\eta_2$. De son c\^ot\'e, la stationnarisation de la partie r\'eelle de $S_{\{\eps,\eta\}}$ va contraindre les signes $\eta_1, \eta_2$. En effet, pour une solution $(\bar{\phi}_1,\bar{\phi}_2)$ de \eqref{6j stationary points}, caract\'eris\'ee par les signes $\eps_1, \eps_2$ et $\eps_{12}$, le signes $\eta_1$ et $\eta_2$ doivent satisfaire :
\be
\sin 2\bigl(\bar{\phi}_e+\eta_e\ \alpha_e\bigr) = 0,\qquad\text{pour}\ e=1,2.
\ee
Ces conditions demandent de prendre : $\eta_1=\eps_2\eps_{12}$, et $\eta_2=\eps_1\eps_{12}$. Cela laisse quatre possiblit\'es, d\'etaill\'ees dans le tableau suivant,
\begin{displaymath}
\begin{array}{c|c|c}
\; & \eta_1=-1 & \eta_1=1 \\
\hline
\eta_2=-1 & \quad \begin{aligned} &\bar{\phi}_1=\alpha_1,\ \text{et}\ \bar{\phi}_2=\alpha_2, \\ &\eps_1=\eps_2=-\eps_{12} \end{aligned} & \quad \begin{aligned} &\bar{\phi}_1=\pi-\alpha_1,\ \text{et}\ \bar{\phi}_2=\alpha_2, \\ &\eps_1=-\eps_2=-\eps_{12} \end{aligned}  \\
\hline
\eta_2=1 & \quad \begin{aligned} &\bar{\phi}_1=\alpha_1,\ \text{et}\ \bar{\phi}_2=\pi-\alpha_2, \\ &-\eps_1=\eps_2=-\eps_{12} \end{aligned} &\quad \begin{aligned} &\bar{\phi}_1=\pi-\alpha_1,\ \text{et}\ \bar{\phi}_2=\pi-\alpha_2, \\ &\eps_1=\eps_2=\eps_{12} \end{aligned}
\end{array}
\end{displaymath}
Les conditions $\eps_{12}^+=\eps_{12}^-=\eps_{12}$ et $\eta_1=\eps_2\eps_{12}$ ainsi que $\eta_2=\eps_1\eps_{12}$ permettent de se d\'ebarrasser de trois sommes dans \eqref{prop exp form}, puis, les configurations pour lesquelles il n'y a pas de points col \'etant exponentiellement supprim\'ees,
\begin{multline}
W_{1122} = \f{1}{32\pi^2\ \calN}\ \f{k_1 k_2}{4\cos^2\thet_t} \sum_{\eps_1, \eps_2, \eps_{12}=\pm 1} \eps_1\,\eps_2\ \int d\phi_1 d\phi_2 du\ f(\phi_1,\phi_2,u)\ e^{\di{t} S_{\{\eps\}}(\phi_1,\phi_2,u)} \\
\times\Bigl( a_{\{\eps\}}(\phi_1,\phi_2)+\f{b_{\{\eps\}}(\phi_1,\phi_2)}{d_{j_t}}+\f{c_{\{\eps\}}(\phi_1,\phi_2)}{d_{j_t}^2}\Bigr),
\end{multline}
et la m\^eme chose pour $\calN$ sans l'insertion de $\f{k_1 k_2}{4\cos^2\thet_t}(a_{\{\eps\}}+b_{\{\eps\}}/d_{j_t}+c_{\{\eps\}}/d_{j_t}^2)$.
Le label $\{\eps\}$ indique simplement la d\'ependance des fonctions en les signes $\eps_1, \eps_2$ et $\eps_{12}$.

\subsection{D\'eveloppement asymptotique aux grandes distances}

Le d\'eveloppement asymptotique de la fonction de corr\'elation se formule comme une s\'erie en puissances de $1/d_{j_t}$, du type
\be
W\,=\, \f1{d_{j_t}} \,\left[w_0+\f1{d_{j_t}}w_1+\f1{d_{j_t}^2}w_2+\dots\right].
\ee
Rappelons que $d_{j_t}$ d\'efinit l'\'echelle de la g\'eom\'etrie du t\'etra\`edre, car $\ell_t/\ell_P = d_{j_t}/2$ et les rapports $k_1, k_2$ avec les longueurs $\ell_1, \ell_2$ sont fix\'es. Puisqu'en 3d la longueur de Planck est $\ell_P = \hbar G$, ce d\'eveloppement rappelle les d\'eveloppements typiques de la th\'eorie quantique des champs pour les fonctions de corr\'elations, en puissance croissantes de $\hbar$ (ou de la constante de couplage $G$), et $\ell_t$ \'etant l'\'echelle.

Notre d\'eveloppement s'obtient en \'etudiant les int\'egrales autour de chacun des quatre points stationnaires du d\'enominateur, $\int f\exp\di{t}S_{\{\eps\}}$, et du num\'erateur, $\int(a_{\{\eps\}}+b_{\{\eps\}}/d_{j_t}+c_{\{\eps\}}/d_{j_t}^2)f\exp d_{j_t}S_{\{\eps\}}$. Plus pr\'ecis\'ement, nous d\'eveloppons l'action $S_{\{\eps\}}$ autour d'un point stationnaire : l'\'evaluation de $S$ en ce point produit un facteur num\'erique, il n'y a \'evidemment pas de termes lin\'eaires, les termes quadratiques d\'efinissent la matrice hessienne $A_{\{\eps\}}$, et finalement tous les termes d'ordres sup\'erieurs (cubiques et plus) sont regroup\'es en un potentiel $\Omega_{\{\eps\}}$. Ce potentiel contient donc l'ensemble des corrections a l'approximation quadratique de l'action. En tant que tel il n'intervient pas \`a l'ordre dominant, mais participe largement aux ordres sup\'erieurs du d\'eveloppement -- le next-to-leading order (NLO), le next-to-next-to-leading order (NNLO), et ainsi de suite. Puis chaque terme de la s\'erie de puissance est \'evalu\'e comme le moment gaussien pour la matrice hessienne $A_{\{\eps\}}$ de termes provenant du d\'eveloppement de $(\exp d_{j_t})f \Omega_{\{\eps\}}$ en puissance de $d_{j_t}$. En g\'en\'eral, de nombreux termes contribuent \`a un m\^eme ordre en puissance de $1/d_{j_t}$. Cette structure complexe est due au fait que le d\'eveloppement de $\exp\di{t}\Omega_{\{\eps\}}$ apporte des puissaances croissantes de $d_{j_t}$, tandis que les moments gaussiens produisent des puissances croissantes de $1/d_{j_t}$.

Par ailleurs une simplification dans notre calcul provient du fait que tous les points stationnaires donnent la m\^eme contribution. C'est une cons\'equence de l'invariance des fonctions impliqu\'ees sous la transformation $\phi_e\rightarrow \pi-\phi_e$. De plus, pour chaque point col, il y a deux configurations des signes $\{\eps\}$ possibles, et les termes correspondants s'av\`erent \^etre complexe-conjugu\'es l'un de l'autre. Leur somme assure donc la r\'ealit\'e du r\'esultat. Nous pouvons donc nous restreindre sans perte de g\'en\'eralit\'e au calcul autour du point col $(\bar{\phi}_1, \bar{\phi}_2,\bar{u}) = (\alpha_1,\alpha_2,0)$, avec le choix de signes $\eps_{12}=1=-\eps_1=-\eps_2$. On a dans ce cas :
\begin{align}\label{a}
a(\phi_1,\phi_2) &= \prod_{e=1,2} \Bigl( \f{1-k_e^2}{2k_e}\sin^2 2(\phi_e- \alpha_e) + 2i \sin2(\phi_e- \alpha_e)\Bigr), \\
b(\phi_1,\phi_2) &= -\f{1}{k_2}\cos 2(\phi_2- \alpha_2)\, \Bigl( \f{1-k_1^2}{2k_1}\sin^2 2(\phi_1- \alpha_1) + 2i \sin2(\phi_1-\alpha_1)\Bigr) + ( e_1\leftrightarrow e_2), \\
c(\phi_1,\phi_2) &= \f{1}{k_1 k_2}\, \cos 2(\phi_1-\alpha_1)\, \cos 2(\phi_2-\alpha_2).
\end{align}
Le potentiel $\Omega$ peut \^etre extrait du d\'eveloppement de Taylor de $S$ au-del\`a de l'ordre trois, $S$ \'etant avec les signes choisis
\be
S(\phi_1,\phi_2,u) = i\bigl(\phi_{12}^+ + \phi_{12}^-\bigr) - \sum_{e=1,2} \bigl(1-k_e^2\bigr)\sin^2(\phi_e-\alpha_e)+\text{termes lin\'eaires}.
\ee
Le d\'eveloppement autour de la g\'eom\'etrie de fond conduit alors \`a la matrice inverse suivante pour la hessienne, \cite{graviton3d-val},
\be \label{Ainv}
A^{-1} = \f{1}{4} \begin{pmatrix} \f{1}{1-k_1^2} & \f{\cos\thet_t}{k_1 k_2}\ e^{i\thet_t} & 0 \\
                                                        \f{\cos\thet_t}{k_1 k_2}\ e^{i\thet_t} & \f{1}{1-k_2^2} & 0 \\
                                                        0 & 0 & \f{2i\ \tan\thet_t}{1-(k_1^2+k_2^2)} \end{pmatrix}.
\ee
La notation suivante exprimant les moments gaussiens de $A$ va se r\'ev\`eler tr\`es utile pour appliquer le th\'eor\`eme de Wick,
\be
A^{-1}_{\vec{\beta}} = \sum_{\substack{\mathrm{appariements} \\ \mathrm{de}\ (\beta_1,\dots,\beta_{2N})}} A^{-1}_{\beta_{i_1}\beta_{i_2}}\dots A^{-1}_{\beta_{i_{2N-1}}\beta_{i_{2N}}},
\ee
pour $\vec{\beta}\in\{1,2,3\}^{2N}$. Le d\'eveloppement asymptotique recherch\'e prend alors la forme :
\be \label{prop final expansion}
W_{1122} = \f{k_1 k_2}{4\cos^2\thet_t}\,\f{\f{\sqrt{1-k_1^2}\sqrt{1-k_2^2}}{2d_{j_t}}\sum_{i,j=1,2}\pp^2_{ij}a\ A^{-1}_{ij} + \sum_{P\geq2}
\f{W_P}{d_{j_t}^P}}{\sum_{P\in\mathbb{N}} \f{N_P}{d_{j_t}^P}}\, .
\ee
Le premier terme du num\'erateur correspond \`a l'ordre dominant, en $1/d_{j_t}$, comme nous le discutons plus bas. Il provient enti\`erement de la fonction $a$ dans \eqref{a}. En fait, $a$ et $b$ s'annulent tous deux au point col, de m\^eme que le gradient de $a$, de sorte que le d\'eveloppement de $a, b$ et $c$ est domin\'e par le terme quadratique de $a$.

L'ensemble des termes d'ordres sup\'erieurs a \'et\'e rassembl\'e dans les sommes. Les coefficients $N_P$ et $W_P$ sont donn\'es par des sommes finies, et s'expriment en vertu du th\'eor\`eme de Wick comme :
\beq \label{coeff norm}
N_P = \sum_{n=0}^{2P} \sum_{\vec{\beta}\in\{1,2,3\}^{2(P+n)}}
\f{1}{(2(P+n))!\ n!}\, \Re\Bigl(i\ e^{-i(2d_{j_t}-\f{1}{2})\thet_t}\, \pp^{2(P+n)}_{\vec{\beta}} \bigl(f\Omega^n\bigr)\ A^{-1}_{\vec{\beta}}\Bigr)_{|\phi_1=\alpha_1,\phi_2=\alpha_2,u=0},
\ee
et
\begin{multline} \label{coeff prop}
W_P = \sum_{n\geq0}\f{1}{n!}\Bigl\{ \sum_{\vec{\beta_a}\in\{1,2,3\}^{2(P+n)}} \f{1}{(2(P+n))!}\ \Re\Bigl(ie^{-i(2d_{j_t}-\f{1}{2})\thet_t}\ \pp^{2(P+n)}_{\vec{\beta_a}}\bigl(af\Omega^n\bigr)\ A^{-1}_{\vec{\beta_a}}\Bigr) \\
+ \sum_{\vec{\beta_b}\in\{1,2,3\}^{2(P+n-1)}} \f{1}{(2(P+n-1))!}\ \Re\Bigl(ie^{-i(2d_{j_t}-\f{1}{2})\thet_t}\ \pp^{2(P+n-1)}_{\vec{\beta_b}}\bigl(bf\Omega^n\bigr)\ A^{-1}_{\vec{\beta_b}}\Bigr) \\
+ \sum_{\vec{\beta_c}\in\{1,2,3\}^{2(P+n-2)}} \f{1}{(2(P+n-2))!}\
\Re\Bigl(ie^{-i(2d_{j_t}-\f{1}{2})\thet_t}\ \pp^{2(P+n-2)}_{\vec{\beta_c}}\bigl(cf\Omega^n\bigr)\
A^{-1}_{\vec{\beta_c}}\Bigr) \Bigr\}_{|\phi_1=\alpha_1,\phi_2=\alpha_2,u=0}.
\end{multline}
Les trois lignes de \eqref{coeff prop} correspondent aux trois contributions distinctes des insertions de $a, b$ et $c$. La somme sur $n$ d\'efinissant $W_P$ est bien finie pour chacune de ces contributions : $n$ est born\'e respectivement par $(2P-2)$, $(2P-3)$ et $(2P-4)$ pour $a$, $b$ et $c$. Les plus hautes d\'eriv\'ees de $\Omega$ impliqu\'ees dans $W_P$ sont respectivement d'ordre $2P$, $2P-1$ et $2P-2$, et d'ordre $2(P+1)$ pour $N_P$, correspondant toutes \`a $n=1$.

La structure complexe de ces contributions avait \'et\'e anticip\'ee en d\'ebut de section. Malgr\'e leur lourdeur apparente, elles restent des expressions alg\'ebriques simples \`a \'evaluer.

\subsubsection*{Les premiers ordres dominants}

Nous utilisons maintenant \eqref{coeff norm} et \eqref{coeff prop} pour obtenir explicitement les premiers ordres du d\'eveloppement. L'ordre dominant (leading order, LO) et le premier sous-dominant (NLO) avait d\'ej\`a \'et\'e calcul\'es diff\'eremment dans \cite{graviton3d-corrections}. Nous les retrouvons ici rapidement. Le calcul du deuxi\`eme terme sous-dominant (NNLO) est en revanche tout nouveau. En particulier, ce terme re\c{c}o\^it des contributions provenant de corrections du symbole 6j \`a la formule asymptotique de Ponzano et Regge (comme il est remarrqu\'e dans \cite{graviton3d-corrections}), qui sont naturellement prises en compte par notre m\'ethode.

Le LO est obtenu en \'evaluant au point col $f_0=f(\alpha_1,\alpha_2,0)$, et la d\'eriv\'ee seconde de $a$ au num\'erateur,
\begin{align}
W^{LO}_{1122} &= \f{k_1 k_2}{4\cos^2\thet_t}\ \f{\f{1}{d_{j_t}}
\Re\bigl(ie^{-i(2d_{j_t} - \f{1}{2})\thet_t}\pp^2_{\phi_1,\phi_2}a\ A^{-1}_{12}\bigr)}{f_0\Re\bigl(ie^{-i(2d_{j_t} - \f{1}{2})\thet_t}\bigr)}, \\
 &= -\f{1}{d_{j_t} \cos\thet_t}\ \f{\sin (2d_{j_t}-\f{3}{2})\thet_t}{\sin (2d_{j_t} - \f{1}{2})\thet_t}. \label{propLO}
\end{align}
Cela reproduit bien le comportement en $1/\ell_t$ attendu. La diff\'erence avec le coefficient calcul\'e dans \cite{graviton3d-corrections} provient du choix diff\'erent de l'\'etat de bord. En particulier, alors que $\thet_t(k_1,k_2)$ reste constant, la d\'ependance en $d_{j_t}$ de la seconde fraction produit d'\'etonnantes oscillations. Celles-ci peuvent \^etre r\'eabsorb\'ees dans l'\'etat de bord, en rempla\c{c}ant $\sin d_{j_e}(\phi+\eta\ \alpha_e)$ dans $\widehat{\Psi}_e$ par : $\sin
\bigl(d_{j_e}(\phi+\eta\ \alpha_e) + \eta\  d_{j_t}\thet_t\bigr)$. La transform\'ee de Fourier devient alors
\be \label{simplified bound state}
\Psi_e(j) = \f{e^{-\gamma_e/2}}{N}\ \Bigl[ I_{\lvert j-j_e\rvert}(\f{\gamma_e}{2})\,
\cos(d_j\alpha_e+d_{j_t}\thet_t) - I_{j+j_e+1}(\f{\gamma_e}{2})\ \cos(d_j\alpha_e-d_{j_t}\thet_t)\Bigr],
\ee
ce qui n'affecte pas le comportement gaussien asymptotique. Avec ce changement nous reproduisons exactement le r\'esultat de \cite{graviton3d-corrections},
\be \label{real LO}
W_{1122}^{LO} = -\f{1}{d_{j_t}\cos\thet_t}\ \f{\sin \f{3}{2}\thet_t}{\sin \f{1}{2}\thet_t}.
\ee
M\^eme si le LO est bien celui de \cite{graviton3d-corrections}, les ordres suivants seront diff\'erents du fait de l'utilisation d'un \'etat de bord diff\'erent.

\begin{figure}[t] \begin{center}
\includegraphics[width=5cm]{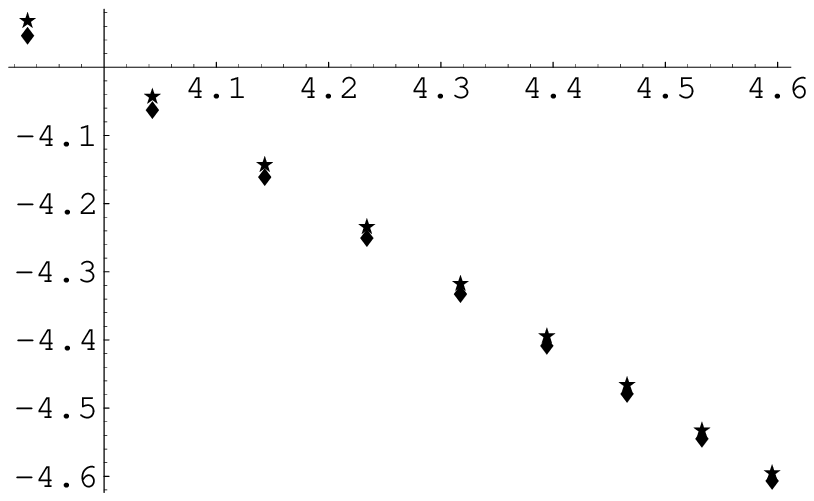}\qquad
\includegraphics[width=5cm]{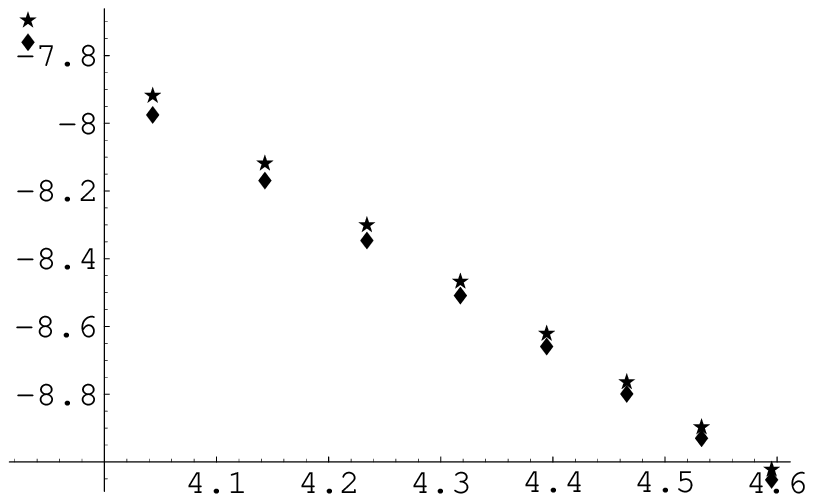}\quad
\includegraphics[width=5cm]{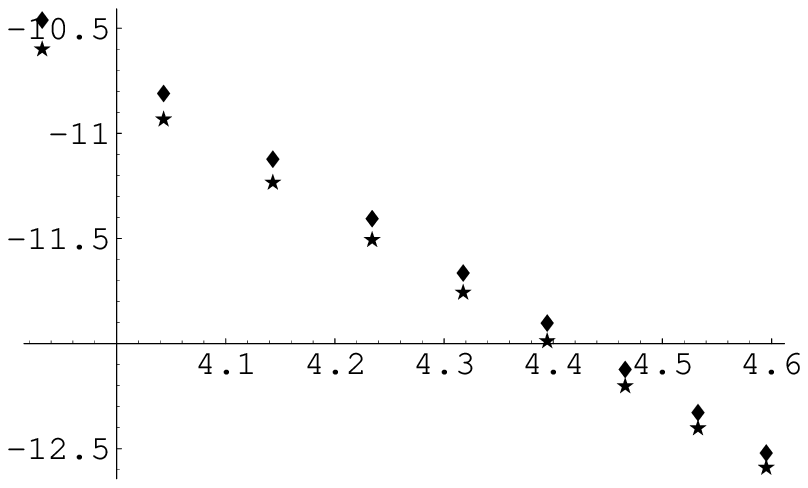}\quad
\caption{ \label{numerics LO NLO NNLO} Graphes en log-log comparant les simulations num\'eriques aux r\'esultats analytiques. A gauche : une simulation de \eqref{exact propagator} (symbole diamand) compar\'e \`a l'ordre dominant de \eqref{equilateral NLO} (symbole \'etoile). Au milieu : le premier ordre sous-dominant de \eqref{equilateral NLO} (symoble \'etoile), et le num\'erique (symbole diamand). A droite : le deuxi\`eme ordre sous-dominant issu de \eqref{equilateral NNLO}.}
\end{center}
\end{figure}

Pour la simplicit\'e de la pr\'esentation, nous nous pla\c{c}ons d\'esormais dans le cas d'un t\'etra\`edre \'equilat\'eral, caract\'eris\'e par $k_1 = k_2= 1/2$, et $\thet_t = \arccos-\f{1}{3}$, les expressions g\'en\'eriques en terme de $k_1, k_2$ \'etant en effet assez lourdes. Le NLO est ensuite obtenu \`a partir de coefficients $N_1$ et $W_2$. Pour garder des expressions compactes, nous adoptons la notation symbolique suivante pour les contractions des d\'eriv\'ees avec les moments gaussiens. Pour des fonctions $f$ et $h$ de $\phi_1, \phi_2$ et $u$, on d\'efinit : $(f_n\, h_m\, A^{-1}_{n+m}) := \f{1}{n!m!}\sum_{i_1,\cdots,i_n=1,2,3}\sum_{j_1,\cdots,j_m=1,2,3}
\pp^n_{i_1,\cdots,i_n}f\ \pp^m_{j_1,\cdots,j_m}h\ A^{-1}_{(i_1,\cdots,i_n,j_1,\cdots,j_m)}$, \'evalu\'e au point col. Par exemple le LO de \eqref{prop final expansion} s'\'ecrit : $\f{1}{d_{j_t}}a_2 A^{-1}_2$. Dans $N_1$, trois puissances de $\Omega$ apparaissent, $\Omega^0$, $\Omega$ and $\Omega^2$. En utilisant l'\'etat de bord \eqref{simplified bound state}, nous avons
\be
N_1 = \Re\Bigl(i\, e^{\f{i}{2}\thet_t}\ \Bigl[f_2\ A^{-1}_2 + \bigl(f_1\ S_3 + f_0\ S_4\bigr) A^{-1}_4 + \f{f_0}{2}\ S_3\ S_3\ A^{-1}_6\Bigr]\Bigr).
\ee
De m\^eme pour les trois contributions \`a $W_2$,
\begin{multline}
W_2 = \Re\Bigl(i\, e^{\f{i}{2}\thet_t}\ \Bigl[ \bigl(a_2\ f_2 + a_3\ f_1 + f_0\ a_4\bigr)A^{-1}_4 + \bigl(a_2\ f_1\ S_3 + f_0 (a_3\ S_3 + a_2\ S_4)\bigr)A^{-1}_6 + \f{f_0}{2}\ a_2\  S_3\ S_3\ A^{-1}_8 \\
+ \bigl(f_0\ b_2 + b_1\ f_1\bigr)A^{-1}_2 + f_0\ b_1\ S_3\ A^{-1}_4 + f_0\ c_0\Bigr]\Bigr).
\end{multline}
Apr\`es un peu d'alg\`ebre, le premier ordre sous-dominant, en $1/d_{j_t}^2$, se laisse calculer :
\be \label{equilateral NLO}
W^{NLO}_{1122} = \f{1}{d_{j_t}} - \f{511}{432\ d_{j_t}^2}.
\ee
Ces r\'esultats pour le LO et le NLO sont bien confirm\'es num\'eriquement, comme on l'observe en figure \ref{numerics LO NLO NNLO}. Un accord avec $0,58\%$ d'erreur pour le LO, et $1,7\%$ d'erreur pour le NLO, est atteint entre les simulations num\'eriques et les coefficients ci-dessus, pour $d_{j_t}=201$ (soit la repr\'esentation de spin $j=100$).

La recette ci-dessus donne acc\`es \`a tous les ordres du d\'eveloppement. De ce point de vue, le NNLO n'offre pas de sp\'ecificit\'es. Nous avons besoin de d\'evelopper l'action \`a l'ordre 6. La normalisation $\calN$ n\'ecessite des corr\'elateurs $A\mone_{\vec{\beta}}$ d'ordre 12, et d'ordre respectivement 14, 10 et 6 pour les insertions de $a, b$ et $c$.
\begin{equation} \begin{split}
N_2 = \Re\Bigl(i\ e^{\f{i}{2}\thet_t}\ \Bigl[ &f_4\ A^{-1}_4 + \bigl(f_3\ S_3 + f_2\ S_4 + f_1\ S_5 + f_0\ S_6\bigr)A^{-1}_6 \\
&+ \f{1}{2}\bigl(f_2\ S_3\ S_3 + 2f_1\ S_3\ S_4 + f_0\ S_4\ S_4 + 2f_0\ S_3\ S_5\bigr)A^{-1}_8 \\
&+ \f{1}{3!}\bigl(f_1\ S_3\ S_3\ S_3 + 3f_0\ S_3\ S_3\ S_4\bigr)A^{-1}_{10} + \f{f_0}{4!}S_3\ S_3\ S_3\ S_3\ A^{-1}_{12}\Bigr]\Bigr).
\end{split}
\end{equation}
Nous \'ecrivons ensuite $W_3 = W_3^{(a)}+W_3^{(b)}+W_3^{(c)}$, avec :
\begin{gather} \begin{split}
W_3^{(a)} =& \Re\Bigl(i\ e^{\f{i}{2}\thet_t}\ \Bigl[\bigl(f_0\ a_6 + f_1\ a_5 + f_2\ a_4 + f_3\ a_3 + f_4\ a_2\bigr)A^{-1}_6 \\
+ \bigl( (f_0\ a_5 + &f_1\ a_4 + f_2\ a_3 + f_3\ a_2)S_3 + (f_0\ a_4 + f_1\ a_3 + f_2\ a_2)S_4 + (f_0\ a_3 + f_1\ a_2)S_5 + f_0\ a_2\ S_6\bigr)A^{-1}_8 \\
&+ \f{1}{2}\bigl( (f_0\ a_4 + f_1\ a_3 + f_2\ a_2)S_3\ S_3 + 2(f_0\ a_3 + f_1\ a_2)S_3\ S_4 + f_0\ a_2(S_4\ S_4 + 2S_3\ S_5)\bigr)A^{-1}_{10} \\
&+ \f{1}{3!}\bigl((f_0\ a_3 + f_1\ a_2)S_3\ S_3\ S_3 + 3f_0\ a_2\ S_3\ S_3\ S_4\bigr)A^{-1}_{12} \\
&+ \f{f_0}{4!}a_2\ S_3\ S_3\ S_3\ S_3\ A^{-1}_{14}\Bigr]\Bigr),
\end{split} \\
\begin{split}
W_3^{(b)} = \Re\Bigl(i\ e^{\f{i}{2}\thet_t}\ \Bigl[ \bigl( &f_0\ b_4 + f_1\ b_3 + f_2\ b_2 + f_3\ b_1\bigr)A^{-1}_4 \\
&+ \bigl( (f_0\ b_3 + f_1\ b_2 + f_2\ b_1)S_3 + (f_0\ b_2 + f_1\ b_1)S_4 + f_0\ b_1\ S_5\bigr)A^{-1}_6 \\
&+ \f{1}{2}\bigl( (f_0\ b_2 + f_1\ b_1)S_3\ S_3 + 2f_0\ b_1\ S_3\ S_4\bigr)A^{-1}_8 \\
&+ \f{f_0}{3!}b_1\ S_3\ S_3\ S_3\ A^{-1}_{10}\Bigr]\Bigr), \end{split} \\
\begin{split}
W_3^{(c)} = \Re\Bigl(i\ e^{\f{i}{2}\thet_t}\ \Bigl[ \bigl(f_0\ c_2 + f_1\ c_1 + f_2\ c_0\bigr)A^{-1}_2 + \bigl( (f_0\ c_1 + f_1\ c_0)S_3 + &f_0\ c_0\ S_4\bigr)A^{-1}_4 \\
&+ \f{f_0}{2}c_0\ S_3\ S_3\ A^{-1}_6\Bigr]\Bigr).
\end{split}
\end{gather}
On obtient ainsi le NNLO :
\be \label{equilateral NNLO}
W^{NNLO}_{1122} = \f{1}{d_{j_t}} - \f{511}{432\ d_{j_t}^2} + \f{520507}{157464\ d_{j_t}^3}
\ee
Ce r\'esultat est lui aussi confirm\'e par les simulations num\'eriques, voir figure \ref{numerics LO NLO NNLO}. Un accord avec $11,3\%$ d'erreur et obtenu pour le coefficient de $1/d_{j_t}^3$ avec $d_{j_t} = 201$. L'erreur peut \^etre r\'eduite en poussant les simulations \`a de plus grandes valeurs du spin.

\section{D\'eveloppement asymptotique du 6j isoc\`ele} \label{sec:asymp6jiso}

La proc\'edure d\'ecrite \`a la section pr\'ec\'edente peut en fait s'appliquer directement au 6j isoc\`ele ! Cela reproduit la formule de Ponzano-Regge, mais fournit en plus l'ensemble des corrections. Ceci est int\'eressant pour un certain nombre de raisons. Les corrections apport\'ees par le 6j \`a la formule de Ponzano-Regge sont en effet des diff\'erences cl\'es en comparaison avec la m\^eme quantit\'e consid\'er\'ee en calcul de Regge quantique. Le symbole 6j est par ailleurs l'unique \'etat physique de la gravit\'e 3d pour une topologie triviale et une triangulation minimale d'un seul t\'etra\`edre, et il est donc aussi int\'eressant \`a ce titre de comprendre la diff\'erence avec l'exponentielle (ou plut\^ot le cosinus) de l'action de Regge. Les mod\`eles de mousses de spins en 4d font appara\^itre des symboles $\SU(2)$-invariants qui peuvent s'exprimer \`a l'aide de symboles 6j, celui-ci \'etant bien s\^ur plus simple. Ainsi, au regard de nombreux aspects de la gravit\'e quantique, il est essentiel d'acqu\'erir une meilleure connaissance du symbole 6j, au-del\`a de l'asymptotique de Ponzano et Regge. C'est la t\^ache que nous entreprenons dans cette section, en effectuant le d\'eveloppement asymptotique du 6j isoc\`ele \`a partir de l'expression exacte \eqref{6jiso so3}.

\'Ecrit comme \eqref{6j exp form}, le 6j isoc\`ele est la fonction de partition d'une th\'eorie d\'efinie par l'action $\Phi_{\{\eps\}}$ et la mesure d'int\'egration $f$. L'action de Regge, $S_{\rm R}=\sum d_{j_e}\f{\thet_e}{2}$, appara\^it naturellement comme l'\'evaluation de $\Phi_{\{\eps\}}$ au point stationnaire,
\be \label{regge action 6jiso}
\left\{ \begin{aligned} &e^{id_{j_t}\Phi_{\{\eps\}}(\bar{\phi}_1,\bar{\phi}_2,0)} = \eps_1\eps_2\ e^{-i\eps_1\eps_2\eps_{12}S_{\rm R}}, \\
&S_{\rm R} = d_{j_t}\bigl(2\thet_t + 2k_1\alpha_1 + 2k_2\alpha_2\bigr). \end{aligned} \right.
\ee
Nous proc\'edons ensuite comme pour le propagateur, sachant que pour chaque configuration des signes le 6j estpiqu\'e sur la g\'eom\'etrie classique du t\'etra\`edre. Le d\'eveloppement autour de cette g\'eom\'etrie plate est donn\'e en termes des moments gaussiens de la hessienne $H_{\{\eps\}}$ de $\Phi_{\{\eps\}}$, par le th\'eor\`eme de Wick. Contrairement aux \'etudes pr\'ec\'edentes de l'asymptotique du 6j, nous utilisons comme variable d'\'echelle la dimension $d_{j_t}$, avec les rapports $k_1, k_2$ fixes, et non le spin $j_t$.

Comme pour le propagateur, les quatre points stationnaires fournissent la m\^eme contribution, et les deux configurations de signes possibles pour chacun sont reli\'ees par conjugaison complexe. Nous pouvons ainsi aboutir \`a des formules assez explicites. Introduisons $\omega$ comme $\Phi_{\{\eps\}}$ priv\'e de son d\'eveloppement jusqu'\`a l'ordre 2, autour de $(\alpha_1,\alpha_2,0)$, avec $\eps_{12}=-\eps_1=-\eps_2=1$. Soit $H\mone$ l'inverse de la hessienne $H_{\{\eps\}}$,
\be
H^{-1} = \f{1}{2} \begin{pmatrix} \f{1}{1-k_1^2}\cot\thet_t & -\f{1}{\sqrt{1-k_1^2}\sqrt{1-k_2^2}\sin\thet_t} & 0 \\
                                  -\f{1}{\sqrt{1-k_1^2}\sqrt{1-k_2^2}\sin\thet_t} & \f{1}{1-k_2^2}\cot\thet_t & 0 \\
                                  0 & 0 & \f{\tan\thet_t}{1-(k_1^2+k_2^2)} \end{pmatrix}.
\ee
Nous avons aussi besoin du volume du t\'etra\`edre, qui intervient dans les int\'egrales gaussiennes de $H$,
\be \label{tetrahedron volume}
V_t = \f{d_{j_t}^3}{12}\ k_1 k_2\sqrt{1-(k_1^2+k_2^2)}.
\ee
Le d\'eveloppement du 6j isoc\`ele est alors donn\'e par (plus de d\'etails dans \cite{graviton3d-val}) :
\be \label{6jiso expansion}
\begin{Bmatrix}
j_1 & j_t & j_t \\
j_2 & j_t & j_t
\end{Bmatrix} = \f{1}{\sqrt{1-k_1^2}\sqrt{1-k_2^2}\sqrt{12\pi V_t}}\ \sum_{p\geq 0}(-1)^p
\Bigl(\f{C_{2p}}{d_{j_t}^{2p}}\ \cos\bigl(S_{\rm R}+\f{\pi}{4}\bigr)\ +\ \f{C_{2p+1}}{d_{j_t}^{2p+1}}\
\sin\bigl(S_{\rm R}+\f{\pi}{4}\bigr)\Bigr),
\ee
les coefficients $C_P$, pour $P=2p,2p+1$, \'etant donn\'es par des sommes finies,
\be \label{coeff 6jiso}
C_P = \sum_{n=0}^P \f{(-1)^n}{(2(P+n))! n!}\ \sum_{\vec{\beta}\in\{1,2,3\}^{2(P+n)}}
\pp^{2(P+n)}_{\vec{\beta}}\bigl(f\omega^n\bigr)_{|(\alpha_1,\alpha_2,0)}\ H^{-1}_{\vec{\beta}}.
\ee
Ainsi, tous les ordres paires sont en phase avec l'ordre dominant l'asymptotique, i.e. la formule originale de Ponzano et Regge, en $\cos(S_{\rm R}+\pi/4)$. Ce terme dominant est ici facilement retrouv\'e en calculant le coefficient $C_0$, avec $f(\alpha_1,\alpha_2,0)=\sqrt{1-k_1^2}\sqrt{1-k_2^2}$,
\be \label{6jiso LO}
\begin{Bmatrix}
j_1 & j_t & j_t \\
j_2 & j_t & j_t
\end{Bmatrix}\,\underset{LO}{\sim}\,\f{1}{\sqrt{12\pi V_t}}\ \cos\bigl(S_R+\f{\pi}{4}\bigr).
\ee
Quant aux ordres impaires, ils sont touts en quadrature avec celui-ci. Si nous avions utilis\'e le spin, $j_t$, comme variable d'\'echelle au lieu de $(2j_t+1)$, le d\'eveloppement asymptotique n'aurait pas mis en \'evidence une structure si simple, et cosinus et sinus auraient \'et\'e m\'elang\'es \`a tous les ordres (sauf le premier) !

Ce d\'eveloppement asymptotique pour le t\'etra\`edre isoc\`ele montre que la seule fr\'equence pertinente est l'action de Regge. Ces calculs ont \'et\'e confirm\'es et largement poursuivis par Dupuis et Livine, \cite{maite-etera-6jcorr, maite-etera-6jasym}, en utilisant d'autres m\'ethodes, elles aussi int\'eressantes, faisant notamment appel aux limites asymptotiques des relations de r\'ecurrence que nous regarderons plus tard !

Le coefficient d'un ordre donn\'e se calcule simplement par les contractions des d\'eriv\'ees de $\omega^nf$ avec les moments gaussiens. Pour un ordre donn\'e $P$, les d\'eriv\'ees de $\omega$ apparaissent jusqu'\`a l'ordre $2(P+1)$, pour $n=1$. Par exemple, le NLO s'obtient en regardant $P=1$. Avec les notations de la section pr\'ec\'edente, nous avons :
\be
C_1 = f_2\ H^{-1}_2 - (f_1\ \omega_3 + f_0\ \omega_4)H^{-1}_4 + \f{f_0}{2}\omega_3\ \omega_3\ H^{-1}_6.
\ee
Et en introduisant le volume r\'eduit, $v = V/d_{j_t}^3$, il vient
\be \label{6jiso NLO}
\begin{Bmatrix}
j_1 & j_t & j_t \\
j_2 & j_t & j_t
\end{Bmatrix}\,\underset{NLO}{\sim}\, \f{1}{\sqrt{12\pi V_t}}\ \cos\bigl(S_{\rm R}+\f{\pi}{4}\bigr) - \f{\cos^2\thet_t}{d_{j_t} \sqrt{12\pi V_t}}\f{P_1(k_1,k_2)}{48(12v)^3}\ \sin\bigl(S_{\rm R}+\f{\pi}{4}\bigr),
\ee
o\`u $P(k_1, k_2)$ est un polyn\^ome sym\'etrique en $k_1$ et $k_2$,
\begin{multline} 
P_1(k_1,k_2) = 3(1-k_1^2)^2(1-2k_1^2) + 3(1-k_2^2)^2(1-2k_2^2) - 3 + 46k_1^2k_2^2 + 25k_1^4k_2^4 \\
- 44(k_1^4k_2^2+k_1^2k_2^4) + 10(k_1^6k_2^2+k_1^2k_2^6).
\end{multline}
Ce polyn\^ome n'est pas reli\'e de mani\`ere \'evidente au volume, et nous ne lui avons pas trouv\'e d'interpr\'etation g\'eom\'etrique naturelle. Cela serait pourtant tr\`es souhaitable pour interpr\'eter physiquement les corrections au propagateur.

Ce polyn\^ome se simplifie pour les valeurs extr\`emes de $k_1$. Pour $k_1=0$, en effet, $P_1(0,k)=3(1-k^2)^2(1-2k^2)$ ; et pour $k_1 =1$, $P_1(1,k)=-4k^4(1-k^2)$, avec des racines \'evidentes en 0 et 1. N\'eanmoins, $k_1$ n'atteint jamais ces valeurs, ses bornes d\'ependant de la repr\'esentation $j_t$ (du fait du recouplage des produits tensoriels de repr\'esentations) :
\beq
\f1{2d_{j_t}}\le k_e \le 1-\f1{2d_{j_t}}.
\ee
Lorsque $k_1$ atteint ces valeurs extr\`emes, les coefficients de $P_1$ sont alors polynomiaux en $1/d_{j_t}^2$.
\footnotemark
\footnotetext{Par exemple lorsque l'ar\^ete $e_1$ a une longueur minimale, $j_1=0$ ou $k_1=\f{1}{2d_{j_t}}$, les coefficients de $P_1$ se lisent :
\beq
P_1(\f{1}{2\di{t}},k_2) = 3\big(1-\f{1}{\di{t}^2} + \f{5}{16\di{t}^4} -
\f{1}{32\di{t}^6}\big)+\big(-12+\f{23}{2\di{t}^2}-\f{11}{4\di{t}^4}+\f{5}{32\di{t}^6}\big)k_2^2 +
\big(15-\f{11}{\di{t}^2}+\f{25}{16\di{t}^4}\big)k_2^4 + \big(-6+\f{5}{2\di{t}^2}\big)k_2^6.
\ee}

\begin{figure}[t] \begin{center}
\includegraphics[width=5cm]{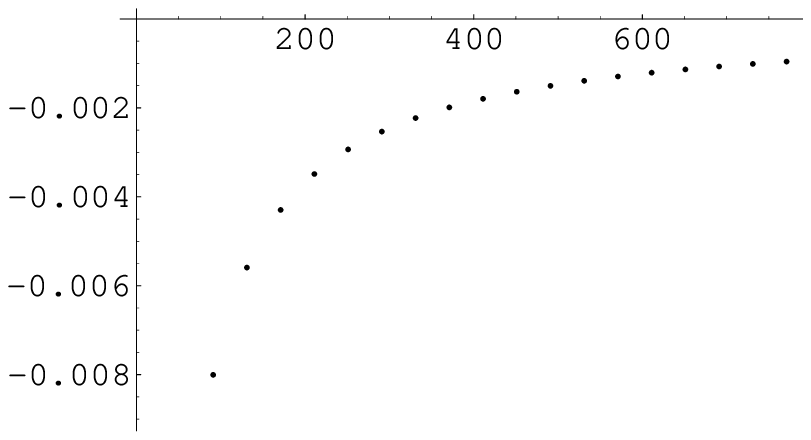}
\includegraphics[width=5cm]{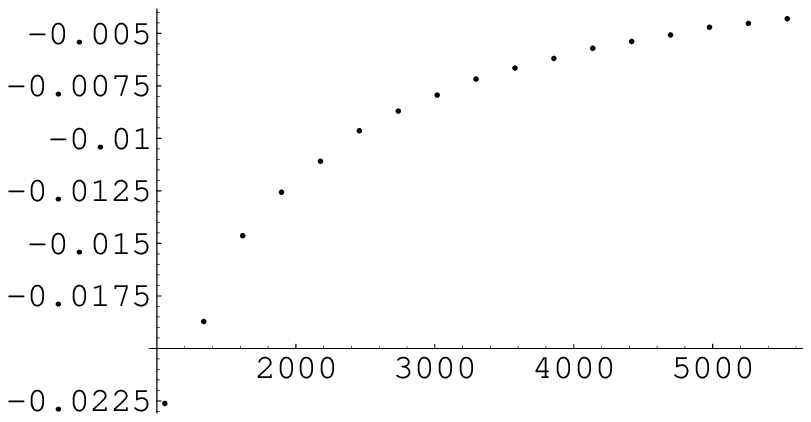}
\includegraphics[width=5cm]{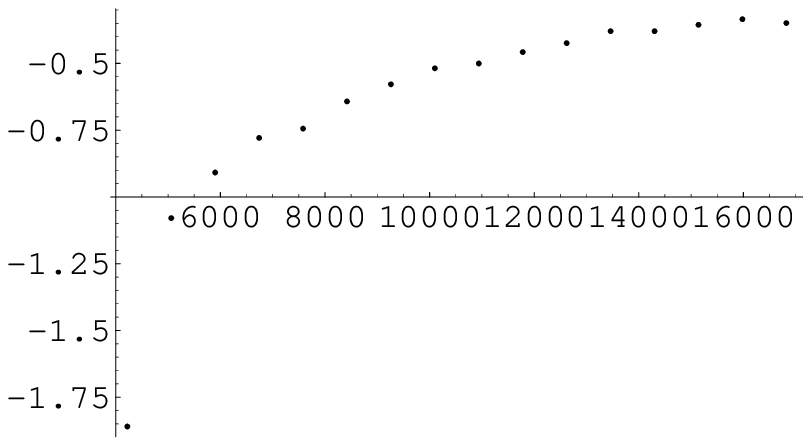}
\caption{ \label{plot 6jNLO} Les diff\'erences entre le symbole 6j et le d\'eveloppement asymptotique \eqref{6jiso NLO}, pour trois paires $(k_1,k_2)=(k,k)$ : de gauche \`a droite : $k=1/2$, $k=3/14$, $k=3/42$. L'axe des abscisses repr\'esente la dimension $d_{j_t} = N d_{j_0}$, o\`u $d_{j_0}$ est fix\'e respectivement \`a 1, 7, et 21, tandis que $N$ va de 200 \`a 800.}
\end{center}
\end{figure}

Le r\'esultat \eqref{6jiso NLO} est bien confirm\'e num\'eriquement, figure \ref{plot 6jNLO}. Ces graphes repr\'esentent des calculs num\'eriques du 6j moins la formule analytique \eqref{6jiso NLO}, pour trois paires $(k_1, k_2)$. Nous avons pris le cas particulier $k_1=k_2=k$, pour lequel
\be
P_1(k,k) = (1-k^2)\Bigl(3-21k^2+55k^4-45k^6\Bigr),
\ee
dont la seule racine dans $[0,1]$ est $k=\f{1}{45}\sqrt{15[(10(81\sqrt{310}+1450))^{\f{1}{3}} +
(10(81\sqrt{310}-1450))^{\f{1}{3}}+55]}\approx 0.8248$. Pour mieux mettre en valeur la comparaison, nous avons multipli\'e les expressions par $d_{j_t}^{5/2}$, pour appr\'ecier la fa\c{c}on dont le coefficient du NLO est approch\'e, et nous avons supprim\'e les oscillations en divisant par celles du NNLO, $\cos(S_{\rm R}+\f{\pi}{4})$. Le num\'erique confirme ainsi le coefficient et la phase.

Notons que dans le cas \'equilat\'eral, $k_1=k_2=1/2$ repr\'esente le (demi-)rapport des longueurs, mais aussi des spins. On peut donc se ramener facilement au d\'eveloppement usuel en $1/j_t$,
\begin{align}
\{6j_t\}^{NLO} &= \f{2^{5/4}}{\sqrt{\pi d_{j_t}^3}}\cos\bigl(S_{\rm R}+\f{\pi}{4}\bigr) - \f{31}{72\cdot 2^{1/4}\sqrt{\pi d_{j_t}^5}}\sin\bigl(S_{\rm R}+\f{\pi}{4}\bigr), \\
&= \f{\cos\bigl(S_{\rm R}+\f{\pi}{4}\bigr)}{2^{1/4}\sqrt{\pi j_t^3}} - \f{1}{2^{9/4}\sqrt{\pi j_t^5}}\Bigl[3\cos\bigl(S_{\rm R}+\f{\pi}{4}\bigr) + \f{31\cdot 2^{5/2}}{576}\sin\bigl(S_{\rm R}+\f{\pi}{4}\bigr)\Bigr].
\end{align}
Ce point de vue montre bien qu'il est plus judicieux d'utiliser les inverses des longueurs plut\^ot que des spins. Finalement signalons que notre d\'eveloppement en termes de cosinus et sinus de l'action de Regge ne tient que pour des valeurs \og moyennes\fg{} de $k_1, k_2$, et n'est plus valable \`a proximit\'e des valeurs extr\`emes 0 et 1. En effet, pour $k_2=0$ on sait que l'asymptotique est donn\'e par des fonctions d'Airy, tandis que pour $k_2=1$, il s'agit d'une exponentielle r\'eelle de l'action de Regge \cite{schulten-gordon2}.

\part{Contraintes hamiltoniennes et relations de r\'ecurrence} \label{sec:recurrence}

On sait, depuis Perez et Noui \cite{noui-perez-ps3d} et plus particuli\`erement Ooguri \cite{ooguri-3d} que le mod\`ele de Ponzano-Regge peut \^etre consid\'er\'e comme un projecteur sur les \'etats physiques (connexions plates), agissant naturellement sur la base des r\'eseaux de spins. On imagine donc qu'il impl\'emente d'une certaine mani\`ere une version quantifi\'ee de l'\'equation de Wheeler-DeWitt en dimension 3. Rechercher un tel lien, plus particulier\`erement en dimension 4, est essentiel pour comprendre comment des mod\`eles de mousses de spins peuvent r\'ealiser quantiquement la contrainte hamiltonienne, ou l'\'equation de Wheeler-DeWitt qui reste une \'equation mal comprise, dont la quantification canonique, par des m\'ethodes standards, pr\'esente de nombreuses \'ecueils. En dimension 4, les mod\`eles de mousses de spins pour la gravit\'e quantique dont nous disposons sont suffisamment compliqu\'es et sp\'eculatifs pour que l'on s'int\'eresse dans un premier temps \`a des situations mieux comprises par ailleurs, telle que la gravit\'e 3d. Cela va nous permettre d'\'elaborer une m\'ethode au potentiel int\'eressant, fond\'ee sur la d\'erivation de relations de r\'ecurrence pour les mousses de spins, dans le but de traquer les liens entre ces mod\`eles en tant que projecteurs hypoth\'etiques sur les \'etats physiques de la th\'eorie, et la contrainte hamiltonienne, ou l'\'equation de Wheeler-DeWitt au niveau quantique.

Quelle sont les diff\'erentes strat\'egies envisageables ? Pour l'heure, les amplitudes de mousses de spins sont formul\'ees de mani\`ere g\'en\'erique via une somme sur les triangulations de l'int\'erieur d'une vari\'et\'e topologique. Une quantit\'e importante est le poids donn\'e \`a chaque simplexe de plus haute dimension -- du point de vue dual, ces derniers \'etant des vertexes, on parle de d\'eveloppement en vertexes. Le poids assign\'e aux simplexes est calcul\'e alg\'ebriquement \`a partir de la th\'eorie des repr\'esentations du groupe de structure consid\'er\'e. Comme r\'esultat de la construction, on s'attend \`a ce que les variables de bord soient les donn\'ees caract\'erisant les r\'eseaux de spins de la th\'eorie. En relativit\'e g\'en\'erale, cela sugg\`ere naturellement de s'int\'eresser de mani\`ere directe aux repr\'esentations des contraintes hamiltoniennes, scalaire et vectorielle, sur les r\'eseaux de spins. Mais la contrainte vectorielle est prise en compte en LQG de mani\`ere totalement diff\'erente (par moyenne sur les diff\'eomorphismes 3d). Quant \`a la contrainte scalaire, malgr\'e la quantification explicite de Thiemann et les r\'esultats associ\'es, impressionnants en soi, un tel lien reste hors de port\'ee. On pourra n\'eanmoins consulter dans une optique similaire la r\'ef\'erence \cite{alesci-noui-sardelli}, dans laquelle les auteurs s'attachent \`a touver des op\'erateurs dont l'action sur les r\'eseaux de spins reproduit, via le produit scalaire, les amplitudes de diff\'erents mod\`eles de mousses de spins, dans le cas d'un seul 4-simplexe certes. Le point innovant de cette approche est que les auteurs parviennent \`a n'utiliser que les r\'eseaux de spins de la LQG, dont le groupe de structure est $\SU(2)$, tandis que les mousses de spins sont plus naturellement formul\'ees avec $\Spin(4)$. L'aspect plus obscur de leur construction est le manque de transparensce des op\'erateurs ainsi d\'efinis quant \`a leur interpr\'etation g\'eom\'etrique et leur sens naturel du point de vue hamiltonien.

Une direction diff\'erente pour interpr\'eter les contraintes hamiltoniennes se dessine \`a partir de la g\'eom\'etrie cod\'ee par les variables de bord, les r\'eseaux de spins. Cela fut d'abord sugg\'er\'e par les analyses semi-classiques des mod\`eles de mousses de spins, dans lesquelles les amplitudes sont donn\'ees par des exponentielles de l'action de Regge. En particulier, les donn\'ees du bord fournissent dans ces cas des g\'eom\'etries de Regge, et on sait que cela peut se g\'en\'eraliser gr\^ace aux g\'eom\'etries tordues. De plus, notre analyse des mod\`eles de mousses de spins comme int\'egrales de chemins sur r\'eseau renforce de telles interpr\'etations, y compris en dehors du r\'egime semi-classique. Ces interpr\'etations g\'eom\'etriques orientent donc vers une version hamiltonienne du calcul de Regge. Une telle construction n'a \'et\'e men\'ee que tr\`es r\'ecemment, par Dittrich et Bahr, en 3 et 4 dimensions, et il n'y a pas de propositions claires pour les contraintes hamiltoniennes, \`a l'exception tout de m\^eme tr\`es int\'eressante des g\'eom\'etries plates. En fait, \`a la vue de ces travaux, il semble qu'il n'y ait g\'en\'eriquement pas de contraintes exactes, mais plut\^ot des pseudo-contraintes, i.e. n'impliquant pas qu'une seule tranche spatiale de la vari\'et\'e. Autrement dit les sym\'etries sont g\'en\'eriquement bris\'ees ! Dittrich et Bahr montrent en 3d dans le cas d'une courbure homog\`ene que les sym\'etries peuvent \^etre restor\'ees par {\it coarse graining}. Cela dirige naturellement vers les mouvements de Pachner dont on verra qu'ils jouent un r\^ole pr\'epond\'erant dans nos \'equations d\'ecrivant les sym\'etries au niveau quantique. Au del\`a des mouvements de Pachner, mes travaux (au niveau quantique) partagent avec ceux de Dittrich et Bahr (au niveau classique) de grandes similitudes qui restent \`a explorer plus profond\'ement.

Une des difficult\'es pour relier de telles constructions canoniques en calcul de Regge aux mousses de spins tient au passage du classique au quantique. En effet, les g\'eom\'etries de Regge aux bords des mousses de spins sont repr\'esent\'ees par des quantit\'es discr\`etes comme des demi-entiers ! Cela nous sugg\`ere deux choses :
\begin{itemize}
\item Se contenter de regarder les sym\'etries au niveau \emph{quantique}, directement dans les mousses de spins.
\item Chercher \`a caracteriser ces sym\'etries par des \'equations \emph{aux diff\'erences}, et non des \'equations diff\'erentielles !
\end{itemize}
En effet, de telles \'equations aux diff\'erences sont assez naturellement susceptibles d'intervenir, comme cons\'equence de la th\'eorie des repr\'esentations du groupe local consid\'er\'e.

Nous allons donc regarder des \'equations aux diff\'erences sur l'amplitude de vertex de diff\'erents mod\`eles, en particulier sur les symboles 6j, 15j, et un peu plus tard sur le 10j de Barrett-Crane. Au moins pour les mod\`eles topologiques, ces objets sont en fait le d\'eveloppement d'\'etats physiques (satisfaisant toutes les contraintes) pour les th\'eories consid\'er\'ees dans la base des r\'eseaux de spins sur le bord d'un simplexe. Ainsi, ces \'equations aux diff\'erences sont des \'equations d\'efinissant (en partie ou compl\`etement) les \'etats physiques, et nous d\'ecrirons au possible de quelle mani\`ere on peut les voir comme des actions de contraintes hamiltoniennes dans la repr\'esentation des \'etats en r\'eseaux de spins. Bien s\^ur, pour obtenir des \'equations similaires sur des \'etats physiques \`a la Hartle-Hawking, sur un bord quelconque, il faudrait \'etudier les sym\'etries pour les amplitudes de mousses de spins globales, et pas seulement l'amplitude d'un simplexe.


Ce type d'id\'ee -- utiliser des \'equations aux diff\'erences -- a d\'ej\`a \'et\'e mis en oeuvre partiellement dans deux situations. La premi\`ere est en cosmologie quantique \`a boucles (LQC) (voir \cite{ashtekar-bojowald-math-lqc, bojowald-living-rev-lqc, ashtekar-quantum-bigbang}). La r\'eduction du nombre de degr\'es de libert\'e par des sym\'etries sp\'eciales (isotropie, homog\'en\'eit\'e) permet dans ce cadre de contruire pr\'ecis\'ement un op\'erateur hamiltonien agissant exactement par diff\'erences finies. Il a de plus \'et\'e r\'ecemment montr\'e \cite{ashtekar-lqc-vertex-expansion} que l'op\'erateur d'\'evolution, i.e. l'exponentielle du hamiltonien, admet un d\'eveloppement en vertexes, \`a la mani\`ere d'une somme sur des mousses de spins \og jouets\fg.

Le second exemple est le mod\`ele de Ponzano-Regge. Des relations de r\'ecurrence exactes pour le symbole 6j sont en effet bien connues \cite{varshalovich-book, yutsis-book, schulten-gordon1}. Elles sont \`a la base de calculs tr\`es efficaces de l'asymptotique \cite{schulten-gordon2, maite-etera-6jcorr}, et sont utilis\'ees pour les \'evaluations num\'eriques. Le lien entretenu avec l'\'equation de Wheeler-DeWitt a d\'ej\`a \'et\'e approch\'e, notamment dans \cite{barrett-crane-wdw}. Ici nous obtenons une connexion plus pr\'ecise en utilisant l'\'ecriture de la contrainte de courbure nulle en termes d'angles dih\'edraux (cet argument n'a pas encore \'et\'e soumis \`a publication) et sachant que le 6j est l'unique \'etat physique sur le bord d'un t\'etra\`edre d\'evelopp\'e dans la base des r\'eseaux de spins, \ref{sec:6j move23}. Je vais pr\'esenter diff\'erentes m\'ethodes pour d\'eriver des relations de r\'ecurrence, certaines connues mais aussi de nouvelles. L'une de ces m\'ethodes est standard et utilise l'invariance topologique, mais n'avait jusque l\`a \'et\'e appliqu\'ee qu'en dimension 3. Je montre ici \ref{sec:15jrecurrence} que cette propri\'et\'e d'invariance contient des relations de r\'ecurrence sur le symbole 15j de $\SU(2)$, qui est l'unique \'etat physique de la th\'eorie BF sur le bord d'un 4-simplex, et la brique de construction du mod\`ele d'Ooguri en dimension 4. J'insiste par ailleurs sur l'interpr\'etation g\'eom\'etrique de ces relations, dans la perspective d'y voir l'action d'un op\'erateur hamiltonien agissant sur les r\'eseaux de spins. Cela permet de d\'egager une interpr\'etation en termes de mouvements \'el\'ementaires sur des simplexes.

Une seconde m\'ethode est plus originale : elle est fond\'ee sur l'action des op\'erateurs d'holonomies sur les fonctionnelles des r\'eseaux de spins, \'evalu\'ee sur les connexions plates, \ref{sec:recurrence tentmoves}. Cette m\'ethode est tr\`es \'el\'egante et construit naturellement des analogues quantiques des mouvements \og tente\fg{}, qui sont les mouvements \'el\'ementaires  utilis\'es par Dittrich et Bahr dans leur formulation canonique du calcul de Regge ! Par ailleurs, j'obtiens ainsi des relations de r\'ecurrence pour des r\'eseaux de spins arbitraires, et qui sont naturellement associ\'ees aux cycles de ceux-ci.

J'insiste d'ores et d\'ej\`a sur le fait que la m\'ethode utilisant l'invariance topologique est \emph{intrins\`equement dynamique}, car elle prend en compte les sommes sur les spins internes des mod\`eles. Je dis cela par opposition \`a la simple donn\'ee du vertex du mod\`ele, i.e de l'amplitude \'el\'ementaire de chaque simplexe, qui d\'etermine la structure des \'etats de bord et serait ainsi plut\^ot de nature \emph{cin\'ematique}. La seconde m\'ethode, utilisant l'op\'erateur d'holonomie, est plus subtile de ce point de vue, en ce qu'un op\'erateur d'holonomie n'est qu'une des composantes de Fourier de la contrainte de courbure nulle sur r\'eseau. Cela montre que les r\'eseaux de spins rendent particuli\`erement simple la description des contraintes du type $F=0$.

Pour les mod\`eles topologiques, l'invariance sous les mouvements de Pachner contient la dynamique g\'en\'er\'ee par les sym\'etries classiques. Ces sym\'etries sont telles que pour une topologie fix\'ee, toutes les triangulations contribuent avec la m\^eme amplitude (apr\`es r\'egularisation) \`a la somme sur les mousses de spins. Ainsi, la projection sur le noyau des contraintes est r\'ealis\'ee \`a l'aide d'une seule triangulation, disons celle avec le moins de simplexes possibles. Mais ce n'est plus le cas pour les mod\`eles non-topologiques. Dans ces cas, on s'attend \`a ce que les sym\'etries classiques soient restor\'ees, et ainsi la projection sur le noyau des contraintes r\'ealis\'ee par les mousses de spins, uniquement en sommant sur les triangulations pour un bord fix\'e. Mais nous ne savons pas \`a l'heure actuelle obtenir des expressions simples et ferm\'ees pour des triangulations non-triviales.

Ce n'est n\'eanmoins pas la fin de l'histoire, et il est certainement int\'eressant de regarder des \'equations aux diff\'erences sur une triangulation fix\'ee. Si l'on imagine formellement le projecteur physique comme l'exponentiation des contraintes (de diff\'eomorphismes et scalaire),  et la somme sur les mousses comme une somme sur les histoires des r\'eseaux de spins, alors on s'attend typiquement \`a ce que l'amplitude de vertex d'un mod\`ele soit reli\'ee aux \'el\'ements de matrices des contraintes. Puisque ceux-ci ne sont pas vraiment accessibles dans le formalisme canonique, il peut \^etre utile de sonder leurs propri\'et\'es g\'eom\'etriques de base via des relations de r\'ecurrence. C'est dans cette perspective que j'insiste sur l'interpr\'etation g\'eom\'etrique des relations obtenues en termes de d\'eformations \'el\'ementaires de simplexes, susceptibles de pr\'efigurer l'action d'un bon op\'erateur hamiltonien sur les r\'eseaux de spins. Par exemple, dans les mod\`eles topologiques mentionn\'es ci-dessus, on trouve que les r\'ecurrences sont g\'en\'er\'ees par le collage de simplexes aplatis \`a des simplexes initiaux, ce qui de leur point de vue correspond au d\'eplacement de sommets produisant ainsi des modifications des longueurs ou des aires.

J'ai donc avec E. Livine et S. Speziale \'egalement cherch\'e et obtenu des relations de r\'ecurrence pour des symboles ne donnant pas lieu \`a des mod\`eles topologiques. Au chapitre \ref{sec:recurrence-6jiso}, je pr\'esente une relation sur un symbole 6j isoc\`ele, qui peut se d\'eriver de diff\'erentes mani\`eres. En particulier, on montre qu'elle est issue de sommes et successions des mouvement topologiques, ce qui autorise \`a l'interpr\'eter comme une condition d'invariance sous des \emph{combinaisons} des d\'eformations \'el\'ementaires du mod\`ele de Ponzano-Regge. Plus tard, \ref{sec:10jrecurrence}, je d\'ecris une relation de r\'ecurrence sur le symbole 10j du mod\`ele, non-topologique, de Barrett-Crane, qui s'interpr\`ete naturellement comme la fermeture d'un 4-simplexe au niveau quantique.



\chapter{Contrainte hamiltonienne sur le 6j : mouvement de Pachner 2-3} \label{sec:6j move23}

\section{Le 6j comme \'etat physique}


Rappelons tout d'abord le sens du symbole 6j en tant qu'\'etat physique de la gravit\'e 3d. Consid\'erons $S^2$ triangul\'ee comme le bord d'un t\'etra\`edre. C'est notre surface canonique. Avec cette topologie et cette triangulation fix\'ees, on sait qu'il n'existe qu'un seul \'etat satisfaisant la contrainte de courbure nulle. L'espace de Hilbert des \'etats invariants de jauge est engendr\'e par les r\'eseaux de spins vivant sur le graphe dual \`a la triangulation (voir figure \ref{fig:6j}) qui est aussi de forme t\'etra\'edrique. En repr\'esentation holonomie, cet espace des \'etats cin\'ematiques est constitu\'e des fonctions $L^2$ pour les 6 \'el\'ements de $\SU(2)$ vivant sur les ar\^etes duales, et invariantes par translation \`a chacun des 4 vertexes duaux. Sans surprise, l'unique \'etat physique n'est pas normalisable et est donn\'e formellement comme $\prod_{(\mathrm{cycles\ ind})}\delta(g_c)$, o\`u $g_c = \prod_{e}g_e$ est l'holonomie le long d'un cycle (ferm\'e) form\'e par des liens du graphe dual. De mani\`ere \'equivalente on peut consid\'erer \`a la place des cycles les faces duales aux sommets de la triangulation. Le produit sur ces cycles ou faces doit se faire pour des cycles ou faces \emph{ind\'ependants}. Ici cela ne pose aucune difficult\'e : imposer que les holonomies autour des quatre faces soient triviales conduit typiquement \`a une divergence \'evidente, du fait qu'une des quatre contraintes est redondante. On fait donc un choix de trois cycles/faces ind\'ependants, tel que
\beq
\psi_{\operatorname{phys}}(g_1,\dotsc,g_6) = \delta(g_4 g_5 g_6)\,\delta(g_1 g_6 g_2\mone)\,\delta(g_2 g_4 g_3\mone).
\ee
Pour conna\^itre le d\'eveloppement de cet \'etat sur les r\'eseaux de spins, on prend son produit scalaire avec une fonctionnelle de r\'eseau de spins $s_{\{j_e\}}(g_1,\dotsc,g_6)$,
\begin{align}
\psi_{\operatorname{phys}}(j_1,\dotsc,j_6) &= \int \prod_{e=1}^6 dg_e\ s_{\{j_e\}}(g_1,\dotsc,g_6)\ \psi_{\operatorname{phys}}(g_1,\dotsc,g_6),\\
&= \begin{Bmatrix} j_1 &j_2 &j_3 \\j_4 &j_5 &j_6 \end{Bmatrix}.
\end{align}
Ce r\'esultat provient simplement du fait que l'\'etat physique en repr\'esentation holonomie force l'\'evaluation de l'int\'egrale sur les holonomies triviales (\`a transformations de jauge pr\`es), et qu'alors : $s_{\{j_e\}}(\mathbbm{1})=\{6j_e\}$.

\section{L'identit\'e de Biedenharn-Elliott et la contrainte hamiltonienne}

Nous savons par ailleurs que l'invariance topologique du mod\`ele de Ponzano-Regge r\'eside dans l'identit\'e de Biedenharn-Elliott satisfaite par le 6j, que nous rappelons,
\begin{equation} \label{bied-elliott}
\begin{Bmatrix} j &h &g\\ k &a &b \end{Bmatrix}\,\begin{Bmatrix} j &h &g\\ f &d &c \end{Bmatrix} = \sum_l (-1)^{S+l}(2l+1)\,\begin{Bmatrix} k &f &l\\ d &a &g\end{Bmatrix}\,\begin{Bmatrix} a &d &l\\ c &b &j\end{Bmatrix}\,\begin{Bmatrix} b &c &l\\ f &k &h\end{Bmatrix},
\end{equation}
o\`u $S$ repr\'esente la somme des neuf spins impliqu\'es. En prenant la valeur $f=1$, la somme du membre de droite se r\'eduit \`a trois termes, $l=k-1,k,k+1$. En \'evaluant explicitement les symboles avec $f=1$, on parvient \`a une relation de r\'ecurrence d'ordre 2 sur le symbole 6j,
\be \label{rec6j}
A_{+1}(j)\,\begin{Bmatrix} j_1+1 &j_2 &j_3 \\ j_4 &j_5 &j_6 \end{Bmatrix} +
A_{0}(j)\,\begin{Bmatrix} j_1 &j_2 &j_3 \\ j_4 &j_5 &j_6 \end{Bmatrix} +
A_{-1}(j)\,\begin{Bmatrix} j_1-1 &j_2 &j_3 \\ j_4 &j_5 &j_6 \end{Bmatrix} = 0.
\ee
Les coefficients $A_{\pm 1}$ prennent la forme : $A_{+1}(j) = j_1 E(j_1+1)$, et $A_{-1}(j) = (j_1-1) E(j_1)$, pour
\be
E(j_1) = \Bigr[\bigl((j_2+j_3+1)^2-j_1^2\bigr) \bigl(j_1^2-(j_2-j_3)^2\bigr) \bigl((j_5+j_6+1)^2-j_1^2\bigr) \bigl(j_1^2-(j_5-j_6)^2\bigr)\Bigr]^{\f12}.
\ee
Le coefficient $A_0$ est donn\'e par :
\begin{multline}
A_0(j) = \bigl(2j_1+1\bigr)\Bigl\{2\bigl[j_2(j_2+1)j_5(j_5+1)+j_6(j_6+1)j_3(j_3+1)-j_1(j_1+1)j_4(j_4+1)\bigr] \\
- \bigl[j_2(j_2+1)+j_3(j_3+1)-j_1(j_1+1)\bigr]\bigl[j_5(j_5+1)+j_6(j_6+1)-j_1(j_1+1)\bigr]\Bigr\}.
\end{multline}
L'\'equation \eqref{rec6j} permet \`a l'aide d'une condition initiale de d\'eterminer l'ensemble des valeurs du 6j. Comme en outre, cette relation, caract\'erisant compl\`etement le 6j, donc l'\'etat physique $\psi_{\operatorname{phys}}$, provient directement de l'invariance topologique, il est tentant de se demander si elle peut s'interpr\'eter comme une \'equation de Wheeler-DeWitt sur le bord d'un t\'etra\`edre.

On se place pour cela dans une situation o\`u les coefficients se simplifient et prennent un sens g\'eom\'etrique, en regardant la limite homog\`ene des grands spins, $j_e\rightarrow N j_e$, pour $N\rightarrow\infty$. On d\'efinit $f(j_e) = \sqrt{12\pi V(\ell_e)}\,\{6j_e\} $, o\`u $V(\ell_e)$ est le volume du t\'etra\`edre dont les longueurs sont donn\'ees par $\ell_e = j_e+\f12$, et poss\'edant des angles dih\'edraux $\theta_e(j_e)$. La relation de r\'ecurrence se r\'e\'ecrit alors
\beq \label{Hlink}
\left[\Delta_{j_1} + 2(1-\cos\theta_1(j_l)) \right]\ \f 1{\sqrt{\sin\theta_1(j_l)}}f(j_l)= 0.
\ee
Ici $\Delta$ est l'op\'erateur de diff\'erences d'ordre 2, d\'efini par : $\Delta f(x) = f(x+1)+f(x-1)-2f(x)$, et $\theta_1$ est l'angle dih\'edral entre les deux triangles se rencontrant sur le lien qui porte le spin $j_1$, calcul\'e en fonction des longueurs $\ell_e$. Schulten et Gordon ont montr\'e \cite{schulten-gordon2} que cette \'equation est intimement reli\'ee \`a une \'equation diff\'erentielle d'ordre 2 qui en fournit des solutions via l'approximation WKB. Cela conduit \`a l'asymptotique d\'esormais bien connue du 6j en terme de l'action de Regge\footnote{En g\'en\'eral, la solution est une combinaison de fonctions d'Airy, qui se r\'eduit \`a \eqref{6jasymp-repeat} quand $V(\ell_e)^2>0$.} $S_{\rm R}$ \cite{PR, schulten-gordon2},
\be \label{6jasymp-repeat}
\begin{Bmatrix} j_1 &j_2 &j_3 \\ j_4 &j_5 &j_6 \end{Bmatrix} \sim
\f1{\sqrt{12\pi V(\ell_l)}}\cos\left( S_{\rm R}[\ell_l] + \f\pi4\right).
\ee
Le r\'egime des grands spins fournit ainsi une notion de limite semi-classique dans laquelle les amplitudes de mousses de spins sont approch\'ees par une forme de calcul de Regge quantique \cite{graviton3d-simone, graviton3d-corrections, graviton3d-val}.

Montrons maintenant que l'\'equation \eqref{Hlink} peut s'obtenir de mani\`ere simple et na\"ive, directement comme quantification de la contrainte hamiltonienne. Rappelons-nous de l'espace des phases pour la gravit\'e sur une triangulation d'une surface 2d. Il y a une paire canonique sur chaque ar\^ete, form\'ee par $(l_e,\theta_e)$ o\`u $l_e$ est la longueur de l'ar\^ete et $\theta_e$ une variable angulaire conjugu\'ee. La contrainte de courbure nulle de la th\'eorie BF se traduit dans ces variables par \eqref{flat dynamics 3d scalar},
\beq \label{Hregge}
\cos\theta_e = \cos\thet_e(l),
\ee
c'est-\`a-dire que le moment $\theta_e$ est en fait l'angle dih\'edral $\thet_e(l)$ calcul\'e gr\^ace aux longueurs.

Tentons alors une quantification tout ce qu'il y a de plus na\"ive, en prenant des fonctions des moments $\theta_e$ sur le bord d'un t\'etra\`edre. $\cos\theta_e$ agit basiquement par multiplication, et les longueurs par d\'erivation sur ces fonctions. On exprime ensuite l'action de \eqref{Hregge} sur les modes de Fourier $\psi(m_1,\dotsc,m_6)$, en d\'eveloppant $\psi(\theta) = \sum_{\{m_e\}}\psi(m_1,\dotsc,m_6) \sum_{e=1}^6\exp( i\sum_{e=1}^6 m_e\theta_e)$. Sur ces modes de Fourier, la contrainte \eqref{Hregge} devient \eqref{Hlink} !
\beq
\left[\Delta_{m_1} + 2(1-\cos\theta_1(m_l)) \right]\ \psi(m_1,\dotsc,m_6)= 0.
\ee
Bien s\^ur, cette \'equation ne d\'efinit pas proprement $\psi$ du fait qu'il manque le facteur $1/{\sqrt{\sin\theta_1(j_l)}}$. De m\^eme, nous n'avons pas pr\^et\'e attention au sens des variables $m_e$ et leur lien avec les longueurs $l_e$ ou $\ell_e$. N\'eanmoins, cet argument montre que la relation de r\'ecurrence sur le 6j \eqref{rec6j} n'est rien d'autre qu'une quantification propre de la contrainte de courbure nulle sur le bord d'un t\'etra\`edre !

\subsection*{Interpr\'etation g\'eom\'etrique}

\begin{figure}
\begin{center}
\includegraphics[width=11.5cm]{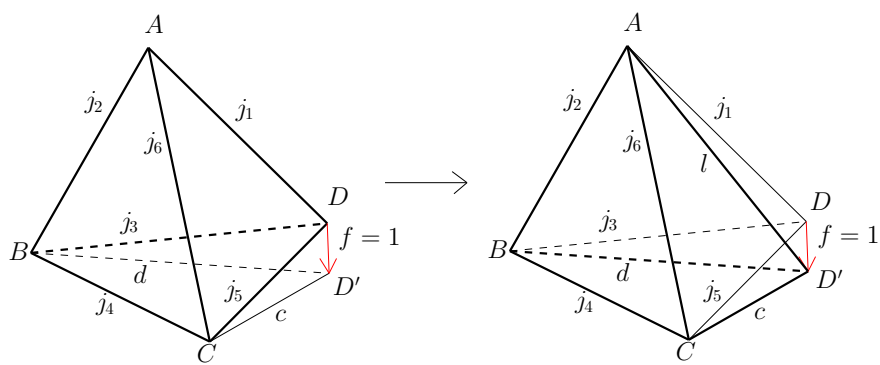}
\end{center}
\caption{ La figure repr\'esente l'identit\'e \eqref{rec6j}. A gauche, le t\'etra\`edre $(ABCD)$, dont les longueurs sont $\ell_e =(j_e+\f12)$, $e=1,\dotsc,6$, est coll\'e le long du triangle $(BCD)$ \`a un t\'etra\`edre aplati, $(BCDD')$, dont le spin sur $(DD')$ est $f=1$. A droite, les trois t\'etra\`edres $(ACDD')$, $(ABDD')$, $(ABCD')$, partagent l'ar\^ete $(AD')$ dont le spin prend les valeurs $l=j_1-1, j_1, j_1+1$. Les t\'etra\`edres $(ACDD')$ and $(ABDD')$ peuvent \^etre consid\'er\'es comme aplatis du fait qu'ils poss\`edent l'ar\^ete $f=1$. Ainsi, du point de vue du t\'etra\`edre initial $(ABCD)$, le mouvement peut \^etre vu comme un l\'eger d\'eplacement du point $D$ vers $D'$, dont il r\'esulte des changements de longueur pour les ar\^etes touchant $D$. Cependant, il est possible de faire ce mouvement de telle sorte que les spins $j_3, j_5$ soient inchang\'es : $c=j_5, d=j_3$.
 \label{move2-3} }
\end{figure}

Avec une telle conclusion, il est d\'esirable de se donner une vision plus intuitive et g\'eom\'etrique de la relation \eqref{rec6j}. Ceci est illustr\'e en figure \ref{move2-3}. (Remarquons que trois liens d'un r\'eseau de spins se rencontrant en un vertex, comme ici $j_1, j_2, j_3$, forment du point de vue de la triangulation un triangle du t\'etra\`edre). L'id\'ee est de coller le long du t\'etra\`edre $ABCD$ un deuxi\`eme t\'etra\`edre (c'est le sens du membre de gauche de l'identit\'e de BE) dont une des longueurs est \og petite\fg{}, $\ell_f = f+1/2$ (c'est le sens de la restriction au cas $f=1$, nous aurions aussi pu prendre $f=1/2$). On peut qualifier ce deuxi\`eme t\'etra\`edre d'aplati. Dans le membre de droite, le t\'etra\`edre qui nous int\'eresse est $ABCD'$, de sorte que l'on peut visualiser l'ensemble du mouvement comme un l\'eger d\'eplacement du point $D$ en $D'$ ! Naturellement, cela modifie les longueurs des trois segments qui se rencontrent initialement en $D$, pour un choix g\'en\'erique de $f$. Avec le choix sp\'ecifique $f=1$ (et en fait avec tout $f\in\N$), on peut choisir de laisser invariant les longueurs des segments $BD$ et $CD$, ce qui correspond \`a faire pivoter le triangle $BCD$ autour de $BC$. Ainsi, seul la longueur de l'ar\^ete $AD$ est modifi\'ee, ce qui donne une relation de r\'ecurrence sur un seul spin !

\chapter{Invariance topologique et relations de r\'ecurrence en 4d} \label{sec:15jrecurrence}

En passant du symbole 6j aux symboles de Wigner $\{3nj\}$, pour $n\geq3$, la litt\'erature devient nettement plus \'evasive, \`a l'exception notable du 9j \cite{varshalovich-book}. Ces symboles \'etant globalement pertinents dans le cadre des mousses de spins pour la gravit\'e quantique (par exemple, le 9j intervient dans les coefficients de fusion du nouveau mod\`ele EPR, voir section \ref{sec:resume models}), il est int\'eressant de voir quelles m\'ethodes utiliser pour les \'etudier.

Consid\'erons un symbole $\{3nj\}$ arbitraire, repr\'esent\'e par un graphe ferm\'e form\'e de vertexes trivalents et d'ar\^etes portant des repr\'esentations irr\'eductibles de $\SU(2)$ que l'on note collectivement $\{j_i\}$. L'\'evaluation du graphe conduit \`a une valeur num\'erique obtenue en suivant des r\`egles conventionnelles, qui assignent des coefficients de Clebsch-Gordan (ou des coefficients $(3mj)$) aux vertexes. Pour \'ecrire ces symboles sans ambiguit\'e \`a l'aide de ces graphes de type r\'eseaux de spins, nous utilisons les conventions de \cite{varshalovich-book} pour l'orientation des ar\^etes et vertexes.

\begin{figure}
\begin{center}
\includegraphics[width=11cm]{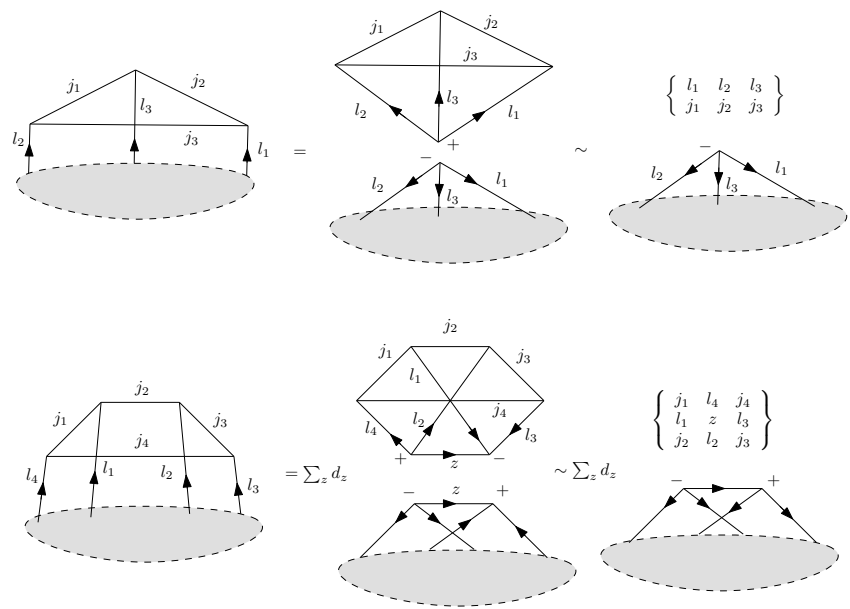}
\end{center}
\caption{ \label{reducible} La figure du haut montre un symbole r\'eductible : le cycle $(j_1 j_2 j_3)$ se factorise du fait qu'il n'existe qu'un unique entrelaceur entre $\calH_{l_1}\otimes \calH_{l_2}\otimes \calH_{l_3}$ et $\C$. Il est alors possible d'utiliser les relations de r\'ecurrence sur le 6j pour g\'en\'erer des translations des spins $(j_i)$. Pour un cycle de quatre ar\^etes, il faut sommer sur les entrelaceurs $z$ entre $\calH_{l_4}\otimes \calH_{l_2}\rightarrow \calH_{l_1}\otimes \calH_{l_3}$. On peut alors appliquer des relations de r\'ecurrence connues agissant sur les $(j_i)$, si les coefficients ne d\'ependent pas de $z$ mais tout au plus des spins s'attachant au cycle.}
\end{figure}

Un moyen direct pour obtenir des relations de r\'ecurrence est d'exploiter la d\'ecomposition des symboles $\{3nj\}$ en symboles plus petits. Cette d\'ecomposition s'effectue en isolant une partie du graphe contenant un cycle, comme sur la figure \ref{reducible}. Le cas le plus simple consiste en un cycle de trois ar\^etes, de sorte que le symbole se d\'ecompose alors en : $\{3nj\} = \{6j\} \{3(n-1)j\}$. Pour un cycle de quatre liens, il faut introduire une somme sur des spins interm\'ediaires, $\{3nj\} = \sum_i \{9j\}_i \{3(n-1)j\}_i$, et ainsi de suite.

Cette proc\'edure que nous allons d\'etailler fonctionne bien pour des 3-cycles et 4-cycles, mais devient plut\^ot lourde et maladroite pour les cycles plus grands. Nous nous int\'eresserons donc \`a une autre m\'ethode pour le cas particulier $n=5$, qui est le plus pertinent pour les mod\`eles de gravit\'e quantique en dimension 4. Cette m\'ethode fond\'ee sur une version r\'egularis\'ee du mouvement de Pachner 4-2 a de plus l'avantage de fournir une interpr\'etation g\'eom\'etrique simple. Nous verrons aussi que les relations de r\'ecurrence obtenues s'appliquent aussi bien \`a tout symbole ayant des cycles identiques \`a ceux des 15j \'etudi\'es.

\section{Relations entre les symboles 3nj}

La situation la plus simple consiste en un symbole $\{3nj\}$ contenant un cycle de trois ar\^etes, comme sur le haut de la figure \ref{reducible}. Le symbole est alors dit \emph{r\'eductible}, du fait qu'il s'\'ecrit comme le produit d'un plus petit symbole $\{3(n-1)j\}$  par un 6j. Puisque les spins du 6j se d\'ecouplent totalement du reste du symbole, nous pouvons assez trivialement utilis\'e les relations de r\'ecurrence du 6j pour en d\'eduire d'identiques sur le $\{3nj\}$. Par exemple,
\be
A_{+1}(j)\begin{Bmatrix} l_1 &l_2 &l_3 \\ j_1+1 &j_2 &j_3 \\ \hdotsfor[2]{3}\end{Bmatrix} +
A_0(j)\begin{Bmatrix} l_1 &l_2 &l_3 \\ j_1 &j_2 &j_3 \\ \hdotsfor[2]{3}\end{Bmatrix} +
A_{-1}(j)\begin{Bmatrix} l_1 &l_2 &l_3 \\ j_1-1 &j_2 &j_3 \\ \hdotsfor[2]{3}\end{Bmatrix} = 0.
\ee

Consid\'erons ensuite un symbole contenant un cycle de quatre ar\^etes, comme au bas de la figure \ref{reducible}. Nous pouvons toujours utilis\'e la th\'eorie du recouplage ({\it recoupling theory}, en anglais) pour \'ecrire un tel symbole en termes de plus petits, mais d\'esormais une somme sur les plus petits symboles appara\^it : le symbole initial n'est pas r\'eductible. Cela ne nous emp\^eche pas n\'ecessairement d'appliquer les relations de r\'ecurrence des plus petits symboles. Mais une attention particuli\`ere doit alors \^etre port\'ee pour \'eviter que ces relations n'impliquent des spins qui sont par ailleurs somm\'es dans la d\'ecomposition du symbole initial. Sur l'exemple de la figure, nous pouvons d\'eriver des relations pour le symbole de gauche \`a partir des relations sur le 9j \`a droite, qui n'impliquent pas le spin $z$ entrela\c{c}ant les repr\'esentations $\{l_i\}$. Typiquement,
\begin{multline}
\Bigl[\f{(j_2+l_2+j_3+1)(j_2+j_3-l_2)(-j_4+j_3+l_3)(j_4+l_3-j_3+1)}{(j_1+l_1+j_2+2)(j_1-l_1+j_2+1)}\Bigr]^{\f{1}{2}} \begin{Bmatrix} j_1+\f{1}{2} &l_4 &j_4+\f{1}{2} \\ l_1 &z &l_3 \\ j_2-\f{1}{2} &l_2 &j_3-\f{1}{2} \\ \hdotsfor[2]{3} \end{Bmatrix} \\
+ \Bigl[\f{(j_2-j_3+l_2)(-j_2+j_3+l_2+1)(j_4+j_3+l_3+2)(j_4+j_3-l_3+1)}{(j_1+l_1+j_2+2)(j_1-l_1+j_2+1)}\Bigr]^{\f{1}{2}}\begin{Bmatrix} j_1+\f{1}{2} &l_4 &j_4+\f{1}{2} \\ l_1 &z &l_3 \\ j_2-\f{1}{2} &l_2 &j_3+\f{1}{2} \\ \hdotsfor[2]{3} \end{Bmatrix} \\
+
\Bigl[\f{(j_2+l_2+j_3+2)(j_2+j_3-l_2+1)(j_4+j_3+l_3+2)(j_4+j_3-l_3+1)}{(j_1+l_1-j_2+1)(-j_1+l_1+j_2)}\Bigr]^{\f{1}{2}}\begin{Bmatrix} j_1+\f{1}{2} &l_4 &j_4+\f{1}{2} \\ l_1 &z &l_3 \\ j_2+\f{1}{2} &l_2 &j_3+\f{1}{2} \\ \hdotsfor[2]{3} \end{Bmatrix} \\
-
\Bigl[\f{(j_2-j_3+l_2+1)(-j_2+j_3+l_2)(-j_4+j_3+l_3)(j_4+l_3-j_3+1)}{(j_1+l_1-j_2+1)(-j_1+l_1+j_2)}\Bigr]^{\f{1}{2}}\begin{Bmatrix} j_1+\f{1}{2} &l_4 &j_4+\f{1}{2} \\ l_1 &z &l_3 \\ j_2+\f{1}{2} &l_2 &j_3-\f{1}{2} \\ \hdotsfor[2]{3} \end{Bmatrix} \\
= (2j_2+1)(2j_3+1)\Bigl[\f{(j_1+l_4+j_4+2)(j_1+l_4-j_4+1)}{(j_1+l_1-j_2+1)(-j_1+l_1+j_2)(j_1+l_1+j_2+2)(j_1-l_1+j_2+1)}\Bigr]^{\f{1}{2}}\begin{Bmatrix} j_1 &l_4 &j_4 \\ l_1 &z &l_3 \\ j_2 &l_2 &j_3 \\ \hdotsfor[2]{3} \end{Bmatrix}
\end{multline}
D'autres relations peuvent ainsi \^etre obtenues \`a partir de r\'ecurrences sur le 9j donn\'ees dans \cite{varshalovich-book}.

Notons que les coefficients de telles relations de r\'ecurrence ne d\'ependent que des spins du cycle et des spins sur les liens directement attach\'es au cycle. Cette propri\'et\'e tient aussi bien pour les cycles de 5 ou 6 liens. N\'eanmoins, la proc\'edure devient alors plus lourde et n\'ecessite de conna\^itre \`a l'avance des relations sur des symboles $12j$ ou plus. Plus bas, nous d\'erivons une alternative plus fructueuse pour les cycles plus grands.

Mentionnons finalement une autre m\'ethode utilis\'ee par Jang \cite{jang}. Un symbole auxiliaire $\{3(n+1)j\}$ est exprim\'e en terme d'une somme de produits d'un symbole $\{3nj\}$ et typiquement de symboles 6j. Puis, en utilisant des sym\'etries sp\'ecifiques du $\{3(n+1)j\}$, il est possible d'obtenir des relations non-triviales entre les symboles $\{3nj\}$, avec des 6j comme coefficients. Cette m\'ethode fonctionne bien pour les symboles dit du premier type ({\it of the first kind}, en anglais, dans la terminologie de \cite{yutsis-book}), mais il n'est pas clair qu'elle soit efficace pour d'autres types de symboles ayant des sym\'etries moins fortes. De plus, cela ne fournit pas d'interpr\'etation g\'eom\'etrique, en termes de mouvements \'el\'ementaires sur des spimplexes, par contraste avec la m\'ethode que nous allons pr\'esenter.

\section{Par le mouvement de Pachner 4-2} \label{sec:move4-2}

Fixons maintenant $n=5$. Le symbole de Wigner correspondant, le 15j, n'est en fait pas unique, \`a la diff\'erence des 6j et 9j. Suivant la terminologie de \cite{yutsis-book}, on peut d\'eterminer cinq types de symboles irr\'eductibles. Selon leur type, diff\'erents cycles seront pr\'esents, et les m\'ethodes d\'evelopp\'ees plus haut s'appliquent \'evidemment. Toutes les ar\^etes des symboles 15j du premier et du second types peuvent \^etre prises en compte dans des cycles de quatre liens, tandis que des cycles de cinq liens sont n\'ecessaires pour toutes les ar\^etes du 15j du cinqui\`eme type. Les symboles du troisi\`eme et quatri\`eme types n\'ecessitent des cycles de cinq liens seulement pour certaines ar\^etes (et de quatre liens pour les autres).

Tout comme le 6j en tant qu'\'etat physique sur le bord d'un t\'etra\`edre, le 15j admet une interp\'etation physique, mais en dimension 4. Consid\'erons en effet le bord d'un 4-simplexe. Le graphe dual accueille un espace de Hilbert, engendr\'e par les r\'eseaux de spins form\'es sur les 5 vertexes et 10 liens duaux. Comme en gravit\'e 3d, le seul \'etat physique sur cette triangulation consiste en des fonctions delta sur les cycles ind\'ependants du graphe dual, ou sur les faces \emph{ind\'ependantes} duales aux ar\^etes de la triangulation, qui imposent aux holonomies le long de ces cycles d'\^etre triviales. Ainsi pour \'eviter les redondances, l'\'etat physique s'\'ecrit en choisissant 6 faces ind\'ependantes parmi les 10. Il est tel qu'\`a des transformations de jauge pr\`es, il impose \`a chacune des 10 holonomies le long des liens duaux d'\^etre triviale -- ce qui est fondamentalement d\^u au fait que le groupe fondamental $\pi_1$ de cette triangulation est trivial. Cet \'etat non-normalisable se d\'eveloppe sur la base des r\'eseaux de spins. En 4d, une telle base s'obtient en choisissant un appariement \`a chaque vertex dual entre les quatre spins $j_e$ qui s'y rencontrent, et en y associant un entrelaceur not\'e par exemple $\lvert (j_1j_2),(j_3j_4);i\rangle$ dont le spin interm\'ediaire $i$ doit courir sur toutes les valeurs admissibles. Alors un calcul similaire au cas 3d conduit aux coefficients de cet \'etat dans la base des r\'eseaux de spins qui se r\'ev\`elent \^etre des symboles 15j !
\beq
\psi_{\operatorname{phys}}(j_1,\dotsc,j_{10};i_1,\dotsc,i_5) = \{15j_e\},
\ee
du fait que l'\'evaluation d'un r\'eseau de spins sur l'identit\'e pour chaque ar\^ete est exactement un symbole 15j. Bien s\^ur la forme pr\'ecise de ces 15j d\'epend des appariements choisis pour chacun des cinq entrelaceurs.

Tout comme le 6j dans le mod\`ele de Ponzano-Regge, le symbole 15j est la brique de construction du mod\`ele d'Ooguri, mod\`ele de type BF en dimension 4 pour le groupe $\SU(2)$, souvent pris comme point de d\'epart pour des mod\`eles de gravit\'e quantique comme nous le verrons. A ce titre ce mod\`ele est construit comme une fonction de partition pour un syst\`eme de connexions discr\`etes plates, et la somme sur les mousses de spins en est une r\'e\'ecriture en modes de Fourier $\SU(2)$. Les int\'egrales sur les connexions discr\`etes se font \`a l'aide de la formule que nous rappelons,
\be \label{intg4}
\int_{\SU(2)}dg_t\ \bigotimes_{f=1}^4 D^{(j_f)}(g_t) = \id_{\Inv}.
\ee
Sur une vari\'et\'e triangul\'ee, $g_t$ repr\'esente le transport parall\`ele entre deux 4-simplexes \`a travers un t\'etra\`edre. Les quatre repr\'esentations $j_f$ sont ainsi attach\'ees une \`a une aux quatre triangles de ce t\'etra\`edre. Ces triangles apparaissent bien s\^ur dans d'autres t\'etra\`edres avec les m\^emes repr\'esentations. L'espace invariant sur le t\'etra\`edre $t$ est le sous-espace $\SU(2)$-invariant du produit tensoriel des quatre repr\'esentations se rencontrant en $t$. Une base orthonormale de tels entrelaceurs quadrivalents est obtenue par d\'eveloppement selon un arbre, i.e. un d\'eveloppement en vertexes trivalents. Il faut pour cela choisir un des trois appariements possibles entre les quatre repr\'esentations $j_f$ et sommer sur une repr\'esentation interm\'ediaire. L'identit\'e est alors donn\'ee par :
\be \label{pairing}
\id_{\Inv} = \sum_i\ \lvert (j_1j_2),(j_3j_4);i\rangle\ \langle (j_1j_2),(j_3j_4);i\rvert
\ee
Chaque t\'etra\`edre re\c{c}oit donc un spin entrelaceur $i_t$. La contraction de tels entrelaceurs le long des triangles donnent naissance \`a des symboles 15j comme amplitude de chaque 4-simplexe. Le lecteur attentif aura not\'e que la structure de chaque 15j d\'ependra du choix des appariements dans \eqref{pairing} pour chacun des cinq entrelaceurs d'un 4-simplexe. Il est n\'eanmoins possible d'\'ecrire le mod\`ele en n'utilisant qu'un seul type de 15j, mais cela requiert g\'en\'eralement des choix d'appariements diff\'erents pour les deux entrelaceurs entrant dans \eqref{pairing}. L'invariance de jauge est alors garantie par l'apparition de symboles 6j appropri\'es \`a chaque t\'etra\`edre pour recoupler correctement les deux appariements diff\'erents.

Une question naturelle \`a laquelle nous n'avons pas trouv\'e de r\'eponses dans la litt\'erature est de savoir si l'exploration de tous les choix d'appariements permet de construire les cinq types diff\'erents de 15j. Il est assez facile de voir qu'elle engendre bien les 15j des premier, troisi\`eme, quatri\`eme et cinqui\`eme types, ainsi que des symboles r\'eductibles. Mais nous n'avons pas \'et\'e en mesure de reproduire le 15j du second type de cette mani\`ere, sans preuve toutefois qu'il ne peut appara\^itre.

La sp\'ecificit\'e du mod\`ele d'Ooguri est naturellement son invariance topologique, i.e. qu'il ne d\'epend pas explicitement de la triangulation choisie, mais seulement de la topologie de la vari\'et\'e. Les preuves les plus simples \cite{ooguri4d, perez-diagrammatic} utilisent les mouvements de Pachner qui relient des triangulations de deux vari\'et\'es hom\'eomorphes. En dimension 4, il existe trois mouvements ind\'ependants, discut\'es en d\'etails dans \cite{kauffman-4dmodel} : 3-3, 4-2 et 5-1. Les mouvements 4-2 et 5-1 engendrent en fait des divergences dans le mod\`ele d'Ooguri, ce qui rend l'invariance topologique un peu subtile. Mais comme le mouvement 4-1 dans le mod\`ele de Ponzano-Regge, ces divergences proviennent simplement de fonctions delta redondantes dans la d\'efinition du mod\`ele en termes de connexions discr\`etes plates.

Puisque le 15j donne naissance \`a un mod\`ele topologique, il semble naturel qu'il existe des relations sur le 15j codant cette invariance topologique ou du moins en d\'ecoulant, tout comme pour l'identit\'e de Biedenharn-Elliott et les relations de r\'ecurrence sur le 6j. Nous allons montrer que c'est bien le cas : des relations de r\'ecurrence s'obtiennent \`a partir du mouvement de Pachner 4-2 (r\'egularis\'e). Comme ce mouvement de Pachner est une composante de l'invariance du mod\`ele sous les hom\'eomorphismes, ces relations de r\'ecurrence peuvent \^etre vues comme des \'equations aux diff\'erences contribuant \`a l'impl\'ementation de cette sym\'etrie, ou de mani\`ere \'equivalente : comme une version discr\`ete des contraintes classiques de la th\'eorie BF sur le bord d'un 4-simplex dont la solution est l'unique \'etat physique $\psi_{\operatorname{phys}}(j_1,\dotsc,j_{10};i_1,\dotsc,i_5)$.

\subsection{R\'egularisation du mouvement 4-2}

\begin{figure}
\begin{center}
\includegraphics[width=5cm]{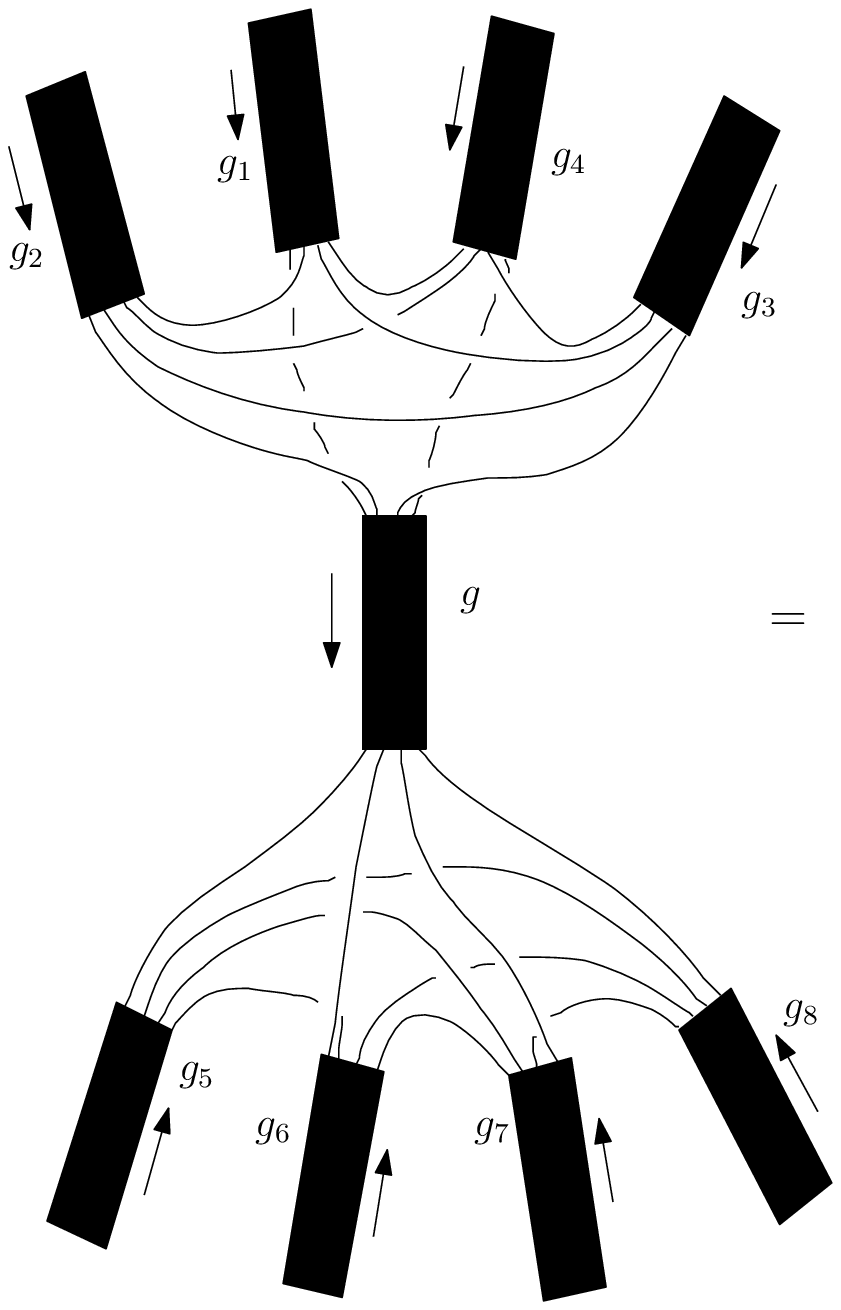}\qquad\qquad\includegraphics[width=8cm]{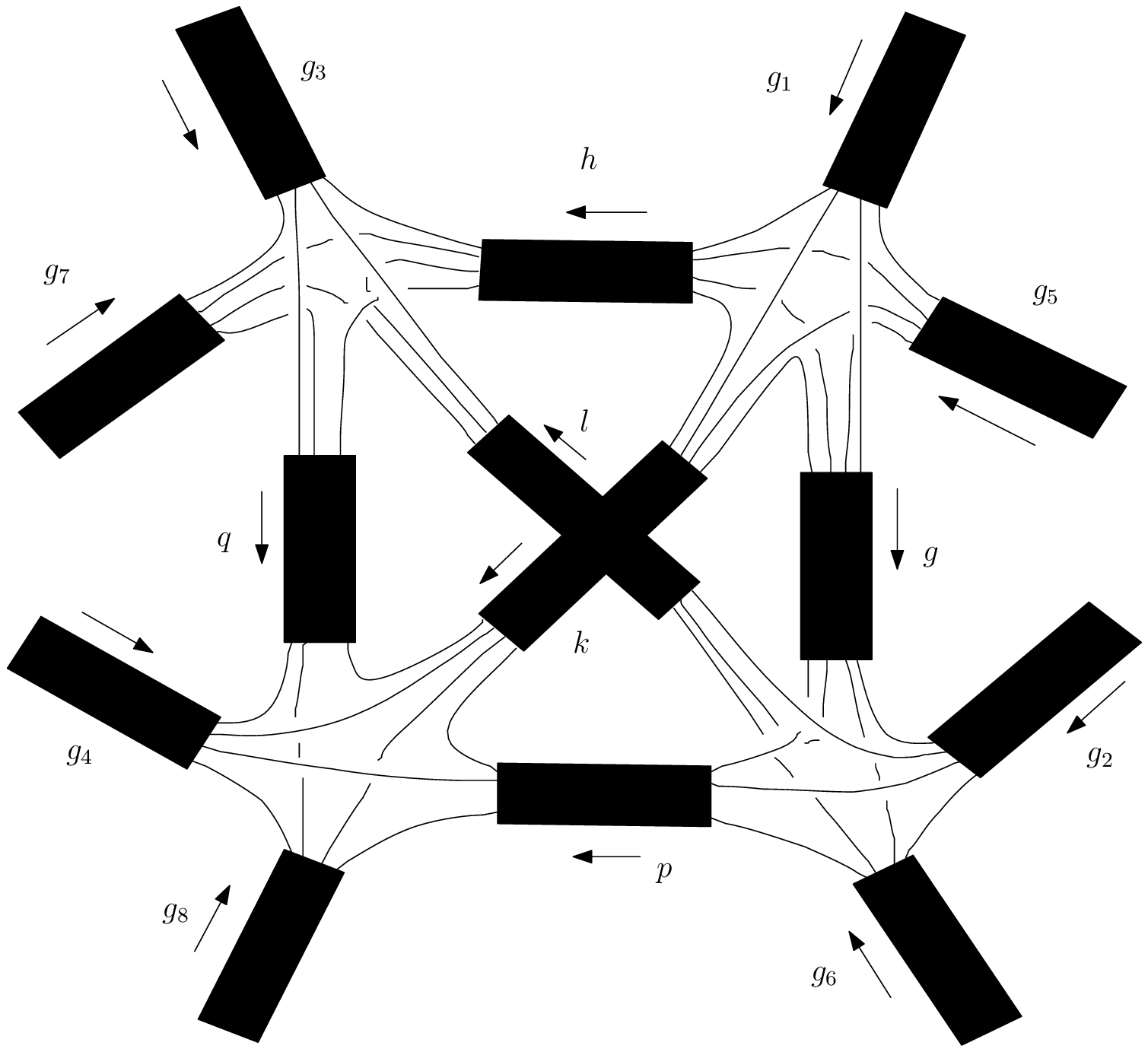}
\end{center}
\caption{ \label{move4-2gft} Ces figures d\'ecrivent le mouvement 2-4 du point de vue du 2-squelette du complexe cellulaire dual \`a la triangulation. Les t\'etra\`edres du bord sont les 8 bo\^ites externes. A gauche il y a deux 4-simplexes coll\'es par un t\'etra\`edre (dual \`a la bo\^ite du milieu), tandis que le graphe de droite donne la structure du collage des quatre 4-simplexes reli\'es deux \`a deux par des t\'etra\`edres tous distincts. Les fl\`eches donnent l'orientation des liens duaux, et les bo\^ites signifient aussi qu'il faut int\'egrer sur les \'el\'ements de groupe port\'es par ces liens duaux. Une fois que le membre de droite est correctement r\'egularis\'e, les deux configurations produisent la m\^eme amplitude BF. Pour effectuer cette r\'egularisation n\'ecessaire, nous avons enlev\'e la condition de courbure nulle pour la face interne passant par $l, q, p$. Par cons\'equent, puisqu'il n'y plus que trois liens \`a passer \`a travers $l, q$, et $p$, l'amplitude de trois 4-simplexes va se r\'eduire \`a des symboles 12j, et nous allons obtenir un seul symbole 15j dans ce membre de droite.}
\end{figure}

Il s'agit donc d'un mouvement reliant l'amplitude associ\'ee \`a deux 4-simplexes coll\'es le long d'un t\'etra\`edre \`a celle de quatre 4-simplexes \'egalement reli\'es deux \`a deux par des t\'etra\`edres distincts, comme sur la figure \ref{move4-2}. Puisque le mod\`ele est d\'efini intrins\`equement sans choix d'appariements, mais plut\^ot en termes d'int\'egrations sur $\SU(2)$, nous regardons le mouvement en utilisant cette repr\'esentation des amplitudes, qui se dessine plus naturellement sur les 2-complexes duaux aux triangulations concern\'ees, suivant la figure \ref{move4-2gft}. Les bo\^ites noires repr\'esentent \`a la fois les liens duaux aux t\'etra\`edres et les int\'egrales sur les \'el\'ements de groupe correspondants $g_t$, tandis que les lignes forment les bords des faces duales aux triangles. Chaque bo\^ite est travers\'ee par quatre lignes, correspondant aux quatre triangles de chaque t\'etra\`edre. Dans cette vision, un 4-simplexe est form\'e de cinq bo\^ites noires reli\'ees deux \`a deux. Un triangle interne se rep\`ere par une plaquette duale enti\`ere, i.e. dont le bord est une ligne ferm\'e, alors que les lignes ouvertes correspondent \`a des triangles du bord.

Dans les deux configurations du mouvement, le bord est constitu\'e de huit t\'etra\`edres portant les \'el\'ements de groupe $(g_i)_{i=1,\dots,8}$. Dans la configuration de gauche (sur la figure \ref{move4-2gft}), le t\'etra\`edre interne, partag\'e par les deux 4-simplexes, porte un \'el\'ement not\'e $g$. En suivant les orientations donn\'ees sur la figure, l'amplitude de gauche se calcule par :
\begin{multline} \label{move4-2lhs}
\int \prod_{t=1}^8 dg_t\,D^{(j_{12})}(g_1 g_2\mone)\,D^{(j_{13})}(g_1 g_3\mone)\,D^{(j_{14})}(g_1 g_4\mone)\,D^{(j_{23})}(g_2 g_3\mone)\,D^{(j_{23})}(g_2 g_3\mone)\,D^{(j_{24})}(g_2 g_4\mone)\\
D^{(j_{34})}(g_3 g_4\mone)\,D^{(j_{56})}(g_5 g_6\mone)\,D^{(j_{57})}(g_5 g_7\mone)\,D^{(j_{58})}(g_5 g_8\mone)\,D^{(j_{67})}(g_6 g_7\mone)\,D^{(j_{68})}(g_6 g_8\mone)\,D^{(j_{78})}(g_7 g_8\mone)\\
\int dg\,D^{(j_{15})}(g_1\,g\, g_5\mone)\,D^{(j_{26})}(g_2\,g\, g_6\mone)\,D^{(j_{37})}(g_3\,g\, g_7\mone)\,D^{(j_{48})}(g_4\,g\, g_8\mone).
\end{multline}
Des produits tensoriels sont entendus entre les matrices $D^{(j)}$. Apr\`es int\'egration sur les huit \'el\'ements $g_t$ des t\'etra\`edres du bord, les indices libres des entrelaceurs, qui correspondent aux extr\'emit\'es libres des lignes ouvertes, peuvent \^etre contract\'es avec huit entrelaceurs choisis et fix\'es. Cela implique notamment le choix d'appariements pour d\'efinir les spins interm\'ediaires/virtuels. L'int\'egration sur $g$ produit la somme \eqref{pairing}, et cela conduit pr\'ecis\'ement \`a la partie gauche de la figure \ref{move4-2graph}.

Sur la partie droite du mouvement, six \'el\'ements de groupe, correspondant aux six t\'etra\`edres internes, doivent \^etre int\'egr\'es. Les lignes ouvertes correspondent aux triangles du bord, et les lignes ferm\'ees aux triangles internes, qui sont partag\'es par exactement trois t\'etra\`edres (et trois 4-simplexes). Chaque triangle interne contribue \`a l'amplitude par un facteur $\delta(\prod_{t \supset f} g_t )$. Apr\`es inspection, il s'av\`ere qu'une de ces fonctions delta est redondante, causant une divergence du type $\delta(\mathbbm{1})$, exactement comme pour le mouvement 4-1 dans le mod\`ele de Ponzano-Regge. Cette divergence peut \^etre \'evit\'ee en supprimant brutalement une des quatre conditions de courbure nulle, disons  $\delta(lqp\mone)$, puisque la contrainte $lqp\mone=\mathbbm{1}$ est garantie par les autres fonctions delta. L'amplitude qui r\'esulte de ce choix est \emph{finie} et se lit
\begin{multline}
\int \prod_{t=1}^8 dg_t\,D^{(j_{15})}(g_1 g_5\mone)\,D^{(j_{26})}(g_2 g_6\mone)\,D^{(j_{37})}(g_3g_7\mone)\,D^{(j_{48})}(g_4 g_8\mone)\ \int dg\,dh\,dk\,dl\,dp\,dq\\
\delta\bigl(g\,l\,h\mone\bigr)\,\delta\bigl(g\,p\,k\mone\bigr)\,\delta\bigl(h\,q\,k\mone\bigr)\ D^{(j_{12})}(g_1\,g\, g_2\mone)\,D^{(j_{56})}(g_5\,g\, g_6\mone)\ D^{(j_{57})}(g_5\,h\, g_7\mone)\,D^{(j_{13})}(g_1\,h\, g_3\mone)\\
D^{(j_{58})}(g_5\,k\, g_8\mone)\,D^{(j_{14})}(g_1\,k\, g_4\mone)\,D^{(j_{23})}(g_2\,l\, g_3\mone)\,D^{(j_{67})}(g_6\,l\, g_7\mone)\\
D^{(j_{24})}(g_2\,p\, g_4\mone)\,D^{(j_{68})}(g_6\,p\, g_8\mone)\,D^{(j_{34})}(g_3\,q\, g_4\mone)\,D^{(j_{78})}(g_7\,q\, g_8\mone).
\end{multline}
Les int\'egrales sur l'int\'erieur se font de fa\c{c}on usuelle, en d\'eveloppant les fonctions delta selon la formule de Plancherel et en usant de mani\`ere r\'ep\'et\'ee de \eqref{intg4}. A nouveau, cette \'etape n\'ecessite un choix d'appariements. Notons que le r\'esultat ne fait appara\^itre qu'un seul symbole 15j : les \'el\'ements de groupe $l$, $q$ et $p$ n'interviennent plus qu'en trois endroits du fait de la fonction delta \'elimin\'ee. Ainsi les amplitudes des trois 4-simplexes partageant le triangle correspondant se r\'eduisent \`a de plus petits symboles 12j.

Les deux expressions coincident apr\`es r\'egularisation, puisque nous n'avons fait que supprimer une fonction delta redondante. Cela peut aussi de voir explicitement \`a l'aide de la m\'ethode graphique de \cite{perez-diagrammatic}. L'id\'ee est la suivante. Observons d'abord que $g$, sur la partie gauche, peut \^etre r\'eabsorb\'e \`a droite des \'el\'ements $g_1$, $g_2$, $g_3$ et $g_4$, en usant de la propri\'et\'e d'invariance par translation des mesures de Haar. D\'ecrivons bri\`evement comment la partie droite se r\'eduit \`a la m\^eme amplitude. La m\^eme m\'ethode, dite de fixation de jauge, permet d'\'eliminer par exemple $h$ et $q$, for\c{c}ant ainsi $k$ \`a \^etre l'identit\'e. $g$ peut aussi \^etre r\'eabsorb\'e, de sorte que $l=\mathbbm{1}$. Finalement, l'int\'egrale sur $p$ est triviale car les simplifications pr\'ec\'edentes ont g\'en\'er\'e une fonction $\delta(p)$.

Comme pr\'eliminaires \`a une m\'ethode que nous introduirons plus bas en lien avec le mouvement de Pachner 4-1 en gravit\'e 3d, imaginons qu'au lieu de supprimer la fonction $\delta(lqp\mone)$, nous gardions la composante, disons de spin $J$, de son d\'eveloppement de Fourier, $\chi_J(lqp\mone)$, dans l'int\'egrant. Apr\`es avoir effectu\'e les int\'egrales et choisi des appariements, la configuration de droite est maintenant constitu\'ee de quatre symboles 15j, avec une d\'ependance en $J$. Cela se traduit dans l'amplitude de gauche de la fa\c{c}on suivante. Puisque les autres fonctions delta (du membre de droite) forcent l'identit\'e $lqp\mone=\mathbbm{1}$, le caract\`ere $\chi_J(lqp\mone)$ vaut pr\'ecis\'ement et simplement $(2J+1)$. Cette quantit\'e peut \'evidemment se factoriser. Ainsi l'amplitude de gauche se trouve tout juste multipli\'ee par un facteur $(2J+1)$ ! Au lieu d'\'ecrire explicitement de telles formules pour le mouvement 4-2, nous \'etudierons cela en d\'etails en gravit\'e 3d, ce qui donnera la formule analogue \eqref{rechol1}.

Nous montrons maintenant comment ce mouvement r\'egularis\'e permet de d\'eriver des relations de r\'ecurrence connues mais aussi de nouvelles.

\subsection{Choix des donn\'ees de bord}

\begin{figure}
\begin{center}
\includegraphics[width=10cm]{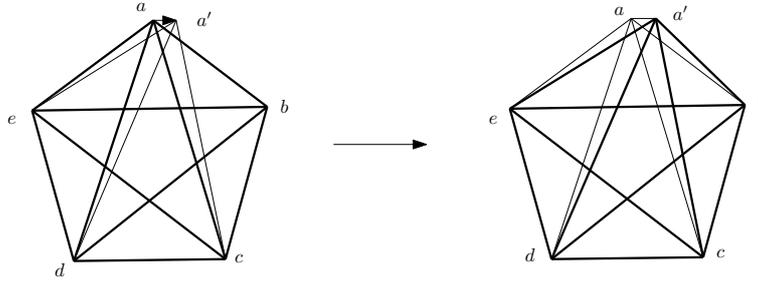}
\end{center}
\caption{ \label{move4-2} Les deux configurations du mouvement 2-4. A gauche, deux 4-simplexes, $(abcde)$ et $(aa'cde)$, sont coll\'es le long du t\'etra\`edre $(acde)$. A droite, nous avons quatre 4-simplexes, $(a'bcde)$, $(a'bacd)$, $(a'bade)$ et $(a'bace)$. Cette configuration a quatre triangles internes, $(a'bc)$, $(a'bd)$, $(a'be)$, $(aa'b)$, chacun partag\'e par trois 4-simplexes, et six t\'etra\`edres internes form\'es par les points $a', b$ et tout choix de deux autres sommets. Quand les spins des triangles $(aa'e)$ et $(aa'c)$ sont pris \`a z\'ero, nous pouvons consid\'erer $a'$ comme tr\`es proche de $a$. Ainsi, les aires du 4-simplexe initial $(abcde)$ de gauche et celles de $(a'bcde)$ \`a droite ne diff\`erent que par de petites variations pour les trois triangles partageant l'ar\^ete $(ad)$.}
\end{figure}

En suivant la d\'erivation de la relation \eqref{rec6j} \`a partir de l'identit\'e de Biedenharn-Elliott, nous regardons le mouvement 4-2 r\'egularis\'e avec certains spins fix\'es de mani\`ere sp\'ecifique sur le bord. Le choix doit \^etre fait correctement, afin de toujours garder dans le membre de droite un symbole 15j. La situation est illustr\'ee figure \ref{move4-2}. Nous consid\'erons un 4-simplexe de sommets $a,b,c,d,e$, et le long du t\'etra\`edre $(acde)$ nous collons un autre 4-simplexe, $(aa'cde)$. Cela forme la partie gauche du mouvement. Chaque triangle est colori\'e par un spin de $\SU(2)$ fix\'e, et chaque t\'etra\`edre sauf $(acde)$ par un entrelaceur fix\'e : ce sont les donn\'ees du bord. Le t\'etra\`edre $(acde)$, partag\'e par les deux 4-simplexes, porte une somme sur les entrelaceurs (due \`a l'int\'egrale sur $g$ dans \eqref{move4-2lhs}). La partie droite du mouvement est form\'ee des quatre 4-simplexes respectivement obtenus en enlevant les points $a,c,d$ et $e$.

Pour \'ecrire les indices des spins et entrelaceurs, nous allons utiliser la notation duale standard : les indices correspondant \`a un simplexe sont les points qui ne sont \emph{pas} des sommets de ce simplexe. Typiquement, la repr\'esentation du triangle $(cde)$ est $j_{aa'b}$ et celle du t\'etra\`edre $(bcde)$ est $i_{aa'}$. Pour r\'egulariser le mouvement de Pachner, nous avons besoin de supprimer une fonction delta sur une des faces internes dans le membre de droite ; nous choisissons celle du triangle $(aa'b)$. Ainsi seule l'amplitude du 4-simplexe qui ne partage pas ce triangle, i.e. $(a'bcde)$, est effectivement un 15j, tandis que les trois autres 4-simplexes deviennent des symboles 12j, comme le montre la figure \ref{move4-2graph}.

Nous fixons maintenant deux spins du bord \`a z\'ero : $j_{bcd}=j_{bde}=0$. Les liens correspondant sur les graphes des r\'eseaux de spins sont repr\'esent\'es avec des points et des tirets sur la figure \ref{move4-2graph}. Ce choix force l'\'egalit\'e $j_{abd}=j_{a'bd}$ sur les spins au bord. En cons\'equence, le 15j sur $(aa'cde)$ dans le membre de gauche se r\'eduit \`a un symbole 9j, \'eventuellement r\'eductible. Il s'av\`ere que pour tous les choix d'appariements pour l'entrelaceur du bord $i_{ab}$, il est possible de d\'evelopper l'entrelaceur interne (et donc somm\'e) $i_{a'b}$ de sorte que ce 9j se r\'eduise encore \`a un simple 6j ou au moins \`a un produit de deux symboles 6j. De cette fa\c{c}on, le membre de gauche de notre mouvement contient un seul 15j et un ou deux 6j.

Une simplification similaire se produit dans le membre de droite. En particulier, parmi les spins qui sont somm\'es, deux deviennent fix\'es : $j_{acd}=j_{a'cd}$ et $j_{ade}=j_{a'de}$. Les trois symboles 12j se r\'eduisent \'egalement \`a des produits de 6j, dont la forme pr\'ecise d\'epend du choix des appariements pour les entrelaceurs au bord.

Nous sommes alors en position d'\'ecrire des \'equations impliquant dans chaque membre un 15j, et avec des 6j pour coefficients. Il est aussi facile de voir que tous ces 6j d\'ependent du spin $\lambda\equiv j_{bce}$. Ce spin param\`etre \'egalement les domaines sur lesquels les autres spins sont somm\'es. Typiquement, on voit sur la figure \ref{move4-2graph} que $\lvert j_{a'ce}-\lambda\rvert\leq j_{ace}\leq j_{a'ce}+\lambda$. Ainsi, tout comme des relations de r\'ecurrence pour le 6j sont obtenues \`a partir de l'\'equation de Biedenharn-Elliott en sp\'ecialisant un des spins du bord \`a la valeur 1 (ou $1/2$ pour avoir des diff\'erences de spins $\pm1/2$), nous pouvons donner de petites valeurs d\'etermin\'ees \`a $\lambda$ et obtenir de la sorte des relations de r\'ecurrence sur des 15j. Notons que les conditions $j_{bcd}=j_{bde}=0$ induisent des relations entre les spins au bord qui d\'ependent aussi de $\lambda$, \`a savoir : $j_{abc}=j_{a'bc}+\alpha$ et $j_{abe}=j_{a'be}+\beta$, pour $\alpha$ et $\beta$ tels que : $-\lambda\leq \alpha,\beta\leq\lambda$.

\subsection{Interpr\'etation g\'eom\'etrique} \label{sec:interpretation-move4-2}

Avant de donner les formules g\'en\'erales, il serait bon de donner une interpr\'etation g\'eom\'etrique de ce que nous avons fait. Une interpr\'etation naturelle des spins sur les triangles et des spins entrelaceurs aux t\'etra\`edres provient de la quantification g\'eom\'etrique d'un t\'etra\`edre (isol\'e) et de la similarit\'e entre la th\'eorie BF $\SU(2)$ et la gravit\'e quantique \`a boucles au niveau \emph{cin\'ematique}, voir section \ref{sec:operatorsLQG} et plus sp\'ecifiquement \ref{sec:quantum tet}. Cela nous invite \`a voir les spins $j$ comme \'etant reli\'es aux aires des triangles, et les spins entrelaceurs comme les aires de parall\'elogrammes coupant en deux chaque t\'etra\`edre (les trois appariements possibles correspondent \`a autant de fa\c{c}on de proc\'eder \`a cette coupe d'un t\'etra\`edre) \cite{baez-bf-spinfoam}.

Alors, le choix $j_{bcd}=j_{bde}=0$ signifie que les triangles $(aa'e)$ et $(aa'c)$ sont \og tr\`es petits\fg{}, de sorte que le 4-simplexe $(aa'cde)$ se trouve \og aplati\fg. Du point de vue du 4-simplexe $(abcde)$, le collage du deuxi\`eme 4-simplexe, aplati, peut ainsi \^etre consid\'er\'e comme un l\'eger d\'eplacement du point $a$ en $a'$, comme la figure \ref{move4-2} l'illustre. D'une part, les relations $j_{acd}=j_{a'cd}$ et $j_{ade}=j_{a'de}$ signifient que les triangles $(a'be)$ et $(a'bc)$ ont la m\^eme aire que les triangles initiaux $(abe)$ et $(abc)$. D'autre part, les aires des triangles partageant l'ar\^ete $(ad)$ sont modif\'ees lorsque le sommet $a$ bouge en $a'$ : comme nous l'avons dit, $j_{abc}=j_{a'bc}+\alpha$ et $j_{abe}=j_{a'be}+\beta$, tandis que $j_{a'ce}$ est \'egalement chang\'e, d'une quantit\'e qui d\'epend de $\lambda$. Faire varier $\lambda$ s'interpr\`ete comme un d\'eplacement de $a'$, de sorte que l'aire du triangle $(aa'd)$ change, en m\^eme temps que les aires de $(aa'c)$ et $(aa'e)$ restent fixes et \og petites\fg.

\subsection{Les formules pour diff\'erents cycles} \label{sec:formule-move4-2}

Le fait que les spins entrelaceurs changent ou non lorsque $a$ bouge en $a'$ (par exemple, si $i_{ab}=i_{a'b}$ ou non) d\'epend des appariements choisis. Plus pr\'ecis\'ement, regardons le d\'eveloppement en arbre d'un entrelaceur du point de vue du graphe du r\'eseau de spins. Si deux spins modifi\'es se rencontrent en un vertex commun avec un spin entrelaceur, alors ce spin n'est pas touch\'e. Mais tout lien ins\'er\'e entre deux liens dont les spins sont modifi\'es verra aussi son spin \^etre touch\'e (a priori). De cette fa\c{c}on, nous obtenons des relations de r\'ecurrence agissant sur des cycles qui peuvent avoir un nombre de liens diff\'erents, en changeant simplement les choix d'appariements des entrelaceurs.

\begin{figure}
\begin{center}
\includegraphics[width=17cm]{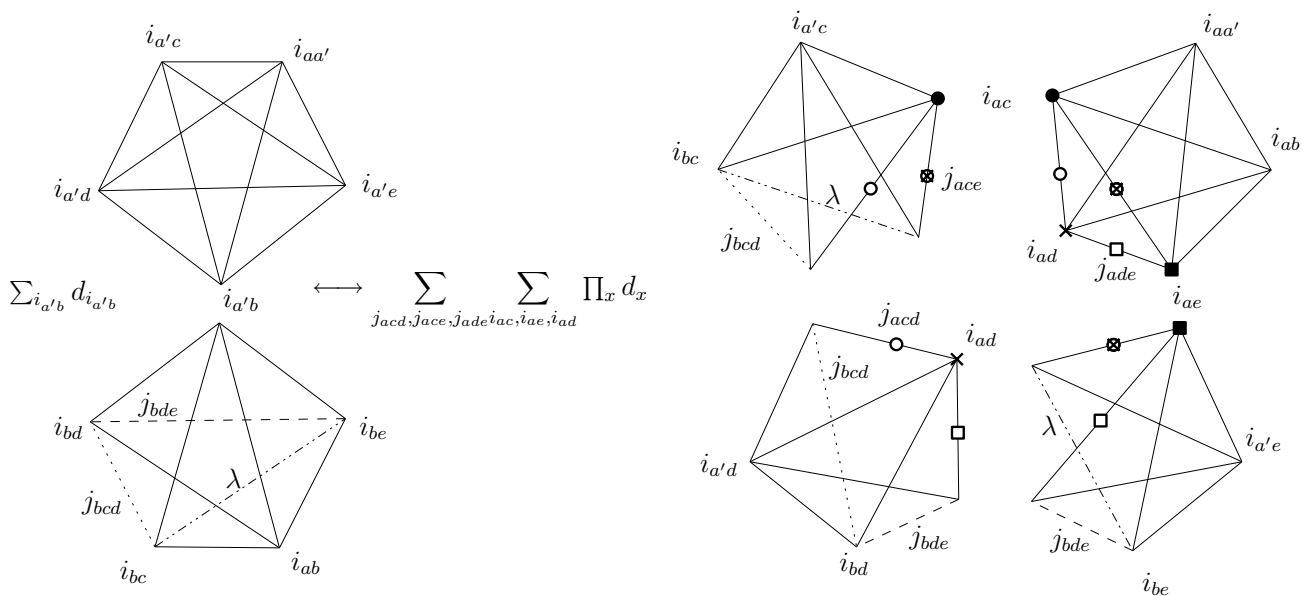}
\end{center}
\caption{ \label{move4-2graph} Le mouvement de Pachner 2-4 r\'egularis\'e, exprim\'e dans le langage des mousses de spins. Seul un des graphes du membre de droite est un symbole 15j du fait que la condition de courbure nulle a \'et\'e supprim\'ee pour le triangle $(aa'b)$. La notation $x$ repr\'esente l'ensemble des spins qui sont somm\'es, et les liens et vertex (dont les entrelaceurs n'ont pas \'et\'e explicitement d\'evelopp\'es) correspondant sont marqu\'es. Par exemple, les trois liens portant un rond vide sont colori\'es par le spin $j_{acd}$. Pour obtenir des formules n'impliquant qu'un seul 15j dans chaque membre du mouvement, nous avons pris $j_{bcd}=j_{bde}=0$, repr\'esent\'es par des lignes en tirets et pointill\'es, de sorte que plusieurs graphes se r\'eduisent \`a de simples produits de symboles 6j. Ensuite, des relations de r\'ecurrence explicites sont obtenues en regardant de petites valeurs du param\`etre $\lambda\equiv j_{bce}$.}
\end{figure}

Les relations que nous obtenons pour des cycles form\'es de trois ou quatre liens sont pr\'ecis\'ement celles sur le 6j et sur le 9j, sans v\'eritable surprise. Pr\'esentons maintenant de nouvelles relations adapt\'ees aux cycles de cinq ou six liens. Il y a deux fa\c{c}ons d'ins\'erer un lien avec un spin entrelaceur entre deux autres liens pour obtenir un 5-cycle. Nous pouvons en effet ins\'erer ou non $i_{a'b}$ entre $j_{a'bc}$ et $j_{a'be}$ dans le membre de gauche. Puis, lorsque $i_{a'b}$ est ins\'er\'e, il est \'equivalent d'ins\'erer $i_{a'e}$ ou $i_{a'c}$. Nous avons donc deux relations diff\'erentes :
\begin{multline} \label{5-cycle 1}
(-1)^{R}\sum_{i_{a'b}=\lvert i_{ab}-\lambda\rvert}^{i_{ab}+\lambda} d_{i_{a'b}} \begin{Bmatrix} \lambda &i_{a'b} &i_{ab} \\ j_{a'bd} &j_{a'bc}+\alpha &j_{a'bc} \end{Bmatrix} \begin{Bmatrix} \lambda &i_{a'b} &i_{ab} \\ j_{aa'b} &j_{a'be}+\beta &j_{a'be} \end{Bmatrix} \\
\times \begin{Bmatrix} i_{a'b} &{} &i_{a'e} &{} &j_{a'bc} \\ {} &j_{a'be} &{} &j_{a'ce} &{} \\ \hdotsfor[2]{5}\end{Bmatrix} \\
= \sum_{j_{ace},i_{ae}} d_{i_{ae}}d_{j_{ace}} \begin{Bmatrix} i_{a'c} &j_{a'bc} &j_{a'ce} \\ \lambda &j_{ace} &j_{a'bc}+\alpha \end{Bmatrix} \begin{Bmatrix} \lambda &i_{a'e} &i_{ae} \\ j_{a'de} &j_{abe}+\beta &j_{a'be} \end{Bmatrix} \begin{Bmatrix} \lambda &i_{a'e} &i_{ae} \\ j_{aa'e} &j_{ace} &j_{a'ce} \end{Bmatrix}\\
\times \begin{Bmatrix} i_{ab} &{} &i_{ae} &{} &j_{a'bc}+\alpha \\ {} &j_{a'be}+\beta &{} &j_{ace} &{} \\ \hdotsfor[2]{5}\end{Bmatrix},
\end{multline}
o\`u $\alpha$ et $\beta$ sont des demi-entiers entre $-\lambda$ et $\lambda$, $d_j\equiv 2j+1$ et $R=j_{a'de}+j_{a'bd}-j_{aa'e}-j_{aa'b}-i_{ac}+2j_{a'bc}+\alpha+\lambda$,
\begin{multline} \label{5-cycle 2}
\begin{Bmatrix} i_{ab} &j_{a'bc} &j_{a'be} \\ \lambda &j_{a'be}+\beta &j_{a'bc}+\alpha \end{Bmatrix} \begin{Bmatrix} j_{a'bc} &{} &i_{a'e} &{} &i_{a'c} \\ {} &j_{a'be} &{} &j_{a'ce} &{} \\ \hdotsfor[2]{5}\end{Bmatrix} = \sum_{i_{ae},j_{ace},i_{ac}} (-1)^S d_{i_{ae}}d_{j_{ace}}d_{i_{ac}} \\
\times \begin{Bmatrix} \lambda &i_{a'e} &i_{ae} \\ j_{a'de} &j_{a'be}+\beta &j_{a'be} \end{Bmatrix}  \begin{Bmatrix} \lambda &i_{a'e} &i_{ae} \\ j_{aa'e} &j_{ace} &j_{a'ce} \end{Bmatrix} \begin{Bmatrix} \lambda &i_{a'c} &i_{ac} \\ j_{a'cd} &j_{a'bc}+\alpha &j_{a'bc} \end{Bmatrix} \\
\times \begin{Bmatrix} \lambda &i_{a'c} &i_{ac} \\ j_{aa'c} &j_{ace} &j_{a'ce} \end{Bmatrix} \begin{Bmatrix} j_{a'bc}+\alpha &{} &i_{ae} &{} &i_{ac} \\ {} &j_{a'be}+\beta &{} &j_{ace} &{} \\ \hdotsfor[2]{5}\end{Bmatrix},
\end{multline}
pour $S= i_{ab}+i_{ac}-i_{a'c}+2i_{a'e}+2j_{a'bc}+\alpha+j_{ade}+j_{aa'e}+j_{acd}+j_{aa'c}$.

\begin{figure}
\begin{center}
\includegraphics[width=14cm]{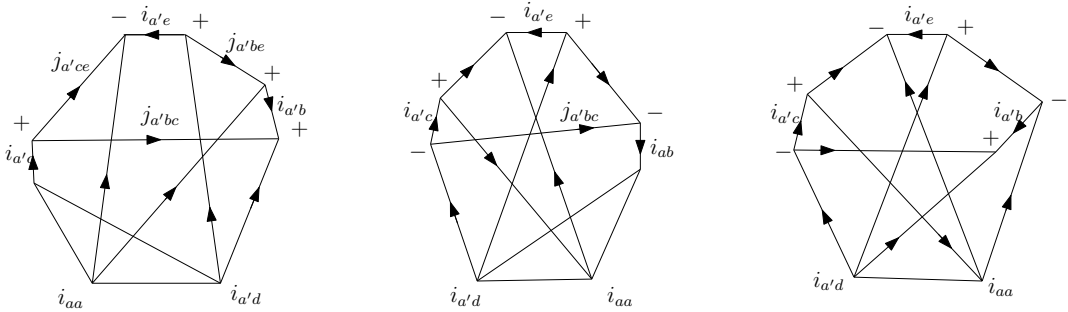}
\end{center}
\caption{ \label{fifj-move4-2} Cette figure repr\'esente les symboles 15j apparaissant respectivement en \eqref{5-cycle 1}, \eqref{5-cycle 2}, donnant des r\'ecurrences sur des cycles de 5 liens, et en \eqref{6-cycle} sur un cycle de six liens. Remarquons que seules les orientations des vertex et des liens de ces cycles et des liens qui les joignent sont n\'ecessaires.}
\end{figure}

Il n'y a qu'une seule fa\c{c}on dans notre situation de cr\'eer un cycle de six liens, donnant la relation suivante :
\begin{multline} \label{6-cycle}
\sum_{i_{a'b}=\lvert i_{ab}-\lambda\rvert}^{i_{ab}+\lambda} d_{i_{a'b}} \begin{Bmatrix} \lambda &i_{a'b} &i_{ab} \\ j_{a'bd} &j_{a'bc}+\alpha &j_{a'bc} \end{Bmatrix} \begin{Bmatrix} \lambda &i_{a'b} &i_{ab} \\ j_{aa'b} &j_{a'be}+\beta &j_{a'be} \end{Bmatrix}\\
\times \begin{Bmatrix} i_{a'b} &{} &i_{a'e} &{} &i_{a'c} &{} \\ {} &j_{a'be} &{} &j_{a'ce} &{} &j_{a'bc} \\ \hdotsfor[2]{6}\end{Bmatrix} \\
= \sum_{i_{ae},j_{ace},i_{ac}} (-1)^T d_{i_{ae}}d_{j_{ace}}d_{i_{ac}} \begin{Bmatrix} \lambda &i_{a'e} &i_{ae} \\ j_{a'de} &j_{a'be}+\beta &j_{a'be} \end{Bmatrix} \begin{Bmatrix} \lambda &i_{a'c} &i_{ac} \\ j_{a'cd} &j_{a'bc}+\alpha &j_{a'bc} \end{Bmatrix} \\
\times \begin{Bmatrix} \lambda &i_{a'e} &i_{ae} \\ j_{aa'e} &j_{ace} &j_{a'ce} \end{Bmatrix}  \begin{Bmatrix} \lambda &i_{a'c} &i_{ac} \\ j_{aa'c} &j_{ace} &j_{a'ce} \end{Bmatrix} \begin{Bmatrix} i_{ab} &{} &i_{ae} &{} &i_{ac} &{} \\ {} &j_{a'be}+\beta &{} &j_{ace} &{} &j_{a'bc}+\alpha \\ \hdotsfor[2]{6}\end{Bmatrix},
\end{multline}
avec $T= 2j_{a'bc}+2\lambda+\alpha+i_{a'c}+i_{ac}+j_{a'bd}+j_{aa'b}+j_{aa'c}+j_{aa'e}+j_{a'de}+j_{a'cd}$. Les coefficients de cette relation sont naturellement ceux du membre de gauche de \eqref{5-cycle 1}, avec ceux du membre de droite de \eqref{5-cycle 2}. Les 15j d\'ecrits dans ces \'equations, \eqref{5-cycle 1}, \eqref{5-cycle 2} et \eqref{6-cycle}, sont repr\'esent\'es en figure \ref{fifj-move4-2}.

\begin{figure}
\begin{center}
\includegraphics[width=17cm]{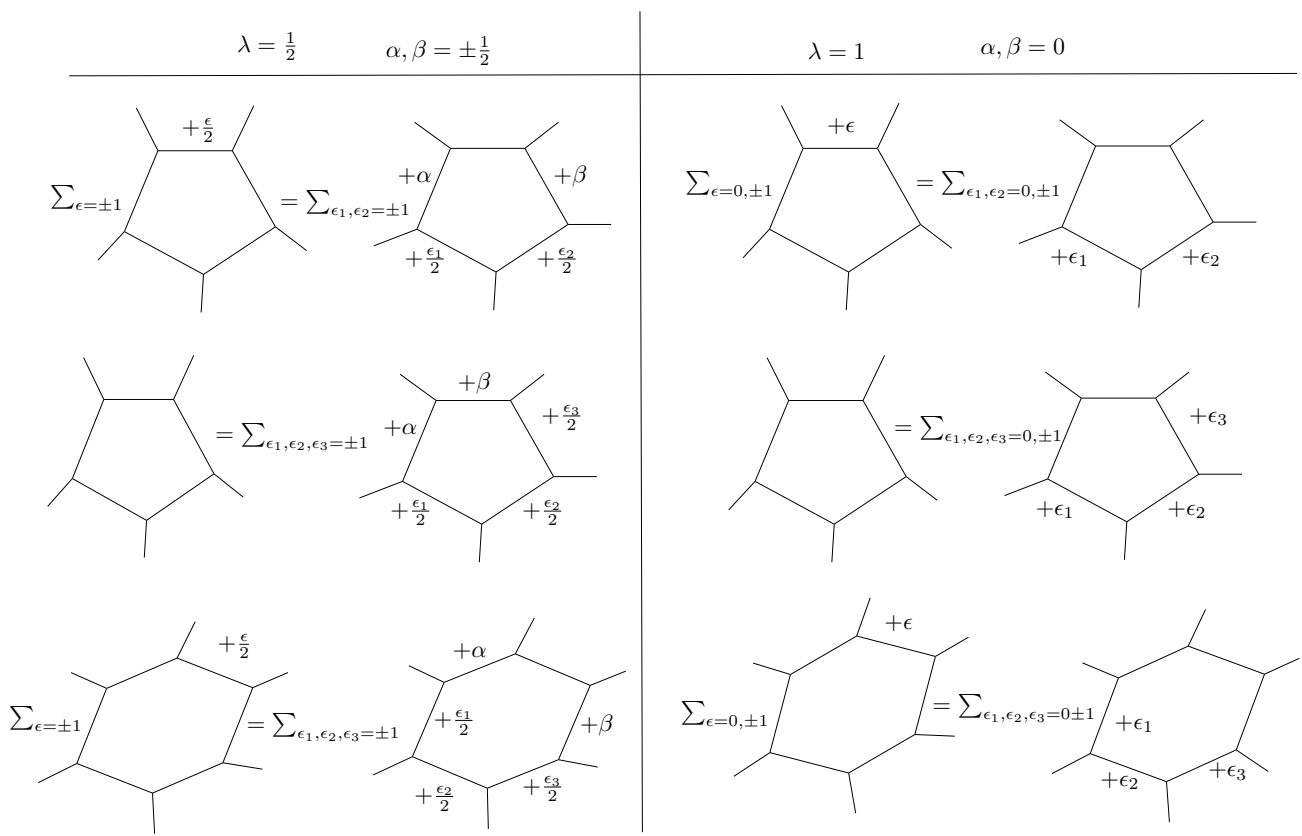}
\end{center}
\caption{ \label{fifj-summary} Ce tableau r\'esume les r\'esultats obtenus pour des cycles de r\'eseaux de spins, form\'es de cinq ou six liens, avec des variations des arguments de $\pm\f12$ ou $\pm1$. Par simplicit\'e, les coefficients de ces relations de r\'ecurrence ont \'et\'e omis, mais ils peuvent \^etre explicitement \'evalu\'es, \'etant simplement des symboles 6j avec des spins $1/2$ ou $1$. Ces relations agissent sur des 5-cycles et des 6-cycles de n'importe quels graphes, et pas seulement des 15j. Pour $\lambda=\f12$, on peut choisir les valeurs des param\`etres libres $\alpha,\beta=\pm\f 12$, et tous les liens d'un cycle sont touch\'es. Pour $\lambda=1$, le choix $\alpha=\beta=0$ des param\`etres libres permet de simplifier les \'equations en laissant invariants certains spins du cycle concern\'e.}
\end{figure}

Ces relations impliquent des symboles 15j dont les spins subissent des diff\'erences de demi-entiers, entre $-\lambda$ et $\lambda$. Pour obtenir des r\'ecurrences plus explicitement, le choix le plus simple est de prendre $\lambda=1/2$. Alors, les param\`etres libres sont $\alpha,\beta=\pm1/2$, et les spins de tous les liens des cycles sont affect\'es. Dans chacune des trois formules, exactement trois triangles voient leur spin touch\'e, puisque les cycles impliqu\'es ne diff\`erent que par les choix d'appariements pour les spins entrelaceurs. Le choix $\lambda=1$ est aussi tr\`es int\'eressant. Dans ce cas, on est libre de choisir $\alpha,\beta=0,\pm1$. Le choix $\alpha,\beta=0$ a le m\'erite de simplifier les \'equations car il laisse invariant deux spins sur chaque cycle. G\'eom\'etriquement, cela signifie que l'aire d'un seul triangle, portant $j_{a'ce}$, est modifi\'ee par le mouvement (avec deux ou trois entrelaceurs). Ces cas sont r\'esum\'es par la figure \ref{fifj-summary}.

Dans ces formules, seuls les spins le long du graphe d'un r\'eseau de spins sont affect\'es. Je souhaite insister sur le fait que les coefficients de ces relations ne d\'ependent que des spins port\'es par le cycle consid\'er\'e et ceux des liens attach\'es aux sommets du cycle. En cons\'equence, ces relations s'appliquent \`a tout symbole $\{3nj\}$ contenant de tels cycles. Cela nous donne donc une g\'en\'eralisation de la m\'ethode d\'ecrite pr\'ec\'edemment pour les plus petits symboles \`a des symboles arbitraires contenant des cycles de cinq ou six liens.

\medskip

En r\'esum\'e, nous sommes partis du mouvement de Pachner 4-2 r\'egularis\'e, et avons obtenu des relations de r\'ecurrence en choisissant des donn\'ees de bord particuli\`eres, comme en gravit\'e quantique 3d avec l'identit\'e de Biedenharn-Elliott. Nous allons dans la suite proposer une nouvelle m\'ethode pour obtenir des relations de r\'ecurrence sur les mod\`eles topologiques de type BF, par action de l'op\'erateur d'holonomie sur les r\'eseaux de spins. Cette m\'ethode, que nous allons pr\'esenter en 3d s'appliquerait tout aussi bien en 4d, en relation avec le mouvement 5-1.

\medskip

Bien que nos relations de r\'ecurrence s'appliquent \`a tout symbole $\SU(2)$-invariant, l'interpr\'etation g\'eom\'etrique provenant du mouvement 4-2 ne tient que pour les symboles 15j, du fait de leur relation \`a la th\'eorie BF 4d. On pourrait n\'eanmoins imaginer qu'une partie de cette interpr\'etation survive dans les mod\`eles visant \`a d\'ecrire la gravit\'e quantique 4d. En effet, les \'etats quantiques de la LQG sont construits avec les r\'eseaux de spins $\SU(2)$, dont les spins et entrelaceurs sont les valeurs propres des op\'erateurs d'aire et de volume. Si un mod\`ele de mousses de spins parvient \`a fournir la dynamique de ces r\'eseaux de spins, on est en droit d'attendre que des relations de r\'ecurrence sur les amplitudes de mousses de spins traduisent les contraintes (i.e. les sym\'etries) classiques de la th\'eorie au niveau quantique. Ainsi des diff\'erences de spins sur les triangles seraient des variations finies d'aires, et il est raisonnable de penser leur description en termes de mouvements \'el\'ementaires simpliciels tels que ceux ici pr\'esent\'es.

\chapter{Contrainte hamiltonienne et mouvement \og tente\fg{} en 3d} \label{sec:recurrence tentmoves}

\section{Contrainte hamiltonienne sur le 6j : mouvement de Pachner 1-4}

Nous nous pla\c{c}ons \`a nouveau dans le cadre de la gravit\'e 3d, afin de pr\'esenter une nouvelle m\'ethode, qui a l'avantage de poss\'eder une interpr\'etation g\'eom\'etrique simple et directement en lien avec l'approche du formalisme hamiltonien en calcul de Regge de Dittrich et Bahr. On consid\`ere donc une triangulation de $S^2$ par le bord d'un t\'etra\`edre, et l'espace de Hilbert sur le graphe dual. Cet espace est engendr\'e par les fonctionnelles des r\'eseaux de spins $s_{\{j_e\}}(g_1,\dotsc,g_6)$, qui sont telles que leur \'evaluation sur la connexion discr\`ete triviale est simplement le 6j ayant les couleurs correspondantes,
\be \label{flat spin network}
\left.s_{\{j_e\}}(g_1,\dotsc,g_6)\right|_{g_e=\mathbbm{1}} = \begin{Bmatrix} j_1 &j_2 &j_3 \\j_4 & j_5 & j_6\end{Bmatrix}.
\ee
Le graphe t\'etra\'edral sous-jacent est le graphe dual \`a la triangulation. Les liens portant les spins $j_4, j_5$ et $j_6$ forment une plaquette triangulaire.

Rappelons l'identit\'e formelle satisfaite par le 6j sous l'effet du mouvement de Pachner 4-1 dans le mod\`ele de Ponzano-Regge :
\begin{multline} \label{div14}
\Bigl(\sum_{l_4}d_{l_4}^2\Bigr)
\begin{Bmatrix} j_1 & j_2 & j_3 \\ j_4 & j_5 & j_6 \end{Bmatrix} =
\sum_{l_1 \ldots l_4} (-)^{\sum_{i=1}^6 j_i+\sum_{j=1}^4 l_j} \biggl(\prod_{i=1}^4 d_{l_i}\biggr)
\begin{Bmatrix} j_1 & j_2 & j_3 \\ l_1 & l_2 & l_3  \end{Bmatrix}
\begin{Bmatrix} j_6 & j_5 & j_1 \\ l_2 & l_3 & l_4  \end{Bmatrix} \\
\times \begin{Bmatrix} j_4 & j_2 & j_6 \\ l_3 & l_4 & l_1  \end{Bmatrix}
\begin{Bmatrix} j_3 & j_5 & j_4 \\ l_4 & l_1 & l_2  \end{Bmatrix}.
\end{multline}
Les coefficients $d_k\equiv (2k+1)$ sont les dimensions des repr\'esentations de spins $k$. Cette relation peut se d\'eriver \`a partir de l'indentit\'e de BE et des relations d'orthogonalit\'e des 6j. Elle est n\'eanmoins formelle du fait que le membre de gauche diverge tr\`es grossi\`erement (et donc le membre de droite aussi). Cela fut corrig\'e par Ponzano et Regge en introduisant un cut-off sur les spins, et le mod\`ele qui en r\'esulte est sens\'e \^etre ind\'ependant de la triangulation dans la limite o\`u le cut-off est envoy\'e \`a l'infini (voir \`a ce propos une discussion d\'etaill\'ee dans \cite{barrett-PR}, et surtout le chapitre consacr\'e aux divergences dans cette th\`ese !).

On souhaite ici se donner un point de vue canonique sur ce mouvement de Pachner, un peu \`a la mani\`ere de \cite{ooguri-3d}. Aussi, les \'equations qui viennent ne sont pas toutes nouvelles, mais la perspective que nous en donnons permet d'aboutir \`a de jolies g\'en\'eralisations et interpr\'etations. La contrainte de courbure nulle sur la plaquette portant les spins $j_4, j_5$ et $j_6$ s'\'ecrit : $\delta(g_4 g_5 g_6)$. Pour la faire agir sur les r\'eseaux de spins, nous la d\'eveloppons selon : $\delta(g_4 g_5 g_6) = \sum_j d_j \chi_j(g_4 g_5 g_6)$. Chaque caract\`ere a une action bien d\'efinie, par multiplication sur les r\'eseaux de spins. Nous pouvons alors effectuer du recouplage entre les \'el\'ements de matrices d'un m\^eme \'el\'ement de groupe, \`a l'aide de coefficients de Clebsch-Gordan. Nous obtenons le r\'esultat suivant :
\begin{multline} \label{constrainthol}
\delta\left(g_4 g_5 g_6\right)\,s_{\{j_l\}}(g_1,\dotsc,g_6) = \sum_{k_4,k_5,k_6,j} (-1)^{j_1+j_2+j_3+j_4+j_5+j_6+k_4+k_5+k_6 + j} d_{k_4} d_{k_5} d_{k_6} d_j\\
\times \begin{Bmatrix} k_4& j_4 & j \\ j_6& k_6 & j_2\end{Bmatrix}
\begin{Bmatrix}k_5& j_5 & j \\ j_4& k_4 & j_3 \end{Bmatrix}
\begin{Bmatrix}k_6& j_6 & j \\ j_5& k_5 & j_1\end{Bmatrix}\,
s_{\{j_1,j_2,j_3,k_4,k_5,k_6\}}(g_1,\dotsc,g_6).
\end{multline}
Il est alors clair que le mouvement 4-1 \eqref{div14} r\'esulte de l'action de la contrainte de courbure nulle sur une face, suivi de l'\'evaluation sur les connexions plates, ce qui explique la divergence. Celle-ci provient formellement de : $\delta(\mathbbm{1}) = \sum_j d_j^2$.

\begin{figure}
\begin{center}
\includegraphics[width=7cm]{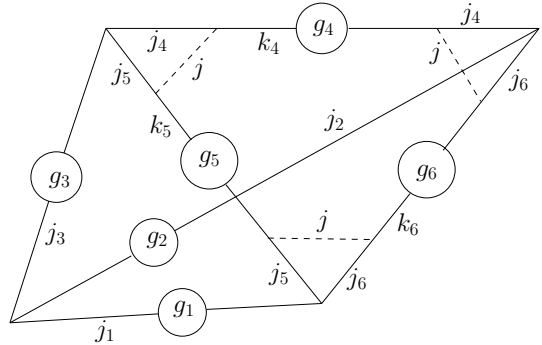}
\end{center}
\caption{\label{picturerechol} La fonctionnelle de r\'eseau de spins $s_{\{j_e\}}$ peut \^etre repr\'esent\'ee par un graphe t\'etra\'edrique, dont les liens sont colori\'es par les spins $(j_e)$ et portent les holonomies $(g_e)$, $e=1,\dotsc,6$. L'action du caract\`ere $\chi_{j}(g_4g_5g_6)$, qui est une composante de Fourier de la contrainte de courbure nulle sur les cycle $(456)$, se traduit apr\`es recouplage en la fonctionnelle dessin\'e ci-dessus, pour laquelle les intersections des liens donnent des symboles 3mj, et les spins $k_1, k_2, k_3$ \'etant somm\'es. Les lignes en pointill\'es portent le spin $j$ fix\'e.}
\end{figure}

Si cela nous fournit un lien \'evident entre les contraintes classiques et les sym\'etries quantiques, nous aimerions affiner et \'eviter les divergences ! Cela est sugg\'er\'e par le fait que la divergence en $\delta(\mathbbm{1})$ se factorise dans le membre de gauche. Puisque les relations quantiques, i.e. sur les symboles 6j, s'obtiennent apr\`es \'evaluation des fonctionnelles $s_{\{j_e\}}$ sur les connexions plates en utilisant \eqref{flat  spin network}, la solution est naturellement de ne pas agir avec $\delta(g_4 g_5 g_6)$, mais seulement avec une des ses composantes de Fourier, i.e. un caract\`ere $\chi_{j}(g_4g_5g_6)$ ! L'\'evaluation de celui-ci est en effet bien d\'efinie en l'identit\'e. Nous pouvons alors recoupler les \'el\'ements de matrices de $\chi_{j}(g_4g_5g_6)$ avec les \'el\'ements de matrices de $g_4, g_5$ et $g_6$ d\'ej\`a pr\'esents dans $s_{\{j_e\}}$, ce qui fait aboutir \`a
\begin{multline} \label{rechol}
\chi_j(g_4g_5g_6)\,s_{\{j_l\}}(g_1,\dotsc,g_6) = \sum_{k_4,k_5,k_6} (-1)^{j_1+j_2+j_3+j_4+j_5+j_6+k_4+k_5+k_6 + j} d_{k_4} d_{k_5} d_{k_6} \\
\times \begin{Bmatrix} k_4& j_4 & j \\ j_6& k_6 & j_2\end{Bmatrix}
\begin{Bmatrix}k_5& j_5 & j \\ j_4& k_4 & j_3 \end{Bmatrix}
\begin{Bmatrix}k_6& j_6 & j \\ j_5& k_5 & j_1\end{Bmatrix}
\,s_{\{j_1,j_2,j_3,k_4,k_5,k_6\}}(g_1,\dotsc,g_6).
\end{multline}
Le recouplage effectu\'e est pr\'esent\'e figure \ref{picturerechol}. En \'evaluant les deux membres de cette \'egalit\'e sur la connexion plate triviale, il vient
\begin{multline} \label{rechol1}
(2j+1)\begin{Bmatrix}j_1& j_2 & j_3 \\ j_4& j_5 & j_6\end{Bmatrix} =
\sum_{k_4,k_5,k_6} (-1)^{j_1+j_2+j_3+j_4+j_5+j_6+k_4+k_5+k_6 + j} d_{k_4} d_{k_5} d_{k_6}
\\ \times \begin{Bmatrix} k_4& j_4 & j \\ j_6& k_6 & j_2\end{Bmatrix}
\begin{Bmatrix}k_5& j_5 & j \\ j_4& k_4 & j_3 \end{Bmatrix}
\begin{Bmatrix}k_6& j_6 & j \\ j_5& k_5 & j_1\end{Bmatrix}
\begin{Bmatrix}j_1& j_2 & j_3 \\ k_4& k_5 & k_6\end{Bmatrix}.
\end{multline}
De telles relations de r\'ecurrence sur le 6j sont d\'ej\`a connues \cite{varshalovich-book}. Remarquons alors que le mouvement 4-1 avec un cut-off sur les spins peut \^etre construit simplement en multipliant \eqref{rechol1} par $d_j$ et en sommant sur $j$. Lorsque le cut-off est envoy\'e \`a l'infini, la somme suppl\'ementaire dans le membre de droite diverge, et le membr de gauche reproduit bien le facteur divergent $\sum_j d_j^2$. La relation de r\'ecurrence \eqref{rechol1} fournit donc une r\'egularisation alternative du mouvement 4-1, dans laquelle le cut-off sur les spins est remplac\'e par une valeur fix\'ee du spin $j$. Cette r\'egularisation peut \^etre vue comme une fixation de jauge partielle \`a a mani\`ere de \cite{freidel-louapre-diffeo}. La r\'egularisation qui y est propos\'ee revient au choix $j=0$, qui donne ici une identit\'e triviale. Ce n'est en fait rien d'autre que la r\'egularisation que nous avons utilis\'ee pour le mouvement de Pachner 4-2 dans le mod\`ele d'Ooguri \`a la section pr\'ec\'edente. Inversement, \eqref{rechol1} appara\^it comme l'\'egalit\'e des deux multi-s\'eries impliqu\'ees dans le mouvement formel pour tout spin $l_4$. Dans ce sens, notre relation de r\'ecurrence est bien une fixation de jauge de la valeur du spin responsable du facteur divergent.

La chose int\'eressante avec cette m\'ethode utilisant l'op\'erateur d'holonomie est qu'elle peut \^etre facilement g\'en\'eralis\'ee \`a des r\'eseaux de spins plus g\'en\'eraux.

\section{Interpr\'etation comme mouvement \og tente\fg}

Il s'agit ici de 
\begin{itemize}
\item g\'en\'eraliser la relation \eqref{rechol} de la section pr\'ec\'edente \`a des r\'eseaux de spins arbitraires,
\item tout en trouvant une interpr\'etation g\'eom\'etrique ad\'equate,
\item et une interpr\'etation en termes de mousses de spins.
\end{itemize}
Pour cela, regardons la construction que nous avons effectu\'ee pour obtenir \eqref{rechol} du point de vue de la triangulation, et non plus du point de vue du graphe dual. Les ar\^etes duales aux liens portant les spins $j_4, j_5$ et $j_6$ ont pour longueur $\ell_e = j_e+\f12$, et sont attach\'ees \`a un sommet commun not\'e $s$. Le caract\`ere $\chi_j(g_4 g_5 g_6)$ agit alors via une boucle autour de ce sommet. On peut imaginer, pour commencer \`a g\'en\'eraliser, que les autres sommets, en liens avec $j_1, j_2$ et $j_3$, sont reli\'es \`a des sommets suppl\'ementaires formant au total une certaine triangulation 2d (donc les ar\^etes duales \`a $j_1, j_2$ et $j_3$ ne supportent pas n\'ecesairement un triangle). Il est assez clair que notre relation \eqref{rechol} s'av\`ere inchang\'ee car le recouplage ne concerne pas les ar\^etes suppl\'ementaires ; seule la fonctionnelle $s_{\{j_e\}}$ doit \^etre adapt\'ee au graphe dual de la triangulation 2d.

\begin{figure}
\begin{center}
\includegraphics[width=10cm]{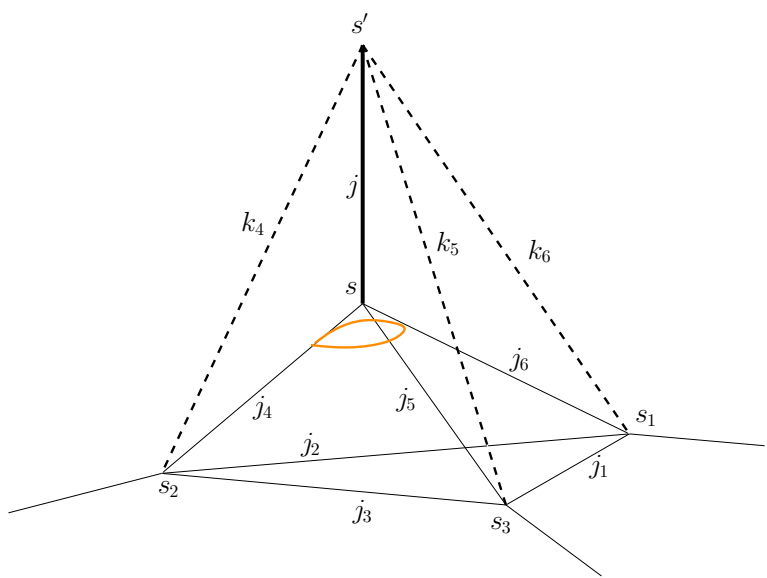}
\end{center}
\caption{ \label{fig:tent3valent} Cette figure pr\'esente le membre de droite de \eqref{rechol}, i.e. l'action de l'op\'erateur d'holonomie autour d'un sommet $s$, apr\`es recouplage, du point de vue de la triangulation d'une surface canonique 2d. Nous avons un morceau d'une surface initiale 2d, form\'e de 3 triangles ayant pour sommets $s, s_1, s_2, s_3$. Nous agissons autour du sommet $s$ avec une composante de Fourier, $\chi_j(g_4g_5g_6)$, de la contrainte de courbure nulle. Par l'\'equation \eqref{rechol},  cela revient \`a construire une nouvelle triangulation 2d en faisant \'evoluer $s$ en $s'$, par un segment de longueur $j+\f12$. Entre ces deux triangulations canoniques, nous avons un morceau de triangulation 3d, contenant trois t\'etra\`edres qui se rencontrent en $(ss')$. Alg\'ebriquement, l'action de l'op\'erateur autour de $s$ se r\'e\'ecrit comme l'amplitude de mousses de spins de Ponzano-Regge pour ce morceau de triangulation 3d, en sommant sur les spins $k_e$.}
\end{figure}

Int\'eressons nous maintenant \`a ce qui se passe dans le membre de droite de \eqref{rechol}, i.e. apr\`es recouplage, en essayant d'en donner une vision en mousses de spins. Dans le cadre du mod\`ele de Ponzano-Regge, les trois symboles 6j correspondent \`a trois t\'etra\`edres positionn\'es comme sur la figure \ref{fig:tent3valent} Autrement dit, le mouvement s'interpr\^ete en ajoutant un sommet $s'$, ensuite reli\'e \`a $s$ et aux autres sommets avec lesquels $s$ est en contact. On peut voir en cela un processus d'\'evolution de la triangulation initiale 2d, dans lequel $s$ est remplac\'e par $s'$, d\'efinissant une nouvelle triangulation identique \`a la premi\`ere. Ce mouvement est appel\'e mouvement \og tente\fg{} pour des raisons \'evidentes, le support de la tente \'etant l'ar\^ete $(ss')$ ! L'avantage d'un tel mouvement sur les mouvements de Pachner est qu'il ne change pas le nombre de variables entre les deux triangulations. C'est pourquoi il se pr\`ete bien \`a la d\'erivation d'un formalisme hamiltonien du calcul de Regge, comme dans \cite{bahr-dittrich-broken-symmetries}. Nous avons donc une vision g\'eom\'etrique en termes d'un mouvement \og tente\fg.

Les deux morceaux de triangulations 2d impliqu\'es peuvent \^etre vus comme le bord d'une triangulation 3d, form\'ee de trois t\'etra\`edres partageant l'ar\^ete $(ss')$. Le mod\`ele de Ponzano-Regge associe \`a ces t\'etra\`edres des symboles 6j, et notre relation de r\'ecurrence, \eqref{rechol}, ne fait que calculer l'\'evolution de la fonction d'onde $s_{\{j_e\}}$ sous cette transformation via le mod\`ele de Ponzano-Regge ! Il s'agit donc d'un cas particulier -- un mouvement \og tente\fg{}-- des amplitudes de transition g\'en\'eriques calcul\'ees par Ooguri dans \cite{ooguri-3d}. En fait, la v\'eritable amplitude de Ponzano-Regge, divergente, n\'ecessiterait d'\'evaluer les fonctionnelles sur la connexion triviale, et de sommer sur le spin $j$ qui repr\'esente la longueur $\ell_{(ss')} = j+\f12$ de l'ar\^ete $(ss')$. Ainsi, le fait d'avoir s\'electionn\'e un mode de Fourier $\chi_j$ est une fixation de jauge du mouvement, qui consiste simplement \`a fixer la longueur de l'ar\^ete $(ss')$ support de la \og tente\fg{} ! Cela donne l'interpr\'etation du mouvement en termes de mousses de spins, et permet de remplir les trois objectifs \'enonc\'es !

\section{Mouvements \og tente\fg{} pour des sommets de valences quelconques}

\begin{figure}
\begin{center}
\includegraphics[width=12cm]{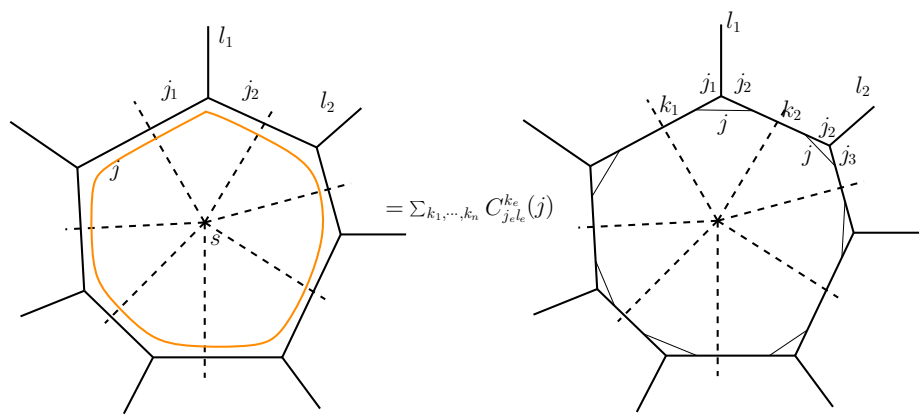}
\end{center}
\caption{ \label{dualtent-nvalent} Cette figure pr\'esente graphiquement la relation \eqref{recholgen}. Nous avons un sommet $s$ d'une triangulation 2d, duale \`a une plaquette. Nous appliquons dans le membre de gauche sur le bord de cette derni\`ere une composante $\chi_j$ de l'op\'erateur de courbure nulle. Le membre de draoite correspond \`a la situation apr\`es recouplage. De chaque vertex dual (trivalent), on peut extraire un symbole 6j, et l'on doit sommer sur les coloriages $k_1,\dotsc,k_n$ des liens duaux.}
\end{figure}

Nous avons donc donn\'e une g\'en\'eralisation de \eqref{rechol} pour des sommets trivalents d'une triangulation 2d, en termes de mouvements \og tente\fg{}. Il en va exactement de m\^eme pour un sommet $s$ de valence $n$ quelconque ! La face duale \`a ce sommet a un bord form\'ee de $n$ liens, portant des spins $(j_e)$ pour $e=1$ \`a $n$. On s'int\'eresse \`a une fonctionnelle de r\'eseaux de spins sur le graphe dual, et en particulier sur la partie repr\'esent\'ee figure \ref{dualtent-nvalent}. L'action de l'op\'erateur d'holonomie $\chi_j(g_1\dotsm g_n)$ sur le r\'eseau de spins se dessine comme une boucle le long de la face duale, dans la repr\'esentation $j$. On peut alors recoupler les \'el\'ements de matrices de ce caract\`ere avec les \'el\'ements de matrices de $g_1,\dotsc,g_n$ d\'ej\`a pr\'esents dans la fonction d'onde $s_{\{j_e\}}(g_e)$. Cela a pour effet de factoriser un symbole 6j \`a chaque vertex dual, jusqu'\`a obtenir
\begin{multline} \label{recholgen}
\chi_j(g_1\dotsm g_n)\,s_{\{j_e\}}(g_e) = \sum_{k_1,\dotsc,k_n}  (-)^{j+\sum_{e=1}^n j_e+k_e+l_e}\prod_{e=1}^n d_{k_e}  \\
\times \begin{Bmatrix} k_2& j_2 & j \\ j_1& k_1 & l_1\end{Bmatrix} 
\begin{Bmatrix} k_3& j_3 & j \\ j_2& k_2 & l_2\end{Bmatrix} \dotsm
\begin{Bmatrix}k_1& j_1 & j \\ j_n& k_n & l_n\end{Bmatrix}\,
s_{\{j_e\}}(g_e).
\end{multline}
Ce processus est repr\'esent\'e du point de vue dual, figure \ref{dualtent-nvalent}.

\begin{figure}
\begin{center}
\includegraphics[width=8cm]{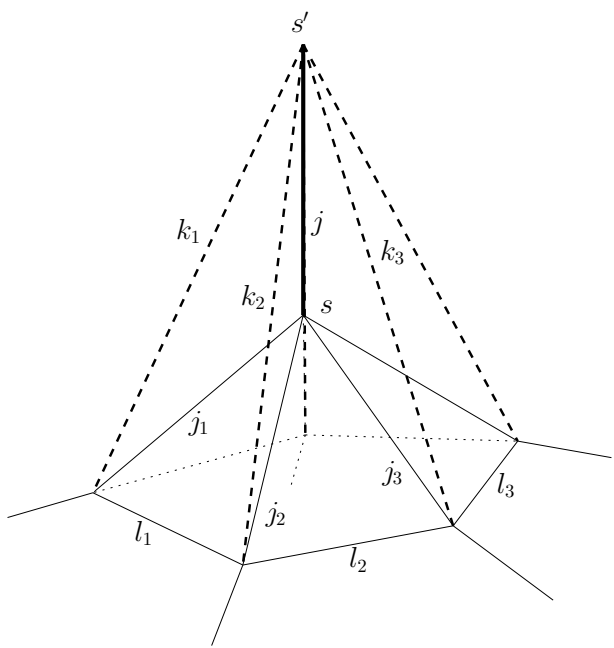}
\end{center}
\caption{ \label{tentnvalent} Interpr\'etation g\'eom\'etrique de l'action de l'holonomie autour du sommet $s$ comme un mouvement \og tente\fg. L'\'evolution de $s$ en $s'$ se fait par une ar\^ete de longueur $j+\f12$, d\'efinissant ainsi une nouvelle triangulation 2d, identique \`a la premi\`ere. Par ailleurs, toutes deux d\'elimitent un morceau de triangulation 3d, contenant $n$ t\'etra\`edres qui se rencontrent le long du pole de la tente. Le membre de droite de \eqref{recholgen} correspond \`a l'\'evaluation de l'amplitude de Ponzano-Regge pour cette triangulation, avec la longueur du pole de la tente fix\'ee, et en sommant sur les longueurs des ar\^etes $k_1,\dotsc, k_n$.}
\end{figure}

L'interpr\'etation est imm\'ediate en regardant la figure \ref{tentnvalent}. On fait \'evoluer la triangulation initiale par l'ar\^ete $(ss')$ dont on fixe la longueur $\ell_{(ss')} = j+\f12$. Les deux morceaux de triangulation 2d forment le bord d'une triangulation 3d, constistu\'ee de $n$ t\'etra\`edres. On calcule alors l'\'evolution de la fonction d'onde $s_{\{j_e\}}(g_e)$ \`a travers cette triangulation 3d, via le mod\`ele de Ponzano-Regge, avec pour fixation de jauge la valeur fix\'ee de $j$. En \'evaluant \eqref{recholgen} sur la connexion plate (triviale sur cette topologie), on obtient l'amplitude de transition pour les r\'eseaux de spins,
\begin{multline}
(2j+1)\  s(j_e) = \sum_{k_1,\dotsc,k_n}  (-)^{j+\sum_{e=1}^n j_e+k_e+l_e}\prod_{e=1}^n d_{k_e}  \\
\times \begin{Bmatrix} k_2& j_2 & j \\ j_1& k_1 & l_1\end{Bmatrix} 
\begin{Bmatrix} k_3& j_3 & j \\ j_2& k_2 & l_2\end{Bmatrix} \dotsm
\begin{Bmatrix}k_1& j_1 & j \\ j_n& k_n & l_n\end{Bmatrix}\
s(j_e).
\end{multline}

\chapter{Une relation non-topologique pour le 6j isoc\`ele} \label{sec:recurrence-6jiso}

Le 6j dit isoc\`ele a plusieurs spins \'egaux. Il a \'et\'e notamment utilis\'e dans des calculs de corr\'elations de type graviton en 3d \cite{graviton3d-simone, graviton3d-val}, du fait qu'il admet une repr\'esentation int\'egrale,
\be\label{6jisoint}
\int_{\SU(2)^2} dg dh\, \chi_j(g)\chi_k(h)\chi_J(gh)\chi_K(gh^{-1}) \,=\, (-1)^{2j} \begin{Bmatrix} j &J &K \\ k &J &K\end{Bmatrix},
\ee
Notons que $(-)^{2j} = (-)^{2k}$ du fait de la condition sur les deux triplets de repr\'esentations, $(j+J+K)\in\N$ et $(k+J+K)\in\N$.

Gr\^ace aux sym\'etries de l'int\'egrant, ces deux int\'egrales sur $\SU(2)$ peuvent se r\'eduire \`a une int\'egrale sur trois variables r\'eelles. Il est judicieux de la consid\'erer en termes de quatre variables angulaires sujettes \`a une contrainte. Notons $\alpha$ et $\beta$ les angles de classe de $g$ et $h$, et $\phi^\pm$ ceux des produits $gh^{\pm}$. Alors, \eqref{6jisoint} se r\'e\'ecrit comme une int\'egrale sur ces quatre angles, avec la contrainte
\be\label{6jIsoC}
\cos\alpha\,\cos\beta - \f12(\cos\phi^++\cos\phi^-) = 0
\ee
Cette contrainte peut s'exprimer \`a l'aide des caract\`eres de la repr\'esentation fondamentale,
\be
\chi_{\f12}(g)\chi_{\f12}(h) - \chi_{\f12}(gh)+\chi_{\f12}(gh^{-1}) = 0.
\ee
En particulier, son insertion dans l'int\'egrale \eqref{6jisoint} donne \'evidemment z\'ero. Par ailleurs, nous disposons d'une formule de recouplage entre un spin $j$ quelconque et la repr\'esentation fondamentale,
\be
\chi_j(g)\,\chi_{\f12}(g) \,=\, \chi_{j-\f12}(g)+\chi_{j+\f12}(g).
\ee
Ainsi, l'insertion de \eqref{6jIsoC} dans \eqref{6jisoint} produit une somme de diff\'erents 6j isoc\`eles. Cela donne la relation de r\'ecurrence suivante :
\be \label{recurrenceiso6j}
\sum_{\eps=\pm} \begin{Bmatrix} j &J+\f\eps2 &K \\k &J +\f\eps2 &K\end{Bmatrix}+\sum_{\eps'} \begin{Bmatrix} j &J &K+\f{\eps'}2 \\k &J &K+\f{\eps'}2\end{Bmatrix} + \sum_{\eta,\eta'} \begin{Bmatrix} j+\f\eta2 &J &K \\k+\f{\eta'}2 &J  &K\end{Bmatrix} = 0.
\ee

Remarquons la simplicit\'e avec laquelle nous avons d\'eriv\'e cette relation. Comme nous le savons, les propri\'et\'es du 6j sont cod\'ees dans l'identit\'e de BE, et donc cette relation en particulier, comme nous le montrons plus bas. Mais il est remarquable qu'une repr\'esentation int\'egrale \`a notre disposition permette de court-circuiter l'identit\'e de BE pour d\'eriver de telles relations de r\'ecurrence.

A la diff\'erence des relations de r\'ecurrence usuelles, que l'on trouve dans la litt\'erature, celle-ci, toute nouvelle \`a notre connaissance, se distingue par ses coefficients triviaux. Le prix \`a payer pour ce b\'en\'efice est le fait que tous les spins sont affect\'es, et l'accroissement du nombre de termes de la relation, habituellement trois ou quatre contre huit ici. Il s'agit d'une \'equation aux diff\'erences d'ordre 2, car nous avons \`a faire \`a la somme et non la diff\'erence de termes impliquant des variations de la m\^eme variable.

\section{Lien avec l'identit\'e de Biedenharn-Elliott}

Nous avons vu auparavant que l'identit\'e de BE, \eqref{bied-elliott}, donne naissance, en sp\'ecifiant la valeur d'un spin du bord, \`a une quantification de la contrainte hamiltonienne de la gravit\'e 3d. Toujours dans la perspective de relier les \'equations quantiques aux contraintes classiques, nous d\'erivons maintenant la relation \eqref{recurrenceiso6j} \`a partir de l'identit\'e de BE. Bien que la d\'erivation soit un peu longue, elle met en \'evidence une intrepr\'etation g\'eom\'etrique simple, comme combinaison des mouvements \'el\'ementaires associ\'es \`a la contrainte $F=0$.

Nous avions pris dans l'identit\'e de BE,
\begin{equation*}
\begin{Bmatrix} j &h &g\\ k &a &b \end{Bmatrix}\,\begin{Bmatrix} j &h &g\\ f &d &c \end{Bmatrix} = \sum_l (-1)^{S+l}(2l+1)\,\begin{Bmatrix} k &f &l\\ d &a &g\end{Bmatrix}\,\begin{Bmatrix} a &d &l\\ c &b &j\end{Bmatrix}\,\begin{Bmatrix} b &c &l\\ f &k &h\end{Bmatrix},
\end{equation*}
le spin $f=1$, et interpr\'et\'e ce choix comme le fait de coller un t\'etra\`edre aplati sur un autre t\'etra\`edre, simulant ainsi un l\'eger d\'eplacement d'un sommet de ce dernier. Nous choisissons maintenant de prendre $f=1/2$, l'interpr\'etation restant donc identique.

Il vient comme cons\'equence de ce choix, $d=g+\beta$ et $c=h+\alpha$, pour $\alpha,\beta=\pm \f{1}{2}$. La somme sur $l$ se r\'eduit simplement \`a deux termes, $l=k\pm\f{1}{2}$, soit :
\begin{multline} \label{recurrencebe}
\begin{Bmatrix} j &h &g \\ \f{1}{2} &g+\beta &h+\alpha\end{Bmatrix} 
\begin{Bmatrix} j &h &g \\k &a &b\end{Bmatrix}\\
= \sum_{l=k\pm\f{1}{2}} (-1)^{S+l}(2l+1)\,\begin{Bmatrix} a &g &k \\ \f{1}{2} &l &g+\beta\end{Bmatrix}\,\begin{Bmatrix} b &h &k \\ \f{1}{2} &l &h+\alpha\end{Bmatrix}\, \begin{Bmatrix} j &h+\alpha &g+\beta \\l &a &b\end{Bmatrix}.
\end{multline}
Une vision g\'eom\'etrique est donn\'ee en figure \ref{move2-3}. Consid\'erons les deux premiers symboles 6j du membre de droite : ils correspondent aux t\'etra\`edres $ABCD'$ et $ACDD'$, coll\'es le long du triangle $ACD'$ dont les ar\^etes portent les spins $(a, g+\beta, l=k\pm\f12)$. Nous appliquons \`a nouveau le mouvement, sous la forme \eqref{recurrencebe}, mais d\'esormais sur le nouveau t\'etra\`edre $ABCD'$. G\'eom\'etriquement, un t\'etra\`edre aplati $ACC'D'$ est coll\'e le long du triangle $ACD'$, et le mouvement 2-3 transforme $ABCD'$ en $ABC'D'$, induisant des variations de spins sur les liens qui se rencontrent en $C$.

Nous choisissons les diff\'erences sur les liens $AC'$ et $C'D'$ comme : $\alpha'=\alpha$ et $\beta'=-\beta$. L'effet de ce choix est d'\'eliminer la variation du spin $g$ due au premier mouvement, tandis que $a$ est modifi\'e d'une quantit\'e $\alpha$, tout comme $h$ sur l'ar\^ete oppos\'ee. Finalement, chacun des deux termes de droite dans \eqref{recurrencebe} donne naissance \`a deux termes, $BC$ devenant $BC'$ avec les deux possibilit\'es $j\pm\f{1}{2}$.

Ces deux mouvements \'el\'ementaires successifs conduisent \`a une relation sur cinq termes :
\be
\begin{Bmatrix} j &h &g \\k &a &b\end{Bmatrix},\qquad\text{and}\qquad \begin{Bmatrix} j+\f{\eta_j}{2} &h+\alpha &g \\k+\f{\eta_k}{2} &a+\alpha &b\end{Bmatrix},\quad \text{pour}\ \ \eta_j,\eta_k=\pm\f{1}{2},
\ee
et pour un $\alpha$ fix\'e. Les coefficients sont g\'en\'eriquement compliqu\'es. Mais ils se simplifient consid\'erablement dans le cas isoc\`ele de la section pr\'ec\'edente ! Prenons \`a cette fin $h=a$ et $g=b$. Red\'efinissons par ailleurs $a\rightarrow a-\alpha$. Nous obtenons pour $\alpha =\f12$ :
\begin{multline} \label{alpha1/2}
d_j d_k \begin{Bmatrix} j &a-\f{1}{2} & b \\k &a-\f{1}{2} & b\end{Bmatrix} \\
=\bigl(k+b-a+\f{1}{2}\bigr)\bigl(a+b-j+\f{1}{2}\bigr) 
\begin{Bmatrix} j-\f{1}{2} &a &b \\k-\f{1}{2} &a &b\end{Bmatrix} - \bigl(k+b-a+\f{1}{2}\bigr)\bigl(a+b+j+\f{3}{2}\bigr) \begin{Bmatrix}j+\f{1}{2} &a &b \\k-\f{1}{2} &a &b\end{Bmatrix} \\
+ \bigl(k+a-b+\f{1}{2}\bigr)\bigl(a+b-j+\f{1}{2}\bigr) \begin{Bmatrix} j-\f{1}{2} &a &b \\k+\f{1}{2} &a &b\end{Bmatrix} - \bigl(k+a-b+\f{1}{2}\bigr)\bigl(a+b+j+\f{3}{2}\bigr) \begin{Bmatrix}j+\f{1}{2} &a &b \\k+\f{1}{2} &a &b\end{Bmatrix},
\end{multline}
et pour $\alpha=-\f12$,
\begin{multline} \label{alphamoins1/2}
d_j d_k \begin{Bmatrix} j &a+\f{1}{2} & b \\k &a+\f{1}{2} & b\end{Bmatrix} \\
= -\bigl(a+b+k+\f{3}{2}\bigr)\bigl(j+a-b+\f{1}{2}\bigr) \begin{Bmatrix} j-\f{1}{2} &a &b \\k-\f{1}{2} &a &b\end{Bmatrix} - \bigl(a+b+k+\f{3}{2}\bigr)\bigl(j+b-a+\f{1}{2}\bigr) \begin{Bmatrix}j+\f{1}{2} &a &b \\k-\f{1}{2} &a &b\end{Bmatrix} \\
+ \bigl(a+b-k+\f{1}{2}\bigr)\bigl(j+a-b+\f{1}{2}\bigr) \begin{Bmatrix}j-\f{1}{2} &a &b \\k+\f{1}{2} &a &b\end{Bmatrix} + \bigl(a+b-k+\f{1}{2}\bigr)\bigl(j+b-a+\f{1}{2}\bigr) \begin{Bmatrix} j+\f{1}{2} &a &b \\k+\f{1}{2} &a &b\end{Bmatrix}.
\end{multline}

Par sym\'etrie, nous obtenons deux relations similaires avec des diff\'erences sur $b$ \`a la place de $a$. Maintenant, la somme de ces quatre relations est pr\'ecis\'ement \eqref{recurrenceiso6j} ! Notons d'abord que les membres de gauche des relations ci-dessus contiennent exactement la moiti\'e des termes attendus dans \eqref{recurrenceiso6j}, tandis que les membres de droite contiennent les autres termes. Il s'av\`ere ensuite que leur somme, sym\'etris\'ee dans l'\'echange $(j\leftrightarrow k)$, conduit bien \`a des coefficients triviaux.

Nous avions vu que l'identit\'e de BE prise avec la valeur $f=1$ peut \^etre vue comme une quantification de la contrainte g\'eom\'etrique $\cos\theta_e = \cos\thet_e(l)$. Il est naturel de se demander si un tel ph\'enom\`ene a lieu avec $f=1/2$. Nous n'avons cependant pas eu le temps d'\'etudier davantage cela. N\'eanmoins, il s'av\`ere que l'\'equation de r\'ecurrence \eqref{rec6j} peut s'obtenir uniquement \`a partir de \eqref{recurrencebe}, en l'it\'erant plusieurs fois un peu comme nous venons de le faire dans ce paragraphe. Il en r\'esulte que \eqref{rec6j} et \eqref{recurrencebe} sont en fait \emph{\'equivalentes}.

Ainsi, les \'equations \eqref{alpha1/2} et \eqref{alphamoins1/2} g\'en\`erent deux de ces mouvements successivement, ce qui d\'eforme le t\'etra\`edre initial par l'ajout de deux t\'etra\`edres aplatis. La relation \eqref{recurrenceiso6j}, qui est plus faible car elle provient de la somme de \eqref{alpha1/2} et \eqref{alphamoins1/2}, a donc la charge d'impl\'ementer une invariance sous une \emph{combinaison} sp\'ecifique de ces mouvements, permettant des compensations entre eux. On voit ainsi comment le symbole 6j isoc\`ele, qui ne donne \emph{pas} naissance \`a un mod\`ele topologique, satisfait une \'equation d'invariance sous une combinaison des mouvements fondamentaux codant la contrainte hamiltonienne au niveau quantique.


\part{Mod\`eles de mousses de spins pour la gravit\'e quantique} \label{part:newmodel}
\chapter{De la th\'eorie BF discr\`ete au calcul de Regge} \label{sec:bf-regge}

Le cadre de ce chapitre est l'\'etude au niveau classique du passage de la th\'eorie BF discr\'etis\'ee en 4d comme \`a la section \ref{sec:bf-spinfoam} \`a une forme discr\`ete de la th\'eorie de Plebanski pour la relativit\'e g\'en\'erale. Le point de d\'epart est donc la discr\'etisation de la th\'eorie BF qui permet de construire le mod\`ele d'Ooguri, l'id\'ee \'etant de s'appuyer sur ce mod\`ele et les contraintes de la th\'eorie de Plebanski pour au final imaginer des mod\`eles de mousses de spins pour la gravit\'e quantique. Cette discr\'etisation de la th\'eorie BF est naturelle puisqu'elle aboutit dans le calcul de la fonction de partition \`a un mod\`ele topologique, i.e. avec une mesure correctement d\'etermin\'ee. De plus, les \'etats de bord du mod\`ele correspondent aux r\'eseaux de spins ! Autrement dit, nous utilisons une discr\'etisation telles que les variables de bord coincident avec l'alg\`ebre des holonomies-flux sur une triangulation ! Cela sugg\`ere l'existence d'un pont au niveau quantique entre le mod\`ele topologique $\Spin(4)$ et la LQG. Il nous faut donc au pr\'ealable clarifier la situation au niveau classique.

Pour cela, en plus de la th\'eorie BF, nous devons discr\'etiser les contraintes de simplicit\'e du continu pour en obtenir une forme qui sera adapt\'ee aux r\'eseaux de spins du bord. Cela est fait suivant l'approche standard, voir \cite{freidel-plebso4} par exemple. Nous donnons ici une preuve originale du fait que ces contraintes de simplicit\'e permettent \`a partir des variables de la th\'eorie BF sur r\'eseau de construire des g\'eom\'etries de Regge. Cela est \`a mettre en parall\`ele avec la discussion \'equivalente au niveau canonique, section \ref{sec:twisted-simplicial}, o\`u des contraintes de simplicit\'e ayant la m\^eme forme n'\'etaient pas suffisantes pour cr\'eer des g\'eom\'etries de Regge. Il en va donc diff\'eremment au niveau discret covariant (4-dimensionnel). Mais notre preuve donne plus que cela ! Elle offre un formalisme duquel les informations g\'eom\'etriques sont extraites tr\`es facilement, comme les corr\'elations entre les directions orthogonales \`a des t\'etra\`edres voisins, et les relations entre angles dih\'edraux 3d et 4d. Cela permet de montrer que la discr\'etisation des variables de Plebanski utilis\'ee est surtout proche de la formulation de calcul de Regge aires - angles 3d. En particulier, les contraintes apparaissant dans cette formulation et qui poss\`edent une forme compliqu\'ee, \eqref{constraints4dangles-aarc}, en prenne ici une tr\`es simple, comme produit d'\'el\'ements de $\SU(2)$ !

Un ingr\'edient essentiel pour \'etablir la correspondance entre la th\'eorie de Plebanski discr\'etis\'ee et la formulation aires-angles 3d du calcul de Regge est une prise en compte attentive des relations de transport parall\`ele sur le r\'eseau. Ces relations sont fondamentales pour d\'ecrire correctement les corr\'elations entre t\'etra\`edres voisins, obtenir les contraintes \eqref{constraints4dangles-aarc}, et ainsi assurer la consistence de la g\'eom\'etrie entre ces t\'etra\`edres. C'est donc ce qui permet d'obtenir la continuit\'e de la m\'etrique de Regge et de passer d'une g\'eom\'etrie tordue \`a une de Regge. Il me semble que l'importance de ces relations de transport parall\`ele avait jusqu'ici \'et\'e un peu n\'eglig\'ee, l'accent ayant \'et\'e mis sur l'amplitude de vertex, et la construction de celle-ci ayant notamment comme influence les travaux sur le t\'etra\`edre quantique dans lesquels les holonomies (et donc le transport parall\`ele) ne jouent pas de r\^ole.

Notre analyse est relativement proche des analyses semi-classiques effectu\'ees dans \cite{barrett-asymEPR, freidel-conrady-semiclass} pour l'asymptotique des plus r\'ecents mod\`eles de gravit\'e quantique. N\'eanmoins, il faut garder \`a l'esprit que notre formalisme d\'ecrit ici la \emph{formulation classique de la th\'eorie sur r\'eseau}. Avoir un cadre classique le plus clair possible et le mieux compris est essentiel avant la quantification, et ce formalisme sera utilis\'e dans les chapitres suivants pour d\'eriver les mod\`eles connus de gravit\'e quantique en calculant \emph{explicitement, de mani\`ere exacte,} la fonction de partition associ\'ee \`a cette formulation sur r\'eseau. Un point cl\'e est l'utilisation d'\'el\'ements de groupe pour coder les contraintes de simplicit\'e, particuli\`erement bien adapt\'ee \`a la d\'erivation de mod\`eles de mousses de spins, comme on l'a vu sur le mod\`ele topologique.

\section{Formulation classique sur r\'eseau et contraintes de simplicit\'e} \label{sec:classical sf}

Nous reprenons le cadre de d\'efinition du mod\`ele d'Ooguri, avec quelques ingr\'edients suppl\'ementaires. Pour constuire les mod\`eles nous proc\'edons \`a une discr\'etisation standard des variables, fixant en premier lieu une triangulation de la vari\'et\'e 4d d'espace-temps. L'avantage de ce choix est de pouvoir consid\'erer sur le graphe dual \`a la triangulation induite sur le bord d'un ensemble de 4-simplexes l'espace de Hilbert engendr\'e par les r\'eseaux de spins. Nous consid\'erons des rep\`eres locaux sur chaque 4-simplexe, comme dans le mod\`ele d'Ooguri, et en plus sur chaque 3-simplexe, i.e. t\'etra\`edre. La courbure est alors concentr\'ee autour des 2-simplexes, les triangles. Ce cadre de travail, inspir\'e du calcul de Regge, sera tout particuli\`erement adapt\'e pour \'etablir des liens pr\'ecis entre mousses de spins pour la th\'eorie BF, pour la gravit\'e quantique et calcul de Regge. Nous utiliserons le complexe dual \`a la triangulation de mani\`ere assez syst\'ematique. Dans ce point de vue, les triangles sont duaux aux faces, tous deux not\'es $f$, les t\'etra\`edres aux liens, not\'es $t$, et les 4-simplexes aux vertexes $v$. Le bord d'une face duale est form\'e de liens et de vertexes duaux aux t\'etra\`edres et 4-simplexes qui partagent le triangle $f$. Les orientations des t\'etra\`edres et des triangles induisent des orientations sur les liens et faces duaux.

Le groupe pertinent est $G=\Spin(4)$ que l'on verra comme deux copies de $\SU(2)$. La connection se discr\'etise comme dans le mod\`ele d'Ooguri, par des \'el\'ements de $\Spin(4)$ autorisant le transport parall\`ele entre diff\'erents rep\`eres locaux, \`a ceci pr\`es qu'il sera tr\`es pratique de pouvoir s'arr\^eter dans les rep\`eres des t\'etra\`edres. Nous prenons donc des \'el\'ements de $\Spin(4)$, $G_{vt}$, entre les t\'etra\`edres et les 4-simplexes. Du point de vue dual, ces \'el\'ements sont attach\'es \`a des demi-liens, et le transport entre deux 4-simplexes adjacents, i.e. \`a travers leur t\'etra\`edre commun, est donn\'e par $G_{vv'} = G_{vt}G_{tv'}$, avec $G_{tv'}= G_{v't}\mone$, si $v$ et $v'$sont les vertexes source et arriv\'ee du lien dual \`a $t$. La courbure autour d'un triangle est ainsi mesur\'ee par le produit orient\'e de ces \'el\'ements de groupe le long du bord de sa face duale $f$, en partant d'un vertex (ou d'un t\'etra\`edre) de r\'ef\'erence $v^\star$, $G_f(v^\star) = \prod_{t\subset \pp f} G_{vt} G_{v't}\mone$. La condition de courbure nulle se lit alors : $G_f(v)=\unit$, pour chaque $f$.


Dans la th\'eorie, topologique, BF nous construisions la fonction de partition simplement en imposant la contrainte de courbure nulle. Mais en relativit\'e g\'en\'erale le champ $B$ de Plebanski est soumis aux contraintes de simplicit\'e. Nous avons donc besoin de discr\'etiser \'egalement celui-ci. En tant que 2-forme, comme mentionn\'e en introduction, il se discr\'etise naturellement par des variables de l'alg\`ebre $\spin(4)$ sur les triangles, appel\'ees \emph{bivecteurs}, dans des r\'ef\'erentiels associ\'es \`a des 4-simplexes $v$ ou \`a des t\'etra\`edres $t$, et not\'ees $B_f(v)$ ou $B_f(t)$. Des expressions de $B_f$ dans des r\'ef\'erentiels diff\'erents sont reli\'ees par transport parall\`ele le long du bord de $f$, ce qui se comprend comme une discr\'etisation de l'\'equation de torsion nulle, $d_A\,B=0$ :
\beq \label{transport bivectors vt}
B_f(v) = \Ad\bigl(G_{vt}\bigr)\ B_f(t) = G_{vt}\,B_f(t)\,G_{vt}\mone,
\ee
pour chaque 4-simplexe $v$ dont le bord contient $t$ (qui contient $f$). Cette \'equation est plus importante qu'elle peut le para\^itre dans la th\'eorie BF, car c'est elle qui permettra la consistence de la g\'eom\'etrie dans les mod\`eles pour la gravit\'e quantique.

Les transformations de jauge se comportent comme dans la th\'eorie BF, suivant la m\^eme r\`egle que l'\'equation pr\'ec\'edente. Une transformation de jauge $K$ est une famille d'\'el\'ements du groupe, $(K_t, K_v)$ agissant par :
\begin{align} \label{gauge transfo spin4}
&K\,\cdot\, G_{vt} = K_v\,G_{vt}\,K_t\mone, \\
&K\,\cdot\, B_f(t) = K_t\,B_f(t)\,K_t\mone,
\end{align}
en notation matricielle. Comme les variables $B_f(t)$ discr\'etisent la 2-forme $B$ sur les triangles, et que comme en calcul de Regge, nous avons concentr\'e la courbure \'egalement sur les triangles, nous pouvons former une action invariante de jauge de type BF discr\`ete ! Elle s'exprime sous la forme d'une somme sur les triangles, comme en calcul de Regge,
\beq \label{discreteBFaction}
S_{\mathrm{BF}}\bigl(B_f(t),G_{vt}\bigr) = \sum_f \Tr\bigl(B_f(t)\,G_f(t)\bigr),
\ee
dans laquelle les variables $B_f(t)$ apparaissent comme multiplicateurs de Lagrange pour la courbure $G_f$, \cite{fk-action-principle}. La trace sur $\Spin(4)\simeq \SU(2)\times\SU(2)$ est prise comme la somme de la trace sur la partie $\SU(2)$ self-duale, $\tr(b_{+f}(t)\,g_{+f}(t))$, et celle sur la partie $\SU(2)$ anti-self-duale, not\'ee $\tr(b_{-f}(t)\,g_{-f}(t))$. Du fait des r\`egles simples de transport parall\`ele, l'action est naturellement ind\'ependante du t\'etra\`edre de r\'ef\'erence choisi sur chaque triangle pour d\'efinir $B_f(t)$ et $G_f(t)$. On peut ainsi choisir comme variables ind\'ependantes les \'el\'ements $G_{vt}$ et un $B_f(t)$ par face.


Les choses ne sont plus si simples lorsque l'on introduit des contraintes de simplicit\'e. On s'attend \`a ce que les $B_{f}(t)$ ne soient plus de simples multiplicateurs. En effet, les contraintes de simplicit\'e discr\'etis\'ees \`a partir des contraintes du continu vont permettre un changement de variables entre les variables $B_f(t)$ et des 4-vecteurs $E^I_\ell(t)$ qui d\'ecrivent la direction (dans l'espace interne) des ar\^etes $\ell$ de la triangulation. Ces 4-vecteurs vont former une discr\'etisation consistente de la cot\'etrade. Leurs normes donnent acc\`es aux longueurs des ar\^etes, qui d\'efinissent une g\'eom\'etrie de Regge. De mani\`ere analogue au continu, les variables $B_f(t)$ vont s'\'ecrire : $(\star B_f(t))^{IJ} = E_{\ell_1}^{[I} E_{\ell_2}^{J]}$, si $\ell_1, \ell_2$ sont deux ar\^etes du triangle $f$ (et les crochets indiquant l'antisym\'etrisation). On dit alors que les variables $B_f^{IJ}$ sont des bivecteurs \emph{simples}. Comme tous les ensembles de 10 bivecteurs (pour les dix triangles d'un 4-simplexe) ne d\'ecrivent pas n\'ecessairement la g\'eom\'etrie d'un 4-simplexe plat, nous avons besoin de contraintes de simplicit\'e. Celles-ci proviennent du th\'eor\`eme de Barrett et Crane \cite{BCpaper}, ou bien de la discr\'etisation des contraintes de simplicit\'e du continu sur un 4-simplexe,
\begin{equation*}
\eps_{IJKL}\,B^{IJ}_{\mu\nu}\,B^{KL}_{\lambda\sigma} \propto \eps_{\mu\nu\lambda\sigma},
\end{equation*}
o\`u $\mu,\nu,\lambda,\sigma$ sont des indices d'espace-temps. Lorsque les deux exemplaires de $B$ ci-dessus sont int\'egr\'es sur un m\^eme triangle ou sur deux triangles partageant une ar\^ete, cela donne les contraintes de simplicit\'e dite diagonale et crois\'ee, exactement comme dans la description canonique, section \ref{sec:twisted-simplicial}. Lorsque $\mu, \nu, \rho$ et $\sigma$ sont pris selon quatre directions ind\'ependantes, la contrainte exprime que la quantit\'e $(\star B_f(v))_{IJ} B_{f'}^{IJ}(v)$ est ind\'ependante de la paire de triangles consid\'er\'ee dans le 4-simplexe $v$, pourvu que ceux-ci ne se rencontrent qu'en un sommet, et elle repr\'esente alors le 4-volume de $v$,
\beq \label{volume constraint}
\Tr\,\bigl(\star B_f(v)\,B_{f'}(v)\bigr) \propto \calV_v.
\ee
Une telle contrainte impliquant la g\'eom\'etrie 4d du 4-simplexe n'\'etait \'evidemment pas pr\'esente dans la description des contraintes de simplicit\'e au niveau canonique, section \ref{sec:twisted-simplicial}. Mais nous avions vu qu'alors des contraintes de recollement, \eqref{edge simplicity}, ressemblant plus ou moins aux contraintes secondaires de la th\'eorie continue, devaient se rajouter aux contraintes diagonales et crois\'ees ! On peut donc imaginer qu'en imposant les contraintes primaires, i.e. diagonales et crois\'ees, sur \emph{tous} les t\'etra\`edres de la triangulation 4d, nous pourrons nous passer de ces contraintes secondaires et oublier du m\^eme coup la contrainte sur le 4-volume \eqref{volume constraint}. C'est effectivement le cas, et la contrainte \eqref{volume constraint} appara\^it bien comme une contrainte secondaire \cite{new-model-ls}, si l'on utilise pleinement la relation de fermeture des t\'etra\`edres comme une des contraintes de simplicit\'e ! Nous prenons donc :
\begin{alignat}{2}
&\text{Simplicit\'e diagonale}&&\qquad\qquad\qquad \Tr\,\bigl(\star B_f(t)\, B_f(t)\bigr) = 0, \label{diag}\\
&\text{Simplicit\'e crois\'ee}&&\qquad\qquad\qquad \Tr\,\bigl(\star B_f(t)\, B_{f'}(t)\bigr) = 0, \label{off-diag}\\
&\text{Relation de fermeture}&&\qquad\qquad\qquad \sum_{f\subset \pp t} \eps_{ft}\,B_f(t) = 0, \label{closure}
\end{alignat}
o\`u $\eps_{ft}=\pm1$ selon l'orientation relative de la face duale $f$ et du lien dual $t$. A la seconde ligne, les triangles $f$ et $f'$ sont des triangles d'un m\^eme t\'etra\`edre (de mani\`ere \'equivalente, partageant une ar\^ete dans un 4-simplexe). Par ailleurs, l'op\'erateur $\star$ de Hodge est d\'efini comme dans les sections pr\'ec\'edentes, $(\star B)^{IJ} = \f12\eps^{IJ}_{\phantom{IJ}KL}\,B^{KL}$. La premi\`ere contrainte, diagonale, exprime le fait que chaque bivecteur doit \^etre simple, i.e. le produit antisym\'etris\'e de deux 4-vecteurs, $B_f^{IJ} = u^{[I}v^{J]}$ (et de m\^eme pour son dual de Hodge). La deuxi\`eme contrainte, crois\'ee, demande que la somme de deux bivecteurs d'un t\'etra\`edre soit aussi simple. Finalement, la relation de fermeture implique que les t\'etra\`edres sont bien ferm\'es. Nous gagnons clairement au change en passant de \eqref{volume constraint} \`a la relation de fermeture, d'autant plus que celle-ci g\'en\`ere via les crochets de Poisson les transformations de jauge $\Spin(4)$. Dans les proc\'ed\'es de quantification en mousses de spins, cette contrainte est donc simplement impl\'ement\'ee en s'assurant de l'invariance de jauge des amplitudes.

Ces contraintes, qui sont formul\'ees pour chaque t\'etra\`edre, permettent de reconstruire la g\'eom\'etrie de chacun d'eux. Il est tout \`a fait remarquable qu'elles soient en fait suffisantes pour reconstruire la g\'eom\'etrie de chaque 4-simplexe ! Cela simplement en les imposant pour les cinq t\'etra\`edres d'un 4-simplexe, et en d\'efinissant les bivecteurs d'un 4-simplexe par transport parall\`ele entre les t\'etra\`edres et le 4-simplexe consid\'er\'e, avec $G_{vt}$. Du point de vue canonique, section \ref{sec:twisted-simplicial}, i.e. avec seulement une triangulation 3d de bord, nous avions vu que m\^eme pour reconstruire le bord d'un 4-simplexe \`a la Regge, des contraintes suppl\'ementaires de recollement \eqref{edge simplicity} (ou \eqref{2dconstraints-aarc}, ou \eqref{constraint 3dangle-area}) sont n\'ecessaires. Ce n'est pas le cas ici, et il s'agit clairement d'un effet du changement de dimension : du passage du bord \`a l'int\'erieur d'une triangulation 4d. Plus pr\'ecis\'ement, nous verrons que les contraintes de simplicit\'e \eqref{diag}, \eqref{off-diag} et la condition de transport parall\`ele des bivecteurs,
\beq \label{transport bivectors}
B_f(t) = \Ad\bigl(G_{tt'}\bigr)\ B_f(t')  
\ee
entre t\'etra\`edres adjacents, tir\'ee de \eqref{transport bivectors vt}, assurent ensemble que des contraintes \'equivalentes \`a celles de recollement \eqref{2dconstraints-aarc} sont bien satisfaites. Le point crucial qui permet ceci est le fait d'\^etre autoris\'e dans le cadre 4d \`a placer tous les bivecteurs d'un 4-simplexe dans un r\'ef\'erentiel commun, celui du 4-simplexe. Autrement dit, les variables ind\'ependantes sur le bord sont des holonomies entre t\'etra\`edres, $G_{tt'}$, alors qu'avec une triangulation 4d authentique, ce sont des holonomies entre t\'etra\`edres et 4-simplexes, $G_{tv}$. En particulier, dans la condition de transport parall\`ele \eqref{transport bivectors}, le fait d'avoir la d\'ecomposition : $G_{tt'} = G_{vt}\mone G_{vt'}$, sera fondamental pour nous. Pour un 4-simplexe, il y a 5 \'el\'ements $G_{vt}$, mais 10 holonomies $G_{tt'}$ sur le bord, d'o\`u les contraintes suppl\'ementaires si l'on consid\`ere les holonomies du bord comme variables ind\'ependantes !


Ces contraintes poss\`edent les m\^emes ambiguit\'es que dans le continu : elles impliquent que les bivecteurs d'un 4-simplexe, ou leurs duaux de Hodge, sont les produits antisym\'etris\'es de 4-vecteurs sur les ar\^etes, \`a un signe pr\`es\footnotetext{Cette ambiguit\'e de signe peut \^etre \'elimin\'ee gr\^ace \`a une condition de positivit\'e sur le volume des t\'etra\`edres, comme dans \cite{BCpaper}. On ne sait n\'eanmoins pas du tout comment traiter cette condition dans les mod\`eles de mousses de spins.}, $\eps=\pm$. L'ambiguit\'e entre les bivecteurs et leurs duaux de Hodge est bien s\^ur celle qui dans le continu fait passer du secteur gravitationnel, l'action de Palatini-Hilbert, au secteur topologique, i.e. le terme topologique ajout\'e pour former l'action de Holst. En pr\'esence du param\`etre d'Immirzi, nous avons vu que le passage d'un secteur \`a l'autre dans la r\'esolution des contraintes de simplicit\'e ne fait que changer le param\`etre d'Immirzi en son inverse. N\'eanmoins il est important de pouvoir distinguer quel secteur nous prenons comme r\'ef\'erence (notamment pour comprendre les limites $\gamma\rightarrow 0$ et $\gamma\rightarrow\infty$). Cela peut se faire d'une mani\`ere extr\`emement g\'eom\'etrique, et c'est ce qui a permis de s'extraire du mod\`ele de Barrett-Crane pour construire le mod\`ele EPR \cite{epr-short} ! L'id\'ee est que les contraintes de simplicit\'e formul\'ees dans un t\'etra\`edre nous disent que les triangles (ou ar\^etes) de ce t\'etra\`edre n'engendrent qu'un sous-espace de dimension 3 de l'espace interne $\R^4$. Autrement dit, il existe un 4-vecteur unitaire $N^I_t$ d\'efinissant la direction orthogonale au t\'etra\`edre de sorte que
\beq
\eps_{IJKL}\,B_f^{IJ}(t)\,N_t^K = 0, \label{off-diag grav}
\ee
pour le secteur gravitationnel, ou
\beq
B_f^{IJ}(t)\,N_{tI} =0, \label{off-diag topo}
\ee
pour le secteur topologique. Comme il semble bien que la quantification propos\'ee en mousses de spins pour le secteur topologique ne donne pas vraiment un mod\`ele topologique (!), i.e ind\'ependant de la triangulation, nous parlerons souvent, dans le cadre des mousses de spins, plut\^ot des secteurs \emph{g\'eom\'etrique} et \emph{non-g\'eom\'etrique}. Le choix de \eqref{off-diag grav} ou \eqref{off-diag topo} remplace alors la contrainte de simplicit\'e crois\'ee. Remarquons que dans le secteur g\'eom\'etrique, le 4-vecteur $B_f^{IJ}(t)\,N_{tI}$ donne la direction orthogonale au triangle au sein du sous-espace 3d orthogonal \`a $N_t$.

Nous avons dit que ces contraintes permettent de reconstuire la g\'eom\'etrie plate de chaque 4-simplexe \`a partir des t\'etra\`edres. Se pose alors la question du collage des simplexes pour reconstruire la g\'eom\'etrie de la triangulation enti\`ere. En fait, il suffit \'egalement que les bivecteurs d\'efinis dans les rep\`eres de t\'etra\`edres se transportent de mani\`ere consistente pour chaque triangle, suivant \eqref{transport bivectors}, comme le prouvent Freidel et Conrady dans \cite{freidel-conrady-semiclass} (en utilisant plut\^ot les rep\`eres des 4-simplexes). (Remarquons alors que l'ambiguit\'e de signe $\eps$ est globale).

\section{La g\'eom\'etrie \`a partir des bivecteurs et des normales} \label{sec:geom normales}

On notera ici : $b_{\pm f} = \vec{b}_{\pm f}\cdot \vec{\tau}\in\su(2)$, pour des g\'en\'erateurs $(\tau_i)_{i=1,2,3}$ anti-hermitiens satisfaisant $[\tau_i, \tau_j] = -\eps_{ij}^{\phantom{ij}k}\tau_k$. Ainsi, les composantes $b_{\pm f}^i$ peuvent \^etre vues comme celles d'un 3-vecteur r\'eel. De plus, la trace est telle que : $\vec{a}\cdot\vec{b} = -2\tr(ab) $.

Puisque les bivecteurs discr\'etisent une 2-forme \`a valeurs dans l'alg\`ebre $\spin(4)$ sur les triangles, comme dans la section \ref{sec:twisted-simplicial}, ils donnent acc\`es aux m\^emes grandeurs g\'eom\'etriques, aires et angles dih\'edraux 3d, soit :
\beq
A_{f}^2 = -\Tr\bigl(B_{f}^2(t)\bigr) = \vec{b}_{+f}^2(t),
\ee
pour les aires, qui ne d\'ependent bien s\^ur pas du r\'ef\'erentiel. Comme dans l'analyse canonique, le secteur self-dual est suffisant du fait des contraintes de simplicit\'e diagonale. Quant \`a l'angle dih\'edral 3d $\phi^t_{ff'}$ entre les triangles $f$ et $f'$ dans le t\'etra\`edre $t$, il s'obtient par le produit scalaire des 4-vecteurs $N_{tJ}\,B_{ft}^{IJ}$ et $N_{tJ}\,B_{f't}^{IJ}$ qui repr\'esentent respectivement les directions orthogonales aux triangles $f$ et $f'$ dans le sous-espace 3d engendr\'e par $t$. On montre \cite{epr-long} que cela revient \`a calculer la trace $\Tr (B_{ft} B_{f't})$, donc :
\beq \label{3d angle}
\cos \phi_{ff'}^t = -\eps_{ff'}\ \f{\vec{b}_{+f}(t)\cdot \vec{b}_{+f'}(t)}{A_f\,A_{f'}},
\ee
o\`u $\eps_{ff'}=\eps_{ft}\eps_{f't}$ repr\'esente l'orientation relative des faces duales \`a $f$ et $f'$.

Puisque nous avons introduit de nouvelles variables, les normales $N_t$, il est judicieux de regarder quelles quantit\'es g\'eom\'etriques permettent-elles de d\'ecrire. Nous donnons ainsi un premier coup d'oeil aux variables g\'eom\'etriques (scalaires) et aux relations qui existent entre les bivecteurs et les normales aux t\'etra\`edres. Cela nous sera tr\`es utile dans la suite, pour comprendre les d\'efauts du mod\`ele de Barrett-Crane, ou encore pour extraire le calcul de Regge de ces contraintes. Cette discussion est issue de \cite{BCpaper-val, bf-aarc-val} (voir aussi des r\'esultats similaires, ind\'ependamment, dans \cite{barrett-asymEPR}).

On \'ecrit les contraintes de simplicit\'e crois\'ee \`a l'aide de la formulation \eqref{off-diag grav}. En utilisant les parties self-duales et anti-self-duales, $B_{f} = b_{+f}\oplus b_{-f}$, cela donne en notations matricielles :
\beq \label{simplicity-selfdual}
b_{-f}(t) +s\,\Ad\bigl( N_t\mone\bigr)\ b_{+f}(t) =0.
\ee
La variable $N_t$ n'est autre que le 4-vecteur $N_t^I$ unitaire et orthogonal au t\'etra\`edre, vu comme un \'el\'ement de $\SU(2)$ par l'isomorphisme entre $\SU(2)$ et $S^3$. De mani\`ere \'equivalente, la rotation $N_t$ est d\'efinie de sorte que la rotation (4d) $(N_t, \unit)\in\Spin(4)$ envoie le vecteur de r\'ef\'erence $N^{(0)}= (1,0,0,0)$ sur la normale $N_t^I$, par la repr\'esentation standard de $\SO(4)$. Le signe $s=\pm$ indique la possibilit\'e de choisir comme r\'ef\'erence le secteur topologique, $s=-1$, ou le secteur gravitationnel, comme nous le ferons dans la suite, avec $s=1$. Remarquons que pour des bivecteurs ind\'ependants, la normale $N_t$ ne souffre aucune ambiguit\'e, except\'e un changement d'orientation.

Puisque la simplicit\'e crois\'ee indique l'existence d'une normale au t\'etra\`edre, on peut imaginer que la simplicit\'e diagonale \eqref{diag} s'interpr\`ete aussi en termes de vecteurs $N_f$ orthogonaux au triangle $B_{f}$, \`a savoir :
\beq \label{diag normales}
\eps_{IJKL}\,B_f^{IJ}\,N_f^K = 0.
\ee
C'est bien s\^ur le cas, comme nous le montrons. Dans les variables pr\'ec\'edentes, la contrainte \eqref{diag} donne : $\vec{b}_{+f}^2 = \vec{b}_{-f}^2$, ind\'ependamment du r\'ef\'erentiel choisi. Autrement dit, il existe une rotation $N_f\in\SU(2)$ qui envoie la partie self-duale sur la partie anti-self-duale,
\be \label{diag sol}
b_{-f} + \Ad\bigl(N_f^{-1}\bigr)\ b_{+f} = 0.
\ee
Naturellement, la rotation $N_f$ d\'efinit aussi une normale via l'action de $(N_f,\unit)\in\Spin(4)$ sur $N^{(0)}$. De plus, un bivecteur simple dans $\R^4$ poss\`ede un \emph{plan} orthogonal. Il n'y a donc pas un choix unique pour $N_f$. On voit en fait qu'\'etant donn\'e un choix de r\'ef\'erence, toutes les rotations de la forme $\exp(\varphi\,\hat{b}_{+f})N_f$, o\`u $\hat{b}_{+f}= b_{+f}/\lvert\vec{b}_{+f}\rvert \in\su(2)$ repr\'esente le vecteur unitaire dans la direction de $b_{+f}$, satisfont aussi \eqref{diag sol}. Cela signifie que le sous-groupe $\U(1)$ engendr\'e par $\hat{b}_{+f}$ permet de parcourir l'ensemble des normales du plan orthogonal \`a $B_f$ ! Plus pr\'ecis\'ement encore, si $\exp(2\varphi\,\hat{b}_{+f})$ relie deux normales $N_f$ et $\tl{N}_f$, alors $\varphi$ n'est autre que l'angle entre ces deux normales ! En effet, un tel angle est donn\'e par le produit scalaire des 4-vecteurs, $N_f\cdot\tl{N}_f$, qui se r\'e\'ecrit avec les \'el\'ements de groupe comme :
\beq
N_f\cdot\tl{N}_f = \f12\ \tr\,\bigl( N_f\mone\,\tl{N}^{\phantom{}}_f\bigr)  = \f12\ \tr\,\exp(2\varphi\,\hat{b}_{+f}) = \cos \varphi.
\ee

La contrainte de simplicit\'e crois\'ee \'etablit alors qu'il existe une normale commune $N_t$ aux quatre triangles d'un t\'etra\`edre. Pour chaque triangle, cette normale s'obtient en tournant une normale de r\'ef\'erence $N_f$ dans le plan orthogonal \`a $\star B$, d'un angle $\varphi_{ft}$, ce qui dans notre langage se formule avec le sous-groupe $\U(1)_{\hat{b}_{+f}}$ stabilisant $b_{+f}$ de sorte que :
\beq
N_t = \exp( 2\varphi_{ft}\,\hat{b}_{+f}(t))\ N_f(t).
\ee
Notons que $N_t$ se laisse d\'efinir le plus naturellement dans le rep\`ere local du t\'etra\`edre $t$. Nous pouvons d\'eduire de cette \'equation, du fait que $N_f$ ne d\'epend que du triangle $f$, \`a un changement de r\'ef\'erentiel pr\`es par transport parall\`ele, que les normales \`a deux t\'etra\`edres $t_1$ et $t_2$ coll\'es le long de $f$ sont reli\'ees : il existe un angle $\varphi_{t_1t_2}$ tel que
\beq \label{coupling normals}
N_{t_1}(v)\,N_{t_2}(v)\mone = \exp(2\varphi_{t_1t_2}\,\hat{b}_{+f}(v)).
\ee
Nous avons \'ecrit toutes les quantit\'es dans un r\'ef\'erentiel commun, disons celui du 4-simplexe $v$ dont le bord contient $t_1$ et $t_2$. La notation vient du fait que la paire $(f,v)$ caract\'erise alors la paire de t\'etra\`edres. L'interpr\'etation de cette relation est imm\'ediate. Puisque $t_1$ et $t_2$ partagent un triangle, leurs normales respectives ne sont pas ind\'ependantes : elles sont toutes deux contenues dans le plan orthogonal au triangle, et s\'epar\'ees d'un angle $\varphi_{t_1t_2}$ ! Ainsi, en pr\^etant attention \`a ce sous-groupe stabilisant $b_{+f}$, on peut esp\'erer obtenir l'angle dih\'edral interne 4d $\thet_{t_1 t_2}$ entre les t\'etra\`edres, simplement en suivant la formule :
\beq \label{4d-angles def}
\cos\thet_{t_1 t_2} = - N_{t_1}(v)\cdot N_{t_2}(v) = -\cos\varphi_{t_1t_2}.
\ee
Ces relations, et en particulier \eqref{coupling normals}, ne sont pas apparues tout de suite dans la litt\'erature, qui visait surtout \`a exploiter la formulation \eqref{off-diag grav} au niveau quantique. C'est pourtant en regardant le d\'etail des structures expos\'ees ici que j'ai pu traduire les variables des mousses de spins en calcul de Regge classiquement, et d\'eriver les mod\`eles quantiques par int\'egrales sur les g\'eom\'etries du r\'eseau.

Pour transporter explicitement les normales $N_t$, nous devons conna\^itre leurs propri\'et\'es de transformations. Sous une transformation de jauge $K$, on d\'eduit de \eqref{simplicity-selfdual},
\beq
K\, \cdot\, N_t \,=\, k_{+t}\,N_t\,k_{-t}\mone.
\ee
Cette loi de transformation particuli\`ere implique que l'on peut toujours choisir la jauge \og temps\fg, dans laquelle $N_t = \unit$, ou de mani\`ere \'equivalente $N_t^I = N^{(0)I}$, pour tous les t\'etra\`edres. Dans les discussions, le sous-groupe de $\Spin(4)$ isomorphe \`a $\SU(2)$ qui laisse invariant la normale $N_t$ interviendra r\'eguli\`erement. Dans la jauge temps, il s'agit simplement du sous-groupe $\SU(2)$ diagonal, dont les \'el\'ements sont de la forme $(k, k)$. Pour une normale quelconque, ce sont les rotations du sous-espace 3d engendr\'e par le t\'etra\`edre $t$ dans $\R^4$,
\begin{align}
\SU(2)_{N_t} &= \bigl\{ K\in\Spin(4), K\, \cdot\, N_t \,=\, N_t \bigr\},\\
&= \bigl\{ K = \bigl(N_t^{\phantom{}}\,k_-\,N_t\mone, k_-\bigr),\ k_-\in\SU(2) \bigr\},\\
&= \bigl\{ K = \bigl(k_+, N_t\mone\,k_+\,N_t^{\phantom{}}\bigr),\ k_+\in\SU(2) \bigr\}.
\end{align}
On peut aussi identifier les \og boosts\fg, pour lesquels la direction \og temps\fg{} est bien s\^ur celle du vecteur $N_t^I$, et qui sont de la forme : $K = (N_t\,k\,N_t\mone, k\mone)$ pour $k\in\SU(2)$.

Le transport parall\`ele s'effectue donc de la m\^eme mani\`ere, permettant de d\'efinir la normale au t\'etra\`edre $t_2$ dans le r\'ef\'erentiel du t\'etra\`edre voisin $t_1$ : $N_{t_2}(t_1) = g_{+t_1t_2}\,N_{t_2}\,g_{-t_1t_2}\mone$. Cette formule permet de v\'erifier explicitement le contenu de \eqref{coupling normals} en montrant que $N_{t_1}\,N_{t_2}(t_1)\mone$ pr\'eserve $b_{+f}(t_1)$. Les ingr\'edients \`a  utiliser sont les contraintes de simplicit\'e crois\'ee pour $t_1, t_2$ et le transport parall\`ele de $B_{f}(t_1)$ en $t_2$ :
\begin{alignat}{2}
\Ad \Bigl(N_{t_1}\,N_{t_2}(t_1)\mone\Bigr)\ b_{+f}(t_1) &= &&\Ad\bigl(N_{t_1}\,g_{-t_1t_2}\,N_{t_2}\mone\bigr)\ b_{+f}(t_2),\\
&=&\, - &\Ad\bigl(N_{t_1}\,g_{-t_1t_2}\bigr)\ b_{-f}(t_2),\\
&=&\, - &\Ad\bigl(N_{t_1}\bigr)\ b_{-f}(t_1),\\
&=&&\ b_{+f}(t_1).
\end{alignat}
Nous avons pr\'esent\'e naturellement les angles dih\'edraux comme les angles entre les normales, exprim\'ees dans un r\'ef\'erentiel commun,
\beq
\cos\thet_{t_1 t_2} = -\f12\,\tr\,\bigl(g_{+t_1 t_2}^{\phantom{}}\,N_{t_2}^{\phantom{}}\,g_{-t_1t_2}\mone\,N_{t_1}\mone\bigr).
\ee
Ainsi, dans la jauge temps, $N_t=\unit$, les angles dih\'edraux se retrouvent contenus dans les holonomies uniquement ! Munis des angles dih\'edraux, nous pouvons essayer d'extraire la courbure, concentr\'ee autour des triangles, et caract\'eris\'ee par les angles de d\'eficit, $\vareps_f = 2\pi-\sum_{(tt')}\thet_{tt'}$. Comme nous le montrons pr\'ecis\'ement \`a la section \ref{sec:gluing-aarc}, les angles de d\'eficit s'obtiennent par la courbure vue sur le groupe $G_f(t)$ comme :
\beq \label{deficithol}
g_{+f}(t)\,N_t\,g_{-f}\mone(t)\,N_t\mone = \exp\bigl( 2\,\vareps_f\,\hat{b}_{+f}(t)\bigr),
\ee
$\vareps_f$ ne d\'ependant pas du choix du t\'etra\`edre de r\'ef\'erence. En particulier, si $G_f(t)$ est dans le sous-groupe $\SU(2)$ qui pr\'eserve la normale $N_t$, i.e. une rotation 3d du t\'etra\`edre dans le sous-espace orthogonal \`a $N_t^I$, alors l'angle de d\'eficit est nul,
\beq \label{zero deficithol}
\vareps_f \,=\, 0 \quad \Leftrightarrow \quad G_f(t) = \bigl(N_t^{\phantom{}}\,g_{-f}(t)\,N_t\mone, g_{-f}(t)\bigr)\,\in\,\SU(2)_{N_t}.
\ee

Nous sommes ici dans un formalisme du premier ordre, o\`u holonomies et bivecteurs sont initialement des variables ind\'ependantes. N\'eanmoins, on s'attend bien s\^ur \`a ce que les contraintes de simplicit\'e, la relation de fermeture des t\'etra\`edres et le transport parall\`ele des bivecteurs (tout ce qui permet de reconstruire une cot\'etrade discr\`ete) nous fassent passer dans un formalisme du second ordre, dans lequel holonomies et angles dih\'edraux sont des fonctions des vecteurs de la cot\'etrade discr\`ete -- c'est ce que nous montrons en d\'etails \`a la section \ref{sec:gluing-aarc}. Ainsi, bien que les \'equations compl\`etes de la dynamique n'aient pas \'et\'e \'ecrites dans les variables holonomies/bivecteurs, on s'attend {\it in fine} \`a pouvoir les formuler d'une mani\`ere \'equivalente aux \'equations de Regge, approchant les \'equations d'Einstein sur la triangulation. Nous souhaitons ici faire une remarque plus modeste qu'un tel objectif, qui justifie \eqref{deficithol} et d\'ecrit l'action g\'en\'erique de l'holonomie de courbure sur les normales aux t\'etra\`edres.

Des \'equations du mouvement de la th\'eorie continue, on tire facilement que : 
\beq
d_A^2\,B \,=\, [F(A), B] \,=\, 0.
\ee
Sur la triangulation, une telle \'equation devient :
\beq
\Ad\bigl(G_f(t)\bigr)\,B_f(t) \,=\, B_f(t),
\ee
qui est bien s\^ur v\'erifi\'ee du simple fait des relations de transport parall\`ele \eqref{transport bivectors vt} tout le long du bord des faces duales. Il existe deux sous-groupes $\U(1)$ pr\'eservant un bivecteur donn\'e, celui g\'en\'er\'e par $B$ et celui g\'en\'er\'e par $\star B$, ou de mani\`ere \'equivalente, g\'en\'er\'es par $b_+$ et par $b_-$. En introduisant les angles $\theta^+_f$ et $\theta^-_f$ correspondants, et en utilisant les contraintes de simplicit\'e, cela donne :
\beq
G_f(t) \,=\, \Bigl(\exp\bigl(2\,\theta^+_f\,\hat{b}_{+f}(t)\bigr),\, N_t\mone\,\exp\bigl( 2\,\theta^-_f\,\hat{b}_{+f}(t)\bigr)\,N_t^{\phantom{}}\Bigr).
\ee
En utilisant \eqref{deficithol}, on relie ces angles \`a l'angle de d\'eficit,
\beq
\vareps_f \,=\, \theta^+_f \,-\,\theta^+_f,
\ee
et on v\'erifie imm\'ediatment que $\theta^+ = \theta^-$ correspond bien \`a $G_f(t)\in\SU(2)_{N_t}$. De mani\`ere g\'en\'erique, on comprend donc que l'holonomie de courbure $G_f(t)$ ne laisse pas la normale invariante, mais la transforme \`a la place en un autre vecteur toujours contenu dans le plan orthogonal \`a $\star B_f(t)$,
\beq
\eps_{IJKL}\ \bigl[G_f(t)\,\cdot\,N_t\bigr]^J\,B_f^{KL}(t) \,=\, 0.
\ee
\'Ecrit ainsi, il devient \'evident que l'angle entre les deux vecteurs unitaires $N_t$ et $G_f(t)\cdot N_t$ dans le plan orthogonal \`a $\star B_f(t)$ n'est autre que l'angle de d\'eficit $\vareps_f$ autour du triangle $f$, ce qui est pr\'ecis\'ement le contenu de l'\'equation \eqref{deficithol} (dans le langage $\SU(2)$) !

\section{Sur le collage des simplexes} \label{sec:collage}

Comme nous allons le d\'etailler dans les sections suivantes, les mod\`eles de mousses de spins sont construits en imposant ces contraintes de simplicit\'e sur chaque simplexe/t\'etra\`edre. Se pose ensuite la question du collage (consistent !) de ces simplexes g\'eom\'etriques. Au niveau classique, ce collage n'est autre que les relations de transport parall\`ele \eqref{transport bivectors} qui assurent la consistence de la g\'eom\'etrie, et contient des informations cruciales telles que les relations entre les normales des t\'etra\`edres adjacents \eqref{coupling normals}. Dans les mousses de spins, les m\'ethodes utilis\'ees ne sont pas explicitement mentionn\'e, et il y a eu peu d'\'etudes d\'etaill\'ees concernant l'influence de diff\'erents proc\'ed\'es (voir \cite{oriti-gluing}, et r\'ecemment \cite{freidel-conrady-pathint}, et plus sp\'ecifiquement \cite{regoli-face}). Si l'on peut penser que le collage ne va pas vraiment jouer sur la structure alg\'ebrique de l'amplitude que le mod\`ele associe \`a chaque 4-simplexe, il est clair qu'il influence le choix des donn\'ees de bord (c'est-\`a-dire quelles variables sont somm\'ees pour \'evaluer les amplitudes), et les amplitudes associ\'ees aux triangles (et donc comment les amplitudes des 4-simplexes sont somm\'ees).

Les m\'ethodes utilis\'ees reposent en g\'en\'eral sur la fa\c{c}on dont s'effectue le collage dans le mod\`ele d'Ooguri. A ce titre, une m\'ethode souvent employ\'ee (voir \cite{freidel-conrady-pathint} et \cite{oriti-gluing}) dans la perspective de d\'eriver des mod\`eles \`a partir d'int\'egrales de chemins sur r\'eseau, est d\'evelopp\'ee dans \cite{fk-action-principle}, pour traiter d'\'eventuelles perturbations autour du mod\`ele topologique. Cette m\'ethode, que nous utiliserons aussi et d\'ecrirons dans le cadre du mod\`ele BC, chapitre \ref{sec:BC} (tir\'e de \cite{BCpaper-val}), repose sur un \'eclatement de la triangulation en 4-simplexes ou t\'etra\`edres isol\'es, suivi d'un processus de collage implicite et indirect impliquant des holonomies de bord sur chaque 4-simplexe. Si celui-ci fonctionne bien dans le mod\`ele topologique, il s'av\`ere qu'il n'est plus efficace en pr\'esence des contraintes de simplicit\'e, comme nous le montrons en chapitre \ref{sec:BC} (du moins s'il est pris tel quel, car \cite{freidel-conrady-pathint} s'en accomode, avec d'autres ingr\'edients).

Pour situer les enjeux, prenons cette perspective d'int\'egrales sur les g\'eom\'etries du r\'eseau. Nous avons donc a priori un seul bivecteur par face, $B_f$, d\'efini disons dans le r\'ef\'erentiel d'un t\'etra\`edre, $B_f(t)$. Partout ailleurs o\`u il intervient, du transport parall\`ele doit \^etre {\it a priori} (implicitement) utilis\'e, selon \eqref{transport bivectors vt}. Ainsi, pour calculer la fonction partition, nous devons utiliser des int\'egrales contraintes par la simplicit\'e, du type:
\beq
\int \prod_f dB_f\ \int \prod_t dN_t\ \prod_{(f,t)} \delta\big(b_{-f}(t)+\Ad\bigl(N_t^{-1}\bigr)\ b_{+f}(t)\bigr),
\ee
o\`u les contraintes sont impos\'ees pour chaque face, avec des normales $N_t$ diff\'erentes dans chaque t\'etra\`edre. Cela contraste pourtant avec les formules que l'on peut trouver dans la litt\'erature. En effet, il est plut\^ot d'usage d'imposer les contraintes sur des bivecteurs $B_{ft}$ \emph{ind\'ependants} dans chaque t\'etra\`edre. On peut imaginer cela par un \'eclatement de la triangulation initiale en simplexes disjoints, de sorte qu'il y a autant de copies du triangle $f$ que de t\'etra\`edres l'ayant dans son bord. La relation de transport parall\`ele \eqref{transport bivectors} n'est ensuite pas prise explicitement en compte, ce qui donne des int\'egrales du type :
\be \label{broken up measure}
\int \prod_{(f,t)} dB_{ft} \int \prod_t dN_t\ \prod_{(f,t)} \delta\big(b_{-ft}+\Ad\bigl(N_t^{-1}\bigr)\ b_{+ft}\bigr).
\ee
L'inconv\'enient d'une telle approche appara\^it sur les corr\'elations entre simplexes. En effet, supposons ici que $N_{t_1}$ et $N_{t_2}$ sont deux normales pour des t\'etra\`edres voisins, donc respectivement orthogonaux \`a $B_{ft_1}$ et $B_{ft_2}$, pour le triangle $f$ en commun. Mais si ces deux bivecteurs ne sont {\it a priori} pas corr\'el\'es, alors $N_{t_1}$ et $N_{t_2}$ sont aussi ind\'ependants. Cela signifie notamment que les corr\'elations exprim\'ees par le sous-groupe $\U(1)_{\hat{b}_{+f}}$ dans \eqref{coupling normals} ne sont {\it a priori} pas d\'ecrites au niveau quantique dans un tel contexte. Une mesure telle que \eqref{broken up measure}, accompagn\'ee du processus de collage mentionn\'e plus haut seront les ingr\'edients de notre d\'erivation du mod\`ele BC au chapitre \ref{sec:BC}.


\section{Traduction en calcul de Regge aires-angles} \label{sec:gluing-aarc}

Nous avons maintenant deux descriptions de la g\'eom\'etrie sur une triangulation : par une th\'eorie de jauge $\Spin(4)$ avec des contraintes de simplicit\'e, et par le calcul de Regge. De plus, nous avons \'etabli une sorte de dictionnaire entre les variables des deux approches \`a la section \ref{sec:geom normales}. Celui-ci s'exprime plus naturellement en termes des aires, angles dih\'edraux 3d et 4d, plut\^ot qu'avec les longueurs, ce qui sugg\`ere que les bivecteurs et normales sont assez proches de la proposition de calcul de Regge par les variables aires-angles.

Cela nous ouvre plusieurs pistes int\'eressantes. Typiquement, on peut se demander comment les contraintes de simplicit\'e traitent les contraintes de recollement du calcul de Regge aires-angles. Il s'agit d'une question primordiale, qui vise notamment \`a comprendre la diff\'erence entre le formalisme canonique pr\'esent\'e \`a la section \ref{sec:twisted-simplicial} et le formalisme utilis\'e dans les mousses de spins, i.e. du point de vue espace-temps sur r\'eseau. Nous avons vu en effet \`a la section \ref{sec:twisted-simplicial} que les contraintes de simplicit\'e crois\'ee que nous utilisons en mousses de spins n'assurent pas une g\'eom\'etricit\'e \`a la Regge sur toute une triangulation du point de vue canonique. Elles peuvent alors \^etre compl\'et\'ees par les contraintes \eqref{edge simplicity}. Or ces derni\`eres induisent en variables aires-angles les contraintes propos\'ees en calcul de Regge aires-angles par Dittrich et Speziale. Puisque du point de vue espace-temps, les contraintes \eqref{edge simplicity} ne sont pas n\'ecessaires, essayons de voir comment les contraintes de recollement aires-angles sont prises en compte, gr\^ace \`a notre dictionnaire de la section \ref{sec:geom normales}.

Un autre point de vue sur la question est celui du passage d'un formalisme initial du premier ordre, i.e. avec bivecteurs et holonomies ind\'ependants, \`a un formalisme du second ordre dans lequel les longueurs doivent d\'eterminer toutes les grandeurs g\'eom\'etriques. Nous savons en effet que les normales $N_t$ et les holonomies contiennent les angles dih\'edraux 4d et les angles de d\'eficit. En calcul de Regge standard, ces grandeurs ne sont pas ind\'ependantes : par d\'efinition ce sont des fonctions des longueurs -- c'est un formalisme du deuxi\`eme ordre. Dans la th\'eorie de Plebanski, qui est par contraste un formalisme du premier ordre, le passage s'effectue via l'\'equation $d_AB=0$, qui se transforme, si $B$ est une 2-forme simple, en l'\'equation de Cartan : $d_A e=0$, pourvu que $e$ soit une cot\'etrade non-d\'eg\'en\'er\'ee. Alors cette \'equation admet une unique solution, $A(e)$, la connexion compatible avec la cot\'etrade, de torsion nulle. Parall\`element, dans notre formalisme (du premier ordre) sur r\'eseau, nous avons discr\'etis\'e l'\'equation $d_AB=0$ par la relation de transport parall\`ele \eqref{transport bivectors} sur les bivecteurs. Il est donc naturel de se demander ici comment les relations de transport parall\`ele et les contraintes de simplicit\'e peuvent permettre d'exprimer les angles dih\'edraux (issus des holonomies) en fonction de la g\'eom\'etrie des bivecteurs, et o\`u intervient le crit\`ere de non-d\'eg\'en\'erescence.

Notons que Barrett \cite{first-order-RC-barrett} a insist\'e sur le fait que le calcul de Regge peut \^etre d\'efini dans un formalisme du premier ordre, i.e. avec longueurs et angles dih\'edraux ind\'ependants, \`a condition que les angles dih\'edraux soient ceux d'une g\'eom\'etrie (quelconque). La g\'eom\'etrie naturellement accessible ici est celle cod\'ee par les bivecteurs. Nous avons vu, section \ref{sec:geom normales}, que ces bivecteurs d\'eterminent les aires et angles 3d. On peut donc esp\'erer obtenir une relation donnant les angles 4d directement comme fonctions des angles 3d, comme le propose par exemple le calcul de Regge aires-angles \cite{dittrich-speziale-aarc}. 

Cette section contient quatre r\'esultats importants, initialement expos\'es dans \cite{bf-aarc-val}. Nous savons que la contrainte de simplicit\'e crois\'ee \eqref{simplicity-selfdual} et la relation de fermeture \eqref{closure} permettent de reconstruire la g\'eom\'etrie de chaque t\'etra\`edre ind\'ependamment. En introduisant des variables g\'eom\'etriques scalaires (angles dih\'edraux), nous montrons que :
\begin{itemize}
\item Les contraintes de simplicit\'e crois\'ee et les relations de transport parall\`ele \eqref{transport bivectors} assurent que l'on peut coller de mani\`ere consistente les t\'etra\`edres, car les angles dih\'edraux 3d satisfont bien les contraintes de recollement propos\'ees en calcul de Regge \og aires - angles 3d\fg.
\item Il est possible d'extraire les angles dih\'edraux 4d, entre t\'etra\`edres, et de montrer qu'ils sont contraints \`a s'exprimer comme fonctions des angles 3d. C'est la r\'ealisation du passage implicite dans un formalisme du premier ordre.
\item Les relations des deux premiers points sont des relations trigonom\'etriques assez \'evolu\'ees. Il est remarquable qu'elles se formulent ici simplement par des produits d'\'el\'ements des groupes $\SU(2)$ et $\U(1)$. Cette structure alg\'ebrique sera particuli\`erement adapt\'ee \`a la construction de mod\`eles de mousses de spins, en utilisant les transform\'ees de Fourier sur ces groupes, voir chapitre \ref{sec:pathint fk}.
\item L'introduction des variables $\U(1)$ suppl\'ementaires nous conduit \`a formalisme qui m\'elange les variables usuelles de la th\'eorie topologique sur r\'eseau et des variables angulaires du calcul de Regge. Cette interpr\'etation g\'eom\'etrique nous sera utile pour construire des mod\`eles de mousses de spins comme int\'egrales sur les g\'eom\'etries simplicielles de la triangulation, voir chapitre \ref{sec:pathint fk}.
\end{itemize}
Naturellement, les mod\`eles annonc\'es aux points 3 et 4 coincident.

\bigskip

Dans une th\'eorie de jauge $\Spin(4)$ sur r\'eseau, il est clair que la norme et la direction des bivecteurs jouent des r\^oles diff\'erents (le transport parall\`ele n'affectant que les directions). De plus, nous allons devoir extraire minutieusement des phases issues des changement des directions par transport parall\`ele. Une bonne param\'etrisation est donc la suivante :
\beq \label{B parametrisation}
b_{\pm ft} = \f{i}{2}A_f\,\hat{b}_{\pm ft}\cdot\vec{\sigma} ,\quad\text{et}\qquad \hat{b}_{\pm ft}\cdot\vec{\sigma} = \pm \Ad\bigl(n_{\pm ft}\bigr)\ \sigma_z,
\ee
pour $n_{\pm ft}\in\SU(2)$. La matrice $\sigma_z$ est la matrice de Pauli $\diag(1, -1)$, et $\vec{\sigma}$ le 3-vecteur form\'e par les matrices de Pauli. Avec ces conventions, $A_f$ repr\'esente l'aire du triangle $f$. La direction de $b_{\pm ft}$ est cod\'ee dans les \'el\'ements de groupe $n_{\pm ft}$ : ces rotations envoient la direction de r\'ef\'erence $\hat{z}$ sur la direction $\hat{b}_{\pm ft}$. Remarquons que seuls deux param\`etres de ces variables sont pertinents, puisque l'on peut changer $n_{\pm ft}\rightarrow n_{\pm ft}\,e^{-\f{i}{2}\lambda_\pm\sigma_z}$ sans affecter $\hat{b}_{\pm ft}$. Cela se comprend bien dans la param\'etrisation d'Euler de $\SU(2)$ (ou la projection de Hopf de $\SU(2)$ sur $S^2$) : $n=e^{-\f{i}{2}\alpha\sigma_z} e^{-\f{i}{2}\beta\sigma_y} e^{-\f{i}{2}\gamma\sigma_z}$, o\`u l'angle $\gamma$ ne joue aucun r\^ole. 

Avec la param\'etrisation \eqref{B parametrisation}, les contraintes dont nous avons besoin pour reconstruire les vecteurs de la cot\'etrade sur les ar\^etes d'un t\'etra\`edre $t$ se r\'eduisent \`a :
\begin{align} \label{simplicity directions}
&\hat{b}_{-ft} + \Ad\bigl(N_t\mone\bigr)\,\hat{b}_{+ft} \,=\,0,\\
&\sum_{f\in\pp t} \eps_{ft}\ A_f\, \hat{b}_{+ft} \,=\,0. \label{closure area-directions}
\end{align}
La premi\`ere \'equation \eqref{simplicity directions} est simplement la contrainte de simplicit\'e crois\'ee de laquelle nous avons \'elimin\'ee l'aire $A_f$ (ayant explicitement r\'esolu la contrainte de simplicit\'e diagonale). En fait cette contrainte est suffisante pour qu'il existe un \emph{authentique t\'etra\`edre, ferm\'e}, dans $\R^4$, dont les directions des bivecteurs, simples, sont donn\'ees par $\hat{b}_{\pm ft}$ ! En effet, les quatre vecteurs $\hat{b}_{+ft}\in S^2$ d'un t\'etra\`edre ne peuvent \^etre lin\'eairement ind\'ependants, ce qui se traduit par une relation de type fermeture, avec certains coefficients $x_{ft}$ r\'eels : $\sum_{f\in\pp t} x_{ft}\,\hat{b}_{+ft}=0$. Alors ces coefficients $x_{ft}$ s'interpr\`etent comme les aires de chacun des quatre triangles.  Mais ces coefficients d\'ependent en g\'en\'eral du t\'etra\`edre et surtout n'ont a priori rien \`a voir avec les variables $A_f$ ! Autrement dit, la relation de de fermeture \eqref{closure area-directions} n'a d'autre but que de forcer les variables $A_f$, a priori quelconques, \`a \^etre les aires du t\'etra\`edre d\'ej\`a d\'etermin\'e par les $\hat{b}_{+ft}$ et $N_t$ !

Une id\'ee cl\'e est maintenant de remplacer les variables de configuration $\hat{b}_{\pm ft}\in S^2$ par les rotations $n_{\pm ft}$.
Nous ne souhaitons pas ici nous donner de prescription pour cela, mais plut\^ot consid\'erer qu'il y a une ambiguit\'e $\U(1)\times \U(1)$ sur chaque paire $(f, t)$. Nous allons maintenant r\'e\'ecrire les relations de transport parall\`ele \eqref{transport bivectors} et les contraintes de simplicit\'e crois\'ee \eqref{simplicity directions}, avec ces nouvelles variables. Pour cela nous devons introduire de nouvelles variables angulaires $\theta_{fv}^\pm$ et $\psi_{ft}$ indispensables pour g\'erer les ambiguit\'es $\U(1)$ sur la d\'efinition de $n_{\pm ft}$. Les relations de transport parall\`ele expriment que $G_{tt'}$ envoie les 3-vecteurs $\hat{b}_{\pm ft'}$ sur les 3-vecteurs $\hat{b}_{\pm ft}$. Cela se r\'esout pour les rotations $n_{\pm ft}$ d'un r\'ef\'erentiel en fonction de celles d'un autre :
\beq \label{gluing spin4}
n_{+ft} = g_{+tt'}\,n_{+ft'}\,e^{\f{i}{2}\theta^+_{fv}\sigma_z}, \qquad\text{et}\qquad n_{-ft} = g_{-tt'}\,n_{-ft'}\,e^{\f{i}{2}\theta^-_{fv}\sigma_z}.
\ee
Une fa\c{c}on \'equivalente de voir ces relations est de dire que les angles $\theta_{fv}^\pm$ permettent de r\'esoudre les \'equations de transport parall\`ele pour les holonomies $G_{tt'}$ en fonction des directions des bivecteurs :
\beq \label{solveholspin4}
g_{+tt'} = n_{+ft}\,e^{-\f{i}{2}\theta^+_{fv}\sigma_z}\,n_{+ft'}\mone, \qquad\text{et}\qquad g_{-tt'} = n_{-ft}\,e^{-\f{i}{2}\theta^-_{fv}\sigma_z}\,n_{-ft'}\mone.
\ee
Insistons tout de m\^eme sur le fait que les angles $\theta^\pm_{fv}$ ne fixent pas tous les degr\'es de libert\'e des holonomies $G_{vt}$, car ces relations ne donnent que $G_{tt'} = G_{vt}\mone  G_{vt'}$, et $f$ est le seul triangle commun \`a $t$ et $t'$.

Passons aux contraintes de simplicit\'e crois\'ee qui vont relier les rotations des secteurs self-dual et anti-self-dual $n_{\pm ft}$ pour chaque triangle dans chaque t\'etra\`edre. Nous introduisons des angles $\psi_{ft}$ qui permettent de compenser des choix a priori arbitraires de phase dans $n_{+ft}$ et $n_{-ft}$ :
\beq \label{normale-phase}
n_{-ft} = N_t\mone\,n_{+ft}\ e^{\f{i}{2}\psi_{ft}\sigma_z}.
\ee
Une deuxi\`eme formulation permet de consid\'erer \'egalement $\psi_{ft}$ comme le degr\'e de libert\'e restant dans la r\'esolution des contraintes pour exprimer les normales $N_t$ en fonction des variables $n_{\pm ft}$.
\beq \label{solvenormals}
N_t = n_{+ft}\ e^{\f{i}{2}\psi_{ft}\sigma_z}\,n_{-ft}\mone.
\ee

Ainsi, le passage des variables $\hat{b}_{\pm ft}$ aux variables $n_{\pm ft}$ s'accompagne d'une r\'e\'ecriture des relations de la th\'eorie discr\`ete impliquant de nouvelles variables suppl\'ementaires, $\theta^\pm_{fv}, \psi_{ft}$. Nous pouvons bien s\^ur consid\'erer ces nouvelles variables comme les param\`etres libres dans la r\'esolution des contraintes (auquel cas il n'y a plus de contraintes). Au lieu de cela, il est aussi possible de consid\'erer que d\'esormais ces nouvelles variables s'ajoutent aux variables initiales en tant que \emph{variables de configuration a priori ind\'ependantes}. Ces deux approches sont naturellement \'equivalentes, mais pour \'eviter toute confusion dans la discussion, nous fixons le statut de ces variables au deuxi\`eme choix. Les contraintes de la th\'eorie sont alors donn\'ees par \eqref{gluing spin4} et \eqref{normale-phase} en fonction des variables $(n_{\pm ft}, g_{\pm vt}, N_t, \theta^\pm_{fv}, \psi_{ft})$.

L'\'elargissement de l'espace des configurations implique de prendre en compte une sorte de sym\'etrie de jauge $\U(1)\times \U(1)$ additionnelle, agissant \`a droite de $(n_{+ft}, n_{-ft})\in\Spin(4)$. En effet, les angles d'Euler $\gamma$ dans la d\'ecomposition de $n_{\pm ft}$ ne sont pas physiques (pour $n=e^{-\f{i}{2}\alpha\sigma_z} e^{-\f{i}{2}\beta\sigma_y} e^{-\f{i}{2}\gamma\sigma_z}$), et les grandeurs $G_{tt'}, N_t$ ne doivent pas d\'ependre des choix faits localement pour ces angles. On voit \`a partir de \eqref{solveholspin4} que l'invariance des holonomies $G_{tt'}$ sous de telles transformations entra\^ine certaines lois de transformations sur les angles $\theta^\pm_{fv}$, et de m\^eme pour $\psi_{ft}$. La d\'efinition de nos nouvelles variables de configuration s'accompagne donc des propri\'et\'es suivantes sous les transformations de jauge $K$ dans $\Spin(4)$ et $\Lambda$ dans $\U(1)\times\U(1)$,
\begin{alignat}{3} \label{u1u1 gauge transfo}
&(K,\Lambda)\ \cdot\ n_{\pm ft}& &=&\ &k_{\pm t}\,n_{\pm ft}\,e^{\f{i}{2}\lambda_{ft}^\pm\sigma_z}, \\
&(K,\Lambda)\ \cdot\ \theta_{fv}^\pm& &=&\ &\theta_{fv}^\pm +\eps_{tt'}^f\bigl(\lambda_{ft}^\pm - \lambda_{ft'}^\pm\bigr), \\
&(K,\Lambda)\ \cdot\ \psi_{ft}& &=&\ &\psi_{ft}-\bigl(\lambda^+_{ft}-\lambda^-_{ft}\bigr).
\end{alignat}
Les angles $\theta_{fv}^\pm$ constitue une sorte de connexion discr\`ete $\mathfrak{u}(1)\oplus\mathfrak{u}(1)$ sur chaque face duale. L'angle $\psi_{ft}$ se transforme alors de mani\`ere \`a r\'eserver la covariance $\U(1)\times\U(1)$ dans les contraintes de simplicit\'e. On voit en particuier qu'en plus de la jauge temps, $N_t=\unit$, il est possible de choisir $\psi_{ft}=0$ partout, de sorte que la contrainte de simplicit\'e crois\'ee se simplifie en : $n_{+ft}= n_{-ft}$. Bien s\^ur les quantit\'es physiques qui nous int\'eressent sont invariantes sous l'ensemble de ces transformations.

Nous pouvons ensuite r\'eexprimer les grandeurs g\'eom\'etriques en fonction des nouvelles variables. L'angle dih\'edral 3d entre les triangles $f, f'$ dans le t\'etra\`edre $t$ est donn\'e par la trace dans la repr\'esentation fondamentale : $\tr(\hat{b}_{+ft}\,\hat{b}_{+f't})$. Avec les rotations $n_{+ft}$, on utilise plus naturellement la repr\'esentation de spin 1 :
\beq \label{3d angle n}
\cos \phi_{ff'}^t = -\eps_{ff'}\ \langle 1,0\lvert n_{+ft}\mone\,n_{+f't}^{\phantom{}}\rvert 1, 0\rangle,
\ee
o\`u $\eps_{ff'}=\eps_{ft}\eps_{f't}$ est l'orientation relative des faces duales \`a $f$ et $f'$.Remarquons que l'\'el\'ement de matrice consid\'er\'e est un simple polynome de Legendre : $\bra 1,0\rv g \lv 1,0\ket = P_1(\cos\beta) = \cos\beta$, pour $g=e^{-\f{i}{2}\alpha\sigma_z} e^{-\f{i}{2}\beta\sigma_y} e^{-\f{i}{2}\gamma\sigma_z}$. Cela nous donne donc l'angle 3d comme l'angle d'Euler $\beta$ dans l'\'ecriture de $n_{+ft}\mone n_{+f't}$.

Nous avons vu \`a la section \ref{sec:geom normales} que le cosinus de l'angle dih\'edral entre les t\'etra\`edres $t$ et $t'$ est donn\'e par l'angle de classe de la rotation dih\'edrale,
\beq
D_{tt'} = N_t^{\phantom{}}\,g_{-tt'}^{\phantom{}}\,N_{t'}\mone\,g_{+tt'}\mone.
\ee
On sait aussi que cette rotation est g\'en\'er\'ee par $b_{+ft}$. En utilisant les expressions de $g_{\pm tt'}$, \eqref{solveholspin4}, et de $N_t$, \eqref{solvenormals}, il est possible d'expliciter son angle en fonction de nos nouvelles variables angulaires ! En effet, nous obtenons :
\beq
D_{tt'} = \exp \biggl(\Bigl(\eps_{tt'}^f(\theta^+_{fv}-\theta^-_{fv})+\psi_{ft}-\psi_{ft'}\Bigr)\,\f{b_{+ft}}{A_f}\biggr).
\ee
La quantit\'e $\eps_{tt'}^f=\pm1$ est positive lorsque le chemin de $t$ \`a $t'$ est orient\'e comme la face duale $f$, et n\'egatif sinon.
La formule \eqref{4d-angles def} nous donne alors les angles dih\'edraux $\thet_{tt'}$ (dont le label est \'equivalent \`a $(fv)$),
\begin{align}
\cos\thet_{tt'} &= -\cos\f{1}{2} \Bigl(\theta_{fv}^+-\theta_{fv}^-+\eps_{tt'}^f \bigl(\psi_{ft}-\psi_{ft'}\bigr)\Bigr), \\
\sin\thet_{tt'} & = \eps\,\eps_{tt'}^f\,\sin\f{1}{2} \Bigl(\theta_{fv}^+-\theta_{fv}^-+\eps_{tt'}^f \bigl(\psi_{ft}-\psi_{ft'}\bigr)\Bigr),
\end{align}
o\`u $\eps$ est pr\'ecis\'ement l'ambiguit\'e de signe sur l'expression des bivecteurs en fonction de la cot\'etrade discr\`ete \cite{barrett-asymEPR}. Remarquons au passage que cela permet de d\'eriver proprement l'\'equation \eqref{deficithol} qui relie l'angle de d\'eficit (d\'efini par les $\thet_{tt'}$ autour d'un triangle) et l'holonomie $G_f(t)$ autour de ce triangle !

\medskip

\begin{figure} \begin{center}
\includegraphics[width=3cm]{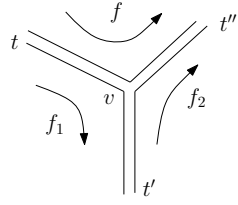}
\caption{ \label{fig:4d-3d angles} Du point de vue dual, les trois t\'etra\`edres $t, t', t''$ d'un 4-simplexe deviennent des liens se rencontrant au vertex $v$. Ces t\'etra\`edres se partagent des triangles deux \`a deux, duaux aux faces $f, f_1, f_2$. Ici, les t\'etra\`edres $t, t''$ se rencontrent sur le triangle dual \`a $f$. L'angle dih\'edral $\thet_{tt''}$ peut se calculer comme fonction des angles 3d en utilisant comme interm\'ediaire trois t\'etra\`edres diff\'erents (puisqu'il y en a cinq se rencontrant en $v$, c'est-\`a-dire que deux ne sont pas dessin\'es). Cela conduit \`a des contraintes \eqref{consistency 3d angles} entre des ensembles d'angles 3d, qui sont \'equivalentes \`a celles propos\'ees en calcul de Regge aire-angle 3d.}
\end{center}
\end{figure}

On \'ecrira donc simplement :
\beq
D_{tt'} = -\cos\thet_{tt'} + i\eps\,\sin\thet_{tt'}\, \Ad\bigl(n_{+ft}\bigr)\, \sigma_z.
\ee
Consid\'erons maintenant trois t\'etra\`edres $t, t'$ et $t''$ sur un 4-simplexe $v$, se rencontrant donc deux \`a deux sur un triangle comme l'illustre la figure \ref{fig:4d-3d angles}. On s'attend \`a ce que nos contraintes permettent d'exprimer les angles 4d entre les t\'etra\`edres du 4-simplexe en termes de angles 3d par la formule standard,
\beq \label{relation 3d-4d angles}
\cos\thet_{tt''} = \f{\cos\phi^{t'}_{f_1f_2}-\cos\phi_{f_1 f}^t\,\cos\phi_{ff_2}^{t''}}{\sin\phi_{f_1 f}^t\,\sin\phi_{ff_2}^{t''}}
\ee
Il est important ici de voir que le t\'etra\`edre $t'$ utilis\'e comme interm\'ediaire pour le calcul, avec les triangles $f_1, f_2$, peut \^etre n'importe lequel des t\'etra\`edres de $v$ (sauf $t$ et $t''$). C'est ce ph\'enom\`ene qui conduit \`a des contraintes sur les angles 3d.

Nous allons montrer que cette relation assez compliqu\'ee est contenue de mani\`ere naturelle dans notre cadre d'\'etude, simplement comme des relations entre variables de $\SU(2)$ n'impliquant que des produits de groupe. Le tout est donc de reconna\^itre dans \eqref{relation 3d-4d angles} la structure de la loi de multiplication de $\SU(2)$. Pour cela, nous exprimons la rotation $D_{tt''}$ en fonction de $D_{tt'}$ et $D_{t't''}$ :
\beq
D_{tt''} = D_{tt'}\,g_{+tt'}\,D_{t't''}\,g_{+tt'}\mone \label{compo D}
\ee
Il s'agit de notre \'equation cl\'e. Du point de vue des \'el\'ements de groupe, elle est relativement triviale, et exprime le fait que le transport parall\`ele de $t$ \`a $t''$ (\`a travers le triangle $f$) peut s'effectuer en utilisant $t'$ comme pivot interm\'ediaire (\`a travers les triangles $f_1$ et $f_2$ \`a la place). Cette possibilit\'e provient de :
\beq
g_{+tt''} = g_{+vt}\mone\,g_{+vt'}^{\phantom{}}\,g_{+vt'}\mone\,g_{+vt''}^{\phantom{}} = g_{+tt'}\,g_{+t't''}.
\ee
Autrement dit, elle provient de l'existence d'un r\'ef\'erentiel commun, celui de $v$, que traversent tous les chemins entre les t\'etra\`edres ! Celui-ci n'\'etait bien s\^ur pas disponible dans le formalisme canonique de la section \ref{sec:twisted-simplicial}, et c'est ce qui explique basiquement que des contraintes de simplicit\'e suppl\'ementaires \eqref{edge simplicity} \'etaient n\'ecessaires pour reconstruire la g\'eom\'etrie de bord d'un 4-simplexe par exemple.

Maintenant, si les rotations $D_{tt'}$ sont \'ecrites comme des \'el\'ements du sous-groupe $\U(1)$ qui stabilise $b_{+ft}$, la relation \eqref{compo D} n'est plus du tout triviale, car elle implique alors les variables $b_{+ft}$ des triangles $f_1, f_2$. En utilisant le transport le long de $f_1$ entre $t$ et $t'$ et en prenant la trace, il vient :
\beq
-\cos \thet_{tt'} = \f{1}{2}\tr\Bigl[\bigl(-\cos\thet_{tt'}+i\eps\,\eps_{tt'}^{f_1}\,\sin\thet_{tt'}\,\Ad\bigl(n_{+f_1t'}\bigr)\sigma_z \bigr) \bigl(-\cos\thet_{t't''}+i\eps\,\eps_{t't''}^{f_2}\,\sin\thet_{t't''}\,\Ad\bigl(n_{+f_2t'}\bigr)\sigma_z \bigr)\Bigr].
\ee
Il s'ensuit :
\beq
\cos \thet_{tt'} = - \cos\thet_{tt'}\ \cos\thet_{t't''} + \sin\thet_{tt'}\ \sin\thet_{t't''}\ \cos\phi_{f_1f_2}^{t'}
\ee
Pour des configurations non-d\'eg\'en\'er\'ees telles que $\sin\thet_{tt'}\neq0$, cette relation donne les angles 3d en fonction des angles 4d. De plus, en l'\'ecrivant pour les trois paires de t\'etra\`edres, elle peut \^etre invers\'ee, reproduisant exactement \eqref{relation 3d-4d angles} ! C'est-\`a-dire que nous avons montr\'e que les angles 4d sont d\'etermin\'es par la g\'eom\'etrie issue des bivecteurs, de le m\^eme mani\`ere que dans \cite{dittrich-speziale-aarc}.


Nous avons maintenant presque prouv\'e que les contraintes de recollement propos\'ees dans \cite{dittrich-speziale-aarc} sont bien satisfaites dans le cadre classique de base des mousses de spins. Il nous suffit pour cela de remarquer que notre d\'efinition de l'angle dih\'edral, disons $\cos\thet_{t_1t_2}$, par $D_{t_1t_2}$ ne d\'epend pas du choix du t\'etra\`edre interm\'ediaire utilis\'e dans \eqref{compo D}. Ainsi en reproduisant le raisonnement pr\'ec\'edent en passant par un autre t\'etra\`edre (un quatri\`eme diff\'erent) de $v$, nous obtenons une autre expression pour $\cos\thet_{t_1 t_2}$ ayant la m\^eme structure. L'\'egalit\'e des diff\'erentes expressions de l'angle revient \`a imposer des contraintes sur les angles 3d,
\beq \label{consistency 3d angles}
\f{\cos\phi_{f_1f_2}^t - \cos\phi_{f_1f}^{t_1}\cos\phi_{ff_2}^{t_2}}{\sin\phi_{f_1f}^{t_1}\sin\phi_{ff_2}^{t_2}}
 = \f{\cos\phi_{f'_1f'_2}^{t'} - \cos\phi_{f'_1f}^{t_1}\cos\phi_{ff'_2}^{t_2}}{\sin\phi_{f'_1f}^{t_1}\sin\phi_{ff'_2}^{t_2}}.
\ee
Il s'agit bien d'un ensemble de contraintes propos\'e dans \cite{dittrich-speziale-aarc} pour assurer la consistence du collage des t\'etra\`edres en calcul de Regge aires-angles, mais aussi dans \cite{dittrich-ryan-simplicial-phase} comme des contraintes qu'il est n\'ecessaire d'ajouter aux contraintes de simplicit\'e du point de vue canonique. Disons aussi qu'il est remarquable que de telles \'equations soient si simplement prises en compte par nos relations \eqref{gluing spin4}, \eqref{normale-phase}, qui n'impliquent que des produits d'\'el\'ements de groupe -- il semble que la structure de $\SU(2)$ soit bien adapt\'ee pour construire la g\'eom\'etrie de la triangulation !

\chapter{Mod\`eles pour la gravit\'e quantique}

\section{La quantification de Engle-Pereira-Rovelli-Livine} \label{sec:epr-quantisation}

Si le mod\`ele de mousses de spins correspondant est commun\'ement appel\'e EPR, il est vrai que ses trois auteurs, Engle, Pereira et Rovelli, ne pr\'esent\`erent d'abord que le mod\`ele pour $\gamma\rightarrow0$, et $\gamma\rightarrow\infty$ (qui est alors le mod\`ele d\'ej\`a connu de Barrett-Crane) \cite{epr-short, epr-long}, puis finalement le mod\`ele pour $\gamma$ fini vint avec Livine \cite{eprl}. Je pr\'esente ici la d\'erivation originale du mod\`ele; pour une pr\'esentation plus r\'ecente, et peut-\^etre mieux murie, je renvoie \`a \cite{rovelli-new-look-lqg}.

L'id\'ee est ici d'imposer les contraintes de simplicit\'e sur les \'etats de bord de chaque 4-simplexe de la triangulation, et de les imposer faiblement, celles-ci \'etant g\'en\'eriquement de seconde classe. Pour cela on consid\`ere la structure symplectique sur le graphe dual au bord d'un 4-simplexe, d\'ej\`a pr\'esent\'ee en \ref{sec:twisted-simplicial}. Nous avons donc un lien dual \`a chaque triangle, et un vertex dual \`a chaque t\'etra\`edre. Les liens portent chacun une holonomie $G_{tt'}$, effectuant le transport parall\`ele entre les t\'etra\`edres $t$ et $t'$, et deux bivecteurs $B_f(t)$ (un pour chaque vertex du lien dual), reli\'es entre eux par transport parall\`ele. La structure symplectique sur $T^*\Spin(4)$ dont nous avons besoin est telle que les moments conjugu\'es $J_f(t)$ aux holonomies sont d\'eform\'es par le param\`etre d'Immirzi,
\beq \label{moment 4s}
J_f(t) = B_f(t) +\f1\gamma \star B_f(t) \quad \Leftrightarrow \quad B_f(t) = \f{\gamma^2}{\gamma^2-1}\Bigl( J_f(t) -\f1\gamma \star J_f(t)\Bigr),
\ee
par rapport au cas de l'action de Palatini-Hilbert. Les cas limites $\gamma\rightarrow0$ et $\gamma\rightarrow\infty$ donnent respectivement :
\beq
J_f(t) = \gamma\mone\star B_f(t), \quad \text{et}\qquad J_f(t) = B_f(t).
\ee
Les deux structures symplectiques associ\'ees, remarqu\'ees dans \cite{baez-barrett-quantum-tet}, correspondent respectivement aux secteurs topologique et gravitationnel, reli\'es par un \og flip\fg{} provoqu\'e par l'op\'erateur de dualit\'e de Hodge. L'origine de ces diff\'erentes structures possibles, d\'ej\`a discut\'ee en section \ref{sec:twisted-simplicial}, tient \`a la d\'ecomposition particuli\`ere $\spin(4) = \su(2)\oplus \su(2)$. Comme on le verra, la structure symplectique \og gravitationnelle\fg{} m\`ene dans la quantification EPR au mod\`ele d\'ej\`a connu de Barrett-Crane (BC). Dans la d\'erivation initiale du mod\`ele EPR, en l'absence du param\`etre d'Immirzi, c'est donc l'utilisation de la structure \og flipp\'ee\fg{} qui a permis d'obtenir un autre mod\`ele que celui de BC. L'argument pour ce choix est de dire que le terme topologique est essentiel pour former les variables d'Ashtekar-Barbero, et par suite dans la LQG (dont les mousses de spins doivent fournir la dynamique). Bien s\^ur, on sait maintenant relier les deux secteurs au niveau classique, et le mod\`ele EPR au mod\`ele BC au niveau quantique, par le param\`etre d'Immirzi (mis \`a part le fait que les cas $\gamma=\pm1$ sont exclus des diff\'erents mod\`eles).

On peut alors r\'eexprimer les contraintes en termes des variables canoniques,
\beq
\bigl(1+\gamma^{-2}\bigr) \Tr\,\bigl(\star J_f\, J_f\bigr) - 2\gamma\mone\,\Tr\,\bigl(J_f\, J_f\bigr) = 0,
\ee
pour la simplicit\'e diagonale. Notons que $\Tr(\star J_f\, J_f)$ et $\Tr(J_f\, J_f)$ sont le Casimir et le pseudo-Casimir de $\Spin(4)$, dont les spectres s'expriment en termes des Casimir des parties $\su(2)$ self-duale et anti-self-duale. Contrairement \`a la contrainte diagonale qui est de prmi\`ere classe, la simplicit\'e crois\'ee ne forme pas un ensemble qui commutent pour les crochets de Poisson. De ce fait, les auteurs proc\`edent subtilement, utilisant une m\'ethode de type \og master constraint\fg, dans l'esprit d\'evelopp\'e dans \cite{thiemann-qsd-master-constraint}. L'id\'ee est d'abord d'utiliser la jauge de type \og temps\fg, comme dans le continu, $N_t=(1,0,0,0)$, de sorte que les t\'etra\`edres sont tous dans l'espace 3d orthogonal, l'invariance de jauge \'etant r\'etablie par moyenne sur le groupe, apr\`es avoir impos\'e les contraintes. La simplicit\'e crois\'ee est ensuite mise au carr\'e pour chaque face,
\beq
\bigl(\vec{J}_f-\gamma\mone\vec{K}_f\bigr)^2 = 0,
\ee
pour $J^i = \f12\eps^{0i}_{\phantom{0i}jk}J^{jk}$, g\'en\'erateur des rotations $\SO(3)$ laissant $N_t$ invariant, et $K^i = J^{0i}$, g\'en\'erateur des \og boost\fg. Cette \'equation se simplifie gr\^ace \`a la simplicit\'e diagonale en :
\beq
\Tr\,\bigl(\star J_f\, J_f\bigr) = 4\gamma\,\vec{J}^2.
\ee
Ces contraintes sont bien s\^ur \`a r\'esoudre ind\'ependamment dans chaque t\'etra\`edre \cite{eprl}.

\begin{figure} \begin{center}
\includegraphics[width=7cm]{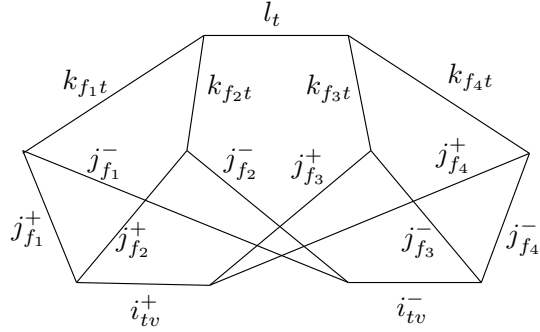}
\caption{ \label{fusion} L'\'evaluation de ce graphe est le coeficient de fusion. Chaque spin $l_t$, $i^+_{tv}$ et $i^-_{tv}$ entrelace 4 repr\'esentations associ\'ees aux 4 faces du t\'etra\`edre $t$. La d\'efinition compl\`ete n\'ecessite des orientations sur les noeuds et les liens, qui doivent \^etre choisies de mani\`ere consistente avec celles des symboles 15j et autres coefficients de fusion dans le mod\`ele.}
\end{center}
\end{figure}

On proc\`ede maintenant \`a la quantification sur le graphe dual au bord du 4-simplex. L'espace de Hilbert, non-invariant de jauge, consiste en les fonctions $L^2$ sur $\Spin(4)=\SU(2)\times\SU(2)$, engendr\'e par les \'el\'ements de matrices des \'el\'ements $G_{tt'}=(g_{+tt'},g_{-tt'})$ vivant sur les liens dans les repr\'esentations irr\'eductibles. Ces derni\`eres correspondent \`a un coloriage $(j_f^+, j_f^-)$ des liens du graphe, o\`u les $j^\pm_f$ sont des demi-entiers. Une base de cet espace est donc form\'ee par les vecteurs :
\beq \label{base magnetic}
s^{\{j^\pm_f\}}_{\{m^\pm_f,n^\pm_f\}}(G_{tt'}) = \prod_f D^{(j^+_f)}_{m^+_f n^+_f}\bigl(g_{+tt'}\bigr)\ D^{(j^-_f)}_{m^-_f n^-_f}\bigl(g_{-tt'}\bigr).
\ee
Ensuite, les moments $J_f(t)$ sont promus au statut d'op\'erateurs, agissant comme des d\'erivations \`a gauche si $t$ est dual au vertex source du lien dual $f$, et \`a droite si $t$ est le vertex d'arriv\'ee. Il existe toujours au niveau quantique des ambiguit\'es d'ordering (donnant naissance \`a des corrections d'ordre $\hbar$ dans les spectres des observables). Il est ici pr\'ef\'erable d'utiliser un spectre en $j^2$ au lieu de $j(j+1)$ pour les Casimir de $\SU(2)$. La contrainte de simplicit\'e diagonale fixe alors le rapport entre les spins self-duaux et anti-self-duaux des vecteurs de base ci-dessus, $j^+_f/j^-_f = \lvert (\gamma+1)/(\gamma-1)\rvert$. En particulier, pour $\gamma>0$, on a $j^+>j^-$ ; et pour $\gamma<0$, $j^->j^+$. Regardons maintenant le cas $\gamma>0$. L'action du sous-groupe $\SU(2)$ diagonal de $\Spin(4)$, qui laisse la normale $N_t=(1,0,0,0)$ invariante, permet de d\'ecomposer l'espace $\calH_{(j^+,j^-)}$ portant la repr\'esentation $(j^+,j^-)$ de $\Spin(4)$ sur les repr\'esentations irr\'eductibles de ce sous-groupe selon
\beq \label{decomposition EPR}
\calH_{(j^+,j^-)} = \bigoplus_{k=\lvert j^+-j^-\rvert}^{j^++j^-} \calH_k.
\ee
Dans cette \'equation, $k^2$ s'interpr\`ete comme la valeur du Casimir $\vec{J}^2$, de sorte que les solutions de la simplicit\'e crois\'ee sont
\beq
k = \begin{cases} j^++j^- &\text{si $0<\gamma<1$}, \\ j^+-j^- &\text{si $\gamma>1$}. \end{cases}
\ee
Ainsi, $\gamma<1$ s\'electionne la plus haute repr\'esentation du sous-groupe $\SU(2)$ diagonal dans la d\'ecomposition de $\calH_{(j^+,j^-)}$, tandis que $\gamma>1$ choisit la plus petite. En r\'esum\'e, $k$ d\'etermine $j^\pm$ par :
\beq \label{contraintes spin epr}
j^\pm_f = \gamma_\pm\,k_f, \quad \text{pour}\qquad \gamma_\pm = \f{\lvert 1\pm\gamma\rvert}{2}.
\ee
En termes de r\'eseaux de spins, cela signifie que les repr\'esentations self-duale et anti-self-duale sur chaque lien s'entrelacent par la repr\'esentation $k_f$. Il nous faut moyenner cette proc\'edure sur $\Spin(4)$ au niveau de chaque t\'etra\`edre (i.e. chaque noeud du r\'eseau). Des int\'egrales sur $\Spin(4)$ produisent alors des entrelaceurs self-duaux et anti-self-duaux entre les quatre liens se rencontrant \`a chaque noeud, vivants dans $\Inv_{\Spin(4)}(\calH_{(j^+_1,j^-_1)}\otimes\dotsm\otimes\calH_{(j^+_4,j^-_4)})$, selon les formules \eqref{intg4} et \eqref{pairing}. Ces int\'egrales peuvent aussi passer \`a travers l'entrelacement des liens self-duaux et anti-self-duaux par $k_f$, pour produire des entrelaceurs du sous-groupe $\SU(2)$ diagonal entre les quatre repr\'esentations $k_f$ impliqu\'ees \`a chaque noeud. On obtient ainsi des r\'eseaux de spins invariants de jauge. De cette fa\c{c}on, l'entrelacement par $k_f$ permet de projeter ces entrelaceurs $\Spin(4)$ sur des entrelaceurs $\SU(2)$ de $\Inv_{\SU(2)}(\calH_{k_1}\otimes\dotsm\otimes\calH_{k_4})$. De mani\`ere \'equivalente, on d\'efinit
\beq
f : \Inv_{\SU(2)}(\calH_{k_1}\otimes\dotsm\otimes\calH_{k_4}) \rightarrow \Inv_{\Spin(4)}\bigl(\calH_{(j^+_1,j^-_1)}\otimes\dotsm\otimes\calH_{(j^+_4,j^-_4)}\bigr),
\ee
qui associe \`a un entrelaceur $\SU(2)$, d\'efini par un choix d'appariement en les quatre liens et un spin interm\'ediaire $l$, un entrelaceur $\Spin(4)$. La projection de ce dernier dans la base utilisant les m\^emes appariements et les spins interm\'ediaires $i^+, i^-$ donne les coefficients dit \emph{de fusion}, repr\'esent\'es figure \ref{fusion},
\beq
f : \iota(k_a,l) \mapsto \sum_{(i^+,i^-)} f^l_{i^+ i^-}\bigl(j^+_a,j^-_a,k_a\bigr)\ \iota(j^+_a,i^+)\,\iota(j^-_a,i^-).
\ee

On obtient ainsi un espace de Hilbert engendr\'e par les r\'eseaux de spins au bord du 4-simplexe, ayant des entrelaceurs de la forme $f(\iota)$, et des coloriages soumis aux contraintes \eqref{contraintes spin epr},
\begin{multline}
s_{\rm EPR}^{\{k_f,\iota_t(k_f,l_t)\}}(g_1,\dotsc,g_{10}) \,=\, \sum_{\{m^\pm_e, n^\pm_e\}} \prod_{f=1}^{10} \langle j^+_f, m^+_f\lv g_{+tt'}\rv j^+_f, n^+_f\rangle\ \langle j^-_f, m^-_f\lv g_{-tt'}\rv j^-_f, n^-_f\rangle\\
\prod_{t=1}^5 \sum_{i^+_t, i^-_t} f_{i^+_t i^-_t}^{l_t}\bigl(j^+_f,j^-_f,k_f\bigr)\ \langle \otimes_{f\,{\rm in}}\, j^+_f,m^+_f \lv i^+_t \rv \otimes_{f\,{\rm out}}\, j^+_f, n^+_f\rangle\ \langle \otimes_{f\,{\rm in}}\, j^-_f,m^-_f \lv i^-_t \rv \otimes_{f\,{\rm out}}\, j^-_f, n^-_f\rangle.
\end{multline}
Nous avons utilis\'e les parties self-duales et anti-self-duales des holonomies, $G_{tt'}=(g_{+tt'}, g_{-tt'})$. Le fait que ces \'etats sont des solutions faibles des contraintes signifie que les \'el\'ements de matrices des contraintes entre ces \'etats sont nuls, ce qui est explicitement v\'erifi\'e dans \cite{ding-rovelli-volume}.

Le mod\`ele EPR d\'efinit alors l'amplitude d'un 4-simplexe par restriction des coloriages somm\'es dans le mod\`ele d'Ooguri $\Spin(4)$ \`a ceux des \'etats EPR. Il s'agit donc de la contraction d'un symbole 15j de $\Spin(4)$, portant des spins $(j^+_f, j^-_f)$ d\'etermin\'es par $k_f$, avec des entrelaceurs $f(\iota)(k_f, l_t)$. De mani\`ere \'equivalente, l'amplitude d'un 4-simplexe dans ce mod\`ele correspond \`a l'\'evaluation des r\'eseaux de spins construits, soit :
\beq \label{vertex epr}
W_v^{\mathrm{EPR}}(j^+_f,j^-_f,k_{f},l_t) = \sum_{\{i^+_t,i^-_t\}} \mathrm{15j}\bigl(j^+_f,j^-_f;i^+_t,i^-_t\bigr)\, \prod_{t\subset v} d_{i^+_t}d_{i^-_t}\,f_{i^+_t i^-_t}^{l_t}\bigl(j^+_f,j^-_f, k_{f}\bigr),
\ee
o\`u les contraintes \eqref{contraintes spin epr} tiennent. Le collage des simplexes, comme le sugg\`ere la notation, se fait simplement en associant la m\^eme repr\'esentation $k_f$ (et donc $j^\pm_f$) \`a la face $f$ dans tous les t\'etra\`edres de la triangulation o\`u elle intervient, et le m\^eme spin $l_t$ pour les deux entrelaceurs du sous-groupe $\SU(2)$ diagonal, intervenant dans les deux 4-simplexes coll\'es le long de $t$.

Avant de d\'ecrire le mod\`ele de mousses de spins en lui-m\^eme, il est int\'eressant de pouvoir comparer cette proc\'edure de quantification avec celle de Freidel et Krasnov.

\section{La quantification de Freidel-Krasnov} \label{sec:fk-quantisation}

Si les id\'ees de base sont naturellement les m\^emes, la m\'ethode, et le r\'esultat pour $\gamma>1$, sont diff\'erents. On cherche \`a imposer les contraintes de simplicit\'e crois\'ee sur chaque t\'etra\`edre au niveau quantique, mais en utilisant des \emph{\'etats coh\'erents} \cite{new-model-fk}. Le but est bien s\^ur de rendre les \'equations des contraintes transparentes au niveau quantique ! Les \'etats coh\'erents de $\SU(2)$ (ou $\Spin(4)$) employ\'es ont \'et\'e introduits dans le contexte des mousses de spins par Livine et Speziale, \cite{livine-coherentBF}. Ils ont alors reformuler les mousses de spins BF en termes d'entrelaceurs coh\'erents, autorisant un contr\^ole g\'eom\'etrique des bivecteurs au niveau quantique. Cela a \'et\'e directement appliqu\'e dans \cite{new-model-ls}, o\`u Livine et Speziale montrent que la quantification EPR peut se r\'e\'ecrire en termes d'\'etats coh\'erents ! Cette surprise incita les auteurs EPR \`a consid\'erer le mod\`ele comme robuste. Mais, Livine et Speziale y remarquent aussi que cette m\'ethode laisse la place \`a un autre mod\`ele, un peu diff\'erent. Celui-ci \'etait dans le m\^eme temps d\'evelopp\'e en d\'etails par Freidel et Krasnov \cite{new-model-fk} !

Les \'etats coh\'erents qui nous int\'eressent sont d\'efinis comme $\lvert j,\hat{n}\rangle$, dans la repr\'esentation de spin $j$, et de direction $\hat{n}\in S^2$, par l'action d'une rotation $n\in\SU(2)$ sur le vecteur de r\'ef\'erence $\lvert j, m=j\rangle$,
\beq \label{def coherent state}
\lvert j,\hat{n}\rangle = D^{(j)}(n)\,\lvert j,j\rangle.
\ee
La rotation $n$ envoie l'axe de r\'ef\'erence $z$ (utilis\'e pour d\'efinir les \'etats de moment magn\'etique $\lvert j,m\rangle$) sur la direction $\hat{n}$. Il y a bien s\^ur une ambiguit\'e dans cette formulation, du fait que $n$ et $n\exp(i\theta\sigma_z)$ envoient tous deux l'axe $z$ sur la m\^eme direction. Leurs actions se diff\'erencient par une phase dans la d\'efinition des \'etats coh\'erents, de sorte qu'il faut plut\^ot les d\'efinir par l'action de $G/H$, pour $G=\SU(2)$ et $H=\U(1)$ le sous-groupe engendr\'e par $\sigma_z$. Cette ambiguit\'e ne joue aucun r\^ole dans la d\'efinition du mod\`ele. N\'eanmoins, disons tout de suite que c'est en en prenant soin que l'on pourra appr\'ecier la g\'eom\'etrie 4d cod\'ee dans le mod\`ele, comme le montrent \cite{barrett-asymEPR, bf-aarc-val, new-model-val}. Pour fixer les id\'ees, prenons pour cette partie la forme $n = \exp(-i\alpha\sigma_z/2)\exp(-i\beta\sigma_y/2)$. Les \'etats coh\'erents jouissent alors de propri\'et\'es avantageuses, formant notamment une famille g\'en\'eratrice (non-orthogonale),
\begin{align}
\id_{\calH_j} &= \sum_{j=-m}^m \lvert j, m\rangle\,\langle j, m\rvert,\\
&= d_j \int_{G/H} d^2\hat{n}\ \lvert j,\hat{n}\rangle\,\langle j,\hat{n}\rvert. \label{id coherent}
\end{align}
Si $\vec{J}$ est le 3-vecteur form\'e par les g\'en\'erateurs hermitiens de $\SU(2)$, satisfaisant $[J^i,J^j] = i\eps^{ij}_{\phantom{ij}k} J^k$, la formule suivante est la cl\'e de l'interpr\'etation de ces \'etats,
\beq
\langle j,\hat{n}\lvert \vec{J}\rvert j,\hat{n}\rangle = j\,\hat{n}.
\ee
Cela signifie que $\vert j,\hat{n}\rangle$ d\'ecrit en moyenne un vecteur de $\R^3$ de norme $j$ et de direction $\hat{n}$. La propri\'et\'e de coh\'erence vient du fait que l'incertitude est minimale,
\beq
\Delta \vec{J}^2 = j,
\ee
tandis qu'en utilisant un \'etat de r\'ef\'erence g\'en\'erique $\lvert j, m\rangle$ \`a la place de $\lvert j, j\rangle$, nous aurions trouv\'e $\Delta \vec{J}^2 = j+j^2-m^2$. Cela indique tout de m\^eme la possibilit\'e de former des \'etats coh\'erents similaires avec l'\'etat de r\'ef\'erence $\lvert j, -j\rangle$. Ils sont reli\'es aux pr\'ec\'edents de la mani\`ere suivante. Introduisons la matrice $\eps = i\sigma_y = \bigl(\begin{smallmatrix} 0&-1\\1 &0\end{smallmatrix}\bigr)$. Ses \'el\'ements de matrice dans la repr\'esentation (du groupe) de spin $j$ sont : $D^{(j)}_{mm'}(\eps) = (-1)^{j-m'}\delta_{m,-m'}$, de sorte que : $\lvert j, -j\rangle = D^{(j)}(\eps)\lvert j, j\rangle$. Il nous faut donc interpr\'eter la multiplication par $\eps$ sur le groupe. On calcule facilement que : $\eps\,\sigma_z\eps\mone = -\sigma_z$. Donc si $n\in\SU(2)$ am\`ene le vecteur $\hat{z}$ sur $\hat{n}$, alors $n\eps$ l'envoie sur $-\hat{n}$. Soit :
\beq
n\,\sigma_z\,n\mone = \vec{\sigma}\cdot \hat{n} \quad \Rightarrow \quad D^{(j)}(n)\lvert j, -j\rangle = D^{(j)}(n\eps)\lvert j, j\rangle = \lvert j, -\hat{n}\rangle.
\ee
Par ailleurs, nous utiliserons le fait que la conjugaison matricielle par $\eps$ est la conjugaison complexe des matrices $\SU(2)$, $\eps g\eps\mone = \eps\mone g\eps = \bar{g}$. Ainsi : $\langle j, -j\lvert g\rvert j, -j\rangle = \overline{\langle j, j\lvert g\rvert j, j\rangle}$.

Nous pouvons aussi utiliser ces \'etats pour construire des invariants sous l'action du groupe, autrement dit des entrelaceurs. Cela se fait naturellement en moyennant sur $\SU(2)$ des produits tensoriels d'\'etats coh\'erents. Comme nous sommes en 4d (ou bien comme les t\'etra\`edres ont quatre triangles$\dotsc$), nous sommes surtout int\'eress\'es par des entrelaceurs 4-valents,
\beq
\lvert \iota(j_a, \hat{n}_a)\rangle = \int_{\SU(2)} dg\ \bigotimes_{a=1}^4 D^{(j_a)}(g)\lvert j_a,\hat{n}_a\rangle.
\ee
Livine et Speziale ont montr\'e dans \cite{livine-coherentBF} que la norme de tels entrelaceurs est piqu\'ee sur les configurations satisfaisant la relation de fermeture quantique, $\sum_{a=1}^4 j_a\hat{n}_a = 0$. Pour un noeud d'un r\'eseau de spins, il est possible d'utiliser une base d'entrelaceurs coh\'erents, formant une famille compl\`ete mais pas orthogonale. On peut alors reformuler le mod\`ele d'Ooguri pour la th\'eorie BF 4d en changeant la repr\'esentation de l'amplitude de vertex : on \'ecrit la r\'esolution de l'identit\'e sur les entrelaceurs $\iota_t$, provenant de l'int\'egrale \eqref{int g4}, associ\'es aux t\'etra\`edres, dans la base des \'etats coh\'erents :
\begin{align}
\int_{\SU(2)} dg\ \bigotimes_{f=1}^4 D^{(j_f)}(g) &=\  \id_{\Inv(\calH_{j_1}\otimes\dotsm\otimes\calH_{j_4})},\\
&= \int \prod_{f=1}^4 (2j_f+1)\,d^2\hat{n}_f\ \lvert \iota(j_a, \hat{n}_a)\rangle\ \langle \iota(j_a, \hat{n}_a) \rvert. \label{identity coherent intertwiners}
\end{align}
Le symbole 15j devient alors un symbole d\'etermin\'e par les 10 spins sur les triangles et \`a la place des cinq spins virtuels aux t\'etra\`edres cinq entrelaceurs coh\'erents.

Revenons maintenant \`a $G=\Spin(4)$. Contrairement aux variables utilis\'ees dans la version discr\`ete de la th\'eorie BF, nous ne d\'efinissons pas un seul bivecteur par triangle, mais une collection $(B_{ft})$ de bivecteurs associ\'es aux t\'etra\`edres qui ont ce triangle dans leur bord, comme nous l'avons d\'ej\`a fait remarqu\'e \`a la section \ref{sec:collage}. Ces bivecteurs sont pris comme \emph{ind\'ependants} a priori ! Autrement dit, on ne cherchera pas \`a imposer fortement les relations de transport parall\`ele \eqref{transport bivectors} entre des bivecteurs d\'ecrivant pourtant le m\^eme triangle mais dans des t\'etra\`edres (et donc r\'ef\'erentiels) diff\'erents. Ces relations seront en fait seulement valables aux points stationnaires de l'int\'egrale de chemins sur la triangulation, comme nous le montrons plus tard (voir aussi \cite{freidel-conrady-semiclass}).

On \'ecrit donc les contraintes de simplicit\'e crois\'ee \`a l'aide de la formulation \eqref{off-diag grav}. En utilisant les parties self-duales et anti-self-duales, $B_{ft} = b_{+ft}\oplus b_{-ft}$, cela donne en notations matricielles :
\beq 
b_{-ft} +s\,\Ad\bigl( N_t\mone\bigr)\ b_{+ft} =0.
\ee
La variable $N_t$ n'est autre que le 4-vecteur $N_t^I$ unitaire et orthogonal au t\'etra\`edre, vu comme un \'el\'ement de $\SU(2)$ par l'isomorphisme entre $\SU(2)$ et $S^3$. De mani\`ere \'equivalente, la rotation $N_t$ est d\'efinie de sorte que la rotation (4d) $(N_t, \unit)\in\Spin(4)$ envoie le vecteur de r\'ef\'erence $N^{(0)}= (1,0,0,0)$ sur la normale $N_t^I$, par la repr\'esentation standard de $\SO(4)$. Le signe $s=\pm$ indique la possibilit\'e de choisir comme r\'ef\'erence le secteur topologique, $s=-1$, ou le secteur gravitationnel, comme nous le ferons dans la suite, avec $s=1$.

Puisqu'apr\`es avoir impos\'e les contraintes, nous moyennerons sur $\Spin(4)$ pour former des entrelaceurs, nous pouvons sans souci travailler dans la jauge \og temps\fg, $N_t = \unit$ comme dans la quantification EPR. Nous exprimons les r\'eseaux de spins vivants au bord du 4-simplexe dans la base des \'etats coh\'erents, \`a la place des nombres magn\'etiques comme dans \eqref{base magnetic},
\beq
s^{\{j^\pm_f\}}_{\{\hat{n}_{\pm ft}\}} (G_{tt'}) = \prod_f \langle j^+_f, \hat{n}_{+ft}\lvert g_{+tt'}\rvert j^+_f, \hat{n}_{+ft'}\rangle\, \langle j^-_f, \hat{n}_{-ft}\lvert g_{-tt'}\rvert j^-_f, \hat{n}_{-ft'}\rangle.
\ee
Le moment conjugu\'e aux holonomies est toujours, selon \eqref{moment 4s}, $J_{ft} = B_{ft} + \gamma\mone \star B_{ft}$, qui devient au niveau quantique une d\'erivation \`a gauche si $t$ est dual au vertex source du lien dual $f$, et \`a droite si $t$ est le vertex d'arriv\'ee. On peut alors exprimer la valeur moyenne de la contrainte \eqref{simplicity-selfdual} comme :
\beq
\langle b_{+ft}+b_{-ft}\rangle = j^+_f\bigl(1-\gamma\mone\bigr) \hat{n}_{+ft} + j^-_f\bigl(1+\gamma\mone\bigr) \hat{n}_{-ft} = 0.
\ee
Consid\'erons le cas $\gamma>0$, puisque $\gamma<0$ se traite alors en inversant les r\^oles des parties self-duales et anti-self-duales. Cette \'equation fournit comme dans la quantification EPR, $j^+_f/j^-_f = \lvert (\gamma+1)/(\gamma-1)\rvert$. La suite diff\`ere de la quantification EPR, si ce n'est que nous devons encore distinguer les cas $\gamma>1$ et $\gamma<1$. Dans ce dernier cas, on a $\hat{n}_{+ft}=\hat{n}_{-ft}$, tandis que pour $\gamma>1$, un signe change, $\hat{n}_{+ft}=-\hat{n}_{-ft}$. Cela se traduit sur les \'el\'ements de groupe d\'efinissant les \'etats coh\'erents par :
\beq \label{solution fk}
n_{-ft} = \begin{cases} n_{+ft}\,e^{i\psi_{ft}\sigma_z} &\text{si $0<\gamma<1$}, \\ n_{+ft}\,e^{i\psi_{ft}\sigma_z}\,\eps &\text{si $\gamma>1$}. \end{cases}
\ee
Le degr\'e de libert\'e $\U(1)$, $e^{i\psi_{ft}\sigma_z}$, provient de l'ambiguit\'e d\'ej\`a mentionn\'ee sur la d\'efinition des \'etats coh\'erents. Dans l'amplitude du 4-simplexe, il ne g\'en\`ere qu'une phase, qui se trouve \^etre supprim\'ee lorsque l'on colle les 4-simplexes de la triangulation. Nous oublions donc cette phase dans ce qui suit. Il se trouve aussi que le cas $\gamma<1$ reproduit pr\'ecis\'ement le mod\`ele EPR ! Nous regardons donc mainntenant plus en d\'etails l'autre situation, $\gamma>1$ qui donne le mod\`ele FK. On forme pour cela \`a chaque noeud du graphe un entrelaceur,
\beq \label{fk intertwiner}
\int dG\ \bigotimes_{a=1}^4 D^{(j^+_a)}(g_+)\lvert j^+_a, \hat{n}_{a}\rangle\otimes D^{(j^-_a)}(g_-)\lvert j^-_a, -\hat{n}_{a}\rangle.
\ee
Les repr\'esentations sont choisies selon
\beq \label{contraintes spin fk}
j^\pm_f = \lvert\gamma_\pm\rvert j_f, \qquad\text{pour}\quad \f{\gamma_+}{\gamma_-} = \f{\gamma+1}{1-\gamma},
\ee
dans lequel $\gamma_\pm$ sont les plus petits entiers (en valeur absolue) satisfaisant la seconde relation et $\gamma_+>0$, et $j_f$ est un demi-entier.

L'amplitude du 4-simplexe dans le mod\`ele FK est alors obtenue en rempla\c{c}ant dans le mod\`ele d'Ooguri $\Spin(4)$ les deux copies de la r\'esolution de l'identit\'e sur les entrelaceurs coherents \eqref{identity coherent intertwiners} par une restriction aux entrelaceurs \eqref{fk intertwiner} munis des contraintes \eqref{contraintes spin fk}. De mani\`ere \'equivalente, l'amplitude du 4-simplexe $v$ est obtenue en r\'eunissant les entrelaceurs ci-dessus accroch\'es aux cinq t\'etra\`edres du bord, et en les contractant, selon la structure du 4-simplexe, comme pour le symbole 15j de $\SU(2)$ :
\beq \label{vertexfk gamma>1}
\int \prod_{t} dG_{vt}\ \prod_{f}\,\bra j^+_f,\hat{n}_{ft}\rv\,g_{+vt}\mone\, g_{+vt'}\,\lv j^+_f,\hat{n}_{ft'}\ket\ \overline{\bra j^-_f,\hat{n}_{ft}\rv\,g_{-vt}\mone\,g_{-vt'}\,\lv j^-_f,\hat{n}_{ft'}\ket},
\ee
pour $\gamma>1$. La conjugaison complexe de la partie anti-self-duale est vraiment due \`a la matrice $\eps$ dans \eqref{solution fk}, elle-m\^eme due \`a la condition $\hat{n}_{+ft}=-\hat{n}_{-ft}$. Ainsi, pour $\gamma<1$ nous pouvons simplement enlever cette conjugaison,
\beq \label{vertexfk gamma<1}
\int \prod_{t} dG_{vt}\ \prod_{f}\,\bra j^+_f,\hat{n}_{ft}\rv\,g_{+vt}\mone\, g_{+vt'}\,\lv j^+_f,\hat{n}_{f't}\ket\ \bra j^-_f,\hat{n}_{ft}\rv\,g_{-vt}\mone\,g_{-vt'}\,\lv j^-_f,\hat{n}_{f't}\ket.
\ee

Pour \'ecrire l'amplitude dans le langage des mousses de spins, on fait d'abord les int\'egrales sur $G_{vt}$, suivant les formules usuelles \eqref{intg4} et \eqref{pairing}. Remarquons que $n_{ft}$ appara\^it \`a la fois dans la partie self-duale et dans la partie anti-self-duale. Ainsi, le fait que les deux directions de $b_{\pm ft}$ soient reli\'ees par les contraintes de simplicit\'e induit un entrelacement des repr\'esentations $j^+_f$ avec $j^-_f$ \`a travers des repr\'esentations $k_{ft}$ du sous-groupe $\SU(2)$ diagonal. Mais, pour $\gamma>1$, contrairement \`a la prescription EPR que l'on retrouve ici pour $\gamma<1$, l'ensemble de la d\'ecomposition \eqref{decomposition EPR}, $\calH_{(j^+,j^-)} = \bigoplus_{k=\lvert j^+-j^-\rvert}^{j^++j^-} \calH_k$, intervient, avec des poids que nous pr\'eciserons. Lorsque l'on colle les 4-simplexes, on identifie les \'el\'ements $n_{ft}$ des triangles pour des 4-simplexes adjacents coll\'es de long de $t$, et l'on int\`egre sur toutes les configurations, trivialement puisque cela reproduit l'identit\'e \eqref{id coherent}. Comme dans la quantification EPR, les int\'egrales sur $\Spin(4)$ peuvent passer les entrelacements des liens \`a travers $k_{ft}$, pour produire des entrelaceurs $\SU(2)$ de $\Inv_{\SU(2)}(\calH_{k_1}\otimes\dotsm\otimes\calH_{k_4})$. Ceux-ci se d\'eveloppent bien s\^ur \`a l'aide d'un choix d'appariement entre ces spins et d'un spin interm\'ediaire $l_t$. Tout cela fait appara\^itre \`a nouveau les coefficients de fusion, et une amplitude :
\beq
W_v^{\mathrm{FK}\gamma>1}(j^+_f,j^-_f,k_{ft},l_t) = \sum_{\{i^+_t,i^-_t\}} \mathrm{15j}\bigl(j^+_f,j^-_f;i^+_t,i^-_t\bigr)\, \prod_{t\subset v} d_{i^+_t}d_{i^-_t}\,f_{i^+_t i^-_t}^{l_t}\bigl(j^+_f,j^-_f, k_{ft}\bigr).
\ee

L'amplitude d'un 4-simplexe ne suffit naturellement pas \`a d\'efinir un mod\`ele de mousses de spins. Encore faut-il pr\'eciser comment ces amplitudes sont somm\'ees ! Cela demande d'\'expliciter les variables ind\'ependantes sur lesquelles sommer, ce que nous avons fait ci-dessus (et comme les notations le prennent en compte), et d'attribuer des amplitudes aux triangles et t\'etra\`edres, ce que nous faisons dans la section qui suit. Celle-ci se veut un petit r\'esum\'e des mod\`eles, d'une mani\`ere \og unificatrice\fg, inspir\'ee de la pr\'esentation de \cite{freidel-conrady-pathint}.

\section{R\'esum\'e des diff\'erents mod\`eles} \label{sec:resume models}

\subsection{En langage mousses de spins}

Il est possible d'exprimer ces mod\`eles \`a l'aide de donn\'ees de bord communes, et de restriction sur celles-ci, d\'ependantes du mod\`ele, cod\'ees dans l'amplitude des t\'etra\`edres. Les donn\'ees de bord consistent en des spins $j^\pm_f$ reli\'es par le param\`etre d'Immrizi,
\beq \label{quantum diag}
j^\pm_f=\lv\gamma_\pm\rv j_f,
\ee
avec $j_f\in\f{\N}{2}$, et $\gamma_\pm$ des entiers d\'etermin\'es par les prescriptions \eqref{contraintes spin epr} pour EPR et \eqref{contraintes spin fk} pour FK. Nous avons ensuite besoin de spins $k_{ft}$ qui entrelacent les repr\'esentations $j^+_f$ et $j^-_f$, ind\'ependamment \`a chaque t\'etra\`edre, puis d'entrelaceurs $\SU(2)$, de $\Inv_{\SU(2)}(\calH_{k_1}\otimes\dotsm\otimes\calH_{k_4})$, qui se d\'eveloppent bien s\^ur \`a l'aide d'un choix d'appariement entre ces spins et d'un spin interm\'ediaire $l_t$. Le spin $j_f$ s'interpr\`ete naturellement comme l'aire quantifi\'ee du triangle $f$, et $j^\pm_f$ comme le flux quantifi\'e de la norme de $b_{\pm f}(t)$ (ind\'ependant du r\'ef\'erentiel).

On peut alors formuler les mod\`eles par des sommes pond\'er\'ees sur les \'etats de chaque 4-simplexe,
\beq
Z = \sum_{\{j_f\}, \{k_{ft}\}, \{l_t\}} \prod_f W_f \prod_t W_t \prod_v W_v.
\ee
L'amplitude $W_v$ est donn\'ee par :
\beq \label{new vertex}
W_v(j^+_f,j^-_f,k_{ft},l_t) = \sum_{\{i^+_t,i^-_t\}} \mathrm{15j}\bigl(j^+_f,j^-_f;i^+_t,i^-_t\bigr)\, \prod_{t\subset v} d_{i^+_t}d_{i^-_t}\,f_{i^+_t i^-_t}^{l_t}\bigl(j^+_f,j^-_f, k_{ft}\bigr).
\ee
Les coefficients de fusion $f_{i^+_t i^-_t}^{l_t}$ permettent de projeter les entrelaceurs $\Spin(4)$ sur des entrelaceurs $\SU(2)$ \`a chaque t\'etra\`edre, en passant par l'entrelacement $j^+_f\otimes j^-_f\rightarrow k_{ft}$. Bien que cette amplitude joue un r\^ole majeur dans les pr\'esentations usuelles que l'on fait de ces mod\`eles, il faut dire qu'elle n'est absolument pas d\'eterminante ! C'est bien plut\^ot le choix de restrictions sur les donn\'ees et d'amplitudes sur les triangles et t\'etra\`edres qui sont pr\'epond\'erants ! 

Nous verrons en effet que les coefficients de fusion interviennent tr\`es naturellement d\`es lors que l'on s'efforce de construire des mod\`eles avec les contraintes de simplicit\'e qui relient parties self-duale et anti-self-duale des bivecteurs, sans pour autant aboutir au mod\`eles EPR et FK. Par ailleurs, simplement en rel\^achant la contrainte de simplicit\'e diagonale \eqref{quantum diag}, et en gardant l'amplitude $W_v$, \eqref{new vertex}, il est possible de retrouver le mod\`ele d'Ooguri $\Spin(4)$ pour la th\'eorie BF discr\`ete ! Cela se fait en remarquant que les coefficients de fusion peuvent \^etre \og invers\'es\fg{} en sommant sur tous les spins $k_{ft}$. En effet, ils peuvent s'\'ecrire comme de simples produits de symboles 9j \cite{asym-fusion},
\beq
f^l_{i^+ i^-}(j^+_f, j^-_f, k_{ft}) = \begin{Bmatrix} j^+_1 & j^+_2 & i^+ \\ j^-_1 & j^-_2 & i^- \\ k_1 & k_2 & l \end{Bmatrix}
\begin{Bmatrix} j^+_3 & j^+_4 & i^+ \\ j^-_3 & j^-_4 & i^- \\ k_3 & k_4 & l \end{Bmatrix}.
\ee
En utilisant alors l'orthogonalit\'e des 9j, on trouve\footnote{Le coefficient de proportionnalit\'e d\'epend de $i^\pm$ bien s\^ur, et est fondamental pour retrouver un mod\`ele topologique. Nous voulons ici juste regarder comment \'eliminer les coefficients de fusion gr\^ace aux sommes sur $k_{ft}$.}
\beq
\sum_l \sum_{k_1,\dotsc,k_4}\ \Bigl(\prod_{a=1}^4 d_{k_a}\Bigr)\ f^l_{i^+ i^-}(j^+_a, j^-_a, k_a)\,f^l_{i'^+ i'^-}(j^+_a, j^-_a, k_a)\ \propto\ \delta_{i^+,i'^+}\ \delta_{i^-, i'^-}.
\ee
Cela montre qu'on peut reproduire le mod\`ele BF avec une amplitude $W_t$ se comportant en fonction des spins $k_{ft}$ comme
\beq
W_t^{\mathrm{BF}}(k_{ft}) = \prod_{f\subset \pp t} d_{k_{ft}}.
\ee
Si la contrainte diagonale n'est pas relach\'ee, on obtient alors un mod\`ele simplement contraint par \eqref{quantum diag}, o\`u les entrelaceurs $\Spin(4)$ ne sont pas contraints. Ce mod\`ele sera utilis\'e en d\'ebut de la prochaine section.

Dans le mod\`ele FK, $\gamma>1$, les spins $k_{ft}$ ne sont pas contraints. Cela signifie qu'il se distingue en fait du mod\`ele que nous venons de discuter par un choix sp\'ecifique de l'amplitude des t\'etra\`edres, provenant directement de la formule \eqref{vertexfk gamma>1} en termes d'\'etats coh\'erents,
\beq \label{FK weight}
W_t^{\mathrm{FK}\gamma>1} = d_{l_t}\,\prod_{f\subset \pp t}d_{k_{ft}}\Bigg[\begin{pmatrix} j^+_f &j^-_f &k_{ft} \\ j^+_f &-j^-_f &j^-_f-j^+_f \end{pmatrix}\Bigg]^2,
\ee
la quantit\'e entre crochets \'etant un symbole 3mj de Wigner. Celui-ci apporte une pond\'eration non-triviale, qui est piqu\'ee autour de la valeur $k=j^+-j^-$, \cite{freidel-conrady-pathint}. Cette valeur n'est autre que la restriction d\'efinissant les \'etats EPR pour $\gamma>1$ :
\beq \label{EPR weight}
W_t^{\mathrm{EPR}\gamma>1} = d_{l_t}\,\prod_{f\subset t} \delta_{k_{ft},j^+_f-j^-_f}.
\ee
Ainsi, m\^eme si les deux mod\`eles peuvent avoir un comportement proche dans le calcul des amplitudes en mousses de spins, ils sont tr\`es diff\'erents en ce que le mod\`ele FK, $\gamma>1$, n'impose aucune restriction sur les \'etats de bord du 4-simplexe (\`a part \eqref{quantum diag} tout de m\^eme), contrairement au mod\`ele EPR dont les donn\'ees de bord ne sont rien d'autre que celles des r\'eseaux de spins LQG ! Autrement dit, le mod\`ele EPR parvient \`a r\'eduire les r\'eseaux de spins $\Spin(4)$ aux r\'eseaux $\SU(2)$ de la LQG, tandis que les donn\'ees de bord utilis\'ees ici pour le mod\`ele FK d\'ecrivent pr\'ecis\'ement la base des r\'eseaux de spins $\Spin(4)$ dits projet\'es, introduits par Livine et Alexandrov, \cite{alexandrov-clqg-hilbert, alexandrov-livine-clqg, livine-projected}, avec la contrainte \eqref{quantum diag}. Pour une discussion comparant les deux m\'ethodes de quantification en lien avec l'alg\`ebre des contraintes de simplicit\'e, on pourra par ailleurs consulter \cite{pereira-engle-classes}.

Lorsque $\gamma=\infty$, le mod\`ele EPR s\'electionne le mode $k_{ft}=0$, du fait qu'alors $j^+_f = j^-_f$. L'amplitude se r\'eduit alors au mod\`ele longtemps \'etudi\'e avant, de Barrett-Crane (BC) \cite{BCpaper}, que nous \'etudions dans la section suivante.

Pour $\gamma<1$, les deux mod\`eles coincident, en s\'electionnant le plus grand spin dans la d\'ecomposition de $\calH_{(j^+,j^-)}$ en repr\'esentations irr\'eductibles du sous-groupe $\SU(2)$ diagonal,
\beq \label{gamma<1}
W_t^{\gamma<1} = d_{l_t} \prod_{f\subset t} \delta_{k_{ft},j^+_f+j^-_f}.
\ee
Pour $\gamma\rightarrow0$, il s'agit de la proposition initiale EPR \cite{epr-long}.

Du fait des m\'ethodes de quantification employ\'ees, on comprend que l'amplitude des triangles soit relativement ambigue. Cela est bien s\^ur reli\'e au fait qu'on ne sait pas comment les restrictions que l'on impose aux r\'eseaux de spin g\'en\'eriques $\Spin(4)$ doivent \^etre prises en compte dans le produit scalaire cin\'ematique. Tr\`es r\'ecemment, Bianchi, Regoli et Rovelli ont \'etabli ce lien pr\'ecis\'ement et montr\'e, \cite{regoli-face} que l'amplitude des triangles peut \^etre d\'etermin\'ee de mani\`ere unique par : la structure de l'espace de Hilbert au bord, la composition des triangulations, et un troisi\`eme crit\`ere de \og localit\'e\fg. L'amplitude des triangles est alors donn\'ee par : $W_f = d_{j_f}$. Ainsi, on peut prendre pour produit scalaire entre les \'etats de bord satisfaisant les contraintes, qui sont des fonctions $\Psi(j_f, k_{ft}, l_t)$ sur un bord $\Gamma$,
\beq
\langle \Phi \vert \Psi\rangle_\Gamma = \sum_{j_f, k_{ft}, l_t} \prod_f W_f\ \prod_t W_t\quad \overline{\Phi}(j_f, k_{ft}, l_t)\,\Psi(j_f, k_{ft}, l_t),
\ee
et on retrouve dans le cas du mod\`ele EPR le produit des r\'eseaux de spins de la LQG sur un graphe (voir \'egalement \cite{freidel-conrady-pathint}). On sait que dans une th\'eorie poss\'edant des contraintes de seconde classe, comme la relativit\'e g\'en\'erale dans l'action de Holst ou de Plebanski, ces contraintes contribuent de mani\`ere non-triviale \`a la mesure de l'int\'egrale de chemins \cite{henneaux-teitelboim}. En particulier, dans les variables de l'espace des configurations, l'int\'egrale de chemins ne se formule plus simplement avec l'exponentielle de l'action lagrangienne \cite{henneaux-pleb, engle-measure2ndclass}. Ce ph\'enom\`ene, connu certes depuis longtemps comme une des difficult\'es de la gravit\'e quantique, doit \^etre trait\'e d'une certaine mani\`ere dans la quantification en mousses de spins, bien qu'il y fut jusqu'ici ignor\'e en pratique, et intervenir vraisemblablement dans les amplitudes des triangles. Il est donc excellent d'avoir des crit\`eres et m\'ethodes pour fixer ces amplitudes, comme ceux fournis par \cite{regoli-face}. Nous allons maintenant pr\'esenter l'approche poursuivie dans cette th\`ese consistant \`a \'etudier les mousses de spins comme des int\'egrales de chemins pour des g\'eom\'etries discr\`etes. Nous verrons en particulier comment une mesure non-triviale dans ces int\'egrales se traduit dans l'amplitude des triangles en termes de mousses de spins. Il s'agit d'un pr\'eliminaire essentiel pour tenter de comprendre pr\'ecis\'ement la relation entre la mesure due aux contraintes de seconde classe et ces amplitudes de triangles.

\subsection{Avec des int\'egrales plut\^ot que des sommes}

Pour finir ce r\'esum\'e, notons que si les mousses de spins sont int\'eressantes du point de vue th\'eorique, comme quantification background-independent en relation avec les r\'eseaux de spins, elles impliquent des sommes disons \emph{affreuses} en pratique ! Pour calculer, comme nous l'avons vu explicitement dans le mod\`ele jouet en 3d, \ref{sec:3dgraviton}, il peut \^etre plus judicieux de trouver une expression avec des int\'egrales plut\^ot que des sommes. Autrement dit, une formulation dans laquelle les variables ne sont plus des quantit\'es discr\`etes mais des variables continues, de groupe ou d'espace homog\`ene typiquement. C'est d'ailleurs la fa\c{c}on la plus simple de construire et de comprendre les mod\`eles topologiques. Les liens duaux aux $(n-1)$-simplexes $t$ portent des holonomies $g_t$, et chaque face duale $f$ est pond\'er\'ee par 
\beq
A_f^{\rm BF}(g_t) = \sum_{j}\bigl(2j+1\bigr)\,\chi_j\bigl(\prod_{t} g_t^{[f:t]}\bigr) = \delta\Bigl( \prod_{t} g_t^{[f:t]}\Bigr),
\ee
le produit \'etant pris le long du bord de la plaquette $f$, et $[f:t]=\pm1$ \'etant le nombre d'incidence d\'etermin\'e par l'orientation relative, et
\beq
Z_{\rm BF} = \int \prod_t dg_t\ \prod_f A_f^{\rm BF}(g_t).
\ee
Autrement dit, on comprend que : (i) les entrelaceurs proviennent de moyennes sur le groupe, et (ii) on est en mesure de faire explicitement la somme sur les spins attach\'es aux faces.

Pour le mod\`ele EPR/FK, on peut reprendre l'expression donn\'ee en \eqref{vertexfk gamma<1} de l'amplitude d'un 4-simplexe, et d\'evelopper diff\'eremment pour factoriser des amplitudes sur les faces. On colle la formule \eqref{vertexfk gamma<1} sur tous les 4-simplexes d'une triangulation, en donnant le m\^eme spin $j_f$ \`a la face $f$ chaque fois qu'elle intervient, et avec un \'etat coh\'erent d\'etermin\'e par $\hat{n}_{ft}\in S^2$ pour chaque paire triangle-t\'etra\`edre,
\begin{align}
Z_{\gamma<1} &= \sum_{\{j_f\}} \int \prod_{(v,t)} dG_{vt}\,\prod_{(f,t)} d^2n_{ft}\ \prod_{(f,v)} \bra j^+_f,\hat{n}_{ft}\rv\,g_{+vt}\mone\, g_{+vt'}\,\lv j^+_f,\hat{n}_{f't}\ket\ \bra j^-_f,\hat{n}_{ft}\rv\,g_{-vt}\mone\,g_{-vt'}\,\lv j^-_f,\hat{n}_{f't}\ket,\\
&= \int \prod_{(v,t)} dG_{vt}\,\prod_{(f,t)} d^2n_{ft}\ A^{\gamma<1}_f(G_{vt}, n_{ft}),
\end{align}
o\`u l'on a d\'efini le poids par face comme contenant la somme sur le spin pr\'esent \`a chaque face :
\beq
A^{\gamma<1}_f(G_{vt}, n_{ft}) = \sum_{j\in\f{\N}{2}} \prod_{v\in\pp f}\,\bra j^+,\hat{n}_{ft}\rv\,g_{+vt}\mone\, g_{+vt'}\,\lv j^+,\hat{n}_{ft'}\ket\ \bra j^-,\hat{n}_{ft}\rv\,g_{-vt}\mone\,g_{-vt'}\,\lv j^-,\hat{n}_{ft'}\ket,
\ee
pour $j^\pm = \gamma_\pm j$. Le point cl\'e est maintenant que le produit scalaire des \'etats coh\'erents a une d\'ependance simplement exponentielle en le spin, (voir plus de d\'etails dans \cite{livine-coherentBF}),
\beq
\langle j, \hat{n}\vert j, \hat{n}'\rangle = e^{2j\ h(\hat{n}, \hat{n}')},
\ee
c'est-\`a-dire que la fonction $h$ ne d\'epend pas de $j$ mais seulement des directions des \'etats coh\'erents. Comme l'action de $\SU(2)$ sur un \'etat coh\'erent est simplement de faire tourner sa direction, cela signifie que la somme sur le spin peut \^etre effectu\'ee de mani\`ere exacte dans le mod\`ele ! Ce qui donne :
\beq \label{sum spins epr}
A^{\gamma<1}_f(G_{vt}, n_{ft}) = \f1{1-\exp \bigl(\sum_{v\in\pp f} \gamma_+ h(\hat{n}_{ft}, \hat{n}_{ft},g_{+vt}\mone\, g_{+vt'}) + \gamma_- h(\hat{n}_{ft}, \hat{n}_{ft},g_{-vt}\mone\, g_{-vt'})\bigr)}.
\ee
Il est important d'observer que cette amplitude est partout d\'efinie sauf lorsque l'argument de l'exponentielle s'annule. Nous pouvons essayer de tracer un parall\`ele avec le mod\`ele topologique, pour anticiper les probl\`emes potentiels li\'es \`a cette singularit\'e. Dans le mod\`ele topologique, l'amplitude est contruite via des fonctions delta sur les holonomies autour des faces. Bien que de tels objets ne soient pas d\'efinis comme fonctions usuelles (et intuitivement \og divergent\fg{} en l'identit\'e), ils le sont au sens des distributions, et permettent donc (dans une certaine mesure) de faire les int\'egrales sur les holonomies. Les divergences du mod\`ele sont alors li\'ees \`a \og un trop grand nombre de fonctions delta\fg{} \cite{freidel-louapre-diffeo}, i.e. que leur produit ne constitue plus une distribution ! Par l\`a on voit bien que ces divergences ne d\'ecoulent pas directement du fait que $\delta(g)$ n'est pas d\'efini en l'identit\'e. La situation est a priori un peu diff\'erente ici, car on ne sait pas \`a ce stade quelles sont les configurations des variables $(G_{vt}, \hat{n}_{ft})$ qui annulent le d\'enominateur de \eqref{sum spins epr}.

Mais ce rappel sur le mod\`ele topologique nous donne une marche \`a suivre : commencer par d\'efinir l'amplitude $A^{\gamma<1}_f$, au moins au sens des \emph{distributions}. Il s'agit d'une route naturelle vers une d\'efinition propre du mod\`ele, incluant la r\'egularisation, puis vers l'\'etude des divergences. Quant \`a l'interpr\'etation des configurations \`a l'origine des \'eventuelles divergences, nous verrons qu'il s'agit de g\'eom\'etries plates \ref{sec:stationnary geom fk} !

\chapter{Simplicit\'e d'un seul bivecteur et repr\'esentations simples de Spin(4)} \label{sec:single bivector}

Ce chapitre vise \`a expliquer sur un exemple simple ce que l'on souhaite faire ensuite sur des mod\`eles plus r\'ealistes, et \`a introduire quelques ingr\'edients qui nous re-serviront. On peut donc y voir un exercice avant de s'attaquer \`a l'ensemble des contraintes de simplicit\'e. En particulier, la d\'erivation du mod\`ele de Barrett-Crane, qui suivra, utilisera des m\'ethodes tr\`es similaires. Bien que notre mani\`ere de traiter les contraintes soit compl\`etement originale, la discr\'etisation de la partie BF que nous utilisons ici est totalement standard : elle reprend le cadre \'etabli dans \cite{fk-action-principle}. On pourra de la sorte appr\'ecier dans quelle mesure les \'equations classiques de la section \ref{sec:classical sf} sont vraiment prises en compte par cette discr\'etisation. Et c'est en comprenant les probl\`emes qu'elle soul\`eve que nous pourrons aller au-del\`a du mod\`ele de Barrett-Crane, et en particulier rencontrer les mod\`eles EPR/FK.

Mais ce chapitre pr\'esente bien s\^ur un int\'er\^et propre : il \'etudie les contraintes de simplicit\'e diagonale \eqref{diag}, pour $\gamma=\infty$. Le mod\`ele qui en r\'esulte a \'et\'e assez peu consid\'er\'e dans la litt\'erature, surtout du point de vue int\'egrale de chemins sur r\'eseau. Le contenu physique de ces contraintes diagonales est simple : elles imposent \`a chaque bivecteur $B_f$, ou de mani\`ere \'equivalente au dual de Hodge $\star B_f$, d'\^etre simple, i.e. de s'\'ecrire comme le produit antisym\'etris\'e de deux 4-vecteurs, $u^{[I} v^{J]}$. Ce mod\`ele est ainsi authentiquement non-toplogique. Les contraintes diagonales, comme nous l'avons vu, sont d'habitude impos\'ees au niveau quantique gr\^ace \`a des proc\'edures de quantification g\'eom\'etrique issues de \cite{baez-barrett-quantum-tet}. L'interpr\'etation g\'eom\'etrique correspondante, motiv\'ee par l'analyse canonique de l'action BF \cite{epr-long}, permet d'associer les repr\'esentations de $\Spin(4)$ coloriant les triangles, $(j^+_f, j^-_f)$, \`a une quantification des aires de ces triangles. Dans le cas $\gamma\rightarrow\infty$, cette identification conduit, comme en \eqref{contraintes spin fk}, \`a $j^+_f = j^-_f$, qui caract\'erise les repr\'esentations dites simples de $\Spin(4)$.

Notre but est ici de v\'erifier dans quelle mesure cette impl\'ementation est correcte du point de vue des int\'egrales, classiques, sur la triangulation. Rappelons-nous que dans le continu, la 2-forme $B$ est un multiplicateur de Lagrange, \eqref{BFaction}, imposant l'\'equation de courbure nulle $F(A) = 0$, \eqref{flatness}. Dans la formulation discr\`ete, il est aussi naturel de consid\'erer que les bivecteurs peuvent servir de multiplicateurs de Lagrange pour imposer la contrainte de courbure nulle, $G_f=\unit$, sur les holonomies. En effet, les bivecteurs discr\'etisent la 2-forme $B$ sur les triangles, tandis que comme en calcul de Regge, nous avons concentr\'e la courbure \'egalement sur les triangles ! Cela a permis de forger l'action suivante de type BF,
\beq
S_{\mathrm{BF}}\bigl(B_f(t),G_{vt}\bigr) = \sum_f \Tr\bigl(B_f(t)\,G_f(t)\bigr),
\ee
dont on peut v\'erifier\footnotemark qu'elle restitue la fonction de partition attendue,
\begin{align}
Z_{\mathrm{BF}} &= \int \prod_{(t,v)} dG_{vt} \Bigg[\prod_f dB_f(v^\star)\ e^{i\Tr(B_f(v^\star)G_f(v^\star))}\Bigg] \label{sf bf1}\\
&= \int \prod_{(t,v)} dG_{vt}\ \prod_f \delta\bigl(G_f(t)\bigr).
\end{align}
Puisque nous disposons d'une formulation claire des contraintes de simplicit\'e discr\`etes, (voir en section \ref{sec:classical sf}), il est tout naturel de chercher \`a les inserer dans l'int\'egrale ci-dessus, et de tenter alors de la r\'e\'ecrire en mousses de spins. On pourrait ainsi trouver les conditions permettant d'aboutir \`a des mod\`eles connus, tels que ceux d\'ecrits \`a la section pr\'ec\'edente, ou voir quels sont les mod\`eles qui \'emergent le plus naturellement, et s'int\'eresser \`a l'ambiguit\'e sur la mesure. Autrement dit, nous allons travailler sur des int\'egrales du type
\beq
\int \prod_{(t,v)} dG_{vt}\ \prod_f \Bigg[dB_f(v^\star)\ e^{i\Tr(B_f(v^\star)G_f(v^\star))}\Bigg]\ \tl{\delta}\bigl(\operatorname{contrainte}(B_f, G_{vt})\bigr),
\ee
o\`u $\tl{\delta}$ est une fonction plus ou moins piqu\'ee et centr\'ee sur la contrainte \`a imposer$\dotsc$

\footnotetext{Regardons explicitement le cas $G=\SU(2)$, en \'ecrivant $g=\pm\sqrt{1-\vec{p}^2} + i\vec{p}\cdot \vec{\sigma}$, avec $\lvert\vec{p}\rvert<1$, et $b = -i b^i\sigma_i/2$. Alors, il vient :
\beq
\int db e^{i\tr(bg)} = \int d^3b e^{i\vec{b}\cdot\vec{p}} = \delta^{(3)}(\vec{p}).
\ee
On montre \cite{freidel-louapre-PR1} que cette fonction delta est \'egale \`a : $\f12(\delta_{\SU(2)}(g) + \delta_{\SU(2)}(-g)) = \delta_{\SO(3)}(g)$, de sorte que la fonction delta provenant de \eqref{sf bf1} doit \^etre comprise sur $\SO(3)\times\SO(3)$. L'invariance $g\rightarrow -g$ provient bien s\^ur du fait que $\tr(bg)$ ne voit pas le terme $\pm\sqrt{1-\vec{p}^2}$.}

L'impl\'ementation standard de la simplicit\'e diagonale, issue de la quantification g\'eom\'etrique, produit la fonction de partition
\beq
Z_{\mathrm{diag}} = \int \prod_t dG_t\ \prod_f Z_f \left(G_f\right), \quad\text{o\`u}\qquad
Z_f\left(G_f\right) = \sum_{j_f} d_{j_f}^k\ \chi_{j_f}(g_{+f})\ \chi_{j_f}(g_{-f}). \label{Z_f simple rep ansatz}
\ee
L'amplitude des triangles habituellement consid\'er\'ee est $k=2$, et provient de celle du mod\`ele d'Ooguri $\Spin(4)$ avec une fonction delta sur les contraintes, $d_{j^+_f}d_{j^-_f}\delta_{j^+_f,j^-_f}$. Nous allons montrer que le calcul de type int\'egrale de chemins donne plus naturellement ce m\^eme r\'esultat avec $k=0$, et nous verrons que cela se g\'en\'eralise \`a :
\be \label{ansatz simple rep measure}
Z_f\left(G_f\right) = \chi(g_{+f})\ \sum_{j_f} \chi_{j_f}(g_{+f})\ \chi_{j_f}(g_{-f}),
\ee
o\`u $\chi$ est une fonction de classe sur $\SU(2)$, typiquement un caract\`ere, pour assurer l'invariance de jauge. Cette fonction repr\'esente une ambiguit\'e de la mesure de l'int\'egrale, et comme nous le verrons, les mod\`eles du type \eqref{ansatz simple rep measure} partagent les m\^emes propri\'et\'es physiques. Lorsque $\chi$ est un caract\`ere, cela induit de simples \og shifts\fg, i.e. des d\'ecalages dans les repr\'esentations self-duales (la raison pour laquelle cela ne brise pas la sym\'etrie avec le facteur anti-self-dual sera claire plus bas). 

Nous devons trouver une param\'etrisation des contraintes \eqref{diag} suffisamment sympathique pour pouvoir effectuer explicitement les int\'egrales sur les bivecteurs. Les contraintes s'expriment comme : $\lvert \vec{b}_{+f}\rvert^2=\lvert \vec{b}_{-f}\rvert^2$. Comme nous l'avons d\'etaill\'e en section \ref{sec:geom normales}, il existe une rotation $N_f\in\SU(2)$ qui envoie la partie self-duale sur la partie anti-self-duale et qui repr\'esente le choix d'une normale au plan d\'efini par $\star B_f$,
\be 
b_{-f} +\ N_f^{-1}\ b_{+f}\ N_f = 0,
\ee
en notation matricielle. Puisque les bivecteurs sont d\'efinis dans des r\'ef\'erentiels d\'etermin\'es, il faut ici d\'efinir les rotations $N_f$ dans ces m\^emes rep\`eres locaux.

A la place d'imposer ces contraintes directement avec des fonctions delta, nous les introduisons dans l'action \`a l'aide de multiplicateurs de Lagrange $q_f\in\SU(2)$,
\be \label{simple rep action}
S_{\mathrm{diag}} = \sum_f \Tr\bigl(B_f(v) G_f(v)\bigr) + \tr\Bigl[q_f(v) \bigl(b_{-f}(v) + N_f\mone b_{+f}^{\phantom{}} N_f^{\phantom{}}(v)\bigr)\Bigr]
\ee
Le choix de multiplicateurs dans $\SU(2)$ est ici le plus naturel, du fait que les \'equations du mouvement obtenues par variations de cette action par rapport \`a $b_{\pm f}$ leur demandent d'\^etre \'egaux \`a des \'el\'ements de groupe :
\beq \label{eom q}
g_{-f} = q_f,\quad\text{et}\qquad g_{+f} = N_f\,q_f\,N_f\mone.
\ee
Cela implique aussi que $g_{+f}$ et $g_{-f}$ sont dans la m\^eme classe de conjugaison.

La mesure de Haar sur $\SU(2)$ n'\'etant pas la mesure de Lebesgue habituellement utilis\'ee pour des multiplicateurs, nous autorisons ici un poids quelconque $\mu(q_f)$ dans la fonction de partition. Cela se traduit bien s\^ur comme des poids diff\'erents sur la quantit\'e \`a contraindre $(b_{-f}+N_f\mone b_{+f} N_f)$ pour chaque face :
\beq \label{simple rep1}
Z_{f} = \int dB_f\ e^{i\tr(B_f G_f)}\ \int dh_f\ \tl{\delta}_\mu\bigl(b_{-f}+h_f^{-1}b_{+f}h_f\bigr),
\ee
avec
\beq
\tl{\delta}_\mu (x_f) = \int_{\SU(2)} dq_f\ \mu(q_f)\ \exp\big\{i\tr [q_fx_f]\big\}.
\ee
A partir des transformations de jauge sur $b_{\pm f}$, on voit que $N_f$ se transforme avec le secteur self-dual \`a gauche, et le secteur anti-self-dual \`a droite. $q_f$ se transforme par conjugaison, et on prend donc $\mu$ comme une fonction de classe. On peut voir que l'imposition stricte de la contrainte, telle que $\tl{\delta}_\mu = \delta$, se produit\footnotemark pour $\mu(q) = \lvert \tr\ q\rvert$. Nous allons par la suite trouver une relation pr\'ecise entre $\mu$ et la fonction $\chi$ de \eqref{ansatz simple rep measure}.

\footnotetext{La mesure de Haar sur $\SU(2)$ peut \^etre reli\'ee \`a la mesure de Lebesgue sur $\R^3$ \`a l'aide du facteur $\lvert \tr q\rvert$ qui prend en compte la structure de la loi de composition du groupe et sa compacit\'e (voir en appendice \ref{sec:app}). Donc nous n'avons pas \`a strictement parl\'e $\tl{\delta}_\mu=\delta$ dans ce cas, car le domaine d'int\'egration reste compact. Mais cela a le m\^eme effet que la fonction delta, comme on peut le voir en int\'egrant les bivecteurs en premier lieu. Par ailleurs, un poids trivial donne : $\int dq\ e^{i\tr(bq)} = \f{J_1(\lv b\rv)}{\lv b\rv}$, dont le maximum est atteint en $\lv b\rv=0$, et $J_1$ \'etant la fonction de Bessel de premi\`ere esp\`ece d'ordre 1, \`a des coefficients inint\'eressants pr\`es.}

Puisque l'action est lin\'eaire en les bivecteurs, l'int\'egrale sur un bivecteur projette les variables restantes sur les \'equations du mouvement \eqref{eom q}. Cela s'accompagne d'un facteur de mesure calcul\'e en \eqref{int b} :
\be \label{simple rep after b int}
Z_{f} = \f{1}{\lvert \tr(g_{+f})\ \tr(g_{-f})\rvert}\ \int dN_f\, dq_f\ \mu(q_f)\
\delta\big(g_{-f}\, q_f\big)\ \delta\big(g_{+f}\, N_f\, q_f\, N_f^{-1}\big).
\ee
Le multiplicateur de Lagrange a donc pour effet de demander \`a $g_{+f}$ et $g_{-f}$ d'\^etre conjugu\'es par $N_f$,
\be \label{physical step simple rep}
Z_{f} = \f{\mu(g_{-f})}{\lvert \tr(g_{+f})\ \tr(g_{-f})\rvert}\ \int dN_f\
\delta\big(g_{+f}\, N_f\, g_{-f}^{-1}\, N_f^{-1}\big).
\ee
$g_{+f}$ et $g_{-f}$ ont le m\^eme angle de classe, et $\mu$ est une fonction de classe, donc $\mu(g_{-f})=\mu(g_{+f})$. La quantit\'e cl\'e dans \eqref{physical step simple rep}, qui fait de ce mod\`ele un mod\`ele non-toplogique, est bien s\^ur : $\delta(g_{+f} h_f g_{-f}^{-1} h_f^{-1})$, qui concentre l'amplitude sur les configurations classiques \eqref{eom q}. Remarquons que
\beq
g_{+f}\,N_f\,g_{-f}\mone = G_f\cdot N_f
\ee
n'est autre que le r\'esultat du transport parall\`ele de la rotation $N_f$ tout autour de la face $f$, jusqu'\`a revenir au point de d\'epart. Ainsi l'amplitude s\'electionne les holonomies qui laissent le vecteur normal $N_f^I$ invariant ! Cette restriction des degr\'es de libert\'e des holonomies doit certainement \^etre consid\'er\'ee comme la caract\'erisation du mod\`ele, de la m\^eme mani\`ere que l'action de type BF, $\Tr(B_f G_f)$, n'est qu'un principe d'action pour d\'efinir le mod\`ele topologique par la condition de courbure nulle, $\delta(G_f)$.

En utilisant le d\'eveloppement standard, $\delta(g) = \sum_j d_j \chi_j(g)$, et l'orthogonalit\'e de ces repr\'esentations, les int\'egrales sur les rotations $N_f$ cnoduisent \`a :
\be \label{final simple rep}
Z_{f} = \f{\mu(g_{+f})}{\left(\tr \,g_{+f}\right)^2}\ \sum_{j_f} \chi_{j_f}\bigl(g_{+f}\bigr)\ \chi_{j_f}\bigl(g_{-f}\bigr),
\ee
qui n'est autre que le r\'esultat annonc\'e \eqref{ansatz simple rep measure}, avec la relation entre les facteurs de mesure : $\mu(g) =
(\tr\,g)^2\chi(g)$. Si l'on souhaite se d\'ebarrasser de $\chi$ dans l'expression finale, il suffit de choisir $\mu(g) = \tr^2 g$, ce qui correspond \`a imposer les contraintes diagonales non pas avec une fonction delta mais plut\^ot avec
\be \label{bessel measure}
\tl{\delta}_\mu(\vec{x}) = \f{J_1(\lvert \vec{x}\rvert)+J_3(\lvert \vec{x}\rvert)}{\lvert \vec{x}\rvert} \quad \Rightarrow\quad \chi=1.
\ee
Ici, $J_1$ et $J_3$ sont des fonctions de Bessel de premi\`ere esp\`ece, d\'efinies par : $J_n(x) = \f{1}{\pi i^n}\int_0^\pi e^{ix\cos \theta}\ \cos(n\theta) d\theta$. En particulier, la fonction  $(J_1(x)+J_3(x))/x$ est centr\'ee autour de 0 avec une largeur finie. Cela signifie que dse fluctuations autour des contraintes sont autoris\'ees. Mais, ces fuctuations sont tr\`es particuli\`eres car elles sont telles que l'int\'egration des bivecteurs am\`ene d'authentiques fonction delta sur le groupe dans \eqref{physical step simple rep}. Puisqu nous souhaitons plut\^ot construire les mod\`eles de mousses de spins \`a partir d'int\'egrales sur le groupe, cette mesure est un choix naturel.

Plus g\'en\'eralement, l'insertion d'un caract\`ere $\mu(g) = \chi_l(g)$ dans la repr\'esentation spin $l$ se traduit par le choix :
\be
\tl{\delta}_{\chi_l}(\vec{x}) = \f{J_{2l+1}(\lvert \vec{x}\rvert)}{\lvert \vec{x}\rvert}
\ee
De telles fonctions ne sont pas centr\'ees autour de z\'ero (et s'y annulent m\^eme pour tout $l>1$) ! La maximum est \`a la place atteint pour une valeur de $x$ qui d\'epend de $l$. Le fait que nous obtenons tout de m\^eme un mod\`ele du type th\'eorie BF restreinte aux repr\'esentations simples de $\Spin(4)$ peut se comprendre par la forme sp\'eciale de $\tl{\delta}_\mu$, comme transform\'ee de Fourier \'evalu\'ee sur $b_{-f}+N_f^{-1}b_{+f}N_f$. A la suite de quoi les int\'egrales sur les bivecteurs, qui apparaissent lin\'eairement dans l'action, projettent sur les configurations satisfaisant la relation cl\'e du mod\`ele, $g_{+f} = N_f g_{-f} N_f^{-1}$.

Introduisons maintenant une nouvelle fa\c{c}on d'ins\'erer les contraintes dans l'int\'egrale de chemins discrets. Cette fa\c{c}on de proc\'eder se trouve \^etre tr\`es naturelle dans le cadre de base utilis\'e ici et dans la d\'erivation du mod\`ele BC au chapitre suivant. L'id\'ee est la suivante. Puisque l'\'equation \eqref{simple rep after b int} force les variables $q_f\in\SU(2)$ \`a \^etre des holonomies, il serait pr\'ef\'erable d'introduire les contraintes dans l'action d'une mani\`ere qui respecte pleinement la structure de groupe et \'evite les facteurs de mesure du type $(\tr\,g_{-f})^2$. Nous avons en effet proc\'ed\'e plus haut de mani\`ere assez na\"ive. Notre introduction des contraintes dans l'action \eqref{simple rep action} conduit \`a des formules pas tout \`a fait naturelles, puisqu'elle donne pour la partie anti-self-duale par exemple, $\tr[b_{-f}(g_{-f}+q_f)]$. C'est justement le fait d'avoir une telle somme d'\'el\'ements de groupe qui conduit aux facteurs de mesure $\lv \tr g\rv$ dans \eqref{simple rep after b int}. Ces derniers n'ont pas un sens physique clair, et il est vraisemblable qu'ils disparaissent si l'on utilise \`a la place de sommes d'\'el\'ements de groupe la loi de composition naturelle de $\SU(2)$. L'outil bien adapt\'e \`a cela est une addition non-commutative, not\'ee ici $\oplus$, qui multiplie les \'el\'ements de groupe \`a la place de les sommer,
\be \label{nc law}
e^{\tr(bg)\oplus \tr(bh)}\equiv e^{\tr(bgh)} \ne e^{\tr(b(g+h))}.
\ee
Cette op\'eration d\'eforme la composition standard des ondes planes lorsque l'espace des moments est le groupe $\SU(2)$, d'une mani\`ere naturellement compatible avce la multiplication $\SU(2)$. Elle va permettre d'\'eliminer tous les facteurs de mesure, au sens o\`u elle prend en compte implicitement la mesure \eqref{bessel measure}. Cette loi non-commutative est en fait naturelle d\`es que l'on regarde des th\'eories de jauge sur r\'eseau impliquant \`a la fois des \'el\'ements d'un groupe de Lie et de son alg\`ebre. En particulier, elle a d\'ej\`a \'et\'e consid\'er\'ee dans l'\'etude des observables dans le mod\`ele de Ponzano-Regge pour la gravit\'e 3d, \cite{freidel-livine-EQFT, freidel-livine-PR3}, et plus r\'ecemment pour donner une repr\'esentation de l'espace de Hilbert des r\'eseaux de spins en termes des flux de la triade (qui sont des variables non-commutatives) \cite{baratin-oriti-flux-ncrep} ! Une approche \'equivalente \`a la notre mais du point de vue plus g\'en\'eral des th\'eories des champs sur le groupe de Lie $\SU(2)$ (GFT) est pr\'esent\'ee par Baratin et Oriti dans \cite{baratin-oriti-ncgft}.

On impl\'emente donc les contraintes diagonales dans l'action par :
\begin{align}
S^\oplus_{\mathrm{diag}} &= \sum_f \Tr\Bigl(B_f(v)\, G_f(v)\Bigr) \oplus \tr\Bigl[q_f(v) \bigl(b_{-f}(v) + N_f^{-1}\,b_{+f}\,N_f(v)\bigr)\Bigr], \\
 &= \sum_f \tr\bigl(b_{-f}(v)\, g_{-f}(v)\, q_f(v)\bigr) + \tr\bigl(b_{+f}(v)\, g_{+f}(v)\, N_f\,q_f\,N_f^{-1}(v)\bigr). \label{diag action}
\end{align}
Cette forme est maintenant similaire \`a celle de l'action \eqref{discreteBFaction} de type BF, d\'eform\'ee par de faux bivecteurs (i.e. qu'ils ne sont pas dans l'alg\`ebre),
\be
\tl{B}_f(v) = Q_f(v)\,B_f(v),\qquad \qquad \text{with}\qquad Q_f(v) = \bigl(N_f\,q_f\,N_f^{-1}(v),q_f(v)\bigr)
\ee
On montre facilement que cette action reproduit les r\'esultats pr\'ec\'edents, mais en \'evitant proprement les facteurs $(\tr\, g_{-f})^2$. Nous obtenons \`a la place de \eqref{simple rep after b int},
\begin{gather} \label{content simple rep}
Z^\oplus_{\mathrm{diag}} = \int \prod_t dG_t \prod_f Z^\oplus_{f},\\
\begin{split}
 \text{avec}\qquad Z^\oplus_{f} &= \int dN_f\ dq_f\ \mu(q_f)\ \delta\big(g_{-f}\, q_f\big)\ \delta\big(g_{+f}\, N_f\, q_f\, N_f^{-1}\big), \\
 &= \mu(g_{-f})\ \int dN_f\ \delta\big(g_{+f}\, N_f\, g_{-f}^{-1}\, N_f^{-1}\big).
\end{split}
\end{gather}
Ainsi, \eqref{Z_f simple rep ansatz} est reproduite avec un poids trivial, $k=0$, et le choix $\mu=1$. La somme $\oplus$ prend donc en compte de mani\`ere naturelle pour les mousses de spins, et implicitement, la mesure \eqref{bessel measure}, \`a travers la structure du groupe $\SU(2)$.

Pour terminer, il est int\'eressant de traiter les contraintes diagonales dans la formulation \eqref{diag normales} directement, $(\star B_f)_{IJ}N_f^J = 0$. Dans la perpsective de traiter les contraintes de simplicit\'e crois\'ee, on pourra comparer l'effet de cette formule avec le secteur topologique, $(B_f)_{IJ}N_f^J = 0$ (m\^eme si bien s\^ur cela ne change rien dans le r\'esultat final tant que l'on ne regarde que la simplicit\'e diagonale). Cette contrainte \'etant un 4-vecteur, on peut prendre sa norme et l'ins\'erer dans l'action avec un multiplicateur de Lagrange $\lambda_f$. On \'ecrit pour une face l'action BF, $\Tr(B_f G_f)= B_f^{IJ} P_{fIJ}$, et il vient 
\be
Z_{f} = \int_{S^3} dN_f \int dB_f \int d\lambda_f\ e^{i B_f^{IJ} P_{fIJ} + i\lambda_f\ (\star B_f)^{IJ} N_{fJ} (\star B_f)_{IK} N_f^K}
\ee
Gr\^ace \`a la covariance $\Spin(4)$, il nous suffit de comprendre cette int\'egrale pour $N_f=(1,0,0,0)$. Alors, nous avons $B_f^{ij}=0$ pour $i,j=1,2,3$, de sorte que seules les composantes $B_f^{0i}$ sont libres. Les int\'egrales sur ces variables donnent donc la contrainte $P_f^{0i} = 0$, ou de mani\`ere covariante : $N_{fJ} P_f^{IJ}=0$, soit :
\be
Z_{f} = \int dN_f\ \delta\Big(N_{f J}\, P_f^{IJ}\Big)
\ee
qui est bien \'equivalent \`a \eqref{content simple rep}. Par dualit\'e, imposer \`a la place la variante \og topologique\fg, $N_{f J}
B_f^{IJ}=0$, am\`ene \`a la restriction duale : $\eps_{IJKL}\, N^J_f P_f^{KL} = 0$.

\chapter{Le mod\`ele de Barrett-Crane} \label{sec:BC}

Le mod\`ele de Barrett-Crane (BC), pr\'esent\'e dans \cite{BCpaper}, fut le premier candidat comme mod\`ele de mousses pour la gravit\'e quantique, en r\'eduisant les degr\'es de libert\'e des r\'eseaux de spins pour $\Spin(4)$ par des contraintes de simplicit\'e. Ces contraintes sont celles qui permettent de reconstruire un 4-simplexe \`a partir de bivecteurs sur les triangles, et sont impos\'ees au niveau quantique sur les liens et les noeuds du r\'eseau de spins au bord d'un 4-simplexe. Une pr\'esentation dans le cadre de la quantification g\'eom\'etrique de l'espace des phases de cinq t\'etra\`edres de bord fut donn\'ee dans \cite{baez-barrett-quantum-tet}. Le mod\`ele ne d\'efinit en fait que l'amplitude assign\'ee aux 4-simplexes, qui se trouve \^etre un symbole appel\'e 10j, d\'ependant de dix spins qui repr\'esentent les aires des 10 triangles. Celui-ci s'obtient \`a partir du symbole 15j de $\Spin(4)$ en deux \'etapes. D'abord par restriction aux repr\'esentations irr\'eductibles \emph{simples}, i.e. $j^+_f=j^-_f=j_f$, impl\'ementant ainsi les contraintes de simplicit\'e diagonale, puis par restriction \`a un unique entrelaceur pour chacun des cinq t\'etra\`edres, not\'e $\iota_{BC}$. Il s'agit du seul entrelaceur 4-valent dont la d\'ecomposition en vertexes 3-valents ne fasse intervenir que des repr\'esentations simples, ind\'ependamment du choix de base (pour l'appariement entre les repr\'esentations incidentes). C'est ainsi que les contraintes de simplicit\'e crois\'ee sont impos\'ees au niveau de chaque t\'etra\`edre. Le symbole 10j admet la repr\'esentation suivante en termes d'int\'egrales sur $\SU(2)$ :
\beq \label{10j int}
\{10j(j_f)\} = \int \prod_{t=1}^5 dg_t\ \prod_{f=1}^{10} \chi_{j_f}(g_{u(f)}^{\phantom{}}\,g_{d(f)}\mone).
\ee
Autrement dit, un \'el\'ement de groupe $g_t$ est associ\'e \`a chacun des cinq t\'etra\`edres. Un triangle $f$ appara\^it dans le bord d'exactement deux t\'etra\`edres, not\'es $u(f), d(f)$. Puis on \'evalue le caract\`ere $\chi_{j_f}(g_{u(f)}\,g_{d(f)}\mone)$ entre les deux \'el\'ements de groupe correspondants, dans la repr\'esentation de spin $j_f$. Les amplitudes associ\'ees aux t\'etra\`edres et triangles ne sont pas clairement d\'efinies (on pourra consulter \cite{baez-convergenceBC} pour une comparaison explicite de diff\'erents poids en rapport avec la convergence du mod\`ele).

Aujourd'hui, le mod\`ele BC appara\^it comme la quantification EPR pour $\gamma\rightarrow\infty$, i.e. sans le param\`etre d'Immirzi, ou sans le terme topologique de l'action de Holst (dont on sait l'importance en LQG). Plus pr\'ecis\'ement, il \'emerge de la d\'ecomposition de $\calH_{j^+_f}\otimes\calH_{j^-_f}$ sur les repr\'esentations irr\'eductibles du sous-groupe $\SU(2)$ diagonal (qui stabilise la normale au t\'etra\`edre) par le choix de la repr\'esentation triviale. L'entrelaceur $\iota_{BC}$ est donc le seul entrelaceur invariant sous $\SU(2)_{\mathrm{diag}}^4$, et peut \^etre vu comme un entrelaceur sur l'espace homog\`ene $S^3 \simeq \Spin(4)/\SU(2)$.

On sait \cite{alesci-BC1} qu'il ne reproduit pas la limite semi-classique attendue dans le cadre du calcul du propagateur \`a un 4-simplexe selon les id\'ees d\'evelopp\'ees au chapitre \ref{sec:semiclass}. Il poss\`ede n\'eanmoins des propri\'et\'es int\'eressantes au niveau de la g\'eom\'etrie semi-classique. On a notamment pu isoler dans la limite des grands spins du 10j l'action de Regge du 4-simplexe, vue comme fonction des dix aires \cite{steele-asym10j, williams-asym10j, freidel-louapre-6j, baez-asym10j} (mais aussi d'autres termes qui dominent en fait l'asymptotique, provenant notamment des configurations d\'eg\'en\'er\'ees du 4-simplexe).

Dans ce chapitre, nous donnons en suivant les id\'ees pr\'esent\'ees au chapitre pr\'ec\'edent une pr\'esentation du mod\`ele BC comme la fonction de partition d'un syst\`eme de bivecteurs, $B_{ft}$, d'holonomies $G_{vt}$ et de normales aux t\'etra\`edres $N_t$, sur une triangulation. L'action classique de ce syst\`eme est construite \`a partir de l'action pour la th\'eorie BF $\Spin(4)$, et des contraintes de simplicit\'e sous leur forme lin\'earis\'ee faisant intervenir les normales $N_t$. Le calcul exact de la fonction de partition sous forme de sommes sur les mousses de spins est alors possible.

Ayant trouv\'e une formulation du mod\`ele quantique comme int\'egrale de chemins pour un syst\`eme classique, la question est de conna\^itre le comportement classique de celui-ci. Etant donn\'e les variables utilis\'ees, on peut se demander s'il s'agit de g\'eom\'etries de Regge ou non, et quelle en est la dynamique. Ainsi, nous allons montrer que le mod\`ele traite correctement les contraintes de simplicit\'e crois\'ee, bien qu'\'etant plus simple que les nouveaux mod\`eles EPR/FK, ce qui contribue \`a le rendre int\'eressant. Mais nous verrons explicitement qu'il ne parvient pas \`a int\'egrer les corr\'elations entre simplexes adjacents (pour un point de vue diff\'erent techniquement, mais \'equivalent dans cette conclusion, voir \cite{new-model-fk}). Cela tient au fait que les relations de transport parall\`ele entre bivecteurs ne sont pas satisfaites aux points stationnaires ! Or nous connaissons depuis \ref{sec:gluing-aarc} l'importance de ces relations pour reconstruire des g\'eom\'etries de Regge \`a partir des bivecteurs et holonomies. De plus nous discutons les \'equations de la dynamique, qui sont loin d'une version discr\`ete des \'equations de la relativit\'e g\'en\'erale.

La section \ref{sec:actionBC} pr\'esente l'action utilis\'ee pour reproduire le mod\`ele, et le calcul correspondant explicite. Un point important de la construction est de consid\'erer comme variables ind\'ependantes des bivecteurs distincts sur chaque 4-simplexe partageant une m\^eme face. Il faut alors introduire un processus de collage permettant d'identifier aux points stationnaires les bivecteurs d'une m\^eme face. Pour cela nous utilisons des holonomies de bord comme dans \cite{fk-action-principle}, mais nous verrons \`a la section \ref{sec:classicalBC} sur l'analyse des configurations classiques que ce processus \'echoue en pr\'esence des contraintes de simplicit\'e.

La section \ref{sec:generalisationBC} pr\'esente des g\'en\'eralisations du mod\`ele qui mettent en avant le r\^ole des amplitudes de t\'etra\`edres et de triangles dans les mousses de spins. Dans le secteur topologique, l'\'equivalent du mod\`ele BC se trouve donn\'e par un mod\`ele restreint aux repr\'esentations simples sur les faces, avec des entrelaceurs libres, comme au chapitre pr\'ec\'edent, mais avec des poids pour les sous-simplexes diff\'erents. Nous verrons \'egalement comment une mesure non-triviale dans l'int\'egrale de chemins conduit \`a l'apparition des coefficients de fusion. Cela montre que ce sont des objets naturels en pr\'esence des contraintes, et que le point cl\'e se trouve encore \^etre les amplitudes des t\'etra\`edres et triangles.

La section \ref{sec:10jrecurrence} reprend le programme des relations de r\'ecurrence pour les mousses de spins. Nous d\'erivons une relation de r\'ecurrence pour le symbole 10j, que nous interpr\'etons comme codant la propri\'et\'e g\'eom\'etrique de fermeture d'un 4-simplexe. Cette r\'ecurrence est satisfaite dans l'asymptotique pour l'action de Regge fonction des aires du 4-simplexe. Et nous discutons la possibilit\'e de voir cette relation comme une combinaison des mouvements g\'eom\'etriques mis en \'evidence dans les relations de r\'ecurrence sur le symbole 15j \`a la section \ref{sec:15jrecurrence}.


\section{Une action discr\`ete pour le mod\`ele BC} \label{sec:actionBC}

Nous nous concentrons ici sur la construction d'une action au niveau discret correspondant \`a la situation simplifi\'ee d\'ecrite \'equation \eqref{broken up measure}. C'est-\`a-dire que nous souhaitons imposer dans un principe d'action de type BF les contraintes de simplicit\'e crois\'ee sur des bivecteurs $B_{ft}$ tous ind\'ependants les uns des autres (ne satisfaisant aucune relation par transport parall\`ele impos\'ee a priori). Il s'agit de la situation la plus rencontr\'ee dans la litt\'erature (voir notamment \cite{freidel-conrady-pathint}). Notre but est bien s\^ur d'aboutir \`a un lien clair et propre entre un principe d'action et le mod\`ele de mousses de spins qui en d\'ecoule. Il a \'et\'e dit qu'imposer fortement les contraintes de simplicit\'e conduit au mod\`ele BC, alors que le mod\`ele EPR les impl\'emente faiblement (puisqu'elles sont de seconde classe) \cite{epr-short}. On peut aussi lire que le mod\`ele BC n'admet pas d'\'equivalent dans le secteur topologique \cite{new-model-fk} (du point de vue Freidel-Krasnov, car du point de vue EPR, c'est le mod\`ele EPR !). Notre \'etude confirme que ces affirmations doivent \^etre temp\'er\'ees, en insistant sur le fait qu'elles d\'ependent du processus de quantification envisag\'e. En effet, le principal r\'esultat de cette section est que le mod\`ele le plus naturel dans ce cadre est le mod\`ele BC. Il est obtenu en imposant les contraintes d'une mani\`ere qu'on pourrait qualifi\'ee de faible, i.e. selon la mesure \eqref{bessel measure}, ou la loi de composition non-commutative $\oplus$ introduite au chapitre pr\'ec\'edent, qui autorise des fluctuations tr\`es particuli\`eres autour des contraintes. De plus, nous observons bien que le mod\`ele BC correspond \`a une quantification du secteur g\'eom\'etrique, tandis qu'un \'equivalent dans le secteur non-g\'eom\'etrique est donn\'e par un mod\`ele du type d'Ooguri simplement restreint par l'utilisation des seules repr\'esentations simples de $\Spin(4)$, avec des poids particuliers pour les t\'etra\`edres et triangles.

Dans \cite{freidel-conrady-pathint}, Conrady et Freidel ont fourni un formalisme d'int\'egrales de chemins pour reproduire les amplitudes du mod\`ele FK. Leur construction pr\'esente les m\^emes variables que nous (avec la m\^eme interpr\'etation physique), et utilise le m\^eme processus de collage des simplexes que celui que nous allons pr\'esenter. N\'eanmoins, l'action qu'ils construisent est destin\'ee \`a offrir une int\'egrale de chemins reproduisant exactement le mod\`ele FK, et n'est pas une action du type BF. Notre perspective est ici diff\'erente puisque nous voulons consid\'erer des int\'egrales de chemins g\'en\'eriques, fond\'ees sur des actions de type BF discret et sur les contraintes de simplicit\'e, et explorer quels mod\`eles \'emergent naturellement. Il r\'esulte de notre \'etude que le mod\`ele le plus naturel dans ce cadre est le mod\`ele BC.

\begin{figure}
\begin{center}
\includegraphics[width=4cm]{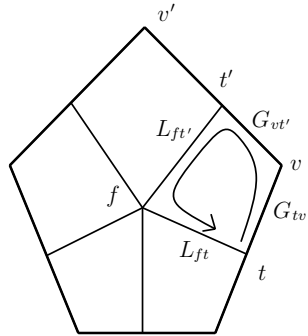}
\end{center}\caption{ \label{fig:wedge} Un wedge $(f,v)$ est la portion d'une face duale $f$ contenue dans un 4-simplexe donn\'e $v$. Du point de vue de la triangulation, cela correspond \`a \'eclater la celle-ci en 4-simplexes disjoints. Pour recoller les diff\'erentes copies d'un m\^eme triangle, on utilise des holonomies $L_{ft}$ le long des bords des wedges, chacune \'etant associ\'ee au t\'etra\`edre commun \`a chaque paire de 4-simplexes.}
\end{figure}

Pour prendre en compte les contraintes sur chaque t\'etra\`edre ind\'ependamment, nous consid\'erons des bivecteurs $B_{ft}$ et des normales $N_t$ telles que, $\eps_{IJKL} B_{ft}^{IJ} N_t^K = 0$. Comme nous ne d\'esirons pas explicitement traiter les relations de transport parall\`ele des bivecteurs, nous allons utiliser le processus de collage d\'evelopp\'e dans \cite{fk-action-principle}. Celui-ci est en fait souvent utiis\'e m\^eme pour traiter le mod\`ele topologique. Il repose sur un d\'ecoupage des faces duales aux triangles en \og wedges\fg, introduit par Reisenberger \cite{}. Un wedge, carat\'eris\'e par la donn\'ee d'un triangle et d'un 4-simplexe, $(f,v)$, est la portion de la face duale contenue dans le 4-simplexe $v$, comme l'illustre la figure \ref{fig:wedge}. Du point de vue de la triangulation, cela correspond \`a un \'eclatement de cette derni\`ere en 4-simplexes disjoints, et \`a consid\'erer ind\'ependamment les unes des autres toutes les copies d'un m\^eme triangle.

Nous avons alors besoin de rajouter des variables de bord sur les 4-simplexes isol\'es, qui doivent permettre d'identifier les copies d'un triangle donn\'e. En plus des holonomies $G_{vt}$, nous consid\'erons donc des holonomies de bord $L_{ft}\in\Spin(4)$, longeant les bords des wedges. Autrement dit, les variables $L_{ft}$ discr\'etisent la connexion sur les liens qui joignent le \og centre\fg{} d'une face duale aux liens formant le bord de la face (voir figure \ref{fig:wedge}). L'orientation de chaque face duale induit une orientation sur chaque wedge qui donne un sens de parcours entre les liens duaux aux t\'etra\`edres $t$ et $t'$. Nous noterons donc un wedge plut\^ot par la paire $(f,t)$ correspondante (car nos bivecteurs et normales sont plut\^ot d\'efinis sur les r\'ef\'erentiels des t\'etra\`edres). Avec les holonomies entre simplexes de dimension 3 et 4, $G_{vt}$, et les holonomies de bord, $L_{ft}$, nous pouvons former une holonomie de wedge $G_{ft}$, se transformant par conjugaison en $t$,
\beq \label{wedgehol}
G_{ft} = L_{ft}\,L_{ft'}\mone\,G_{vt'}\mone\,G_{vt}.
\ee
Ces holonomies permettent de construire une action pour le mod\`ele topologique, en les couplant aux bivecteurs,
\beq \label{wedgeaction}
\sum_{(f,\tau)} \Tr \big( B_{f\tau}\, G_{f\tau}\big).
\ee
Naturellement, cette action est diff\'erente de l'action $\Tr(B_f(t) G_f(t))$, \eqref{discreteBFaction}, puisqu'elle n'utilise m\^eme pas les m\^emes variables. Mais la fonction de partition obtenue en int\'egrant les $B_{ft}, G_{vt}$ et $L_{ft}$ est bien celle du mod\`ele topologique. Les \'equations du mouvement sont int\'eressantes \`a regarder. Les variations des bivecteurs, des holonomies $G_{vt}$ et $L_{ft}$ conduisent respectivement \`a :
\begin{align}
&G_{ft} = \unit,\\
&\sum_{f\in\pp t} \eps_{ft}\,B_{ft} = 0,\\
&B_{ft} = \Ad\bigl(G_{tt'}\bigr)\ B_{ft'},
\end{align}
o\`u dans la deuxi\`eme \'equation $\eps_{ft}=\pm$ code l'orientation relative de la face duale et du lien dual, et dans la troisi\`eme \'equation $t$ et $t'$ sont deux t\'etra\`edres voisins (d'un m\^eme 4-simplexe), et $G_{tt'} = G_{vt}\mone G_{vt'}$. On reconstruit la courbure concentr\'ee autour des triangles par le produit des holonomies de wedge d'une face donn\'ee (conjugu\'ees par $L_{ft}\mone$). Ainsi, la premi\`ere \'equation reproduit bien : $G_f = \unit$. La deuxi\`eme \'equation est la relation de fermeture du t\'etra\`edre $t$, ce qui confirme son interpr\'etation comme la contrainte de Gau\ss{} (elle est obtenue par variation des holonomies $G_{vt}$), engendrant les transformations de jauge au niveau canonique. Quant \`a la troisi\`eme \'equation, c'est en quelque sorte la raison d'\^etre des holonomies de bord : elles permettent classiquement d'obtenir les relations de transport parall\`ele des bivecteurs !

Il est important de souligner que les deux \'equations de fermeture et transport parall\`ele des bivecteurs sont obtenues \emph{apr\`es simplification} par l'\'equation de courbure nulle $G_{ft}=\unit$ ! Il n'est pas \'evident qu'elles puissent toujours tenir en pr\'esence de \og sources\fg{} pour les holonomies$\dotsc$

Nous pourrions alors comme au d\'ebut du chapitre pr\'ec\'edent, introduire les contraintes de simplicit\'e \eqref{simplicity-selfdual}, \`a l'aide de multiplicateurs de Lagrange $q_{ft}\in\SU(2)$ en ajoutant \`a l'action \eqref{wedgeaction} le terme :
\beq
\sum_{(f,t)} \tr\Bigl[ q_{ft}\bigl( b_{-ft} + N_t\mone\,b_{+ft}\,N_t\bigr)\Bigr].
\ee
Le mod\`ele de mousses de spins qui en r\'esulte est pr\'esent\'e dans \cite{BCpaper-val}. En utilisant la mesure de largeur finie \eqref{bessel measure} autour des contraintes, il s'agit du mod\`ele BC. Le choix \og naturel\fg{} de cette mesure tient au fait que l'action totale implique des sommes d\'el\'ements de groupe du type : $\tr[b_{-f\tau}(g_{-f\tau}+q_{f\tau})]$. Comme nous l'avons expliqu\'e au chapitre pr\'ec\'edent, il est possible d'oublier ces subtilit\'es, qui sont alors cod\'ees de mani\`ere naturelle dans la structure du groupe $\SU(2)$, en utilisant la loi de composition \eqref{nc law}, bien adapt\'ee \`a cette structure. L'action correspondante s'\'ecrit :
\beq
S_{\mathrm{BC}}(B_{ft}, G_{vt}, L_{ft}, N_t, q_{ft}) = \sum_{(f,t)} \tr\bigl(b_{-ft}\,g_{-ft}\,q_{ft}\bigr) + \tr\bigl(b_{+ft}\,g_{+ft}\,N_t q_{ft}N_t^{-1}\bigr).
\ee
Les configurations classiques et la proximit\'e de cette action avec les id\'ees de la quantification EPR seront \'etudi\'ees \`a la section suivante \ref{sec:classicalBC}. Nous montrons ici comment cette action reproduit bien le mod\`ele BC. Nous int\'egrons d'abord sur les bivecteurs $B_{ft}$, qui sont rappelons-le tous \emph{ind\'ependants}, selon la formule usuelle :
\beq \label{intB-BC}
\int_{\R^3} d^3b_{+ft}\ e^{i \tr(b_{+ft}\,g_{+ft}\,N_t q_{ft}N_t^{-1})} = \delta\bigl(g_{+ft}\,N_t\,q_{ft}\,N_t^{-1}\bigr),
\ee
et de m\^eme pour le secteur anti-self-dual. Les int\'egrales sur les multiplicateurs $q_{ft}$ s'effectuent comme pour la contrainte diagonale du chapitre pr\'ec\'edent,
\beq
\int_{\SU(2)} dq_{ft}\ \delta\bigl(g_{+ft}\,N_t\,q_{ft}\,N_t^{-1}\bigr)\, \delta\bigl(g_{-ft}\,q_{ft}\bigr) = \delta\bigl(g_{+ft}\,N_t\,g_{-ft}\mone\,N_t^{-1}\bigr).
\ee
Tout comme le mod\`ele topologique est construit sur la contrainte de courbure nulle, $\delta(G_f)$, la fonction delta apparaissant ici peut \^etre vue comme la caract\'erisation ou la d\'efinition du mod`ele BC : les holonomies du wedge, $g_{\pm ft}$, sont reli\'ees par conjugaison par la normale $N_t$, ou de mani\`ere \'equivalente, le transport parall\`ele autour du wedge doit laisser la normal invariante, $g_{+ft}N_t g_{-ft}\mone = N_t$.

L'\'etape cl\'e consiste ensuite \`a int\'egrer les holonomies de bord $L_{ft}$, de mani\`ere \`a recoller les wedges de chaque face (de mani\`ere consistente ou non, cela sera d\'etaill\'e \`a la section suivante). Pour cela, on d\'eveloppe les fonctions delta de la formule ci-dessus sur les caract\`eres des repr\'esentations irr\'eductibles de $\SU(2)$, puis chaque caract\`ere en somme de produits d'\'el\'ements de matrices des variables concern\'ees. Puisque chaque \'el\'ement $L_{ft}$ n'appara\^it que deux fois pour chaque face, sur deux wedges voisins, il suffit d'invoquer la relation d'orthogonalit\'e des \'el\'ements de matrices des repr\'esentations. On voit facilement que ces int\'egrales identifient tous les spins associ\'es aux wedges d'une face \`a un seul spin not\'e $j_f$. L'int\'egrale sur une variable $L_{ft}$ va alors recombiner les deux caract\`eres des wedges voisins en deux caract\`eres diff\'erents, et cela se fait successivement tout autour de la face. En prenant en compte les facteurs de dimension des repr\'esentations, on aboutit pour une face $f$ \`a :
\beq \label{intL}
\int \prod_{t} dL_{ft}\ \prod_{\text{wedges de }f} \delta\bigl(g_{+ft}\,N_t\,g_{-ft}\mone\,N_t^{-1}\bigr) = \sum_{j_f} d_{j_f}^2 \prod_v \f1{d_{j_f}}\,\chi_{j_f}\bigl(g_{+tt'}^{\phantom{}}\,N_{t'}^{\phantom{}}\,g_{-tt'}\mone\,N_t\mone\bigr).
\ee
Le produit sur les vertexes duaux \`a tous les 4-simplexes $v$ partageant $f$ correspond \`a un produit sur les wedges de la face duale \`a $f$. Par ailleurs, $g_{\pm tt'} = g_{\pm vt}\mone g_{\pm vt'}^{\phantom{}}$, o\`u $t, t'$ sont les deux t\'etra\`edres partageant le triangle $f$ dans le 4-simplexe $v$.

On sait (on suspectait) que le mod\`ele BC manque de corr\'elations entre 4-simplexes voisins. Pourtant il semble ici qu'en prenant une normale par t\'etra\`edre nous soyons bien parti (c'est ce que nous demande de faire les contraintes de simplicit\'e), \'etant donn\'e que cette normale est partag\'ee par les deux 4-simplexes $v, v'$ coll\'es le long de $t$. Mais en fait, ceci n'est vrai qu'\`a du transport parall\`ele pr\`es ; il ne faut pas oublier les degr\'es de libert\'e des variables $G_{vt}$. Ainsi une normale $N_t$ doit \^etre transport\'ee diff\'eremment vers les 4-simplexes $v$ et $v'$, et prend alors les formes :
\beq \label{decoupling normals}
N_{tv} = g_{+vt}^{\phantom{}}\,N_t^{\phantom{}}\,g_{-vt}\mone, \quad \text{et}\qquad N_{tv'} = g_{+v't}^{\phantom{}}\,N_t^{\phantom{}}\,g_{-v't}\mone.
\ee
Les 4-vecteurs correspondants s'interpr\`etent bien s\^ur comme la normale \`a $t$ vu dans le r\'ef\'erentiel de $v$. Les caract\`eres qui interviennent dans \eqref{intL} se lisent alors : $\chi_{j_f}(N_{t'v}\,N_{tv}\mone)$, i.e. qu'ils comparent les normales de t\'etra\`edres adjacents dans leur r\'ef\'erentiel commun, celui de $v$. En fait, la d\'efinition \eqref{decoupling normals} se comprend aussi comme un changement de variables sur disons $g_{+vt}$, qui absorbe sur sa droite $N_t g_{-vt}$. Ce qui en fait un changement de variables efficace, c'est que l'holonomie $g_{+vt}$ n'appara\^it que sous cette forme, i.e. uniquement pour transporter la normale $N_t$ dans le r\'ef\'erentiel du 4-simplexe $v$ ! Ainsi, les int\'egrales sur $N_t$ et $g_{-vt}$ sont triviales (avec des mesures de Haar normalis\'ees).

Notons comme rappel\'e dans \cite{livine-speziale-10jgraviton} que les caract\`eres apparaissant dans \eqref{intL} peuvent aussi s'\'ecrire, dans la jauge temps $N_t=\unit$, comme l'\'el\'ement de matrices $\langle J_f 0\lv G_{vt}\,G_{vt'}\mone\rv J_f0\rangle$, o\`u $\rv J_f0\rangle$ est le seul \'etat dans la repr\'esentation $J_f=(j_f,j_f)$ de $\Spin(4)$ qui soit invariant sous le sous-groupe $\SU(2)$ diagonal (sans la jauge temps, sous le sous-groupe $\SU(2)$ qui laisse la normale invariante), C'est bien ce qui caract\'erise le mod\`ele BC.

En rassemblant ces r\'esultats pour chaque triangle, on obtient la fonction de partition dans le langage des mousses de spins, pour lesquelles l'amplitude des 4-simplexes est donn\'ee par le symbole 10j de Barrett-Crane, que l'on retrouve sous forme d'int\'egrales \eqref{10j int} :
\be \label{ZBC final}
Z_{\mathrm{BC}} = \sum_{\{j_f\}}\ \prod_f d_{j_f}^2\ \prod_t \f{1}{\prod_{f\subset \pp t} d_{j_f}}\ \prod_v \Biggl[\int \prod_{t\subset\pp v} dN_{vt}\, \prod_{f\subset\pp v} \chi_{j_f}\bigl(N_{vt}^{\phantom{}}\,N_{vt'}\mone\bigr)\Biggr].
\ee
L'amplitude sur les faces est celle souvent utilis\'ee par convention, i.e. comme provenant du mod\`ele topologique. Mais nous l'obtenons ici diff\'eremment, par un calcul complet, sans trous. L'amplitude des t\'etra\`edres est celle utilis\'ee dans \cite{livine-speziale-10jgraviton}, et propos\'ee dans \cite{oriti-gluing} comme alternative \`a celles de \cite{perez-BC-bubble} et \cite{depietri-BC-gft}.

\section{Quelques g\'en\'eralisations : de l'importance des amplitudes des t\'etra\`edres et triangles} \label{sec:generalisationBC}

Nous gardons le m\^eme cadre d'\'etude, \`a savoir l'utilisation de bivecteurs $B_{ft}$ tous ind\'ependants et de wedges pour imposer les contraintes sur chaque simplexe. Nous pr\'esentons bri\`evement deux g\'en\'eralisations du mod\`ele BC, d\'etaill\'ees dans \cite{BCpaper-val}, obtenues respectivement en passant dans le secteur non-g\'eom\'etrique et en utilisant une mesure quelconque pour les multiplicateurs de Lagrange $q_{ft}$.

En ce qui concerne le secteur topologique, ou plut\^ot non-g\'eom\'etrique puisqu'aucun des mod\`eles propos\'e n'est topologique (!), les diff\'erentes quantifications conduisant aux mousses de spins entra\^inent diff\'erentes conclusions ! Pour la quantification \`a la EPR, l'\'equivalent du mod\`ele BC pour $\gamma\rightarrow0$ n'est autre que le mod\`ele EPR \cite{epr-long}. Pour la quantification \`a la FK, son \'equivalent n'existe tout simplement pas \cite{new-model-fk}. Ici, nous obtenons encore un r\'esultat diff\'erent : l'\'equivalent non-g\'eom\'etrique du mod\`ele BC est un mod\`ele de type Ooguri $\Spin(4)$ dont les repr\'esentations sur les triangles sont simples, comme dans le chapitre \ref{sec:single bivector}, et avec des amplitudes bien particuli\`eres sur les t\'etra\`edres et triangles. Ce r\'esultat est obtenu en ins\'erant les contraintes de simplicit\'e dans l'int\'egrale de chemins sur r\'eseau \emph{avant} l'int\'egration des bivecteurs. Il nous permet donc en comparaison avec le mod\`ele du chapitre \ref{sec:single bivector} d'insister sur l'importance des amplitudes des t\'etra\`edres et triangles dans les mod\`eles, qui peuvent traduire des situations physiques totalement diff\'erentes touten ayant la m\^eme amplitude de 4-simplexe.

Puis nous regardons l'influence de la mesure des multiplicateurs de Lagrange, ce qui r\'ev\`ele l'apparition des coefficients de fusion, ingr\'edients cl\'e des nouveaux mod\`eles ! Cela montre en fait que ces coefficients, qui entrelacent les spins self-duaux et anti-self-duaux sont des objets tr\`es naturels d\`es lors que l'on s'int\'eresse au contraintes de simplicit\'e crois\'ee. Ce calcul nous permet de plus d'insiter \`a nouveau sur l'importance des amplitudes des t\'etra\`edres et triangles ! En effet, notre construction reproduit pr\'ecis\'ement les amplitudes des mod\`eles EPR/FK pour un 4-simplexe, mais les amplitudes des plus petits simplexes de ces mod\`eles sont ici hors de port\'ee !

\subsection{Le secteur non-g\'eom\'etrique et les repr\'esentations simples de Spin(4)}

Les contraintes \`a ins\'erer sont obtenues par changement du signe dans la relation entre les parties self-duale et anti-self-duale des bivecteurs :
\beq
b_{-ft} - \Ad\bigl(N_t\mone\bigr)\ b_{+ft} = 0.
\ee
Pour les prendre en compte dans l'action, l'approche na\"ive est d'ajouter au terme de type BF avec wedge \eqref{wedgeaction} le terme :
\beq
\sum_{(f,t)} \tr\Bigl[ q_{ft}\bigl( b_{-ft} - N_t\mone\,b_{+ft}\,N_t\bigr)\Bigr],
\ee
pour des multiplicateurs de Lagrange $q_{ft}\in\SU(2)$. La question est ensuite de savoir comment le changement de signe se traduit dans l'utilisation de la loi de composition non-commutative $\oplus$, qui permet de combiner les contraintes avec \eqref{wedgeaction} d'une mani\`ere naturelle sur le groupe et l'alg\`ebre de Lie. Cela est tr\`es simple, il suffit de remarquer que : $\tr(bq\mone) = -\tr(bq)$. Ainsi, nous n'avons qu'\`a changer $q_{ft}$ en $q_{ft}\mone$ dans la \emph{partie sef-duale} de l'action conduisant au mod\`ele BC,
\beq
S_{\mathrm{non-geo}}(B_{ft}, G_{vt}, L_{ft}, N_t, q_{ft}) = \sum_{(f,t)} \tr\bigl(b_{-ft}\,g_{-ft}\,q_{ft}\bigr) + \tr\bigl(b_{+ft}\,g_{+ft}\,N_t q_{ft}\mone N_t^{-1}\bigr).
\ee

Les int\'egrales sur les bivecteurs et les multiplicateurs dans la fonction de partition se font comme \`a la section pr\'ec\'edente, et conduisent \`a :
\beq
\prod_{(f, t)} \delta\bigl(g_{+ft}\,N_t\,g_{-ft}\,N_t^{-1}\bigr).
\ee
Par rapport au mod\`ele BC, $g_{-ft}\mone$ a \'et\'e chang\'e en $g_{-ft}$. \'Etant donn\'e les r\`egles de transport parall\`ele des normales $N_t$, la condition impos\'ee par ces fonctions delta est que la normale est laiss\'ee invariante par transport autour du wedge, non plus sous l'action de l'holonomie $G_{ft}$ comme dans le mod\`ele BC, mais sous l'action de :
\beq
G_{\mathrm{non-geo}ft}\ \cdot\ N_t = g_{+\mathrm{non-geo}ft}\,N_t\,g_{-\mathrm{non-geo}ft}\mone = N_t,\quad\text{pour}\qquad G_{\mathrm{non-geo}ft} = (g_{+ft}^{\phantom{}}, g_{-ft}\mone).
\ee

Malheureusement, le mod\`ele obtenu apr\`es int\'egration des holonomies de bord $L_{ft}$ n'est pas bien d\'efini au sens o\`u il d\'epend de la parit\'e du nombre de liens duaux autour de chaque face duale. La solution fournie dans \cite{BCpaper-val} est de d\'ecouper chaque wedge en demi-wedge, et d'imposer les contraintes \`a artir de ces demi-wedges. Du point de vue de la triangulation, ce red\'ecoupage signifie que l'on \'eclate le bord de chacun des 4-simplexes isol\'es en t\'etra\`edres disjoints. Les diff\'erentes copies d'un m\^eme triangle portent alors des bivecteurs ind\'ependants, $B_{ftv}$. Le recollement de tous ces demi-wedges autour d'une face se fait dans la fonction de partition gr\^ace \`a l'introduction d'holonomies de bord suppl\'ementaires $L_{fv}$ sur les t\'etra\`edres disjoints de chaque 4-simplexe. On aboutit par cette proc\'edure au calcul suivant :
\begin{align}
Z_{\mathrm{non-geo}} &= \int\prod_t dG_{t}\ \sum_{\{j_f\}}\ \prod_t \f{1}{\prod_{f\subset\pp t}d_{j_f}^2}\ \prod_f d_{j_f}^2\ \chi_{j_f}\bigl(g_{+f}\bigr)\ \chi_{j_f}\bigl(g_{-f}\bigr), \\
&= \sum_{\{j_f\}}\sum_{\{(i_{+t},i_{-t})\}}\ \prod_f d_{j_f}^2\ \prod_t \f{1}{\prod_{f\subset\pp t}d_{j_f}^2}\ \prod_v 15j_{\mathrm{Spin(4)}}\bigl((j_f,j_f);(i_{+t},i_{-t})\bigr). \label{topo model}
\end{align}
Le mod\`ele s'interpr\`ete donc imm\'ediatement comme un mod\`ele restreint aux repr\'esentations simples de $\Spin(4)$ sur les triangles, mais libre sur les intrelaceurs. Si l'amplitude du 4-simplexe est bien la m\^eme que dans le mod\`ele du chapitre \ref{sec:single bivector}, ces mod\`eles diff\`erent quant aux amplitudes des t\'etra\`edres et triangles. Ainsi, le fait que les mod\`eles soient construits sur des contenus physiques tr\`es diff\'erents, \`a savoir la simplicit\'e des bivecteurs isol\'ement ou la simplicit\'e crois\'ee dans le secteur non-g\'eom\'etrique, ne se traduit \emph{pas} par des amplitudes de 4-simplexes diff\'erentes, mais uniquement dans les amplitudes des plus petits simplexes ! Il me semble que c'est un ph\'enom\`ene peu consid\'er\'e dans la litt\'erature qui insiste plut\^ot sur les informations physiques cod\'ees dans l'amplitude d'un 4-simplexe, et pour lequel nous avons ici un exemple non-trivial.

\subsection{Apparition des coefficients de fusion} \label{sec:fusion coeff BC}

Jusqu'\`a maintenant, nos constructions d'int\'egrales de chemins sur r\'eseau n'ont pas fait appara\^itre les coefficients de fusion, une des caract\'eristiques principales des nouveaux mod\`eles EPR/FK, si ce n'est des coefficients de fusion trivialis\'es par la projection sur le sous-espace invariant sous le sous-groupe $\SU(2)$ diagonal. Pour en obtenir de non-triviaux, il faut entrelacer les repr\'esentations self-duale et anti-self-duale de chaque triangle sous l'effet du sous-groupe $\SU(2)$ diagonal. Pour cela nous ins\'erons un poids non-trivial sur les multiplicateurs $q_{ft}\in\SU(2)$. Nous choisissons donc une repr\'esentation $k_{ft}$ fix\'ee pour chaque paire triangle-t\'etra\`edre et regardons de pr\`es la quantit\'e :
\beq \label{measureBC}
Z_{\{k_{ft}\}} = \int \prod_{(f,t)} dq_{ft}\,dL_{ft}\ \chi_{k_{ft}}\big(q_{ft}\big)\ \delta\bigl(g_{-ft} q_{ft}\bigr)\ \delta\bigl(N_t\mone\,g_{+ft}\,N_t\,q_{ft}\bigr).
\ee
Bien s\^ur la condition d'invariance de la normale $N_t$ sous le transport autour du wedge par $G_{ft}$ tient toujours, comme on le voit :
\beq
Z_{\{k_{ft}\}} = \int \prod_{(f,t)} dL_{ft}\ \chi_{k_{ft}}\big(g_{+ft}\big)\ \delta\bigl(N_t\mone\,g_{+ft}\,N_t\,g_{-ft}\mone\bigr).
\ee
N\'eanmoins, pour garder la sym\'etrie explicite entre les secteurs self-duaux et anti-self-duaux, il est pr\'ef\'erable d'nt\'egrer sur $q_{ft}$ en d\'eveloppant les caract\`eres et fonctions delta de \eqref{measureBC} sur les \'el\'ements de matrices des repr\'esentations de $\SU(2)$. On voit alors en utilisant la formule \eqref{int g3} pour $q_{ft}$ que les repr\'esentations self-duale et anti-self-duale pour chaque wedge sont entrelac\'ees via la repr\'esentation de spin $k_{ft}$. Les int\'egrales de collage des wedges, sur $L_{ft}$ se font toujours gr\^ace \`a l'orthogonalit\'e des matrices de repr\'esentations. On aboutit \`a :
\begin{multline}
Z_{\{k_{ft}\}} = \sum_{\{j^+_f, j^-_f\}} \prod_f d_{j^+_f} d_{j^-_f} \prod_{(f,v)} \begin{pmatrix} j^+_f & j^-_f & k_{ft'} \\a^+_{ft'} &a^-_{ft'} & m_{ft'} \end{pmatrix}\ \begin{pmatrix} j^+_f & j^-_f & k_{ft} \\b^+_{ft} &b^-_{ft} & m_{ft} \end{pmatrix}\\
\langle j^+_f, a^+_{ft'}\lvert\ N_{t'}\mone\,g_{+vt'}\mone\,g_{+vt}^{\phantom{}}\,N_t^{\phantom{}}\ \rvert j^+_f, b^+_{ft}\rangle\ \langle j^-_f, a^-_{ft'}\lvert\ g_{-vt'}\mone\,g_{-vt}^{\phantom{}}\ \rvert j^-_f, b^-_{ft}\rangle.
\end{multline}
Les notations $a^\pm_{ft}, b^\pm_{ft}$ et $m_{ft}$ repr\'esentent des nombres magn\'etiques des repr\'esentations $j^\pm_{f}$ et $k_{ft}$, et sont implicitement somm\'es. Les symboles $3mj$ de Wigner, entre parenth\`eses, r\'ealisent l'entrelacement recherch\'e : $\calH_{j^+_f}\otimes\calH_{j^-_f}\rightarrow \calH_{k_{ft}}$ (\`a une dualisation de $k_{ft}$ pr\`es).

\begin{figure} \begin{center}
\includegraphics[width=10cm]{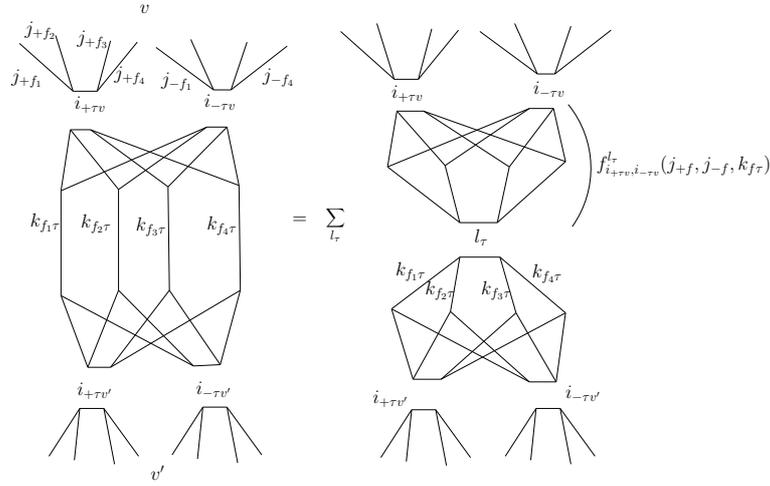}
\caption{ \label{general BCvertex} C'est ce qui r\'esulte des int\'egrales sur les \'el\'ements de groupe de la fonction de partition. Le graphe du milieu \`a gauche peut \^etre vu comme une amplitude de t\'etra\`edre, mais il est pr\'ef\'erable de faire le d\'eveloppement du membre de droite, de sorte que l'on forme les coefficients de fusion et la m\^eme amplitude, structurellement, que dans les nouveaux mod\`eles. Les triangles apparaissent ici comme liens des graphes et portent des repr\'esentations  irr\'eductibles $(j_{+f},j_{-f})$, qui sont entrelac\'ees au niveau des t\'etra\`edres avec les repr\'esentations $k_{ft}$.}
\end{center} \end{figure}

Pour extraire la repr\'esentation en mousses de spins, nous n'avons plus qu'\`a int\'egrer sur les holonomies (les normales $N_t$ pouvant \^etre r\'eabsorb\'ees \`a droite dans les variables $g_{+vt}$, gr\^ace \`a l'invariance de la mesure de Haar). Les int\'egrales sur $g_{\pm vt}$ produisent l'identit\'e sur l'espace invariant $\Inv(\oplus_{f=1}^4\calH_{j^\pm_f})$ du produit tensoriel des quatre spins $j^\pm_f$ se rencontrant au t\'etra\`edre $t$. Cette identit\'e se d\'eveloppe sur les entrelaceurs 4-valents caract\'eris\'es par les spins interm\'ediaires $i^\pm_{vt}$. Ces entrelaceurs se contractent entre les cinq t\'etra\`edres de chaque 4-simplexe, formant des symboles 15j, et se contractent de l'autre c\^ot\'e avec les entrelaceurs $\calH_{j^+_f}\otimes\calH_{j^-_f}\rightarrow \calH_{k_{ft}}$. Il devient \'evident alors que l'on peut faire appara\^itre les coefficients de fusion. Cela se fait en ins\'erant l'identit\'e sur l'espace invariant $\Inv(\oplus_{f=1}^4\calH_{k_{ft}}$ \`a chaque t'etra\`edre, identit\'e qui se d\'eveloppe aussi sur une base orthogonale d'entrelaceurs 4-valents caract\'eris\'es par un spin $l_t$ (voir illustration figure \ref{general BCvertex}). Il vient :
\beq
Z_{\{k_{ft}\}} = \sum_{\{j^+_f, j^-_f\}, \{l_t\}} \prod_f d_{j^+_f} d_{j^-_f}\ \prod_t d_{l_t}\ \prod_v W_v(j^+_f, j^-_f, k_{ft}, l_t),
\ee
o\`u $W_v$ n'est autre que l'amplitude des nouveaux mod\`eles EPR/FK donn\'ee en \eqref{new vertex}.

Puisque l'amplitude des 4-simplexes est la m\^eme que dans les nouveaux mod\`eles, quelques commentaires sont de rigueur. Quelles sont les diff\'erences ? Tout d'abord, les repr\'esentations des triangles ne sont simples, $j^+_f\neq j^-_f$. Par ailleurs, il faut se demander comment sommer ces amplitudes $W_v$ ! En effet, les repr\'esentations $k_{ft}$ sont ici fix\'ees au pr\'ealable, i.e. \emph{avant} le d\'eveloppement sur les spins $j^\pm_f, l_t$. Ainsi, il n'est possible de sommer $Z_{\{k_{ft}\}}$ qu'avec des coefficients ne d\'ependant que des spins $k_{ft}$. En revanche, dans les mod\`eles EPR/FK, les amplitudes des t\'etra\`edres portent des couplages entre $k_{ft}$ et les autres spins somm\'es, tels que \eqref{FK weight} : $\Bigl(\begin{smallmatrix} j^+_f & j^-_f & k_{ft} \\ j^+_f & -j^-_f & j^-_f-j^+_f\end{smallmatrix}\Bigr)$, impossibles \`a obtenir ici.

\section{Le mod\`ele BC au niveau classique : probl\`eme de collage} \label{sec:classicalBC}

Il est int\'eressant de r\'e\'ecrire cette proposition en introduisant une source de courbure $\Spin(4)$ sur chaque wedge, not\'ee $Q_{ft}$, telle que
\beq
S_{\mathrm{BC}} = \sum_{(f, t)} \Tr\bigl( B_{ft}\ G_{ft}\,Q_{ft}\bigr),
\ee
et que l'on contraint \`a \^etre de la forme :
\beq
Q_{ft} = (N_t\,q_{ft}\,N_t\mone, q_{ft}).
\ee
Cela signifie que $Q_{ft}$ est contraint \`a \^etre dans le sous-groupe $\SU(2)$ qui pr\'eserve la normale au t\'etra\`edre $t$ :
\beq
Q_{ft}\ \cdot\ N_t \,=\, q_{+ft}\,N_t\,q_{-ft}\mone \,=\, N_t,
\ee
ou de mani\`ere \'equivalente :
\beq
(Q_{ft})^I_{\phantom{I}J}\,N_t^J = \bigl((N_t, \unit)\,(q_{ft}, q_{ft})\bigr)^I_{\phantom{I}J}\,(1,0,0,0)^J = N_t^I.
\ee
Nous avons utilis\'e la repr\'esentation matricielle standard sur $\R^4$, dans laquelle $(q_{ft}, q_{ft})$ est une rotation de l'espace 3d orthogonal au vecteur $(1, 0, 0, 0)$. Dans la jauge temps, $N_t = \unit$, on voit imm\'ediatement que lorsque $q_{ft}$ parcourt $\SU(2)$, alors $Q_{ft}$ parcourt le sous-groupe $\SU(2)$ diagonal. On comprend ainsi que c'est l'int\'egrale sur $q_{ft}$ dans la fonction de partition qui g\'en\`ere l'invariance $\SU(2)$ diagonal sp\'ecifique au mod\`ele BC. C'est la mani\`ere dont l'int\'egrale du chemins r\'ealise la projection de $\calH_{(j,j)} = \oplus_{k=0}^{2j} \calH_k$ sur le sous-espace invariant $k=0$, du point de vue de la quantification EPR pour $\gamma \rightarrow\infty$.

On peut aussi voir de la sorte pourquoi notre construction dans le secteur non-g\'eom\'etrique ne conduit pas au mod\`ele EPR. Celui-ci s'obtient par projection de $\calH_{(j,j)} = \oplus_{k=0}^{2j} \calH_k$ sur le sous-espace (non-invariant) $k=2j$. Or, dans la jauge temps, notre construction appliqu\'ee au secteur non-g\'eom\'etrique demande \`a la source $Q_{ft}$ d'\^etre non plus une rotation 3d mais un \og boost\fg,
\beq
Q_{\mathrm{non-geo}ft} = (q_{ft}\mone, q_{ft}).
\ee
La moyenne sur ces boosts dans la fonction de partition ne r\'ealise pas du tout de projection sur le sous-espace $k=2j$.

\medskip

On s'attend bien s\^ur \`a ce que les configurations classiques de l'action $S_{\rm BC}$ jouent un r\^ole important dans l'\'evaluation des amplitudes (limite semi-classique, calcul des divergences par exemple), bien que la fa\c{c}on dont cela se fait ne soit pas forc\'ement encore tr\`es claire -- et d\'epende du mod\`ele. Nous nous contentons d'\'etudier ici les \'equations des points stationnaires de l'action $S_{\rm BC}$. L'\'equation la plus simple est celle obtenue par les variations des bivecteurs, \'equation que l'on conna\^it d\'ej\`a car l'action \'etant lin\'eaire en les bivecteurs, l'amplitude du mod\`ele est piqu\'ee par des fonctions delta sur ces configurations \eqref{intB-BC},
\beq \label{BeomBC}
Q_{ft}^{\phantom{}} = G_{ft}\mone \qquad \Rightarrow\qquad G_{ft}\ \cdot\ N_t \,=\, g_{+ft}\,N_t\,g_{-ft}\mone \,=\, N_t.
\ee
La premi\`ere \'equation exprime les multiplicateurs de Lagrange en fonction des holonomies de wedge, et la deuxi\`eme nous dit que la normale $N_t$ est pr\'eserv\'ee par le transport parall\`ele autour du wedge. Comme nous l'avons d\'ej\`a point\'e, cette deuxi\`eme \'equation peut \^etre pens\'ee comme la d\'efinition du mod\`ele BC, \`a l'instar de $\delta(G_f)$ pour le mod\`ele topologique -- auquel cas la discussion sur les bivecteurs et les contraintes permet de justifier la d\'efinition et d'extraire l'interpr\'etation physique. Une formulation \'equivalente est de dire que l'holonomie de wedge est restreinte \`a la forme :
\beq
G_{ft} = (N_t\,g_{-ft}\,N_t\mone, g_{-ft}) \in \SU(2)_{N_t},
\ee
similaire \`a celle de $Q_{ft}$. Dans le secteur non-g\'eom\'etrique l'holonomie de wedge n'est plus une rotation du sous-groupe $\SU(2)$ stabilisant $N_t$, mais en revanche l'\'el\'ement $(g_{+ft}, g_{-ft}\mone)$ l'est ! Ce qui signifie que $G_{ft}$ est plut\^ot un \og boost\fg, la direction \og temps\fg{} \'etant celle de la normale $N_t^I$,
\beq
\text{(secteur non-g\'eom\'etrique)}\qquad G_{ft} = (N_t\,g_{-ft}\mone\,N_t\mone, g_{-ft}).
\ee
Ainsi, nous pouvons appr\'ecier le fait que comme dans le continu les deux secteurs solutions des contraintes de simplicit\'e sont bien reli\'es par la dulait\'e de Hodge sur $\Spin(4)$ (qui envoie les g\'en\'erateurs des rotations sur les boosts et r\'eciproquement) !


Nous discuterons en conclusion du chapitre le fait que l'holonomie de wedge soit dans le sous-groupe $\SU(2)_{N_t}$, et l'utilisons pour le moment pour simplifier les autres \'equations du mouvement. La variation de $q_{ft}$ reproduit bien la contrainte de simplicit\'e crois\'ee \eqref{simplicity-selfdual}. Les variations des holonomies $G_{vt}$ et des normales $N_t$ donnent les relations de fermeture suivantes :
\beq
\sum_{f\subset \pp t} B_{ft} = 0\quad \text{et}\qquad \sum_{f\subset \pp t} G_{ft}\, B_{ft}\, G_{ft}^{-1} = 0
\ee
La premi\`ere relation est bien s\^ur la relation de fermeture des t\'etra\`edres, \eqref{closure}, fortement attendue puisqu'elle est n\'ecessaire \`a la reconstruction de le g\'eom\'etrie simplicielle, en tant que contrainte de simplicit\'e (voir section \ref{sec:classical sf}). Dans la th\'eorie BF discr\`ete, la deuxi\`eme relation est \'equivalente \`a la fermeture du fait qu'alors $G_{ft}=\unit$. Mais ce n'est plus le cas ici \`a cause de la pr\'esence des variables $Q_{ft}$ apparaissant comme sources pour les holonomies de wedge \eqref{BeomBC}. Nous avons donc une \'equation non-triviale qui demande que la fermeture soit toujours valable apr\`es transport parall\`ele des bivecteurs autour de leurs wedges respectifs.

Gr\^ace \`a la relation de fermeture et aux contraintes de simplicit\'e crois\'ee, nous sommes assur\'es de la metricit\'e des bivecteurs pour chaque t\'etra\`edre ind\'ependamment les uns des autres. En fait, puisque nous avons utilis\'e un bivecteur $B_{ft}$ par wedge, i.e. pour un triangle dans un 4-simplexe, nous pouvons consid\'er\'e que la m\'etricit\'e est garantie dans les 4-simplexes ind\'ependamment les uns des autres. Il nous faut donc \'etudier la fa\c{c}on dont les holonomies de bord $L_{ft}$ prennent en charge le recollage des 4-simplexes que nous avons isol\'es les uns des autres par le d\'ecoupage des faces en wedges. Un tel collage est une op\'eration non-triviale, et le fait qu'il fonctionne dans l'action \eqref{wedgeaction} de type BF est fortement li\'e \`a la particularit\'e du mod\`ele topologique d'avoir une courbure triviale. En fait, puisque l'action $S_{\rm BC}$ a la m\^eme structure : $\Tr(B_{ft}\tl{G}_{ft})$, pour $\tl{G} = GQ$, menant en particulier aussi \`a une \'equation de trivialit\'e sur $G_{ft}Q_{ft}$, les variations de $L_{ft}$ conduisent aussi \`a une \'equation impliquant le transport parall\`ele des bivecteurs. Mais celle-ci est d\'eform\'ee par la pr\'esence de la \og sources\fg{} $Q_{ft}$. Si $t_1$ et $t_2$ sont deux t\'etra\`edres adjacents, associ\'es \`a des wedges voisins pour le triangle $f$, nous obtenons en effet :
\beq \label{bad gluing BC}
G_{t_1 t_2}\,B_{ft_2}\,G_{t_1 t_2}\mone = Q_{ft_1}\,B_{ft_1}\,Q_{ft_1}\mone \neq B_{ft_1}.
\ee
Ce qui signifie que le collage des simplexes de se fait pas de mani\`ere consistente, et donc de m\^eme pour la reconstruction de la g\'eom\'etrie \`a partir des bivecteurs ! Ainsi les raisonnements des sections \ref{sec:geom normales} et \ref{sec:gluing-aarc} ne peuvent plus s'appliquer. En particulier, la relation \eqref{coupling normals} entre les normales $N_{t_1}, N_{t_2}$ des t\'etra\`edres adjacents ne tient plus, car la diff\'erence entre les normales, $N_{t_1}(v)N_{t_2}\mone(v)$, n'est plus dans le sous-groupe $\U(1)$ stabilisant $b_{+ft_2}(v)$. G\'eom\'etriquement, cela qui signifie que $N_{t_1}$ n'est pas contenu dans le plan orthogonal au bivecteur $\star B_{ft_2}$. Ainsi, on voit que les corr\'elations entre simplexes adjacents ne sont pas correctement prises en compte dans le mod\`ele BC.


\section{Une relation de r\'ecurrence sur l'amplitude du 4-simplexe} \label{sec:10jrecurrence}

Dans la lign\'ee des id\'ees d\'evelopp\'ees \`a la partie \ref{sec:recurrence}, nous d\'erivons et \'etudions une relation de r\'ecurrence sur le symbole 10j du mod\`ele BC. S'il est loin d'\^etre \'evident, voire peu vraisemblable, que cette relation traduise une sym\'etrie associ\'ee \`a la contrainte hamiltonienne, il n'en reste pas moins qu'elle code une propri\'et\'e g\'eom\'etrique du 4-simplexe que nous identifions comme sa relation de fermeture, en s'assurant qu'elle est v\'erifi\'ee dans l'asymptotique par l'action de Regge fonction des aires du 4-simplexe. Nous proposons \'egalement une interpr\'etation de l'\'equation obtenue comme combinaison des mouvements \'el\'ementaires sur le 4-simplexe \`a l'origine des relations de r\'ecurrence sur le symbole 15j, vus \`a la section \ref{sec:15jrecurrence}.

\subsection{De la matrice de Gram \`a la relation de r\'ecurrence}

Nous d\'ebutons avec l'expression du symbole 10j pour le 4-simplexe euclidien comme une int\'egrale sur cinq copies de $\SU(2)$,
\begin{equation*}
\{10 j\}\,=\,
\int_{\SU(2)^5} \prod_{a=1}^5 dN_a \, \prod_{a<b} \chi_{j_{ab}}(N_a^{-1}\,N_b^{\phantom{}}).
\end{equation*}
C'est une fonction de dix spins $j_{ab}\in\N/2$, un sur chacun des 10 triangles du 4-simplexe. Le caract\`ere $\chi_j(g)$ dans la repr\'esentation de spin $j$ est la trace qui s'\'evalue pour $g = \exp(i\theta\,\hat{n}\cdot\vec{\sigma})$ \`a :
\begin{equation*}
\chi_j(g) = \f{\sin(d_j\theta)}{\sin\theta},
\end{equation*}
$\theta$ \'etant l'angle de classe (i.e. la moiti\'e de l'angle de rotation). Une expression tr\`es utilis\'ee du 10j, \`a la fois num\'eriquement et analytiquement, s'obtient en \'ecrivant l'int\'egrant comme une fonction des dix angles de classe $\theta_{ab}$ des produits $N_a\mone N_b$. Naturellement, ces angles ne sont pas tous ind\'ependants, de sort que le changement de variables implique une mesure non-triviale sur les nouvelles variables angulaires, donn\'ee par une contrainte \cite{freidel-louapre-6j},
\beq \label{10jint-angles}
\{10 j\}\,=\,\f{4}{\pi^6}\int_{[0,\pi[^{10}} \prod_{a<b}d\theta_{ab}\ \delta\bigl(\det G[\theta]\bigr) \ \prod_{a<b}\sin\,
(2j_{ab}+1)\theta_{ab}.
\ee
La matrice de Gram $G[\theta]$ est une matrice sym\'etrique $5\times 5$, d\'efinie par : $G_{ab} = \cos\theta_{ab}$ (et $\cos\theta_{aa}=1$). L'id\'ee, en accord avec notre d\'erivation du mod\`ele BC, est que les variables $N_a\in\SU(2)$ dans l'int\'egrant du 10j repr\'esentent via l'isomorphisme $\SU(2)\simeq S^3$ les vecteurs unitaires orthogonaux \`a chacun des cinq t\'etra\`edres. Comme nous l'avons vu, l'angle de classe de $N_a\mone N_b$ est alors l'angle entre les deux normales, de sorte que $G[\theta]$ est bien la matrice de Gram de ce syst\`eme de normales, $G_{ab} = N_a\cdot N_b$. Puisque nous avons cinq normales dans $\R^4$, celles-ci sont lin\'eairement d\'ependantes, ce qui \'equivaut \`a l'annulation du d\'eterminant de $G[\theta]$. Ainsi, dix angles $\theta_{ab}$ satisfaisant cette contrainte sont bien les angles d'un 4-simplexe euclidien. Cest ce qui permet d'extraire du 10j l'information g\'eom\'etrique n\'ecessaire pour le relier \`a l'action de Regge du 4-simplexe dans l'asymptotique \cite{freidel-louapre-6j, baez-asym10j, williams-asym10j, steele-asym10j, livine-speziale-10jgraviton}.

Pour construire une relation de r\'ecurrence, il suffit de consid\'erer l'identiti\'e trigonom\'etrique suivante :
\beq \label{trigo-10j}
\sin\,(2j+2)\theta\,+\,\sin\,(2j)\theta\,=\, 2\cos\theta\ \sin\,(2j+1)\theta.
\ee
L'id\'ee est d'appliquer cette identit\'e dans l'int\'egrant du 10j \eqref{10jint-angles} pour plusieurs spins $j_{ab}$ en m\^eme temps, de sorte que les facteurs $\cos\theta_{ab}$ provenant du membre de droite de \eqref{trigo-10j} se combinent pour former le d\'eterminant de Gram,
\beq
\det G[\theta] = \sum_{\sigma}\eps(\sigma)\prod_{a=1}^5\cos\theta_{a\sigma(a)}.
\ee
Il s'agit d'une somme sur les permutations $\sigma$ entre cinq \'el\'ements, $\eps(\sigma)$ \'etant la signature de la permutation. De mani\`ere \'equivalente, nous ins\'erons le d\'eterminant de Gram dans l'int\'egrale \eqref{10jint-angles} :
\beq \label{tenj-recurrence}
\int_{[0,\pi[^{10}} \prod_{a<b}d\theta_{ab}\ \delta\bigl(\det G[\theta]\bigr)\ \det(G)\ \prod_{a<b}\sin\,
(2j_{ab}+1)\theta_{ab} = 0.
\ee
Le d\'eterminant ins\'er\'e est alors d\'evelopp\'e sur les permutations, et la formule trigonom\'etrique \eqref{trigo-10j} est utilis\'ee sur chaque facteur en la lisant de droite \`a gauche. Cela reproduit pour chaque facteur $\cos\theta$ du d\'eterminant une combinaison lin\'eaire d'int\'egrants du 10j, mais avec des spins diminu\'es et augment\'es selon \eqref{trigo-10j}.

La permutation triviale n'introduit aucun facteur $\cos\theta$, de sorte qu'elle reproduit le symbole 10j initial sans affecter les repr\'esentations. Il n'y a aucune permutations fixant exactement quatre \'el\'ements. L'\'etape suivante est donc les permutations fixant exactement trois \'el\'ements, disons $a=1,2,3$. Cette permutation des \'el\'ements $a=4, 5$ produit un facteur $-\cos^2\theta_{45}$ qui se transforme par :
\beq
-\cos^2\theta\ \sin\,(2j+1)\theta\,=\,-\f{1}{2}\,\sin\,(2j+1)\theta \,-\, \f{1}{4}\bigl(\sin\,(2j+3)\theta \,+\, \sin\,(2j-1)\theta\bigr).
\ee
Cela produit une combinaison lin\'eaire du symbole 10j initial accompagn\'e de deux autres 10j pour lesquels la repr\'esentation $j_{45}$ est modifi\'ee de $\pm1$. Viennent ensuite les permutations fixant exactement deux \'el\'ements, disons $a=1,2$. L'action sur les \'el\'ements $a=3, 4, 5$ est l'une des deux permutations cycliques de ces trois \'el\'ements. Elles donnent toutes deux le m\^eme facteur $+\cos\theta_{34}\cos\theta_{45}\cos\theta_{35}$, conduisant \`a une combinaison lin\'eaire o\`u les spins $j_{34}, j_{45}$ et $j_{35}$ sont affect\'es de $\pm\f12$. Quant aux permutations fixant exactement un \'el\'ement, $a=1$, elles sont constitu\'ees des permutations cycliques des quatre \'el\'ements restants (6 possibilit\'es), et des permutations sur deux paires (3 possibilit\'es). Les 4-cycles donnent des facteurs du type : $+\cos\theta_{23}\cos\theta_{34}\cos\theta_{45}\cos\theta_{25}$, qui produisent des variations des spins correspondants de $\pm\f12$. Les permutations sur des paires donnent des facteurs du type : $\cos^2\theta_{23}\cos^2\theta_{45}$ qui induisent au final des diff\'erences de $\pm1$ sur les spins correspondants. Finalement, il nous reste les permutations ne fixant aucun \'el\'ement : soit un 5-cycle ($4!$ possibilit\'es), soit une combinaison d'un 3-cycle et d'un 2-cycle ($10\times 2$ possibilit\'es). On v\'erifie alors que : $1+10+(2\times 10)+5\times (6+3) + (4!+20) =5!$.

Cette proc\'edure conduit \`a une relation de r\'ecurrence sur le symbole 10j, impliquant l'amplitude initiale, pour les spins $j_{ab}$, des amplitudes ayant jusqu'\`a cinq spins affect\'es de $\pm\f12$, et/ou des amplitudes ayant aux plus deux spins d\'eplac\'es de $\pm1$ (du fait des 2-cycles). La relation compl\`ete comporte $5! = 120$ termes, de sorte que nous ne l'\'ecrirons pas explicitement ! D'une part, nous sommes conscients qu'une telle formule n'est pas directement utile pour \'evaluer efficacement le symbole 10j (elle requiert notamment une infinit\'e de conditions initiales). Mais elle pourrait par ailleurs \^etre utile dans l'\'etude du comportement asymptotique et des propri\'et\'es g\'eom\'etriques du mod\`ele, et plus g\'en\'eralement pour le comportement des corr\'elations, de type graviton, aire-aire, dans des mod\`eles proches de celui-ci, comme c'est le cas en 3d \cite{maite-etera-6jcorr}.

La relation de r\'ecurrence que nous obtenons est une \'equation aux diff\'erences d'ordre 4, chose que nous commenterons plus bas.

\subsection{La relation de r\'ecurrence comme contrainte de fermeture}

La r\'ecurrence d\'ecrite plus haut est assez complexe, puisqu'elle r\'esulte du d\'eveloppement du d\'eterminant de Gram $5\times5$. Il est donc souhaitable de pouvoir identifier une quantit\'e simple et naturelle qui satisfait aussi cette relation, au moins asymptotiquement. Il y a bien s\^ur les solutions triviales, qui sont des constantes, car : $\sum_\sigma \eps(\sigma) = 0$. Nous cherchons maintenant des solutions pus g\'eom\'etriques.

Dans ce but, insistons sur le fait que la r\'ecurrence fut obtenue en ins\'erant dans une int\'egrale la quantit\'e $\det (\cos\theta_{ab})$. L'annulation de ce d\'eterminant est la condition n\'ecessaire et suffisante pour que les angles $\theta_{ab}$ soient les angles dih\'edraux d'un authentique 4-simplexe plat, i.e. \emph{ferm\'e}. De plus, \`a partir des dix spins $j_{ab}$, interpr\'et\'es comme les aires des dix triangles d'un 4-simplexe plat par la formule : $A_{ab} = (2j_{ab}+1)$, il est possible de calculer les angles dih\'edraux $\bar{\theta}_{ab}(j)$ de ce 4-simplexe, qui satisfont la condition : $\det (\cos\bar{\theta}_{ab})=0$. Dans le r\'egime asymptotique \cite{freidel-louapre-6j, baez-asym10j, williams-asym10j, steele-asym10j, livine-speziale-10jgraviton}, la partie oscillante du 10j est alors donn\'ee par l'action de Regge de ce 4-simplexe, \'evalu\'ee sur les variables d'aires $(2j_{ab}+1)$. Un candidat naturel pour tester notre relation de r\'ecurrence est donc cette action de Regge,
\be
S_{\rm R} = \sum_{a<b}(2j_{ab}+1)\bar{\theta}_{ab}(j),
\ee
Plus pr\'ecis\'ement, nous allons montrer que son cosinus, $\cos\bigl(S_{\rm R}(j_{ab}) + \alpha\bigr)$, pour un phase $\alpha$ quelconque satisfait bien la relation de r\'ecurrence dans l'asymptotique. De plus, l'identit\'e cl\'e pour obtenir ce r\'esultat est la fermeture du 4-simplexe, soit alg\'ebriquement :
\be
\det\, \bigl(\cos\bar{\theta}_{ab}\bigr) = 0.
\ee

La relation de r\'ecurrence touche plusieurs spins \`a la fois, en leur faisant subir des translations de $\pm\f12$. Imaginons d'abord une situation o\`u seul un spin est affect\'e, disons $j_{12}$. Selon l'identit\'e trigonom\'etrique \eqref{trigo-10j}, chaque d\'eplacement de $+\f12$ s'accompagne d'un terme \'equivalent avec un d\'eplacement de $-\f12$. Par ailleurs, en suivant l'analyse asymptotique de Schulten et Gordon \cite{schulten-gordon2} sur le symbole 6j, nous consid\'erons la limite homog\`ene des grands spins et d\'eveloppons $S_{\rm R}$ au premier ordre, en n\'egligeant les d\'eriv\'ees secondes. Il vient alors :
\begin{align}
\f 12\cos\Bigl(S_{\rm R}\bigl(j_{12}+\f 12\bigr)+\alpha\Bigr) + \f 12\cos\Bigl(S_{\rm R}\bigl(j_{12}-\f 12\bigr)+\alpha\Bigr) &= \f 12\sum_{\eta=\pm 1}\cos\Bigl(S_{\rm R}(j_{12})+\alpha + \eta\,\bar{\theta}_{12}(j_{12})\Bigr), \\
&= \cos\Bigl(S_{\rm R}(j_{12})+\alpha\Bigr)\,\cos\bar{\theta}_{12}(j_{ab}).
\end{align}
Pour obtenir ce r\'esultat, il faut utiliser l'identit\'e de Schl\H{a}fli, $\sum_{a<b}(2j_{ab}+1)\delta\bar{\theta}_{ab} = 0$, de sorte que la variation de $S_{\rm R}$ par rapport \`a $j_{12}$ est simplement $2\bar{\theta}_{12}$.

Cette relation s'\'etend facilement au cas d'un nombre quelconque de spins translat\'es. Notons $j_{ij}$, pour $i<j$, les spins touch\'es par des translations dans un terme donn\'e de la relation de r\'ecurrence. Alors nous obtenons le produit du cosinus de l'action de Regge par le produit des cosinus des angles $\bar{\theta}_{ij}$ :
\begin{align}
\f 12 \sum_{\eta_{ij}=\pm 1} \cos\Bigl(S_{\rm R}\bigl(j_{ij}+\eta_{ij}\f 12\bigr)+\alpha\Bigr) &= \f 12\sum_{\eta_{ij}=\pm 1}\cos\Bigl(S_{\rm R}(j_{ab})+\alpha + \sum \eta_{ij}\bar{\theta}_{ij}(j_{ab})\Bigr), \\
&= \cos\Bigl(S_{\rm R}(j_{ab})+\alpha\Bigr)\prod_{i<j} \cos\bar{\theta}_{ij}(j_{ab}).
\end{align}
Nous ne regardons par ailleurs pas des variations arbitraires des spins, mais suivant les permutations de cinq \'el\'ements comme d\'ecrit plus haut, le signe de chaque terme \'etant d\'etermin\'e par la signature de la permutation correspondante. Cela fait aboutir \`a :
\be
\sum_{\sigma} \eps(\sigma)\,\cos\bigl(S_{\rm R}+\alpha\bigr)\,\prod_{a=1}^5 \cos\bar{\theta}_{a\sigma(a)}
= \cos\bigl(S_{\rm R}+\alpha\bigr)\ \det G[\bar{\theta}] = 0,
\ee
du fait que les angles $\bar{\theta}_{ab}$ sont les angles dih\'edraux d\'etermin\'es par les aires.

Ce r\'esultat montre que la r\'currence trouv\'ee sur le 10j est associ\'ee \`a une propri\'et\'e g\'eom\'etrique simple d'un 4-simplexe seul, \`a savoir le fait qu'il est bien ferm\'e, ce qui contraste avec les relations de r\'ecurrence \'etablies sur les mod\`eles topologiques \`a partir des mouvements de Pachner. En effet, le symbole 10j ne d\'efinit pas de mod\`ele invariant sous ces mouvements et n'est donc pas topologique. Il ne correspond pas non plus \`a une quantification de la relativit\'e g\'en\'erale \cite{alesci-BC1}. Il est pourtant en mesure de d\'ecrire quelques propri\'et\'es simples, en se restreignant \`a un seul 4-simplexe isol\'e, et en ne regardant que les observables d'aires \cite{livine-speziale-10jgraviton, speziale-propagation-lqg, christensen-numerical10j}. Par cons\'equent, la relation de r\'ecurrence impl\'emente au niveau quantique les propri\'et\'es de ce mod\`ele \og jouet\fg.

N\'eanmoins, cette jolie interpr\'etation g\'eom\'etrique est obtenue en se concentrant sur la partie \`a la Regge de l'asymptotique du 10j, et en n\'egligeant tous les autres termes non-g\'eom\'etriques \cite{baez-asym10j, steele-asym10j, freidel-louapre-6j}. Ces derniers dominent pourtant le comportement asymptotique : lorsque tous les spins sont multipli\'es par un large param\`etre $\lambda$, la partie oscillante \`a la Regge d\'ecro\^it en $\lambda^{-\f 92}$, tandis que la contribution dominante est en $\lambda^{-2}$ et correspond \`a des configurations d\'eg\'en\'er\'ees. D'autres comportements interm\'ediaires non-g\'eom\'etriques ont \'et\'e mis en \'evidence depuis, \cite{christensen-subleading10j}. La situation est donc beaucoup plus riche qu'avec le 6j, pour lequel les points col \og g\'eom\'etrique\fg{} et \og d\'eg\'en\'er\'e\fg{} contribuant \`a l'asymptotique produisent les m\^emes comportements. Une \'equation de r\'ecurrence du second ordre, \eqref{Hlink}, suffit alors \`a restituer le comportement de ces points col. Ici les diff\'erents comportements asymptotiques du 10j font qu'une relation de r\'ecurrence pour celui-ci sera g\'en\'eralement d'ordre sup\'erieur. Consid\'erons \`a ce propos un exemple simple. Nous prenons le comportement assymptotique suivant,
\be\label{totti}
\{10j\}\sim\,\f{a}{\lambda^\alpha}+\f{1}{\lambda^{\alpha+\beta}}(b\cos \lambda S+ b'\sin\lambda
S)+\dots,
\ee
o\`u $S$ est l'action de Regge ind\'ependante de l'\'echelle $\lambda$, d\'efinissant la fr\'equence d'oscillations du 10j, et $a, b, b'$ sont des param\`etres num\'eriques fix\'es. Cette forme est motiv\'ee par le v\'eritable asymptotique du 10j, duquel les termes interm\'ediaires sont n\'eglig\'es. On peut alors montrer que \eqref{totti} est solution d'une \'equation diff\'erentielle lin\'eaire du trois\`eme ordre :
\be
\lambda(\beta(\beta-1)-aS^2\lambda^2)(\pp_\lambda^3(\lambda^{\alpha+\beta}f)+S^2\pp_\lambda(\lambda^{\alpha+\beta}f))
-\beta((\beta-1)(\beta-2)-aS^2\lambda^2)(\pp_\lambda^2(\lambda^{\alpha+\beta}f)+S^2(\lambda^{\alpha+\beta}f))
=0.
\ee
Des termes suppl\'ementaires dans \eqref{totti} entre le terme dominant en $\lambda^{-\alpha}$ et le r\'egime d'oscillations requiereraient une \'equation d'ordre plus \'elev\'e. Cet exemple justifie partiellement le fait que nous obtenons pour le 10j complet une \'equation aux diff\'erences d'ordre 4. Mais nous n'avons pas \'et\'e capables d'\'etudier l'asymptotique de cette \'equation et d'\'etablir un lien explicite avec les points col contribuant au 10j dans ce m\^eme r\'egime.

Cela sugg\`ere \'egalement la question suivante : en se restreignant aux points col d'origine g\'eom\'etrique, ne serait-il pas possible de s\'electionner des relations de r\'ecurrence plus simples, du second ordre ? Il existe une strat\'egie qui pourrait permettre cela et qui s'est r\'ev\'el\'ee efficace dans les calculs de corr\'elations en mousses de spins \cite{alesci-BC1, livine-speziale-10jgraviton} pour supprimer les contributions des points col non-g\'eom\'etriques dans l'int\'erieur de la triangulation (avec un unique simplexe jusque l\`a !). Elle consiste, comme nous l'avons fait dans le mod\`ele en 3d pour les corr\'elations des fluctuations de longueurs, section \ref{sec:3dgraviton}, \`a introduire un \'etat de bord qui s\'electionne naturellement les configurations g\'eom\'etriques de l'int\'erieur de la triangulation en accord avec la g\'eom\'etrie du bord, comme le montre \cite{modesto-perturbativeregge}.

\subsection{Invariance sous des mouvements g\'eom\'etriques}

Nous proposons ici une interpr\'etation des diff\'erents termes de la relation de r\'ecurrence. Puisque le symbole 10j ne donne pas lieu \`a un mod\`ele topologique, une interpr\'etation directe en termes d'invariance sous des mouvements de Pachner, comme c'est le cas avec le 6j et le 15j, est impossible. Mais nous aimerions profiter de l'interpr\'etation des spins du 10j comme valeurs propres des aires (quantifi\'ees) des triangles d'un 4-simplexe pour identifier des s\'eries de mouvements g\'eom\'etriques engendrant les variations des spins observ\'ees dans les termes de la relation de r\'ecurrence. Dans cette optique l'interpr\'etation des relations de r\'ecurrence sur le 15j obtenues par l'\'etude du mouvement de Pachner 2-4 semble bien appropri\'ee, car nous avions alors mis en \'evidence un mouvement \'el\'ementaire consistant \`a ajouter \`a un 4-simplexe un 4-simplexe aplati. Vu comme le l\'eger d\'eplacement d'un sommet du 4-simplexe initial, cela engendre bien des variations des aires des triangles. Rappelons qu'un tel mouvement g\'eom\'etrique est simplement issu de l'invariance du mod\`ele topologique sous les mouvements de Pachner, ceux-ci repr\'esentant bien les sym\'etries/la dynamique de la th\'eorie au niveau quantique.

Nous voulons faire remarquer dans cette section que la mani\`ere, ou la structure selon laquelle les spins sont affect\'es dans la relation sur le 10j est similaire \`a celle qu'offrirait une s\'erie de mouvements \'el\'ementaires sur des symboles 15j de $\SU(2)$ ! Cela nous fournit une vision qualitative de la r\'ecurrence comme l'affirmation d'une \emph{invariance sous une combinaison} des mouvements \'el\'ementaires pr\'esent\'es \`a a section \ref{sec:interpretation-move4-2}. Pour cela, il nous faut passer outre le fait que les entrelaceurs des 15j sont touch\'es par ces mouvements (puisque le mod\`ele BC a un unique entrelaceur) et nous concentrer sur les spins des triangles.

Nos arguments restent donc ici largement qualitatifs. Mais nous avons vu au chapitre \ref{sec:recurrence-6jiso} qu'il existe une situation similaire en 3d avec le symbole 6j restreint aux configurations isoc\`eles. Si celles-ci ne d\'efinissent pas un mod\`ele topologique, une relation de r\'ecurrence peut \^etre d\'eriv\'ee \`a partir d'une expression du 6j isoc\`ele sous forme d'int\'egrales, exactement comme ici. Elle s'interpr\`ete de plus comme une \'equation d'invariance sous une \emph{combinaison} de mouvements g\'eom\'etriques simples provenant des mouvements de Pachner ; y compris au niveau alg\'ebrique, puisque nous avions pu la d\'eriver \'egalement \`a partir de l'identit\'e de Biedenharn-Elliott. Cet exemple est donc l'exemple \`a suivre pour aller plus loin que la discussion qui suit.

\bigskip

Nous avons d\'ecrit aux sections \ref{sec:interpretation-move4-2}, \ref{sec:formule-move4-2} deux types de mouvements \'el\'ementaires sur les 4-simplexes, tous deux consistant \`a ajouter un 4-simplexe aplati \`a un premier, dont le r\'esultat est un l\'eger d\'eplacement d'un sommet, mais avec des manifestations alg\'ebriques diff\'erentes. D'abord, en bougeant un sommet d'un 4-simplexe, les spins de trois triangles qui se rencontrent sur une ar\^ete sont d\'ecal\'es de $\pm\f12$. C'est le mouvement 1, obtenu en prenant $\lambda=1/2$ dans les formules de la section \ref{sec:formule-move4-2}. Deuxi\`emement, si ce sommet est d\'eplac\'e un peu plus, i.e. en prenant $\lambda=1$, alors il n'est possible de ne toucher au spin que d'un seul triangle, par une translation de $\pm 1$ : c'est le mouvement dit 2. Rappelons-nous par ailleurs que dans le graphe du symbole 10j, les liens repr\'esentent les triangles et les noeuds les t\'etra\`edres.

Regardons d'abord les termes les plus simples, o\`u un seul spin est chang\'e de $0, \pm1$, chose qui survient d\`es lors qu'on prend en compte une permutation de deux \'el\'ements dans \eqref{tenj-recurrence}. Dans ce cas, aucun triangle partageant une ar\^ete avec le triangle concern\'e ne voit son spin modifi\'e. On peut donc consid\'erer que de tels termes s'obtiennent par le mouvement 2. Pour \^etre plus pr\'ecis, nous prenons la situation de la section \ref{sec:move4-2}, avec les m\^emes notations, et en particulier les figures \ref{move4-2} et \ref{move4-2graph}. Le 4-simplexe initial est $(abcde)$, et nous proc\'edons \`a un petit d\'eplacement de $a$ en $a'$ (pour les symboles 15j, nous prenons les spins des triangles $(aa'c)$ et $(aa'e)$ \`a z\'ero, et celui de $(aa'd)$ \`a 1). ALors, nous avons vu que cela affecte l'aire du triangle $(abd)$ (encore une fois, nous ne cherchons \`a \'ecrire des formules alg\'ebriques, et nous oublions ce qui se passe aux entrelaceurs). Ce mouvement \'el\'ementaire peut bien s\^ur \^etre pratiqu\'e ind\'ependamment sur tous les triangles qui ne partagent pas une ar\^ete.

Une interpr\'etation simmilaire peut \^etre formul\'ee pour les permutations cycliques de trois \'el\'ements. Elles g\'en\`erent des variations de spins de $\pm\f12$ sur trois liens formant une boucle sur le graphe du r\'eseau de spins correspondant, soit sur trois triangles partageant une ar\^ete. C'est donc pr\'ecis\'ement le mouvement 1. Par exemple, en prenant le spin du triangle $(aa'd)$ \`a $\f12$ dans le 4-simplexe aplati $(aa'cde)$, nous avions vu pour le 15j que les trois triangles se rencontrant le long de l'ar\^ete $(ad)$ \'etaient touch\'es.

Regardons l'effet d'une permutation cyclique des cinq \'el\'ements dans \eqref{tenj-recurrence}. La strat\'egie est maintenant d'appliquer trois fois le mouvement 1, sur trois boucles diff\'erentes du point de vue du graphe du r\'eseau de spins associ\'e au 10j. Sur ce graphe, la permutation concerne 5 liens formant une boucle. Nous formons alors trois boucles de trois liens chacune, qui \`a toutes les trois passent sur les cinq liens qui nous int\'eressent. Pour cela, on consid\`ere en plus deux liens du graphe, choisis de mani\`ere \`a ce que deux de ces boucles se rencontrent en un noeud et soient chacune form\'ees de deux des liens portant les spins affect\'es par la permutation (tous diff\'erents) et d'un des liens consid\'er\'es en plus. La troisi\`eme boucle est constitu\'ee du dernier des liens formant la boucle de cinq liens initiale et des deux liens consid\'er\'es en plus. Nous appliquons ensuite le mouvement 1 deux fois ind\'ependamment, sur les deux boucles se rencontrant en un noeud, ce qui affecte six spins. On utilise une troisi\`eme fois le mouvement pour compenser les variations induites sur les deux liens consid\'er\'es en plus, tout en affectant le dernier des cinq liens impliqu\'es par la permutation.

Les autres termes de la relation de r\'ecurrence sont analys\'es sous cet angle dans \cite{recurrence-paper}. Cela donne bien une image de la relation de r\'ecurrence comme une \'equation d'invariance sous une \emph{combinaison} des mouvements \'el\'ementaires sous lesquels le mod\`ele topologique est invariant.

\section{Conclusion sur le mod\`ele BC} \label{sec:conclusionBC}

Nous avons donc fournit une action discr\`ete et un calcul de type int\'egrales de chemins sur la triangulation pour le mod\`ele BC. Une telle formulation \'eclaire le contenu du mod\`ele en termes de g\'eom\'etrie simplicielle. En particulier les bivecteurs qui sont toujours habituellement trait\'es comme des op\'erateurs sur l'espace des \'etats du bord des 4-simplexes sont ici des variables classiques que l'on contraint de mani\`ere \`a ce qu'ils d\'ecrivent des t\'etra\`edres et 4-simplexes dans $\R^4$. Le mod\`ele en version mousses de spins est alors obtenu par int\'egrations sur ces bivecteurs classiques et contraints, et sur les holonomies.

Le mod\`ele BC \cite{BCpaper} fut construit par quantification g\'eom\'etrique sur un 4-simplexe isol\'e, en imposant les contraintes de simplicit\'e sur chaque t\'etra\`edre au niveau quantique. \'Etant donn\'e que le mod\`ele n'est pas une quantificaton de la gravit\'e, on faut essayer de comprendre o\`u les probl\`emes se cr\'eent. Du point de vue mousses de spins, le probl\`eme r\'eside dans le manque de libert\'e sur les entrelaceurs, autrement dit, le manque de possibilit\'e de corr\'elations entre simplexes adjacents. Cette interpr\'etation est renforc\'ee par la d\'erivation du mod\`ele selon les m\'ethodes FK, et par notre travail, bien que sous des angles assez diff\'erents. Dans \cite{new-model-fk}, Freidel et Krasnov propose une approche de ce mod\`ele avec des bivecteurs compl\`etement ind\'ependants sur chaque triangle de chaque t\'etra\`edre de chaque 4-simplexe, soit $B_{ftv}$. Ainsi c'est la quantification sur des 4-simplexes \emph{isol\'es} qui est mise en accusation. Le mod\`ele FK \'emerge alors dans cette approche en imposant les contraintes de la m\^eme mani\`ere mais \`a l'\'echelle de la triangulation, i.e en identifiant les bivecteurs de deux simplexes adjacents : $B_{ftv}=B_{ftv'}$ si $v, v'$ sont coll\'es le long de $v$. Dans notre approche, les 4-simplexes sont bien disjoints, mais nos bivecteurs satisfont bien la relation $B_{ft} = B_{ftv}=B_{ftv'}$ ; le probl\`eme se situe alors dans le collage des simplexes.

On pourrait penser que les corr\'elations entre t\'etra\`edres sont amoindries par le fait que le noyau du 10j fasse intervenir les normales aux t\'etra\`edres sous la forme \eqref{ZBC final} :
\beq
\chi_{j_f}\bigl(N_{vt}\mone N_{vt'}^{\phantom{}}\bigr),
\ee
i.e. avec des normales $N_{vt}$ ind\'ependantes pour les deux simplexes $v, v'$ coll\'es par $t$. Mais un tel ph\'enom\`ene se retrouve dans les nouveaux mod\`eles, et s'explique tr\`es clairement ici. Contrairement \`a ce que le r\'esultat, i.e. le 10j, sugg\`ere, nous avons d\'emarr\'e l'analyse et effectu\'e les calculs avec une m\^eme normale $N_t$, quelque soit le 4-simplexe envisag\'e, mais \`a du transport parall\`ele pr\`es. En effet, il ne faut pas oublier que nous sommes dans une th\'eorie de jauge (sur r\'eseau), et donc avec des r\'ef\'erentiels \emph{locaux}. Ainsi, la fonction de partition fait intervenir, avant toute int\'egration sur les holonomies, les normales \emph{transport\'ees}, $N_t(v), N_t(v')$. On voit bien, par exemple en utilisant la jauge temps, que les holonomies ont alors un r\^ole majeur dans l'amplitude. Et ce n'est qu'en int\'egrant sur les holonomies $G_{vt}$ que l'on peut absorber les normales dans de nouvelles variables $N_{vt} = N_t(v)$, voir \'equation \eqref{decoupling normals}, cr\'eant l'illusion de normales d\'ecorr\'el\'es entre 4-simplexes adjacents.

En revanche, la difficult\'e avec ce mod\`ele se situe bien dans le manque de corr\'elations entre les normales de \emph{t\'etra\`edres} adjacents ! En effet, lorsque nous r\'esolvons les contraintes de simplicit\'e dans chaque t\'etra\`edre, nous introduisons des vecteurs normaux \`a chacun d\'eterminant leur direction orthogonale dans $\R^4$. Deux t\'etra\`edres $t_1, t_2$ coll\'es le long du triangle $f$ doivent alors avoir pour normales deux vecteurs tous deux contenus dans le \emph{plan} orthogonal au triangle, i.e \`a $\star B_f$. Mais pour dire cela, nous devons encore comparer des quantit\'es d\'efinies dans des r\'ef\'erentiels diff\'erents, et donc les transporter du r\'ef\'erentiel de $t_1$ \`a celui de $t_2$ disons. Mais nous avons vu qu'aux points stationnaires de l'action, les bivecteurs $B_{ft_1}$ et $B_{ft_2}$ ne satisfont pas la condition de transport parall\`ele, \'equation \eqref{bad gluing BC} !

Nous voyons de la sorte que les corr\'elations induites entre t\'etra\`edres voisins et la consistence des changements de r\'ef\'erentiels sont intimement li\'ees. En particulier, il devient impossible de calculer avec consistence, aux points stationnaires, des angles dih\'edraux ou des angles de d\'eficit si le processus de collage \'echoue. Pour rem\'edier \`a ce probl\`eme du mod\`ele BC, nous prendrons grand soin aux sections suivantes du collage ds simplexes. En particulier, nous abandonnerons les wedges et les holonomies de bord qui n'ont pas donn\'e les r\'esultats escompt\'es, alors m\^eme que nous avons trait\'e les contraintes de mani\`ere conventionnelles ! Pour cela, il nous faudra pr\^eter une attention particuli\`ere aux conditions de transport parall\`ele des bivecteurs, en exploitant notamment le formalisme des sections \eqref{sec:geom normales} et \ref{sec:gluing-aarc}.

Par ailleurs, souvenons-nous que les mousses de spins visent \`a impl\'ementer la dynamique de la th\'eorie ! Cela appara\^it clairement dans les mod\`eles de Ponzano-Regge et d'Ooguri, construits en int\'egrant sur les connexions discr\`etes de courbure nulle. Peut-on essayer de saisir la dynamique propos\'ee dans le mod\`ele BC ? Faisons la remarque suivante qui nous \'eclairera sur ce que l'on attend (ou pas) de la dynamique en termes des variables d'holonomies et des normales. Le mod\`ele impose sur chaque wedge une condition d'invariance de transport pour les normales,
\beq
G_{ft}\, \cdot\, N_t \,=\, g_{+ft}\,N_t\,g_{-ft}\mone \,=\, N_t,
\ee
et une condition duale sur le secteur non-g\'eom\'etrique. L'interpr\'etation n'est pas \'evidente du fait des wedges et holonomies de bord. Supposons que les points stationnaires de l'action rendent correctement compte du transport parall\`ele et que l'on puisse constuire \`a partir des conditions ci-dessus une \'equation similaire pour l'holonomie autour d'un triangle $f$, soit :
\beq
G_f(t)\, \cdot\, N_t \,=\, g_{+f}(t)\,N_t\,g_{-f}\mone(t) \,=\, N_t.
\ee
Alors cette \'equation nous dit une chose simple : l'angle de d\'eficit autour de $f$ est nul ! Comme nous l'avons vu \`a l'\'equation \eqref{deficithol} et dans la discussion qui la suit, nous nous attendons plut\^ot, dans une discr\'etisation consistente, \`a ce que $G_f(t)$ transforme la normale au t\'etra\`edre $N_t$ en un autre vecteur \'egalement contenu dans le plan orthogonal \`a $\star B_{ft}$, l'angle entre ces deux vecteurs \'etant pr\'ecis\'ement l'angle de d\'eficit.



\chapter{Int\'egrales sur les g\'eom\'etries du r\'eseau : le mod\`ele EPR/FK} \label{sec:pathint fk}

Apr\`es avoir aux chapitres pr\'ec\'edents trouv\'e des syst\`emes classiques de g\'eom\'etries sur r\'eseau dont les fonctions de partition reproduisent le mod\`ele BC et celui contraignant les repr\'esentations sur les triangles, nous aimerions :
\begin{itemize}
 \item trouver une pr\'esentation similaire pour le mod\`ele EPR/FK, prenant en compte le param\`etre d'Immirzi,
 \item analyser le comportement classique correspondant, pour s'assurer que les probl\`emes de corr\'elations mis en \'evidence dans le mod\`ele BC sont corrig\'es.
 \item par variations des spins ou aires de l'action trouv\'ee, obtenir une id\'ee de la dynamique engendr\'ee par le mod\`ele, ou plus modestement des configurations qui ont un r\^ole particulier, soit en dominant la fonction de partition, soit en n\'ecessitant une r\'egularisation. En particulier, nous aimerions comprendre le sens physique des configurations provoquant des divergences dans la formulation \eqref{sum spins epr} en termes d'int\'egrales.
 \item dans une optique diff\'erente, utiliser le formalisme des sections \ref{sec:geom normales}, \ref{sec:gluing-aarc} pour construire des fonctions de partition associ\'ees \`a des syst\`emes de g\'eom\'etries discr\`etes et les \'ecrire sous forme de mousses de spins.
\end{itemize}
Si les trois premiers points fixent des objectifs, le quatri\`eme est plut\^ot une m\'ethode dont on ne conna\^it pas le r\'esultat a priori. De mani\`ere remarquable, c'est une m\'ethode qui va nous permettre d'atteindre les objectifs fix\'es, i.e. de construire le mod\`ele FK pour la gravit\'e quantique, pour un param\`etre d'Immirzi fini, de constater que les d\'efauts de corr\'elations du mod\`ele BC sont bien corrig\'es, et de voir que les singularit\'es de la formulation \eqref{sum spins epr} sont associ\'ees aux g\'eom\'etries plates !

Dans \cite{freidel-conrady-pathint}, Conrady et Freidel ont \'egalement fourni un formalisme d'int\'egrales de chemins pour reproduire les amplitudes du mod\`ele FK. Leur construction fait pr\'ecis\'ement appel aux variables que nous avons utilis\'e pour le mod\`ele BC au chapitre pr\'ec\'edent \ref{sec:actionBC} (avec la m\^eme interpr\'etation physique), et utilise le m\^eme processus de collage des simplexes. N\'eanmoins, l'action qu'ils construisent est destin\'ee \`a offrir une int\'egrale de chemins reproduisant exactement le mod\`ele FK, et n'est pas une action du type BF. Notre perspective est ici diff\'erente puisque nous voulons consid\'erer des int\'egrales de chemins g\'en\'eriques, fond\'ees sur des actions de type BF discret et sur les contraintes de simplicit\'e, et explorer quels mod\`eles \'emergent naturellement. Nous avons vu qu'en appliquant telles quelles ces id\'ees, c'est le mod\`ele BC qui se construisait.

Afin d'aller au-del\`a du mod\`ele BC, il nous faut en effet prendre au s\'erieux les probl\`emes de collage de simplexes, et d\'evelopper une approche plus adapt\'ee. Dans cette persective, le formalisme pr\'esent\'e aux sections \ref{sec:geom normales}, \ref{sec:gluing-aarc} est prometteur : il permet de d\'etailler la g\'eom\'etrie, montrant que les contraintes de simplicit\'e autorisent la construction de g\'eom\'etries de Regge, et assure, typiquement, que l'on peut calculer de mani\`ere consistente les angles dih\'edraux 4d en fonction des angles 3d. Cela \'etablit le lien avec le calcul de Regge. De plus, ce formalisme traite de la m\^eme mani\`ere les conditions de transport parall\`ele que les contraintes de simplicit\'e : en les formulant uniquement en termes d'\'el\'ements du groupe $\SU(2)$ (et sans avoir recours \`a des holonomies de bord), ce qui est tout indiqu\'e pour en d\'eveloppement en mousses de spins.

Nous verrons ainsi qu'il est possible de repr\'esenter le mod\`ele FK sous la forme d'une int\'egrale sur des g\'eom\'etries d'une triangulation. Par ailleurs, cela nous renseignera sur la diff\'erence entre la fonction de partition pour le calcul de Regge et le mod\`ele de mousses de spins sur une triangulation fix\'ee, \`a savoir que seules les configurations stationnaires de l'action que nous allons construire sont des g\'eom\'etries de Regge, les fluctuations quantiques \'etant d\'etermin\'ees par le choix de fonctions sp\'ecifiques au mod\`ele. Notons \`a ce propos les travaux de Freidel et ses collaborateurs, \cite{conrady-quantum-tet, freidel-quantum-tet}, qui r\'e\'ecrivent l'int\'egrale de chemins discrets avec des fonctions delta permettant d'avoir une g\'eom\'etrie de Regge sur chaque simplexe (mais sur la triangulation seulement on-shell), en utilisant des propri\'et\'es remarquables des \'etats coh\'erents. Le prix \`a payer est alors une nouvelle mesure particuli\`ere, faisant intervenir des fonctions de corr\'elations d'une th\'eorie des champs conforme.

De plus, en regardant l'ensemble des variations de l'action, en particulier par rapport aux aires, nous pouvons obtenir une \'equation sur la dynamique de ces g\'eom\'etries. Ici, nous verrons que les g\'eom\'etries stationnaires sont plates. N\'eanmoins, cela ne signifie pas que ces configurations dominent la fonction de partition, mais c'est plut\^ot le signe qu'elles n\'ecessitent une r\'egularisation suppl\'ementaire avant de pouvoir discuter la dynamique plus pr\'ecis\'ement.

Mais avant de lancer ce formalisme original tel quel au niveau quantique, il faut nous assurer qu'en l'absence des contraintes de simplicit\'e, nous pouvons effectivement restituer le mod\`ele topologique, avec la bonne mesure. Cela n\'ecessite de construire une action adapt\'ee aux variables de la section \ref{sec:gluing-aarc}, qui m\'elangent les variables usuelles de la th\'eorie BF sur r\'eseau (holonomies), des variables de $\SU(2)$ (codant les bivecteurs), et des variables angulaires, interpr\'et\'ees comme des \'el\'ements d'un sous-groupe $\U(1)$ de $\SU(2)$. Et afin que les amplitudes r\'esultantes soient bien topologiques, il faut d\'eterminer avec pr\'ecision les mesures pour ces variables. Nous faisons tout cela \`a la section \ref{sec:back-topological}, en calculant explicitement la fonction de partition. De plus, nous nous int\'eressons \`a l'insertion d'observables locales, fonctions des bivecteurs, et montrons que l'on peut facilement les traduire en tant qu'insertions dans les mousses de spins. L'insertion de la norme au carr\'e d'un bivecteur conduit comme on s'y attend \`a l'insertion d'un casimir, et celle du cosinus de l'angle entre deux bivecteurs reproduit \`a une normalisation pr\`es le r\'esultat de la quantification canonique sur les r\'eseaux de spins de Major, pr\'esent\'ee en section \ref{sec:quantum tet} !

Puis \`a la section \ref{sec:action eprfk}, nous construisons une action pour des g\'eom\'etries \`a la Regge sur une triangulation, pour laquelle on montre \`a la section \ref{sec:partition eprfk} que la fonction de partition reproduit bien le mod\`ele FK ! L'action s'exprime en termes des variables pr\'esent\'ees \`a la section \ref{sec:gluing-aarc} et contient principalement deux parties : une partie prenant la forme de l'action de Regge compactifi\'ee mais avec le param\`etre d'Immirzi, et une partie qui assure que les relations de transport parall\`ele sont bien satisfaites on-shell. On trouve que les g\'eom\'etries stationnaires (non-d\'eg\'en\'er\'ees) sont des g\'eom\'etries de Regge, satisfaisant les contraintes du calcul de Regge aires-angles 3d, et que le param\`etre d'Immirzi dispara\^it au niveau classique (suite \`a un r\'esultat de \cite{freidel-conrady-semiclass}). De plus, en faisant varier les aires (ou les spins), on constate que les g\'eom\'etries stationnaires sont plates (r\'esultat que je discute en conclusion de la th\`ese).

Le mod\`ele FK est ainsi d\'eriv\'e \`a partir d'une int\'egrale sur des configurations g\'eom\'etriques, en utilisant une fonction sp\'ecifique pour traiter les relations de transport parall\`ele. Cela sugg\`ere de g\'en\'eraliser l'action utilis\'ee. Nous montrons dans ce cas qu'il est toujours possible de construire des g\'eom\'etries de Regge. Mais, g\'en\'eriquement, les variables qui apparaissaient classiquement comme les aires dans la d\'erivation du mod\`ele FK ne peuvent plus \^etre interpr\'et\'ees comme les aires physiques. Ce sont \`a la place des multiplicateurs imposant une condition d'angles de d\'eficit nuls, qui donnent des g\'eom\'etries on-shell plates. Les mod\`eles de mousses de spins obtenues par cette approche, section \ref{sec:partition eprfk}, ont la m\^eme structure que celles des nouveaux mod\`eles EPR/FK, mais les donn\'ees de bord du 4-simplexe sont diff\'erentes, de m\^eme que la fa\c{c}on dont elles sont somm\'ees dans les mod\`eles.

\section{Retour sur les mod\`eles topologiques} \label{sec:back-topological}

\subsection{L'action en variables \og aires - angles dih\'edraux\fg}

Pour les besoins du mod\`ele topologique de type BF, sans contrainte g\'eom\'etrique, il nous suffit de consid\'erer le groupe $\SU(2)$. Donc, nous avons des holonomies $g_{vt}\in\SU(2)$ pour transporter des objets entre les r\'ef\'erentiels locaux des t\'etra\`edres et 4-simplexes. Nous avons aussi des variables de l'alg\`ebre sur les triangles $b_{ft}\in\su(2)$, d\'efinis ind\'ependamment dans les r\'ef\'erentiels de chaque t\'etra\`edre\footnote{On pourrait de la m\^eme fa\c{c}on introduire des variables $b_{fv}$ dans les rep\`eres des 4-simplexes, mais cela ne nous servira pas.} dont le bord contient le triangle $f$. En dimension 4, on parle du mod\`ele d'Ooguri, mais il suffit de changer les notations pour se mettre en dimension quelconque\footnote{En dimension $n$, les holonomies effectuent du transport parall\`ele entre $n$-simplexes et $(n-1)$-simplexes, la courbure \'etant concentr\'ee autour des $(n-2)$-simplexes. Ce sont ces derniers qui portent les variables de l'alg\`ebre, discr\'etisant la $(n-2)$-forme $B$. Le point de vue du 2-squelette dual est unificateur : les holonomies discr\'etisent toujours la connexion sur les liens duaux, et les variables $b_f$ sont associ\'ees aux faaces duales.}, et le formalisme que nous d\'eveloppons fonctionne toujours ! La consistence de la g\'eom\'etrie, au sens des th\'eories de jauge, demande les conditions
\beq \label{parallel transport su2}
b_{ft} - \Ad\bigl(g_{tt'}\bigr)\,b_{ft'} \,=\,0,
\ee
pour $g_{tt'} = g_{vt}\mone g_{vt'}$. Dans la d\'erivation usuelle du mod\`ele topologique BF, \cite{fk-action-principle}, utilisant wedges et holonomies de bord des 4-simplexes, cette \'equation n'a de sens qu'aux points stationnaires de l'action. Mais nous avons vu qu'en pr\'esence des contraintes de simplicit\'e, la situation n'est plus totalement sous contr\^ole quant au collage des t\'etra\`edres, et nous trouvons prudent de chercher \`a imposer ces conditions directement dans la fonction de partition. On peut donc penser \`a utiliser une mesure telle que :
\beq \label{su2 gluing}
\prod_{(t,v)}dg_{vt}\ \prod_{(f,t)} d^3b_{ft}\ \prod_{(f,v)}\delta^{(3)}\bigl(b_{ft} - \Ad\bigl(g_{tt'}\bigr)\,b_{ft'}\bigr),
\ee
qui restreint les configurations \`a int\'egrer \`a celles satisfaisant les conditions de transport parall\`ele de mani\`ere exacte. $dg$ est la mesure de Haar sur $\SU(2)$, et $d^3b$ sera pr\'ecis\'e par la suite. Notons que la donn\'ee $(f,v)$, repr\'esentant un wedge, sp\'ecifie compl\`etement la paire de t\'etra\`edres de $v$ coll\'es le long de $f$. Avec une telle mesure, le recours aux holonomies de bord des wedges, comme utilis\'ees pr\'ec\'edemment, dont les variations dans l'action visent \`a reproduire \eqref{parallel transport su2} aux points stationnaires, est inutile : le collage est d\'esormais explicitement pris en compte et garanti dans la d\'efinition du mod\`ele.

Le groupe de structure \'etant $\SU(2)$, la norme et la direction de $b_{ft}$ jouent des r\^oles diff\'erents. Et comme on le sait les contraintes de simplicit\'e leur assignent aussi des r\^oles diff\'erents. La norme de $b_{ft}$ est invariante de jauge et s'\'ecrit disons $A_f$. On utilise alors la param\'etrisation suivante, similaire \`a \eqref{B parametrisation},
\beq \label{b parametrisation}
b_{ft} = \f{i}{2}\,A_f\ \hat{b}_{ft}\cdot\vec{\sigma} ,\qquad\qquad \hat{b}_{ft}\cdot\vec{\sigma} = \Ad\bigl(n_{ft}\bigr)\,\sigma_z,
\ee
o\`u $n_{ft}\in\SU(2)$ est une rotation qui envoie la direction de r\'ef\'erence $\hat{z}$ de $\R^3$ sur la direction $\hat{b}_{ft}$. On sait que seul deux des trois param\`etres de $n_{ft}$ sont pertinents du fait du sous-groupe $\U(1)$ qui laisse $\hat{z}$ invariant (ou le sous-groupe qui laisse $\hat{b}_{ft}$ invariant, de mani\`ere \'equivalente), voir section \ref{sec:gluing-aarc}. Pour r\'e\'ecrire la mesure en fonction des rotations $n_{ft}$, l'id\'ee principale est qu'en fonction de ces variables, l'\'equation de transport parall\`ele \eqref{parallel transport su2} nous dit que $n_{ft}$ et $g_{tt'}n_{ft'}$ ne coincident qu'\`a un \'el\'ement de $\U(1)$ selon $\hat{z}$ pr\`es. Ainsi, nous commen\c{c}ons par remplacer dans \eqref{su2 gluing} les int\'egrales et mesures sur $b_{ft}$ par une seule int\'egrale par triangle sur la norme, $d\mu(A_f)$ \`a pr\'eciser, et pour chaque triangle dans chaque t\'etra\`edre la mesure usuelle $d^2\hat{b}_{ft}$ sur $S^2$ invariante par rotation,
\beq
\prod_f d\mu\bigl(A_f\bigr)\ \prod_{(f,t)} d^2\hat{b}_{ft}\ \prod_{(f,v)} \delta^{(2)}\bigl(\hat{b}_{ft}-\Ad\bigl(g_{tt'}\bigr)\,\hat{b}_{ft'}\bigr).
\ee
La mesure $d^2\hat{b}_{ft}$ se r\'eexprime comme une mesure sur $\SU(2)$ pour les rotations $n_{ft}$, les ambiguit\'es $\U(1)$ dans leur d\'efinition \'etant prises en compte via les contraintes de collage. Il nous faut en effet int\'egrer sur les solutions des contraintes \eqref{parallel transport su2} caract\'eris\'ees par un \'el\'ement de $\U(1)$ :
\beq \label{gluing measure}
\prod_f d\mu\bigl(A_f\bigr)\ \prod_{(f,t)} dn_{ft}\ \prod_{(f,v)} \int_0^{4\pi}d\theta_{fv}\ \delta_{\SU(2)}\Bigl(n_{ft}\mone\ g_{tt'}\,n_{ft'}\ e^{\f{i}{2}\theta_{fv}\sigma_z}\Bigr).
\ee
En fait les int\'egrales sur les fibres $\U(1)$ qui pr\'eservent $\hat{z}$ pour chaque $n_{ft}$ peuvent \^etre r\'eabsorb\'ees par invariance de la mesure de Haar dans les int\'egrales sur les variables angulaires $\theta_{fv}$. Nous ne sommes pas int\'eress\'es par la correspondance pr\'ecise, incluant les jacobiens, entre les deux expressions ci-dessus, car nous voulons la mesure la plus simple et la plus naturelle au regard des th\'eories de jauge sur r\'eseau, i.e. en utilisant des \'el\'ements de groupe et les mesures de Haar correspondantes le plus possible. L'inconnue restante est la mesure $d\mu(A)$, qui sera fix\'ee \`a la mesure de Lebesgue sur $\R$ par l'exigence de reproduire le mod\`ele topologique attendu.

\medskip

Au d\'epart, nous avons consid\'er\'e que les variables de configuration sont les holonomies $g_{vt}$, les normes $A_f$ et les directions $\hat{b}_{ft}$. En particulier, ce sont les variables dans lesquelles doit s'\'ecrire l'action. L'id\'ee importante, d\'ej\`a expos\'ee \`a la seciton \ref{sec:gluing-aarc} est de maintenant remplacer les directions $\hat{b}_{ft}$ par les rotations $n_{ft}$, et d'ajouter comme variables de base les angles $\theta_{fv}$ qui sont int\'egr\'es dans la mesure \eqref{gluing measure}. D\'esormais l'espace des configurations est form\'e des holonomies $g_{vt}$, des normes $A_f$, des rotations $n_{ft}$ et des angles $\theta_{fv}$. Ces variables sont reli\'ees, au moins au niveau classique si l'on utilise une autre mesure que \eqref{gluing measure}, par la condition de transport parall\`ele qui est devenue :
\beq \label{gluing n}
n_{ft}\mone\,g_{tt'}\,n_{ft'}\,e^{\f{i}{2}\theta_{fv}\sigma_z} \,=\,\mathbbm{1}.
\ee
Notre approche a l'avantage de nous permettre de consid\'erer d\'esormais des fonctions de $n_{ft}$ et $\theta_{fv}$ m\^eme lorsque les contraintes ci-dessus ne tiennent pas, ce qui sera utile pour construire les nouveaux mod\`eles EPR/FK. En particulier, on est en droit d'\'ecrire l'action comme une fonction des variables $g_{vt}, A_f, n_{ft}, \theta_{fv}$.

Il faut bien comprendre que ce sont les angles $\theta_{fv}$ qui assurent l'ind\'ependance vis-\`a-vis du choix de phase local \`a droite des rotations $n_{ft}$. En effet, en r\'e\'ecrivant cette \'equation pour extraire les holonomies : $g_{tt'} = n_{ft}\,e^{-\f{i}{2}\theta_{fv}\sigma_z}\,n_{ft'}\mone$, on voit que l'invariance de celles-ci sous les changements de r\'ef\'erence $\U(1)$ locale des rotations $n_{ft}$ conduit \`a une transformation des angles $\theta_{fv}$. Ainsi, l'\'elargissement de l'espace des configurations s'accompagne d'une sort de sym\'etrie de jauge additionnelle, tout comme en section \ref{sec:gluing-aarc}, agissant sur chaque paaire $(f, t)$ : Chacune d'elles est \'equip\'ee d'un r\'ef\'erentiel local $\U(1)$ qui peut \^etre chang\'e par multiplication \`a droite de $n_{ft}$ et une transformation ad\'equate des angles $\theta_{fv}$. Ces derniers forment ainsi une sorte de connexion pour cette sym\'etrie. Les transformations de jauge $\SU(2)$ et ces nouvelles transformations agissent par des famille d\'el\'ements $k = (k_t, k_v \in\SU(2))$ et $\lambda = (\lambda_{ft}\in\mathfrak{u}(1))$ par :
\begin{alignat}{2} \label{su2 u1 transfo}
&(k,\lambda)\,\cdot\,g_{vt} &\,=\,& k_v\,g_{vt}\,k_t\mone,\\ 
&(k,\lambda)\,\cdot\, n_{ft} &\,=\,& k_t\,n_{ft}\,e^{\f{i}{2}\lambda_{ft}\sigma_z}, \\
&(k,\lambda)\,\cdot\, \theta_{fv} &\,=\,& \theta_{fv} +\eps_{tt'}^f\l(\lambda_{ft} - \lambda_{ft'}\r).
\end{alignat}
Puisque $t, t'$ sont voisins, leurs liens duaux se rencontrent au vertex $v$, le long du bord de la face duale $f$. Alors, $\eps_{tt'}^f=1$ si le chemin de $t$ \`a $t'$ par $v$ est orient\'e comme la face $f$, et $\eps_{tt'}^f=-1$ dans l'autre cas.

Si les variables similaires de la section \ref{sec:gluing-aarc}, $\theta^\pm_{fv}$, se trouvent \^etre directement reli\'es aux angles dih\'edraux physiques, les angles $\theta_{fv}$ contiennent ici l'information pour extraire les angles de d\'eficit de la courbure $\SU(2)$ :
\beq \label{su2discrete curv}
g_f(t) = \exp\Bigl(-\theta_f\,\f{b_{ft}}{A_f}\Bigr),\quad\text{pour}\qquad \theta_f = \sum_{v\in\pp f} \theta_{fv}.
\ee
C'est-\`a-dire que $\theta_f$ est l'angle de classe de l'holonomie autour du triangle $f$, holonomie qui, exprim\'ee dans le r\'ef\'erentiel de $t$, est naturellement restreinte au sous-groupe qui stabilise $b_{ft}$ (c'est un \'equivalent discret de l'\'equation continue : $d_A^2B = [F(A),B] = 0$). Nous pouvons maintenant reconsid\'erer l'action discr\`ete de type BF, \eqref{discreteBFaction}, sans utiliser de d\'ecoupage des face en wedges, mais plut\^ot en couplant un $b_{ft}$ par face \`a l'holonomie $g_f(t)$, le tout \'etant ind\'ependant du r\'ef\'erentiel choisi pour chaque face, gr\^ace aux contraintes de transport parall\`ele que nous allons imposer,
\beq
S_{\rm BF}(g_{vt}, A_f, n_{ft}, \theta_{fv}) \,=\,\sum_f \tr\bigl( b_{ft}\,g_f(t)\bigr).
\ee
L'action on-shell, i.e. lorsque \eqref{gluing n} tient, se met sous la forme \'evocatrice suivante :
\beq \label{BFregge action}
S_{\rm BF}(g_{vt}, A_f, n_{ft}, \theta_{fv}) \,=\,\sum_f A_f\, \sin\l(\f{\theta_f}{2}\r).
\ee
Cette action est invariante sous les transformations d\'efinies plus haut. Si l'on \'etait capable d'interpr\'eter $A_f$ et les angles $\theta_{fv}$ comme les aires des triangles et les angles dih\'edraux, i.e. si ces quantit\'es pouvaient \^etre fonctions d'une m\'etrique discr\`ete, alors l'action serait l'action de Regge dite compactifi\'ee, propos\'ee en \cite{caselle-RC-poincare, kawamoto-nielsen} ! Nous verrons comment obtenir ce r\'esultat, \`a l'aide des contraintes de simplicit\'e, et m\^eme comment y introduire le param\`etre d'Immirzi !

\subsection{Calcul de l'amplitude} \label{sec:bf-regge calcul}

Munis de l'action \eqref{BFregge action} et de la mesure \eqref{gluing measure}, nous veillons \`a faire en sorte que notre nouveau formalisme reproduise bien le mod\`ele topologique. Dans la th\'eorie continue, le champ $B$ est un multiplicateur de Lagrange pour la courbure. Au niveau discret on sait que la th\'eorie consiste \`a int\'egrer sur toutes les connexions discr\`etes telles que les holonomies autour des faces duales soient $\pm\mathbbm{1}\in\SU(2)$, i.e. que leurs projections sur $\SO(3)$ soient l'identit\'e :
\begin{align} \label{ZBF so3}
Z_{\mathrm{BF}} &= \int \prod_{(t,v)} dg_{vt}\ \prod_f \delta_{\SO(3)}\bigl( g_f\bigr), \\
 &= \sum_{\{j_f\in\N\}} \int \prod_{(t,v)} dg_{vt}\ \prod_f d_{j_f}\,\chi\bigl(g_f\bigr).
\end{align}
Les facteurs de dimension sont $d_j = (2j+1)$, et les caract\`eres des repr\'esentations : $\chi_j(g) = \sum_m \langle j, m\lvert g\rvert j, m\rangle = \f{\sin d_j\phi}{\sin\phi}$, pour $g=e^{i\phi\hat{n}\cdot\vec{\sigma}}$.

\medskip

En utilisant les variables introduites pr\'ec\'edemment, nous nous int\'eressons \`a :
\beq \label{def Zbf}
Z_{\mathrm{BF}} = \int \prod_{(t,v)} dg_{vt}\ \prod_{(f,t)} dn_{ft}\ \prod_{(f,v)} \Bigl[d\theta_{fv}\,\delta\bigl(n_{ft}\mone\,g_{tt'}\,n_{ft'}\,e^{\f{i}{2}\theta_{fv}\sigma_z}\bigr)\Bigr] \prod_f d\mu\bigl(A_f\bigr) e^{iA_f\sin\f{\theta_f}{2}},
\ee
o\`u $dn$ est la mesure de Haar sur $\SU(2)$, $d\theta$ la mesure de Haar sur $\U(1)$, et $d\mu(A)$ sera choisie pour reproduire \eqref{ZBF so3} (comme \'etant la mesure de Lebesgue sur $\R$). Il est clair que sur chaque face, les fonctions delta permettent d'\'eliminer tous les $n_{ft}$ sauf un, choisi dans un t\'etra\`edre de r\'ef\'erence, disons $n_{ft^*}$. La fonction delta restante sur chaque face impose l'\'equation \eqref{su2discrete curv} : $\delta(n_{ft^*}\,g_f(t^*)\,n_{ft^*}\mone\,e^{\f{i}{2}\theta_f\sigma_z})$. Par invariance de la mesure $d\theta$ par translation, tous les angles autour d'une face peuvent absorb\'es en un seul angle $\theta_f$ ind\'ependant.

Le formalisme des mousses de spins consiste \`a \'ecrire $Z_{\rm BF}$ comme une fonction de partition, i.e. une somme sur tous les \'etats du syst\`eme au niveau quantique, caract\'eris\'es par des donn\'ees provenant de la th\'eorie des repr\'esentations des groupes concern\'es. Nous devons donc d\'evelopper les ingr\'edients de $Z_{\rm BF}$ sur leurs modes de Fourier pour $\U(1)$ et $\SU(2)$. Tout d'abord, pour chaque face, les conditions de collage conduisent \`a une fonction delta sur la classe de conjugaison $\theta_f$ pour l'holonomie $g_f$,
\beq \label{conj class bf su2}
\int_{\SU(2)} dn\ \delta\bigl(n\,g\,n\mone\,e^{\f{i}{2}\theta\sigma_z}\bigr) = \sum_{j\in\f{\N}{2}} \chi_j\bigl(g\bigr)\,\chi_j\bigl(e^{\f{i}{2}\theta\sigma_z}\bigr)
\ee
En particulier, $\chi_j(e^{\f{i}{2}\theta\sigma_z}) = \sum_{p=-j}^j e^{-ip\theta}$.

Nous pouvons aussi bien prendre les aires $A_f$ sur $\R_+$ ou sur $\R$ (en renversant la direction de $\hat{b}_{ft}$). Si $A_f$ est int\'egr\'e sur $\R$ avec la mesure de Lebesgue, alors $\theta_f$ est clairement contraint \`a \^etre $0$ ou $2\pi$. Il s'ensuit alors que $g_f=\pm\mathbbm{1}$, et donc \eqref{ZBF so3} suit imm\'ediatement. Mais il est plus instructif, et plus dans l'esprit \og mousses de spins\fg, d'utiliser des d\'eveloppements dont les donn\'ees s'interpr\`etent bien et peuvent \^etre consid\'er\'ees comme physiquement pertinentes -- en tant que valeur propre de tel ou tel op\'erateur d'aire, de volume, etc \cite{rovelli-PRTVO}. De plus, nous devons nous pr\'eparer \`a affronter l'ajout des contraintes de simplicit\'e. Pour ces raisons, nous prenons $A_f\in\R_+$, et d\'eveloppons l'exponentielle de $i$ fois l'action sur les modes $\U(1)$ :
\beq \label{bfaction u1modes}
e^{iA\sin\f{\theta}{2}} = \sum_{m\in\Z} J_m(A)\,e^{im\f{\theta}{2}}.
\ee
Il s'agit du d\'eveloppement de Jacobi-Anger, qui d\'efinit en fait les fonctions de Bessel de premi\`ere esp\`ece $J_m$ comme : $J_m(A)=\int_0^\pi \f{d\theta}{\pi\, i^m}\,\cos(m\theta)\,e^{iA\cos\theta}$. Ces fonctions satisfont quelques propri\'et\'es int\'eressantes \cite{abramowitz-stegun}. Leur int\'egrale sur $\R_+$ vaut $1$ pour $m\in\N$, et parmi les sym\'etries : $J_{-m}(A) = J_m(-A) = (-1)^mJ_m(A)$. En combinant tout cela, il vient :
\beq \label{int area bf}
\int_{\R_+}dA\ e^{iA\sin\f{\theta}{2}} = 1 + \sum_{m>0} \bigl(e^{im\f{\theta}{2}} + (-1)^m\,e^{-im\f{\theta}{2}}\bigr).
\ee
Les int\'egrales sur les angles $\theta_f$ se font selon :
\beq
\sum_{m>0}\sum_{p=-j}^j \int_0^{4\pi} \f{d\theta}{4\pi}\ e^{-ip\theta}\bigl(e^{im\f{\theta}{2}} + (-1)^m\,e^{-im\f{\theta}{2}}\bigr) = \sum_{m>0}\sum_{p=-j}^j \delta_{m,2p}\bigl(1+(-1)^m\bigr),
\ee
pour $j\in\N/2$. Cette expression s'annule sauf si $m$ est pair, ce qui imlique que $j$ est entier. Ces conditions satisfaites, nous obtenons pour $m=2m'$ : $2\sum_{m'=1}^j 1 = 2j$. En ce qui concerne le terme $m=0$ dans \eqref{bfaction u1modes}, l'int\'egrale sur $\theta$ est alors triviale, et restreint $j$ \`a \^etre entier. Rassemblons ces r\'esultats :
\beq
Z_{\mathrm{BF}} = \int \prod_{(t,v)} dg_{vt}\ \prod_f \sum_{j\in\N} \bigl(2j+1\bigr)\,\chi_j\bigl(g_f\bigr),
\ee
qui est bien le r\'esultat d\'esir\'e, avec : $\sum_{j\in\N} (2j+1)\,\chi_j(g) = \delta_{\SO(3)}(g)$.

\subsection{Insertions d'observables g\'eom\'etriques locales} \label{sec:insertion obsBF}

Nous pouvons ici profiter de notre formalisme qui d\'ecrit pr\'ecis'ement et de mani\`ere ind\'ependante les normes et directions des $b_{ft}$. Les donn\'ees qui apparaissent dans la somme sur les mousses de spins sont g\'en\'eralement interpr\'et\'ees \`a la vue de la quantification canonique \cite{rovelli-book, baez-bf-spinfoam}. En particulier, les spins $j_f$ sur les faces duales repr\'esentent en 3d la longueur des ar\^etes (c'est l'ansatz de Ponzano et Regge, justifi\'e par la quantification en boucles \cite{rovelli-PRTVO}), et en 4d les aires des triangles quantifi\'ees. Ces r\'esultats doivent pouvoir s'obtenir \'egalement par l'approche covariante, en calculant la valeur moyenne des observables voulues, par insertion dans une int\'egrale fonctionnelle, et en particulier dans l'approche des mousses de spins, en ins\'erant des \'etats dans le mod\`ele pour \og exciter\fg{} les faces duales. C'est ce qui a \'et\'e fait proprement, et assez r\'ecemment dans \cite{livine-ryan-Bobs}, en dimension 3. Nous ne proposons pas un tel traitement ici, notamment du fait que la sym\'etrie de jauge de translation de la th\'eorie topologique est plus subtile en dimensions sup\'erieures (c'est une sym\'etrie r\'eductible \cite{blau-thompson-bf}), et donc de m\^eme pour la fixation de jauge n\'ecessaire.

Nous sommes plut\^ot int\'eress\'es par la mani\`ere dont l'insertion d'observables locales dans l'int\'egrale sur les g\'eom\'etries de la triangulation se traduit dans le langage des mousses de spins. C'est la m\^eme id\'ee que celle d\'evelopp\'ee dans \cite{speziale-grasping-rules} pour le mod\`ele de Ponzano-Regge, mais avec une m\'ethode assez diff\'erente (celle \'etablie dans \cite{fk-action-principle}), nos calculs \'etant plus directs. Les auteurs de \cite{speziale-grasping-rules} \'etablissent des r\`egles dites de grasping pour traduire l'insertion d'observables du champ $B$ dans le mod\`ele de Ponzano-Regge (ce ne sont pas \`a proprement parler des calculs de valeurs moyennes). La structure du mod\`ele d'Ooguri \'etant assez similaire (m\^eme groupe de sym\'etrie $\SU(2)$), nous trouverons des formules analogues.

Consid\'erons d'abord un triangle donn\'e $f^*$ et d\'efinissons $\mathcal{I}(A_{f^*})$ comme $Z_{\mathrm{BF}}$, \eqref{def Zbf}, dans laquelle on a ins\'er\'e la norme au carr\'e $A_{f^*}^2$,
\begin{multline}
\mathcal{I}(A_{f^*}) = \int \prod_{(t,v)} dg_{vt} \prod_{(f,t)} dn_{ft} \prod_{(f,v)} \Bigl[d\theta_{fv}\,\delta\bigl(n_{ft}\mone\,g_{tt'}\,n_{ft'}\,e^{-\f{i}{2}\theta_{fv}\sigma_z}\bigr)\Bigr] \\
\times\prod_{f\neq f^*} d\mu\bigl(A_f\bigr) e^{iA_f\sin\f{\theta_f}{2}} \Bigl[d\mu\bigl(A_{f^*}\bigr)\,A_{f^*}^2\, e^{iA_{f^*}\sin\f{\theta_{f^*}}{2}}\Bigr].
\end{multline}
Le r\'esultat \eqref{ZBF so3} n'est naturellement pas modifi\'e pour les triangles $f\neq f^*$. Seul l'int\'egrale \eqref{int area bf} est affect\'ee pour $f^*$ et produit d\'esormais des facteurs :
\beq
\int_{\R_+} dA\,A^2\,J_m\bigl(A\bigr) = m^2-1,\qquad\text{for}\ m\in2\N
\ee
Le point important, comme dans notre cacul pr\'ec\'edent, est qu'il est possible d'effectuer la somme sur les moments magn\'etiques $m_{f^*}$. En plus du poids standard sur les triangles, $d_{j_f}$, cela revient \`a ins\'erer dans \eqref{ZBF so3} l'aire quantique suivante :
\beq
\hat{A}_j = \f{4}{3}\Bigl[ j\bigl(j+1\bigr)-\f{3}{4} \Bigr] = \f{1}{3}\bigl(d_j^2 - 4\bigr)
\ee
Ce spectre donne une valeur n\'egative pour $j=0$, s'annulerait sur $j=1/2$ (le calcul nous restreint \`a $j$ entier), puis cro\^it comme $\f{4}{3}j(j+1)$. Ce comportement doit \^etre compar\'e \`a la quantification canonique des flux de $B$ de la th\'eorie BF $\SU(2)$ dans le formalisme des r\'eseaux de spins \cite{baez-bf-spinfoam} : ils coincident \`a un facteur num\'erique pr\`es. On s'attend d'ailleurs \`a ce que ce comportement soit reproduit dans les mod\`eles de mousses de spins pour la gravit\'e quantique, puisque la quantification (\`a boucles, par les r\'eseaux de spins) canonique des aires est identique, au param\`etre d'Immirzi pr\`es \cite{rovelli-book}. Par ailleurs, le spectre des flux de $B$ dans la quantification canonique est le m\^eme que celui des longueurs en gravit\'e 3d, ce qui sugg\`ere de comparer notre r\'esultat \`a celui de \cite{livine-ryan-Bobs}. A un coefficient $1/3$ pr\`es, il correspond bien au r\'esultat obtenu dans \cite{livine-ryan-Bobs} en 3d en prenant soin des fixations de jauge. Comme l'explique les auteurs, le r\'esultat d\'epend crucialement des facteurs de mesure et de la mani\`ere dont la courbure est discr\'etis\'ee sur le r\'eseau. Nous avons donc un r\'esultat int\'eressant et prometteur, \'etant donn\'e notre cadre de travail impliquant des variables, des mesures et des calculs diff\'erents.

\medskip

On s'int\'eresse maintenant \`a l'insertion de l'angle 3d, $\cos \phi_{ff'}^t$, entre les directions des deux triangles $\hat{b}_{ft}$ et $\hat{b}_{f't}$ dans le t\'etra\`edre $t$ -- il s'agit bien s\^ur d'un analogue $\SU(2)$ du v\'eritable angle dih\'edral entre les triangles, donn\'e \`a la section \ref{sec:geom normales}. Soit deux triangles $f_1, f_2$ dans un t\'etra\`edre $t^*$. Notre angle est, suivant \eqref{3d angle} et \eqref{3d angle n},
\beq
\cos \phi_{f_1f_2}^{t^*} \,=\, -\eps_{f_1f_2}\ \langle 1,0\lvert n_{f_1t^*}\mone\,n_{f_2t^*}^{\phantom{}}\rvert 1, 0\rangle.
\ee
Nous prendrons dans la suite la m\^eme orientation pour les faces, $\eps_{f_1f_2} =1$. Nous nous attendons naturellement \`a ce que la traduction de cette quantit\'e en mousses de spins fasse intervenir le spin entrelaceur $i_{t^*}$ qui couple $j_1$ et $j_2$ (en choisissant l'appariement entre les quatre spins se rencontrant en $t^*$ de fa\c{c}on ad\'equate), et nous esp\'erons pouvoir comparer notre r\'esultat (avec succ\`es) avec la quantification des angles dih\'edraux 3d en gravit\'e canonique \`a boucles donn\'ee par Major \cite{major-3dangles}. On regarde donc :
\begin{multline}
\mathcal{I}(\cos\phi_{f_1 f_2}^{t^*}) = \int \prod_{(t,v)} dg_{vt} \prod_{(f,t)} dn_{ft} \prod_{(f,v)} \Bigl[d\theta_{fv}\,\delta\bigl(n_{ft}\mone\,g_{tt'}\,n_{ft'}\,e^{-\f{i}{2}\theta_{fv}\sigma_z}\bigr)\Bigr] \prod_{f} d\mu\bigl(A_f\bigr) e^{iA_f\sin\f{\theta_f}{2}}\\
\times\Bigl[D^{(1)}_{00}\bigl(n_{f_1 t^*}\mone\,n_{f_2t^*}\bigr)\Bigr]
\end{multline}
Nous \'ecrivons ensuite : $D_{00}^{(1)}(n_1\mone\,n_2) = \sum_{A=-1}^1 D^{(1)}_{0A}(n_1\mone)\,D^{(1)}_{A0}(n_2)$, en notant les \'el\'ements de matrices des repr\'esentations $D^{(j)}_{mn}(g) = \langle j, m\lvert g\rvert j, n\rangle$. Apr\`es int\'egration sur les normes $A_f$ et les angles $\theta_{fv}$, l'int\'egrant sur $f_1$ est :
\beq \label{insertion 3d angle}
\Biggl\{\sum_{j_1\in\N} d_{j_1} D^{(j_1)}_{00}(n_1\mone g_1n_1) + \sum_{m_1>0}\sum_{j_1\in\f{\N}{2}} d_{j_1}\Bigl( D^{(j_1)}_{\f{m_1}{2}\f{m_1}{2}}(n_1\mone g_1n_1) + (-1)^{m_1} D^{(j_1)}_{-\f{m_1}{2}-\f{m_1}{2}}(n_1\mone g_1n_1)\Bigr)\Biggr\}\,D^{(1)}_{0A}(n_1\mone)
\ee
L'int\'egrale sur les rotations $n_1, n_2$ n'est donc plus trivial comme en \eqref{conj class bf su2}, et se calcule explicitement :
\beq \label{int n 3d angle}
\int_{\SU(2)}dn\ D^{(j)}_{ll}\bigl(n\mone g n\bigr)\ D^{(1)}_{0A}(n\mone) = -(-1)^{j-l}\begin{pmatrix} j & j & 1 \\ -l & l & 0\end{pmatrix} \sum_{a,b} (-1)^{j-a}(-1)^{1-A} \begin{pmatrix} j & j & 1 \\ -a & b & -A \end{pmatrix} D^{(j)}_{ab}(g)
\ee
Les d\'etails du calcul sont fournis dans \cite{bf-aarc-val}. A nouveau, le point important est qu'il est possible de faire les sommes sur les moments magn\'etiques $m_1, m_2$ pour ne laisser subsister que les spins $j_1, j_2$ ! Par ailleurs, ces spins sont ici contraints \`a \^etre dans $\N+\f12$.

Les int\'egrales sur les holonomies produisent toujours une somme sur les entrelaceurs entre les quatre repr\'esentations se rencontrant en $t^*$. On choisit la base telle que $j_1$ et $j_2$ s'apparient via $i_{t^*}$. Il est clair dans l'\'equation ci-dessus que l'insertion provoque une s\'erie d'entrelacement de $j_1$ avec $j_2$ via le spin $1$ selon : $\calH_{j_1} \otimes \calH_{j_1} \rightarrow \calH_{j=1} \rightarrow \calH_{j_2} \otimes \calH_{j_2}$. Au final, nous obtenons en $t^*$ un symbole 6j de recouplage entre la s\'erie d'entrelacements pr\'ec\'edente et : $\calH_{j_1} \otimes \calH_{j_2} \rightarrow \calH_{i_{t^*}} \rightarrow \calH_{j_1} \otimes \calH_{j_2}$,
\beq
\bigl(\widehat{\cos\phi_{f_1f_2}^t}\bigr)_{j_1,j_2,i_t} = -\f{(-1)^{j_1+j_2+i_t}}{4}\,\sqrt{\f{d_{j_1}^3\,d_{j_2}^3}{j_1(j_1+1)j_2(j_2+1)}}\ \begin{Bmatrix} j_1 & j_1 & 1 \\ j_2 & j_2 & i_t\end{Bmatrix},\qquad j_1,j_2\,\in \N+\f{1}{2}.
\ee
Ce symbole 6j correspond pr\'ecis\'ement \`a la r\`egle du double grasping de \cite{speziale-grasping-rules} pour calculer les angles, et est explicitement connu, ce qui conduit \`a :
\beq \label{quantum angle}
\bigl(\widehat{\cos\phi_{f_1f_2}^t}\bigr)_{j_1,j_2,i_t}\ =\ N_{j_1j_2}\ \f{j_1(j_1+1)+j_2(j_2+1)-i_t(i_t+1)}{2\sqrt{j_1(j_1+1)\,j_2(j_2+1)}},
\ee
o\`u $N_{j_1j_2}$ est un coefficient born\'e, ind\'ependant de $i_{t}$, et \'equivalent \`a $1$ dans l'asymptotique :
\beq
N_{j_1j_2} = \f{(2j_1+1)\,(2j_2+1)}{4\sqrt{j_1(j_1+1)\,j_2(j_2+1)}}.
\ee
En fait, $N_{j_1j_2}$ est le rapport de deux des spectres les plus utilis\'es pour les observables d'aires \cite{alekseev-area-entropy}, $(j+\f12)$ et $\sqrt{j(j+1)}$, qui sont bien s\^ur \'equivalents dans la limite semi-classique. A la normalisation $N_{j_1j_2}$ pr\`es, c'est exactement la quantification du cosinus des angles donn\'ee par Major dans \cite{major-3dangles} dans le formalisme de la LQG (voir discussion \`a la section \ref{sec:operatorsLQG}), et d\'ej\`a propos\'ee par Penrose (voir les r\'ef\'erences dans \cite{major-3dangles}) dans son programme d'espace-temps combinatoire ! Etre capable de retrouver de telles expressions en partant des grandeurs purement classiques, quantifi\'ees par l'int\'egrale sur les g\'eom\'etries du syst\`eme, nous permet d'avoir confiance dans le formalisme d\'evelopp\'e et confirme son potentiel en tant que point de d\'epart pour la construction de mod\`eles de gravit\'e quantique.


\section{Action discr\`ete pour les nouveaux mod\`eles} \label{sec:action eprfk}

Nous utilisons maintenant l'ensemble des variables introduites \`a la section \ref{sec:gluing-aarc}, dans le but de regarder quels mod\`eles de mousses de spins \'emergent de l'int\'egrale sur les g\'eom\'etries exprim\'ees dans ces variables. Pour se donner une r\'ef\'erence, nous allons surtout pr\'esenter une action qui conduit au mod\`ele FK (et EPR, $\gamma<1$), et pr\'esenterons par la suite bri\`evement la classe de mod\`eles naturelle qui le contient.

Nos variables sont les suivantes : des holonomies $G_{vt}\in\Spin(4)$, les aires des triangles $A_f$, des rotations $n_{\pm ft}\in\SU(2)$, des variables angulaires sur chaque paire de t\'etra\`edres voisins $\theta^\pm_{fv}$ qui contiennent les angles dih\'edraux physiques correspondants. Pour impl\'ementer les contraintes de simplicit\'e, nous avons aussi les normales aux t\'etra\`edres $N_t\in\SU(2)\simeq S^3$, et des angles $\psi_{ft}$, qui peuvent par ailleurs \^etre fix\'es respectivement \`a l'identit\'e et $0$, ramenant l'invariance de jauge de $\Spin(4)$ au sous-groupe $\SU(2)$ diagonal.

Par rapport au jeu de variables standard pour d\'ecrire les g\'eom\'etries simplicielles \`a partir de la th\'eorie topologique, voir \cite{new-model-fk}, nous avons ajouter les variables angulaires $\theta_{fv}^\pm$ et $\psi_{ft}$. Les relations \`a imposer \emph{classiquement} entre toutes ces variables et leur sens g\'eom\'etrique, expos\'es en section \ref{sec:gluing-aarc}, seront repris plus bas. En particulier, toutes nos variables suppl\'ementaires sont, au niveau classique, reli\'ees aux holonomies et aux rotations $n_{\pm ft}$. L'avantage de notre approche sur les autres est donc de pouvoir consid\'erer des fonctions de ces variables suppl\'ementaires m\^eme \emph{off-shell} ! C'est ce qui sera d\'eterminant pour d\'eriver les nouveaux mod\`eles, car les relations classiques ne sont pas toutes impos\'ees fortement dans l'int\'egrale fonctionnelle sur la triangulation, au niveau quantique.

L'action que nous proposons contient deux parties jouant des r\^oles physiquement distincts :
\beq
S_\gamma = S_{{\rm R}\gamma}(A_f,\theta_{fv}^+,\theta_{fv}^-) + S_{s_+}^{\mathrm{CS}}\bigl(n_{+ft},g_{+vt},\theta^+_{fv}\bigr)  +S_{s_-}^{\mathrm{CS}}\bigl(n_{-ft},g_{-vt},\theta^-_{fv}\bigr).
\ee
L'action $S_{{\rm R}\gamma}$ est une action \`a la Regge, compactifi\'ee, et qui, pour la premi`ere fois \`a notre connaissance, incorpore le param\`etre d'Immirzi $\gamma$ :
\beq \label{regge-gamma action}
S_{{\rm R}\gamma} = - \sum_f A_f\,\sin\Biggl(\sum_{v\supset f}\f{\gamma_+}{2}\,\theta^+_{fv} + \f{\gamma_-}{2}\,\theta^-_{fv}\Biggr).
\ee
Notons tout de suite que c'est la seule partie de l'action qui d\'epend des aires des triangles, et de plus de mani\`ere lin\'eaire ($A_f$ est bien une variable ind\'ependante) !

Les quantit\'es $\gamma_\pm$ sont d\'efinies \`a partir de $\gamma$, et plusieurs choix sont possibles en s'inspirant de la litt\'erature, pour reproduire les diff\'erentes prescriptions \eqref{contraintes spin epr} ou \eqref{contraintes spin fk}. Nous consid\'ererons souvent le choix le plus na\"if, obtenu par identification des coefficients avec la d\'ecomposition self-duale/anti-self-duale de l'action de Holst, dans la limite continue na\"ive de $S_{{\rm R}\gamma}$,
\beq \label{naive presc}
\text{prescription na\"ive :}\qquad \qquad\gamma_\pm = \f{1\pm\gamma}{\gamma}\cdot
\ee
Par comparaison, la prescription de Freidel-Krasnov \eqref{contraintes spin fk} permet de d\'efinir sans souci le cas $\gamma=0$. Bien entendu, nos calculs s'appliquent \`a des choix quelconques de $\gamma_{\pm}$.

La deuxi\`eme partie de l'action est une action de collage des simplexes, qui vise \`a concentrer les amplitudes autour des configurations qui satisfont les relations de transport parall\`ele, \eqref{transport bivectors} ou \eqref{gluing spin4}. Le choix sp\'ecifique $S^{\rm CS}_s$ permet de reproduire le mod\`ele FK, mais d'autres choix sont envisageables. La notation \og CS\fg{} r\'ef\`ere \`a \og coherent states\fg, soit \'etats coh\'erents, car sa forme tr\`es sp\'eciale est la cl\'e pour reproduire la formulation du mod\`ele FK en termes d'\'etats coh\'erents. $S^{\mathrm{CS}}_s$ ne d\'epend pas explicitement du param\`etre d'Immirzi, mais des signes $s_\pm = \operatorname{sign}(\gamma_\pm)$, ind\'ependamment pour les parties anti/self-duales. Elle se construit en fait \`a l'aide de variables suppl\'ementaires, interpr\'et\'ees comme des multiplicateurs de Lagrange \`a valeurs discr\`etes, $j^\pm_{fv}\in\f\N2$,
\beq \label{FK gluing action}
S_s^{\mathrm{CS}}\bigl(n_{ft},g_{vt},\theta_{fv}\bigr) = -2i\sum_{(f,v)} j_{fv}\,\ln\,\tr \left(\f 12 \bigl(\mathbbm{1}+s\,\sigma_z\bigr)\,n_{ft}\mone\,g_{tt'}\,n_{ft'}\,e^{\f{i}{2}\theta_{fv}\sigma_z}\right).
\ee
Nous verrons que les secteurs g\'eom\'etriques et non-g\'eom\'etriques sont reli\'es par $\gamma_-\rightarrow -\gamma_-$. Il faut aussi garder \`a l'esprit que le choix de l'action de collage influe sur les points stationnaires, et $S^{\rm CS}_s$ est tr\`es sp\'ecifique de ce point de vue -- c'est notamment une quantit\'e complexe.

L'action propos\'ee $S_\gamma$ est invariante sous les transformations de jauge $\Spin(4)$ provenant du mod\`ele topologique, et les transformations $\U(1)\times\U(1)$ associ\'ees aux variables suppl\'ementaires. Ces derni\`eres viennent du fait que l'angle d'Euler $\gamma$ dans la d\'ecomposition $n = e^{-\f i2 \alpha\sigma_z}\,e^{-\f i2 \beta\sigma_y}\,e^{-\f i2 \gamma\sigma_z}$ des rotations $n_{\pm ft}$ n'est pas physique ; le r\^ole physique de ces rotations est de d\'eterminer les directions de 3-vecteurs d\'efinis par l'action adjointe (i.e. la repr\'esentation vectorielle) des rotations sur l'axe de r\'ef\'erence $\hat{z}$ :
\beq
\hat{b}_{\pm ft}\cdot \vec{\sigma} \,\equiv\, \pm \Ad\bigl(n_{\pm ft}\bigr)\,\sigma_z.
\ee
Ces transformations agissent par des familles $K = (K_t, K_v\in\Spin(4))$ et $\Lambda = ((\lambda^+_{ft}, \lambda^-_{ft})\in\mathfrak{u}(1)\oplus\mathfrak{u}(1))$ selon :
\begin{alignat}{3} \label{gauge transfo regge-gamma action}
&(K,\Lambda)\ \cdot\ G_{vt}& &=&\ &K_v\,G_{vt}\,K_t\mone,\\
&(K,\Lambda)\ \cdot\ N_{t}& &=&\ &k_{+t}^{\phantom{}}\,N_{t}^{\phantom{}}\,k_{-t}\mone,\\
&(K,\Lambda)\ \cdot\ n_{\pm ft}& &=&\ &k_{\pm t}\,n_{\pm ft}\,e^{\f{i}{2}\lambda_{ft}^\pm\sigma_z}, \\
&(K,\Lambda)\ \cdot\ \theta_{fv}^\pm& &=&\ &\theta_{fv}^\pm +\eps_{tt'}^f\bigl(\lambda_{ft}^\pm - \lambda_{ft'}^\pm\bigr), \\
&(K,\Lambda)\ \cdot\ \psi_{ft}& &=&\ &\psi_{ft}-\bigl(\lambda^+_{ft}-\lambda^-_{ft}\bigr).
\end{alignat}

\subsection{Action de transport parall\`ele pour le collage des simplexes}

La forme de l'action de collage $S^{\rm CS}_s$ permet donc de reproduire sp\'ecifiquement le mod\`ele FK et a des cons\'equences non-triviales sur les \'equations des points stationnaires de l'action compl\`ete. Avant de voir cela, nous regardons dans cette section en quoi elle permet de vraiment coller les simplexes.

Remarquons d'abord que $S^{\rm CS}_s$ a la m\^eme forme fonctionnelle que l'action propos\'ee dans \cite{freidel-conrady-pathint} par Conrady et Freidel, mais avec des variables diff\'erentes et surtout ici un v\'eritable sens g\'eom\'etrique ! Pour le voir, nous traitons momentan\'ement les variables $j^\pm_{fv}$ comme d'authentiques multiplicateurs de Lagrange, \`a savoir des variables r\'eelles. La stationnarit\'e de l'action par rapport \`a ces variables impose au logarithme complexe de \eqref{FK gluing action} de s'annuler pour chaque paire $(f, v)$,
\beq \label{stationarity spins}
\ln\,\tr \left(\f 12 \bigl(\mathbbm{1}+s\,\sigma_z\bigr)\,n_{ft}\mone\,g_{tt'}\,n_{ft'}\,e^{\f{i}{2}\theta_{fv}\sigma_z}\right) \,=\,0.
\ee
On param\`etre : $g=\cos\phi+i\sin\phi\,\hat{u}\cdot\vec{\sigma}\in\SU(2)$. Alors, la fonction qui nous int\'eresse est :
\beq \label{ln tr}
\ln\,\tr \left(\f 12 \bigl(\mathbbm{1}+s\,\sigma_z\bigr)\,g\right) = \ln \Bigl(\cos\phi + is\sin\phi\,u_z\Bigr),
\ee
o\`u $u_z$ est la projection de $\hat{u}\in S^2$ sur la direction $\hat{z} = (0,0,1)$. Elle s'annule uniquement pour $\phi=0$, soit $g=\mathbbm{1}$. Les \'equations du mouvement associ\'ees sont donc pr\'ecis\'ement \eqref{solveholspin4},
\beq \label{solveholspin4 bis}
g_{+tt'} = n_{+ft}\,e^{-\f{i}{2}\theta_{fv}^+\sigma_z}\,n_{+ft'}\mone,\qquad\text{et}\qquad g_{-tt'} = n_{-ft}\,e^{-\f{i}{2}\theta_{fv}^-\sigma_z}\,n_{-ft'}\mone,
\ee
pour des t\'etra\`edres adjacents coll\'es le long de $f$. Nous avons vu \`a la section \ref{sec:gluing-aarc} qu'il s'agit des relations de transport parall\`ele \'ecrites dans les nouvelles variables. En effet, il faut voir $n_{\pm ft}$ comme des rotations d\'efinissant les directions d'un bivecteur $\hat{b}_{+ft}\oplus \hat{b}_{-ft}\in\spin(4)$ en envoyant la direction $\hat{z}$ sur $\Ad(n_{\pm ft}) \hat{z} =\pm \hat{b}_{\pm ft}$. Les \'equations ci-dessus donnent alors :
\beq
\hat{b}_{+ft} \,=\, \Ad\bigl(g_{+tt'}\bigr)\,\hat{b}_{+ft'},\qquad\text{et}\qquad \hat{b}_{-ft} \,=\, \Ad\bigl(g_{-tt'}\bigr)\,\hat{b}_{-ft'}.
\ee

Une caract\'eristique de ce choix de l'action de collage est donc d'imposer une contrainte sur trois variables r\'eelles (car \eqref{solveholspin4 bis} est une \'equation sur $\SU(2)$) \`a partir d'une \'equation sur une seule variable r\'eelle \eqref{stationarity spins}. N\'eanmoins, les variables $j^\pm_{fv}$ ne sont pas r\'eelles mais seulement demi-enti\`eres. Il faut donc penser \`a l'action de collage comme une mesure de l'int\'egrale fonctionnelle qui concentre, avec une largeur finie, l'amplitude autour des configurations satisfaisant les r\`egles correctes de transport parall\`ele. En effet, dans l'int\'egrales de chemins, nous avons \`a faire \`a une somme sur chaque spin $j^\pm_{fv}$, qui se calcule exactement :
\begin{align} \label{gluing function}
\tl{\delta}_s^{\mathrm{cs}}(g) &\equiv \sum_{j\in\f{\N}{2}}e^{2j\,\ln\,(\tr\,\f 12(\mathbbm{1}+s\sigma_z)g)}, \\
&= \sum_{j\in\f{\N}{2}} \bra j,sj\rv\,g\,\lv j,s j\ket, \\
&= \f1{1-\cos\phi- is\,\sin\phi\,u_z},
\end{align}
pour $g=\cos\phi+i\sin\phi\,\hat{u}\cdot\vec{\sigma}$. Nous avons utilis\'e le fait que : $\bra j,\pm j\rv\,g\,\lv j,\pm j\ket = [\tr\,\f 12(\mathbbm{1}\pm\sigma_z)g]^{2j} = (\cos\phi\pm i\sin\phi\,u_z)^{2j}$. Cette fonction est bien s\^ur piqu\'ee autour de l'idenit\'e car elle diverge pour $\phi = 0$. Cette action de collage revient donc \`a consid\'erer l'int\'egrale de chemins pour les variables $G_{vt}, n_{\pm ft}, \theta_{fv}^\pm$ avec un poids $\tl{\delta}_s^{\mathrm{cs}}$ sur triangle de chaque 4-simplexe :
\beq \label{gluing measure fk}
\sum_{\{j^+_{fv},j^-_{fv}\}} e^{iS^{\mathrm{CS}}_{s_+}}\ e^{iS^{\mathrm{CS}}_{s_-}} =
\prod_{(f,v)}\,\tl{\delta}_{s_+}^{\mathrm{cs}}\Bigl(n_{+ft}\mone\,g_{+tt'}\,n_{+ft'}\,e^{\f i2 \theta_{fv}^+\sigma_z}\Bigr)\,\tl{\delta}_{s_-}^{\mathrm{cs}}\Bigl(n_{-ft}\mone\,g_{-tt'}\,n_{-ft'}\,e^{\f i2 \theta_{fv}^-\sigma_z}\Bigr).
\ee
La normalisation de l'amplitude finale des t\'etra\`edres peut \^etre modifi\'ee \`a cette \'etape, en introduisant des puissances des dimensions $d_{j^\pm_{fv}}$ dans la somme \eqref{gluing function}, ce qui pour des puissances positives revient \`a prendre des d\'eriv\'ees de $\tl{\delta}^{\mathrm{cs}}_s$.

\medskip

Les contraintes de simplicit\'e crois\'ee \eqref{simplicity directions} s'incorporent facilement \`a l'action car elles expriment directement disons $n_{-ft}$ en fonction des autres variables, \eqref{normale-phase},
\beq
n_{-ft} = N_t\mone\,n_{+ft}\ e^{\f{i}{2}\psi_{ft}\sigma_z}.
\ee
Nous n'aurons donc qu'\`a remplacer $n_{-ft}$ par cette expression partout. Nous savons (voir par exemple \cite{baez-barrett-quantum-tet, dittrich-ryan-simplicial-phase}) que cette contrainte avec la relation de fermeture \eqref{closure} permet de reconstruire la g\'eom\'etrie de chaque t\'etra\`edre ind\'ependamment. De plus, nous avons vu \`a la section \ref{sec:gluing-aarc} que cette contrainte avec la relation de transport parall\`ele \eqref{solveholspin4 bis} permet d'extraire les angles dih\'edraux 4d comme \'etant
\begin{align}
\cos\thet_{tt'} &= -\f12\,\tr\,\bigl(g_{+t_1 t_2}^{\phantom{}}\,N_{t_2}^{\phantom{}}\,g_{-t_1t_2}\mone\,N_{t_1}\mone\bigr),
&= -\cos\f{1}{2} \Bigl(\theta_{fv}^+-\theta_{fv}^-+\eps_{tt'}^f \bigl(\psi_{ft}-\psi_{ft'}\bigr)\Bigr),
\end{align}
fournissant l'interpr\'etation des angles $\theta^\pm_{fv}, \psi_{ft}$, mais aussi de montrer que ces angles sont d\'etermin\'es par la g\'eom\'etrie des t\'etra\`edres,
\beq
\cos\thet_{tt''} = \f{\cos\phi^{t'}_{f_1f_2}-\cos\phi_{f_1 f}^t\,\cos\phi_{ff_2}^{t''}}{\sin\phi_{f_1 f}^t\,\sin\phi_{ff_2}^{t''}},
\ee
pour $\cos\phi^t_{ff'} = \langle 1, 0\lv n_{+ft}\mone\,n_{+f't}\rv 1, 0\rangle$. De ce fait, nous avons \'egalement pu prouver que les g\'eom\'etries des t\'etra\`edres voisins sont bien consistentes !

\subsection{Action de Regge avec param\`etre d'Immirzi}

Munie de cette description g\'eom\'etrique, nous formons la quantit\'e suivante, qui est on-shell la somme des angles dih\'edraux :
\beq
\thet_f = \sum \thet_{tt'} = \f 12(\theta^+_f-\theta^-_f),\qquad \text{pour}\qquad \theta^\pm_f = \sum_{v\supset f}\theta^\pm_{fv},
\ee
et l'action de Regge compactifi\'ee correspondante :
\beq
S_{{\rm R}\gamma} = -\sum_f A_{f}\,\sin \f 14\Biggl(\bigl(\gamma_+-\gamma_-\bigr)\,2\thet_f + \bigl(\gamma_++\gamma_-\bigr)\bigl(\theta^+_f+\theta^-_f\bigr)\Biggr).
\ee
Avec la prescription na\"ive \eqref{naive presc} pour les coefficients $\gamma_\pm$, cela se r\'eduit \`a :
\beq
S_{{\rm R}\gamma} = -\sum_f A_{f}\,\sin \left(\thet_f + \f1{2\gamma}\bigl(\theta^+_f+\theta^-_f\bigr)\right).
\ee
Cette forme est particuli\`erement \'evocatrice. Tout d'abord, commentons la proposition pour $\gamma=\infty$. Nous pouvons alors introduire l'angle de d\'eficit, $\vareps_f = 2\pi - \thet_f$, et pour des petits angles de d\'eficit, cette action prend excatement la forme de l'action de Regge, $\sum_f A_f\,\vareps_f$. Insistons n\'eanmoins sur le fait qu'ici les aires et les angles sont des variables \emph{ind\'ependantes}, et qu'on ne s'attend \`a ce qu'elles soient fonctions d'une m\'etrique discr\`ete, i.e. des longueurs, uniquement on-shell !

Pour $\gamma$ fini, on avons \`a c\^ot\'e de l'angle de d\'eficit, form\'e par la diff\'erence de $\theta^+_f$ et $\theta^-_f$, un autre angle qui lui vient de la somme de $\thet^+_f$ et $\theta^-_f$, avec un facteur le param\`etre d'Immirzi. Une situation similaire a \'et\'e observ\'ee dans les \'etudes asymptotiques, \cite{freidel-conrady-semiclass, barrett-asymEPR}. Quand les scontraintes sont pleinement satisfaites, il se trouve que \cite{freidel-conrady-semiclass} : $\theta^+_f+\theta^-_f=2\pi n$ pour un entier $n$. Ainsi, l'action on-shell esst ind\'ependante de $(\gamma_++\gamma_-$ d\`es lors que ce dernier est un entier pair. C'est \'egalement une condition suffisante pour que l'asymptotique calcul\'e dans \cite{freidel-conrady-semiclass} ne soit pas nul. Cela rejoint ainsi le fait que les \'equations du mouvement sont ind\'ependantes de $\gamma$ dans la th\'eorie continue !

Lorsque les contraintes de recollement \eqref{solveholspin4 bis} tiennent, et en utilisant la simplicit\'e crois\'ee, on montre que l'action \`a la Regge compactifi\'ee prend la forme d'une action \`a la BF,
\beq \label{bf-like regge-gamma action}
S_{{\rm R}\gamma} = -\sum_f \tr\Bigl(b_{+ft}\ g_{+f}^{\gamma_+}(t)\,N_t^{\phantom{}}\,g_{-f}^{\gamma_-}(t)\,N_t\mone\Bigr), \qquad\text{on-shell.}
\ee
Pour une rotation $g=e^{\f i2\phi\hat{u}\cdot\vec{\sigma}}\in\SU(2)$ d'angle $\phi$ et de direction $\hat{u}$, l'\'el\'ement de groupe $g^\alpha$ est d\'efini comme ayant la m\^eme direction et un angle $\alpha\phi$. Pour chaque triangle, il faut un t\'etra\`edre de r\'ef\'erence, mais la trace ne d\'epend pas de ce choix gr\^ace au transport parall\`ele. L'action \eqref{bf-like regge-gamma action} est bien une discr\'etisation de la th\'eorie BF $\Spin(4)$ avec les contraintes de simplicit\'e. Prenons en effet un syst\`eme de coordonn\'ees dans lequel la longueur typique d'une ar\^ete est d'ordre $\eps$. Quand $\eps$ tend vers zero, nous pouvons utiliser le d\'eveloppement : $g_{\pm f}\approx 1+\epsilon^2 F_{\pm \vert f}$, o\`u $F_{\pm \vert f}$ est la composante de la courbure dans les directions de la face duale au triangle $f$. Ainsi : $g_{\pm f}^{\gamma_\pm}\approx 1+\epsilon^2 \gamma_\pm F_{\pm \vert f}$. Pour chaque triangle, l'action \eqref{bf-like regge-gamma action} se r\'eduit \`a :
\beq
\epsilon^2\,\gamma_+\tr\, \bigl(b_{+ft}\,F_{+\vert f}\bigr) - \epsilon^2\,\gamma_-\tr\, \bigl(b_{-ft}\,F_{-\vert f}\bigr),
\ee
o\`u nous avons utilis\'e : $b_{-ft}=-\Ad(N_t\mone)\, b_{+ft}$. C'est bien la limite continue attendue.

Dans la th\'eorie continue, il est possible de consid\'erer comme r\'ef\'erence aussi bien le secteur topologique que le secteur g\'eom\'etrique, les deux \'etant reli\'es par la transformation $\gamma \rightarrow \gamma\mone$. Regardons comment cela se fait aussi dans le formalisme discret. Nous d\'emarrons avec la prescription na\"ive \eqref{naive presc} pour $\gamma_\pm$ dans le secteur g\'eom\'etrique. Cela ne concerne que la partie \`a la Regge de notre action,
\beq
S_{\mathrm{R}\gamma} = - \sum_f A_f\,\sin\Biggl(\sum_{v\supset f}\f{\gamma_+}{2}\,\theta^+_{fv} + \f{\gamma_-}{2}\,\theta^-_{fv}\Biggr),\qquad \text{pour}\quad \gamma_\pm = \gamma\mone\pm 1.
\ee
L'ambiguit\'e de signes entre les deux secteurs appara\^it originellement dans la contrainte de simplicit\'e crois\'ee : $b_{-ft}=\mp \Ad(N_t\mone)\,b_{+ft}$, et nous cherchons \`a la traduire dans cette action. En observant \eqref{bf-like regge-gamma action} et sa limite continue, on remarque que la transformation revient \`a changer le signe de $\gamma_-$, et donc pour nous l'exposant de l'holonomie anti-self-duale : $g^{\gamma_-}_f\rightarrow g_f^{-\gamma_-}$. Changeons dans le m\^eme temps le param\`etre d'Immirzi en $\gamma\mone$. Cela donne l'action pour le secteur non-g\'eom\'etrique comme r\'ef\'erence,
\beq
S_{\mathrm{R,non-g},\gamma\mone} = - \sum_f A_f\,\sin\Biggl(\sum_{v\supset f}\f{1+\gamma}{2}\,\theta^+_{fv} + \f{1-\gamma}{2}\,\theta^-_{fv}\Biggr),
\ee
ce qui correspond pr\'ecis\'ement \`a la prescription EPR \eqref{contraintes spin epr} pour $\gamma<1$.

\subsection{Les g\'eom\'etries stationnaires} \label{sec:stationnary geom fk}

Nous regardons maintenant les \'equations du mouvement restantes. Nous nous attendons \`a trouver : la relation de fermeture, pour compl\'eter le jeu des contraintes de g\'eom\'etricit\'e, une \'equation exprimant les multiplicateurs de Lagrange $j^\pm_{fv}$ en fonction des autres variables (notamment les aires), et une \'equation d\'ecrivant la \emph{dynamique}. Avec l'interpr\'etation g\'eom\'etrique simplicielle d\'evelopp\'ee jusqu'\`a pr\'esent, on pourrait esp\'erer trouver \`a une \'equation \'equivalentes aux \'equations de Regge, approchant les \'equations d'Einstein sur la triangulation. En fait, nous obtenons plut\^ot une \'equation caract\'erisant des g\'eom\'etries simplicielles plates ! Bien s\^ur la question de savoir si ces g\'eom\'etries, singularis\'ees par l'\'etude des points stationnaires, sont pr\'epond\'erantes au niveau quantique reste une question ouverte~$\dotsc$

Les aires $A_f$ n'interviennent dans $S_\gamma$ que dans la partie \`a la Regge, et de mani\`ere lin\'eaire, ce qui donne l'\'equation de stationnarit\'e suivante,
\beq \label{area variation}
\sin\, \f 14\left(\bigl(\gamma_+-\gamma_-\bigr)\,2\theta_f + \bigl(\gamma_++\gamma_-\bigr)\bigl(\theta^+_f+\theta^-_f\bigr)\right) = 0,
\ee
qui est pour la dynamique l'analogue de l'\'equation de courbure (discr\`ete) nulle dans le mod\`ele topologique. De plus, cette \'equation est revient \`a extr\'emiser la version on-shell \eqref{bf-like regge-gamma action} par rapport aux variables $b_{+ft}$ :
\beq \label{area variation 2}
g_{+f}^{\gamma_+}(t)\,N_t\,g_{-f}^{\gamma_-}(t) \,=\, \pm N_t.
\ee
Comme nous le savons, dans le cas $\gamma=\infty$ (donc $\gamma_+ = -\gamma_- = 1$) et lorsque toutes les contraintes sont satisfaites (simplicit\'e crois\'ee, fermeture v\'erifi\'ee ci-dessous, transport parall\`ele), cela siginifie que la courbure physique, i.e. les angles de d\'eficit, est nulle (voir \eqref{zero deficithol}). Reste \`a ce niveau \`a d\'eterminer l'influence du param\`etre d'Immirzi.

Ensuite nous faisons varier les angles $\theta^\pm_{fv}$ qui couplent la partie \`a la Regge $S_{{\rm R}\gamma}$ et la partie collage $S^{\rm CS}$ :
\beq
\gamma_\pm\,A_f\ \cos\, \f 14\left(\bigl(\gamma_+-\gamma_-\bigr)\,2\theta_f + \bigl(\gamma_++\gamma_-\bigr)\bigl(\theta^+_f+\theta^-_f\bigr)\right) = 2s_\pm\,j^\pm_{fv},
\ee
pour chaque vertex dual $v$ sur le bord de la face duale $f$. Du fait de l'\'equatio \eqref{area variation}, le cosinus se r\'eduit \`a $\pm1$. Cela nous am\`ene au r\^ole des signes $s_\pm=\mathrm{sign}(\gamma_\pm)$. En effet, quand $\gamma$ devient plus grand que $1$, $\gamma_-$ devient n\'egatif. Mais toutes les quantit\'es de l'\'equation ci-dessus sont positives, except\'ee $s_-$ qui change \'egalement de signe. Finalement, le cosinus doit \^etre $+1$, ce qui correspond au signe $+$ dans \eqref{area variation 2}. Nous aboutissons \`a deux \'equations :
\beq \label{preflatness}
\cos\ \f 14\left(\bigl(\gamma_+-\gamma_-\bigr)\,2\theta_f + \bigl(\gamma_++\gamma_-\bigr)\bigl(\theta^+_f+\theta^-_f\bigr)\right) = 1,
\ee
et
\beq
\f{2}{\lv\gamma_+\rv}\ j^+_{fv} \,=\, \f{2}{\lv\gamma_-\rv}\ j^-_{fv} \,=\, A_f. \label{solve j}
\ee
Parmi les cons\'equences, les aires sont quantifi\'ees et prennent des valeurs rationnelles quand $\gamma_\pm$ sont entiers. Les spins sont les m\^emes sur tous les 4-simplexes dont le bord contient le triangle $f$, $j^\pm_{fv} = j^\pm_f$, et v\'erifient la contrainte de simplicit\'e diagonale \eqref{quantum diag}, avec $j_f$ remplac\'e par $A_f/2$ (qui n'est pas forc\'ement demi-entier).

La stationnarit\'e par rapport aux variables $n_{\pm ft}$ ne g\'en\`ere pas de nouvelles \'equations.

Il nous faut v\'erifier que les variations de l'action par rapport aux holonomies conduisent bien \`a la relation de fermeture \eqref{closure},
\beq
\sum_{f\subset t} \eps_{ft}\,b_{\pm ft} \,=\, 0, \qquad\text{pour}\qquad b_{\pm ft} \,=\, \pm A_f\ \Ad\bigl(n_{\pm ft}\bigr)\,\sigma_z,\label{closure bis}
\ee
pour les bivecteurs $B_{ft} = b_{+ft}\oplus b_{-ft}$ de chaque t\'etra\`edre, ce qui n'est g\'en\'eralement pas assur\'e pour une fonction de collage quelconque comme nous allons le discuter. C'est bien s\^ur vrai pour l'action BF sur r\'eseau \eqref{discreteBFaction}, mais la situation est quelque peu diff\'erente ici. L'action \`a la Regge $S_{{\rm R}\gamma}$ n'est du type BF que on-shell, et les holonomies interviennent en fait dans l'action de collage par transport parall\`ele. Remarquons que c'est gr\^ace \`a cette \'equation que les variables $A_f$ s'interpr\`etent bien comme les aires des triangles.

Pour v\'erifier que la relation de fermeture est effectivement produite lorsqu'on choisit le collage sp\'ecifique $S^{\rm CS}$, nous regardons l'effet de variations $\delta g_{+vt}$, qui sont telles que par la forme de Maurer-Cartan, $\xi_{+vt} = g_{+vt}\mone\,\delta g_{+vt}$ est un \'el\'ement de l'alg\`ebre $\su(2)$. En prenant la m\^eme orientation pour les quatre triangles du t\'etra\`edre $t$, il vient :
\beq \label{closure variation}
\delta S^{\mathrm{CS}}_{s_+} = -2i\sum_{f\subset t} j_{fv}^+ \f{\tr\bigl(\f 12 \bigl(\mathbbm{1}+s_+\sigma_z\bigr)\,n_{+ft'}\mone\,g_{+t't}\,\xi_{+vt}\,n_{+ft}\,e^{\f{i}{2}\theta^+_{fv}\sigma_z}\big)}{\tr\bigl(\f 12 \bigl(\mathbbm{1}+s_+\sigma_z\bigr)\,n_{+ft'}\mone\,g_{+t't}\,n_{+ft}\,e^{\f{i}{2}\theta^+_{fv}\sigma_z}\bigr)}.
\ee
On peut alors v\'erifier \cite{bf-aarc-val} que les \'equations du mouvement d\'eriv\'ees pr\'ec\'edemment simplifient cette \'equation en la relation de fermeture \eqref{closure bis}.

Nous voulons conna\^itre l'influence du choix de l'action de collage sur les \'equations des points stationnaires, et comme nous l'avons annonc\'e sur la relation de fermeture des bivecteurs \eqref{closure bis}. C'est-\`a-dire qu'\`a la place d'utiliser la mesure $\tl{\delta}^{\mathrm{cs}}$, \eqref{gluing measure fk}, nous voulons introduire une mesure diff\'erente $\tl{\delta}$ pour piquer l'int\'egrale sur les relations de transport parall\`ele. A titre d'illustration, et parce que c'est le choix le plus naturel, nous d\'etaillons le cas o\`u $\tl{\delta}(g) = \delta(g)$. Au niveau de l'action, cela signifie que nous imposons les relations de collage, \eqref{solveholspin4 bis}, \`a l'aide de multiplicateurs de Lagrange $x_{\pm fv}\in\su(2)$ pour chaque wedge,
\beq \label{strong gluing}
S_\gamma^{\mathrm{strong}} = S_{\mathrm{R}\gamma} + \sum_{(f,v)} \tr\Bigl(x_{+fv}\,g_{+tt'}\,n_{+ft'}\,e^{\f{i}{2}\theta^+_{fv}\sigma_z}\,n_{+ft}\mone\Bigr) + \tr\Bigl(x_{-fv}\,g_{-tt'}\,n_{-ft'}\,e^{\f{i}{2}\theta^-_{fv}\sigma_z}\,n_{-ft}\mone\Bigr).
\ee
Remarquons alors que chaque contrainte, qui est une relation sur $\SU(2)$ donc sur trois variables r\'eelles, est bien impos\'ee via les trois param\`etres r\'eels de chaque $x_{\pm fv}$, contrastant avec le fait que l'action de collage associ\'ee aux nouveaux mod\`eles, $S^{\rm CS}$, n'utilise qu'une variable r\'eelle, $j^\pm_{fv}$, par contrainte $\SU(2)$ !

Les variations de l'action ci-dessus par rapport aux holonomies $g_{\pm vt}$ conduisent naturellement \`a une relation de fermeture sur les t\'etra\`edres, mais pour les multiplicateurs,
\beq \label{fake closure}
\sum_{f\subset t} \eps_{ft}\,x_{\pm fv} =0.
\ee
Il faut donc regarder quel est le lien entre les multiplicateurs et les bivecteurs. Tout comme \eqref{solve j}, cela se fait par stationnarit\'e par rapport aux angles $\theta^\pm_{fv}$. Ces variations conduisent \`a projeter $x_{\pm fv}$ sur les directions $\hat{b}_{\pm ft}$ :
\beq \label{project multiplier}
A_f = \pm\f {i}{2\gamma_+}\,\tr\bigl( x_{+fv}\,\Ad\bigl(n_{+ft}\bigr)\,\sigma_z\bigr) = \pm\f {i}{2\gamma_-}\,\tr\bigl( x_{-fv}\,\Ad\bigl(n_{-ft}\bigr)\,\sigma_z\bigr),
\ee
de sorte que ces projections correspondent aux variables $A_f$. Mais les composantes de $x_{\pm fv}$ orthogonales \`a $\hat{b}_{\pm ft}$ restent libres, et il ne semble pas possible de les \'eliminer de \eqref{fake closure}. La diff\'erence cruciale avec $S^{\rm CS}$ se situe dans l'insertion du projecteur $\f 12(\mathbbm{1}+\sigma_z)$, qui permet d'aboutir \`a la relation de fermeture pour les bivecteurs \eqref{closure bis} (voir \cite{bf-aarc-val}).

Ainsi, pour une action de collage g\'en\'erique, on s'attend \`a ne pas pouvoir interpr\'eter les variables $A_f$ comme les aires des triangles ! Mais cela ne signifie pas qu'on ne puisse pas reconstuire de g\'eom\'etrie simplicielle, \`a la Regge, avec les variables $\hat{b}_{\pm ft}$. Comme nous l'avons vu en commentaire de l'\'equation \eqref{closure area-directions}, les quatre 3-vecteurs $\hat{b}_{+ft}$ d'un t\'etra\`edre satisfont toujours une relation de fermeture, pour laquelle les coefficients s'interpr\`etent comme les v\'eritables aires, \`a un scaling pr\`es.

Mais si les aires ne coincident g\'en\'eralement pas avec les variables $A_f$, quel est le sens de ces derni\`eres ? Puisque l'action est lin\'eaire en les $A_f$, on peut consid\'erer qu'elles visent simplement \`a imposer l'\'equation du mouvement correspondante, \eqref{area variation}. Et comme nous le d\'ecrivons maintenant cette \'equation nous dit que les points stationnaires sont d'authentiques g\'eom\'etries simplicielles plates. A ce titre, remarquons qu'imposer la relation de fermeture des bivecteurs \`a la main, i.e. explicitement, pourrait r\'esoudre simultan\'ement ces deux situations qui sont li\'ees : d'abord, permettre d'interpr\'eter les variables $A_f$ comme les aires. Puis comme toutes les contraintes seraient imposer directement, autoriser le changement de variables vers les longueurs des ar\^etes, restituant ainsi une dynamique proche des \'equations de Regge.

Regardons maintenant les \'equations de la dynamique. La discussion suivante ne d\'epend pas du fait que les variables $A_f$ s'interpr\`etent on-shell comme les aires ou non. Nous avons d\'ej\`a remarqu\'e que pour $\gamma=\infty$ l'\'equation \eqref{area variation 2} d\'ecrit des angles de d\'eficit nuls autour des triangles. Pour traiter le cas d'un param\`etre d'Immirzi fini, nous utilisons un r\'esultat de Conrady et Freidel \cite{freidel-conrady-semiclass} : lorsque toutes les contraintes de g\'eom\'etricit\'e sont satisfaites, l'angle $(\theta^+_f+\theta^-_f)$ en facteur du param\`etre d'Immirzi dans l'action $S_{{\rm R}\gamma}$ est un multiple de $2\pi$. Cela vient du fait que l'on peut assigner de mani\`ere consistente des 4-vecteurs aux ar\^etes de la triangulation, tels que leurs expressions dans diff\'erents r\'ef\'erentiels sont reli\'ees par transport parall\`ele, \`a des signes pr\`es. En cons\'equence, et pourvu que $(\gamma_++\gamma_-)$ soit un entier pair, la somme des angles dih\'edraux $\thet_f$ est :
\beq \label{flatness sf}
(\theta^+_f+\theta^-_f)=0,2\pi\ \mod (4\pi)\quad\Rightarrow\quad\thet_f = \f{4\pi p_f}{\gamma_+-\gamma_-},
\ee
pour $p_f\in\N$. De plus, avec la prescription na\"ive \eqref{naive presc}, $\gamma_+-\gamma_- =2$, cela se r\'eduit \`a :
\beq
\cos\sum_{(t,t')}\thet_{tt'} = 1,
\ee
dont la signification est que la courbure concentr\'ee autour des triangles est nulle. Notons n\'eanmoins que cela ne tient pas dans le secteur non-g\'eom\'etrique $\gamma=0$ ! 


\section{Calcul de l'int\'egrale de chemins sur la triangulation en mousses de spins} \label{sec:partition eprfk}

\subsection{Pour les nouveaux mod\`eles}

Le r\'esultat de cette section est que l'amplitude quantique associ\'ee \`a l'action discr\`ete d\'ecrite \`a la sectoin pr\'ec\'dente est pr\'ecis\'ement celle du mod\`ele FK (et donc aussi EPR pour $\gamma<1$). Soit $Z_\gamma$ d\'efini par :
\begin{multline} \label{path int}
Z_\gamma\equiv \int\prod_{(t,v)}dG_{vt}\,\prod_{(f,t)}dn_{+ft}\,dn_{-ft}\,d\psi_{ft} \prod_t dN_t\ \prod_{(f,t)}\delta\Bigl(n_{-ft}\mone\,N_t\mone\,n_{+ft}\,e^{\f i2 \psi_{ft}\sigma_z}\Bigr)\ \prod_f dA_f\quad e^{iS_{\mathrm{R}\gamma}(A_f,\theta^+_{fv},\theta^-_{fv})} \\
\times \prod_{(f,v)}\int d\theta^+_{fv}d\theta^-_{fv} \prod_{(f,v)} \tl{\delta}^{\mathrm{cs}}_{s_+}\Bigl(n_{+ft}\mone\,g_{+tt'}\,n_{+ft'}\,e^{\f i2\theta^+_{fv}\sigma_z}\Bigr)\,\tl{\delta}^{\mathrm{cs}}_{s_-}\Bigl(n_{-ft}\mone\,g_{-tt'}\,n_{-ft'}\,e^{\f i2\theta^-_{fv}\sigma_z}\Bigr).
\end{multline}
Alors la r\'e\'ecriture de toutes ces int\'egrales en langage \og mousses de spins\fg{} donne :
\beq
Z_\gamma = \sum_{\{j_f,k_{ft},l_t\in\f{\N}{2}\}}\prod_t W_t\bigl(\lv\gamma_+\rv j_f,\lv\gamma_-\rv j_f,k_{ft},l_t\bigr)\quad \prod_v W_v\bigl(\lv\gamma_+\rv j_f,\lv\gamma_-\rv j_f,k_{ft},l_t\bigr),
\ee
o\`u l'amplitude des t\'etra\`edres est donn\'ee par \eqref{FK weight} ou \eqref{gamma<1}, et l'amplitude des 4-simplexes par \eqref{new vertex}. La s\'erie de fonctions delta dans \eqref{path int} prend en compte la simplicit\'e crois\'ee, $S_{{\rm R}\gamma}$ est l'action \`a la Regge, compactifi\'ee, incluant le param\`etre d'Immirzi $\gamma$, et la mesure constituant la seconde ligne de \eqref{path int} concentre l'amplitude sur les configurations satisfaisant les relations de transport parall\`ele, n\'ecessaires \`a la consistence de la g\'eom\'etrie. Aussi, $s_\pm=\mathrm{sign}\gamma_\pm$. Nous donnons maintenant la preuve, expos\'ee initialement dans \cite{new-model-val}, et renvoyons \`a cette publication pour des discussions int\'eressantes sur les propri\'et\'es de l'amplitude notamment sous les changements d'orientations.

L'id\'ee cl\'e des mousses de spins est d'\'ecrire l'amplitude comme une fonction de partition, soit une somme sur les \'etats du syst\`eme caract\'eris\'e au niveau quantique par des donn\'ees au sens g\'eom\'etrique bien d\'efini. En pratique, de tels d\'eveloppements s'obtiennent par transform\'ees de Fourier sur les groupes de Lie consid\'er\'es. Nous d\'eveloppons donc d'abord l'exponentielle de $i$ fois l'action de Regge sur ses modes $\U(1)$. Par le d\'eveloppement de Jacobi-Anger, \eqref{bfaction u1modes}, d\'efinissant les fonctions de Bessel de premi\`ere esp\`ece $J_m$, nous avons pour chaque face :
\beq \label{expand regge-gamma}
e^{-iA_f\sin\f12(\gamma_+\theta^+_f+\gamma_-\theta^-_f)} = \sum_{m_f\in\f{\Z}{2}} J_{2m_f}(A)\, e^{-im_f(\gamma_+\theta^+_f+\gamma_-\theta^-_f)}.
\ee
Comme \`a la section \ref{sec:bf-regge calcul}, nous utilisons : $\int_{\R_+}dA\,J_m(A) = 1$, pour $m\in\N$ et : $J_{-m}(A) = J_m(-A) = (-1)^mJ_m(A)$, ce qui conduit \`a :
\beq \label{int area regge-gamma}
\int_{\R_+}dA_f\ e^{-iA_f\sin\f12(\gamma_+\theta^+_f+\gamma_-\theta^-_f)} = 1 + \sum_{m_f\in\f{\N*}{2}} \Bigl(e^{-i\gamma_+m_f\theta^+_f}\,e^{i\gamma_-m_f\theta^-_f} + (-1)^{2m_f}\,e^{i\gamma_+m_f\theta^+_f}\,e^{-i\gamma_-m_f\theta^-_f}\Bigr).
\ee
Notons que les modes conjugu\'es \`a $\theta^+$ et $\theta^-$ sont :
\beq \label{diag momenta}
m^\pm_f=\gamma_\pm m_f,
\ee
pour $m_f\in\N/2$. Ainsi, ils satisfont une relation de la forme de la contrainte de simplicit\'e diagonale \eqref{quantum diag} qui dans la quantification usuelle relie les spins self-duaux et anti-self-duaux de chaque triangle. Mais ici cette relation tient plus naturellement pour des repr\'esentations de $\U(1)$ qui vues dans $\SU(2)$ s'interpr\`etent comme des nombres magn\'etiques $m^\pm_f$. Cela est d\^u au d\'eveloppement de l'action de Regge compactifi\'ee, et est ind\'ependant du choix de l'action pour le transport parall\`ele.

Viennent ensuite les effets propres aux fonctions $\tl{\delta}^{\rm CS}$. Nous avons recours au d\'eveloppement suivant, sur les \'el\'ements de matrices des repr\'esentations de $\SU(2)$, de plus haut ou plus bas nombres magn\'etiques,
\beq \label{dev gluing action}
\sum_{\{j^+_{fv}\}}e^{iS^{\mathrm{CS}}(n_{+ft},g_{+vt},\theta^+_{fv})} = \prod_{(f,v)}\sum_{j^+_{fv}}\bra j^+_{fv},s_+\,j^+_{fv}\rv n_{+ft}\mone\,g_{+tt'}\,n_{+ft'}\lv j^+_{fv},s_+\,j^+_{fv}\ket\,e^{is_+j^+_{fv}\theta^+_{fv}}.
\ee
Pour un mode $m_f$ donn\'e, les int\'egrales sur les variables angulaires $\theta^\pm_{fv}$ s'effectuent selon :
\begin{multline} \label{theta int}
\prod_{v\supset f}\int d\theta^+_{fv}d\theta^-_{fv}\  e^{is_+j^+_{fv}\theta^+_{fv}+is_- j^-_{fv}\theta^-_{fv}} \Bigl(e^{-i\gamma_+m_f\theta^+_f+i\gamma_-m_f\theta^-_f} + (-1)^{2m_f}\,e^{i\gamma_+m_f\theta^+_f-i\gamma_-m_f\theta^-_f}\Bigr) \\= \prod_{v\supset f} \delta_{j^+_{fv},\lv\gamma_+\rv m_f}\ \delta_{j^-_{fv},\lv\gamma_-\rv m_f}.
\end{multline}
Cela contraint les spins dans le membre de droite de \eqref{dev gluing action} \`a la valeur : $j^\pm_{fv}=\lv\gamma_\pm\rv m_f$. C'est de cette fa\c{c}on que le choix sp\'ecifique $\tl{\delta}^{\rm cs}$ pour la fonction de recollement se manifeste : elle permet d'identifier les modes $\U(1)$ $m_f$ \`a des spins de $\SU(2)$ (\`a $\gamma_\pm$ pr\`es) ! Ainsi, la simplicit\'e diagonale sur les moments magn\'etiques \eqref{diag momenta} devient la relation usuelle sur les spins \eqref{quantum diag}.

Nous prenons la notation $j_f\equiv m_f$, et d\'efinissons : $j^\pm_f=\lv\gamma_\pm\rv j_f$. A ce stade, nous sommes arriv\'es \`a :
\begin{multline}
Z_\gamma = \sum_{\{j_f\in\f{\N}{2}\}} \int\prod_{(t,v)}dG_{vt}\,\prod_{(f,t)}dn_{+ft}\,dn_{-ft}\,d\psi_{ft} \prod_t dN_t\ \prod_{(f,t)}\delta\Bigl(n_{-ft}\mone\,N_t\mone\,n_{+ft}\,e^{\f i2 \psi_{ft}\sigma_z}\Bigr) \\
\times\prod_{(f,v)}\,\bra j^+_f,s_+ j^+_f\rv n_{+ft}\mone\,g_{+tt'}\,n_{+ft'}\lv j^+_f,s_+ j^+_f\ket\,\bra j^-_f,s_- j^-_f\rv n_{-ft}\mone\,g_{-tt'}\,n_{-ft'}\lv j^-_f,s_- j^-_f\ket,
\end{multline}
La contrainte de simplicit\'e crois\'ee s'\'elimine directement en ins\'erant partout o\`u n\'ecessaire : $n_{-ft}=N_t\mone\,n_{+ft}\,e^{\f i2 \psi_{ft}\sigma_z}$. Les rotations $N_t$ peuvent \^etre absorb\'ees \`a droite des \'el\'ements $g_{-vt}$ qui sont aussi int\'egr\'es, ou de mani\`ere \'equivalente, fix\'ees \`a l'identit\'e par invariance de jauge $\Spin(4)$. De m\^eme, les angles $\psi_{ft}$ disparaissent trivialement. Tout cela simplifie l'\'ecriture pr\'ec\'edente en :
\begin{multline} \label{Zgamma}
Z_\gamma = \sum_{\{j_f\in\f{\N}{2}\}} \int\prod_{(t,v)}dG_{vt}\,\prod_{(f,t)}dn_{ft}
\prod_{(f,v)}\,\bra j^+_f,s_+ j^+_f\rv n_{ft}\mone\,g_{+tt'}\,n_{ft'}\lv j^+_f,s_+ j^+_f\ket\\
\times \bra j^-_f,s_- j^-_f\rv n_{ft}\mone\,g_{-tt'}\,n_{ft'}\lv j^-_f,s_- j^-_f\ket.
\end{multline}
Le plus simple est alors d'introduire les \'etats coh\'erents de $\SU(2)$, pour aboutir directement aux expressions de la section \ref{sec:fk-quantisation} ! Dans chaque \'el\'ement de matrices ci-dessus, une rotation $n_{ft}$ agit sur un vecteur de moment magn\'etique bien d\'efini selon l'axe de r\'ef\'erence $\hat{z}$, $\lv j^\pm_f, s_\pm j^\pm_f\rangle$. Ainsi, on d\'efinit les \'etats coh\'erents par la direction $\hat{n}_{ft}\in S^2$ sur laquelle l'axe $\hat{z}$ est envoy\'e par la rotation $n_{ft}$ :
\beq
\lv j^\pm_f, \hat{n}_{ft}\rangle \equiv n_{ft}\,\lv j^\pm_f, j^\pm_f\rangle, \quad\text{\`a une phase pr\`es.}
\ee
L'ambiguit\'e de phase est celle dont nous avons pris soin en introduisant les angles $\theta^\pm_{fv}$ auparavant, de sorte qu'elle n'intervient pas d\`es lors que le t\'etra\`edre $t$ n'est pas au bord de la triangulation (donc qu'il est bien partag\'e par deux 4-simplexes) ! Il nous faut maintenant distinguer diff\'erentes situations selon la valeur du param\`etre d'Immirzi. Sans perte de g\'en\'eralit\'e, prenons $\gamma\geq 0$, de sorte que $s_+=1$. Alors, pour $\gamma<1$, nous avons $s_-=1$ et :
\beq
Z_{\gamma<1} = \sum_{\{j_f\in\f{\N}{2}\}} \int\prod_{(t,v)}dG_{vt}\,\prod_{(f,t)}dn_{ft}
\prod_{(f,v)}\,\bra j^+_f,\hat{n}_{ft}\rv\,g_{+vt}\mone\, g_{+vt'}\,\lv j^+_f,\hat{n}_{f't}\ket\ \bra j^-_f,\hat{n}_{ft}\rv\,g_{-vt}\mone\,g_{-vt'}\,\lv j^-_f,\hat{n}_{f't}\ket,
\ee
ce qui n'est autre que l'expression du mod\`ele EPR/FK pour $\gamma<1$, \eqref{vertexfk gamma<1} ! Le cas $\gamma>1$ produit un signe $s_-=-1$. On se ram\`ene aux \'etats coh\'erents par : $\bra j,-j\rv g \lv j,-j\ket=\overline{\bra j,j\rv g \lv j,j\ket}$, de sorte que :\beq
Z_{\gamma>1} = \sum_{\{j_f\in\f{\N}{2}\}} \int\prod_{(t,v)}dG_{vt}\,\prod_{(f,t)}dn_{ft}
\prod_{(f,v)}\,\bra j^+_f,\hat{n}_{ft}\rv\,g_{+vt}\mone\, g_{+vt'}\,\lv j^+_f,\hat{n}_{f't}\ket\ \overline{\bra j^-_f,\hat{n}_{ft}\rv\,g_{-vt}\mone\,g_{-vt'}\,\lv j^-_f,\hat{n}_{f't}\ket}.
\ee
C'est bien l'expression souhait\'ee, donn\'ee en \eqref{vertexfk gamma>1}. Je renvoie \`a \cite{freidel-conrady-pathint, new-model-val} pour des discussions des propri\'et\'es de ces amplitudes.

\subsection{Avec un collage g\'en\'erique}

Le formalisme d\'evelopp\'e \`a la section \ref{sec:gluing-aarc} et tout au long de ce chapitre permet de contr\^oler g\'eom\'etriquement et analytiquement la fa\c{c}on dont les diff\'erents ingr\'edients -- contraintes de simplicit\'e, relations de transport parall\`ele -- participent aux amplitudes de mousses de spins. En effet, \emph{tous} les ingr\'edients cl\'e de la g\'eom\'etrie de la triangulation sont pris en compte par des relations entre variables du groupe $\SU(2)$ ou $\U(1)$. Nous avons compris en particulier le r\^ole jou\'e par le choix sp\'ecifique de la fonction de collage $\tl{\delta}^{\rm cs}$ dans les nouveaux mod\`eles. Aux points stationnaires de l'int\'egrant, il permet d'interpr\'eter les variables $A_f$ comme les aires des triangles, et au niveau quantique, il semble que l'effet principal soit de pouvoir identifier les modes $\U(1)$ de l'action de Regge compactifi\'ee \`a des spins de $\SU(2)$ sur les triangles. Nous avons \`a la section \ref{sec:stationnary geom fk} discuter le r\^ole des variables $A_f$ classiquement, et de la reconstruction de la g\'eom\'etrie aux points stationnaires. Il nous reste donc \`a \'etudier l'influence d'un changement de la fonction de collage dans l'int\'egrale sur les g\'eom\'etries $Z_\gamma$. Pour voir les effets sur les ampitudes de mousses de spins, nous consid\'erons le d\'eveloppement d'une fonction de collage $\tl{\delta}$ sur $\SU(2)$,
\beq \label{generic gluing function}
\tl{\delta}(g) = \sum_{j\in\f{\N}{2}}\sum_{m=-j}^j c^j_m\ \bra j,m\rv\,g\,\lv j,m\ket.
\ee
Nous restreignons le d\'eveloppement sur les \'el\'ements de matrices diagonaux, pour assurer l'invariance sous les transformations \eqref{gauge transfo regge-gamma action}. Donnons quelques exemples typiques :
\begin{alignat}{2}
&\tl{\delta}(g) =\tl{\delta}^{\mathrm{cs}}_{s=\pm 1} &\qquad& c^j_m = \delta_{m,sj}, \\
&\tl{\delta}(g) = \delta(g) && c^j_m = (2j+1), \\
&\tl{\delta}(g) = \f{1}{N}e^{\f{1}{2\eps}\tr^2(g\,\vec{\sigma})} && c^j_m(\epsilon) = \f{1}{N'}\left(I_j(\epsilon\mone) - I_{j+1}(\epsilon\mone)\right),\\
&\tl{\delta}(g) = K_\eps(g) & & c^j_m (\eps) = (2j+1)\,e^{-4 j(j+1)\eps}.
\end{alignat}
A la troisi\`eme ligne, il s'agit d'une sorte de gaussienne sur le groupe, de largeur $\eps$, celle utilis\'ee dans les calculs semi-classiques de la section \ref{sec:3dgraviton} ($N, N'$ sont des normalisations, et $I_j$ est une fonction de Bessel modifi\'ee, d\'efinie par $I_j(z) = i^j J_j(-iz)$). A la quatri\`eme ligne nous proposons le noyau de la chaleur sur $\SU(2)$, solution de l'\'equation : $(\pp_\eps -\Delta)K_\eps(g) = 0$, avec la condition initiale $\lim_{\eps\rightarrow0}K_\eps(g) = \delta(g)$.

\medskip

A chaque vertex dual de chaque face, nous pouvons alors remplacer dans la d\'efinition de $Z_{\gamma}$, \eqref{path int}, les fonctions de collage $\tl{\delta}^{\rm cs}$ par l'expression g\'en\'erique \eqref{generic gluing function} :
\begin{multline}
\prod_{(f,v)}\ \tl{\delta}\bigl(n_{ft}\mone\,g_{+tt'}\,n_{ft'}\,e^{\f i2\theta^+_{fv}\sigma_z}\bigr)\ \tl{\delta}\bigl(n_{ft}\mone\,g_{-tt'}\,n_{ft'}\,e^{\f i2\theta^-_{fv}\sigma_z}\bigr) 
= \prod_{(f,v)} \sum_{j^+_{fv},j^-_{fv}}\sum_{m^+_{fv},m^-_{fv}} \ e^{im^+_{fv}\theta^+_{fv}}\,e^{im^-_{fv}\theta^-_{fv}}\\
c^{j^+_{fv}}_{m^+_{fv}}\,c^{j^+_{fv}}_{m^+_{fv}}\ \bra j^+_{fv},m^+_{fv}\rv\,n_{ft}\mone\,g_{+tt'}\,n_{ft}\,\lv j^+_{fv},m^+_{fv}\ket\,\bra j^-_{fv},m^-_{fv}\rv\,n_{ft}\mone\,g_{-tt'}\,n_{ft}\,\lv j^-_{fv},m^-_{fv}\ket,
\end{multline}
l'action \`a la Regge, $S_{{\rm R}\gamma}$, \'etant bien s\^ur inchang\'ee. L'exponentielle de cette derni\`ere est toujours d\'evelopp\'ee sur ses modes $\U(1)$, $m_f$. Puisque la fonction de collage ne s\'electionne plus uniquement les \'el\'ements de matrices des repr\'esentations de plus hauts ou plus bas nombres magn\'etiques, l'\'equation pour les int\'egrales sur les angles \eqref{theta int} doit\^etre \'ecrite en rempla\c{c}ant $s_\pm j^\pm_{fv}$ par un nombre magn\'etique $m^\pm_{fv}$. Celui-ci est ainsi contraint \^etre : $m^\pm_{fv}=\gamma_\pm m_f$, ou l'oppos\'e. Quant aux spins issus du d\'eveloppement \eqref{generic gluing function}, ils restent libres, et doivent \^etre somm\'es de mani\`ere ind\'ependante pour chaque wedge $(f,v)$ ! Introduisons donc :
\beq \label{R_m}
R_{\gamma_\pm m_f}\bigl(g_{\pm vt},n_{ft}\bigr) = \prod_{v\supset f} \sum_{j^\pm_{fv}\geq\lv\gamma_\pm m_f\rv} c^{j^\pm_{fv}}_{\gamma_\pm m_f}\,\bra j^\pm_{fv},\gamma_\pm j_f\rv n_{ft}\mone\,g_{\pm tt'}\,n_{ft'}\,\lv j^\pm_{fv},\gamma_\pm m_f\ket.
\ee
La fonction de partition devient :
\begin{multline}
Z_\gamma = \int\prod_{(t,v)}dG_{vt}\,\prod_{(f,t)}dn_{ft} \prod_f \Bigl[R_0\bigl(g_{+ vt},n_{ft}\bigr)\,R_{0}\bigl(g_{- vt},n_{ft}\bigr) \\
+\sum_{m_f\in\f{\N^\star}{2}} \Bigl(R_{\gamma_+ m_f}\bigl(g_{+ vt},n_{ft}\bigr)\,R_{\gamma_- m_f}\bigl(g_{- vt},n_{ft}\bigr) + (-1)^{2m_f}R_{-\gamma_+ m_f}\bigl(g_{+ vt},n_{ft}\bigr)\,R_{-\gamma_- m_f}\bigl(g_{- vt},n_{ft}\bigr)\Bigr)\Bigr],
\end{multline}
qui se r\'eduit bien \`a \eqref{Zgamma} lorsque $c^j_m=\delta_{sj,m}$. Supposons maintenant la sym\'etrie : $c^j_{-m}=c^j_m$ (comme c'est le cas pour : $\tl{\delta}=\tl{\delta}^{\mathrm{cs}}_{s=1}+\tl{\delta}^{\mathrm{cs}}_{s=-1}$). En observant l'effet de la transformation $n_{ft}\arr n_{ft}\eps$, pour la matrice $\eps = i\sigma_y = \bigl(\begin{smallmatrix} 0&-1\\1 &0\end{smallmatrix}\bigr)$ d\'ej\`a utilis\'ee \`a la section \ref{sec:fk-quantisation}, on peut voir que seuls les $m_f$ entiers contribuent. Nous pouvons \'ecrire de la sorte :
\beq
Z_\gamma = \sum_{\{m_f\in\Z\}} \int\prod_{(t,v)}dG_{vt}\,\prod_{(f,t)}dn_{ft} \prod_f R_{\gamma_+ m_f}\bigl(g_{+ vt},n_{ft}\bigr)\,R_{\gamma_- m_f}\bigl(g_{- vt},n_{ft}\bigr).
\ee

La repr\'esentation de l'amplitude en mousses de spins provient du d\'eveloppement des derni\`eres int\'egrales, sur les holonomies $G_{vt}$ et les rotations $n_{ft}$, en sommes sur des entrelaceurs. Les donn\'ees de bord d'un 4-simplexe sont les m\^emes que dans les mod\`eles EPR/FK, except\'e le remplacement des spins $j_f$ en nombres magn\'etiques $m_f\in\Z$. Nous d\'efinissons sur le mod\`ele des contraintes de simplicit\'e diagonale, et tout comme en \eqref{diag momenta} : $m^\pm_f=\gamma_\pm m_f$. L'amplitude du 4-simplexe est donn\'ee par la somme sur les dix spins des triangles de l'amplitude du mod\`ele FK, $W_v$, \eqref{new vertex}, pond\'er\'ee par des coefficients d\'ependant de $m_f$ :
\begin{multline}
\tl{W}_v\bigl(m_f,k_{ft},l_t\bigr) = \sum_{J^\pm_f \geq \lv\gamma_\pm m_f\rv} W_v\bigl(J^+_f,J^-_f,k_{ft},l_t\bigr) \prod_f\Biggl[c^{J^+_f}_{\gamma_+m_f}\,c^{J^-_f}_{\gamma_-m_f}\\ 
\times \prod_t \begin{pmatrix} J^+_f & J^-_f & k_{ft} \\ \gamma_+m_f & \gamma_-m_f & -(\gamma_++\gamma_-)m_f \end{pmatrix} \Biggr].
\end{multline}
Ce n'est certainement pas une surprise de retrouver l'amplitude $W_v$ dans cette expression. En effet, sa principale caract\'eristique r\'eside dans les coeffficients de fusion, \ref{fusion}, dont on sait qu'ils apparaissent naturellement en pr\'esence des contraintes de simplicit\'e crois\'ee, voir section \ref{sec:fusion coeff BC}.

On peut r\'esumer ainsi la principale diff\'erence entre un mod\`ele g\'en\'erique issu de la classe de mod\`eles ici expos\'es et les mod\`eles EPR/FK qui en sont des exemples particuliers : le collage entre les 4-simplexes impliquent des donn\'ees de bord l\'eg\`erement diff\'erentes. En effet, une fois l'amplitude d'un 4-simplexe d\'etermin\'ee dans les quantifications EPR/FK, il faut se donner une prescription pour coller ces amplitudes entre les 4-simplexes de toute la triangulation. L'id\'ee physique derri\`ere le collage est que l'aire d'un triangle est la m\^eme dans tous les 4-simplexes qui le partagent. Donc si les aires sont repr\'esent\'ees/ont pour valeurs propres, au niveau quantique, des spins de $\SU(2)$ (en accord avec la LQG au niveau canonique), cela conduit aux mod\`eles EPR/FK. Mais, dans notre approche, les int\'egrales de chemins, sur les g\'eom\'etries discr\`etes de la triangulation, il semble que les informations sur les aires sont plut\^ot contenues dans des nombres magn\'etiques, les valeurs propres de la projection du spin sur un axe. Le poids relatif du passage des spins aux nombres magn\'etiques est donn\'e par les coefficients $c^j_m$ du d\'eveloppement de la fonction de collage \eqref{generic gluing function}. Dans cette perspective, les mod\`eles EPR/FK sont parmi les mod\`eles les plus simples de notre famille de mod\`eles, caract\'eris\'es par le choix $c^j_m=\delta_{j,\lv m\rv}$, qui s\'electionne les plus bas spins $J^\pm_f$ de la somme sur $J^\pm_f\geq \lv\gamma_\pm m_f\rv$ dans l'expression de $\tl{W}_v$.



{\renewcommand{\thechapter}{}\renewcommand{\chaptername}{}
\addtocounter{chapter}{-1}
\chapter{Conclusion}\markboth{Conclusion}{Conclusion}}


Loop quantum gravity (LQG) has been developped as a rigourous non-perturbative quantization of general relativity, yet to be completed. But it has also provided us with more than LQG: a fully background independent framework to study diffeomorphism invariant gauge theories ! At the canonical level, one has the following picture: fundamental excitations of the space of quantum connections are supported by graphs colored by some data coming from the representation theory of the structure group. They form a basis of kinematical quantum states, called spin network states, which are gauge invariant and can be made diffeomorphism invariant. A set of well-defined geometric operators acts on these states, which in the LQG context correspond to length, area, volume and angle operators with discrete spectra, thus giving a nice interpretation of spin networks as describing quantum 3-geometry.

The dynamics of spin network states has led to the spin foam formalism, which describes their evolution along a 2-complex joining spin network graphs. Thus, it is supposed to yield the analoguous 4-dimensional picture of quantum geometry, and to implement the full 4-dimensional diffeomorphism symmetry. The spin foam quantization is known to be a success for Schwarz-type BF theories which are typically closely related to 2d Yang-Mills theory and 3d gravity. In particular it gives the appropriate integration measure so that their partition functions are topological invariant. In four dimensions, the $\SO(4)$ Plebanski action gives geometricity constraints which can be added to the BF topological action to recover the local degrees of freedom of general relativity. This relation has been exploited to build spin foam models for quantum gravity from the spin foam quantization of the BF theory.

Both spin networks and spin foams are discrete structures emerging from the full, exact theory, and the former describe quantum 3-geometry. A natural question is: what kind of discrete 4-geometries do spin foams describe ? Can we describe them explicitly, taking inspiration from Regge calculus, a nice and well-known approximation of general relativity, and further exhibiting relations to the Regge geometries ? Then, how does the dynamics proposed by spin foam models to spin network states can be expressed in terms of those 4d discrete geometries ?

To answer such questions, I developed some \emph{tools}, especially for BF and Plebanski theories, but which hopefully may be useful to generic spin foam quantization.
\begin{itemize}
 \item A formalism, equivalent to the standard discretization of BF theory, and in which the geometricity constraints can be added, which enables to extract in a clear way the geometric information encoded into discrete BF and Plebanski theories, at the classical level.
 \item A quantum implementation of this formalism. It is designed to exactly rewrite spin foam models as path integrals for systems of discrete geometries on a triangulation of spacetime. This involves the choice of a classical action to reproduce a given spin foam model. This action can then be studied, and in particular its stationnary configurations are expected to be related to some classical formulation of BF-like or Plebanski-like theories.
 \end{itemize}
These items focus on spin foam geometry and consist in a programme to see spin foam models as path integrals for some discrete geometries. A second programme aims at studying algebraic equations of spin foam amplitudes, through:
\begin{itemize}
 \item The use of recurrence relations for spin foam amplitudes as a way to encode the dynamics, or more generally the symmetries at the quantum level. Using the geometric interpretation of the spin foam colorings, a recurrence relation means invariance under some deformations of simplices. For the topological BF models, we propose to derive recurrence relations from the invariance under Pachner moves, or from the action of holonomy operators. In some cases, use can be made of integral representations of the amplitudes.
 \item The formalism of the first item above can be adapted to the canonical framework, to describe more geometrically spin network states on a given graph, as has been done by other authors, \cite{freidel-speziale-twisted-geom}. Then, one can hope to recast the Hamiltonian constraint in geometric terms, typically in a form relating extrinsic to intrinsic curvatures, and finally to fill the gap between the canonical and the covariant frameworks by identifying some recurrence relations on spin foams as a quantization of the Hamiltonian constraint on spin networks.
 \end{itemize}
I have also presented some basic ideas about how to compute correlation functions via spin foams, and have mainly illustrated what can be actually done using a toy model in 3d gravity. The method I have used gives an original access to the semi-classical asymptotics, including in a natural way all corrections to the Ponzano-Regge asymptotics of the 6j symbol. This enables to discuss the choice of the boundary states, and it also gives an example where it is possible to compute to any desired order in the large spin limit.

 \bigskip 
 
I applied these methods to a number of situations, with the following results about specific theories and models. As for the last method, we have obtained the full asymptotics expansion for the correlations of the fluctuations between the lengths of two curves, in the 3d toy model, which is the first time that computations are made beyond the first order (but the interpretation is not straightforward). As a side-product of the analysis, we have got the full asymptotics expansion of an isoceles 6j symbol, thus showing that the Regge action is the only relevant frequency, and making explicit the pattern of oscillations for each order. Such computations are important to show how spin foams go beyond quantum Regge calculus.

Using the discrete path integral programme, I have shown that: 
\begin{itemize}
 \item The classical framework used to set up spin foams involves genuine Regge geometries, and is more closely related to its area-3d angles formulation. In doing so, a dictionary between variables coming from lattice BF theory and those of Regge calculus was derived.
 \item I have also found a discrete action of (compactified) Regge form which includes the Immirzi parameter.
 \item It has been used to reproduce the new spin foam models from path integrals over variables on a triangulation. It is important to notice that the Regge-like action has to be supplemented with an action for parallel transport, crucial so that the stationnary points are Regge geometries.
 \item Using the very same ideas with a generalisation of the action for parallel transport, one arrives at a set of spin foam models, containing the EPR/FK model as one of them, sharing the same geometric properties but with different quantum fluctuations. In the spin foam language, this shows up in terms of different gluings between the basic building amplitudes attached to simplices.
\end{itemize}
This work thus makes explicit the similarities and differences between spin foams on a single triangulation and Regge calculus in any regime (not only in the asymptotics) ! In an analogous way to the continuum, the Immirzi parameter disappears on-shell, which clarifies its role within the spin foam formalism. Applying the same ideas to a derivation of the old-fashioned Barrett-Crane model, it turns out that the parallel transport relations are not correctly reproduced: one can understand this way how the new model corrects this BC model.

It is important to see that the different ways so far proposed to derive spin foam models generically lead different models, which certainly means different views on quantum geometry. Typically, the EPR quantization process gives to the BC model an equivalent in the topological, or non-geometric sector of the Holst action, that is for $\gamma\rightarrow0$, which is the initial EPR model, while for the FK quantization, the BC model does not admit an equivalent in this sector. Using our path integral formalism for discrete geometries, we here concluded that a natural equivalent to the BC model is an Ooguri-like model restricted to simple representations of $\Spin(4)$ on the triangles, giving a third perspective. Similarly, the EPR and FK models, despite their agreement in the $\vert\gamma\vert<1$ case, do differ for a larger Immirzi parameter, when $\vert\gamma\vert>1$. At the end of the day, it is hard to tell which quantization is better, as for instance we do not know how to get a spin foam model for the purely topological sector $\gamma\rightarrow0$ which would really be invariant under changes of the triangulation.

As a first step towards understanding the dynamics implemented by spin foams, the intimate relation found between the spin foam geometry and Regge calculus enables to identify the stationnary points as flat geometries. It is a preliminary result towards a better account for dynamics, which shows the direction to further proceed, and more precisely that flat configurations ask for a specific attention and may be in need for a regularisation. Indeed, we have seen in section \ref{sec:resume models} that the EPR/FK model can be written with integrals instead of sums. In this representation, the sum over each spin on the faces is performed using geometric series, to get \eqref{sum spins epr},
\begin{align}
A_f^{\gamma<1}(G_{vt}, \hat{n}_{ft}) &= \sum_{j\in\f{\N}{2}} \prod_{v\in\pp f}\,\bra j^+,\hat{n}_{ft}\rv\,g_{+vt}\mone\, g_{+vt'}\,\lv j^+,\hat{n}_{ft'}\ket\ \bra j^-,\hat{n}_{ft}\rv\,g_{-vt}\mone\,g_{-vt'}\,\lv j^-,\hat{n}_{ft'}\ket,\\
&= \f1{1-\exp \bigl(\sum_{v\in\pp f} \gamma_+ h(\hat{n}_{ft}, \hat{n}_{ft},g_{+vt}\mone\, g_{+vt'}) + \gamma_- h(\hat{n}_{ft}, \hat{n}_{ft},g_{-vt}\mone\, g_{-vt'})\bigr)}.
\end{align}
Such an amplitude is well-defined as a function everywhere except when the argument of the exponential vanishes. This is in contrast to what happens in the BF model with similar holonomy variables : the amplitude for a face is a discrete version of the flatness constraint, $\delta (\prod_{v\in \pp f} g_{tt'})$. It is thus not defined as a function but as a distribution, which is fine, while that of the new model is not ! When the exponential is 1, it is not clear at all how to interpret the formula and avoid the divergence.

Interestingly, this happens on flat geometries, so that our classical analysis of the action gives the first fruits relative to this phenomenon ! This can be seen by observing first that the norm of the hermitian product of coherent states is always smaller than unity and, $\lvert \langle j, \hat{n}\vert j, \hat{n}'\rangle\rvert =1$ if and only if $\hat{n} = \hat{n}'$. This shows that the parallel transport equations: $\hat{n}_{ft} = \Ad(g_{\pm tt'})\hat{n}_{ft'}$, hold in the divergent configurations. Then, one can introduce phases like in \eqref{solveholspin4 bis}, so that:
\beq
\sum_{v\in\pp f} \gamma_+ h(\hat{n}_{ft}, \hat{n}_{ft},g_{+vt}\mone\, g_{+vt'}) + \gamma_- h(\hat{n}_{ft}, \hat{n}_{ft},g_{-vt}\mone\, g_{-vt'}) = \bigl(\gamma_+-\gamma_-\bigr)\,2\thet_f + \bigl(\gamma_++\gamma_-\bigr)\bigl(\theta^+_f+\theta^-_f\bigr).
\ee
Here the quantity $\thet_f$ is the sum of the dihedral angles around $f$ determined, in non-degenerate configurations, by the triangle directions $\hat{n}_{ft}$ and so functions of the 3d angles, while $(\theta^+_f+\theta^-_f)$ is expected to disappear, all like in section \ref{sec:action eprfk}. Thus, this briefly shows how flat geometries ask for further regularisation of the model after areas/spins have been integrated out.

\medskip 

As for my second programme, that consisting in describing quantum symmetries through recurrence relations, nice results emerged, especially in topological models.
\begin{itemize}
 \item In the 3d case, I found that the flatness constraint geometrically expresses the extrinsic curvature, i.e 3d angles, in terms of the intrinsic geometry, i.e 2d angles.
 \item A naive quantization of this constraint on the boundary of a tetrahedron reproduces the asymptotics of a well-known recurrence relation for the corresponding Ponzano-Regge amplitude. As far as I know, this is the most direct link between the classical Hamiltonian constraint and the spin foam quantization.
 \item A detailed analysis of a Pachner move in 4d, leading to new recurrence relations, on the 15j-symbols which are the basic building blocks of the Ooguri model.
 \item These recurrence relations more generally apply to any evaluation of spin networks, depending on their cycles.
 \item In the 3d context, we have shown that the holonomy operator acting on spin network states generates tent moves, and associated recurrence relations for the Ponzano-Regge model. This also shows how non-trivial gauge-fixing can be involved in getting recurrence relations, and interpreted as fixing the tent pole length.
\end{itemize}
These last items are important since it is commonly believed that the evolution of spin network states, and hence diffeomorphisms at the discrete level, can be described in terms of elementary deformations, typically generated by Pachner moves or tent moves. Such deformations are also to be considered in the process of renormalisation of spin foams, that is evaluating different foams between given spin network states. In particular, upon using Pachner moves, we interpreted the recurrence relations as a statement of invariance under elementary displacements of vertices of simplices. In non-topological situations, some recurrence relations have been found, which capture some important classical features of simplices, and may be understood in the simplest case as invariance under a \emph{combination} of the above elementary deformations. To go beyond, one should try to find new recurrence relations, in particular for the more realistic, new EPR/FK model, in combination with tent move processes for instance.

\bigskip

To conclude this manuscript, and my doctoral work, let me give some overview of possible future developments of loop quantum gravity and spin foams. If the new spin foam models have given a clear kinematical picture of the involved geometries, many aspects remain to be understood, in particular concerning the dynamics. As for that of the BF model, the key point is that we know how to perform the sums over spins. Thus, an interesting direction is to do the same with the new models, along the line of equation \eqref{sum spins epr}. Such efforts seem necessary to go beyond the semi-classical analysis of a single simplex. I think that the recent proposal of spin foam cosmology \cite{vidotto-sfcosmo} gives a good starting point, with a well-posed physical context, to address the question with a larger scope: what are the relevant approximations, necessary to evaluate spin foam amplitudes, and what are the corresponding regimes ?

Since the new models have brought the spin foam formalism closer to LQG, it is certainly time to come back to working on the Hamiltonian constraint on spin network states, trying to fill the gap with spin foams like in \cite{alesci-rovelli-hamiltonian} or using ansatz for some definite physical situations. The beautiful picture of quantum geometry so far obtained could also be improved by progress on the realization of symmetries in lattice gravity, along the lines of the work of Dittrich and collaborators, which could then be translated into the spin foam language via recurrence relations. The field has recently received attention to find new parametrisations, like that of projected spin networks \cite{livine-projected}, twisted geometries \cite{freidel-speziale-twisted-geom}, a flux representation for LQG \cite{baratin-oriti-flux-ncrep}, and also holomorphic \cite{conrady-quantum-tet} or $\U(N)$ \cite{freidel-livine-un} intertwiners, and holomorphic coherent states \cite{bahr-thiemann-cs, bianchi-holomorphic-spinfoam}. Certainly, each of them comes with new algebraic and analytical features and can bring advantages for some issues.

Finally, the direct challenge to come is to study the sum over spin foams, which is supposed to implement the Hamiltonian constraint, to be necessary to recover the degrees of freedom of the full theory, to make spacetime topology an emergent property, for examples. The framework of group field theory (GFT) enables to see spin foam amplitudes as Feynman graphs of a non-local field theory. Thus, the key question is that of its renormalizability, which implies to understand in a precise way the divergences. Notice that it gives to the elementary deformations, like Pachner moves, used to provide evolution schemes the flavour of a coarse-graining process, or a renormalisation group implementation. In this respect, I hope that some recurrence relations may be applied to the GFT context, and that coarse-graining can be translated into a renormalisation flow for the discrete actions we have exhibited in the lattice path integral programme. In agreement with the line of reasoning of my present work, I think taking the GFT scenario seriously implies that further tools should be developed, and typically tested on the topological models, like those presented in \cite{cell-homology}.


\appendix
\chapter{Appendice} \label{sec:app}


Les \'el\'ements de $G=\SU(2)$ peuvent \^etre param\'etr\'es de la fa\c{c}on suivante :
\beq
g=\cos\psi\,\unit+i\sin\psi\,\hat{n}\cdot\vec{\sigma},
\ee
o\`u $\vec{\sigma} =(\sigma_x, \sigma_y, \sigma_z)$ est le 3-vecteur form\'e par les matrices de Pauli,
\beq
\sigma_x =\begin{pmatrix} 0 & 1\\ 1& 0\end{pmatrix},\qquad \sigma_y = \begin{pmatrix} 0 & -i\\i & 0\end{pmatrix},\qquad \sigma_z = \begin{pmatrix} 1 & 0\\0 & -1\end{pmatrix}.
\ee
Celles-ci forment une base de l'alg\`ebre $\su(2)$, et satisfont les relations de commutation fondamentales : $[\sigma_i,\sigma_j]=2i\eps_{ij}^{\phantom{ij}k}\sigma_k$. L'angle de classe de $g$ est $\psi\in[0,\pi]$, invariant par conjugaison sous le groupe, $g\mapsto hgh\mone$ pour tout $h\in\SU(2)$ (et d\'efinit donc les classes de conjugaison), et correspond \`a la moiti\'e de l'angle de rotation. Le vecteur $\hat{n}\in S^2$ est l'axe de la rotation, et se transforme comme un vecteur sous les conjugaisons. La m\'etrique $G$-invariante sur le groupe, obtenue via la forme de Killing-Cartan sur l'alg\`ebre, est :
\beq
ds^2_g = d\psi \otimes d\psi + \sin^2\psi\bigl( d\theta\otimes d\theta + \sin^2\theta\,d\phi\otimes d\phi\bigr),
\ee
si $(\theta, \phi)$ sont les coordonn\'ees polaires de $\hat{n}$. La mesure de Haar sur $\SU(2)$ (non-normalis\'ee) s'en d\'eduit :
\beq
dg = \sin^2\psi\,\sin\theta\ d\psi\,d\theta\,d\phi.
\ee
Il est possible de r\'e\'ecrire cette mesure en utilisant une autre param\'etrisation : $\vec{p} = \sin\psi\ \hat{n}$, qui repr\'esente la projection de $g$ sur les matrices de Pauli. Alors, $g = \pm\sqrt{1-\vec{p}^2}\unit + i\vec{p}\cdot\vec{\sigma}$, avec le signe $+$ pour la partie $\SO(3)$ correspondant \`a $\cos\psi\geq0$. La mesure prend la forme :
\be
dg = \f{d^3p}{\sqrt{1-\vec{p}^2}}.
\ee
C'est-\`a-dire que le facteur $(1-\vec{p}^2)^{\f{1}{2}}$ est la trace de la compacit\'e et de la courbure du groupe par rapport \`a la mesure de Lebesgue $d^3p$. En utilisant des fonctions test, on peut montrer que : $\delta(g h^{-1}) = \sqrt{1-\vec{p}_g^2}\, \delta^{(3)}(\vec{p}_g-\vec{p}_h)$. Si l'on utilise la transform\'ee de Fourier du delta sur $\R^3$, et en assimilant cet espace \`a l'alg\`ebre $\su(2)$, on obtient :
\be \label{int b}
\int d^3b\ e^{i\big[\tr(bg)-\tr(bh)\big]} = \f{1}{\sqrt{1-\vec{p}_g^2}}\Big( \delta_{\SU(2)}\big(g h^{-1}\big)+\delta_{\SU(2)}\big(-g h^{-1}\big)\Big) = \f{1}{\lvert\tr\ g\rvert}\ \delta_{\SO(3)}\big(g h^{-1}\big),
\ee
avec $b=-\f i2 \vec{b}\cdot\vec{\sigma} \in \su(2)$. Notons que cette formule ne se g\'en\'eralise pas \`a plus de 2 \'el\'ements de groupe, car une somme du type $\vec{p}_{h_1}+\vec{p}_{h_2}$ n'est g\'en\'eralement pas la projection d'un \'el\'ement de $\SU(2)$ sur les matrices de Pauli.

\medskip

Les repr\'esentations irr\'eductibles sont indic\'ees par des spins, demi-entiers, et agissent sur des espaces $\calH_j$ de dimension $d_j=(2j+1)$ pour $j\in\f{\N}{2}$. On note $(\vert j, m\rangle)_{m=-j,\dotsc,j}$ les vecteurs formant la base des nombres magn\'etiques, tels que : $\vec{\sigma}^2 \vert j, m\rangle = 4 j(j+1)\vert j, m\rangle$, et : $\sigma_z \vert j, m\rangle = 2m\vert j, m\rangle$. Soit $D^{(j)}_{mn}(g) = \langle j,m\lvert g\rvert j,n\rangle$ l'\'el\'ement de matrice de $g$ dans la repr\'esentation de spin $j$. Alors, l'espace des fonctions de carr\'e sommable sur $\SU(2)$ admet comme base hilbertienne l'ensemble de ces \'el\'ements matriciels,
\beq
L^2(\SU(2),dg) = \Bigl\{ f : g\mapsto \sum_{j,m,n} f^j_{mn}\ \sqrt{d_j}\,D^{(j)}_{mn}(g),\ \sum_{j,m,n}\lvert f^j_{mn}\rvert^2 <\infty\Bigr\},
\ee
et est muni du produit hermitien :
\beq
\int_{\SU(2)} dg\ f^*(g)\,h(g) = \sum_{j,m,n} f^{j*}_{mn}\,h^j_{mn}.
\ee
La relation d'orthogonalit\'e des \'el\'ements matriciels s'\'ecrit :
\beq \label{orthogonality}
\int dg\ D^{(j_1)}_{ab}(g)\ D^{(j_2)*}_{cd}(g) = \f{1}{d_{j_1}}\delta_{j_1j_2}\ \delta_{ac}\delta_{bd}.
\ee
Nous avons aussi une d\'ecomposition de Fourier de la distribution de Dirac sur le groupe, qui n'est autre que la relation de compl\'etude de la base choisie de $L^2(\SU(2),dg)$,
\beq
\delta(g) = \sum_{j\in\f{\N}{2}} d_j\ \chi_j(g),\qquad\text{d\'efinie par :}\qquad \int_{SU(2)}dg\ f(g)\,\delta(g\,h\mone) = f(h).
\ee
Ici, $\chi_j(g)=\sum_{m=-j}^j D^{(j)}_{mm}(g)$ d\'esigne le caract\`ere de la repr\'esentation de spin $j$, qui se calcule explicitement comme :
\beq
\chi_j(g) = \f{\sin d_j\psi}{\sin\psi}.
\ee
Ces caract\`eres ne sont autres que les polynomes de Chebyshev de second esp\`ece en la variable $\cos\psi$, soit $U_{2j}(\cos\psi)$.

La param\'etrisation d'Euler est aussi int\'eressante,
\beq
g = e^{-\f i2 \alpha\sigma_z}\ e^{-\f i2 \beta\sigma_y}\ e^{-\f i2 \gamma\sigma_z},
\ee
en isolant \`a gauche et \`a droite le sous-groupe $\U(1)$ g\'en\'er\'e par $\sigma_z$. La repr\'esentation de spin 1 est utile et s'\'ecrit alors facilement :
\beq
D^{(1)}_{mn}(g) = e^{-i\alpha m}\ d^1_{mn}(\beta)\ e^{-i\gamma n},\quad \text{avec}\quad d^1(\beta) = \begin{pmatrix} \f{1+\cos\beta}{2} & -\f{\sin\beta}{\sqrt{2}} & \f{1-\cos\beta}{2} \\ \f{\sin\beta}{\sqrt{2}} & \cos\beta & -\f{\sin\beta}{\sqrt{2}} \\ \f{1-\cos\beta}{2} & \f{\sin\beta}{\sqrt{2}} & \f{1+\cos\beta}{2} \end{pmatrix}.
\ee

La fonction de partition des mod\`eles de mousses de spins fait intervenir des moyennes de produits d'\'el\'ements matriciels sur le groupe. Donnons quelques formules explicites. En 2d, la formule pertinente est simplement la relation d'orthogonalit\'e \eqref{orthogonality}. Dans le mod\`ele de Ponzano-Regge, \ref{sec:pr model}, nous avons \`a int\'egrer le roduit de trois \'el\'ements matriciels, ce qui se fait d'abord via le produit tensoriel de deux repr\'esentations,
\beq \label{int g3}
D^{(j_1)}_{m_1 n_1}(g)\ D^{(j_2)}_{m_2 n_2}(g) = \sum_{J}\sum_{M,N =-J}^J d_J\ \begin{pmatrix} j_1 & j_2 & J\\ m_1 & m_2 & M\end{pmatrix}\,\begin{pmatrix} j_1 & j_2 & J\\ n_1 & n_2 & N\end{pmatrix}\ D^{(J)*}_{MN}(g),
\ee
puis en utilisant les relations d'orthogonalit\'e, on aboutit \`a :
\beq
\int_{\SU(2)}dg\ D^{(j_1)}_{m_1 n_1}(g)\ D^{(j_2)}_{m_2 n_2}(g)\ D^{(j_3)}_{m_3 n_3}(g) = \begin{pmatrix} j_1 & j_2 & j_3\\ m_1 & m_2 & m_3\end{pmatrix}\ \begin{pmatrix} j_1 & j_2 & j_3\\ n_1 & n_2 & n_3\end{pmatrix}.
\ee
Plus formellement, nous pouvons \'ecrire :
\beq
\int_{\SU(2)}dg\ D^{(j_1)}(g)\otimes D^{(j_2)}(g)\otimes D^{(j_3)}(g) = \vert \iota^{j_1 j_2 j_3}\rangle\,\langle \iota^{j_1 j_2 j_3}\vert,
\ee
o\`u $\vert \iota^{j_1 j_2 j_3}\rangle$ est le seul vecteur invariant du produit tensoriel $\calH_{j_1}\otimes \calH_{j_2} \otimes \calH_{j_3}$, \`a la normalisation pr\`es. Il s'agit donc d'un entrelaceur avec la repr\'esentation triviale : $\otimes_{a=1}^3 \calH_{j_a}\rightarrow \C$. Sa projection sur les \'etats de base de l'espace produit tensoriel d\'efinit le symbole 3mj de Wigner utilis\'e plus haut (et qui est r\'eel) :
\beq
\bigl(\langle j_1, m_1\vert \otimes \langle j_1, m_1\vert \otimes \langle j_1, m_1\vert\bigr)\vert \iota^{j_1 j_2 j_3}\rangle =  \begin{pmatrix} j_1 & j_2 & j_3\\ m_1 & m_2 & m_3\end{pmatrix}.
\ee
Ce symbole est invariant sous les permutations cycliques. Une permutation impaire produit une phase $(-1)^{j_1+j_2+j_3}$, c'est ce qui se passe lorsque l'on change l'orientation d'un vertex trivalent dans le graphe d'une fonctionnelle de r\'eseaux de spins, ou dans l'\'evaluation d'un symbole invariant obtenu par contractions des tels entrelaceurs. La dualisation de la repr\'esentation $j_1$ produit le symbole :
\beq
(-1)^{j_1-m_1}\,\begin{pmatrix} j_1 &j_2 &j_3 \\ -m_1 &m_2 &m_3\end{pmatrix}.
\ee
Ainsi, la dualisation des trois repr\'esentations ne changent pas la valeur du symbole, car
\beq
\begin{pmatrix} j_1 &j_2 &j_3 \\ -m_1 &-m_2 &-m_3\end{pmatrix} = (-1)^{j_1+j_2+j_3}\,\begin{pmatrix} j_1 &j_2 &j_3 \\ m_1 &m_2 &m_3\end{pmatrix},
\ee
et $m_1+m_2+m_3=0$ n\'ecesairement. Ces symboles permettent ensuite de construire des entrelaceurs g\'en\'eriques, et de calculer les moyennes de produits de plus d'\'el\'ements matriciels. Dans les mod\`eles de mousses de spins en 4d, \ref{sec:ooguri model}, par exemple, nous avons besoin d'entrelaceurs 4-valents. Pour cela, il faut choisir un appariement parmi les quatre repr\'esentations incidentes, par exemple $j_1$ avec $j_2$, et choisir un spin virtuel $i$ provenant \`a la fois de la d\'ecomposition de $\calH_{j_1}\otimes\calH_{j_2}$, et de $\calH_{j_3}\otimes \calH_{j_4}$ en repr\'esentations irr\'eductibles. On forme ainsi les quantit\'es :
\begin{align} \label{tree 4-node}
\bigl(\otimes_{a=1}^4\langle j_a, m_a\vert\bigr)\vert \iota^{j_1 j_2, i, j_3 j_4}\rangle &= 
\iota^{j_1 j_2,i,j_3 j_4}_{m_1 m_2 m_3 m_4},\\
&= \sum_{m=-i}^i \begin{pmatrix} j_1 & j_2 & i \\ m_1 & m_2 & m \end{pmatrix} (-1)^{i-m} \begin{pmatrix} i & j_3 & j_4 \\ -m & m_3 & m_4 \end{pmatrix},
\end{align}
que l'on \'ecrira simplement $\iota_{m_a}(j_a,i)$, bien que cela ne fasse pas appara\^itre explicitement le choix de l'appariement. En faisant varier le spin virtuel $i$, nous obtenons une base orthogonale des entrelaceurs entre $\otimes_{a=1}^4\calH_{j_a}$ et $\C$, avec la relation d'orthogonalit\'e suivante :
\beq
\sum_{\{m_a\}} \iota_{m_a}(j_a,i)\, \iota_{m_a}(j_a,i') = \f{\delta_{i,i'}}{d_i}.
\ee
De plus, nous pouvons exprimer l'int\'egrale suivante :
\beq
\int_{\SU(2)} dg\ \prod_{a=1}^4 D^{(j_a)}_{m_a n_a}(g) = \sum_i d_i\ \iota_{m_a}(j_a,i)\,\iota_{n_a}(j_a,i),
\ee
comme une somme sur les entrelaceurs, ce qui compl\`ete l'\'ecriture du mod\`ele topologique en mousses de spins \eqref{oog model}.

\begin{figure} \begin{center}
\includegraphics[width=5cm]{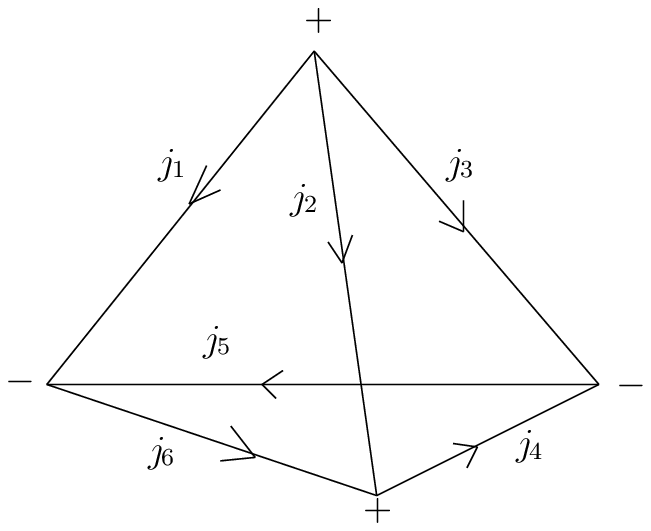}
\caption{ \label{fig:6jdef}}
\end{center}
\end{figure}

Ces symboles 3mj permettent par contractions de former toute une th\'eorie des invariants pour le couplage de spins, voir notamment \cite{varshalovich-book} dont nous utilisons les conventions graphiques. Le symbole 6j, qui repr\'esente l'amplitude d'un t\'etra\`edre dans le mod\`ele de Ponzano-Regge, est d\'efini comme :
\begin{multline}
\begin{Bmatrix}
 j_1 & j_2 & j_3 \\ j_4 & j_5 & j_6
\end{Bmatrix}
= \sum_{m_1,\dotsc,m_6} (-1)^{\sum_{i=1}^6(j_i-m_i)}\begin{pmatrix} j_1 & j_2 & j_3\\ m_1 & m_2 & m_3\end{pmatrix}\, \begin{pmatrix} j_3 & j_4 & j_5 \\ -m_3 & -m_4 & m_5 \end{pmatrix}\\
\begin{pmatrix} j_1 & j_5 & j_6 \\ -m_1 & -m_5 & m_6 \end{pmatrix}\,\begin{pmatrix}j_2 & j_6 & j_4 \\ -m_2 & -m_6 & m_4 \end{pmatrix}.
\end{multline}
Il se repr\'esente graphiquement sur le r\'eseau t\'etra\'edrique dual \`a la triangulation du bord d'un t\'etra\`edre, selon la figure \ref{fig:6jdef}. Il est aussi bien d\'efini par l'identit\'e de Biedenharn-Elliott, \eqref{bied-elliott}, dont un cas particulier est une relation de r\'ecurrence d'ordre 2 sur un seul spin, \eqref{rec6j}, et qui d\'etermine donc tous les coefficients et permet de les \'evaluer num\'eriquement, \cite{schulten-gordon1, schulten-gordon2}.

\bibliographystyle{amsplain}

\bibliography{/home/ubuntu/Papers/bibliographie/biblio}

\end{document}